\documentclass[a4paper,10pt]{report}

\usepackage{graphics}

\newcommand{\ket}[1]{\ensuremath{\left| #1 \right\rangle}}

\newcommand{\br}[1]{\ensuremath{\left\langle #1 \right.}}

\newcommand{\bra}[1]{\ensuremath{\left. \br{#1} \right|}}

\newcommand{\bk}[2]{\br{{#1}}\ket{{#2}}}

\newcommand{\kb}[2]{\ket{{#1}}\bra{{#2}}}

\newcommand{\ip}[3]{\bra{{#1}} {#2} \ket{{#3}}}

\newcommand{\proj}[1]{\kb{{#1}}{{#1}}}

\newcommand{\trace}[1]{\ensuremath{\mathrm{Tr}\left[{#1}\right]}}

\newcommand{\partrace}[2]{\ensuremath{\mathrm{Tr}_{{#2}}\left[{#1}\right]}}

\newcommand{\mean}[1]{\ensuremath{\left\langle
                                    {#1} \right\rangle}}

\newcommand{\magn}[1]{\ensuremath{\left| {#1} \right|^2}}

\newcommand{\com}[2]{\ensuremath{\left[{#1},{#2}\right]_-}}
\newcommand{\acom}[2]{\ensuremath{\left[{#1},{#2}\right]_+}}

\newcommand{\partime}[1]{\ensuremath{
    \frac{\partial {#1}}{\partial t}}}

\newcommand{\Sch}{Schr\"odinger }

\newcommand{\pict}[2]
{
\begin{figure}[htb]
    \resizebox{\textwidth}{!}{
        \includegraphics{#1}
        }
    \caption{#2\label{fg:#1}}
\end{figure}
}

\newcommand{\pictpict}[4]
{
\begin{figure}[htb]
    \resizebox{\textwidth}{!}{
        \includegraphics{#1}
        \includegraphics{#2}
        }
    \caption{#3\label{fg:#4}}
\end{figure}
}

\begin{document}
\title{Information and Entropy in Quantum Theory}
\author{O J E Maroney\\
\\
\\
Ph.D. Thesis \\
\\
\\
Birkbeck College \\
University of London\\
Malet Street \\
London \\
WC1E 7HX}
\date{}

\maketitle

\vspace{1in} \textbf{\LARGE{Abstract}} \vspace{0.5in}
 \label{ch:abstract}

Recent developments in quantum computing have revived interest in
the notion of information as a foundational principle in physics.
It has been suggested that information provides a means of
interpreting quantum theory and a means of understanding the role
of entropy in thermodynamics. The thesis presents a critical
examination of these ideas, and contrasts the use of Shannon
information with the concept of 'active information' introduced by
Bohm and Hiley.

We look at certain thought experiments based upon the 'delayed
choice' and 'quantum eraser' interference experiments, which
present a complementarity between information gathered from a
quantum measurement and interference effects. It has been argued
that these experiments show the Bohm interpretation of quantum
theory is untenable. We demonstrate that these experiments depend
critically upon the assumption that a quantum optics device can
operate as a measuring device, and show that, in the context of
these experiments, it cannot be consistently understood in this
way. By contrast, we then show how the notion of 'active
information' in the Bohm interpretation provides a coherent
explanation of the phenomena shown in these experiments.

We then examine the relationship between information and entropy.
The thought experiment connecting these two quantities is the
Szilard Engine version of Maxwell's Demon, and it has been
suggested that quantum measurement plays a key role in this. We
provide the first complete description of the operation of the
Szilard Engine as a quantum system. This enables us to demonstrate
that the role of quantum measurement suggested is incorrect, and
further, that the use of information theory to resolve Szilard's
paradox is both unnecessary and insufficient. Finally we show
that, if the concept of 'active information' is extended to cover
thermal density matrices, then many of the conceptual problems
raised by this paradox appear to be resolved.

\tableofcontents \listoffigures \listoftables

\chapter{Introduction}\label{ch:intro}

In recent years there has been a significant interest in the idea
of information as fundamental principle in physics\cite[amongst
others]{Whe83,Whe90,Zur90b,Per93,FS95,Fri98,Deu97,Zei99,Sto90,Sto92,Sto97}.
While much of this interest has been driven by the developments in
quantum computation\cite{Gru99,CN01} the issues that are addressed
are old ones. In particular, it has been suggested that:
\begin{enumerate}
\item Information theory must be introduced into physical theories
at the same fundamental level as concepts such as energy;
\item Information theory provides a resolution to the measurement
problem in quantum mechanics;
\item Thermodynamic entropy is equivalent to information, and that
information theory is essential to exorcising Maxwell's Demon.
\end{enumerate}
The concept of information used in these suggestions is
essentially that introduced by Shannon\cite{Sha48} and it's
generalisation to quantum theory by Schumacher\cite{Sch95}. This
concept was originally concerned with the use of different signals
to communicate messages, and the capacity of physical systems to
carry these signals, and is a largely static property of
statistical ensembles.

A completely different concept of information was introduced by
Bohm and Hiley\cite{BH93} in the context of Bohm's interpretation
of quantum theory\cite{Boh52a,Boh52b}. This concept was much more
dynamic, as it concerned the manner in which an individual system
evolves.

In this thesis we will be examining some of these relationships
between information, thermodynamic entropy, and quantum theory. We
will use information to refer to Shannon-Schumacher information,
and active information to refer to Bohm and Hiley's concept. We
will not be examining the ideas of Fisher
information\cite{Fis25,Fri88,Fri89,FS95,Fri98,Reg98}, although it
is interesting to note that the terms that result from applying
this to quantum theory bear a remarkable equivalence to the
quantum potential term in the Bohm approach. Similarly, we will
not be considering the recently introduced idea of {\em total}
information due to Bruckner and Zeilinger\cite{BZ99,BZ00a,BZ00b}.
We will also leave aside the concept of {\em algorithmic}
information\cite{Ben82,Zur89a,Zur89b,Zur90a,Cav93,Cav94}, as this
concept has only been defined within the context of classical
Universal Turing Machines. To be meaningful for quantum systems
this concept must be extended to classify quantum bit strings
operated upon by a Universal Quantum Computer, a task which
presents some considerable difficulties.

The structure of the thesis is as follows.

In Chapter \ref{ch:info} we will briefly review Shannon and
Schumacher information, and the problems for interpreting
information in a quantum measurement. Chapter \ref{ch:interf} will
introduce Bohm and Hiley's concept of active information, and will
examine recent thought experiments\cite{ESSW92} based upon the use
of 'one-bit detectors' which criticises this interpretation. We
will show that this criticism is unfounded.

Chapter \ref{ch:szmd} introduces the relationship between entropy
and information, by reviewing the discussion of Szilard's
Engine\cite{Szi29}. This thought experiment has been used to
suggest that an intelligent being (a Maxwell Demon) could reduce
the entropy of a system by performing measurements upon it. To
prevent a violation of the second law of thermodynamics it has
been argued that the information processing necessary for the
demon to perform it's function must lead to a compensating
dissipation.

Despite the extensive debate surrounding this thought experiment,
we will find that a number a key problems have not been addressed
properly. Of particular concern to us will be an argument by
Zurek\cite{Zur84} that the quantum measurement process plays a key
role in the operation of the Engine. If correct, this would appear
to imply that 'no collapse' theories of quantum mechanics (such as
Bohm's) would be unable to explain why the Engine cannot produce
anti-entropic behaviour. We will show this is not the case.

In Chapters \ref{ch:szqm} to \ref{ch:szsol} we will explicitly
construct a complete quantum mechanical description of the Szilard
Engine, and use it to examine the entropy-information link. We
will find that

\begin{enumerate}
\item The attempts to apply quantum theory to the experiment have
made a fundamental error, which we correct. Wavefunction collapse
then plays no role in the problem;
\item The Engine is not capable of violating the second law of thermodynamics;
\item Information theory is neither necessary nor sufficient to
completely resolve the problems raised by the Szilard Engine;
\end{enumerate}

In Chapters \ref{ch:szmd} and \ref{ch:szsol} we will encounter
Landauer's Principle\cite{Lan61}, which also attempts to directly
link information to entropy. We will examine this Principle in
more depth in Chapter \ref{ch:comp}. Properly interpreted, it is a
physical limitation upon the thermodynamics of computation. It
does not prove that information and entropy are equivalent,
however, as we will demonstrate that there are logically
reversible processes which are not thermodynamically reversible,
and further that there are thermodynamically reversible processes
which are not logically reversible. Although the information
functional and the entropy functional have the same form, their
physical interpretations have critical differences.

Finally in Chapter \ref{ch:active} we will re-examine the concept
of active information to see if it has any relevance to
thermodynamics. We will find that recent developments of the Bohm
interpretation\cite{BH00} suggest that the problems surrounding
the Szilard Engine may be viewed in a new light using the concept
of active information. The fundamental conflict in interpreting
thermodynamics is between the statistical ensemble description,
and the state of the individual system. We will show that, by
extending Bohm's interpretation to include the quantum mechanical
density matrix we can remove this conflict in a manner that is not
available to classical statistical mechanics and does not appear
to be available to other interpretations of quantum theory.

With regard to the three issues raised above, therefore, we will
have found that:
\begin{enumerate}
\item The introduction of information as a fundamental principle in
physics certainly provides a useful heuristic device. However, to
be fruitful a much wider concept of information than Shannon's
seems to be required, such as that provided by Bohm and Hiley;

\item The use of Shannon-Schumacher information in a physical theory
must presume the existence of a well defined measurement
procedure. Until a measurement can be certain to have taken place,
no information can be gained. Information theoretic attempts to
resolve the quantum measurement problem are therefore essentially
circular unless they use a notion of information that goes beyond
Shannon and Schumacher;

\item Although Shannon-Schumacher information and Gibbs-Von Neumann
entropy are formally similar they apply to distinctly different
concepts. As an information processing system must be implemented
upon a physical system, it is bound by physical laws and in an
appropriate limit they become related by Landauer's Principle.
Even in this limit, though, the different nature of the concepts
persists.
\end{enumerate}
\chapter{Information and Measurement}\label{ch:info}

In this Chapter we will briefly review the concept of Shannon
information\cite{Sha48,SW49} and it's application to quantum
theory.

Section 1 reviews the classical notion of information introduced
by Shannon and it's key features. Section 2 looks at the
application of Shannon information to the outcomes of quantum
measurements\cite{Kul59,Per93,Gru99,CN01}. We will be assuming
that a quantum measurement is a well defined process. The Shannon
measure may be generalised to Schumacher information, but the
interpretation of some of the quantities that are constructed from
such a generalisation remains unclear. Finally in Section 3 we
will consider an attempt by \cite{AC97b} to use the quantum
information measures to resolve the measurement problem, and show
that this fails.

\section{Shannon Information}

Shannon information was original defined to solve the problem of
the most efficient coding of a set of signals\cite{SW49,Sha48}. We
suppose that there is a source of signals (or sender) who will
transmit a given message $a$ with probability $P_a$. The message
will be represented by a bit string (an ordered series of 1's and
0's). The receiver will have a decoder that will convert the bit
string back into it's corresponding message. Shannon's theorem
shows that the mean length of the bit strings can be compressed to
a size

\begin{equation}
I_{Sh}=-\sum_a p_a \log_2 p_a \label{eq:shinf}
\end{equation}

without introducing the possibility of errors in the decoded
message\footnote{This assumes there is no noise during
transmission.}. This quantity $I_{Sh}$ is called the Shannon
information of the source. As it refers to the length in bits, per
message, into which the messages can be compressed, then a
communication channel that transmits $I_{Sh}$ bits per message has
a signal capacity of $I_{Sh}$.

This concept of information has no relationship to the meaning or
significance that the sender or the receiver attributes to the
message itself. The information content of a particular signal,
$-\log_2 p_a$, is simply an expression of how likely, or unlikely
the message is of being sent. The less likely the occurrence of a
message, the greater information it conveys. In the limit where a
message is certain to occur ($P_a=1$), then no information is
conveyed by it, as the receiver would have known in advance that
it was going to be received. An extremely rare message conveys a
great deal of information as it tells the receiver that a very
unlikely state of affairs exists. In many respects, the Shannon
information of the message can be regarded as measuring the
'surprise' the receiver feels on reading the message!

The most important properties of the Shannon information, however,
are expressed in terms of conditional $I(\alpha|\beta)$ and mutual
$I(\alpha:\beta)$ information, where two variables $\alpha$ and
$\beta$ are being considered. The probability of the particular
values of $\alpha=a$ and $\beta=b$ simultaneously occurring is
given by $P(a,b)$, and the joint information is therefore
\[
I(\alpha,\beta)=-\sum_{a,b}P(a,b) \log_2 P(a,b)
\]
From the joint probability distribution $P(a,b)$ we construct the
separate probability distributions
\begin{eqnarray*}
P(a) &=& \sum_b P(a,b) \\
P(b) &=& \sum_a P(a,b)
\end{eqnarray*}
the conditional probabilities
\begin{eqnarray*}
P(a|b)&=&\frac{P(a,b)}{P(b)} \\
P(b|a)&=&\frac{P(a,b)}{P(a)}
\end{eqnarray*}
and the correlation
\[
P(a:b)=\frac{P(a,b)}{P(a)P(b)}
\]
This leads to the information terms\footnote{These terms may
differ by the minus sign from the definitions given elsewhere. The
Shannon information as given represents the ignorance about the
exact state of the system.}
\begin{eqnarray*}
I(\alpha) &=& -\sum_{a,b}P(a,b) \log_2 P(a) \\
I(\beta) &=& -\sum_{a,b}P(a,b) \log_2 P(b) \\
I(\alpha|\beta) &=& -\sum_{a,b}P(a,b) \log_2 P(a|b) \\
I(\beta|\alpha) &=& -\sum_{a,b}P(a,b) \log_2 P(b|a) \\
I(\alpha:\beta) &=& -\sum_{a,b}P(a,b) \log_2 P(a:b)
\end{eqnarray*}
which are related by
\begin{eqnarray*}
I(\alpha|\beta) &=& I(\alpha,\beta)-I(\beta) \\
I(\beta|\alpha) &=& I(\alpha,\beta)-I(\alpha) \\
I(\alpha:\beta) &=& I(\alpha,\beta)-I(\alpha)-I(\beta)
\end{eqnarray*}
and obey the inequalities
\begin{eqnarray*}
I(\alpha,\beta) \ge & I(\alpha) & \ge 0 \\
I(\alpha,\beta) \ge & I(\alpha|\beta) & \ge 0 \\
\min\left[I(\alpha),I(\beta)\right] \ge & -I(\alpha:\beta) & \ge 0
\end{eqnarray*}

We can interpret these relationships, and the $\alpha$ and $\beta$
variables, as representing communication between two people, or as
the knowledge a single person has of the state of a physical
system.

\subsection{Communication} If $\beta$ represents the signal states
that the sender transmits, and $\alpha$ represents the outcomes of
the receivers attempt to decode the message, then $P(a|b)$
represents the reliability of the transmission and
decoding\footnote{There are many ways in which the decoding may be
unreliable. The communication channel may be noisy, the decoding
mechanism may not be optimally designed, and the signal states may
be overlapping in phase space}.

The receiver initially estimates the probability of a particular
signal being transmitted as $P(b)$, and so has information
$I(\beta)$. After decoding, the receiver has found the state $a$.
Presumably knowing the reliability of the communication channel,
she may now use Bayes's rule to re-estimate the probability of the
transmitted signals
\[
P(b|a)=\frac{P(a|b)P(b)}{P(a)}
\]
On receiving the result $a$, therefore, the receiver has
information
\[
I(\beta|a)=\sum_b P(b|a) \log_2 P(b|a)
\]
about the signal sent. Her information gain, is
\begin{equation}
\Delta I_a(\beta)=I(\beta|a)-I(\beta) \label{eq:igain}
\end{equation}
Over an ensemble of such signals, the result $a$ will occur with
probability $P(a)$. The mean information possessed by the receiver
is then
\[
\mean{I(\beta|a)}=\sum_a P(a) I(\beta|a) = I(\beta|\alpha)
\]
So the conditional information $I(\beta|\alpha)$ represents the
average information the receiver possesses about the signal state,
given her knowledge of the received state, while the term
$I(\beta|a)$ represents the information the receiver possesses
given a specific outcome $a$. The mean information gain
\[
\mean{\Delta I(\beta|a)}=
    \sum_a P(a) \Delta I_a(\beta)=I(\alpha:\beta)
\]
The mutual information is the gain in information the receiver has
about the signal sent. It can be shown that, given that the sender
is also aware of the reliability of the transmission and decoding
process, that the conditional information $I(\alpha|\beta)$
represents the knowledge the sender has about the signal the
receiver actually receives. The mutual information can then be
regarded as the symmetric function expressing the information both
receiver and sender possess in common, or equivalently, the {\em
correlation} between the state of the sender and the state of the
receiver.

If the transmission and decoding processes are completely
reliable, then the particular receiver states of $\alpha$ will be
in a one-to-one correspondence with the signal states of $\beta$,
with probabilities $P(a|b)=1$. This leads to
\begin{eqnarray*}
I(\alpha) &=& I(\beta) \\
I(\beta|\alpha)=I(\alpha|\beta) &=& 0 \\
I(\alpha:\beta) &=& -I(\alpha)
\end{eqnarray*}
It should be remembered that the information measure of complete
certainty is zero, and it increases as the uncertainty, or
ignorance of the state, increases. In the case of a reliable
transmission and decoding, the receiver will end with perfect
knowledge of the signal state, and the sender and receiver will be
maximally correlated.

\subsection{Measurements}

The relationships above have been derived in the context of the
information capacity of a communication channel. However, it can
also be applied to the process of detecting and estimating a state
of a system. The variable $\beta$ will represent the {\em a
priori} probabilities that the system is in a particular state.
The observer performs a measurement upon the system, obtaining the
result in variable $\alpha$.

The initial states do not have to represent an exact state of the
system. If we start by considering a classical system with a
single coordinate $x$ and it's conjugate momentum $p_x$, the
different states of $\beta$ represent a partitioning of the phase
space of the system into separate regions $b$, and the
probabilities $P(b)$ that the system is located within a
particular partition. The measurement corresponds to dividing the
phase space into a partitioning, represented by the different
states of $\alpha$ and locating in which of the measurement
partitions the system is located.

We now find that the conditional information represents the
improved knowledge the observer has of the initial state of the
system (given the outcome of the measurement) and the mutual
information, as before, represents the average gain in information
about the initial state.

Note that if the measurement is not well chosen, it may convey no
information about the original partitioning. Suppose the
partitioning of $\beta$ represents separating the phase space into
the regions $p_x>0$ and $p_x<0$, with equal probability of being
found in either ($P(p_x>0)=P(p_x<0)=\frac12$ and a uniform
distribution within each region. Now we perform a measurement upon
the {\em position} of the particle, separating the phase space
into the regions $x>0$ and $x<0$. The probabilities are
\begin{eqnarray*}
P(p_x>0|x>0)=&\frac{P(x>0|p_x>0)P(p_x>0)}{P(x>0)}&=\frac12 \\
P(p_x<0|x>0)=&\frac{P(x>0|p_x<0)P(p_x<0)}{P(x>0)}&=\frac12 \\
P(p_x>0|x<0)=&\frac{P(x<0|p_x>0)P(p_x>0)}{P(x<0)}&=\frac12 \\
P(p_x<0|x<0)=&\frac{P(x<0|p_x<0)P(p_x<0)}{P(x<0)}&=\frac12
\end{eqnarray*}

A measurement based upon the partition $x>0$ and $x<0$ would
produce no gain in information. However, it is always possible to
a define a finer grained initial partitioning (such as dividing
the phase space into the four quadrants of the $x,p_x$ axes) for
which the measurement increases the information available, and in
this case would provide complete information about the location of
the  original partition.

If the measurement partition of $\alpha$ coincides with the
partition of $\beta$ then the maximum information about $\beta$
will be gained from the measurement. In the limit, the partition
becomes the finely grained partition where each point $(p_x,x)$ in
the phase space is represented with the probability density
function $\Pi(p_x,x)$.

In classical mechanics the observer can, in principle, perfectly
distinguish all the different states, and make the maximum
information gain from a measurement. However, in practice, some
finite partitioning of the phase space is used, owing to the
physical limitations of measuring devices.

\section{Quantum Information}\label{s:info2}

When attempting to transfer the concept of information to quantum
systems, the situation becomes significantly more complex. We will
now review the principal ways in which the measure and meaning of
information is modified in quantum theory.

The first subsection will be concerned with the generalisation of
Shannon's theorem, on communication capacities. This produces the
Schumacher quantum information measure. Subsection 2 will consider
the Shannon information gain from making measurements upon a
quantum system. Subsection 3 reviews the quantities that have
proposed as the generalisation of the relative and conditional
information measures, in the way that Schumacher information
generalises the Shannon information. These quantities have
properties which make it difficult to interpret their meaning.

\subsection{Quantum Communication Capacity}

The primary definition of information came from Shannon's Theorem,
on the minimum size of the communication channel, in mean bits per
signal, necessary to faithfully transmit a signal in the absence
of noise. The theorem was generalised to quantum theory by
Schumacher\cite{Sch95,JS94}.

Suppose that the sender wishes to use the quantum states $\psi_a$
to represent messages, and a given message will occur with
probability $p_a$. We will refer to $I[\rho]$ as the Shannon
information of the source. The quantum coding theorem demonstrates
that the minimum size of Hilbert space $H$ that can be used as a
communication channel without introducing errors is
\[
\mathrm{Dim}(H)=2^{S[\rho]}
\]
where
\begin{eqnarray}
\rho_a  &=& \proj{\psi_a} \nonumber \\
\rho    &=& \sum_a p_a \rho_a \nonumber \\
S[\rho] &=& -\trace{\rho \log_2 \rho} \label{eq:schinf}
\end{eqnarray}
By analogy to the representation of messages in bits, a Hilbert
space of dimension 2 is defined as having a capacity of 1 qbit,
and a Hilbert space of dimension $n$, a capacity of $\log_2 n$
qbits.

If the signal states are all mutually orthogonal
\[
\rho_a \rho_{a^\prime}=\delta_{aa^\prime} \rho_a^2
\]
then
\[
S[\rho]=-\sum_a p_a \log_2 p_a
\]
If this is the case, then the receiver can, in principle, perform
a quantum measurement to determine exactly which of the signal
states was used. This will provide an information gain of exactly
the Shannon information of the source.

However, what if the signal states are not orthogonal? If this is
the case, then\cite{Weh78}
\[
S[\rho]  <  I[\rho]
\]
It would appear that the signals can be sent, without error, down
a smaller dimension of Hilbert space. Unfortunately, as the signal
states are not orthogonal, they cannot be unambiguously
determined. We must now see how much information can be extracted
from this.

\subsection{Information Gain}
To gain information, the receiver must perform a measurement upon
the system. The most general form of a measurement used in quantum
information is the Positive Operator Valued Measure
(POVM)\cite{BGL95}. This differs from the more familiar von
Neumann measurement, which involves the set of projection
operators $\proj{a}$ for which
$\bk{a}{a^\prime}=\delta_{aa^\prime}$ and
\[
\sum_a \proj{a}=I
\]
is the identity operator. The probability of obtaining outcome
$a$, from an initial state $\rho$ is given by
\[
p_a=\trace{\rho \proj{a}}
\]
This is not the most general way of obtaining a probability
measure from the density matrix. To produce a set of outcomes $a$,
with probabilities $p_a$ according to the formula
\[
p_a=\trace{\rho A_a}
\]
the conditions upon the set of operators $A_a$ are that they be
{\em positive}, so that
\[
\ip{w}{A_a}{w} \ge 0
\]
for all states $\ket{w}$, and that the set of operators sums to
the identity
\[
\sum_a A_a=I
\]
For example, consider a spin-$\frac12$ system, with spin-up and
spin-down states $\ket{0}$,$\ket{1}$ respectively and the
superpositions
$\ket{u}=\frac1{\sqrt{2}}\left(\ket{0}+\ket{1}\right)$
$\ket{v}=\frac1{\sqrt{2}}\left(\ket{0}-\ket{1}\right)$ then the
following operators
\begin{eqnarray*}
A_1 &=& \frac12 \proj{0} \\
A_2 &=& \frac12 \proj{1} \\
A_3 &=& \frac12 \proj{u} \\
A_4 &=& \frac12 \proj{v}
\end{eqnarray*}
form a POVM. A given POVM can be implemented in many different
ways\footnote{The example given here could be implemented by, on
each run of the experiment, a random choice of whether to measure
the 0-1 basis or u-v basis. This will require a correlation to a
second system which generates the random choice. In general a POVM
will be implemented by a von Neumann measurement on an extended
Hilbert space of the system and an auxiliary\cite{Per90,Per93}.},
but will typically require an auxiliary system whose state will be
changed by the measurement.

The signal states $\rho_b$ occur with probability $p_b$. Using the
same expression for information gain as in Equation \ref{eq:igain}
so we can now apply Bayes's rule as before, with
\[
p(a|b)=\trace{A_a \rho_b}
\]
to give the probability, on finding outcome $a$, that the original
signal state was $b$
\begin{equation}
p(b|a)=\frac{p(b)\trace{A_a \rho_b}}{p(a)} \label{eq:qbayes}
\end{equation}
We now define the relative information, information gain and
mutual information as before
\begin{eqnarray*}
I(\beta|a)&=&\sum_b P(b|a) \log_2 P(b|a) \\
\Delta I_a(\beta)&=&I(\beta|a)-I(\beta) \\
\mean{I(\beta|a)}&=&\sum_a P(a) I(\beta|a) = I(\beta|\alpha) \\
\mean{\Delta I(\beta|a)}&=&
    \sum_a P(a) \Delta I_a(\beta)=I(\alpha:\beta)
\end{eqnarray*}
It can be shown that the maximum gain in Shannon information,
known as the Kholevo bound, for the receiver is the Schumacher
information\cite{Kho73,HJSWW96,SW97,Kho98}.
\[
I[\alpha:\beta] \le S[\rho]
\]
So, although by using non-orthogonal states the messages can be
compressed into a smaller volume, the information that can be
retrieved by the receiver is reduced by exactly the same amount.

\subsection{Quantum Information Quantities}
The information quantity that results from a measurement is still
defined in terms of Shannon information on the measurement
outcomes. This depends upon the particular measurement that is
performed. We would like to generalise the joint, conditional, and
mutual information to quantum systems, and to preserve the
relationships:
\begin{eqnarray*}
S[A|B]&=&S[AB]-S[B] \\
S[B|A]&=&S[AB]-S[A] \\
S[A:B]&=&S[AB]-S[A]-S[B]
\end{eqnarray*}
This generalisation\cite[and references
therein]{AC95,Gru99,SW00,CN01} is defined from the joint density
matrix of two quantum systems $\rho_{AB}$.
\begin{eqnarray}
\rho_A &=& \partrace{\rho_{AB}}{B} \nonumber\\
\rho_B &=& \partrace{\rho_{AB}}{A} \nonumber\\
S[AB] &=& -\trace{\rho_{AB} \log_2 \rho_{AB}} \nonumber\\
S[A] &=& -\trace{\rho_{AB} \log_2 (\rho_A \otimes 1_B) } \nonumber\\
         &=& -\trace{\rho_A log_2 \rho_A} \nonumber\\
S[B] &=& -\trace{\rho_{AB} \log_2 (1_A \otimes \rho_B)} \nonumber\\
        &=& -\trace{\rho_B log_2 \rho_B} \nonumber\\
S[A|B]&=& -\trace{\rho_{AB} \log_2 \rho_{A|B}} \nonumber\\
S[B|A]&=& -\trace{\rho_{AB} \log_2 \rho_{B|A}} \nonumber\\
S[A:B]&=& -\trace{\rho_{AB} \log_2 \rho_{A:B}} \label{eq:qinf}
\end{eqnarray}
where the matrices\footnote{Where all the density matrices
commute, then
\begin{eqnarray*}
\rho_{A|B}&=&\rho_{AB} \left(\rho_A \otimes 1_B \right)^{-1} \\
\rho_{A:B}&=&\rho_{AB} \left(\rho_A \otimes \rho_B\right)^{-1}
\end{eqnarray*}
in close analogy to the classical probability functions}
\begin{eqnarray*}
\rho_{A|B} &=&\lim_{n\rightarrow \infty}
    \left[\rho_{AB}^{1/n}\left(1_A \otimes \rho_B \right)^{-1/n}\right]^n \\
\rho_{B|A} &=&\lim_{n\rightarrow \infty}
    \left[\rho_{AB}^{1/n}\left(\rho_A \otimes 1_B \right)^{-1/n}\right]^n \\
\rho_{A:B} &=&\lim_{n\rightarrow \infty}
    \left[\rho_{AB}^{1/n}\left(\rho_A \otimes \rho_B \right)^{-1/n}\right]^n
\end{eqnarray*}

However, these quantities display significantly different
properties from Shannon information. The most significant result
is that it is possible for $S[A]>S[AB]$ or $S[B]>S[AB]$. This
allows $S[A|B],S[B|A]<0$ and $-S[A:B]>S[AB]$ which cannot happen
for classical correlations, and does not happen for the Shannon
information quantities that come from a quantum measurement. A
negative conditional information $S[A|B]<0$, for example, would
appear to imply that, given perfect knowledge of the state of $B$,
one has 'greater than perfect' knowledge of the state of $A$!

The clearest example of this is for the entangled state of two
spin-$\frac12$ particles, with up and down states represented by
$0$ and $1$:
\[
\psi=\frac12 \left(\ket{00}+\ket{11}\right)
\]
This is a pure state, which has
\[
S[AB]=0
\]
The subsystem density matrices are
\begin{eqnarray*}
\rho_A &=& \frac12 \left(\proj{0}+\proj{1}\right) \\
\rho_B &=& \frac12 \left(\proj{0}+\proj{1}\right)
\end{eqnarray*}
so that
\[
S[A]=S[B]=1
\]
The conditional quantum information is then
\[
S[A|B]=S[B|A]=-1
\]

The significance that can be attributed to such a negative
conditional information is a matter of some
debate\cite{AC95,AC97b,SW00}. We have noted above that the Shannon
information of a measurement on a quantum system does not show
such a property. However, the Kholevo bound would appear to tell
us that each of the quantities $S[A]$, $S[B]$ and $S[AB]$ can be
the Shannon information gained from a suitable measurement of the
system.

The partial resolution of this problem lies in the fact that, for
quantum systems, there exist joint measurements which cannot be
decomposed into separate measurements upon individual systems.
These joint measurements may yield more information than can be
obtained for separable measurements even in the absence of
entanglement\cite{GP99,Mas00,BDFMRSSW99,Mar01}. In terms of {\em
measurements} the quantities of $S[AB]$, $S[A]$ and $S[B]$ may
refer to information gains from mutually incompatible experimental
arrangements. There is correspondingly no {\em single}
experimental arrangement for which the resulting {\em Shannon}
information will produce a negative conditional information.

\subsection{Measurement} \label{s:info2.4}

We have so far reviewed the existence of the various quantities
that are associated with information in a quantum system. However,
we have not really considered what we mean by the information
gained from a quantum measurement.

In a classical system, the most general consideration is to assume
a space of states (whether discrete digital messages or a
continuous distribution over a phase space) and probability
distribution over those states.

There are two questions that may be asked of such a system:
\begin{enumerate}
\item What is the probability distribution?
\item What is the state of a given system?
\end{enumerate}

If we wish to determine the probability distribution, the means of
doing this is to measure the state of a large number of
equivalently prepared systems, and as the number of experiments
increases the relative frequencies of the states approaches the
probability distribution. So the measurement procedure to
determine the state of the given system is the same as that used
to determine the probability distribution.

For a quantum system, we must assume a Hilbert space of states,
and a probability distribution over those states. Ideally we would
like to ask the same two questions:
\begin{enumerate}
\item What is the probability distribution?
\item What is the state of a given system?
\end{enumerate}
However, we find we a problem. The complete statistical properties
of the system are given by the density matrix
\[
\rho=\sum_a p_a \rho_a
\]
where the state $\rho_a$ occurs with probability $p_a$. We can
determine the value of this density matrix by an {\em
informationally complete} measurement\footnote{An informationally
complete measurement is one whose statistical outcomes uniquely
defines the density matrix. Such a measurement can only be
performed using a POVM\cite[Chapter V]{BGL95}. A single
experiment, naturally, cannot reveal the state of the density
matrix. It is only in the limit of an infinite number of
experiments the relative frequencies of the outcomes uniquely
identifies the density matrix.}. However, this measurement does
{\em not} necessarily tell us the states $\rho_a$ or $p_a$. The
reason for this is that the quantum density matrix does not have a
unique decomposition. A given density matrix $\rho$ could have
been constructed in an infinite number of ways. For example, the
following ensembles defined upon a spin-$\frac{1}{2}$ system
\paragraph{Ensemble 1}
\begin{eqnarray*}
\rho_1 &=& \proj{0} \\
\rho_2 &=& \proj{1} \\
p_1 &=& \frac12 \\
p_2 &=& \frac12
\end{eqnarray*}
\paragraph{Ensemble 2}
\begin{eqnarray*}
\rho_A &=& \proj{u} \\
\rho_B &=& \proj{v} \\
p_A &=& \frac12 \\
p_B &=& \frac12
\end{eqnarray*}
\paragraph{Ensemble 3}
\begin{eqnarray*}
\rho_1 &=& \proj{0} \\
\rho_2 &=& \proj{1} \\
\rho_A &=& \proj{u} \\
\rho_B &=& \proj{v} \\
p_1 &=& \frac14 \\
p_2 &=& \frac14 \\
p_A &=& \frac14 \\
p_B &=& \frac14
\end{eqnarray*}
with $\ket{u}=\frac1{\sqrt{2}}\left(\ket{0}+\ket{1}\right)$
$\ket{v}=\frac1{\sqrt{2}}\left(\ket{0}-\ket{1}\right)$, all
produce the density matrix $\rho=\frac12 I$, where $I$ is the
identity.

The informationally complete measurement will reveal the value of
an unknown density matrix, but will not even reveal the
probability distribution of the states that compose the density
matrix, unless the different $\rho_a$ states happen to be
orthogonal, and so form the basis which diagonalises the density
matrix (and even in this case, an observer who is ignorant of the
fact that the signal states have this property will not be able to
discover it).

To answer the second question it is necessary to have some {\em a
priori} knowledge of the 'signal states' $\rho_a$. In the absence
of {\em a priori} knowledge, the quantum information gain from a
measurement has no objective significance. Consider a measurement
in the basis $\proj{0}$, $\proj{1}$. With Ensemble 1, the
measurement reveals the actual state of the system. With Ensemble
2, the measurement causes a wavefunction collapse, the outcome of
which tells us nothing of original state of the system, and
destroys all record of it. Without the knowledge of which ensemble
we were performing the measurement upon we are unable to know how
to interpret the outcome of the measurement.

This differs from the classical measurement situation. In a
classical measurement we can refine our partitioning of phase
space, until in the limit we obtain the probability density over
the whole of the phase space. If the classical observer starts
assuming an incorrect probability distribution for the states, he
can discover the fact. By refining his measurement and repeatedly
applying Bayes's rule, the initially subjective assessment of the
probability density asymptotically approaches the actual
probability density. The initially subjective character of the
information eventually becomes an objective property of the
ensemble.

In a quantum system, there is no measurement able to distinguish
between different distributions that combine to form the same
density matrix. The observer will never be able to determine which
of the ensembles was the actual one. If he has assumed the correct
signal states $\rho_a$, then he may discover if his probabilities
are incorrect. However, if his initial assumption about the signal
states going into the density matrix are incorrect, he may never
discover this.

It might be argued that the complete absence of {\em a priori}
knowledge is equivalent to an isotropic distribution over the
Bloch sphere\footnote{The Bloch sphere represents a pure state in
a Hilbert space of dimension 2 by a point on a unit sphere.}. An
observer using such a distribution could certainly devise a
optimal measurement, in terms of information gain\cite{Dav78}.
Although some information might be gained, the {\em a posteriori}
probabilities, calculated from Bayes's rule, would be
distributions over the Bloch sphere, conditional upon the outcome
of the experiments. However, the outcomes of such a measurement
would be same for each of the three ensembles above. The {\em a
posteriori} probabilities continue to represent an assessment of
the observer's knowledge, rather than a property of the ensemble
of the systems.

On the other hand, we are not at liberty to argue that only the
density matrix is of significance. If we {\em are} in possession
of {\em a priori} knowledge of the states composing the density
matrix, we will construct very different measurements to optimise
our information gain, depending upon that knowledge. The optimal
measurement for Ensemble 2 is of the projectors $\proj{u}$ and
$\proj{v}$, while for Ensemble 3 a POVM must be used involving all
four projectors. All of these differ from the optimal measurement
for an isotropic distribution\footnote{Recent
work\cite{BZ99,BZ00a,Hal00,BZ00b} by Bruckner and Zeilinger
criticises the use of Shannon-Schumacher information measures in
quantum theory, on similar grounds. While their suggested
replacement of {\em total information} has some interesting
properties, it appears to be concerned exclusively with the
density matrix itself, rather than the states that are combined to
construct the density matrix.}.

\section{Quantum Measurement}
So far we have made a critical assumption in analysing the
information gained from measurements, namely that measurements
have well defined outcomes, and that we have a clear understanding
of when and how a measurement has occurred. This is, of course, a
deeply controversial aspect of the interpretation of quantum
theory. Information theory has, occasionally, been applied to the
problem\cite[Chapter III, for example]{DG73}, but usually this is
only in the context of a predefined theory of measurement (thus,
in \cite{DG73} the use of information theory is justified within
the context of the Many-World Interpretation).

In \cite{AC97b}, Cerf and Adami argue that the properties of the
quantum information relationships in Equation \ref{eq:qinf} can,
in themselves, be used to resolve the measurement problem. We will
now examine the problems in their argument.

Let us start by considering a measurement of a quantum system in a
statistical mixture of orthogonal states $\proj{\psi_n}$ with
statistical weights $w_n$, so that
\[
\rho=\sum_n w_n \proj{\psi_n}
\]
In this case, the density matrix is actually constructed from the
$\ket{\psi_n}$ states, rather than some other mixture leading to
the same statistical state. We now introduce a measuring device,
initially in the state $\ket{\phi_0}$ and an interaction between
system and device
\begin{equation}
\ket{\psi_n \phi_0} \rightarrow \ket{\psi_n \phi_n}
\label{eq:meas1}
\end{equation}
This interaction leads the joint density matrix to evolve from
\[
\rho_n \otimes \proj{\phi_0}
\]
to
\begin{equation}
\rho^\prime=\sum_n w_n \proj{\psi_n \phi_n} \label{eq:meas2}
\end{equation}
We can now consistently interpret the density matrix $\rho^\prime$
as a statistical mixture of the states $\ket{\psi_n \phi_n}$
occurring with probability $w_n$. In particular, when the
measuring device is in the particular state $\ket{\psi_n}$ then
the observed system is in the state $\ket{\phi_n}$. The
interaction in \ref{eq:meas1} above is the correct one to measure
the quantity defined by the $\ket{\psi_n}$ states.

Unfortunately, the linearity of quantum evolution now leads us to
the measurement problem when the initial state of the system is
not initial in a mixture of eigenstates of the observable.
Supposing the initial state is
\[
\ket{\Psi}=\sum_n \alpha_n \ket{\psi_n}
\]
(where, for later convenience, we choose $\magn{\alpha_n}=w_n$),
then the measurement interaction leads to a state
\begin{equation}
\ket{\Psi \Phi}=\sum_n \alpha_n \ket{\psi_n \phi_n}
\label{eq:meas3}
\end{equation}
This is a pure state, not a statistical mixture. Such an entangled
superposition of states cannot be interpreted as being in a
mixture of states, as there are observable consequences of
interference between the states in the superposition.

To complete the measurement it is necessary that some form of
non-unitary projection takes place, where the state \ket{\Psi
\Phi} is replaced by a statistical mixture of the \ket{\psi_n
\phi_n} states, each occurring randomly with probability
$\magn{\alpha_n}=w_n$.

\paragraph{Information}
From the point of view of information theory, the density
matrix in Equation \ref{eq:meas2} has a information content of
\begin{eqnarray*}
S_1[\phi]=S_1[\psi]=S_1[\phi,\psi] &=& -\sum_n w_n \log_2 w_n =S_0\\
S_1[\phi|\psi] =S_1[\psi|\phi] &=& 0 \\
S_1[\phi:\psi] &=& -S_0
\end{eqnarray*}
The conditional information being zero indicates that, given the
knowledge of the state of the measuring apparatus we have perfect
knowledge of the state of the measured system, and the mutual
information indicates a maximum level of correlation between the
two systems.

For the superposition in Equation \ref{eq:meas3}, the information
content is
\begin{eqnarray*}
S_2[\phi,\psi] &=& 0\\
S_2[\phi]=S_2[\psi]&=& S_0 \\
S_2[\phi|\psi] =S_2[\psi|\phi] &=& -S_0 \\
S_2[\phi:\psi] &=& -2S_0
\end{eqnarray*}
We now have situation where the knowledge of the state of the
combined system is perfect, while, apparently, the knowledge of
the individual systems is completely unknown. This leads to a
negative conditional information - which has no classical meaning,
and a correlation that is twice the maximum that can be achieved
with classical systems.

\cite{AC95} do not attempt to interpret these terms. Instead they
now introduce a third system, that 'observes' the measuring
device. If we represent this by $\ket{\xi}$, this leads to the
state
\begin{equation}
\ket{\Psi \Phi \Xi}=\sum_n \alpha_n \ket{\psi_n \phi_n \xi_n}
\label{eq:meas4}
\end{equation}
Now, it would appear we have simply added to the problem as our
third system is part of the superposition. However, by
generalising the quantum information terms to three systems,
\cite{AC95} derive the quantities
\begin{eqnarray*}
S_3[\xi]=S_3[\phi]=S_3[\xi,\phi] &=& -\sum_n w_n \log_2 w_n =S_0\\
S_3[\xi|\phi] =S_3[\phi|\xi] &=& 0 \\
S_3[\xi:\phi] &=& -S_0
\end{eqnarray*}
This shows the same relationships between the second 'observer'
and the measuring device as we saw initially between the measuring
device and the observed system {\em when the system was in a
statistical state}. This essentially leads \cite{AC95} to believe
they can interpret the situation described after the second
interaction as a classical correlation between the observer and
the measuring device.

\cite{AC95} do not claim that they have introduced a non-unitary
wavefunction collapse, nor do they believe they are using a
'Many-Worlds' interpretation. What has happened is that, by
considering only two, out of three, subsystems in the
superposition, they have traced over the third system (the
original, 'observed' system), and produced a density matrix
\begin{equation}
\partrace{\proj{\Psi \Phi \Xi}}{\psi}=
    \sum_n w_n \proj{\phi_n \xi_n} \label{eq:meas5}
\end{equation}
which has the same form as the classically correlated density
matrix. They argue that the original, fundamentally quantum
systems $\ket{\Psi}$ are always unobservable, and it is only the
correlations between ourselves (systems $\ket{\Xi}$) and our
measuring devices (systems $\ket{\Phi}$) that are accessible to
us.

They argue that there is no need for a wavefunction collapse to
occur to introduce a probabilistic uncertainty into the unitary
evolution of the \Sch equation. It is the occurrence of the
negative conditional information
\[
S_3[\psi|\phi,\xi]=-S_0
\]
that introduces the randomness to quantum measurements. This
negative conditional information allows the $\Phi,\Xi$ system to
have an uncertainty (non-zero information), even while the overall
state has no uncertainty
\[
S_3[\psi,\phi,\xi]=S_3[\psi|\phi,\xi]+S_3[\phi,\xi]=0
\]

The basic problem with this argument is the assumption that when
we have an apparently classically correlated density matrix, such
as in Equation \ref{eq:meas2} above, we can automatically
interpret it as actually {\em being} a classical correlation. In
fact, we can only do this if we know that it is actually
constructed from a statistical ensemble of correlated states. As
we have seen above, the quantum density matrix does not have a
unique decomposition and so could have been constructed out of
many different ensembles. These ensembles may be constructed with
superpositions, entangled states, or even, as with the density
matrix in Equation \ref{eq:meas5}, without involving ensembles at
all.

What \cite{AC95} have shown is the practical difficulty of finding
any observable consequences of the entangled superposition, as the
results of a measurement upon the density matrix in Equation
\ref{eq:meas5} are identical to those that would occur from
measurements upon a statistical mixture of classically correlated
states. However, to even make this statement, we have to have
assumed that we know when a measurement has occurred in a quantum
system, and this is precisely the point at issue\footnote{Their
argument is essentially a minimum version of the decoherence
approach to the measurement problem\cite{Zur91}. For a
particularly sharp criticism of why this approach does not even
begin to address the problem, see \cite[Chapter 4, footnote
16]{Alb92}}.

When applying this to \Sch's cat, treating $\Phi$ as the cat and
$\Xi$ as the human observer, they say
\begin{quotation}
The observer notices that the cat is either dead or alive and thus
the observer's own state becomes classically correlated with that
of the cat, although in reality, the entire system (including atom
\ldots the cat and the observer) is in a {\em pure entangled
state}. It is {\em practically} impossible, although not in
principle, to undo this observation i.e. to resuscitate the cat
\end{quotation}
Unfortunately this does not work. The statement that the observer
notices that the cat is either alive or dead must {\em presume}
that it is actually the case that the cat is either alive or dead.
That is, in each experimental realisation of the situation there
is a matter of fact about whether the cat is alive or dead.
However, if this was the case, that the cat is, in fact, either
alive or dead, then the system would not described by the
superposition at all. It is {\em because} a superposition cannot
readily be interpreted as a mixture of states that the measurement
problem arises in the first place.

\cite{AC97b}'s resolution depends upon their being able make the
assumption that a superposition does, in fact, represent a
statistical mixture of the cat being in alive and dead states,
with it being a matter of fact, in each experimental realisation,
which state the cat is in. Only then can we interpret the reduced
density matrix (\ref{eq:meas5}) as a statistical correlation.

There are, in principle, observable consequences of the system
actually being in the superposition, that depend upon the
co-existence of all branches of the superposition\footnote{We will
be examining some of these in more detail in Chapter
\ref{ch:interf}.}. Although these consequences are, in practice,
very difficult to observe, we cannot simply trace over part of the
system, and assume we have a classical correlation in the
remainder. Indeed, the 'resuscitation' of the cat alluded to
requires the use of all branches of the superposition. This
includes the branch in which the observer sees the cat alive as
well as the branch in which the observer sees the cat as dead. If
both branches of the superposition contribute to the resuscitation
of the cat, then both must be equally 'real'.

To understand the density matrix (\ref{eq:meas5}) as a classical
correlation, we must interpret it as meaning that, in each
experiment, the observer actually sees a cat as being alive or
actually sees the cat as being dead. How are we then to understand
the status of the unobserved outcome, the other branch of the
superposition, that enables us to resuscitate the cat, without
using the Many-Worlds interpretation? To make the situation even
more difficult, we need only note that, not only can we
resuscitate the $\Phi$ cat, we can also, in principle at least,
restore the $\Psi$ system to a reference state, leaving the system
in the state
\[
\psi_0 \phi_0 \sum_n \alpha_n \xi_n
\]
The observer is now effectively in a superposition of having
observed the cat alive and observed the cat dead (while the cat
itself is alive and well)! Now the superposition of the states of
the observer is quite different from a statistical mixture. We
cannot assume the observer either remembers the cat being alive or
remembers the cat being, nor can we assume that the observer must
have 'forgotten' whether the cat was alive or dead. The future
behaviour of the observer will be influenced by elements of the
superposition that depend upon his having remembered both.
\cite{AC95} must allow states like this, in principle, but offer
no means of understanding what such a state could possibly mean.

\section{Summary}

The Shannon information plays several different roles in a
classical system. It derives it's primary operational significance
 as a measure of the capacity, in bits, a communication channel
must have to faithfully transmit a ensemble of different messages.
Having been so defined, it becomes possible to extend the
definition to joint, conditional and mutual information. These
terms can be used to describe the information shared between two
different systems - such as a message sender and message receiver
- or can be used to describe the changes in information an
observer has on making measurements upon a classical system. In
all cases, however, the concept essentially presupposes that the
system is in a definite state that is revealed upon measurement.

For quantum systems the interpretation of information is more
complex. Within the context of communication, Schumacher
generalises Shannon's theorem to derive the capacity of a quantum
communication channel and the Kholevo bound demonstrates that this
is the most information the receiver can acquire about the message
sent.

However, when considering the information of unknown quantum
states the situation is less clear. Unlike the classical case
there is no unique decomposition of the statistical state (density
matrix) into a probability distribution over individual states. A
measurement is no longer necessarily revealing a pre-existing
state. In this context, finally, we note that the very application
of information to a quantum system presupposes that we have a
well-defined measuring process.
\chapter{Active Information and Interference}\label{ch:interf}

In Chapter \ref{ch:info} we reviewed the status of information
gain from a quantum measurement. This assumed that measurements
have outcomes, a distinct problem in quantum theory.

We now look at the concept of 'active information' as a means of
addressing the measurement problem within the Bohm approach to
quantum theory. This approach has been recently criticised as part
of a series of though experiments attempting to explore the
relationship between information and interference. These thought
experiments rely upon the use of 'one-bit detectors' or
'Welcher-weg' detectors, in the two slit interference experiment.
In this Chapter we will show why these criticisms are invalid, and
use the thought experiment to illustrate the nature of active
information. This will also clarify the relationship between
information and interference.

Section \ref{s:interf1} will introduce the Bohm interpretation and
highlight it's key features. This will introduce the concept of
active information. The role of active information in resolving
the measurement problem will be briefly treated.

Section \ref{s:interf2} analyses the which-path interferometer. It
has been argued that there is a complementary relationship between
the information obtained from a measurement of the path taken by
an atom travelling through the interferometer, and the
interference fringes that may be observed when the atom emerges
from the interferometer.  As part of the development of this
argument, a quantum optical cavity has been proposed as a form of
which path, or 'welcher-weg' measuring device. The use of this
device plays a key role in 'quantum eraser' experiments and in the
criticism of the Bohm trajectories. We will therefore examine
carefully how the 'welcher-weg' devices affect the interferometer.

Finally, in Section \ref{s:interf3} we will argue that the manner
in which the term 'information' has been used in the which path
interferometers is ambiguous. It is not information in the sense
of Chapter \ref{ch:info}. Rather, it appears to be assuming that a
quantum measurement reveals deeper properties of a system than are
contained in the quantum description, and this is the information
revealed by the measurement.

We will show that this assumption is essential to the
interpretation of the 'welcher-weg' devices as reliable which path
detectors. However, it will be shown that the manner in which this
interpretation is applied to the 'welcher-weg' devices is not
tenable, and this is the reason they are supposed to disagree with
the trajectories of the Bohm approach. By contrast, the concept of
active information, in the Bohm interpretation, does provide a
consistent interpretation of the interferometer, and this can
clarify the relationship between which path measurements and
interference.

\section{The Quantum Potential as an Information Potential}
\label{s:interf1}

The Bohm interpretation of quantum
mechanics\cite{Boh52a,Boh52b,BH87,BHK87,BH93,Hol93,Bel87} can be
derived from the polar decomposition of the wave function of the
system, $\Psi=Re^{iS}$, which is inserted into the \Sch
equation\footnote{We set $\hbar =1$}

\[
i\frac{\partial \Psi}{\partial t}= \left(-\frac{\nabla
^2}{2m}+V\right)\Psi
\]
yielding two equations, one that corresponds to the conservation
of probability, and the other, a modified  Hamilton-Jacobi
equation:

\begin{equation}
-\frac{\partial S}{\partial t}= \frac{(\nabla S)^2}{2m}+V-
\frac{\nabla ^2R}{2mR} \label{eq:qhj}
\end{equation}

This equation can be interpreted in the same manner as a classical
Hamilton- Jacobi, describing an ensemble of particle trajectories,
with momentum $p=\nabla S$, subject to the classical potential V
and a new quantum potential $Q=-\frac{\nabla ^2R}{2mR}$.  The
quantum potential, Q, is responsible for all the non-classical
features of the particle motion.  It can be shown that, provided
the particle trajectories are distributed with weight $R^2$ over a
set of initial conditions, the weighted distribution of these
trajectories as the system evolves will match the statistical
results obtained from the usual quantum formalism. It should be
noted that although the quantum Hamilton-Jacobi equation can be
regarded as a return to a classical deterministic theory, the
quantum potential has a number of the non-classical features that
make the theory very different from any classical theory.  We
should regard Q as being a new quality of global energy that
augments the kinetic and classical potential energy to ensure the
conservation of energy at the quantum level. Of particular
importance are the properties of non-locality and form-dependence.

\subsection{Non-locality}

Perhaps the most surprising feature of the Bohm approach is the
appearance of non-locality. This feature can be clearly seen when
the above equations are generalised to describe more than one
particle.  In this case the polar decomposition of
$\Psi(x_1,x_2,\cdots ,x_N)=R(x_1,x_2,\cdots
,x_N)e^{iS(x_1,x_2,\cdots ,x_N)}$ produces a quantum potential,
$Q_i$, for each particle given by:

\[
Q_i=-\frac{\nabla ^2_iR(x_1,x_2,\cdots ,x_N)}{2mR(x_1,x_2,\cdots
,x_N)} \]

This means that the quantum potential on a given particle $i$
will, in general, depend on the instantaneous positions of all the
other particle in the system.  Thus an external interaction with
one particle may have a non-local effect upon the trajectories of
all the other particles in the system.  In other words groups of
particles in an entangled state are, in this sense, non-separable.
In separable states, the overall wave function is a product of
individual wave functions.

For example, when one of the particles, say particle 1, is
separable from the rest, we can write $\Psi(x_1,x_2,\cdots
,x_N)=\phi(x_1)\xi(x_2,\cdots ,x_N)$.  In this case
$R(x_1,x_2,\cdots ,x_N)= R_1(x_1)R_{2\cdots N}(x_2,\cdots ,x_N)$,
and therefore:

\[
Q_1=-\frac{\nabla ^2_1R_1(x_1)R_{2\cdots N}(x_2,\cdots ,x_N)}
{2mR_1(x_1)R_{2\cdots N}(x_2,\cdots ,x_N)} =-\frac{\nabla
^2_1R_1(x_1)}{2mR_1(x_1)}
\]

In a separable state, the quantum potential does not depend on the
position of the other particles in the system.  Thus the quantum
potential only has non-local effects for entangled states.

\subsection{Form dependence}

We now want to focus on one feature that led Bohm \& Hiley
\cite{BH93} to propose that the quantum potential can be
interpreted as an `information potential'.  As we have seen above
the quantum potential is derived from the R-field of the solution
to the appropriate \Sch equation.  The R-field is essentially the
amplitude of the quantum field $\Psi$ . However, the quantum
potential is not dependant upon the amplitude of this field (i.e.,
the intensity of the R-field), but only upon its {\em form}.  This
means that multiplication of R by a constant has no effect upon
the value of Q.  Thus the quantum potential may have a significant
effect upon the motion of a particle even where the value of R is
close to zero.  One implication of this is that the quantum
potential can produce strong effects even for particles separated
by a large distance.  It is this feature that accounts for the
long- range EPRB-type correlation upon which teleportation relies.

It is this form-dependence (amongst others things) that led Bohm
\& Hiley \cite{BH84,BH93}  to suggest that the quantum potential
should be interpreted as an information potential. Here the word
`information' signifies the {\em action} of forming or bringing
order into something. Thus the proposal is that the quantum
potential captures a dynamic, self-organising feature that is at
the heart of a quantum process.

For many-body systems, this organisation involves a non-local
correlation of the motion of all the bodies in the entangled
state, which are all being simultaneously organised by the
collective R-field.   In this situation they can be said to be
drawing upon a common pool of information encoded in the entangled
wave function. The informational, rather than mechanical, nature
of this potential begins to explain why the quantum potential is
not definable in the 3-dimensional physical space of classical
potentials but needs a 3N-dimensional configuration space. When
one of the particles is in a separable state, that particle will
no longer have access to this common pool of information, and will
therefore act independently of all the other particles in the
group (and vice versa).  In this case, the configuration space of
the independent particle will be isomorphic to physical space, and
its activity will be localised in space-time.

\subsection{Active, Passive and Inactive Information}

In order to discuss how and what information is playing a role in
the system, we must distinguish between the notions of active,
passive and inactive information.  All three  play a central role
in our discussion of teleportation.  Where a system is described
by a superposition $\Psi(x)=\Psi_a(x)+\Psi_b(x)$, and $\Psi_a(x)$
and $\Psi_b(x)$ are non-overlapping wavepackets, then
\[
\Psi_a(x)\Psi_b(x) \approx 0
\]
for all values of $x$. We will refer to this as {\em
superorthogonality}. The actual particle position will be located
within either one or the other of the wavepackets. The effect of
the quantum potential upon the particle trajectory will then
depend only upon the form of the wavepacket that contains the
particle. We say that the information associated with this
wavepacket is active, while it is passive for the other packet. If
we bring these wavepackets together, so that they overlap, the
previously passive information will become active again, and the
recombination will induce complex interference effects on the
particle trajectory.

Now let us see how the notion of information accounts for
measurement in the Bohm interpretation.  Consider a two-body
entangled state, such as
$\Psi(x_1,x_2)=\phi_a(x_1)\xi_a(x_2)+\phi_b(x_1)\xi_b(x_2)$, where
the active information depends  upon the simultaneous position of
both particle 1 and particle 2.  If the $\phi_a$ and $\phi_b$ are
overlapping wave functions, but the $\xi_a$ and $\xi_b$ are
non-overlapping, and the actual position of particle 2 is
contained in just one wavepacket, say $\xi_a$, the active
information will be contained only in $\phi_a(x_1)\xi_a(x_2)$, the
information in the other branch  will be passive.  Therefore only
the $\phi_a(x_1)$ wavepacket will have an active effect upon the
trajectory of particle 1. In other words although $\phi_a$ and
$\phi_b$ are both non-zero in the vicinity of  particle 1,  the
fact that particle 2 is in $\xi_a(x_2)$ will mean that only
$\phi_a(x_1)\xi_a(x_2)$ is active, and thus particle 1 will only
be affected by $\phi_a(x_1)$.

If  $\phi_a(x_1)$ and $\phi_b(x_1)$  are separated, particle 1
will always be found within the location of $\phi_a(x_1)$. The
position of particle 2 may therefore be regarded as providing an
accurate measurement of the position of particle 1. Should the
$\phi_a$ and $\phi_b$ now be brought back to overlap each other,
 the separation of the wavepackets of particle 2 will continue to ensure that
only the information described by $\phi_a(x_1)\xi_a(x_2)$ will be
active.  To restore activity to the passive branches of the
superposition requires that both $\phi_a(x_1)$ and $\phi_b(x_1)$
and $\xi_a(x_2)$ and $\xi_b(x_2)$ be simultaneously brought back
into overlapping positions.  If the $\xi(x_2)$ represents a
thermodynamic, macroscopic device, with many degrees of freedom,
and/or interactions with the environment, this will not be
realistically possible. If it is never possible to reverse all the
processes then the information in the other branch may be said to
be inactive (or perhaps better still `deactivated'), as there is
no feasible mechanism by which it may become active again. This
process replaces the collapse of the wave function in the usual
approach. For the application of these ideas to the problem of
teleportation in quantum information, see Appendix \ref{ap:telep}
and \cite{HM99}.

Rather than see the trajectory as a particle, one may regard it as
the 'center of activity' of the information in the wavefunction.
This avoids the tendency to see the particle as a wholly distinct
object to the wavefunction. As the two feature can never be
separated from each other, it is better to see them as two
different aspects of a single process.

In some respects the 'center of activity' behaves in a similar
manner to the 'point of execution' in a computer program. The
'point of execution' determines which portion of the computer code
is being read and acted upon. As the information in that code is
activated, the 'point of execution' moves on to the next portion
of the program. However, the information read in the program will
determine where in the program the point of execution moves to. In
the quantum process, it is the center of activity that determines
which portion of the information in the wavefunction is active.
Conversely, the activity of the information directs the movement
of the 'center'.

The activity of information, however, differs from the computer in
two ways. Firstly, the wavefunction itself is evolving, whereas a
computer program is unlikely to change it's own coding (although
this is possible). Secondly, when two quantum systems interact,
this is quite unlike any interaction between two computer
programs. The sharing of information in entangled systems means
that the 'center of activity' is in the joint configuration space
of both systems. The movement of the center of activity through
one system depends instantaneously upon the information that is
active in the other system, and vice versa. This is considerably
more powerful than classical parallel processing and may well be
related to the increased power of quantum
computers\cite{Joz96,Joz97}.

\section{Information and interference} \label{s:interf2}

In a series of papers\cite{ESSW92,ESSW93,Scu98}, the Bohm
interpretation has been criticised as
'metaphysical','surrealistic' and even 'dangerous', on the basis
of a thought experiment exploiting 'one-bit' welcher-weg, or
which-way, detectors in the two slit interference
experiment\footnote{Similar criticisms were raised by
\cite{Gri99a} in the context of the Consistent Histories
interpretation of quantum theory. A full examination of Consistent
Histories lies outside the scope of this thesis. However, an
analysis of Griffiths argument, from\cite{HM00} is reproduced in
Appendix \ref{ap:chba}.}. Although these criticisms have been
partially discussed elsewhere\cite{DHS93,DFGZ93,AV96,Cun98,CHM00},
there are a number of features to this that have not been
discussed. The role of information, and active information has
certainly not been discussed in this context. The thought
experiment itself arises in the context of a number of similar
experiments in quantum optics \cite[Chapter 20]{SZ97} which
attempt to apply complementarity to information and interference
fringes\cite{WZ79} and the 'delayed choice' effect\cite{Whe82} in
the two-slit interference experiment.  It is therefore useful to
examine how the problems of measurement, information and active
information are applied to this situation.

To properly consider the issues raised by this thought-experiment,
it will be necessary to re-examine the basis of the two-slit
experiment. This will be considered in Subsection
\ref{s:interf2.1}. The role of information in destroying the
interference effects will be reviewed in Subsection
\ref{s:interf2.2}. The analysis of this is traditionally based
upon the exchange of momentum with a detector destroying the
interference. We will find that the quantum optics welcher-weg
devices, which we will discuss in Subsection \ref{s:interf2.3} do
not exhibit such an exchange of momentum, but still destroy the
interference. Subsection \ref{s:interf2.4} then examines the Bohm
trajectories for this experiment, and shows why \cite{ESSW92}
regard them as 'surreal'.

\subsection{The basic interferometer} \label{s:interf2.1}
We will now describe the basic interferometer arrangement in
Figure \ref{fg:inter1}. \pict{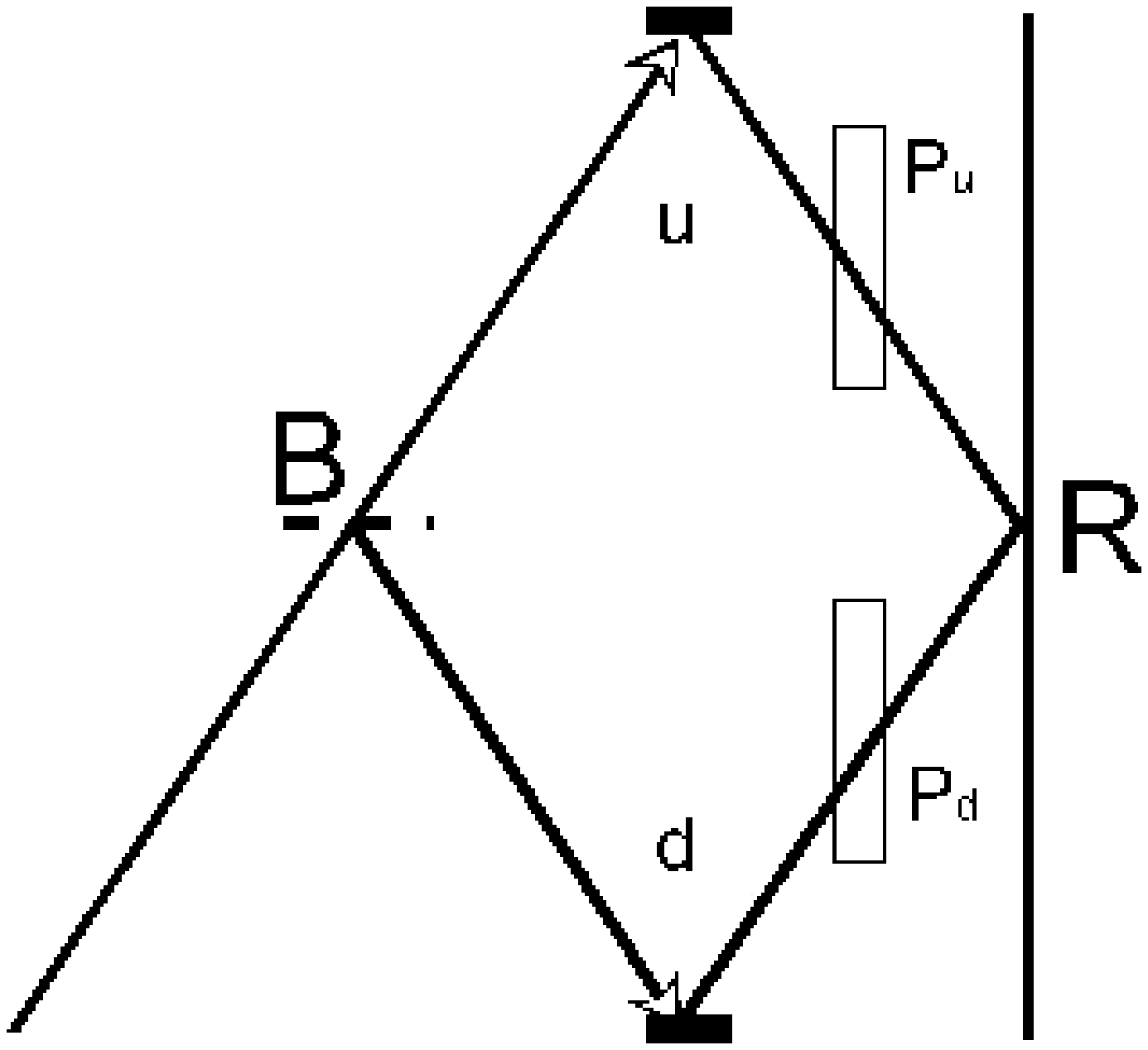}{Basic Interferometer} An
atom, of position co-ordinate $x$, is described by the narrow
wavepacket
\[\psi(x,t)\]. At time $t=t_0$, it is in the initial state
\[\psi(x,t_0)\] and passes through a beam splitter at B, and at
$t=t_1$ has divided into the states
\[
\psi(x,t_1)=\frac{1}{\sqrt2}\left(\psi_u(x,t_1)+\psi_d(x,t_1)\right)
\]
where $\psi_u(x)$ is the wavepacket travelling in the upper branch
of the interferometer, and $\psi_d(x)$ is the wavepacket in the
lower branch.

After $t=t_1$, the wavepackets are reflected so that at $t=t_2$
they are moving back towards each other
\[
\psi(x,t_2)=\frac{1}{\sqrt2}\left(\psi_u(x,t_2)+\psi_d(x,t_2)\right)
\]
They recombine at $t=t_3$, in the region R, where the atoms
location is recorded on a screen. The probability distribution
across the screen is then
\[
\magn{\psi(x,t_3)}=\frac12
\left(\magn{\psi_u(x,t_3)}+\magn{\psi_d(x,t_3)}
    +\psi_u(x,t_3)^*\psi_d(x,t_3)+\psi_u(x,t_3)\psi_u(x,t_3)^*\right)
\]
In Figure \ref{fg:inter1} we have also included phase shifters at
locations $P_u$ and $P_d$, in the two arms of the interferometer.
These may be controlled to create a variable phase shift of
$\phi_u$ or $\phi_d$ in the respective wavepacket. The settings of
these phase shifters will play an important role in the later
discussion, but for the moment, they will both be assumed to be
set to a phase shift of zero, and thus have no effect upon the
experiment.

If we apply the polar decomposition $\psi=Re^{iS}$ to this, we
obtain
\[
\magn{\psi(x,t_3)}=\frac12 \left(R_u(x,t_3)^2+R_d(x,t_3)^2
    +2R_u(x,t_3)R_d(x,t_3)\cos(S_u(x,t_3)-S_d(x,t_3))\right)
\]
We can simplify this by assuming the beam splitter divides the
wavepackets equally, so that in the center of the interference
region
\[
R_u(x,t_3)=R_d(x,t_3)=R(x,t_3)
\]
and
\[
\magn{\psi(x,t_3)}=R(x,t_3)^2
    \left(1+\cos(\Delta S(x,t_3))\right)
\]
where $\Delta S(x,t_3)=S_u(x,t_3)-S_d(x,t_3)$.

The cosine of the phase produces the characteristic interference
fringes. Had we blocked one of the paths ($u$, for example) we
would have found the probability distribution was $R(x,t_3)^2$.
The probability distribution is not simply the sum of the
probability distributions from each path. The superposition of
states given by $\psi(x,t_3)$ cannot be simply interpreted as half
the time the atom goes down the $u$ path, and half the time going
down the $d$ path.

Now let us consider the addition to the interferometer of the
phase shifters in each of the paths. These could be implemented by
simply fine tuning the length of each arm. The $u$ path is shifted
by a phase $\phi_u$ and the $d$ path by $\phi_d$. The effect on
the interference pattern is simply to modify the cosine term to
\[
\cos\left(\Delta S(x,t_3)+(\phi_u-\phi_d)\right)
\]
Now we have
\[
\magn{\psi(x,t_3)}=R(x,t_3)^2
    \left(1+\cos\left(\Delta S(x,t_3)+(\phi_u-\phi_d)\right)\right)
\]
At the points $x_n$, where
\[
\Delta S(x_n,t_3)+(\phi_u-\phi_d)=\frac{\pi}{2}+n\pi
\]
then the value of $\magn{\psi(x_n,t_3)}=0$ ie. there is no
possibility of the atom being located at that point. The important
point to note is that the values of $x_n$ are determined by the
values of both $\phi_u$ {\em and} $\phi_d$, that is by the setting
of the phase shifters in {\em both} arms of the interferometer.

This emphasises the point that we are unable to regard the
superposition of states in $\psi(x,t_1)$ as simply representing a
situation where, in half the cases the atom travels the d-path,
and in half the cases the u-path. Not only is the interference
pattern not simply the sum of the probability distribution from
each of the two paths, but critically, the location of the nodes
in the interference pattern depends upon the settings of
instruments in {\em both} paths.

A simplistic way of stating this is in terms of what the atom
'knows' it should do when reaching the screen. If the atom
proceeds down one path, and the other path is blocked, it can
arrive at locations that are forbidden if the other path is not
blocked. How does the atom 'know' whether the other path is
blocked or not? The phase shifters demonstrate that, not only must
the atom 'know' whether or not the paths are blocked, but even if
they are not blocked, the very locations which are forbidden to it
depend upon the atom 'knowing' the values of the phase shifts in
{\em both} arms. If the atom only travels down one path or the
other, how is it to 'know' the phase shift in the other path?

This is a generic property of superpositions. We cannot interpret
these as a statistical mixture as this implies that in each
experiment either one or the other possibility is realized while
we can always exhibit interference effects which depend upon both
of the elements of the superposition.

\subsection{Which way information} \label{s:interf2.2}
We now turn to the attempts to measure which way the atom went.
The interference pattern builds up from the ensemble of individual
atoms reaching particular locations of the screen. If we could
know which path the atom takes, we could separate the ensemble of
all the atoms that travelled down the u-branch from the atoms
travelling down the d-branch, and this might shed light upon the
questions raised by the introduction of the phase shifters.

As is well known, however, the attempt to measure the path taken
by the atom destroys the interference pattern recorded on the
screen. The paradigm explanation\cite[Chapter 37]{Fey63},
originally due to Heisenberg, involves scattering a photon from
the atom, to show it's location. To be able to determine which
path the atom takes, the wavelength of the photon must be less
than the separation of the paths. However, this scattering changes
the momentum of the atom, according to the uncertainty
relationship $\Delta x \Delta p \ge \hbar$. This random addition
to the wavefunction of the atom destroys the phase coherence of
the two branches of the superposition and so destroys the
interference. The measurement of the atoms location changes the
quantum system from the pure state $\psi(x,t_1)$ to the
statistical density matrix
\[
\rho=\frac12
\left(\proj{\psi_u(x,t_1)}+\proj{\psi_d(x,t_1)}\right)
\]
where $\proj{\psi_u(x,t_1)}$ is correlated to the measurement
outcome locating the atom in the u-path, and
$\proj{\psi_d(x,t_1)}$ is correlated to the atom located in the
d-path. The values of the phase shifters is now irrelevant, and no
interference occurs in the region $R$. We will not now find any
inconsistency in treating the system as a statistical mixture.

\paragraph{Quantity of information}
The information obtained from the position measurement above is
'all or nothing'. We either do not measure the path, and get an
interference pattern, or we measure it, and lose the interference
pattern. This often leads to a tendency to adopt the language
where the quantum object is said to behave in a 'particlelike'
manner, when the which path information is measured, and in a
'wavelike' manner when the interference is observed.

In \cite{WZ79} the experiment is refined by varying the certainty
one has about the path taken by the atom. There are several
different methods proposed for this, but the most efficient
suggested is equivalent to changing the beam splitter in Figure
\ref{fg:inter1}, such that the atomic beam emerges with state
\[
\psi^\prime(x,t_2)=\alpha \psi_u(x,t_2)+\beta \psi_d(x,t_2)
\]
where $\magn{\alpha}+\magn{\beta}=1$. Wootters and Zurek deem the
information 'lacking' about the path of the atom to be
\begin{equation}
I_{WZ}=-p_u \log_2 p_u-p_d \log_2 p_d \label{eq:wzinf}
\end{equation}
where $p_u=\magn{\alpha}$ and $p_d=\magn{\beta}$.

The resulting interference pattern on the screen is given by
\[
\magn{\psi^\prime(x,t_3)}=R(x,t_3)^2
    \left(1+2\sqrt{p_u p_d}\cos(\Delta S(x,t_3)+(\phi_u-\phi_d)+\theta)\right)
\]
where $\theta$ is the relative phase between the complex numbers
$\alpha$ and $\beta$. If the value of $p_u$ approaches zero or
one, then the atom will always go down one arm or the other.
$I_{WZ}$ goes to zero, so there is no information lacking about
the path of the atom, but the interference term disappears. The
largest interference term occurs when $p_u=p_d=\frac12$, for which
$I_{WZ}=-\log_2 2$ represents a maximum lack of information. It is
noticeable that this experiment does not actually involve a
measurement at all. However, Wootters and Zurek show that, for a
given size of the interference term, the information that can be
obtained from any measurement is no more than $I_{WZ}$. In this
respect, the complementarity between the interference and $I_{WZ}$
is equivalent to the equality in the uncertainty relationship
$\Delta x \Delta p \ge \hbar$. What is significant here is that in
Wootters and Zurek's view it is not the momentum transfer that
destroys the interference effects, rather it is the information we
have about the path of the atom.

Finally we can consider Wheeler's delayed choice
experiment\cite{Whe82} where the screen may be removed from Figure
\ref{fg:inter1} and detectors are placed at $D_1$ and $D_2$, as in
Figure \ref{fg:inter2}. Now the wavepackets continue through the
interference region, and become separate again at $t=t_4$
\[
\psi(x,t_4)=\left(\psi_u(x,t_4)+\psi_d(x,t_4)\right)
\]
A detection at $D_1$ of the wavepacket $\psi_d(x,t_4)$ is
interpreted as detecting that the atom went through the d-path in
the interferometer. \pict{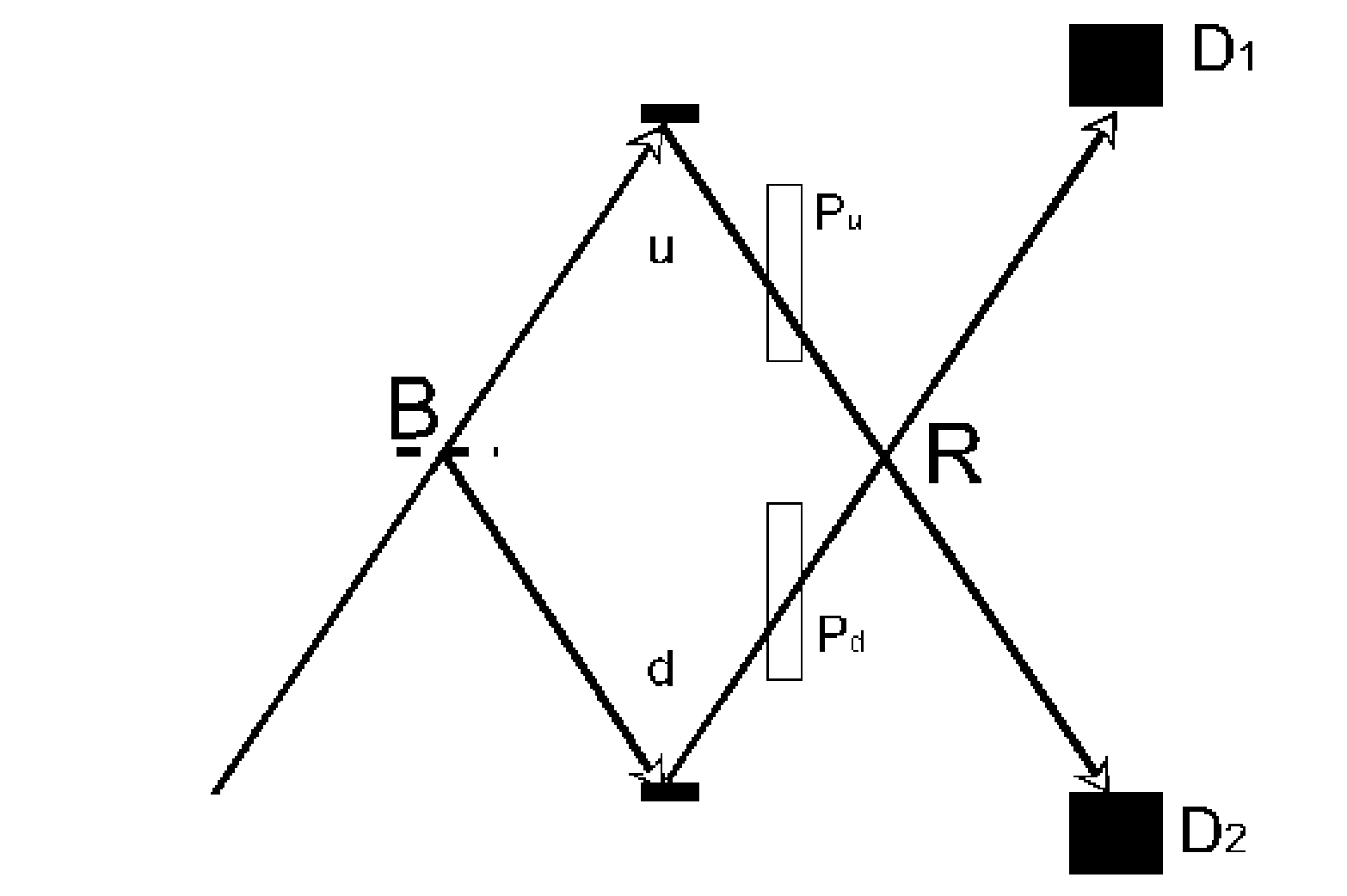}{Which-path delayed choice} Now,
the choice of whether to insert the screen can be made after the
wavepackets have entered the interferometer arms (and even passed
the phase shifters). The choice as to whether we obtain
interference (the atom is a wave in both arms of the
interferometer) or information about which path the atom took (the
atom is a particle in one branch of the interferometer) is delayed
until after the quantum system has actually entered the
interferometer.

\subsection{Welcher-weg devices}\label{s:interf2.3}
In a series of articles\cite[and references
within]{ESW91,ESSW92,SZ97}, it has been suggested that the
which-path information can be measured by using certain quantum
optical devices, which we will follow the authors of these papers
in referring to as 'welcher-weg' (German for 'which way') devices.
These devices do not make a random momentum transfer to the atom
and so it is argued they represent an advance in the understanding
of the which path interferometer. It is the use of these devices
that is essential to understanding the 'quantum eraser'
experiments and the criticism of the Bohm interpretation.

There are three key physical processes that are involved in these
experiments, all involving a two-level circular Rydberg atom. This
is an atom whose outer shell contains only a single electron, the
state of which can be treated effectively as in a hydrogen atom.
The two levels refers to the ground ($\ket{g}$) and first excited
($\ket{e}$) state of the outer shell electron, which differ by the
energy $\Delta E_R$. The processes to which this atom is subjected
are:

\begin{itemize}
\item  Timed laser pulses producing Rabi oscillations.
\item Interaction with a single mode micromaser cavity.
\item Selective ionization
\end{itemize}

Full details of these processes can be found
in\cite{AE74,MW95,SZ97}. We will describe only their essential
features here.

\paragraph{Rabi oscillations}
The atom rapidly passes through an intense electromagnetic field,
oscillating at a single frequency. This can be achieved using a
pulsed laser, and the intensity of the electromagnetic field
allows it to be treated as a semiclassical perturbation on the
atomic states.

The frequency $\omega_R$ of the laser is tuned to the energy gap
between the ground and first excited state of the atom $\Delta
E_R=\hbar \omega_R$. The effect upon the atomic state is to
produce a superposition of ground and excited states

\begin{equation}
\alpha(t)\ket{g}+\beta(t)\ket{e} \label{eq:rabi}
\end{equation}

whose equation of motion is

\begin{eqnarray*}
\frac{d\alpha(t)}{dt}&=& \imath \frac{R}{2} \beta(t) \\
\frac{d\beta(t)}{dt}&=& \imath \frac{R}{2} \alpha(t)
\end{eqnarray*}

where $R$ is the Rabi oscillation term. This factor is a constant,
whose exact value is a function of the overlap integral between
the \ket{g} and \ket{e} states under the influence of perturbation
field of the laser.

The solutions to these coupled equations are
\begin{eqnarray*}
\alpha(t)&=& \alpha(0) \cos\left(\frac{Rt}2\right)
    +\imath \beta(0) \sin\left(\frac{Rt}2\right)\\
\beta(t)&=& \beta(0) \cos\left(\frac{Rt}2\right)
    +\imath \alpha(0) \sin\left(\frac{Rt}2\right)
\end{eqnarray*}

If we time the length of the pulse carefully, we can manipulate
the excitation of the atom. Of particular importance is the $\pi$
pulse, where $Rt=\pi$, as this has the effect of flipping the
atomic state so that $\ket{e}\rightarrow \imath \ket{g}$ and
$\ket{g} \rightarrow \imath \ket{e}$.

\paragraph{Single Mode Cavity}
The Rabi oscillations are produced from an intense, semiclassical
electromagnetic field. The single mode cavity involves the
interaction of the atom with a field with very few photon states
excited. The operation is essentially based upon the
Jaynes-Cumming model\cite{CJ63}.

Instead of using a laser pulse, the circular Rydberg atom is sent
through a high quality microwave cavity, which is tuned to have
the same fundamental resonant frequency $\omega_R$ as the atom. We
will describe the state of the electromagnetic field in the cavity
using the Fock state basis, giving the number of photons excited
in the cavity at the fundamental frequency. Where there are $n$
photons in the cavity, it's quantum state is described as \ket{n}.

If the length of time the atom spends in the cavity is carefully
controlled, there are only three interactions we need to consider
for the purposes of the experiments involved:
\begin{eqnarray}
\ket{g0}\rightarrow \ket{g0} \nonumber \\
\ket{g1}\rightarrow \ket{e0} \nonumber \\
\ket{e0}\rightarrow \ket{g1} \label{eq:cavity}
\end{eqnarray}
If an excited atom goes through an unexcited cavity, it decays to
the ground state, and the $\hbar \omega_R$ energy excites the
first photon state of the cavity. If the atom in the ground state
goes through a cavity with a single photon excitation, the energy
is absorbed, exciting the atom and de-exciting the cavity. If
neither atom nor cavity are excited, then no changes can take
place.

The most important property of these devices is that, if an
excited atom passes through the cavity, it deposits its energy
into the photon field with certainty. As we shall see, it is this
that leads \cite{ESSW92} to describe them as 'welcher-weg' devices
\footnote{A second property of interest is that the interaction of
the atom and cavity has negligible effect upon the momentum of the
atomic wavepacket.}.

\paragraph{Selective Ionization}

State selective field ionization passes the atom through a
electric field that is sufficiently strong to ionize the atom when
the electron in the excited state, but insufficiently strong to
ionize the atom with the electron in the ground state. The ionized
atom and electron are then detected by some amplification process.
For completeness, the ionization of the excited state may be
followed by a second selective ionization and detection, capable
of ionizing the ground state. As long as the first ionization is
very efficient, a reliable measurement of the ground or first
excited state will have taken place.

\cite{ESSW92} now proposed the experiment where a welcher-weg
cavity is placed in each arm of the delayed choice interferometer,
as shown in Figure \ref{fg:inter3}.\pict{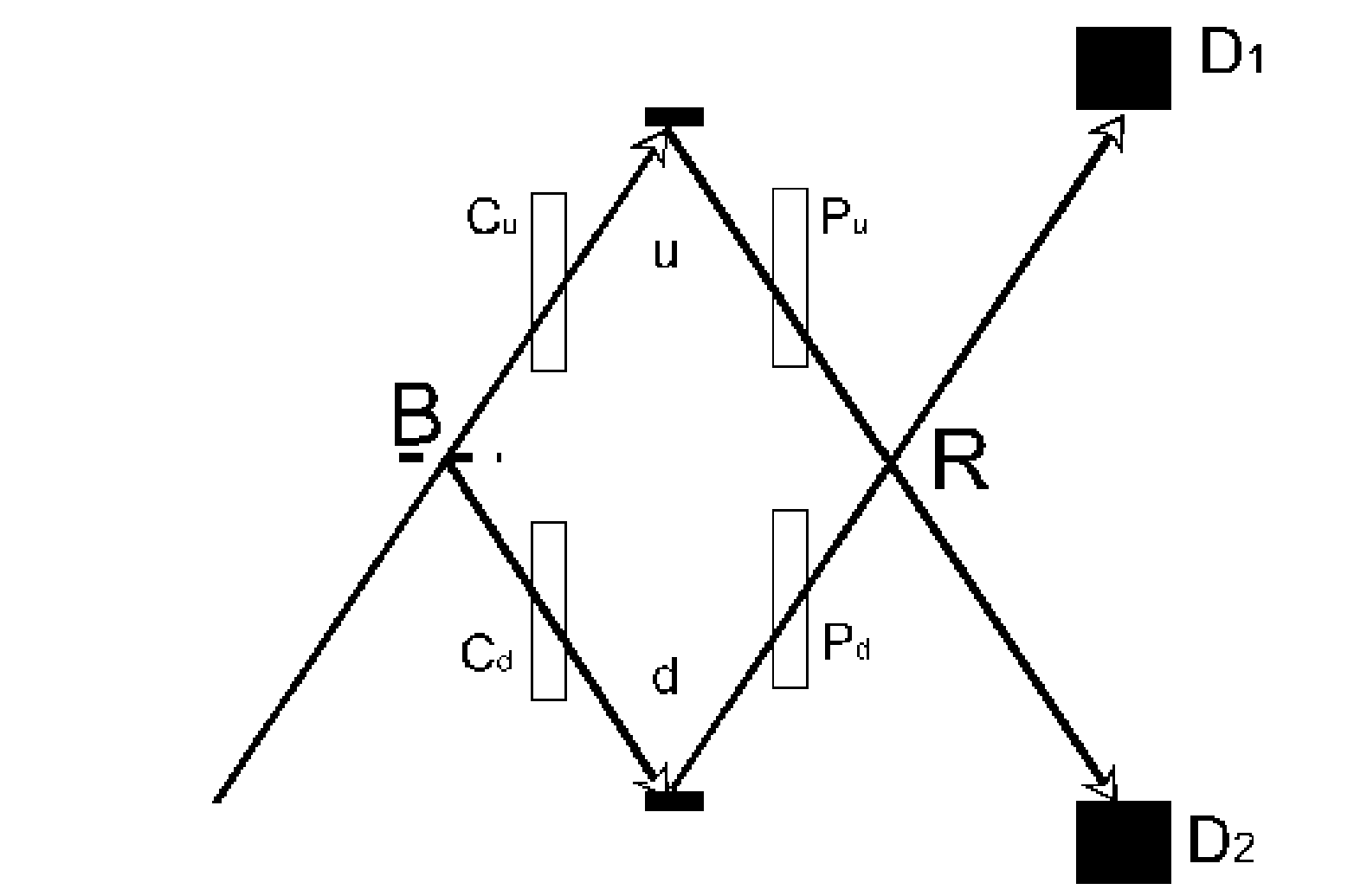}{Welcher-weg
cavities} The atomic wavepackets, initially in the ground state,
are given a $\pi$ pulse just before entering the interferometer.
The electron excitation is passed on to the cavity field mode,
leaving the cavity excited. With the screen missing, the atomic
wavepacket is then detected at either $D_1$ or $D_2$. The location
of the photon, in the upper or lower cavity, is detected by
sending another ('probe') atom, initially in the ground state,
through the cavity and performing a state selective ionization
upon it.

If we follow the quantum evolution of this system, we have:
\begin{enumerate}
\item At $t=t_0$, the atom has not yet encountered the beam
splitter, but is $\pi$ pulsed into the excited state \ket{e},
while the u-path and d-path cavities are in the ground state
($n=0$).
\[
\ket{\Psi(t_0)}=\ket{\psi(t_0),e,0_u,0_d}
\]

\item The atom passes into the interferometer and the wavepacket
is split into the two arms:
\[
\ket{\Psi(t_1)}=\frac{1}{\sqrt2}
    \left(\ket{\psi_u(t_1),e,0_u,0_d}
    +\ket{\psi_d(t_1),e,0_u,0_d}\right)
\]

\item The wavepackets encounter the welcher-weg cavities. The
excited electron energy is deposited in the photon field of the
relevant cavity
\[
\ket{\Psi(t_2)}=\frac{1}{\sqrt2}
    \left(\ket{\psi_u(t_2),g,1_u,0_d}
    +\ket{\psi_d(t_2),g,0_u,1_d}\right)
\]

\item The wavepackets pass through the interference region. The
triggering of the measuring device $D_1$ collapses the state to
\[
\ket{\psi_d(t_4),g,0_u,1_d}
\]
while triggering $D_2$ produces
\[
\ket{\psi_u(t_4),g,1_u,0_d}
\]
\item Probe atoms are sent through the welcher-weg cavities. If
$D_1$ was triggered, then the d-path probe atom will absorb a
photon and be detected by the selective ionization, while a $D_2$
detector triggering will be accompanied by the u-path probe atom
absorbing a photon and being ionized.
\end{enumerate}
This certainly appears to confirm Wheeler's interpretation of the
delayed choice path measurement. If the atom travels down the
d-path, it deposits the energy in the d-cavity, passes through the
interference region and is detected by $D_1$. Conversely, if the
atom travels down the u-path, it deposits the energy in the
u-cavity, passes through the interference region and is detected
by $D_2$.

If we place the screen back in the interference region, what
pattern do we see? The answer is now
\[
\magn{\bk{x}{\Psi(t_3)}}=R(x,t_3)^2
\]
There is no interference term. The reason the interference
disappears is due to the orthogonality of the {\em welcher-weg
cavity} states \ket{1_u,0_d} and \ket{0_u,1_d}.

\cite{ESSW92} interpret this situation as the location of the
photon in one or the other cavity representing a measurement of
the path of the atom. If we had found an interference pattern in
\ket{\Psi(t_3)}, we could still have sent our probe atoms through
the cavities, and discovered which way the atom went. This would
violate the information-interference complementarity relationship.
The welcher-weg cavities are therefore 'one-bit' detectors,
recording and storing the information about the path the atom
took. It is important to notice that it is now the absence of
interference that is being taken to imply that a measurement has
taken place.

\subsection{Surrealistic trajectories}\label{s:interf2.4}

As we saw in Section \ref{s:interf1}, the Bohm interpretation
describes a set of trajectories for the location of the atom.
 In \cite{ESSW92} these
trajectories are calculated and produce the results shown in
Figure \ref{fg:inter4}\footnote{As is shown in
\cite{DHS93,Cun98,CHM00}, trajectories equivalent to Figure
\ref{fg:inter3} will also occur. However, the fact remains that
{\em some} trajectories will still behave in the manner of Figure
\ref{fg:inter4}, which was not appreciated by \cite{Cun98}}. The
atom that travels down the u-path in the interferometer deposits
the excitation energy in the $C_u$ cavity, but it's trajectory
reverses in the region $R$ and it proceeds to be detected at the
$D_1$ detector. \pict{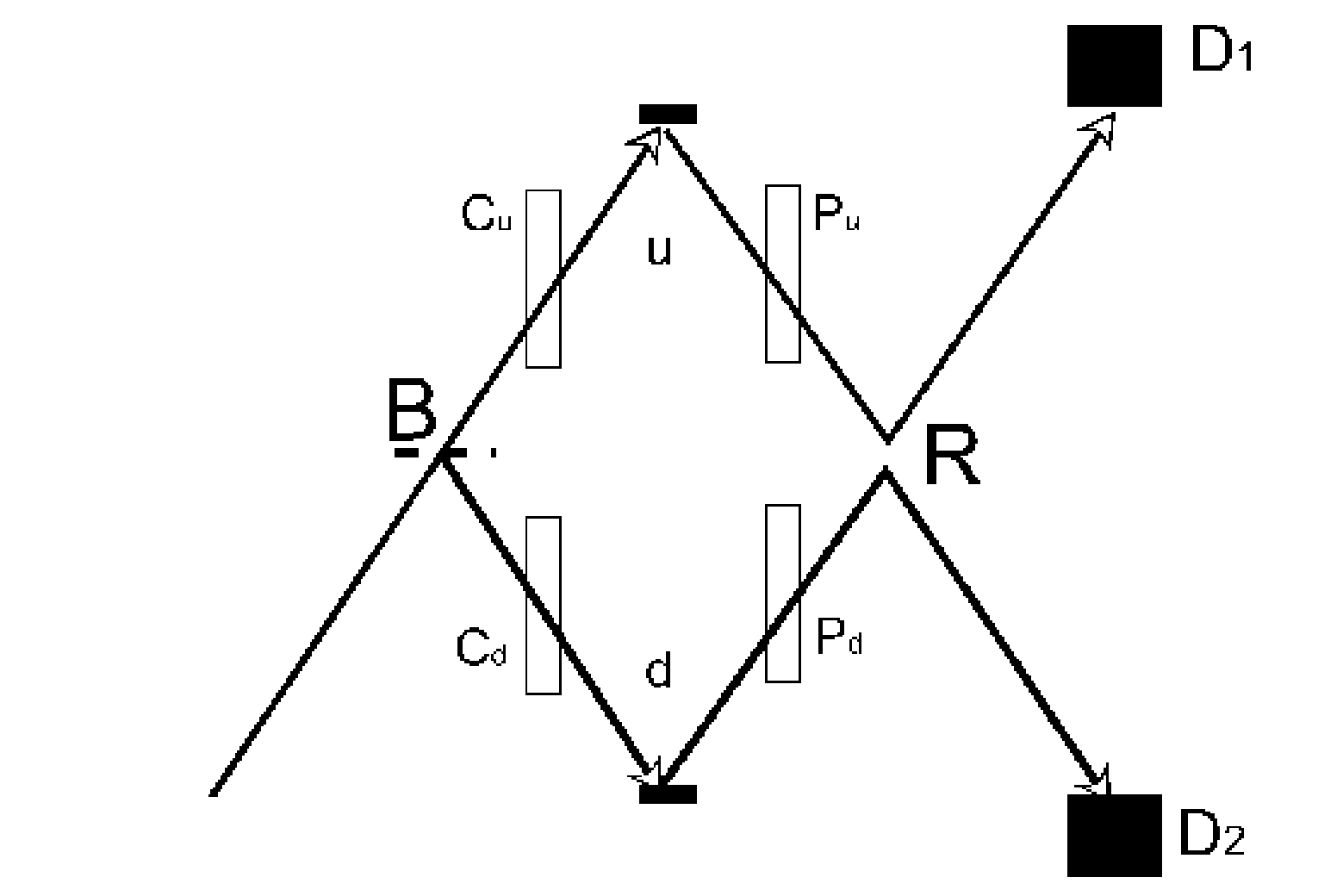}{Surrealistic Trajectories}
Similarly, the atom travelling down the d-path deposits energy in
the $C_d$ cavity, reverses direction in the region $R$ and ends up
in the $D_2$ detector. These results might at first appear to
contradict the experimentally verifiable predictions made in
Subsection \ref{s:interf2.3} and so produce an experimental test
of the Bohm interpretation. However, no such contradiction occurs,
as the Bohm interpretation also predicts that, when the atom is
detected in $D_1$, it is the probe atom going through $C_d$,
rather than $C_u$, that is ionized in the excited state, and vice
versa!

To understand how this occurs we must analyse  why these
trajectories occur with the welcher-weg devices. For conventional
measurements, the trajectories behave as in Figure
\ref{fg:inter3}. We must consider how the single mode cavity
differs from a conventional measuring device, and what effect this
has upon the Bohm trajectories in the various versions of the
interferometer discussed above.

\subsubsection{Delayed choice trajectories}
Let us first note that trajectories of the kind shown in Figure
\ref{fg:inter4} have long been known in the Bohm interpretation,
and discussed in the context of the Wheeler delayed choice
experiment\cite{DHP79,Bel87}. However, these discussions of the
delayed choice experiment suggested that the effect occurs only
when the path of the atom is not measured in the arm of the
interferometer. If detectors are placed in the interferometer
arms, then the result should be the trajectories shown in Figure
\ref{fg:inter3}. It is then argued that the detection of an atom
at $D_1$ in the arrangement of Figure \ref{fg:inter2} cannot be
taken to imply the atom actually travelled down the d-path, except
through the application of a 'naive classical
picture'\cite[Chapter 14]{Bel87} and the possibility of observing
the interference fringes in the region $R$ undermine any such
picture.

By adding their welcher-weg devices \cite{ESSW92} appear to
destroy this position. Two properties emerge. Firstly, the
location of the atom in the detectors coincides with the location
of the photon in the cavity, in the manner shown in Figure
\ref{fg:inter3}. This is taken to confirm Wheeler's assumption
that atom did indeed pass down the d-path when detected in the
$D_1$ detector, and the u-path when detected in the $D_2$
detector. Secondly, the Bohm trajectories still are able to behave
in the manner shown in Figure \ref{fg:inter4} despite the
measurement of the atom's path by the welcher-weg devices.
\cite{ESSW92} conclude that "the Bohm trajectory goes through one
[path], but the atom [goes] through the other", the Bohm
trajectories are "at variance with the observed track of the
particle" and are therefore "surrealistic". In \cite{ESSW93} they
say
\begin{quotation}
If the trajectories \ldots have no relation to the phenomena, in
particular to the detected path of the particle, then their
reality remains metaphysical, just like the reality of the ether
of Maxwellian electrodynamics \end{quotation} and emphasise
\begin{quotation}
this trajectory can be macroscopically at variance with the
detected, actual way through the interferometer
\end{quotation}

We will consider the basis of \cite{ESSW92}'s arguments in detail
in the next Section. Before we do this, however, we will need to
examine in more detail how the Bohm trajectories behave in the
interferometer, and how the ionization of the probe atoms become
correlated to the detectors.

\subsubsection{The cavity field}
The treatment of the field theory in the Bohm interpretation is
developed in \cite{BHK87,BH93,Hol93,Kal94}. In essence, while the
particle theory given in Section \ref{s:interf1} has a particle
position co-ordinate $x$, guided by the wavefunction, the field
theory supposes that there is an actual field, whose evolution is
guided by a wavefunctional. This wavefunctional is the same as the
probability amplitude for a particular field configuration in the
standard approach to quantum field theory.

For a single mode cavity, such as the welcher-weg devices, this
takes a particularly simple form and has been examined in great
detail in \cite{DL94a,DL94b}. The Bohm field configuration can be
represented by a single co-ordinate (the field mode co-ordinate
for the resonant cavity mode) and the wavefunctional reduces to a
wavepacket representing the probability amplitude for the field
mode co-ordinate. As long as one remembers that the 'beable' is
field mode co-ordinate representing a distribution of an actual
field, rather than a localised position co-ordinate, the single
mode cavity may be treated in much the same manner as the particle
theory in Section \ref{s:interf1}.

For the cavity $C_u$, therefore, we need only introduce a mode
co-ordinate $q_u$, the wavefunctional for the cavity mode ground
state \ket{0_u} and for the first excited state \ket{1_u}.
Similarly, for the cavity $C_d$ we introduce $q_d$, \ket{0_d} and
\ket{1_d}. It is important to note that, although the states
\ket{0} and \ket{1} are orthogonal, they are {\em not}
superorthogonal.

\subsubsection{Basic interferometer}

We now review the evolution of the Bohm trajectories in the
experimental arrangements in Figures \ref{fg:inter1} and
\ref{fg:inter2}

As in Subsection \ref{s:interf2.1}, the atomic wavefunction, in
state $\psi(x,t_1)$ divides at the beam splitter. The trajectory
of the atom will move into one or the other of the wavepackets
$\psi_u(x,t_2)$ or $\psi_d(x,t_2)$. As the wavepackets move
through the interferometer arms, the information in only one
wavepacket is active and the other is passive. However, when the
interference region is reached, the two wavepackets begin to
overlap and the previously passive information becomes active once
more. Now the information from both arms of the interferometer is
active upon the particle trajectory. This allows the phase shift
information $\phi_u$ and $\phi_d$ from both phase shifters to
guide the path of the trajectory, and the interference pattern can
show nodes at locations dependant upon the setting of both
devices.

If the screen is not present, the wavepackets separate again. As
both wavepackets were active in the interference region, there is
no guarantee that the trajectory emerges in the same wavepacket in
which it entered. In fact, for the simplest situations, the
trajectory will never be in the same wavepacket! The trajectories
follow the type of paths in Figure
\ref{fg:inter4}\cite{DHP79,Bel80}.

\subsubsection{Which way measurement}
We now add conventional measuring devices to the arms of the
interferometer. These will be described by a co-ordinate ($y_u$ or
$y_d$) and a wavefunction, initially in state $\xi_0(y)$. When the
wavepacket of the atom moves through the arm of the
interferometer, it interacts with the measuring device to change
it's state to $\xi_1(y)$:
\begin{eqnarray*}
\ket{\psi_u(t_2)\xi_0(y_u)\xi_0(y_d)} &\rightarrow&
    \ket{\psi_u(t_2)\xi_1(y_u)\xi_0(y_d)} \\
\ket{\psi_d(t_2)\xi_0(y_u)\xi_0(y_d)} &\rightarrow&
    \ket{\psi_d(t_2)\xi_0(y_u)\xi_1(y_d)}
\end{eqnarray*}
The states $\xi_0$ and $\xi_1$ are superorthogonal and represent
macroscopically distinct outcomes of the measurement (such as
pointer readings). We will assume further that the measuring
device has large number of constituents and interacts with the
environment, in such a manner as to destroy any phase coherence
between the $\xi_0$ and $\xi_1$ states.

Now, the state of the atom and measuring devices after the
interaction is
\[
\frac{1}{\sqrt2} \left(\ket{\psi_u(t_2)\xi_1(y_u)\xi_0(y_d)}
    +\ket{\psi_d(t_2)\xi_0(y_u)\xi_1(y_d)}\right)
\]
As described in Section \ref{s:interf1}, if the atom trajectory is
located in the u-path of the interferometer, then only the
information in $\psi_u(x,t_2)$ is active. The $y_u$ co-ordinate
moves into the $\xi_1$ wavepacket and the $y_d$ co-ordinate
remains in the $\xi_0$ wavepacket. We describe the information in
the other half of the superposition as passive. Had the atom
trajectory initially entered the d-path, $y_d$ would have entered
the $\xi_1$ wavepacket.

When the atomic wavepackets encounter the interference region, the
$\psi_u(x,t_3)$ and $\psi_d(x,t_3)$ begin to overlap. However the
measuring device states are still superorthogonal. The information
in the other branch of the superposition does not become active
again. Consequently, the atom trajectory continues to be acted
upon only by the wavepacket it entered at the start of the
interferometer. No interference effects occur in the $R$ region,
and, if the screen is not present, the u-path trajectory passes
through the interference region to encounter the detector at $D_2$
while the d-path trajectory goes through to the detector at $D_1$.
The superorthogonality of the measuring devices ensures that the
trajectories do not reflect in the interference region, and the
results of the measuring devices in the arms of the interferometer
agree with the detectors at $D_1$ and $D_2$ that the atom has
followed the paths indicated in Figure \ref{fg:inter3}.

Although it is the superorthogonality that plays the key role in
producing the measurement outcome, we will now say a few words
about the role of the loss of phase coherence. As the macroscopic
$\xi$ states interact with the environment, further entangled
correlations build up with large numbers of environmental
particles. This leads to habitual decoherence in the macroscopic
states. From the point of view of active information, however,
what is most significant is that if even a {\em single} one of the
environmental particles is correlated to the measuring device
states in superorthogonal states, then the passive information in
the measuring device states cannot be made active again. As an
example, if the measuring device at $\xi_1$ leads to the
scattering of an atom in the air to a different place than if the
device had been at $\xi_0$, then the passive information in
$\xi_0$ cannot be made active unless the atom in the air is also
brought back into overlapping states. As, for all practical
purposes, the interaction with the environment makes this
impossible, we can describe the information in the 'empty'
wavepacket as inactive, or deactivated.

\subsubsection{Welcher weg devices}
We are now in a position to examine the {\em experimentum crucis}
of \cite{ESSW92}. In place of the measuring devices above, we have
optical cavities in the paths of the interferometer. At $t=t_2$
the wavefunction is
\[
\ket{\Psi(t_2)}=\frac{1}{\sqrt2}
    \left(\ket{\psi_u(t_2),g,1_u,0_d}
    +\ket{\psi_d(t_2),g,0_u,1_d}\right)
\]
Now if the atom trajectory is in the u-path, then in cavity $C_u$
the information in \ket{1_u} is active, and the field mode
co-ordinate $q_u$ will behave as a single photon state. In cavity
$C_d$, it is \ket{0_d} that is active, so $q_d$ behaves as a
ground state. Had the atom trajectory been in the d-path, the
situation would be reversed.

Now, unlike the measurement above, the welcher-weg states are not
superorthogonal, and undergo no loss of phase coherence. When the
atomic wavepackets enter the overlap region $R$, all the
wavepackets in the state
\[
\ket{\Psi(t_3)}=\frac{1}{\sqrt2}
    \left(\ket{\psi_u(t_3),g,1_u,0_d}
    +\ket{\psi_d(t_3),g,0_u,1_d}\right)
\]
are overlapping. The trajectory co-ordinates for $x$, $q_u$ and
$q_d$ are in non-zero portions of the wavefunction for both
branches of the superposition. The previously passive information
becomes active again. It is this that allows the atomic
trajectories to become reflected in $R$ and emerge from this
region in the opposite wavepacket to the one they entered, as in
Figure \ref{fg:inter4}.

If the atom trajectory emerges from $R$ in the wavepacket
$\psi_u(x,t_4)$, then the information in the d-path wavepacket
becomes passive again. This includes the activity of the $q_u$ and
$q_d$ field mode co-ordinates, so only the \ket{1_u} information
is active for $q_u$ and the \ket{0_d} information is active for
$q_d$. The $C_u$ cavity therefore appears to hold the photon,
while the $C_d$ cavity appears empty. {\em This will be the case
even if the atom trajectory originally passed through the $C_d$
cavity.}

Finally, the atom trajectory encounters the detector either at
$D_1$ or $D_2$ and the probe atoms are sent through the cavities.
The probe atom that is sent through the cavity for which the
\ket{1} information is active will be excited, and ionized, and
the correlation between the excited state ionization and the atom
detectors will appear to be that of Figure \ref{fg:inter3}. This
shows how, despite having trajectories of the form in Figure
\ref{fg:inter4}, the Bohm approach produces exactly the same
experimentally verifiable predictions as quantum theory.

\subsection{Conclusion}

The Bohm interpretation clearly provides an internally consistent
means for describing the interference experiments, and produces
all the same observable predictions as 'standard' quantum
mechanics. Nevertheless, \cite{ESSW92,ESSW93,Scu98} argue that the
trajectories followed by the atom in the Bohm interpretation are
\begin{quotation}
macroscopically at variance with the detected, actual way through
the interferometer
\end{quotation}
The claim is that the location of the photon in the welcher-weg
device, after the atomic wavepackets have left the region $R$ tell
us the way the atom actually went. If this claim is true the Bohm
trajectories cannot be an accurate representation of what actually
happened. As we have established the internal consistency of the
Bohm interpretation, we must now examine the internal consistency
of \cite{ESSW92}'s interpretation of their welcher-weg devices.
This examination should not be from the point of view of the Bohm
interpretation, but rather from the point of view of 'standard'
quantum mechanics.

It should be clear from the discussion above that the essential
difference between the standard measuring device, for which the
Bohm trajectories behave as in Figure \ref{fg:inter3}, and the
welcher-weg devices, is that in the cavities there is a coherent
overlap between the excited and ground states throughout the
experiment. This is the property of the welcher-weg devices that
allows the Bohm trajectories to reverse in the region $R$ and
produce the effect that \cite{ESSW92} call 'surrealistic'. If, for
example, the probe atoms were sent though the cavities and ionized
{\em before} the interference region was encountered, then the
ionization and detection process would lead to a loss of phase
coherence, or in the Bohm approach a deactivation of information
in the passive wavepacket. In this case the Bohm trajectories
could not reverse, and the trajectories would follow the paths in
\ref{fg:inter3}. We must therefore investigate the consequences of
the persistence of phase coherence in standard quantum theory, to
see how this affects our understanding of the welcher-weg devices.

\section{Information and which path measurements}
\label{s:interf3}

First we will examine the nature of the which-path 'information'
obtained in the conventional measurement. This, it turns out, is
not information in the sense we encountered it in Chapter
\ref{ch:info}, although it is related to the Shannon information
from a measurement. The information can be interpreted in two
ways: as a strictly operational term, referring to the observable
consequences of a conventional measurement, or as revealing a
pre-existing situation or property of the object being measured.
The second interpretation implicity assumes that there is a deeper
level of reality than that provided by the quantum mechanical
description of a system.

We will then consider the quantum cavity "welcher-weg" devices.
These do not fulfil the criteria of a conventional measuring
device and there are observable consequences of this. The
interpretation \cite{ESSW92} place upon the information derived
from their "welcher-weg" devices is that of revealing pre-existing
properties of the atom, namely it's location. To make this
interpretation, they must implicitly make two assumptions - that
quantum objects, such as atoms or photons, possess an actual
location, beyond the quantum description, and that the atom can
only interact with the welcher-weg devices if the actual location
of the atom is within the device.

However, we will demonstrate that the continued existence of phase
coherence between the welcher-weg states does allow the
observation of interference effects, and these make the
combination of these two assumptions untenable. The welcher-weg
devices cannot be interpreted as providing a reliable measurement
of the location of the atom. This conclusion will be from the
perspective of 'standard' quantum mechanics. We will therefore
find that \cite{ESSW92}'s argument that the location of the
ionized electron reveals the actual path taken by the atom (and
contradicting the Bohm trajectories) is not supported by standard
quantum mechanics, and cannot be consistently sustained. Finally,
we will show how the interference effects observed can be
naturally explained within the context of active information.

\subsection{Which path information}

In \cite{WZ79} it is suggested that it is not the momentum
transfer of a scattered photon that destroys interference fringes,
but rather the gathering of information about the path taken by
the atom. This would appear to be supported by the welcher-weg
devices, as these do not significantly affect the momentum of the
atom. However, we need to consider what we mean by the information
gathered. We will assume the beam splitter can be adjusted, as in
Subsection \ref{s:interf2.2}, to produce the state
\[
\psi^\prime(x,t_2)=\alpha \psi_u(x,t_2)+\beta \psi_d(x,t_2)
\]

The information term $I_{WZ}$ in Equation \ref{eq:wzinf}, although
expressed as a Shannon information, does not correspond to the
quantum information terms in Chapter \ref{ch:info}. The atom is
initially in the pure state $\psi(x,t_0)$. It continues to be in a
pure state after it has split into two separate beams in the
interferometer. The Schumacher information of the atomic state is
zero. This represents a complete knowledge of the system. If we
calculate the information gain from a conventional measurement of
the path taken by the atom, we find that it is always zero. The
initial state is $\psi(x,t_0)$ with probability one. The
measurement of the location of the particle has outcomes $u$ and
$d$ with probabilities \magn{\alpha} and \magn{\beta}, so Bayes's
rule (Equation \ref{eq:qbayes}) produces the trivial result
\begin{eqnarray*}
p(\psi|u)=&\frac{\magn{\alpha} 1}{\magn{\alpha}}&=1 \\
p(\psi|d)=&\frac{\magn{\beta} 1}{\magn{\beta}}&=1
\end{eqnarray*}

We saw this in Subsection \ref{s:info2.4}. The information gain
from a measurement relates to the selection of particular state
from a statistical mixture of states. As this particular situation
is not described by a mixture\footnote{Or, equivalently, is
described by the trivial mixture, for which $p(\psi)=1$} but by a
pure state, there is no uncertainty. Information revealed by the
measurement is {\em not} a gain of information about the {\em
quantum} properties of the system.

From the perspective of information gain, only if the wavepacket
\[
\psi^\prime(x,t_1)=\alpha \psi_u(x,t_1)+\beta \psi_d(x,t_1)
\]
was replaced by the statistical mixture
\[
\rho^\prime=\magn{\alpha} \proj{\psi_u(x,t_1)}
    +\magn{\beta} \proj{\psi_d(x,t_1)}
\]
of \ket{\psi_u(t_1)} and \ket{\psi_d(t_1)} states, would there be
an information gain $I_{WZ}$ from a measurement, but in this case
there would be no interference.

\paragraph{Information about the measurement}
How can we understand $I_{WZ}$ when the initial state is a pure
state? There are two possible ways of doing this. The first method
is to note that $I_{WZ}$ does represent the Shannon uncertainty
about the outcome of the measurement. Let us be very careful what
we mean here. We are proposing that the measuring device is a
conventionally defined, macroscopic object, with an observable
degree of freedom, such as the pointer on a meter. $I_{WZ}$
represents our prior ignorance of the state the pointer will be in
when the measurement is concluded. Naturally, this assumes the
measurement problem is solved so that it is meaningful to talk
about the pointer being in a state, and the measurement being
concluded.

This remains a controversial topic in the interpretation of
quantum theory. However, it is generally accepted, and is
certainly part of the 'standard' approach to quantum theory, that
such a measurement involves an amplification of the quantum state
to macroscopic levels that is, for all practical purposes,
irreversible, and is accompanied by an irretrievable loss of phase
information between the different measurement outcomes. At the end
of such a process, the entangled state between the measuring
device and the measured object can be replaced by a statistical
mixture{\em without in any way affecting the future evolution of
the experiment}. It more or less follows that it can only be
applied to the kind of macroscopically large objects for which a
classical description is valid.

At the end of the measurement, we would know what state the
quantum object was in, as a result of the correlation to the
measuring device. However, we could not infer from this that the
quantum object was in that state {\em prior} to making the
measurement. If we had considered making a complementary
measurement before our path measurement, we could have observed
the kind of interference effects that preclude the assumption that
the measured object was in one or the other state, but that the
state was unknown to us.

In this respect we would be viewing the experiment in the manner
Bohr\cite{Boh58} appears to recommend:
\begin{quotation}
all unambiguous use of space-time concepts in the description of
atomic phenomena is confined to the recording of observations
which refer to marks on a photographic plate or to similar
practically irreversible amplification effects
\end{quotation}
From this point of view, the quantity $I_{WZ}$ refers to the
properties of the macroscopically observable measuring device
outcomes in the particular experimental arrangement. It does {\em
not} represent a statement of the ignorance of the properties of
the atom itself. Our knowledge of the state of the atom, as a
quantum object, is already complete (it is in a pure state). It is
only the future states of the measuring device of which we are
uncertain.

\paragraph{Information about the atom}
The second way of viewing $I_{WZ}$ is to suppose that the
measuring device does precisely what it was intended to do - that
is, measure the actual location of the atom. This must assume that
the atom does indeed have an actual location, and the measurement
reveals that location. This involves the attribution to the atom
of a property (well defined location) which goes beyond the {\em
quantum} description of the object.

When we have only the either/or options of designing an
interference experiment to test the wave nature of the quantum
object, or a which path experiment to test the particle nature of
the quantum object, the tendency is to talk loosely of the quantum
object as being a particle or a wave depending upon the
experimental arrangement. However, the intermediate cases
introduced by \cite{WZ79} make this more difficult, as the object
is supposedly manifesting both particlelike and wavelike
properties in the one arrangement:
\begin{quotation}
The sharpness of the interference pattern can be regarded as a
measure of how wavelike the [object] is, and the amount of
information we have obtained about the [object]'s trajectories can
be regarded as a measure of how particlelike it is
\end{quotation}
The problem here is the talk of our possessing information about
the trajectory taken. The normal meaning of this sentence would be
clear: it would mean that the object had a well-defined
trajectory, and we had some probabilistic estimate of which path
was taken in any given experiment. This meaning applies even when
the ignorance of the path is maximal. This would be the case where
$I_{WZ}=1$. In this case, the consistent use of the word {\em
information} must be taken to mean that the atom follows the
u-path half the time and the d-path the other half the time.

Unfortunately, this is exactly the situation considered in the
basic interferometer (Subsection \ref{s:interf2.1}). The
proponents of an information-interference complementarity would
argue the interference fringes appear because we lack information
about which path was taken. To consistently understand the meaning
of the word information here, we must assume that the atom does,
in fact follow a particular path, it is just that we ourselves are
ignorant of which one. However, the settings of the phase shifters
demonstrates that the ultimate location of the atom in the
interference region depends upon the phase shift in {\em both}
arms of the interferometer. This leads to the exact situation
Bohr\cite{Boh58} warns against, where
\begin{quotation}
we would, thus, meet with the difficulty: to be obliged to say, on
the one hand, that the [atom] always chooses {\em one} of the two
ways and, on the other hand, that it behaves as if it had passed
{\em both} ways.
\end{quotation}

\subsection{Welcher-weg information}

We have seen that the interpretation of which-path information in
the context of a conventional quantum measurement is not without
it's problems. We will now consider the welcher-weg devices.

As we have seen, these devices maintain phase coherence between
the u- and d-branches of the superposition, and this phase
coherence is essential to produce the 'surrealistic' behaviour of
the Bohm trajectories. Such phase coherence is a property that a
conventional measuring device must not possess. It is only when
the state selective ionization takes place that a conventional
measurement can be said to have taken place. This must be after
the atoms have traversed the interference region $R$.

When considering the 'which-path' measurement above, the
destruction of phase coherence in the measurement prevented the
occurrence of interference fringes in the region $R$. With the
welcher-weg devices in place, we similarly lose interference
fringes. If we add the phase shifters to the welcher-weg
experiment, this leads to the state at $t=t_3$
\[
\ket{\Psi(t_3)^{\prime\prime}}=\frac{1}{\sqrt2}
    \left(e^{\imath \phi_u}\ket{\psi_u(t_3),g,1_u,0_d}
    +e^{\imath \phi_d}\ket{\psi_d(t_3),g,0_u,1_d}\right)
\]
The probability distribution in the interference region turns out
to be
\[
\magn{\bk{x}{\Psi(t_3)^{\prime\prime}}}=R(x,t_3)^2
\]
The values of $\phi_u$ and $\phi_d$ have no effect upon the
pattern that emerges if a screen is placed in the region $R$.

The reason for this is that the atom is not, in itself, in a pure
state. It is in an entangled superposition with the photon states
of the fields in the two micromaser cavities. If one traces over
the entangled degrees of freedom, one obtains the density matrix
\[
\frac{1}{2}
    \left(\proj{\psi_u(t_3)}+\proj{\psi_d(t_3)}\right)
\]
which is the same result one would have obtained if there had been
a statistical mixture of atomic wavepackets travelling down one
path or the other. As all the observable properties of a system
are derivable from the density matrix there is no way, {\em from
measurements performed upon the atom alone}, to distinguish
between the state \ket{\Psi(t_3)} and the statistical mixture.

It might therefore seem unproblematical to argue, as \cite{ESSW93}
do, that, although the welcher-weg devices are not {\em
conventional} measurement devices, they are still reliable
\begin{quotation}
Perhaps it is true that it is "generally conceded that \ldots [a
measurement] \ldots requires a \ldots device which is more or less
macroscopic" but our paper disproves this notion because it
clearly shows that one degree of freedom per detector is quite
sufficient. That is the progress represented by the quantum
optical which-way detectors.
\end{quotation}
To \cite{ESSW92,SZ97} the absence of the interference terms
demonstrates information {\em has} been gathered, and that
correspondingly a measurement must have taken place
\begin{quotation}
As long as no information is available about which alternative has
been realized, interference may be observed. On the other hand, if
which-path information is stored in the cavities, then
complementarity does not allow for interference \cite[pg574]{SZ97}
\end{quotation}

However, the tracing over the cavity states does not mean we can
simply replace the entangled superposition with the density
matrix, nor does it mean that we can interpret the entangled
superposition as a statistical mixture. Although interference
properties can no longer be observed from operations performed
upon a single subsystem, we {\em can} observe interference effects
from correlated measurements upon the entire system because, {\em
unlike in a conventional measurement}, phase coherence still
exists.

\paragraph{Interference}
We will now demonstrate how to observe interference effects, by
operations performed upon the probe atom, {\em after} the atomic
wavepacket has reached the region $R$ and {\em after} the probe
has left the cavity. The location of the photon excitation energy
is determined by the selective ionization of a probe atom sent
through the cavity. The probe atom is initially in the ground
state \ket{g_P}. The evolution is
\begin{eqnarray*}
\ket{g_P0} \rightarrow \ket{g_P0} \\
\ket{g_P1} \rightarrow \ket{e_P0}
\end{eqnarray*}

The state of the system becomes
\[
\ket{\Psi(t_4)}=\frac{1}{\sqrt2}
    \left(e^{\imath \phi_u}\ket{\psi_u(t_3),g,e_{P_u},g_{P_d}}
    +e^{\imath \phi_d}\ket{\psi_d(t_3),g,g_{P_u},e_{P_d}}\right)\ket{0_u,0_d}
\]
where \ket{g_{P_u}} represents the ground state of the u-cavity
probe atom etc. The ionization measurement of the probe atoms
leads to the states:
\begin{eqnarray*}
\ket{e_{P_u},g_{P_d}} \Rightarrow \ket{\psi_u(x,t_4)} \\
\ket{g_{P_u},e_{P_d}} \Rightarrow \ket{\psi_d(x,t_4)}
\end{eqnarray*}
which appears to give us a measurement of the atomic position.

We should remember that this is a measurement of the atomic
position {\em after} the atomic wavepackets have left the
interference region $R$, and for which there is no disagreement
between the Bohm trajectories and \cite{ESSW92}'s interpretation
of the location of the atom.

Let us consider what happens if the screen had been placed in the
interference region $R$. Each experiment would lead to a
scintillation at some point on the screen. By correlating the
detected position of the atom in the interference region with the
outcomes of the probe atom ionizations, we would select two
subensembles, which would each have a distribution of
$R(x,t_3)^2$. No interference would be visible.

Now we consider the modification necessary to observe
interference. Before ionizing the probe atoms, let us pass them
each through a pulsed laser beam, producing Rabi oscillations, as
in Equation \ref{eq:rabi}. The size of the pulse should now be
$Rt=\frac12 \pi$. This produces the rotation
\begin{eqnarray*}
\ket{g} \rightarrow \frac1{\sqrt2}
    \left(\ket{g}+\imath \ket{e}\right) \\
\ket{e} \rightarrow \frac1{\sqrt2}
    \left(\imath \ket{g}+ \ket{e}\right)
\end{eqnarray*}
and the state of the system (ignoring the now irrelevant cavity
modes) is
\begin{eqnarray*}
\ket{\Psi(t_3)^\prime} &=&\frac12
    \left(e^{\imath \phi_u}\left(\ket{\psi_u(t_3),e_{P_u},g_{P_d}}
    +\imath \ket{\psi_u(t_3),g_{P_u},g_{P_d}} \right. \right. \\
&&  \left. +\imath \ket{\psi_u(t_3),e_{P_u},e_{P_d}}
    -\ket{\psi_u(t_3),g_{P_u},e_{P_d}} \right) \\
&&    +e^{\imath \phi_d}\left(\ket{\psi_d(t_3),g_{P_u},e_{P_d}}
    +\imath \ket{\psi_d(t_3),e_{P_u},e_{P_d}} \right. \\
&&  \left. \left. +\imath \ket{\psi_d(t_3),g_{P_u},g_{P_d}}
    -\ket{\psi_d(t_3),e_{P_u},g_{P_d}}
    \right)\right)
\end{eqnarray*}
which can be rewritten as
\begin{eqnarray*}
\ket{\Psi(t_3)^\prime}&=&\left(e^{\imath \phi_u}\ket{\psi_u(t_3)}
        -e^{\imath \phi_d}\ket{\psi_d(t_3)}
    \right)\frac{\ket{e_{P_u},g_{P_d}}-\ket{g_{P_u},e_{P_d}}}{2} \\
&&  +\imath \left(e^{\imath \phi_u}\ket{\psi_u(t_3)}
        +e^{\imath \phi_d}\ket{\psi_d(t_3)}
    \right)\frac{\ket{g_{P_u},g_{P_d}}+\ket{e_{P_u},e_{P_d}}}{2}
\end{eqnarray*}
Now when the probe atoms are ionized the atomic wavefunction is
either
\[
\ket{\Psi_a(t_3)}=\frac{1}{\sqrt2}
    \left(e^{\imath \phi_u}\ket{\psi_u(t_4)}
    -e^{\imath \phi_d}\ket{\psi_d(t_4)} \right)
\]
or
\[
\ket{\Psi_b(t_3)}=\frac{1}{\sqrt2}
    \left( e^{\imath \phi_u}\ket{\psi_u(t_4)}
    +e^{\imath \phi_d}\ket{\psi_d(t_4)} \right)
\]
The probability distribution in the interference region is now
either
\[
\magn{\bk{x}{\Psi_a(t_3)}}=\frac{R(x,t_3)^2}{2}
    \left(1+\cos\left(\Delta S(x,t_3)+(\phi_u-\phi_d)\right)\right)
\]
or
\[
\magn{\bk{x}{\Psi_b(t_3)}}=\frac{R(x,t_3)^2}{2}
    \left(1-\cos\left(\Delta S(x,t_3)+(\phi_u-\phi_d)\right)\right)
\]
Both of these exhibit interference patterns in the region $R$ and,
critically for our understanding of the situation, the location of
the nodes of this interference pattern will be dependant upon the
phase shifts $\phi_u$ and $\phi_d$ in {\em both} arms of the
interferometer. Had the cavities been conventional measuring
devices, no such interference patterns could have been observed.
The mixture of the two distributions loses the interference
pattern. It is only when the results of the probe atom
measurements are correlated to the ensemble of atomic locations
that the interference effects can be observed. This is
characteristic of entangled systems, where the interference can
only ever be seen using correlated or joint
measurements\footnote{If interference effects could be seen
without such correlations, they could be used to violate the
no-signalling theorem, and send signals faster than light.}.

It is important to note that the choice of whether or not to pulse
the probe atoms with the $\frac12 \pi$ pulse can be made {\em
after} the atomic wavepacket has entered into the region $R$ and
had it's location recorded on a screen. The information about the
phase shift settings must somehow be present in the atom position
measurements {\em before} we choose whether to pulse the probe
atoms or not.

\paragraph{Quantum erasers}
The arrangement considered here is similar to the quantum eraser
experiments\cite{ESW91,SZ97}. It may be argued that, by pulsing
the probe atom, we are 'erasing' the which path information and so
restoring the interference. The problem is that this implicitly
assumes that there is a matter of fact about which path the atom
took, and that the interference appears only because the
information as to which path the atom took is not stored anywhere.

Thus we read in \cite{SZ97}
\begin{quote}
As long as no information is available about which alternative has
been realized, interference may be observed \end{quote}

This ignores the fact that it is not simply the existence of
interference that is the problem. It is also a problem that the
location of the nodes in the interference pattern so clearly
depend upon the settings of the phase shifters in both arms of the
interferometer. If there is a matter of fact about which path the
atom took ("which alternative has been realized"), that is if we
understand the term 'information' in it's normal usage, then we
cannot account for the fact that the atom is able to avoid
locations that depend upon the configuration of both phase
shifters. There is a fundamental ambiguity in \cite{SZ97}'s
description of the quantum 'eraser': is it only the {\em
information} about which path the atom took that is erased, or is
it the very fact that the atom {\em did} take one or the other
path? We are forced, as Bohr warned, to say the atom travels down
one path, but behaves as if it has travelled down both.

\subsection{Locality and teleportation}

We have established that the welcher-weg devices are not
conventional measuring devices and that there are observable
consequences of this. We will now examine what affect this has
upon \cite{ESSW92,ESSW93,Scu98}'s criticism of the Bohm
interpretation.

The essence of the argument is that when the photon is found in
the cavity the atom must have travelled down that arm of the
interferometer
\begin{quotation}
we do have a framework to talk about path detection: it is based
upon the local interaction of the atom with the \ldots resonator,
described by standard quantum theory with its short range
interactions only \cite{ESSW93}
\end{quotation}
The local interaction between the atom and photon, in terms of the
Hamiltonian interaction in the \Sch equation, is here being taken
to mean that the atom can deposit a photon in the cavity only if
it actually passed through the cavity.

We can identify two key assumptions that are necessary for the
interpretation of the welcher-weg devices as reliable indicators
of the {\em actual} path of the atom:
\begin{enumerate}
\item  This storage of information is a valid measurement, even though it is
not a conventional quantum measurement. The atom can only interact
with the welcher-weg device, and deposit a photon in it, if the
actual path of the atom passes through the device.
\item The reason the interference pattern initially disappears is because
the cavity stores information about the path of the atom. The
storage of information implies that there is a matter of fact,
which may be unknown, about which path the atom took, in all
realizations of the experiment.
\end{enumerate}

\paragraph{Local interactions}
Let us consider why these two assumptions are necessary. The first
assumption is based upon the local interaction Hamiltonian between
the atom and the cavity field. However, when the atom is in a
superposition, as in the interferometer, the effect of this
Hamiltonian is to produce an entangled correlation between the
atom and the cavity mode wavefunctions. Part of the atomic {\em
wavefunction} interacts with {\em each} cavity wavefunction. If we
took the wavefunction to be a physically real entity, we could not
say that the atom in the interferometer interacts with only one
cavity, we would have to say that the atom interacts with both
cavities, in all experiments. If this were the case, then could
draw no conclusions about the path taken by the atom from the
location of the photon. To reach \cite{ESSW92}'s conclusion we
must argue, as is standard, that the wavefunction is not
physically real but
\begin{quote}a tool used by theoreticians to
arrive at probabilistic predictions\end{quote}
If one is
consistently to take this view, however, one must also apply it to
the Hamiltonian interaction, which acts upon the wavefunctions.
Consequently, the first assumption is not based upon the
\begin{quote}local interaction of the atom with the \ldots
resonator, described by standard quantum theory with its short
range interactions only\end{quote}
In \cite{Scu98}, it is stated
that
\begin{quote}the photon emission process is always
(physically and calculationally) driven locally by the action of
the cavity field on the atom\end{quote}
While the emission process
can be said to be {\em calculationally} driven by the local
Hamiltonian acting upon the wavefunction, to say that it is also
{\em physically} local is to attribute reality to something deeper
than the quantum level of description. The assumption that finding
the photon in one cavity implies the atom actually passed through
that cavity is an addition to 'standard' quantum theory.

In \cite{Scu98}, this is made particularly clear. To defend his
interpretation of the experiment, Scully wishes to rule out the
transfer of the photon from one cavity to the other, as the atom
traverses the interference region. He argues that the transfer of
the photon from one micromaser cavity to the other, in the Bohm
approach, represents a teleportation of energy. This teleportation
of energy is 'qualitatively different' and a 'stronger type' of
non-locality to that found in EPR
correlations\footnote{\cite{SZ97,Scu98} appears to state that EPR
correlations can be attributed to 'common cause' and there is
'nothing really shockingly non-local here'. It is precisely
because EPR correlations violate the Bell inequalities that this
point of view encounters considerable
difficulties\cite{Red87,Bel87}.}.

However, the non-locality of entangled photon states in micromaser
cavities has been studied and has even been suggested to be used
in quantum teleportation experiments\cite{BDHMR93,BDHRZ94,CP94}.
In Appendix \ref{ap:telep} and \cite{HM99} we can see that the
welcher-weg interferometer involves exactly the same processes as
in EPR entanglement and quantum teleportation, whether one uses
the Bohm interpretation or 'standard' quantum mechanics.
Consequently, Scully's argument that finding the photon in the
cavity after the interference region has been passed implies that
the photon {\em must} have been in the cavity before the
interference region was encountered is, again, an argument that is
not part of standard quantum mechanics, and rests upon the
assumptions above.

\paragraph{Actual paths of atoms}
The second assumption is necessary to understand the use of the
term 'information'. If the welcher-weg device stores information
about the actual path of atom, this implies that there is a matter
of fact about which path the atom actually takes. The erasure of
such information would simply affect our, real or potential,
knowledge of which path the atom took, but would not affect the
actual reality of which path the atom took.

Can we deny this point without losing the interpretation of the
welcher-weg devices as reliable measuring devices? It would seem
not, as if we do deny this we find ourselves contradicting the
first assumption. Suppose we interpret the atom having a path only
in the experiments where the probe atoms are not pulsed, but not
having a path when the probe atoms are pulsed (and interference is
observed).  The problem lies in the fact that the cavities are
themselves simply two level quantum systems. The location of the
photon in the cavity, which is taken to represent the information
about the path the atom travelled, is a quantum state of the
optical field. If there is no matter of fact about whether the
atom is taking one path or the other, before the measurement is
performed, there is equally no matter of fact about which cavity
contains the photon. The interaction of the atom with the cavity
does not create a matter of fact about whether the atom took one
path or the other, so cannot be said to represent a measurement of
the atoms location.

So when would the measurement take place that determines whether
there is a matter of fact about the path of the atom? The answer
is only when the probe atom is ionized. In other words, when a
{\em conventional} quantum measurement takes place. It is not the
welcher-weg devices that are measuring the path of the atom at
all. There is no matter of fact about whether the atom travelled
down one path or the other, or any matter of fact about which
cavity contains the photon, until the probe atom is ionized, which
cannot take place until {\em after} the interference region has
been traversed.

It is in the interference region that the atom changes wavepackets
and the excitation of the cavity modes switches from one cavity to
the other in the Bohm interpretation. In other words, if we deny
the second assumption, the 'surrealistic' behaviour of the Bohm
trajectories will take place only if there is no matter of fact
about which path the atom took and which cavity contains the
photon. In which case we cannot conclude that the Bohm
trajectories are at variance with the actual path taken by the
atom, as it is not meaningful to talk about the actual path of the
atom. Without the second assumption the addition of the
welcher-weg devices to Wheeler's delayed choice experiment has had
no effect on it's interpretation.

This demonstrates that these two assumptions are essential to the
interpretation \cite{ESSW92} wish to place upon the welcher-weg
devices, and further that neither assumption can be considered
part of 'standard' quantum theory.

\paragraph{Phase coherence}
As we have seen, to contradict the Bohm trajectories it is
essential that the welcher-weg devices maintain phase coherence in
the entangled superposition. However, this allows us to display
interference effects in the location of the atom that depend upon
the settings of phase shifters in both arms of the interferometer.
Such a result seems to undermine both of these assumptions
necessary for \cite{ESSW92}'s interpretation of the welcher-weg
devices.

We can emphasise this by removing the phase shifter from one arm
and the cavity from the other. Firstly, let us consider the
results of ionizing an unpulsed probe atom. If the unpulsed probe
atom is measured to be in the excited state, we would assume that
the atom passed down the arm of the interferometer containing the
cavity, while if the probe atom is measured in the unexcited
state, we would assume that the atom passed down the other arm.
These would each occur with a 50\% probability. In other words,
half of the atoms could not have interacted with the phase
shifter, and the other half could not have interacted with the
cavity.

Now let us consider what happens if we pulse the probe atom. We
separate the pattern the atom makes upon the screen in the
interference region $R$ into subensembles based upon the outcome
of the ionized probe atom measurement. These subensembles each
display the full interference pattern, the location of whose
maxima and minima are determined by the phase shifter. Now, if we
are to assume that the atom did, in fact, travel down only one
path or the other, and could only interact with the device in the
path it travelled through we cannot consistently interpret these
results.

Consider the atom that hypothetically travelled down the arm with
the cavity. This deposited a photon in the cavity, and encountered
the screen. Neither cavity nor atom interact locally with the
phase shifter. However if we pulse the probe atom, before
ionization, the location of the atom in the interference region
shows fringes which depend upon the setting of the phase shifter,
which neither atom nor cavity interacted with.

If we consider the atom that hypothetically travels down the arm
with the phase shifter, we find the situation even worse. Now the
cavity does not interact with the atom and is left empty. If we
send the probe atom through this empty cavity, then pulse and
ionize it, the result of this ionization is to produce
interference patterns, with minima at different locations. If the
cavity never interacted with the atom, how can the result of
measuring the probe atom possibly be correlated to the location of
the forbidden zones in the interference patterns?

\subsection{Conclusion}

It seems to consistently interpret these results we must either
abandon the notion that there is a matter of fact about which path
the atom takes or abandon the idea that the atom can only interact
with the cavity (or phase shifter) if it actually passes down the
same arm of the interferometer. If either of these concepts are
abandoned, however, the interpretation \cite{ESSW92} place upon
the welcher-weg devices is untenable. We are therefore forced to
conclude that the welcher-weg devices do not have the properties
necessary to be interpreted as detectors.

If we abandon the second assumption, and we apply the information
term (\ref{eq:wzinf}) strictly to the outcomes of experiments, we
can make no inference at all about the actual path taken by the
atom. This takes us to the interpretation urged by
Bohr\cite{Boh58} and to 'standard' quantum theory. Here only the
outcomes of macroscopic measurements can be meaningfully
discussed. The macroscopic phenomena emerges, but cannot be
interpreted in terms of microscopic processes. In the case of the
experiments above, the interference effects are predicted by the
quantum algorithm, but no explanation is offered, nor can be
expected, as to how they arise. In particular, the single mode
cavities are normal quantum devices, and so cannot be interpreted
as reliable measuring devices.

If we abandon the first assumption, how do we understand an atom
travelling down one path, but acting as if it travels down both?
We can interpret this in terms of the active information in the
Bohm approach.  A trajectory travels down one path, but a
wavepacket travels down both paths. The wavepackets interact with
the cavity or phase shifter, according to the local Hamiltonian,
regardless of which path the atomic trajectory actually takes.

Now the entangled state means that the information on the setting
of the phase shifter is part of the common pool of information
that guides both the atomic trajectory and the cavity field mode.
When the atom enters the interference region, all the branches of
the superposition become active. The behaviour of the atom is now
being guided by the information from both wavepackets and so can
be influenced by the phase information from both arms of the
interferometer. However, the field modes are also being guided by
this common pool of information.

If the atom encounters the screen at some location $x$ in the
interference region, this is amplified in some, practically
irreversible process, that renders all the other information in
the entangled quantum state inactive. The non-local quantum
potential connects the motion of the atomic trajectory to the
motion of the cavity field mode, so now the excitation of the
cavity field is correlated to the position at which the atom was
detected. If the atom is detected at the specific location $X$,
the active wavefunction for the cavity field modes is now
proportional to
\[
\psi_u(X)\ket{1_u,0_d}+\psi_d(X)\ket{0_u,1_d}
\]
where $\psi_u(X)$ and $\psi_d(X)$ are just the complex numbers
corresponding to the probability amplitudes for the actually
detected location of the atom at $X$. This demonstrates how the
information active upon the cavity field modes is correlated to
the measured location of the atom through the non-locality of the
quantum potential.

When the probe atom is sent through the cavity, and pulsed, this
can be rewritten as
\begin{eqnarray*}
&&\left(e^{\imath \phi_u}\psi_u(X,t_3)
        -e^{\imath \phi_d}\psi_d(X,t_3)
    \right)\frac{\ket{e_{P_u},g_{P_d}}-\ket{g_{P_u},e_{P_d}}}{2} \\
&&+\imath \left(e^{\imath \phi_u}\psi_u(X,t_3)
        +e^{\imath \phi_d}\psi_d(X,t_3)
    \right)\frac{\ket{g_{P_u},g_{P_d}}+\ket{e_{P_u},e_{P_d}}}{2}
\end{eqnarray*}
The probabilities of detection of the states of the probe atoms
are therefore
\begin{eqnarray*}
\ket{e_{P_u},g_{P_d}},\ket{g_{P_u},e_{P_d}} & \Rightarrow &
    \frac{\magn{e^{\imath \phi_u}\psi_u(X,t_3)
        -e^{\imath \phi_d}\psi_d(X,t_3)}}{R(X,t_3)^2} \\
\ket{g_{P_u},g_{P_d}},\ket{e_{P_u},e_{P_d}} & \Rightarrow &
    \frac{\magn{e^{\imath \phi_u}\psi_u(X,t_3)
        +e^{\imath \phi_d}\psi_d(X,t_3)}}{R(X,t_3)^2}
\end{eqnarray*}
We can express this as the conditional probabilities
\begin{eqnarray*}
P(ee,gg|X)=\frac12 \left(1+\cos\left(\Delta S(X,t_3)
    +(\phi_u-\phi_d)\right)\right) \\
P(eg,ge|X)=\frac12 \left(1-\cos\left(\Delta S(X,t_3)
    +(\phi_u-\phi_d)\right)\right)
\end{eqnarray*}
Correlating the ionisation state back to the location of the atom,
using Bayes's rule, reveals the interference fringes
\begin{eqnarray*}
P(X|ee,gg)=R(X,t_3)^2 \left(1+\cos\left(\Delta S(X,t_3)
    +(\phi_u-\phi_d)\right)\right) \\
P(X|eg,ge)=R(X,t_3)^2 \left(1-\cos\left(\Delta S(X,t_3)
    +(\phi_u-\phi_d)\right)\right)
\end{eqnarray*}
The interference exists as a correlation between the entangled
systems. It is usual to regard this as the probe atom ionization
leading to the selection of subensembles of the atomic position
which display interference. As we can see here, we may equally
well have regarded the location of the atom on the screen as
selecting interference subensembles in the ionization of the probe
atom. The phase shifts, $\phi_u$ and $\phi_d$, do not act upon a
single subsystem, rather they form part of the common pool of
information which guides the joint behaviour of both systems.

\paragraph{Information}
We can modify this to produce a POVM measure of the which-path
information suggested by Wootters and Zurek. Suppose that the
resonance between the atomic beam and the cavities are adjusted,
by speeding up the atoms. The transition is no longer
\[
\ket{e0} \rightarrow \ket{g1}
\]
but becomes
\[
\ket{e0} \rightarrow \alpha \ket{g1} + \beta \ket{e0}
\]
We then send the probe atoms through the cavities, and ionise them
{\em while the atomic wavepacket is still in the interferometer}.
The ionisation of the probe atom can now represent a measurement
of the atom's location. The POVM is
\begin{eqnarray*}
A_u&=& \frac12 \magn{\alpha}\proj{\phi_u} \\
A_d&=& \frac12 \magn{\alpha}\proj{\phi_d} \\
A_0&=& \magn{\beta}\mathbf{I}
\end{eqnarray*}

If we represent the location of the Bohm trajectory in the
u-branch by $X_u$ and in the d-branch by $X_d$, then the initial
probabilities are
\begin{eqnarray*}
P(X_u)&=&\frac12 \\
P(X_d)&=&\frac12
\end{eqnarray*}
giving an initial information of $I(X)=1$. The probability of the
measurement outcomes are
\begin{eqnarray*}
P(u)&=& \frac12 \magn{\alpha} \\
P(d)&=& \frac12 \magn{\alpha} \\
P(0)&=& \magn{\beta}
\end{eqnarray*}
where $P(u)$ is the probability of the u-probe atom ionising,
$P(d)$ the d-probe atom ionising, and $P(0)$ neither ionising.

If either probe atom ionises, the wavepacket in the other branch
is deactivated and the correlated ensemble of atoms in the region
$R$ displays no interference. If neither ionises, both wavepackets
become active again and a full interference pattern occurs. The
total pattern is
\[
R(X,t_3)^2\left(1+\magn{\beta}
    \cos\left(\Delta S(X,t_3)+(\phi_u-\phi_d)\right)\right)
\]
The conditional probabilities after the measurement are
\begin{eqnarray*}
P(X_u|u)&=& 1 \\
P(X_d|d)&=& 1 \\
P(X_u|0)&=& \frac12 \\
P(X_d|0)&=& \frac12
\end{eqnarray*}
so the conditional information on the path $(X)$ taken by the atom
after the measurement $(M)$ is
\[
I(X|M)= \magn{\beta}
\]
which represents the remaining ignorance of the path taken. The
gain in information is
\[
I(X:M)= \magn{\alpha}
\]
The size of the interference fringes are given by
$\magn{\beta}=1-\magn{\alpha}$. As we gain more information about
the path, we reduce the size of the interference pattern.

The concept of active information, in the Bohm interpretation,
thus provides a natural way to understand the interference effects
in the experiments considered.

\section{Conclusion}.

We have considered in detail the relationship between information
and interference proposed in a series of thought experiments. We
have found that the concept of 'information' being used, although
quantified by a Shannon information term (\ref{eq:wzinf}) is not
the same as information used in the sense of Chapter
\ref{ch:info}. Shannon information represents a state of ignorance
about an actual state of affairs. The measurement in a quantum
system cannot, in standard quantum theory, be interpreted as
revealing a pre-existing state of affairs. If we can interpret the
term $I_{WZ}$ at all, in standard quantum theory, it is as our
ignorance of the outcome of a particular measurement. It cannot be
used to make inference about the existence of actual properties of
quantum objects.

The measurements that must be used, in standard quantum theory,
involve macroscopic devices, for which the phase coherence between
the different measurement outcomes is, for all practical purposes,
destroyed. This allows us to replace the entangled pure state with
a statistical density matrix, without in any way affecting the
future behaviour of the system. The welcher-weg devices suggested
by \cite{ESSW92,SZ97} do not have this essential feature. It is
entirely because they do not have this feature that they produce
the effects in the quantum eraser experiments\cite{ESW91} and that
appear to contradict the Bohm trajectories. However, the
interpretation \cite{ESW91,ESSW92,SZ97} placed upon the
welcher-weg devices is not consistent with standard quantum
theory, precisely because they lack this feature, and it seems
difficult how this interpretation can be sustained.

The concept of active information, by contrast, provides a natural
way of interpreting these results. If we measure the path taken by
the trajectory, we render the information in the other wavepacket
inactive, because of the superorthogonality of the measuring
device states. When the atom encounters the interference region it
is guided only by the information in the one wavepacket, and so
cannot display interference effects that depend upon phase
differences between both branches of the superposition. If we do
not measure the path taken, then both wavepackets are active when
the interference region is encountered, and the atomic trajectory
is guided by information from both arms of the interferometer.

Active information is clearly different from that given by
$I_{WZ}$. Here we are not talking about our ignorance of a
particular state of affairs ('information-for-us'), but rather a
dynamic principle of how the experimental configuration acts upon
the constituent parts of the quantum system ('informing the
behaviour of the object'). Nevertheless, it connects to our
measurements as, when we gather information-for-us from a
measurement, the dynamic information in the other wavepackets
becomes inactive. This explains why, in the interference
experiments, as we increase our 'information-for-us' about the
path measurements, we increase the {\em deactivation} of the
information about the phase shifts in the arms of the
interferometer, and this leads to the attenuation of the
interference fringes. The Bohm interpretation provides a coherent
means of understanding the information-interference
complementarity in experiments such as\cite{WZ79}, while
welcher-weg devices do not.
\chapter{Entropy and Szilard's Engine} \label{ch:szmd}

In this part of the thesis we will examine the role of information
in thermodynamics. We will be particularly interested in the
quantitative connections suggested between the Shannon/Schumacher
measure of information and the thermodynamic entropy. This will
require us to analyse in detail the quantum mechanical version of
Szilard's thought experiment \cite{Szi29} relating entropy to the
information gained by a measurement. This thought experiment has
been made the paradigm argument to demonstrate the information
theoretic explanation of entropy\cite[for example]{LR90} but it
continues to be strongly criticised\cite{BS95,EN98,EN99,She99b}.

The structure of this is as follows:
\begin{itemize}
\item  Chapter \ref{ch:szmd} will review the attempts that have been made
to make a quantitative link between information and entropy, based
upon Maxwell's Demon and the Szilard Engine. This will be in some
detail, in order to clarify the points that are at issue, and to
motivate the analysis in subsequent Chapters. This will allow us
to construct a modified, and quantum mechanical, form of the
"demonless" Szilard Engine, which will be used to examine the
validity of the various 'resolutions'.

\item In Chapter \ref{ch:szqm} we will make a careful and detailed
description of the quantum mechanical operation of all stages of
the Szilard Engine. The only physical restriction we place upon
this Engine is that it must be consistent with a unitary time
evolution.

\item Chapter \ref{ch:szsm} adds statistical mechanics to the
microscopic motion, by introducing canonical heat baths and
ensembles. No other thermodynamic concepts (such as entropy or
free energy) will be used at this stage. The behaviour of the
Engine will then be shown to quite consistent with the statistical
mechanical second law of thermodynamics.

\item Thermodynamic concepts are introduced and justified in
Chapter \ref{ch:szth}. It will be shown that the entropy of the
Szilard Engine never decreases. In Chapter \ref{ch:szsol} the
behaviour of the Engine is generalised to give a complete
explanation of why Maxwell's Demon cannot produce anti-entropic
behaviour. We then show how the other resolutions suggested, where
they are correct, are contained within our analysis.

\end {itemize}

Our analysis will show that both the information theoretic
resolution, and it's criticisms, are incomplete, each
concentrating on only part of the problem. When we complete this
analysis, we will show that, despite the formal similarity between
Shannon/Schumacher information and Gibbs/Von Neumann entropy,
information theory is both unnecessary and insufficient to give a
complete resolution of the issues raised by the Szilard Engine.

We will now consider the general arguments for a relationship
between entropy and information. Section \ref{s:szmd1} will review
one of the issues raised by statistical mechanics, and why this
may be taken to identify entropy with information. Section
\ref{s:szmd2} then considers the Szilard Engine version of
Maxwell's demon. This has been used as the paradigm thought
experiment to demonstrate the relationship between the entropy of
a system and the information gained from performing measurements
on the system. The final Subsection will consider a 'demonless'
version of the thought experiment, used to deny the role of
information in understanding the problem. Finally, in Section
\ref{s:szmd3} we review what we believe are the key points of
contention in Section \ref{s:szmd2}, and how we propose to address
them in Chapters \ref{ch:szqm} to \ref{ch:szsol}.

\section{Statistical Entropy} \label{s:szmd1}

The attempts to derive the phenomenological laws of thermodynamics
from classical mechanics lead to the identification of entropy
with a statistical property of a system, rather than an intrinsic
property. Unlike other intensive thermodynamic variables (such as
mass or energy) the statistical entropy is not expressed as the
average over some property of the microstates, but is a property
of the averaging process itself.  The unfortunate consequence of
this is that there may not appear to be a well-defined entropy of
an individual system. So, the Boltzmann entropy of a microstate
$S_B=klnW$ depends upon a particular (and possibly arbitrary)
partitioning of phase space, while the Gibbs entropy $S_G=-k\int
p\ln p$ depends upon the inclusion of the microstate in a
'representative' (and possibly arbitrary) ensemble. If we were to
choose to describe the partition of phase space differently, or
include the same microstate in a different ensemble, we would
ascribe a different entropy to it.

Attempting to understand how something as fundamental as entropy
could be so apparently arbitrary has lead many to suggest that
entropy, and it's increase, represents a measure of our ignorance
about the exact microstate of the individual system:

\begin{quote}
the idea of dissipation of energy depends on the extent of our
knowledge \ldots [it] is not a property of things in themselves,
but only in relation to the mind which perceives them\cite[pg 3,
quoting Maxwell]{DD85}
\end{quote}

\begin{quote}
irreversibility is a consequence of the explicit introduction of
ignorance into the fundamental laws \cite{Bor49}
\end{quote}

\begin{quote}
The entropy of a thermodynamic system is a measure of the degree
of ignorance of a person whose sole knowledge about its microstate
consists of the values of the macroscopic quantities \ldots which
define its thermodynamic state \cite{Jay79}
\end{quote}

\begin{quote}
What has happened, and this is very subtle, is that my knowledge
of the possible locations of the molecule has changed \ldots the
less information we have about a state, the higher the entropy
\cite{Fey99a}
\end{quote}

How this ignorance arises, whether it is a subjective or objective
property, and why or how it increases with time have been argued
in many ways. For example, it is often suggested that the
ignorance arises because of the large number of microstates
available to macroscopic bodies, and the difficulty of physically
determining exactly which microstate the body is in. Similarly,
the growth of entropy with time is then identified with the
difficulty of following the exact trajectories of a large number
of interacting bodies.

A frequent criticism that is raised against this interpretation is
that it seems to be implying that the large number of irreversible
processes that surround us (gas diffuses, ice melts, the Sun
shines) are illusory and occur only because of our lack of
detailed knowledge of the exact microstate of the gas, ice cube,
or star:
\begin{quote}
it is clearly absurd to believe that pennies fall or molecules
collide in a random fashion {\em because we do not know} the
initial conditions, and that they would do otherwise if some demon
were to give their secrets away to us \cite{Pop56}
\end{quote}

The discussions and criticisms of this point of view is too large
to fully review here
\cite{Pop57,Pop74,LT79,DD85,LR90,Red95a,Bri96}. Nor will we be
dealing with the problem of the origin of irreversibility
\cite{HMZ94,Alb94,Uff01}. Instead we will concentrate on a
quantitative link between knowledge (information) and entropy. In
particular we will be considering the issues raised by the
following problem:
\begin{quote} If entropy is a measure of ignorance, and
information is a measure of lack of ignorance, how is it that
entropy increases with time, while our information, or knowledge,
also increases with time?\end{quote}

If we cannot follow the exact microstates of a system, it may
appear that our information about the system is decreasing. The
knowledge we have about a system, at some given point in time,
{\em when defined in terms of coarse-grained 'observational
states'}\cite{Pen70}, will provide less and less information about
the system as time progresses, due to coarse-grained 'mixing'.
This decrease in information will be identical (with a sign
change) to the increase in the {\em coarse-grained} entropy of the
system.

On the other hand, the problem arises as we are constantly
increasing our knowledge, or information, by observing the world
around us. Each observation we make provides us with new
information that we did not possess at the earlier time. Does this
process of acquiring {\em new} information reduce the entropy of
the world, and should this be regarded as an apparent violation of
the second law of thermodynamics? This is the key paradox which
needs to be investigated.

We will quantify our knowledge by using the Shannon-Schumacher
measure of information obtained from measurements we perform. The
Gibbs-von Neumann entropy is identical in form to this measure,
and so will be used for the thermodynamic entropy (we will avoid
using 'coarse-grained' entropy as we will be dealing with
microscopic systems for which 'observational states' cannot be
sensibly defined). We now need to consider how the gain in
information from a measurement can be related to the change in
entropy of the system that is measured.

\section{Maxwell's Demon} \label{s:szmd2}

 When we measure a system, we only gain information about it if
it was possible for the measurement to have had several different
outcomes. In the case of a thermodynamic ensemble, the measurement
amounts to the selection of subensembles. The potentially
anti-entropic nature of such a selection was first suggested by
Maxwell\cite[and references therein]{LR90} when he proposed a
sorting demon that would, by opening and closing a shutter at
appropriate times, allow a temperature difference to develop
between two boxes containing gases initially at the same
temperature. Once such a temperature difference develops heat can
be allowed to flow back from the hotter to the colder, via a
Carnot cycle, turning some of it into work in the process. As
energy is extracted from the system, in the form of work, the two
gases will cool down. The result would be in violation of the
Kelvin statement of the second law of thermodynamics:
\begin{quote}
No process is possible whose sole result is the complete
conversion of heat into work.
\end{quote}

There have been many variations upon this theme, and attempts to
resolve the apparent 'paradox'. 'Demonless' versions like
Smoluchowski's trapdoor, or Feynman's ratchet \cite{Fey63}
emphasise the manner in which thermal fluctuations develop in the
physical mechanism designed to effect the sorting, and prevent the
mechanism from operating. A quite different approach was started
by Szilard\cite{Szi29} which will concern us here.

The Szilard Engine (Figure \ref{fg:szeng}) consists of a single
atom (G) confined within a tube of volume $V$. The tube is in
constant contact with a heat bath at temperature $T_G$, providing
a source of energy for the random, thermal kinetic motion of the
atom. At some point a piston (P) is inserted in the center of the
tube, trapping the atom upon one side or the other, or confining
it to a volume $V/2$. If we now attach a pulley and weight (W) to
the piston, we may use the collision of the atom against the
piston to assist us in moving the piston and lifting the weight.
\pict{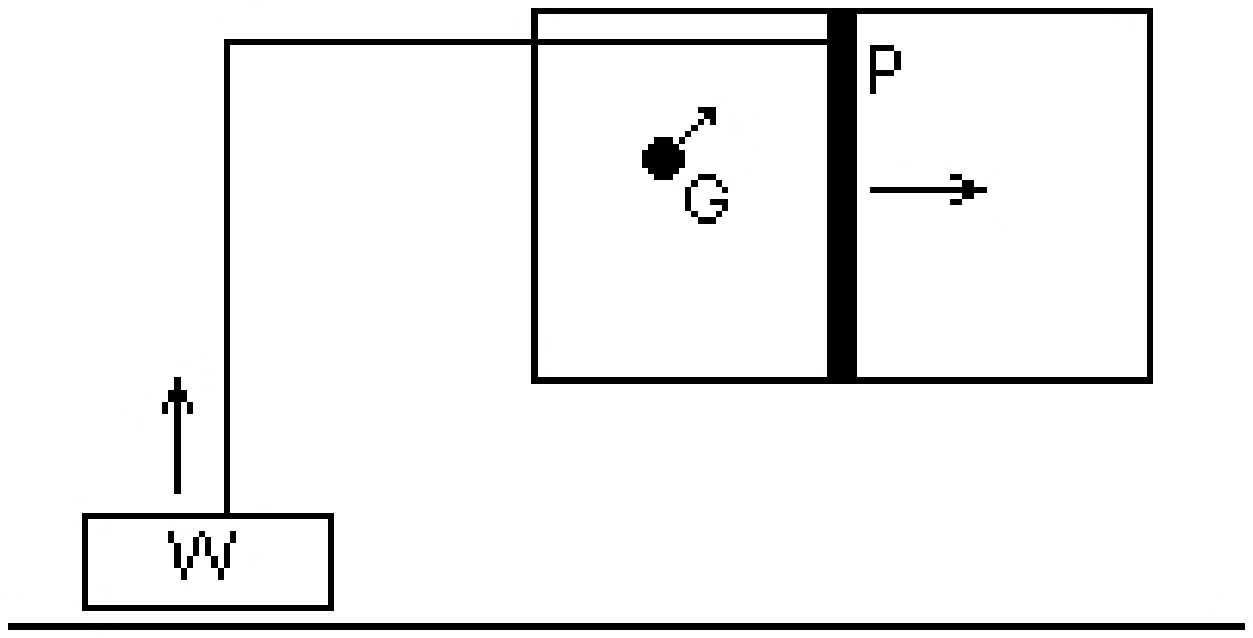}{The Szilard Engine}
 If we consider this as the
expansion of a gas from a volume $V/2$ to $V$ then the isothermal
work which may be extracted in this manner is $k T_G \ln 2$. At
the end of the procedure the atom again occupies the full volume
of the tube $V$ and the piston may be reinserted into the center.
It appears we have extracted work from heat, in violation of the
second law of thermodynamics. This is the essence of the Szilard
Paradox.

Szilard argued that the problem lay in determining upon which side
of the piston the atom was located. Without this information, the
pulley and weight cannot be connected up to the piston in the
correct manner. Having eliminated all other sources of a
compensating entropy increase, he concluded that the act of making
a measurement must be responsible for an increase in entropy. Thus
a 'demon' cannot decrease the entropy of a system by acquiring
information about it, without creating at least as much entropy
when performing the measurement necessary to acquire the
information.

We will now examine the developments of Szilard's idea, and their
criticisms.

\subsection{Information Acquisition}
The next major development of Szilard's
argument\cite{Bri51,Gab51,Bri56}(referred to as [GB]) tried to
quantify the link between the information gained from a
measurement and the entropy decrease implied by that measurement.
The essence of their development was to demonstrate situations in
which the process of acquiring information required a dissipation
of energy. The amount of this dissipation more than offset any
gain in energy that could be achieved by decreasing the entropy of
the system.

 Although [GB]'s arguments are no longer supported by
the main proponents of an information-entropy link, their physical
models are (rather ironically) often still supported by opponents
of that link \cite[for example]{DD85,EN99} so we will need to give
consideration to them here.

[GB] were able to make a quantitative statement of the information
gained from a measurement based upon Shannon's work. They then
went on to produce models to show that at least as much entropy
was created by the physical process by which the information was
acquired. Their analysis was based upon the need for the 'demon'
to {\em see} the location of the atom, and that this required the
atom to scatter at least one photon of light. The tube containing
the atom at temperature $T_G$ would, in thermal equilibrium,
already be filled with photons with a blackbody spectrum. In order
to locate the atom accurately, the scattered photon must be
reliably distinguishable from a photon whose source was the
blackbody radiation. This requires the photon to be of a frequency
$\hbar \omega \gg kT_G$. Brillouin later refined this to argue
that the minimum frequency for a 50\% reliable observation was
given by $\hbar \omega = kT_G \ln 2$. This photon would be
absorbed by a photo-detector, and the energy in the photon would
be lost. This would represent an increase in entropy of the
environment of $\frac{\hbar \omega}{T_G} \geq k \ln 2$ which
compensates for the entropy decrease in the state of the one atom
gas.

Both Gabor and Brillouin generalised from this basic result to
claim that any measurement that yielded information would require
a dissipation of energy, with an entropy increase at least as
large as the information gained. Brillouin, in particular
developed a theory of information as negentropy \cite{Bri56},
essentially based upon the equivalence of the Shannon and Gibbs
formula.

However, it is easy to argue that this equivalence can be ignored,
and with it the information link, and instead concentrate upon the
physical process involved. We note there are two steps in the
above argument: firstly that an information increase occurs when
an entropy decrease occurs; and secondly that this information
increase requires an entropy expenditure. Given the identical form
of the Shannon and Gibbs formulas, this first step may be regarded
as an almost trivial relabelling exercise. If we dispense with
this relabelling as superfluous, we are still left with the second
step, now as an argument that the entropy reducing measurement
must involve entropy increasing dissipation, without reference to
information at all. This approach is essentially that advocated by
\cite[Section 5.4]{DD85} and \cite[Appendix 1]{EN99}.

There are other criticisms of this resolution, however, that rest
upon the question of how universal the measurement procedure used
by [GB] is. We will examine these next, and will return to the
arguments of [GB] in Section \ref{s:szsol3}

\subsection{Information Erasure}

The current principal advocates of the Szilard Engine as the
paradigm of a quantitative information-entropy link no longer
accept the arguments of [GB]
\cite{Ben82,Zur84,Zur89a,Zur90a,Cav90,LR90,Cav93,LR94,Sch94,Lef95,Fey99a}.
Instead they focus upon the need to restore the Engine, and demon,
to their initial states to make a cyclic operation. This, they
argue, requires the demons memory to be 'erased' of the
information gained from the measurement, and that this erasure
requires the dissipation of energy.

The origin of the information erasure argument comes from work on
the thermodynamics of computation. The work of Gabor and Brillouin
was rapidly developed into an assumption that, for each logical
operation, such as the physical measurement or transmission of 1
bit of information, there was a minimum dissipation of $kT \ln 2$
energy. This was challenged by Landauer\cite{Lan61} who, analysing
the physical basis of computation, argued that most logical
operations can be performed reversibly and have no minimal
thermodynamic cost. The only operation which requires the
dissipation of energy is the {\em erasure} of a bit of
information, which loses $kT \ln 2$ energy. This has become known
as Landauer's Principle. Given the importance attached to this
principle, we shall now present a simplified version of Landauer's
argument (see also \cite{Lan92})
\begin{figure}[htb]
\resizebox{\textwidth}{!}
    {\includegraphics{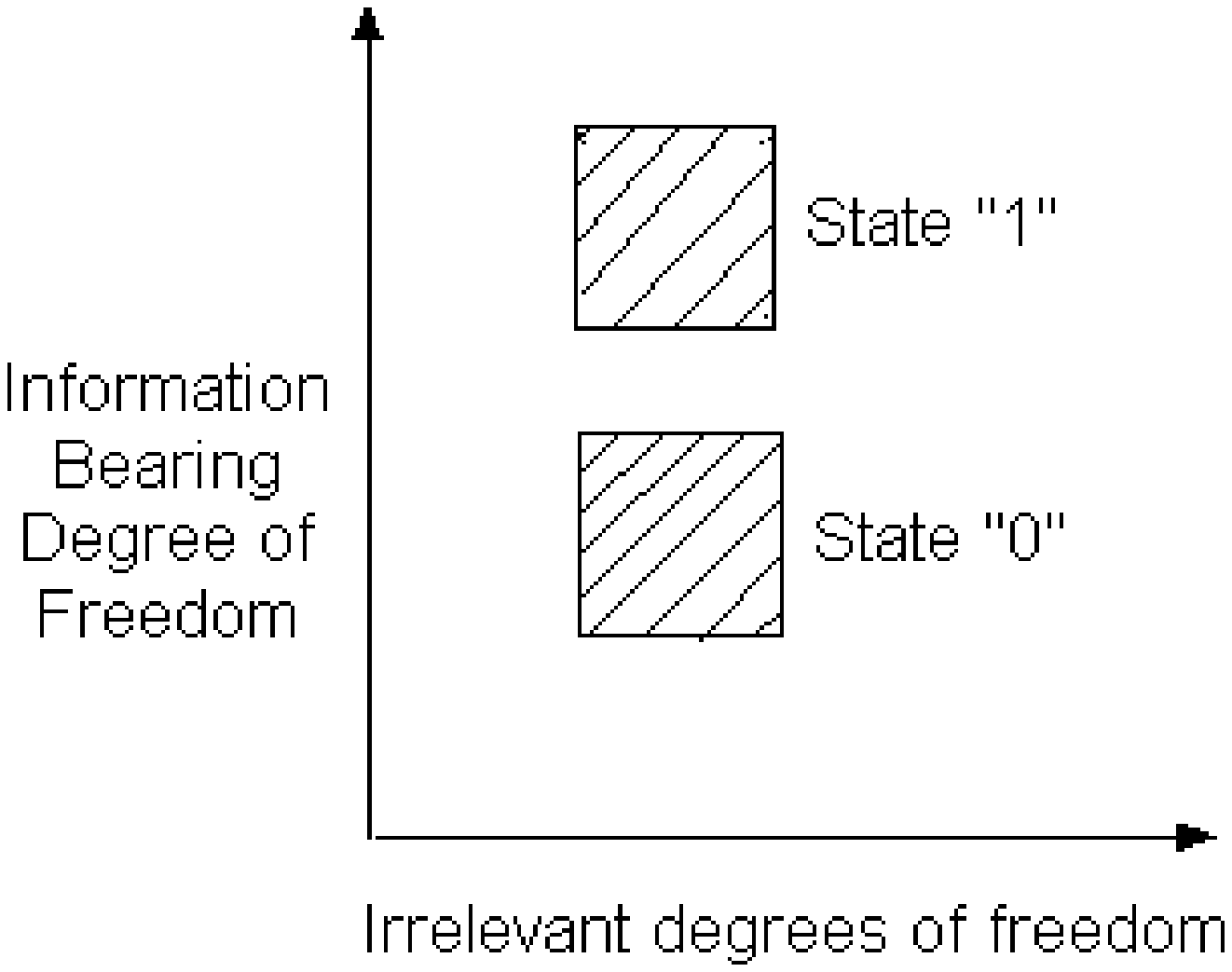}
    \includegraphics{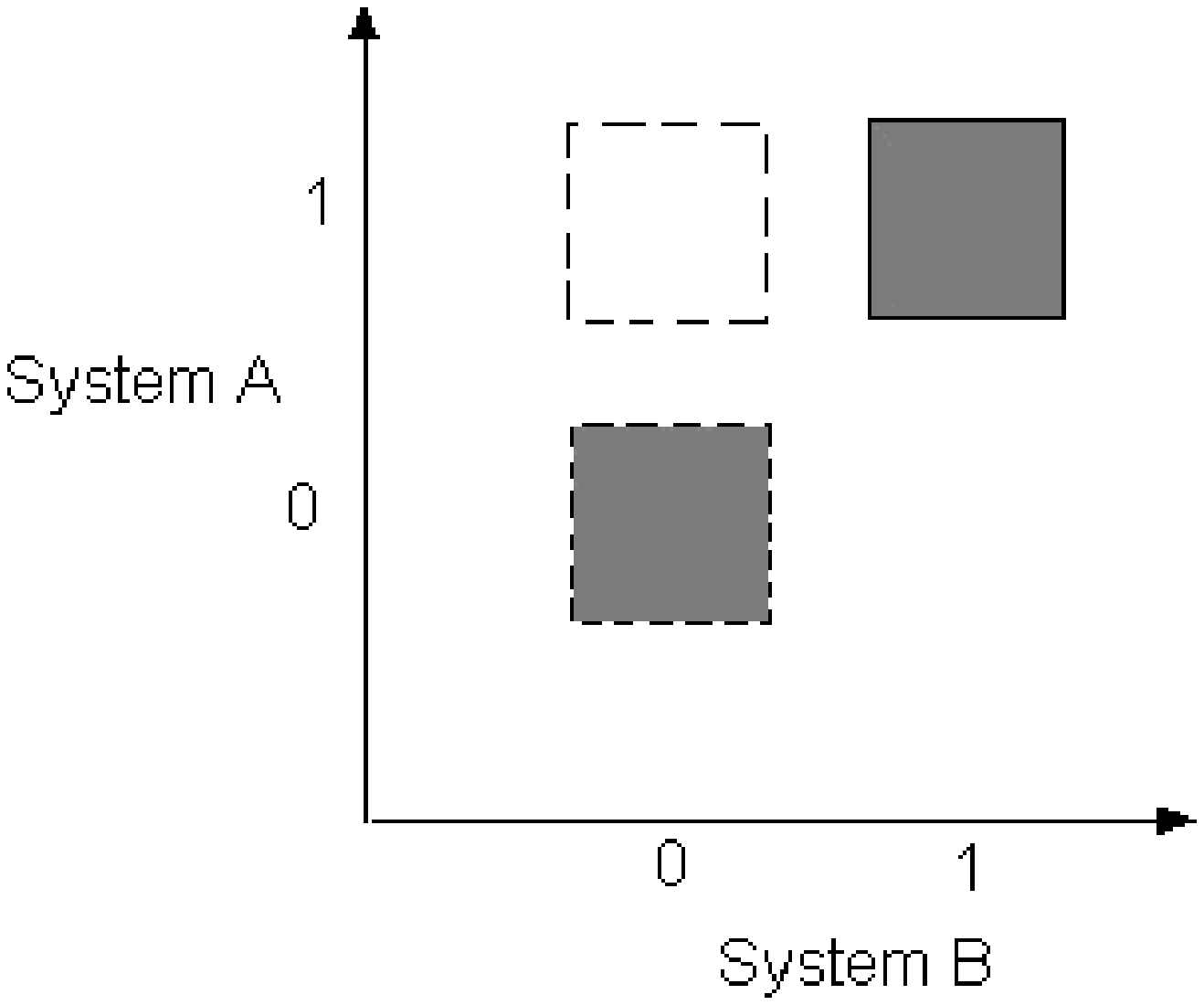}}
\caption{Landauer Bit and Logical Measurement\label{fg:ld1}}
\end{figure}
We shall assume that each logical state of a system has one
relevant (information bearing) degree of freedom, and possibly
many irrelevant (internal or environmental) degrees of freedom. We
will represent this by a diagram such as Figure \ref{fg:ld1}(a)
where the marked areas represent the area of phase space occupied
by the physical representation of the logical state.

A measurement can be represented, in logical terms, by a
Controlled-Not (CNOT) gate (Table \ref{tb:cnot}), where some
System B is required to measure the state of some System A. System
A is in one of two possible states, 0 or 1, while System B is
initially in the definite state 0 (represented by the areas
bounded by dotted lines in Figure \ref{fg:ld1}(b) - the
'irrelevant' degrees of freedom now occupying a third axis). After
System B interacts with System A, through a CNOT interaction, it
moves into the same state as A (the states of the two systems are
now represented by the shaded areas). System B has 'measured'
System A. The operation is completely reversible. If we allow the
systems to interact by the CNOT operation again, they return to
their initial states.

The essential point argued by Landauer is that both before and
after there are only two possible logical states of the combined
system, and the area of phase space occupied by the combined
system has not changed. As the entropy is a function of the
accessible area of phase space, then the entropy has not
increased. The operation is both logically {\em and}
thermodynamically reversible.

\begin{table}
    \begin{tabular}{cc||cc}
    \multicolumn{2}{c}{Input} &
    \multicolumn{2}{c}{Output} \\
    A & B & A & B \\
    \hline \hline
    0 & 0 & 0 & 0 \\
    0 & 1 & 0 & 1 \\
    1 & 0 & 1 & 1 \\
    1 & 1 & 1 & 0 \\
    \end{tabular}
    \caption{The Controlled Not Gate\label{tb:cnot}}
\end{table}

The development of [GB]'s work, to argue that each logical
operation required a minimal dissipation of energy, is shown to be
invalid. A measurement may be performed, and reversed, without any
dissipation. \pict{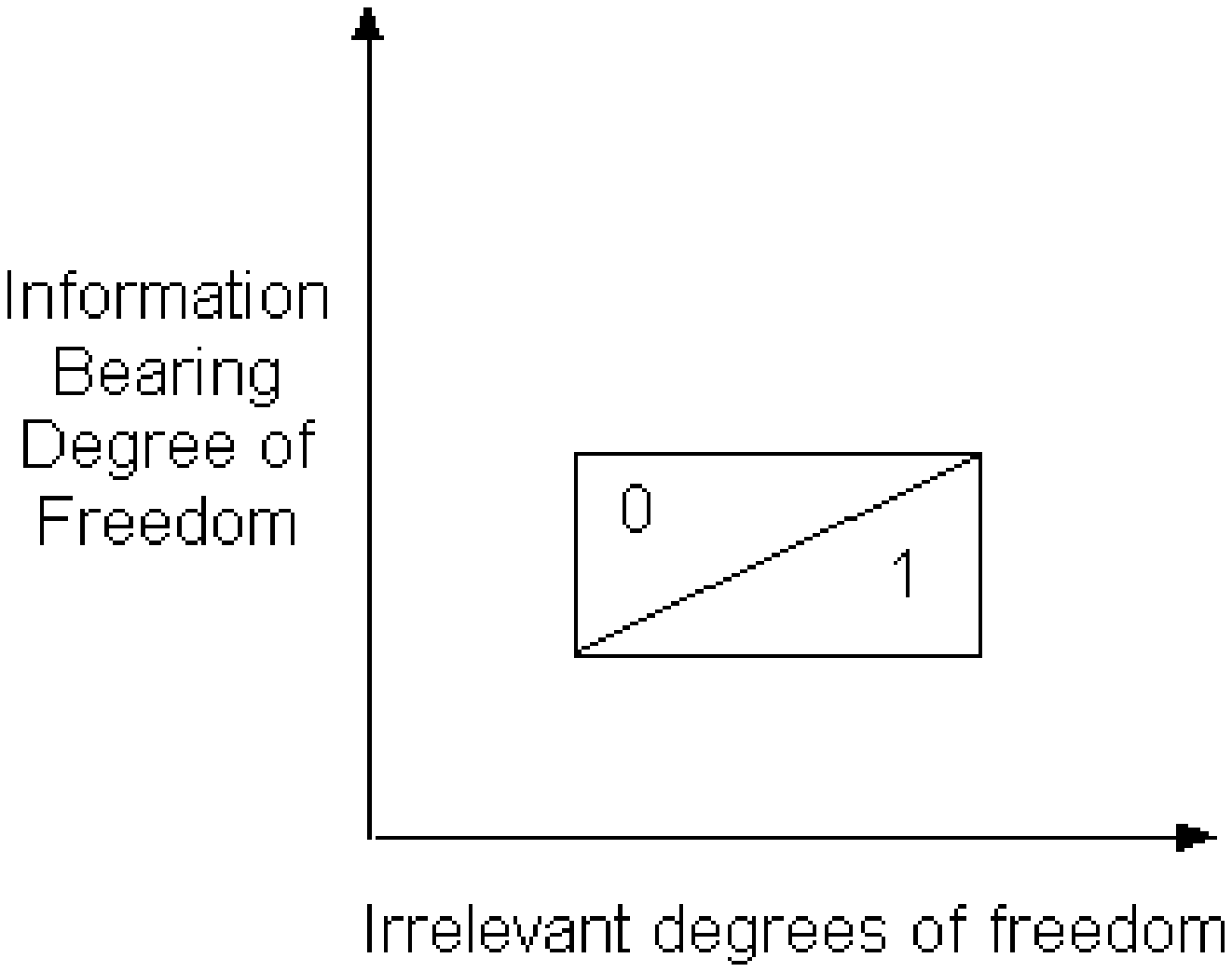}{Bit Erasure} Landauer did identify a
logical procedure which is inherently dissipative. This was called
RESTORE TO ZERO. This operation requires the logical bit,
initially in one of the two states as in Figure \ref{fg:ld1}(a),
to be set to the state zero, regardless of it's initial state,
leading to Figure \ref{fg:ld3}. The triangles represent the
location of the original microstate in Figure \ref{fg:ld1}. The
"width" of phase space occupied by the information bearing degree
of freedom has been reduced from the width of the 0 and 1 states
to the width of the 0 state. To satisfy Liouville's theorem, the
"width" occupied by the {\em non-information} bearing degrees of
freedom must be doubled. This amounts to the increase of entropy
of the environment by a factor of $k \ln 2$. If the environment is
a heat bath at temperature $T$, then we must dissipate at least
$kT \ln 2$ energy into the heat bath.

Landauer was not principally concerned with issues such as the
Szilard Engine, and it was left to Bennett\cite{Ben82} to
re-examine the exorcism of Maxwell's demon. Bennett's analysis
accepted that the Demon did not have to dissipate energy to
measure the location of the atom. Instead, he argues the demon has
acquired one bit of information, and that bit of information must
be stored somewhere in the demon's memory. After the demon has
extracted $kT_G \ln 2$ energy from the expansion of the one atom
gas, the demon is left with a memory register containing a record
of the operation. In order to complete the cycle, the demon's
memory must be restored to it's initial, clear, state. This
requires a RESTORE TO ZERO operation, which, by the Landauer
Principle, will dissipate $kT_G \ln 2$ energy. This exactly
compensates for the energy gained from the expansion of the gas. A
similar conclusion was reached by \cite[Chapter VI]{Pen70}.

This then forms the basis of the argument forging a quantitative
link between entropy and information theory. We will summarise it
as follows:
\begin{itemize}
    \item Entropy represents a state of ignorance about the actual state
    system;
    \item When an observer makes a measurement upon a system, she gains
    information about that system, and so reduces her ignorance;
    \item This does indeed reduce the entropy of the observed
    system, by an amount equal to the gain in Shannon information
    from the measurement;
    \item However, she must store this information in her memory;
    \item To perform this operation cyclically the memory must be
    erased;
    \item By Landauer's Principle, the erasure must dissipate
    energy equal to the temperature times the Shannon information
    erased, compensating for the entropy gain due to the measurement.
\end{itemize}

Perhaps the clearest problem in this 'resolution' of Maxwell's
Demon is the circularity of the argument. Landauer's Principle,
that the 'erasure' of a bit of information costs $kT \ln 2$
energy, was derived by Landauer {\em on the assumption that the
second law is true}. It's use by Bennett to prove that the second
law is {\em not} violated is appealing to the truth of the very
point which is in doubt. This is what Earman and Norton\cite{EN98}
refer to as the "sound vs. profound" dilemma of the information
theoretic resolution, and undermines confidence in its
universality.

We will now review the main counter-example to the
information-entropy link using Szilard's Engine.

\subsection{"Demonless" Szilard Engine}

In this Subsection we will examine the question, first raised by
Popper, of whether it is possible to construct a "demonless"
version of Szilard's Engine. The issues raised by this will form
the basis of the analysis of Szilard's Engine in the subsequent
Chapters.

The "demonless" Engine has been suggested many times by critics of
the information-entropy link\cite{Pop57,Fey66,JB72,Cha73,Pop74},
to demonstrate that a measurement is unnecessary to understand the
operation of the Engine. Unfortunately, both the consequence of
these modifications, and their criticism, have been poorly thought
out, and leave the question of a violation of the second law of
thermodynamics unanswered.

We will present a simple modification of the "demonless" Engine to
answer criticisms that have been made of this approach, and which
appears to lead to a systematic entropy reduction. The detailed
analysis of this version of the Engine, and showing where and how
it fails, will occupy the following three Chapters, and will be
used to critically examine the resolution of the Maxwell's Demon
problem. \pict{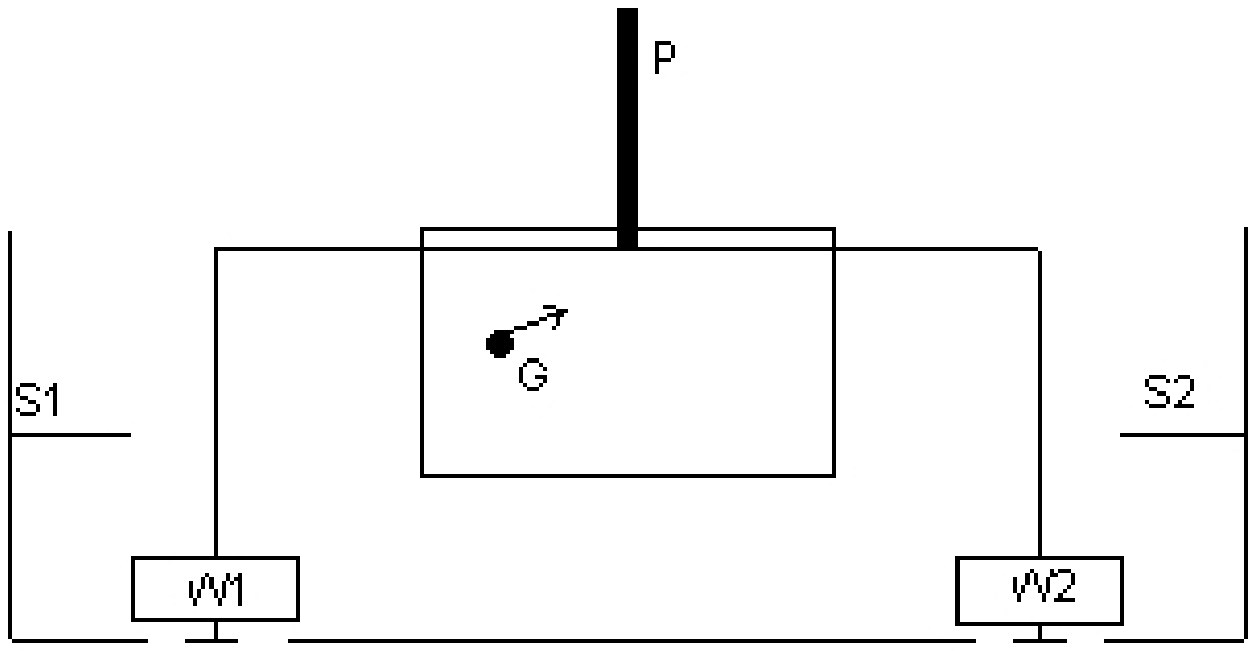}{The Popper version of Szilard's Engine} The
simplest version of the "demonless" engine is described by
Feyerabend\cite{Fey66} (Figure \ref{fg:szpop}). The essence of
this is that weights are attached on each side of the partition,
and rest upon a floor. If the atom, G, is located on the left when
the piston, P, is inserted, then the piston will move to the
right, raising the left weight, W1, and leaving the right weight,
W2, on the floor. If G is located to the right, then W2 will be
raised and W1 will remain upon the floor. The height that a weight
of mass M can be raised through is $\frac{kT_G}{Mg} \ln 2$. The
result is that heat has apparently been used to lift a weight
against gravity, without the need for a demon to perform a
measurement, dissipative or not.

It is very unclear whether this version should be taken as a
violation of the second law. Feyerabend certainly takes the
situation at face value and claims this is a perpetual motion
machine. Popper \cite{Pop74} argues that the machine works only
because it contains only a single atom, and that the atom only
occupies a small fraction of the volume of the cylinder at any one
time, so it's entropy is not increasing. Only if the gas were
composed of many atoms would it make sense to describe it as
expanding. Similarly, Chambadal\cite{Cha73} argues that
thermodynamic concepts are only applicable to many-body systems,
so that the Szilard Engine has nothing to do with entropy, and
Jauch and Baron\cite{JB72} claim the example is invalid because
inserting the partition violates the ideal gas laws
\footnote{Jauch and Baron earlier state that a demon is unable to
operate a Szilard Engine because of thermal fluctuations, but give
no explanation of how these thermal fluctuations enter into their
actual analysis of the Engine later}.

The logic of these arguments is hard to follow. They seem to
accept that heat can be used to lift a weight, and may continue to
do so, without any compensating dissipation. If this is the case,
the fact that a single atom gas has been used is irrelevant: the
Kelvin statement of the second law of thermodynamics has been
violated. The fact that the amount of energy obtained in this way
is small is also irrelevant. Advances in nanotechnology and
quantum computing develop technologies that allow the manipulation
of the states of individual atoms. It is conceivable that, in the
not-too-distant future, it would be possible to construct an
engine consisting of a macroscopically large number of microscopic
Popper-Szilard Engines. As long as each engine could reliably
transfer a small amount of heat to work per cycle, we would be
able to extract significant amounts of work directly from the
temperature of the environment.

Unfortunately many objections to the Popper-Szilard Engine are
equally obscure. \cite{CT74,Rot79} appear to argue that it is the
design of the engine that now embodies the 'information' that
balances the entropy reduction. However, this can hardly be
supported, as such 'structural negentropy' is a one-off cost,
while the engine, once built, could extract unlimited energy.
Others \cite{Bri96}\cite[page74]{SB98} appear to confuse the
Engine with Feyerabend and Popper's opinions on Brownian
motion\cite{Pop57,Pop74,Fey74}.

However, there are two objections to the Popper-Szilard Engine
which do require consideration. These are due to Leff and
Rex\cite[pages 25-28]{LR90} and to Zurek\cite{Zur84} and
Biedenharn and Solem\cite{BS95}.

Leff and Rex offer an argument based upon Landauer's Principle.
They argue that, at the end of the cycle, when one of the weights
has been raised, the location of the piston and pulleys serves as
a memory of the location of the atom. In order to commence the new
cycle, the piston must be removed from either end of the
container, and reinserted in the center. This constitutes an
'erasure' of the memory and must be accompanied by a $kT_G \ln 2$
dissipation.

It is certainly the case that the analysis of the Popper-Szilard
Engine leaves out how this restoration is to take place without
having to perform a measurement of the position of the piston. In
order to see if Leff and Rex's criticism is justified, we will now
suggest a method by which the restoration may take place.

In Figure \ref{fg:szpop} there are two shelves, S1 and S2, on the
left and right of the Engine, at a height $\frac{kT_G}{Mg} \ln 2$
above the floor. When the gas has expanded, these shelves emerge
on both sides of the Engine. This will support whichever weight
has been raised. There is now a correlation between the location
of the weights and the position of the piston. By means of the
reversible CNOT interaction (Table \ref{tb:cnot} and Figure
\ref{fg:ld1}(b)) we can use the location of the raised weights as
System A and the piston as System B. The correlation of the
logical states "0" and "1" is equivalent to that between the
states of the piston and weights. If W1 is raised the piston is to
the right while if W2 is raised, the piston is to the left. This
should allow us to conditionally remove the piston from whichever
end of cylinder it is in and move it to the central position
outside the cylinder. This would appear to be in complete
agreement with Landauer's Principle, without having to perform an
external measurement, or dissipate energy.

Of course, it may be argued that now we have the weight to restore
to it's unraised position before we have truly 'completed' a
cycle\footnote{It could be objected that raising the weight is
precisely the 'work' that we were trying to achieve. To demand
that all weights be restored to their initial conditions appears a
vacuous way of ensuring that 'work' cannot be extracted. This
shows that even the concept of 'work' needs to be clarified.}. An
obvious way of doing this is to pull the shelves back and allow
the raised weight to fall inelastically to the floor, dissipating
the $kT_G \ln 2$ energy required to raise it. This appears to
confirm the resolution based upon Landauer's Principle. However,
this is deceptive.

To dissipate the raised energy, the weights must be in contact
with an environment at some temperature (we will assume a heat
bath located below the floor). Nothing so far has required that
the heat bath of the weight need be the same as the heat bath of
the one atom gas (we will also assume that the partition and
pulleys are perfect insulators). Consider what happens if the heat
bath into which the weight dissipates it's energy is at a {\em
higher} temperature than $T_G$. Now we appear to have completed
the cycle, to the satisfaction of everyone, and have apparently
satisfied the Landauer Principle. Unfortunately, we have also
reliably transferred energy from a colder heat bath to a hotter
one, and can continue to do so. Such a state of affairs would
still constitute a violation of the second law of thermodynamics,
according to the Clausius version:
\begin{quote}
No process is possible whose sole result is the transfer of heat
from a colder to a hotter body
\end{quote}
 We
could now attach a small Carnot engine between the two heat baths
and allow the same small amount of energy to flow back by
conventional means, extracting some of it as work in the process.
It is far from clear that information theory is of any use in
identifying where the argument above must fail.

The second objection, due to Zurek\footnote{This objection was
endorsed by \cite{BS95} although they disagree with the
information interpretation of entropy}, is more subtle. Zurek
argues that quantum measurement plays a role in preventing the
demonless Engine from operating. A classical atom is trapped on
one side or other of the piston, when it is inserted. The
demonless Engine seeks to exploit this without making a
measurement, to prove that the "'potential to do work' [is]
present even before \ldots a measurement is
performed"\cite{Zur84}.

For a quantum object, the situation is more complex:

\begin{quotation}
The classical gas molecule, considered by Szilard, as well as by
Jauch and Baron, may be on the unknown side of the piston, but
cannot be on 'both' sides of the piston. Therefore intuitive
arguments concerning the potential to do useful work could not be
unambiguously settled in the context of classical dynamics and
thermodynamics. Quantum molecule, on the other hand, can be on
'both' sides of the potential barrier, even if its energy is far
below the energy of the barrier top, and it will 'collapse' to one
of the two potential wells only if [it] is 'measured' \cite{Zur84}

This is non-intuitive \ldots but quantum mechanics is unequivocal
on this point \ldots the objections of Popper and Jauch and Baron
- that the Szilard engine could extract energy without requiring
any observation - is clearly wrong. Even with the shutter closed,
the single-molecule gas has both sides available for its thermal
wave function. Observation is require to isolate it on one side or
the other. \cite{BS95}
\end{quotation}

If true, this would certainly invalidate the arguments of Jauch
and Baron, Popper and Feyerabend, and would make the act of
quantum measurement a fundamental part of reducing the entropy of
an ensemble by gaining information about it's microstate. The
attempt to connect 'wavefunction collapse' with entropy changes is
widespread\cite{Neu55,WZ83,Lub87,Par89a,Par89b,Alb94}, although it
is usually associated with an entropy {\em increase}. If Zurek's
argument here holds good, this calls into question how
'no-collapse' versions of quantum theory, such as Bohm's or the
Many-Worlds Interpretation could explain the Szilard Engine.
Unfortunately, neither Zurek nor Biedenharn and Solem actually
demonstrate that the piston does not move.

Zurek calculates the Free Energies, based upon the quantum
partition function, to justify the argument that the gas can only
lift a weight if it is completely confined to one side or the
other. This requires us to assume that the statistical Free Energy
is a valid measure of the 'potential to do work'. A little thought
should show that this will {\em only} be the case if the second
law of thermodynamics is known to be valid, and this is precisely
the point which is under contention.

Biedenharn and Solem simply state that "the pressure on both sides
of the shutter is the same, the piston remains stationary" without
showing their calculations. They proceed to argue that the act of
observation must perform work {\em upon} the gas, and it is this
work which is extracted in the subsequent expansion. Again,
however, they do not provide a convincing demonstration of how
this work is performed.

This leaves the quantum superposition argument an intriguing
possibility to block the operation of the modified Popper-Szilard
Engine, but essentially incomplete. We will address this by
constructing an explicitly quantum mechanical version of the
Popper-Szilard Engine in the next Chapter.

\section{Conclusion} \label{s:szmd3}

The thorough analysis of the points of contention regarding the
Szilard Engine has lead us to construct a modified version of it
which, aside from the question of quantum superposition, appears
to be capable of producing anti-entropic behaviour. The operation
of this Engine is summarised in Figure \ref{fg:raise}
\pict{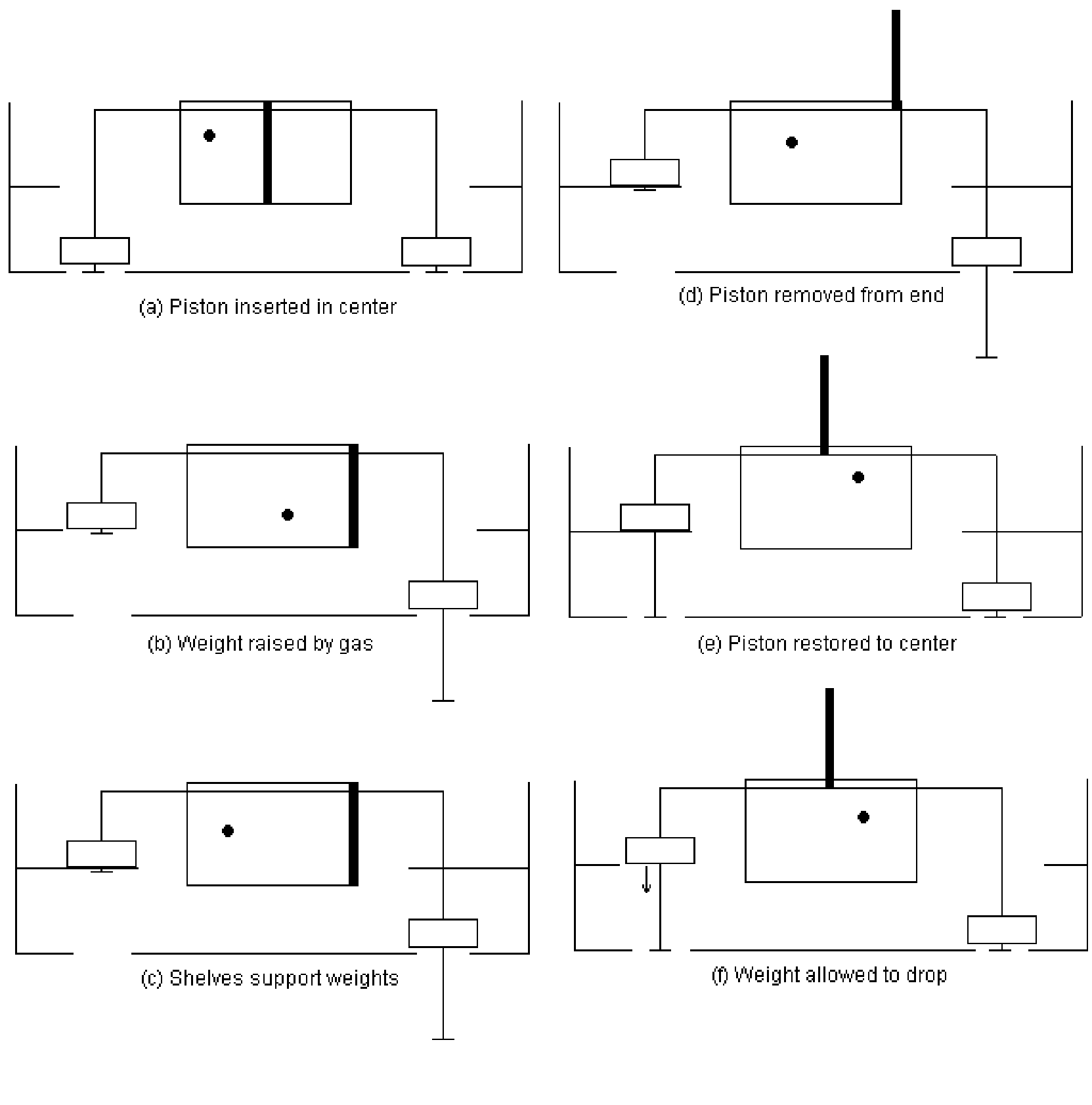}{The Cycle of the Popper-Szilard Engine} In Stage (a),
the piston is inserted into the box, which contains a single atom
in contact with a heat bath.  Stage (b) shows how the pressure of
the atom on the piston, from the left, causes the lefthand weight
to be lifted. The righthand weight remains at rest upon the floor.
In Stage (c), moveable shelves come out on both sides, and support
whichever weight has been raised. Stage (d) removes the piston
from the box. In this case it is on the righthand side. It's
position outside the box is correlated to the position of the
raised weight. Stage (e) uses this correlation to reset the
piston, by means of a Controlled-NOT type interaction. The
'information' as to which side of the box originally contained the
atom is now recorded in the location of the raised weight. If we
now remove both shelves, whichever weight is raised will fall to
the floor. This dissipates the energy used to raise it, and
restores the machine to it's initial state. However, if the weight
is in contact with a {\em higher} temperature heat bath than the
atom, then heat has been transferred from a colder to a hotter
heat bath, in apparent violation of the second law of
thermodynamics.

A detailed analysis of the physics of this cycle will pursued in
Chapters \ref{ch:szqm} and \ref{ch:szsm}. We will not assume any
thermodynamic relationships which depend upon the second law for
their validity. We will start by examining the interactions
between the microscopic states of the Engine. When we have
thoroughly analysed the time evolution of the system at the level
of individual quantum states, we will introduce a statistical
ensemble of these states, by means of density matrices. This will
enable us to calculate the mean, and long term, behaviour of the
Engine, and show that, in the long term, it is not capable of
producing heat flows which violate the Clausius statement.

The central issues that must be addressed, when constructing the
quantum mechanical Popper-Szilard Engine, are:

\begin{enumerate}
\item  What is involved in the process of 'confining' the particle
to one side of the box? Does this require only the inserting of a
potential barrier in the center of the box or must there also be a
'measurement' upon the position of the particle?

\item Does this 'confining' require an input of energy to the system?
This input of energy may come through perturbing existing
eigenvalues, or by a transition between eigenstates. The effect on
energy expectation values of both of these processes must be
calculated.

\item  Can a piston in the center of the box move, when the gas is still
in a superposition of being on both sides of the box?

\item  Can this movement be coupled to a pulley, to lift a weight?
Two weights may be involved.

\item  Can the partition be restored to the center of the box without
making an external measurement?
\end{enumerate}

Only after we have done this will we introduce the concepts of
entropy and free energy, in Chapter \ref{ch:szth}. Our
introduction of these concepts will be justified on the basis of
the analysis of the previous chapters, rather than the reverse. We
will show that these concepts are valid, even for single atom
systems, and that the entropy of the Engine is always increasing.

Finally, in Chapter \ref{ch:szsol} we will use the thermodynamic
concepts to generalise the resolution beyond the specific case of
the Popper-Szilard Engine. We will show that this generalisation
resolves the problems found in our discussion of the Szilard
Engine and Maxwell's Demon above, and provides a complete answer
to the Szilard Paradox. This will show that the information
theoretic resolutions are both unnecessary and insufficient. The
Szilard Engine is unsuccessful as a paradigm of the
information-entropy link.
\chapter{The Quantum Mechanics of Szilard's Engine}\label{ch:szqm}

In Chapter \ref{ch:szmd} we reviewed the historical analysis
 of the Maxwell's Demon, and Szilard Engine thought experiments.
In particular the question was raised of whether information
 processing or quantum measurement was an essential part of understanding these
 problems.

In this Chapter we will analyse the quantum mechanics of the
operation of the Szilard Engine. We are particularly interested in
whether the arguments of \cite{Zur84} or \cite{BS95} regarding the
role of quantum measurements are valid. To complete the analysis
of the Szilard Engine, the machine must be connected up to
statistical mechanical heat reservoirs. The effects of the
resulting statistical considerations will be examined in Chapter
\ref{ch:szsm}.

We can summarise the two issues that need to be assessed in each
stage of the operation of the quantum Szilard Engine as:

\begin{enumerate}
\item Can the operation proceed without an external agent ('demon')
needing to acquire and make conditional use of knowledge of the
systems microstate?

\item Can the transformation be achieved without making a significant
alteration in the internal energy of the Engine? In other words,
does it require work upon the system in order to drive its
operation?
\end{enumerate}

This Chapter will be primarily concerned with the first question,
although it will also calculate changes in internal energy of
specific microstates. The complete answer to the second question
will need consideration of the statistics of thermal ensembles in
Chapter \ref{ch:szsm}.

In order to analyse the questions above, it will, of course, be
necessary to make a number of abstractions and idealisations. All
motion is, as usual, considered to be frictionless. In the absence
of thermal heat baths, the systems are not decoherent so pure
states will evolve into pure states, not density matrices. In
Appendix \ref{ap:szun} we argue that the requirement that no
measurements are performed upon the system by external agents
('Demons' and the like), is equivalent to the requirement that a
single unitary operator is capable of describing the evolution of
the system. Rather than attempt to construct explicit Hamiltonians
for the interaction between parts of the Szilard Engine, we will
focus upon the question of how to describe the evolution of the
engine in terms of unitary operators. If the required evolution is
unitary, then there is {\em some} Hamiltonian that, in principle,
could be used to construct a suitable Engine. This approach will
enable us to make more general conclusions than if we were to
attempt to solve a particular Hamiltonian. We nevertheless will
show that the essential properties of our idealised unitary
evolution operators are the same as those that would result from a
more realistically constructed Hamiltonian.

The evolution of the quantum states of the Szilard Engine will be
studied in six sections. We will avoid introducing any external
measuring devices, and will concentrate upon the constraints that
unitarity imposes upon the evolution of the system. The sections
are:

\begin{enumerate}
\item  The unperturbed eigenstates of the particle in a box of width $2L$
. This is a standard quantum mechanical problem. Hereafter, the particle in
the box will be referred to as a 'gas';
\item  The perturbation of these eigenstates as a potential barrier of
width $2d$ $(d \ll L)$ is raised in the center of the box, up to
an infinite height. This must be considered in detail as
\cite{JB72} have pointed out the gas laws cannot be relied upon
for a single atom. The adiabatic transition was analysed
essentially correctly by \cite{Zur84,BS95}, but more detail is
presented here. Further, an error in the asymptotic form of the
energy eigenvalues given by Zurek is examined and corrected;
\item  The barrier is replaced by a moveable piston, also treated as a
quantum system. The effect of the interaction pressure from the gas is
analysed on both sides of the piston, and then combined into a single time
evolution operator;
\item  The quantum state of the weight to be lifted against gravity is
analysed. Again, this is a standard problem, with solutions given by Airy
functions. An evolution operator is constructed to connect the weight,
partition and gas;
\item  The problem of restoring the piston to the center of the box is
analysed in terms of unitary operators, which will be shown to
require correlating the movement of the piston to the final state
of the raised weights. However, it is found that the quantum state
of the weight leads to an uncertainty in the operation of the
resetting mechanism. This uncertainty leads to the possibility of
the Engine going into reverse. The effects of this reversal will
be evaluated in Chapter \ref{ch:szsm};
\item The conclusion of Sections \ref{s:szqm3} and \ref{s:szqm4} is
that, if the gas is capable of raising a weight when the gas is
confined to one side of the piston (which is generally accepted),
then it can still raise a weight when the single-atom gas is in a
superposition on both sides of the piston. This is contrary to the
analysis of \cite{Zur84,BS95} and calls into question the role
that the demon is alleged to play in either of their analysis.
Some of the objections of \cite{Pop74,Pop56,Fey66,JB72,Cha73} are
therefore shown to be valid in the quantum domain. This
constitutes the main result of this Chapter. However, the problem
of restoring the system, including piston, to it's initial state
has only been partially resolved and can only be fully evaluated
in the next Chapter.
\end{enumerate}
\section{Particle in a box} \label{s:szqm1}
We start by analysing the eigenstates of the one atom gas in the
engine, before any potential barrier or piston is inserted. The
one atom gas occupies the entire length of the Szilard Box, as in
Figure \ref{fg:szeng}. The Hamiltonian for the atom in the box is
then
\begin{eqnarray} \label{eq:szh0}
 H_{G0}\Psi_n=\left( -\frac {\hbar^2} {2m}\frac{\partial ^2}{\partial
x^2}+V\left( x\right) \right) \Psi _n
\end{eqnarray}
with
\[
V\left( x\right) =\left\{
\begin{array}{cc}
\infty & \left( x<-L\right) \\
 0 & \left( -L<x<L\right) \\
\infty
& \left( x>L\right)
\end{array}
\right\}
\]
This is the standard particle in an infinite square well
potential, with integer $n$ solutions of energy
\[
E_n=\frac{\hbar ^2\pi ^2}{8mL^2}n^2
\]
It will be easier to divide these into odd $(n=2l)$ and even
$(n=(2l-1))$ symmetry \footnote{Unfortunately odd symmetry
solutions have even values of $n$ and vice-versa. Odd and even
will exclusively be used to refer to the {\em symmetry}
properties.} solutions and make the substitutions
\begin{eqnarray*}
    K_n &=& L \frac {\sqrt{2mE_n}}{\hbar}
 \\ X &=&\frac {x}{L}
 \\ \epsilon &=& \frac{\hbar^2 \pi^2}{8mL^2}
\end{eqnarray*}
\subsubsection{Odd symmetry solutions}
\begin{eqnarray}\label{eq:psiodd1}
    \psi_l &=& \frac {1}{\sqrt{L}}\sin \left( K_lX\right)
    \\ E_l &=& 4\epsilon l^2 \nonumber
\end{eqnarray}
\subsubsection{Even symmetry solutions}
\begin{eqnarray}\label{eq:psieven1}
     \psi _l &=& \frac {1}{\sqrt{L}} \cos \left( K_lX\right)
  \\ E_l &=& 4\epsilon \left(\frac{2l-1}{2}\right) ^2 \nonumber
\end{eqnarray}
\section{Box with Central Barrier}\label{s:szqm2}
We now need to consider the effect of inserting the partition into
the Szilard Engine (Figure \ref{fg:raise}(a)). It will be simplest
to follow Zurek, and treat this as a potential barrier of width
$2d$ $(d \ll L)$, and variable height $V$, in the center of the
box:
\begin{eqnarray*}
V\left( x\right) &=& \left\{
    \begin{array}{cc}
        \infty & \left( x<-L\right)
         \\ 0 & \left( -L<x<-d\right)
         \\ V & \left( -d<x<d\right)
         \\ 0 & \left( d<x<L\right)
         \\ \infty & \left(L<x\right)
    \end{array}
\right\}
\end{eqnarray*}
Initially the barrier is absent, $V=0$. As the partition is
inserted, the barrier rises, until, when the partition is fully
inserted, dividing the box in two, the barrier has become
infinitely large, $V=\infty$.  This is a time dependant
perturbation problem as the barrier height $V$ is a function of
time. The instantaneous Hamiltonian, for a barrier height $V$, can
be written in terms of the instantaneous eigenstates and
eigenvalues as:
\[
H_{G1}(V)=\sum_l\left\{ E_l^{odd}(V)
    \kb{\Psi _l^{odd}(V)}{\Psi_l^{odd}(V)}
        +E_l^{even}(V)
    \kb{\Psi_l^{even}(V)}{\Psi _l^{even}(V)} \right\}
\]
The adiabatic theorem (see \cite[chapter 17]{Mes62b} and Appendix
\ref{ap:szun}) shows that if the barrier is raised sufficiently
slowly, the $n$'th eigenstate will be continuously deformed
without undergoing transitions between non-degenerate eigenstates.
The unitary evolution operator for the rising barrier is then
approximated by
\begin{equation}
U_{G}(t) \approx \sum_l\left\{
\begin{array}{l}
    e^{\frac i\hbar \int^t E_l^{odd}(\tau) d\tau}
        \kb{\Psi_l^{odd}(V)}{\Psi_l^{odd}(0)}
 \\ +e^{\frac i\hbar \int^t E_l^{even}(\tau) d\tau}
        \kb{\Psi_l^{even}(V)}{\Psi _l^{even}(0)}
\end{array}
\right\}
\label{eq:barrier}
\end{equation}
As this is from a time dependant Hamiltonian, it is not energy
conserving. In agreement with Zurek, and Biedenharn and Solem, we
will not regard this as a problem, as long as the change in energy
caused by inserting the potential barrier can be shown to be
negligible when compared to the energy extracted by the engine
(this will be shown in Chapter \ref{ch:szsm}).

The problem of raising the potential barrier is now that of
solving the stationary \Sch equation for an arbitrary barrier
height $V$. This is analysed in detail in Appendix \ref{ap:szpb}.
It is shown (see Figure \ref{fg:szpb1}) that the energy
eigenvalues and eigenstates change continuously from the zero
potential barrier to the infinitely high barrier.

The main results of Appendix \ref{ap:szpb} are now summarised, for
the limit of a high potential barrier, $V \gg E$ and $p=d/L \ll
1$.
\subsubsection{Odd Symmetry}
\begin{eqnarray} \label{eq:psiodd2}
\Psi &\approx&
 \left\{\begin{array}{lc}
        \frac{1}{\sqrt{L(1-p)}} \sin(K_{al}(1+X)) & (-1<X<-p)
     \\ \frac{(-)^l}{\sqrt{L}} \left(\frac{K_{al}}{K_{cl}}\right)
        \frac{e^{-K_{cl}(p-X)}-e^{-K_{cl}(p+X)}}{\sqrt{(1-p)}}
            & (-p<X<p)
     \\ - \frac{1}{\sqrt{L(1-p)}} \sin (K_{al}(1-X))
            & (p<X<1)
     \end{array} \right.
 \\ K_{al} &\approx& \frac{l\pi }{(1-p)}
        \left( 1-\frac{(1-2e^{-2K_{cl}p})}{K_{cl}(1-p)}\right)
        \nonumber
 \\ E_l &\approx &\epsilon \left( \frac{2l}{(1-p)}\right)^2
        \left(1-2\frac{(1-2e^{-2K_{cl}p})}{K_{cl}(1-p)}\right)\nonumber
 \\ K_{cl}p &\approx& \frac{d \sqrt{2mV}}{\hbar} \gg 1 \nonumber
\end{eqnarray}
\subsubsection{Even Symmetry}
\begin{eqnarray}\label{eq:psieven2}
    \Psi &\approx& \left\{ \begin{array}{lc}
             \frac{1}{\sqrt{L(1-p)}} \sin (K_{al}(1+X)) & (-1<X<-p)
          \\ \frac{(-)^l}{\sqrt{L}}
             \left(\frac{K_{al}}{K_{cl}}\right)
             \frac{e^{-K_{cl}(p-X)}+e^{-K_{cl}(p+X)}}{\sqrt{(1-p)}}
               & (-p<X<p)
          \\ \frac{1}{\sqrt{L(1-p)}} \sin (K_{al}(1-X)) & (p<X<1)
        \end{array} \right.
 \\ K_{al} &\approx& \frac{l\pi }{(1-p)}
        \left(1-\frac{(1+2e^{-2K_{cl}p})}{K_{cl}(1-p)}\right)
        \nonumber
 \\ E_l &\approx &\epsilon \left( \frac{2l}{(1-p)}\right)^2
        \left(1-2\frac{(1+2e^{-2K_{cl}p})}{K_{cl}(1-p)}\right)
        \nonumber
 \\ K_{cl}p &\approx& \frac{d \sqrt{2mV}}{\hbar} \gg 1 \nonumber
\end{eqnarray}
The $l^{th}$ odd and even eigenstates become
degenerate\footnote{The question of whether the asymptotic
degeneracy of the odd and even solutions represents a problem for
the application of the adiabatic theorem can be answered by noting
that, as the perturbing potential is symmetric, then the
probability of transition between odd and even solutions is always
zero.} in the limit, with energy levels $E_l=\epsilon \left(
\frac{2l}{1-p}\right)^2$.

As the adiabatic theorem shows we can insert the barrier without
inducing transitions between states, the only energy entering into
the system when inserting the partition is the shift in
eigenvalues. From the above results the energy level changes are
\[
\begin{array}{ccccc}
    & V=0 & V=E & V=\infty
 \\ \mathrm{Odd} & \epsilon\left( 2l\right) ^2 & \epsilon \left( 2l\right) ^2
     & \epsilon \left( \frac{2l}{1-p}\right) ^2
 \\ \mathrm{Even} & \epsilon \left( 2l-1\right) ^2 & \epsilon \left( \frac{2l-1}{1-p}\right) ^2
     & \epsilon \left( \frac{2l}{1-p}\right) ^2
\end{array}
\]
The fractional changes in odd and even symmetry energies,
respectively, are
\begin{eqnarray*}
\frac{E(\infty)-E(0)}{E(0)} &=& \left\{
\begin{array}{ccl}
    \frac{p(2-p)}{(1-p)^2} &\approx& 2p
 \\ \frac{p(2-p)}{( 1-p)^2}+\frac{4l-1}{(1-p)^2(2l-1)^2}
     &\approx& 2p+\frac{1+2p}{l}
\end{array} \right.
\end{eqnarray*}
where the approximations assume $p \ll 1$ and $l \gg 1$ . In both
cases it can be seen that the energy added is a small fraction of
the initial energy. However, for low energy even states, where $l
\gg 1$ is not valid, relatively large amounts of energy must be
added even when $p \ll 1$. For example $l=1$ leads to $\Delta
E\approx 3 E(0)$. Some work must be done upon the gas to insert
the partition. The size of this work required will be evaluated in
Section \ref{s:szsm2} as part of the statistical mechanics of the
system.

These results can be best understood in terms of the wavelength of
the eigenstate in the region where the potential barrier is zero
\[
\lambda_l=2 \pi K_{al} L
\]
The number of nodes within the box is $2 L/\lambda_l$, as the box
is of width $2L$. The energy of the eigenstate is directly related
to the {\em density of nodes} within the box.

The odd symmetry wavefunctions are simply expelled from the region
of the barrier, without changing the number of nodes. The same
number of nodes are therefore now confined in a volume reduced by
a factor $1-p$. The wavelength must decrease by this factor,
leading to an increase in energy levels.

Even symmetry wavefunctions must, in addition, become zero in the
center of the box, as the barrier becomes high. This requires an
additional node, increasing their number to the same as the next
odd symmetry wavefunction. The wavelength must decrease
sufficiently so that the original number of nodes, plus one, is
now confined to the reduced volume. This is a higher increase in
density of nodes than the corresponding odd symmetry, but as the
original number of nodes increases, the effect of the additional
node becomes negligible.

In the limit of very high barriers, the wavefunctions become
\begin{eqnarray*}
    \Psi_l^{even} \approx \Psi_l^{odd} \approx
     & \frac{1}{\sqrt{L(1-p)}} \sin \left( l \pi \frac{1+X}{1-p} \right)
     &(-1<X<-p)
 \\ \Psi_l^{even} \approx \Psi_l^{odd} \approx
    & 0 & (-p<X<p)
 \\ \Psi_l^{even} \approx -\Psi_l^{odd} \approx
    & \frac{1}{\sqrt{L(1-p)}} \sin \left( l \pi \frac{1-X}{1-p} \right)
    & (p<X<1)
\end{eqnarray*}
\pict{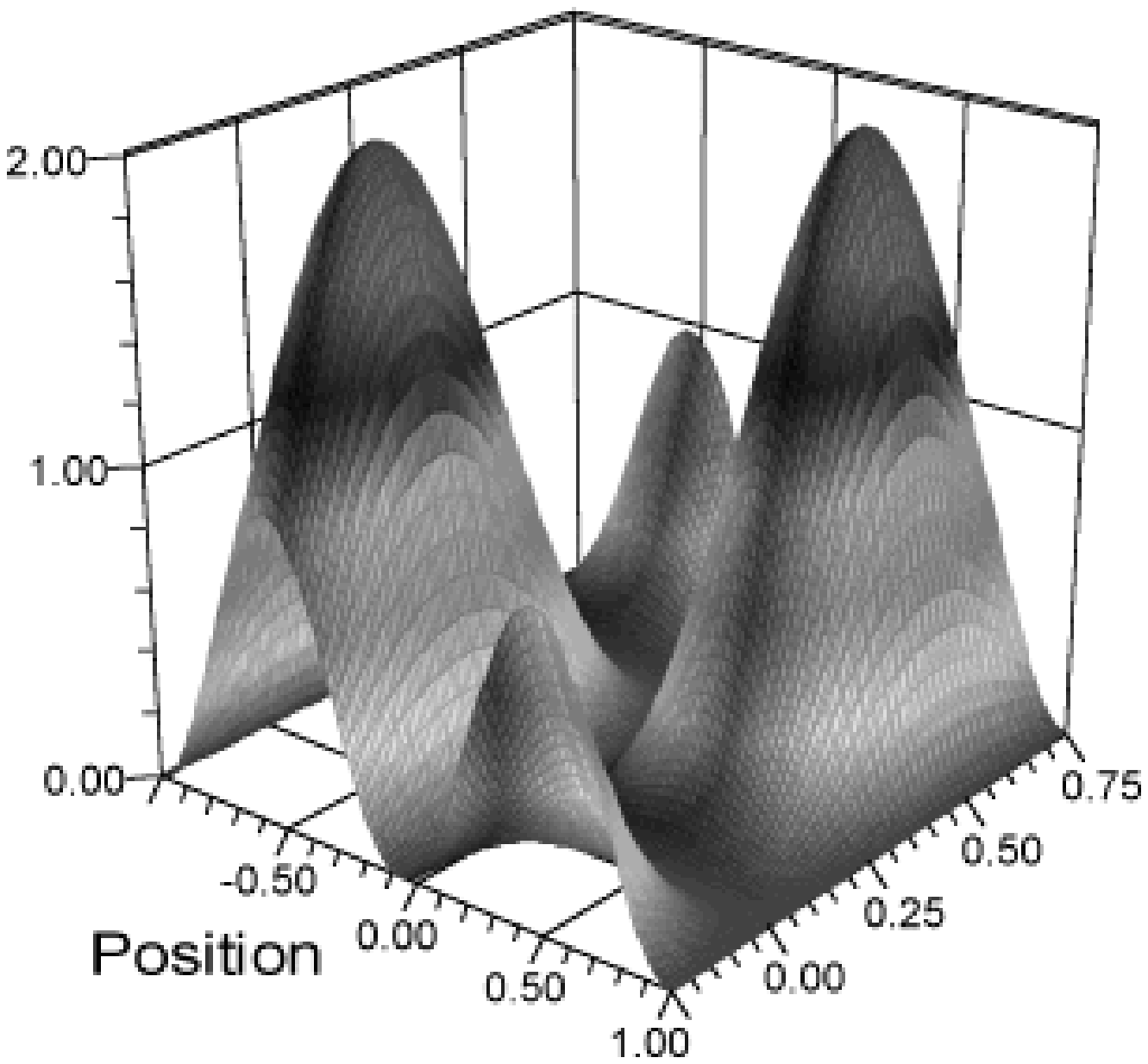}{Superpositions of odd and even symmetry states}
As
these are degenerate, we may form energy eigenstates from any
superposition of these states
\begin{eqnarray*}
\Psi_l(r,\alpha)&=& r e^{i\alpha}\Psi_l^{even}
                    +\sqrt{1-r^2}e^{-i\alpha}\Psi_l^{odd}
\end{eqnarray*}
Figure \ref{fg:szqm1} shows the probability density $ \left|
 \Psi_1(\frac{1}{\sqrt{2}},\alpha)\right|^2$ as $\alpha$ varies
between $-\pi /4$ and $3 \pi /4$. Of particular interest are the
pair of orthogonal states that occur when $\alpha=0$ and
$\alpha=\pi/2$
\begin{eqnarray*}
    \Psi_l^\lambda &=&\frac {1}{\sqrt{2}}\left( \Psi_l^{even}-\Psi_l^{odd}\right)
 \\ &=& \left\{ \begin{array}{lr}
        \sqrt{\frac{2}{L(1-p)}} \sin \left( l \pi \frac{1+X}{1-p} \right)
        & (-1<X<-p)
     \\ 0 & (-p<X<1)
     \end{array} \right.
 \\ \Psi_l^\rho &=& \frac {1}{\sqrt{2}}\left(\Psi_l^{even}+\Psi _l^{odd}\right)
 \\  &=& \left\{ \begin{array}{lr}
         0 & (-1<X<p)
      \\ \sqrt{\frac{2}{L(1-p)}} \sin \left( l \pi \frac{1-X}{1-p} \right)
        & (p<X<1)
      \end{array} \right.
\end{eqnarray*}
These represent situations where the one atom gas is located
entirely on the left or the right of the partition, respectively.
When we consider the system with the partition fully inserted, the
natural inclination is to describe the Hilbert space by a basis in
which the one-atom gas is confined to one side or the other. The
$\Psi_l^\lambda$ and $\Psi_l^\rho$ provide this basis and allow us
to write the final Hamiltonian in the form:
\begin{equation}
H_{G1}=\frac{4\epsilon}{(1-p)^2} \sum_l l^2\left(
        \kb{\Psi_l^\lambda}{\Psi_l^\lambda}
         +\kb{\Psi_l^\rho}{\Psi_l^\rho}
     \right)
     \label{eq:szh1}
\end{equation}
We can now start to consider Zurek's argument that the one-atom
gas must be measured to be confined to one side or the other of
the Szilard Engine. Suppose the gas is initially in an even
symmetry eigenstate $\Psi_l^{even}(0)$, with no barrier. As the
barrier is gradually inserted this eigenstate is deformed
continuously through $\Psi_l^{even}(V)$ until in the limit it
reaches
$\frac{1}{\sqrt{2}}\left(\Psi_l^\lambda+\Psi_l^\rho\right)$. The
single atom is not confined, or in a mixture of states, but is in
a superposition of being on both sides of the barrier. The same
will be true if we had started with an odd symmetry eigenstate.

It is worth noting, though, that if we had started with a
superposition of energy eigenstates\footnote{Ignoring a trivial,
time dependant phase factor that arises between the odd and even
symmetry states as their energy levels change by different
quantities}
\[
\Psi=\frac {1}{\sqrt{2}}\left(
\Psi_l^{even}(0)-\Psi_l^{odd}(0)\right)
\]
the adiabatic insertion of the potential barrier leads to the
state $\Psi_l^\lambda$. This {\em is} confined entirely to the
left of the barrier. A similarly constructed initial state leads
to the one-atom gas being confined entirely to the right of the
barrier. In order to draw a conclusion about the effect of the
quantum superposition upon the Szilard Engine we will need to
explicitly construct the full interaction between the one-atom gas
and the piston itself. This will be performed in Section
\ref{s:szqm3}, below.
\subsection{Asymptotic solutions for the HBA, $V \gg E $}
In this subsection we will briefly investigate a
discrepancy between Zurek's results, and those given above. The
expressions derived for energy eigenvalues in Appendix
\ref{ap:szpb} differ from those presented in \cite{Zur84}. We will
compare these two expressions with the numerical solutions to the
eigenvalue equations, and show that the HBA solutions are a closer
match to the numerical results.

In the High Barrier Approximation (HBA), the eigenvalues differ
only by an energy splitting:
\begin{eqnarray*}
E_l^{even} &\approx &4\epsilon \left( \frac{l}{1-p}\right)^2
    \left(1-2\frac{1+2e^{-2K_{cl}p}}{K_{cl}(1-p}\right) =E_l-\Delta _l
 \\ E_l^{odd} &\approx &4\epsilon \left(\frac {l}{1-p}\right)^2
    \left(1-2\frac{1-2e^{-2K_{cl}p}}{K_{cl}(1-p)}\right) =E_l+\Delta _l
\end{eqnarray*}
where
\begin{eqnarray*}
E_l &=&\epsilon\left(\frac{2l}{1-p}\right)^2
    \left(1-\frac{2}{K_{cl}(1-p)}\right)
 \\ \Delta_l &=&\epsilon \left(\frac{4l}{1-p}\right)^2
    \frac{e^{-2K_{cl}p}}{K_{cl}(1-p)}
\end{eqnarray*}
For comparison, in \cite{Zur84} Zurek appears to be suggesting the
following results (after adjusting for different length scales):
\begin{eqnarray*}
    E_{Zl} &=&\epsilon\left(\frac{2l}{1-p}\right) ^2
 \\ \Delta _{Zl} &=&\frac{\epsilon}{\pi}
    \left(\frac{4}{1-p}\right)^2 e^{-2K_{cl}p}
\end{eqnarray*}
Notice, that this would imply that the odd symmetry energy levels
are {\em falling} slightly for very high barrier heights, despite
initially being {\em lower} than the limiting value.
\begin{figure}
 \resizebox{\textwidth}{!}
    {\includegraphics{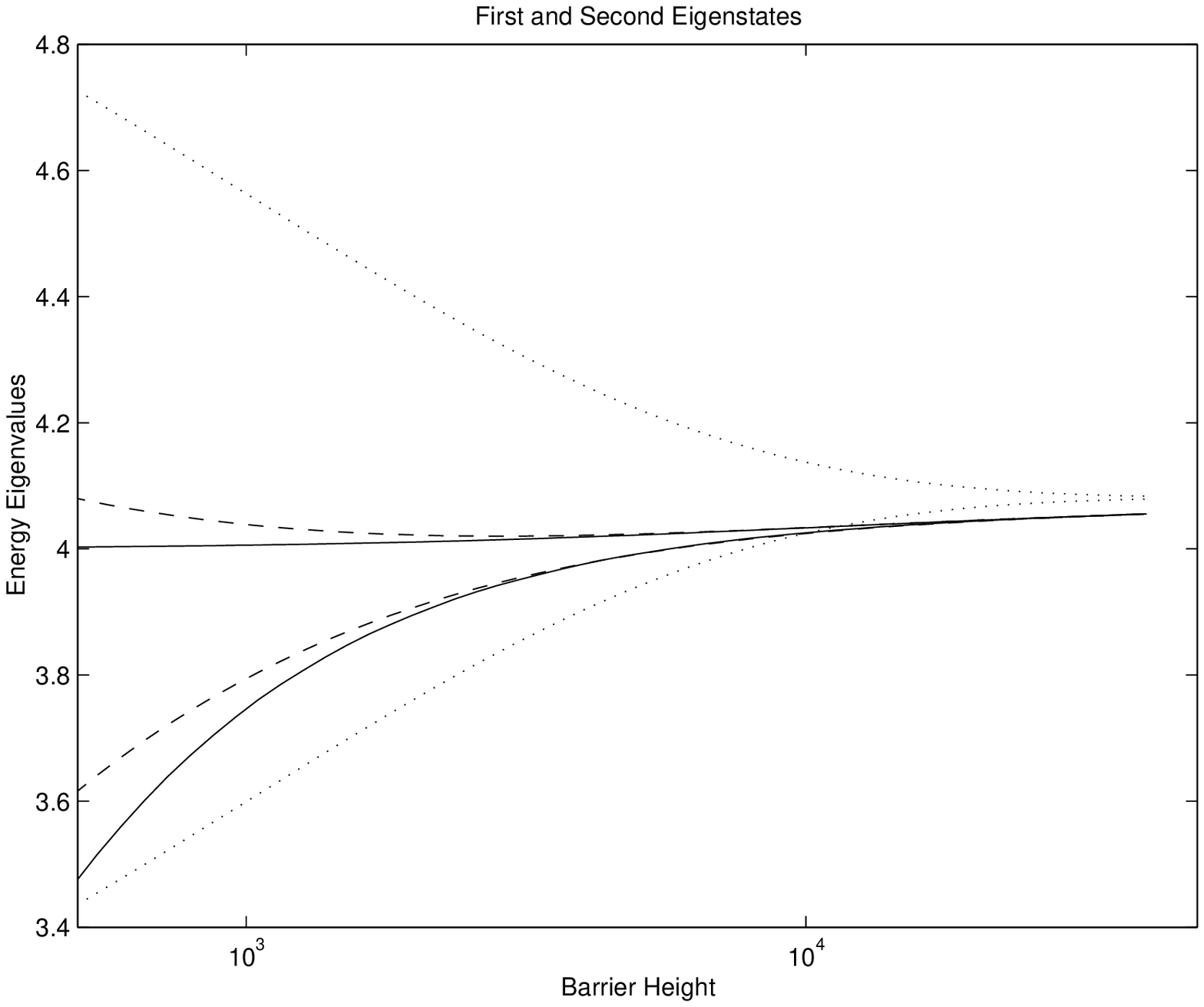}
     \includegraphics{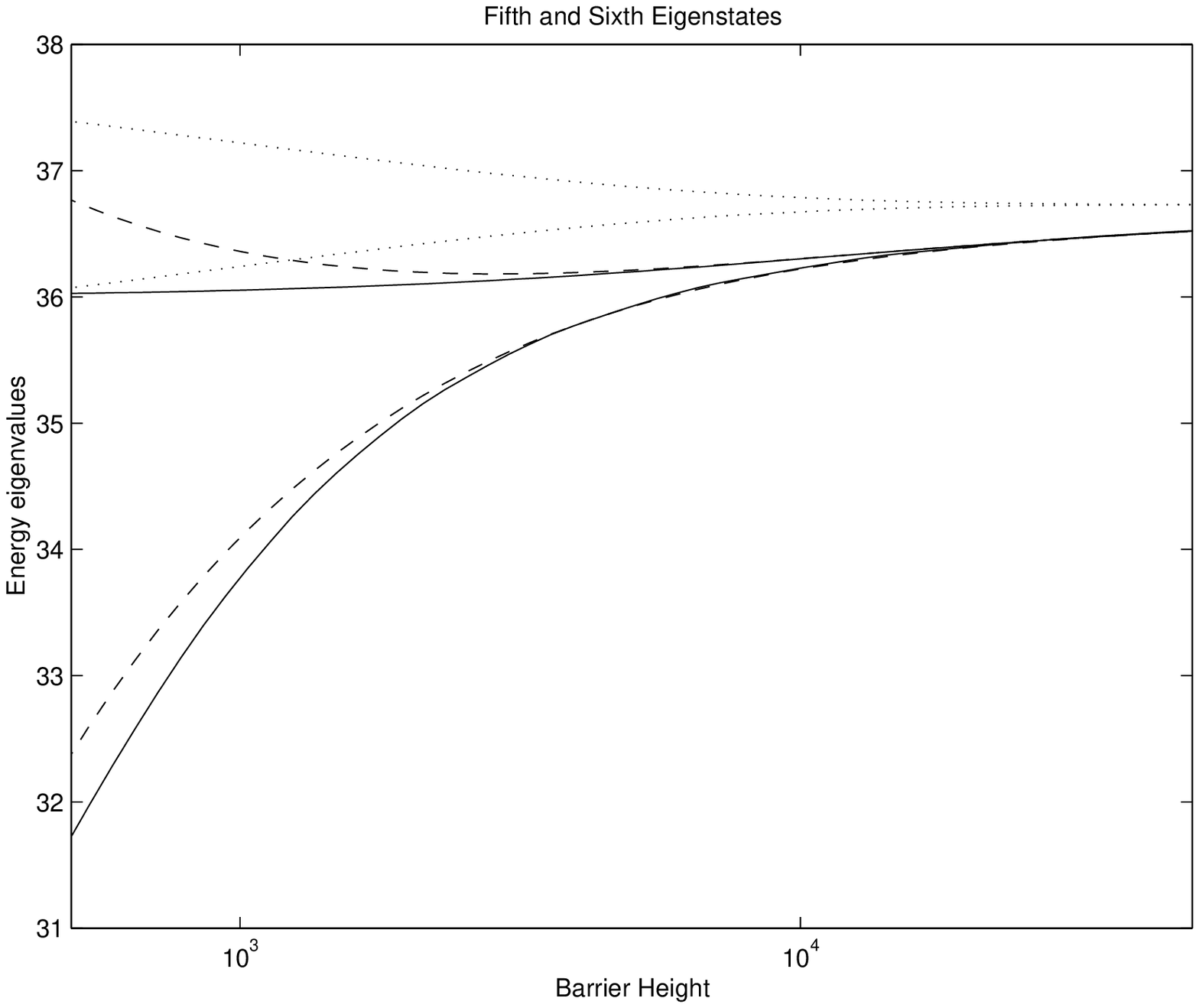}}
 \caption{Asymptotic Values of Energy Levels\label{fg:szqm2}}
\end{figure}
Numerical analysis of the eigenvalue equations (Appendix
\ref{s:apszpb3}) leads to Figure \ref{fg:szqm2}. This shows the
results for the first and third pairs of eigenstates. The dotted
lines are Zurek's solution, while the dashed lines are the HBA
approximations. Finally the unbroken lines give the numerical
solution, for which the energy splitting becomes less than the
difference between the limiting energy and the mean energy. The
odd and even numerical solutions approach degeneracy faster than
they approach the limiting value and the odd symmetry eigenvalues
are always less than the limit.

The HBA results closely match the numerical solution while Zurek's
results are too high, and his splitting is too large. The reason
for this is unclear, as Zurek gives no explanation for his
approximation. However, it is very similar to the central
potential barrier problem considered by Landau and Lifshitz
\cite[chapter 5]{LL77}. Landau and Lifshitz give a formula for the
energy splitting, which matches Zurek's $\Delta_{Zl}$, but no
formula for the mean energy - which Zurek appears to assume to be
equal to the limiting value. This assumption, that the mean energy
approaches the limiting value much faster than the energy levels
become degenerate, is clearly incorrect in this instance. As the
energy splitting formula of Landau and Lifshitz does not agree
with either the asymptotic approximation calculated here, or the
numerical solutions to the equations, it is also unclear that the
semi-classical approximation they use is applicable to this
situation.
\section{Moveable Partition}\label{s:szqm3}
In Section \ref{s:szmd2} one of the key arguments against the
operation of the Popper-Szilard Engine was that of
Zurek\cite{Zur84}, and Biedenharn and Solem\cite{BS95}, that in
the quantum case the partition does not move when the particle is
in a superposition of being on both sides of the partition.

However, neither actually provide a description of the interaction
between the one atom gas and the piston. Instead, both refer to
thermodynamic concepts to justify their arguments. Zurek, somewhat
confusingly, goes on to concede that
\begin{quote}..one can almost equally well maintain that this ...
describes a molecule which is on an 'unknown but definite' side of
the partition\end{quote} There is as much reliance upon
'intuitive' arguments as the classical analysis they criticise. To
improve on this situation it is necessary to analyse the actual
interaction between the piston and the one-atom gas, in terms of
unitary evolution operators. Only when this has been completed can
the effect on a statistical ensemble be calculated, and the
validity of thermodynamic concepts evaluated.

There are two main issues that need to be considered:
\begin{itemize}
\item The description of the moveable partition (piston). We will
need to treat the piston as a quantum object. To do this
rigorously would require dealing with some very subtle
difficulties regarding Hilbert spaces with continuous parameters
and localised states (e.g. see \cite[Chapter 4]{Per93}). However,
these difficulties are not relevant to the problem considered
here. Instead we will construct a fairly simple Hilbert space,
with a basis that corresponds to the minimum properties a piston
is required to possess.
\item The interaction between the piston and the one atom gas.
Before dealing with the problem of the gas in a superposition, we
shall analyse the situation where the gas is already confined to
one side of the piston. In this situation it is generally agreed
that the gas is capable of expanding, and pushing the piston in
doing so. If it were not the case, then it would be impossible to
extract any energy from an expanding one atom gas even when a
demon had knowledge of its location, and the entire debate over
Szilard's Engine would be redundant.
\end{itemize}

We will therefore assume only those properties of the piston state
that are necessary to be able to describe the expansion of the gas
when it is known to be confined to one side or another. We will
then use these properties, and the description of the expansion of
the gas, to examine the situation when the gas is in a
superposition of both sides of the piston. We will not attach a
weight to the piston until Section \ref{s:szqm4}.
\subsection{Free Piston}
The first problem we need to solve is to find a suitable
description of a piston as a quantum system. We will start by
defining a simple Hilbert space, without taking the gas into
account, with an appropriate unitary evolution operator for a
frictionless piston.

We will consider the piston to be an object, centered at some
point $-(1-p)>Y>(1-p)$ , with a width $2p \ll 1$. The quantum
state for a piston located at $Y$ will be $\ket{\Phi(Y)}$. The
width $p$ represents the width of the 'hard sphere repulsion'
potential that the piston will have for the gas. This corresponds
to an effective potential for the gas of
\[
V(X,Y) =\left\{
\begin{array}{cc}
    \infty & (X<-1)
 \\ 0 & (-1<X<Y-p)
 \\ \infty & (Y-p<X<Y+p)
 \\ 0 & (Y+p<X<1)
 \\ \infty & (X>1)
\end{array}
\right.
\]
It is important to note that $p$ is not the spread (or quantum
uncertainty) in the position co-ordinate $Y$. If the piston is a
composite object, $Y$ would be a collective co-ordinate describing
the center of the object. For a reasonably well localised object,
the spread in the co-ordinate $Y$, denoted by $\delta$, is
expected to be much smaller than the extent of the object,
represented by $p$. \pictpict{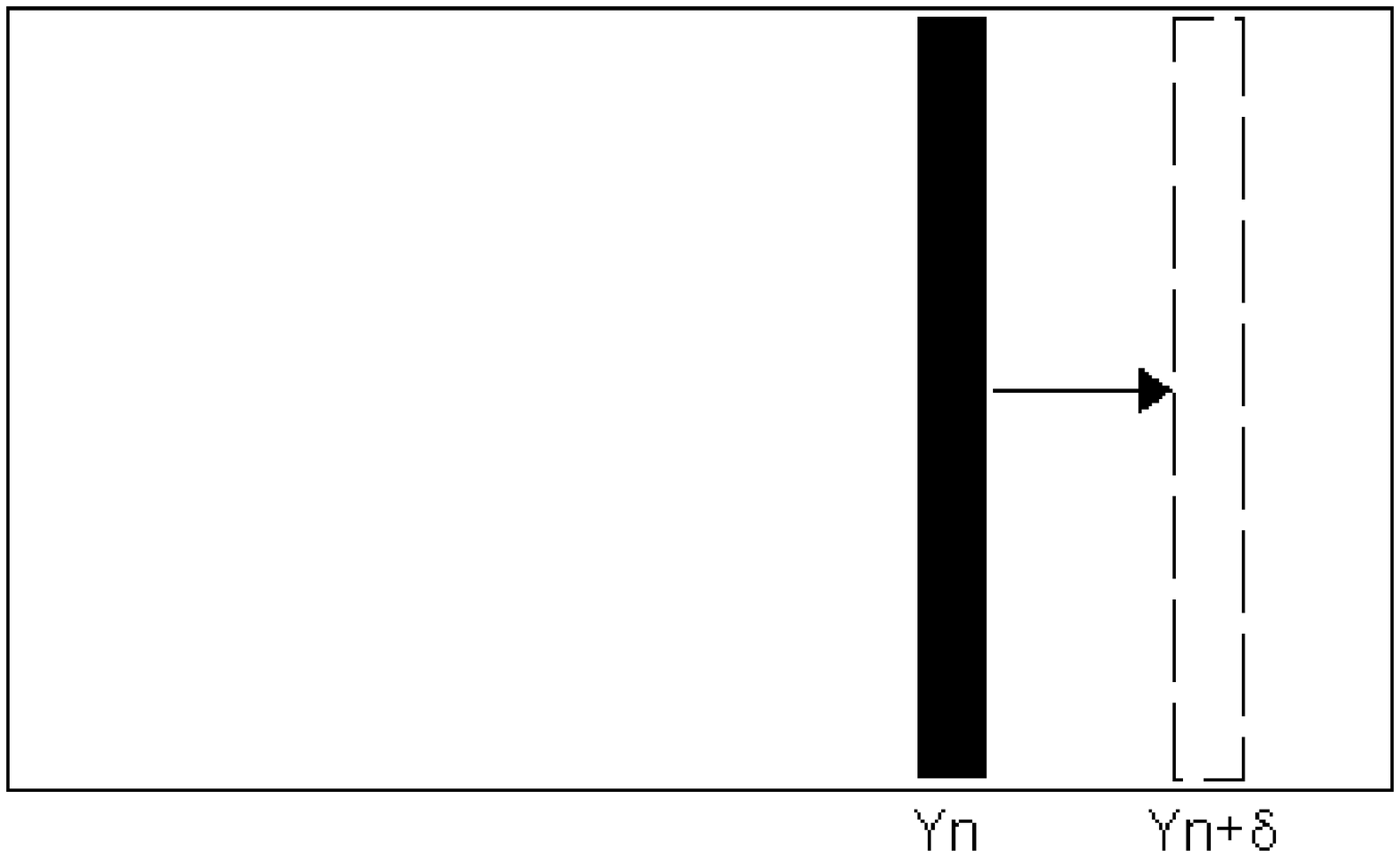}{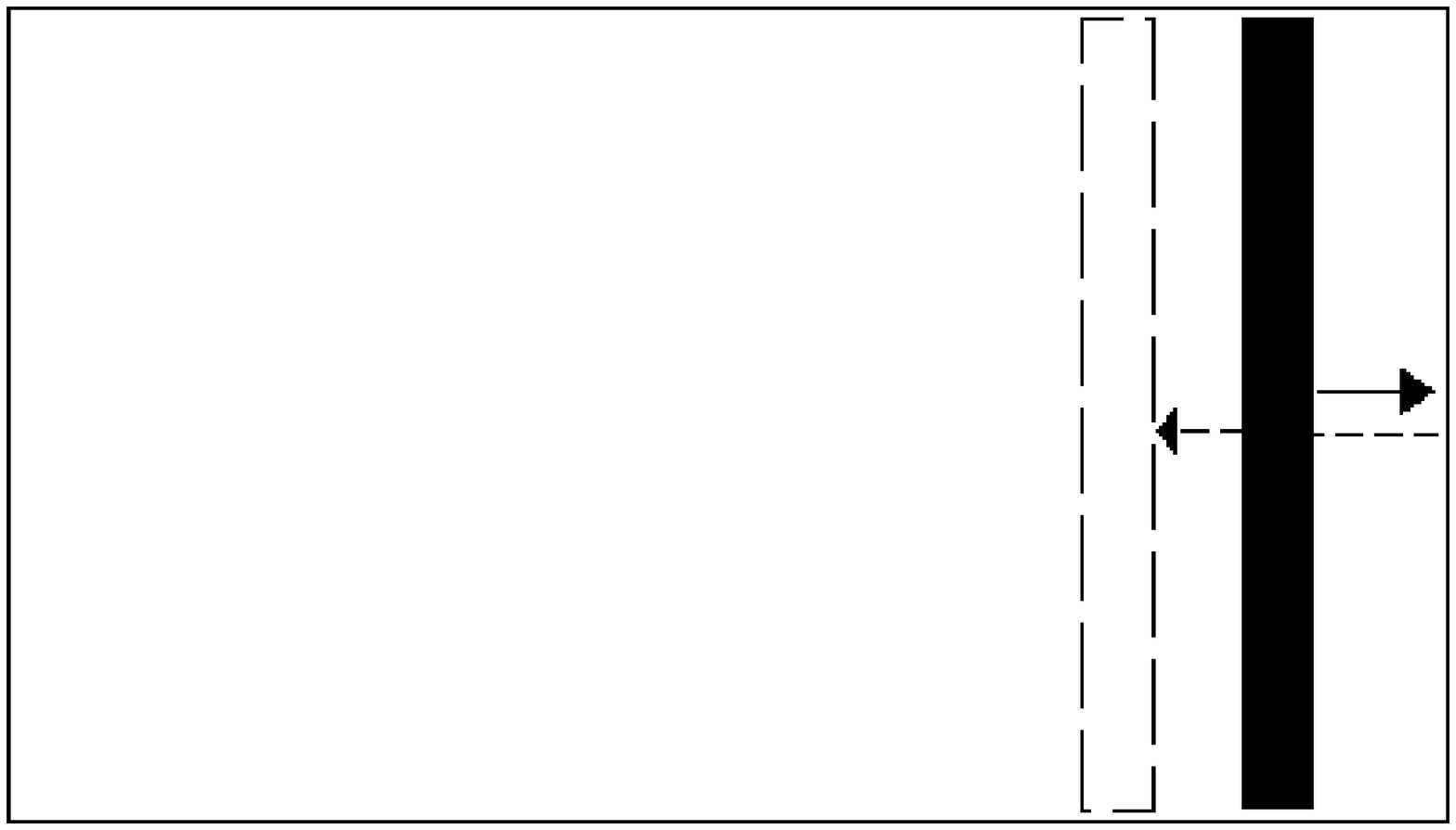}{Motion of
Piston}{szqm3} Now consider the behaviour required of the
frictionless piston in the absence of the gas. If the piston is
initially in state $\ket{\Phi(Y)}$, and is moving to the right,
then after some short period $\tau$ it will have advanced to the
state $\ket{\Phi(Y+\delta)}$ (see Figure \ref{fg:szqm3}(a) where
the distance $\delta$ has been exaggerated to be larger than $p$).
We will assume that two piston states separated by a distance
greater than $\delta$ are non-overlapping and therefore
orthogonal:
\[
\bk{\Phi(Y)}{\Phi(Y^\prime)}
 \approx 0 ; (\left|Y-Y^\prime\right| \geq \delta)
\]
The motion to the right must be described by a unitary operation
\[
U(\tau)\ket{\Phi(Y)}=\ket{\Phi(Y+\delta)}
\]

When the piston reaches the end of the Szilard Box
$(\ket{\Phi(1)}$ it cannot come to a complete halt as this would
require an evolution operator of
\begin{eqnarray*}
    U(\tau)\ket{\Phi(1-\delta)}&=&\ket{\Phi(1)}
 \\ U(\tau)\ket{\Phi(1)}&=&\ket{\Phi(1)}
\end{eqnarray*}
and a mapping of orthogonal onto non-orthogonal states is not
unitary. Instead the piston must collide elastically with the edge
of the box and start moving uniformly to the left (Figure
\ref{fg:szqm3}(b) ).

We now have to distinguish left from right moving piston states,
so that
\begin{eqnarray*}
    U(\tau)\ket{\Phi_L(Y)} &=& \ket{\Phi_L(Y-\delta)}
 \\ U(\tau)\ket{\Phi_R(Y)} &=& \ket{\Phi_R(Y+\delta)}
\end{eqnarray*}
Without this distinction we would need a left moving evolution
\[
    U(\tau)\ket{\Phi(Y)} = \ket{\Phi(Y-\delta)}
\]
and a right moving evolution
\[
    U(\tau)\ket{\Phi(Y)} = \ket{\Phi(Y+\delta)}
\]
and again, this would not be unitary, as the same state
$\ket{\Phi(Y)}$ is mapped to different states.

Left and right moving states are automatically required to be
orthogonal, even if they are spatially overlapping, owing to the
fact that inner products are invariant under unitary evolution, so
that
\begin{eqnarray*}
    \ip{\Phi_L(Y)}{U^\dag(\tau)U(\tau)}{\Phi_R(Y)} &=&
        \bk{\Phi_L(Y-\delta)}{\Phi_R(Y+\delta)}
 \\     \bk{\Phi_L(Y)}{\Phi_R(Y)} &=& 0
\end{eqnarray*}

From this, we can now construct a Hilbert space spanned by a set
of $N=2(2j+1)$ states, each centered on $Y_n=n\delta ,$
$n=-j,...j$ where $j=\frac{1-p}\delta $. The required evolution
operator is:
\begin{eqnarray}\label{eq:up1}
U_{P1}(\tau) &=& \sum_{n=-j}^{j-1}
    \kb{\Phi_R(Y_{n+1})}{\Phi_R(Y_n)}
    +\kb{\Phi_L(Y_j)}{\Phi_R(Y_j)} \nonumber
 \\&&+\sum_{n=-j+1}^j
    \kb{\Phi_L(Y_{n-1})}{\Phi_L(Y_n)}
    +\kb{\Phi_R(Y_{-j})}{\Phi_L(Y_{-j})}
\end{eqnarray}
The first line represents a piston moving to the right, and
reversing direction at $n=j$, while the second line is the piston
moving to the left, and reversing at $n=-j$. Movement is with a
fixed speed $\omega =\frac \delta \tau $, so that over the
characteristic period of time $\tau $ it has moved exactly one
'step' to the left or right.

This operator will be unitary, providing
\begin{eqnarray} \label{eq:szdel}
\bk{\Phi_A(Y_n)}{\Phi_B(Y_m)}&=&\delta_{AB}\delta_{nm}
\end{eqnarray}
It is possible to construct a Hilbert space and unitary evolution
satisfying these conditions, by adapting the quantum clock system
\cite{Per80}. It is important to note that the moving piston
states above are \textit{not} eigenstates of the Hamiltonian
associated with $U_{P1}(\tau)$, and so do not have well defined
energies. This is necessary to ensure that they are {\em moving}
states. States with well defined energies would necessarily be
stationary.
\subsection{Piston and Gas on one side}
Having defined our piston states, we can now start to consider the
interaction between the piston and the single atom gas. This
requires us to define a unitary evolution operator that acts upon
the joint space of the piston and gas states. The key question
that has been raised is whether the piston will move when the gas
is in a superposition of being on both sides of the Szilard Box.
We must not prejudice this question by {\em assuming} the
evolution does (or does not) produce this result, so we need to
find some other basis for constructing our unitary evolution
operator. We will approach this problem by analysing situations
where there is general agreement about how the piston and gas
interact. As we have noted before, there is general agreement
that, when the one atom gas is confined entirely to one side of
the piston, it is capable of exerting a pressure upon the piston
and causing the piston to move (see for example\cite{BBM00}). We
will therefore proceed by analysing the situation where the gas is
located entirely on one side of the piston, and construct a
suitable unitary evolution operator to describe this.

We will start with the one-atom gas on the left of the piston
(once this has been solved we will be able to transfer the results
to the one-atom gas on the right by a simple symmetry operation).
As noted above, the piston acts as a potential barrier of width
$2p$, centered upon $Y_n$. A basis for this subspace of the
Hilbert space of the gas is given by the states
$\ket{\Psi_l^\lambda (Y_n)}$ where
\begin{equation}\label{eq:psileft}
\Psi_l^\lambda(Y_n,X)=\bk{X}{\Psi_l^\lambda (Y_n)}
    =\sqrt{\frac{2}{L(Y_n+1-p)}}
    \sin\left(l\pi \frac{1+X}{Y_n+1-p}\right)
\end{equation}
 and $-1<X<Y_n-p$. We will use the superscript $\lambda $ to represent a gas state on
the left of the piston, and $\rho$ for states of the gas on the
right of the piston.

The left gas states and the piston states are combined to define a
joint basis:
\[\ket{\Psi_l^\lambda(Y_n)\Phi_B(Y_n)}\]
First we will define the internal energy of the gas subsystem,
then we will construct an evolution operator for the joint system,
including the interaction between the gas and piston.

The internal energy of the gas state $\ket{\Psi_l^\lambda(Y_n)}$
is $4\epsilon\left(\frac {l}{Y_n+1-p}\right)^2$ so the Hamiltonian
for the one-atom gas subsystem's internal energy is given by
\begin{eqnarray}
H_{G2}^\lambda &=&\sum_{n=0}^{j}\rho(Y_n)H_{G2}^\lambda(Y_n)
    \label{eq:szhg2} \\
H_{G2}^\lambda(Y_n) &=&\sum_l 4\epsilon
    \left(\frac{l}{Y_n+1-p}\right)^2
    \kb{\Psi_l^\lambda(Y_n)}{\Psi_l^\lambda(Y_n)} \nonumber
\end{eqnarray}
It is important to be clear about the role played by the operators
$\rho(Y_n)=
\kb{\Phi_L(Y_n)}{\Phi_L(Y_n)}+\kb{\Phi_R(Y_n)}{\Phi_R(Y_n)}$. This
does not imply that the piston is part of the gas subsystem, or
that this particular Hamiltonian includes an interaction energy
between the gas and piston. The $H_{G2}^\lambda(Y_n)$ represent
the internal energy states of the gas, given a particular position
of the piston. The combined Hamiltonian $H_{G2}^\lambda$ includes
$\rho(Y_n)$ to project out the position of the piston. The
parameter $Y$ is an external parameter of the gas, describing an
external configuration, or boundary condition, upon the gas, as
opposed to $X$ which is an internal parameter. It is the motion
associated with $X$ that generates the internal energy in
$H_{G2}$, not $Y$.

Details of the internal energy of the piston would depend upon
it's construction as a composite system, so we will simply include
a term $H_P$ to represent this, and assume that there is no
interaction between the internal piston states and it's external
position, or the gas states.

Neither $H_{G2}$ nor $H_P$ represent the interaction between the
gas and piston properly, as they give only internal energies for
each subsystem. A Hamiltonian consisting of $H=H_{G2}+H_P$ would
not lead to a moving piston at all. Instead we must construct an
idealised evolution operator to describe the expansion of the gas,
pushing the piston. When the piston reaches the end of the box, it
will collide elastically, as before, and as it's direction
reverses it will compress the gas. For simplicity we assume that
when the piston reaches the center of the box, it is not capable
of compressing the gas any further, and will reverse back to it's
original direction\footnote{This assumption will be more realistic
when the attached weight is included in the system, in the next
Section.}. This motion can be described by the unitary operator:
\begin{eqnarray}
U_{P2}^\lambda(\tau)&=&\sum_l \{
 \sum_{n=1}^{j-2}
    \kb{\Psi_l^\lambda(Y_{n+1})\Phi_R(Y_{n+1})}
       {\Psi_l^\lambda(Y_n)\Phi_R(Y_n)}\nonumber
\\ &&+ \sum_{n=2}^{j-1}
    \kb{\Psi_l^\lambda(Y_{n-1})\Phi_L(Y_{n-1})}
       {\Psi_l^\lambda(Y_n)\Phi_L(Y_n)}\nonumber
\\ &&+\kb{\Psi_l^\lambda(1-p)\Phi(1-p)}
        {\Psi_l^\lambda(Y_{j-1})\Phi_R(Y_{j-1})}\nonumber
\\ &&+\kb{\Psi_l^\lambda(Y_{j-1})\Phi_L(Y_{j-1})}
        {\Psi_l^\lambda(1-p)\Phi(1-p)}\nonumber
\\ &&+\kb{\Psi_l^\lambda(0)\Phi(0)}
        {\Psi_l^\lambda(Y_{1})\Phi_L(Y_{1})}\nonumber
\\ &&+\kb{\Psi_l^\lambda(Y_{1})\Phi_R(Y_{1})}
        {\Psi_l^\lambda(0)\Phi(0)}\label{eq:szqm1}
\}
\end{eqnarray}
The first and second lines represent the piston moving to the
right (gas expanding) and the left (gas compressing) respectively.
The third and fourth lines represent the right moving piston
reaching the end of the box, coming to an instantaneous halt in
the state $\ket{\Phi(1-p)}$, and reflecting to the left, starting
to recompress the gas. The fifth and sixth lines, similarly,
represents the piston, reaching the maximum compression of the gas
in the center of the box, coming to a halt in $\ket{\Phi(0)}$,
before starting to move back to the right under pressure from the
gas\footnote{This operator assumes the expansion does not cause
transitions between internal states of the gas. As long as the
expansion period $\tau$ is sufficiently long, this will be
consistent with the adiabatic theorem (Appendix \ref{ap:szun}).}.

The eigenstates of $U_{P2}^\lambda(\tau)$ are superposition of all
the $Y_n$ states:
\begin{eqnarray*}
\ket{\Lambda_{al}}
    &=&\sum_{n=1}^{j-1}
    \left\{e^{ina}\ket{\Psi_l^\lambda(Y_n)\Phi_R(Y_n)}
    +e^{-ina}\ket{\Psi_l^\lambda(Y_n)\Phi_L(Y_n)}\right\}
 \\ &&+\ket{\Psi_l^\lambda(0)\Phi(0)}
    +e^{ija}\ket{\Psi_l^\lambda(1-p)\Phi(1-p)}
 \\ U_{P2}^\lambda(\tau)\ket{\Lambda_{al}}&=&e^{ia}\ket{\Lambda_{al}}
\end{eqnarray*}
Continuity at $\ket{\Psi_l^\lambda(1-p)\Phi(1-p)}$ requires that
$e^{-ija}=e^{ija}$ . This imposes a periodic boundary condition
upon the system, and gives a discrete set of eigenstates
$\ket{\Lambda_{al}}$ that satisfy $ja=\pi m$, $m=-j+1, \ldots ,j$

The Hamiltonian that drives the unitary evolution
$U_{P2}^\lambda(\tau)$ is
\[
H_{\tau 2}^\lambda=\frac {1}{\tau}
    \sum_{a,l}a\kb{\Lambda_{al}}{\Lambda _{al}}
\]
This does not offer any simple interpretation in terms of an
internal energy $H_{G2}$ of the gas plus an interaction term
representing the pressure of the gas upon the piston. The simplest
way to take into account the internal energy of the gas, and also
any internal states of the piston system, is with a total
Hamiltonian:
\[
H_{T2}^\lambda =(1-h(t))H_{G2}^\lambda +h(t)H_{\tau 2}^\lambda
+H_{P}
\]
The time dependant function $h(t)$ allows the 'switching on' and
'switching off' of the pressure interaction between the piston and
the gas. It is equal to one when the piston is present in the box,
and zero when the piston is absent\footnote{It may be objected
that $H_{T2}$ is unrealistic as it appears to requires the
internal energy of the gas to be 'switched off' during the
expansion phase. An obvious, if woefully contrived, way to correct
this is to have $H_{G2}$ at all times, but to 'switch on' an
interaction Hamiltonian $H_{I2}=(H_{T2}-H_{G2})$. That more
realistic Hamiltonians will ultimately produce the same result is
argued later.}. While $h(t)$ is one, the interaction of gas and
piston drives the system through the evolution
$U_{P2}^\lambda(t)=e^{iH_{\tau 2}^\lambda t}$, causing the gas to
expand, with the piston moving to the right, or to compress, with
the piston moving to the left, in a cyclic motion.

If the interaction is 'switched on' for just long enough to expand
the gas to it's full extent, and then 'switched off', the final
states will be at a lower energy than they were before the
expansion\footnote{The Hamiltonian $H_{T2}$ is time dependant}.
The excess energy will have been stored in the interaction between
the gas and piston, and the combination of 'switching on' and
'switching off' of the interaction requires energy to be deposited
in, or drawn from, a work reservoir.

We have now constructed a suitable Hamiltonian, and a unitary
evolution operator, that encapsulates the expected behaviour of
the gas and piston system, when the gas is confined to one side of
the piston. We now turn to the case where the gas can be in a
superposition.
\subsection{Piston with Gas on both sides}\label{s:szqm3:3}
This subsection will demonstrate one of the main results of this
Chapter, that the superposition of gas states does not lead to a
stationary piston.

We will extend the results of the previous subsection to include
the situation where the gas is confined entirely to the right. The
combination of the left and right unitary evolution operators will
then be shown to produce a unitary evolution operator that acts
upon the entire space of the gas and piston system, including
situations where the gas is in a superposition of being on the
left and right side of the piston. Applying this unitary operator
to the superposition of gas states and shows that, rather than
staying in the center, the piston moves into an entangled
superposition of states, contrary to the arguments of Zurek and of
Biedenharn and Solem. We will then show how this result
generalises beyond the specific unitary evolution operator
constructed here. Finally we will examine how this evolution
affects the internal energy of the one atom gas.

It is evident that had we considered the situation where the gas
was confined entirely to the right of the piston, we would have
obtained the Hamiltonians:
\begin{eqnarray*}
    H_{\tau 2}^\rho &=&\sum_l \frac {1}{\tau} \sum_{a,l}a \kb{R_{al}}{R_{al}}
 \\ H_{G2}^\rho &=& \sum_{n=-j}^{0}\rho(Y_n)H_{G2}^\rho(Y_n)
\end{eqnarray*}
with
\begin{eqnarray*}
 H_{G2}^\rho(Y_n) &=& \sum_l
     4\epsilon \left(\frac{l}{1-p-Y_n}\right)^2
     \kb{\Psi_l^\rho(Y_n)}{\Psi_l^\rho(Y_n)}
\\ \ket{R_{al}}&=&\sum_{n=-j+1}^{-1}\left\{
    e^{ina}\ket{\Psi_l^\rho(Y_n)\Phi_R(Y_n)}
    +e^{-ina}\ket{\Psi_l^\rho(Y_n)\Phi_L(Y_n)}
    \right\}
\\ && +\ket{\Psi_l^\rho(0)\Phi(0)}
    +e^{ija}\ket{\Psi_l^\rho(-1+p)\Phi(-1+p)}
\end{eqnarray*}
and the gas state $\ket{\Psi_l^\rho(Y_n)}$ represents the gas
confined entirely to the right of the piston $(Y_n+p<X<1)$, with
wavefunction
\[
\Psi_l^\rho(Y_n,X)=\bk{X}{\Psi_l^\rho(Y_n)}
    =\sqrt{\frac{2}{L(1-p-Y_n)}}
    \sin\left(l \pi \frac{1-X}{1-p-Y_n}\right)
\]
During an interaction period, in which $H_{\tau 2}^\rho$ is
'switched on', the unitary evolution operator is
\begin{eqnarray}
U_{P2}^\rho(\tau) &=& \sum_l\{
    \sum_{n=-j+1}^{-2}\kb{\Psi_l^\rho(Y_{n+1})\Phi_R(Y_{n+1})}
        {\Psi_l^\rho(Y_n)\Phi_R(Y_n)} \nonumber
\\ && +\sum_{n=-j+2}^{-1}\kb{\Psi_l^\rho(Y_{n-1})\Phi_L(Y_{n-1})}
        {\Psi_l^\rho(Y_n)\Phi_L(Y_n)}\nonumber
\\ &&+\kb{\Psi_l^\rho(0)\Phi(0)}{\Psi_l^\rho(Y_{-1})\Phi_R(Y_{-1})}
\nonumber
\\ && +\kb{\Psi_l^\rho(Y_{-1})\Phi_L(Y_{-1})}{\Psi_l^\rho(0)\Phi(0)}
\nonumber
\\ && +\kb{\Psi_l^\rho(-1+p)\Phi(-1+p)}{\Psi_l^\rho(Y_{-j+1})\Phi_L(Y_{-j+1})}
\nonumber
\\ && +\kb{\Psi_l^\rho(Y_{-j+1})\Phi_R(Y_{-j+1})}{\Psi_l^\rho(-1+p)\Phi(-1+p)}
\label{eq:szqm2} \}
\end{eqnarray}

We now need to construct a Hamiltonian and corresponding unitary
time evolution operator that acts upon the Hilbert space for the
gas particle on either (or both) sides of the piston. The natural
assumption would be to use:
\[
H_{T2} = h(t) \left[ H_{\tau 2}^\lambda+H_{\tau 2}^\rho \right]
    +(1-h(t)) \left[ H_{G2}^\lambda+H_{G2}^\rho \right] + H_P
\]
where $h(t)$ is again a time dependant function, zero when the
pressure interaction between the piston and gas is 'switched off'
and one otherwise. The question is whether the left and right
Hamiltonians can be added without changing the resultant unitary
evolution. We will be able to answer this affirmatively from the
fact that left and right Hamiltonians, and their respective
unitary evolution operators, act upon disjoint subspaces of the
joint gas-piston Hilbert space.

Firstly, we must prove that the addition of the Hamiltonians leads
to an operator that acts upon the whole of the joint system
Hilbert space. This will be the case if the states
\ket{\Psi_l^\alpha(Y_n)\Phi_B(Y_n)} form an orthonormal basis for
the joint Hilbert space.

Consider the inner product:
\begin{eqnarray}
\bk{\Psi_k^\alpha(Y_m)\Phi_A(Y_m)}
    {\Psi_l^\beta(Y_n)\Phi_B(Y_n)}
=\delta_{nm}\delta_{\alpha \beta }\delta_{kl}\delta_{AB}
\label{eq:szdel2}
\end{eqnarray}
\begin{itemize}
\item $\delta_{nm}$ and $\delta_{AB}$ come from the orthonormality of the different
piston states (Equation \ref{eq:szdel}).
\item $\delta_{\alpha \beta}$ clearly holds if the
wavefunctions of the $\alpha$ and $\beta$ gas states have no
overlap. A right gas wavefunction is non-zero only to the right of
the piston position. Similarly a left gas wavefunction is non-zero
only to the left of the piston position. The right and left gas
wavefunctions can therefore only be overlapping if their
respective piston states are to the left and right of the other.
If this is the case, then $Y_n \neq Y_m$ and then $\delta_{nm}$
guarantees orthogonality, so the joint states are orthogonal.
\item $\delta_{kl}$ is certainly true for wavefunctions where $\alpha$
and $\beta$ are the same. The $\delta_{\alpha \beta}$ term then
automatically prevents interference between these states in the
combined Hilbert space.
\end{itemize}

For any given piston position, the combination of left and right
gas states will span the subspace of the gas states, and the
piston states span the piston subspace, so the above states form
an orthonormal basis for the joint space. This basis splits into
two disjoint subspaces, corresponding to the gas on the left or
right of the piston.

Now let us consider a general property of unitary operators acting
upon subspaces. If $U_a$ acts entirely upon the subspace $S_a$ and
$U_b$ acts upon $S_b$, each unitary operator can be extended to
act upon the entire space $S^T=S_a\oplus S_b$ by means of:
\begin{eqnarray*}
U_a^T &=&U_a\oplus I_b \\ U_b^T &=&I_a\oplus U_b
\end{eqnarray*}
where $I_a$ and $I_b$ are the identity operators upon $S_a$ and
$S_b$ respectively. It is therefore possible to form the joint
operator
\[
U^T=U_a\oplus U_b=U_a^TU_b^T=U_b^TU_a^T
\]
The commutativity implies that, with a unitary operator written in
the form $U=e^{iK}$, where $K$ is a Hermitian operator
\[
U^T=e^{iK^T}=e^{iK_a}e^{iK_b}=e^{i\left(K_a \oplus K_b\right) }
\]
Applying this back to the equation of motion,
\[
i\hbar \frac{\partial U}{\partial t}=HU
\]
it is deducible that if $H_a$ and $H_b$ are Hamiltonians defined
upon disjoint subspaces, and $U_a$ and $U_b$ are their associated
evolution operators, then the joint Hamiltonian $H^T=H_a+H_b$ has
an associated evolution operator given by $U^T$. This proves that
the solutions for the separate cases of the gas confined to the
left and right side of the piston can be combined into a single
unitary evolution operator for the combined Hilbert space.
\subsubsection{Combined Evolution Operator}
We have now shown that the complete unitary evolution operator for
the combined gas piston system, with the interaction 'switched
on', is
\begin{eqnarray*}
U_{T2}(\tau)&=&U_{P2}^\rho(\tau) \oplus U_{P2}^\lambda(\tau)
\end{eqnarray*}
To study the properties of this evolution we will simplify the
operator in two ways. Firstly, we will allow the interaction to
run for exactly the time necessary for the gas wavefunction to
completely expand or compress. This will take
$j=\frac{1-p}{\delta}$ steps, and will result in a unitary
evolution $U_{T2}(j\tau)=(U_{T2}(\tau))^j$.

Secondly, we will start with only those states for which the
piston is in the central position and only look at those states
that occur from $U_{T2}(j\tau)$ acting upon this initial subspace.

With these two simplifications, the evolution operator becomes
\begin{eqnarray*}
U_{T2}&=& \sum_l
    \kb{\Psi_l^\rho(-1+p)\Phi(-1+p)}{\Psi_l^\rho(0)\Phi(0)}
\\ && +\kb{\Psi_l^\rho(0)\Phi(0)}{\Psi_l^\rho(-1+p)\Phi(-1+p)}
\\ && +\kb{\Psi_l^\lambda(1-p)\Phi(1-p)}{\Psi_l^\lambda(0)\Phi(0)}
\\ && +\kb{\Psi_l^\lambda(0)\Phi(0)}{\Psi_l^\lambda(1-p)\Phi(1-p)}
\end{eqnarray*}

If we apply this evolution operator to an initial state, where the
gas is in a superposition of being on both sides of the piston:
\[
\ket{\chi_{initial}}=\left(\alpha \ket{\Psi_l^\rho(0)}
    +\beta\ket{\Psi_m^\lambda(0)}\right)\ket{\Phi(0)}
\]
this state will evolve into
\[
\ket{\chi_{final}}=\alpha\ket{\Psi_l^\rho(-1+p)\Phi(-1+p)}
    +\beta\ket{\Psi_m^\lambda(1-p)\Phi(1-p)}
\]
This demonstrates the central result of this Section. Guided only
by the argument that the {\it confined} one-atom gas is capable of
pushing the piston, we have shown that the condition of unitarity
leads to an evolution operator which does {\em not} leave the
piston stationary when the gas is initially in a superposition.
This is contrary to the arguments of Zurek and of Biedenharn and
Solem. However, it is also the case that the piston is now in an
entangled quantum superposition, so the situation is still quite
different from the classical case.

We have examined the piston gas interaction in considerable
detail, in order to carefully demonstrate that the evolution
operator $U_{T2}$ can be derived from a continuous expansion of
the gas states and is consistent with the agreed behaviour of the
one atom gas when it is confined. The unitary operator, however,
was not derived from a particularly realistic interaction
Hamiltonian.  We will now present a simple argument that a less
idealised Hamiltonian would produce the same result.

The key property is that the confined one atom gas can expand
adiabatically against the piston. If the gas is initially on the
right of the piston, this expansion is given by some unitary
operation $U$
\[
U\ket{\Psi_l^\rho(0)}\ket{\Phi(0)}
    =\ket{\Psi_l^\rho(-1+p)}\ket{\Phi(-1+p)}
\]
while if the gas is initially to the left, the expansion is
\[
U\ket{\Psi_l^\lambda(0)}\ket{\Phi(0)}
    =\ket{\Psi_l^\lambda(1-p)}\ket{\Phi(1-p)}
\]
These equations\footnote{up to a phase factor} must be derivable
from {\it any} interaction Hamiltonian $H$ that, over a
sufficiently long period, allows the adiabatic expansion of a one
atom gas. Provided the two expansions can be combined into a
single unitary operator, and we have shown that they can, it
follows from the linearity of $U$ that a superposition of gas
states leads to the same entangled superposition of piston and gas
states as we reached with $U_{T2}$ above. The piston state will
not be stationary, even with a more realistically derived
Hamiltonian.
\subsubsection{Expansion of the Gas States}
We will now examine the effect of the expansion upon the internal
energy states of the one atom gas. It is assumed that, as long as
$\tau$ is sufficiently large, or equivalently, that the expansion
takes place sufficiently slowly, the adiabatic theorem will apply,
and there will be no transitions between eigenstates. However, the
internal energy eigenstates and eigenvalues continuously change as
the piston position $Y_n$ changes. This forms the basis of the
'work' that will be extracted from the expansion of the gas.

For an initial, odd symmetry state, $\ket{\Psi_l^{odd}}$ the
insertion of the piston makes negligible change upon the energy,
but splits the wavefunction into a superposition of left and right
wavefunctions $\Psi_l^\lambda(0)$ and $\Psi_l^\rho(0)$. The energy
of this state is approximately $4\epsilon l^2$. As the piston
moves into a superposition, the energies of the left and right
states go down, until at the end of the expansion, the internal
energy of the gas state is approximately $\epsilon l^2$.

The reason for this can be seen from the wavelength, and node
density of the gas wavefunction. The wavefunction for a left gas
state is
\[
\Psi_l^\lambda(Y_n,X)=\sqrt{\frac {2}{L(Y_n+1-p)}}
    \sin\left(l\pi \frac{1+X}{Y_n+1-p}\right)
\]
The number of nodes in this wavefunction is constant, and equal to
half the number of nodes in the initial odd symmetry wavefunction.
When the expansion has finished, these nodes are spread over twice
the volume, so the density of nodes has decreased by a factor of
two, and the energy decreased by a factor of four.

The same is true for the right gas wavefunctions. In fact, at the
end of the expansion stages, the wavefunctions are
\begin{eqnarray*}
\Psi_l^\lambda(1-p,X)&=&\frac {1}{\sqrt{L(1-p)}}
    \sin\left(\frac{\pi l}{2}\left(\frac{1+X}{1-p}\right)\right);(-1<X<1-2p)
\\ \Psi_l^\rho(-1+p,X)&=&\frac {1}{\sqrt{L(1-p)}}
    \sin\left(\frac{\pi l}{2}\left(\frac{1-X}{1-p}\right)\right);(-1+2p<X<1)
\end{eqnarray*}
These differ by, at most, a sign change and a shift in position of
order $2p \ll 1$:
\begin{equation}
\Psi_l^\lambda(1-p,X) \approx \Psi_l^\rho(-1+p,X)\approx
    \left\{ \begin{array}{lc}
    \psi_{l/2} & l \ \mathrm{even}
 \\ \psi_{(l+1)/2} & l \ \mathrm{odd}
 \end{array} \right\}
 \label{eq:gas}
\end{equation}
where $\psi_l$ are the unperturbed wavefunctions given in Section
\ref{s:szqm1}. The value of $l$ is approximately halved during the
expansion.

For an initial even symmetry wavefunction, the same analysis
applies, only now a single node is inserted in the center of the
wavefunction, as the piston is inserted, requiring some work. This
corresponds, neglecting terms of order $p$, to an energy input and
output of:
\[
\begin{array}{cccc}
\mathrm{Symmetry}&\mathrm{Input}&\mathrm{Output}&\mathrm{Net}
\\ \mathrm{Odd}&0&3\epsilon l^2&3\epsilon l^2
\\ \mathrm{Even}&\epsilon (4l-1)&3\epsilon l^2&\epsilon
(l-1)(3l-1)
\end{array}
\]

The net energy extracted is always positive, with the single
exception of the ground state, which is the even symmetry $l=1$
state. In this case one node is added, when the barrier is
inserted, and one node is removed, when the wavefunction expands,
so the energy input exactly matches the energy output. So on each
cycle of the Szilard Engine, some energy is extracted, as the
number of the eigenstate is approximately halved, and the gas is
left in a lower energy state than it started. This continues until
the ground state is reached, at which point no more energy can be
extracted, and the work output during the expansion phase is the
work done upon the system when the barrier is inserted.

There are two points that can be drawn from this. Firstly, this
shows that energy could be extracted from the operation of the
Szilard Engine, if all the other stages of the Engine operate as
required. This energy is not energy that is inserted into the
system by performing a measurement.

Secondly, the state of the one atom gas will fall to the ground
state, at which point no further energy can be extracted. In
Chapter \ref{ch:szsm} the gas will be brought into contact with a
heat bath. This will allow energy to flow back into the gas,
restoring the energy extracted by the expansion.

\section{Lifting a weight against gravity}
\label{s:szqm4} In the previous Section it was shown that the
single atom gas can be made to expand against a piston, and that
this expansion is associated with a reduction in the internal
energy of the gas. We now need to incorporate the manner in which
that internal energy is converted into work. The paradigm of work
being performed is taken to be the raising of a weight.

In the Popper version of the Szilard engine, it is the connection
of a weight on either side of the engine that is supposed to allow
work to be extracted without a measurement of the position of the
gas particle (Figure \ref{fg:raise}(b)). However, when the one
atom gas is initially in a superposition of left and right gas
states, the quantum Popper-Szilard Engine becomes a superposition
of left moving and right moving piston states. To include the
piston raising a weight, we must include the weights themselves in
the quantum mechanical description of the system.

A quantum weight, of mass $M_w$, resting upon a floor at height
$h$, in a gravitational field $g$ is described by the \Sch
equation

\begin{equation}
H_W(h)A_n(z,h)=\left(-\frac{\hbar}{2M_w}\frac{\partial
^2}{\partial z^2}+V(z,h)\right)A_n(z,h) \label{eq:airyh}
\end{equation}

with

\[
V(z,h) =\left\{
\begin{array}{cc}
    \infty & (z \leq h) \\
    M_wg(z-h) & (z>h)
\end{array}
\right\}
\]

The solution to this equation is derived from the Airy function
$A(z)$ (see \cite{AS70,NIST}) by applying the requirements that
the wavefunction $A_n(z,h)$ be normalised, and the boundary
condition $A_n(h,h)=0$. This leads to wavefunction solutions

\begin{equation}
 A_n(z,h)=\left\{
 \begin{array}{cc}
    \frac{A\left(\frac{z-h}{L}+a_n\right)}
        {\sqrt{H}A^{\prime }(a_n)} &(z>h) \\
    0 & (z \leq h)
 \end{array}
 \right\}
 \label{eq:airy1}
\end{equation}

with a characteristic height, depending upon the strength of the
gravitational field and the mass of the weight

\[
H=\left(\frac{\hbar ^2}{2M_w^2g}\right)^{\frac{1}{3}}
\]

and an energy eigenvalue

\[
E_n=(h-a_nH)M_wg
\]

\begin{figure}[htb]
\resizebox{\textwidth}{!}
    {\includegraphics{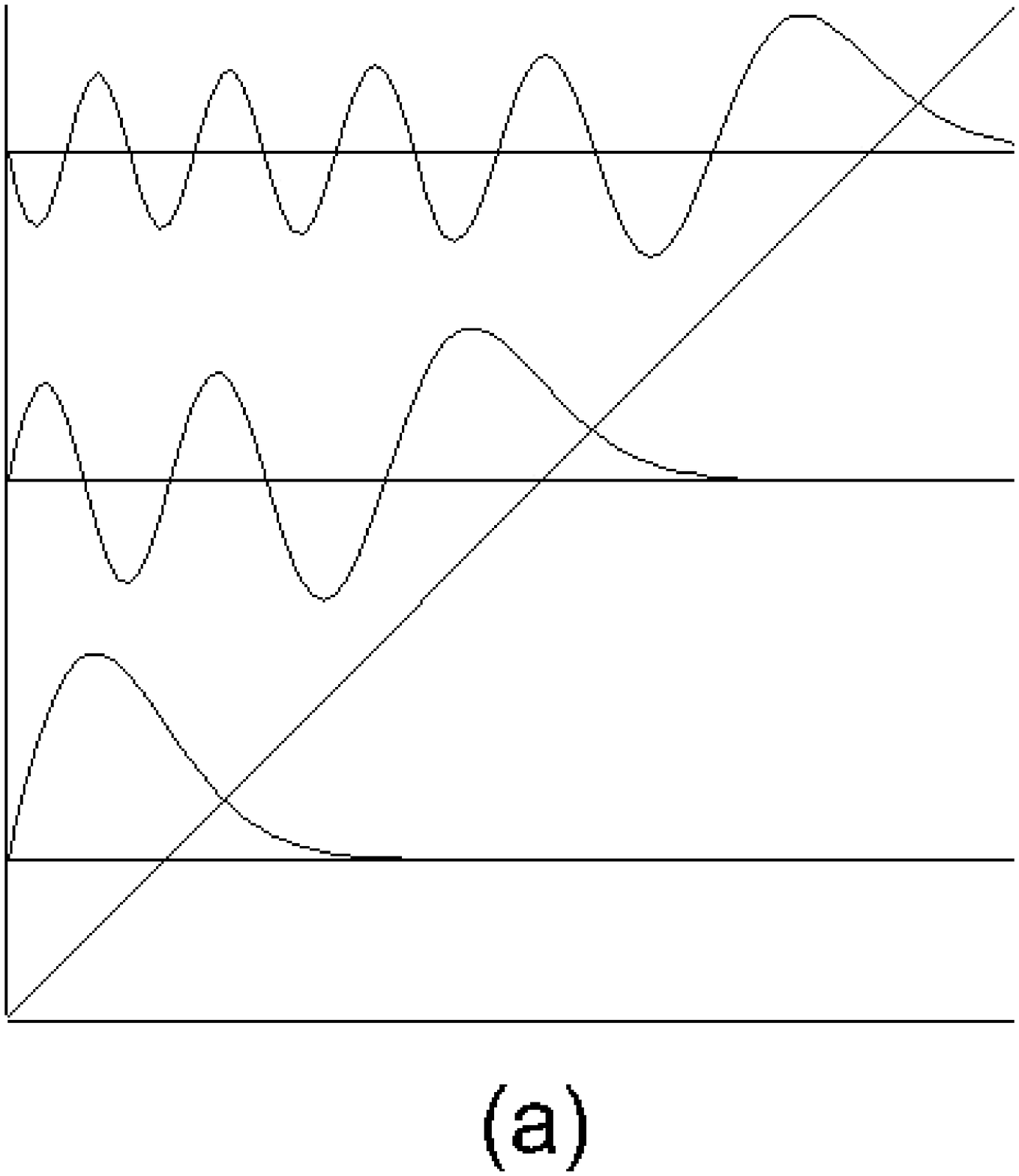}
    \includegraphics{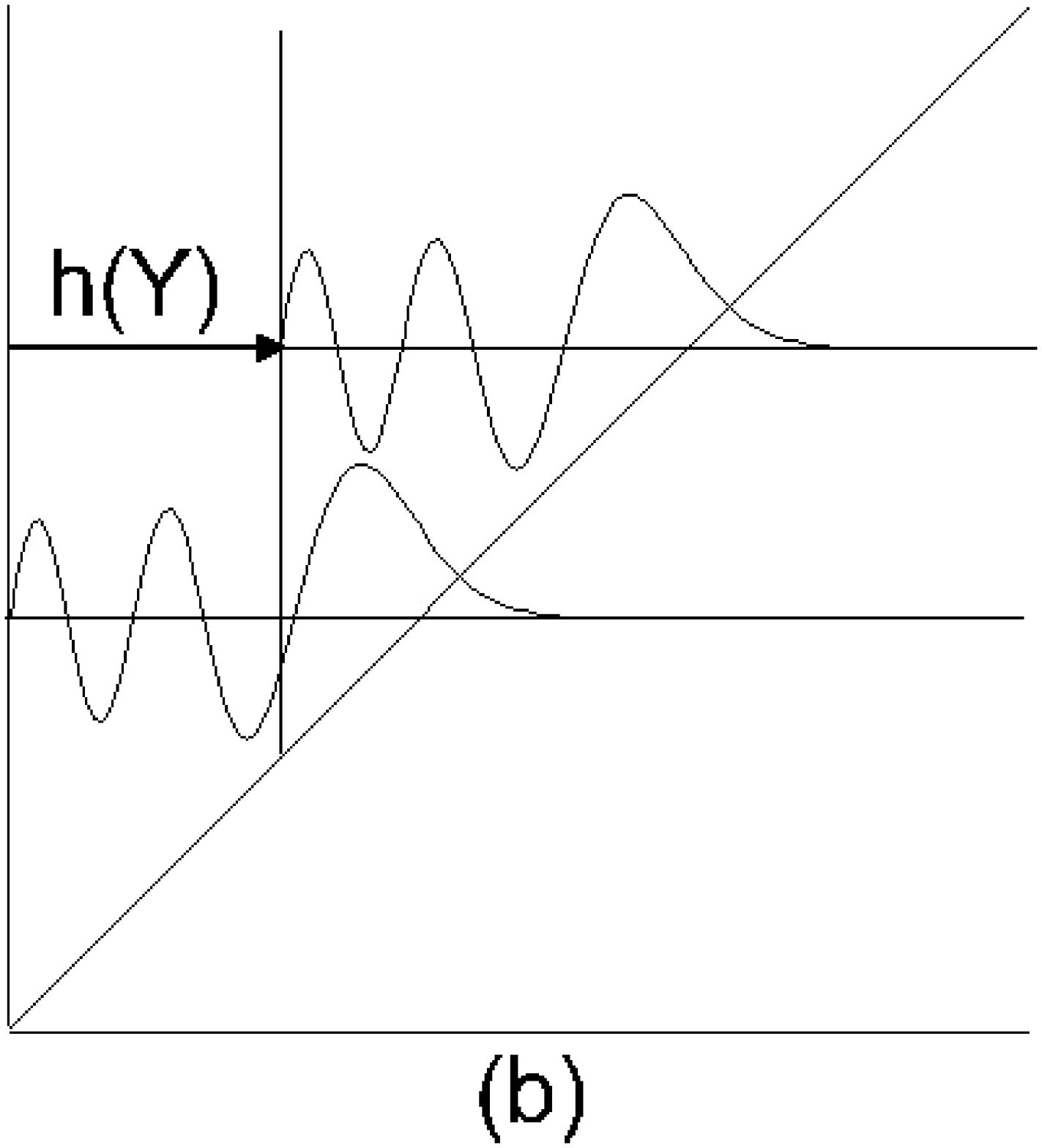}}
\caption{Airy Functions for a Mass in Gravitational
Field\label{fg:airy}}
\end{figure}

The values $a_n$ correspond to the values of $z$ for which the
Airy function $A(z)=0$. These values are always negative, and
become increasingly negative as $n$ increases. For large $n$ they
have the asymptotic form $a_n=-\left(\frac{3 \pi
n}{2}\right)^{\frac{2}{3}}$. $A^{\prime}(z)$ is the first
derivative of the Airy function. Note that $A_n\left(z,h\right)
=A_n(z-h,0).$ The first, fifth and tenth eigenstates are shown in
Figure \ref{fg:airy}(a).
We will proceed as before, by considering the gas on one side of
the piston (the left), and lifting a weight attached to that side,
by raising the floor below it. From now on, when referring to the
piston, or it's position, we will be referring to the entire
system of piston, pulleys, and 'pan' supporting the weight.

If the floor is raised through a distance $\delta h$ the change in
energy will be $\delta E=M_w g \delta h$ (which is independant of
the eigenstate\footnote{The old set of eigenstates $A_n(x)$ will
transform into new eigenstates $A_n(x-\delta h)$. If the floor is
raised sufficiently slowly, then by the adiabatic theorem, there
will be no transitions between states.}). By contrast, when the
piston expands through a distance $\delta Y,$ the change in
internal energy of the n'th eigenstate of the gas will be $\delta
E_n=-\frac{8\epsilon n^2}{(1-p+Y)^3}\delta Y$. If the expansion of
the gas is to exactly supply the energy to lift the weight, a
gearing mechanism that raises the weight through a different
distance than that moved by the piston is required, so that
$h=h(Y)$ and

\[
\frac{\partial h}{\partial Y}=\frac{8\epsilon n^2}{M_wg(1-p+Y)^3}
\]

However, the height raised should not be dependant upon the
specific eigenstate of the gas as there will be a statistical
ensemble of gas states. We cannot arrange for pulley connecting
the piston to the weight to have a different gearing ratio for
different states of the gas. Instead a mean gearing ratio must be
used, such as

\[
\frac{\partial h}{\partial Y}=\frac \alpha {\left( 1-p+Y\right) ^3}
\]

The exact form of the function $h(Y)$ can only be determined when
we know the statistical ensemble, in Section
\ref{s:szsm4}\footnote{The insensitivity of $h(Y)$ to $n$ means
that there will be a difference between the energy extracted from
the expanding gas and the energy put into raising the weight. This
will have to be drawn from a work reservoir. Fortunately it will
be shown, in Section \ref{s:szsm4}, that the energy drawn from the
work reservoir can be made negligible.}. For now we will simply
represent the gearing by the function $h(Y)$. The final height of
the floor of the raised weight is $h_T=h(1-p)$ and we will assume
$h(0) =0$. We will simplify the Dirac notation by dropping the
$h$, so that the wavefunction $A_n(z,h(Y))=\bk{z}{A_n(Y)}$. Figure
\ref{fg:airy}(b) shows the effect upon the fifth eigenstate
$A_5(z,h)$ as the floor height is raised.

Following the same procedure as in Section \ref{s:szqm3} above,
the subsystem internal energy for the lefthand weight is given by
the Hamiltonian

\begin{equation}
H_{W2}^\lambda=\sum_n \rho(Y_n) H_W(h(Y_n))
\label{eq:hw2}
\end{equation}

where
$\rho(Y_n)=\kb{\Phi_R(Y_n)}{\Phi_R(Y_n)}+\kb{\Phi_L(Y_n)}{\Phi_L(Y_n)}$
and we can write
\[
H_W(h(Y_n))=\sum_m (h(Y_n)-a_mH)M_wg
    \kb{A_m^\lambda(Y_n)}{A_m^\lambda(Y_n)}
\]

We now need to construct a 'raising weight' unitary operator
$U_{W3}(t)$ to describe the joint motion of the combined gas,
piston and weights. If we look at the situation where the gas is
located on the left, and only include the description of the
lefthand weight, the appropriate unitary operator is

\begin{eqnarray*}
U^\lambda_{W3}(\tau)&=& \sum_{l,m} \{
 \sum_{n=1}^{j-2}
    \kb{A_m^\lambda(Y_{n+1}) \Psi_l^\lambda(Y_{n+1})\Phi_R(Y_{n+1})}
       {A_m^\lambda(Y_n)\Psi_l^\lambda(Y_n)\Phi_R(Y_n)} \\
 &&+ \sum_{n=2}^{j-1}
    \kb{A_m^\lambda(Y_{n-1})\Psi_l^\lambda(Y_{n-1})\Phi_L(Y_{n-1})}
       {A_m^\lambda(Y_n)\Psi_l^\lambda(Y_n)\Phi_L(Y_n)} \\
 &&+\kb{A_m^\lambda(1-p)\Psi_l^\lambda(1-p)\Phi(1-p)}
       {A_m^\lambda(Y_{j-1})\Psi_l^\lambda(Y_{j-1})\Phi_R(Y_{j-1})} \\
 &&+\kb{A_m^\lambda(Y_{j-1})\Psi_l^\lambda(Y_{j-1})\Phi_L(Y_{j-1})}
       {A_m^\lambda(1-p)\Psi_l^\lambda(1-p)\Phi(1-p)} \\
 &&+\kb{A_m^\lambda(0)\Psi_l^\lambda(0)\Phi(0)}
        {A_m^\lambda(Y_1)\Psi_l^\lambda(Y_1)\Phi_L(Y_1)} \\
 &&+\kb{A_m^\lambda(Y_1)\Psi_l^\lambda(Y_1)\Phi_R(Y_1)}
        {A_m^\lambda(0)\Psi_l^\lambda(0)\Phi(0)}
\}
\end{eqnarray*}

This operator expresses the same behaviour as the operator
$U_{P2}^\lambda(\tau)$, in Equation \ref{eq:szqm1}, but now
includes the lifting of the weight. The first line represents the
piston moving to the right, the gas state on the left of the
piston expanding slightly, and the lefthand weight rising from
$h(Y_n)$ to $h(Y_{n+1})$. The second line gives the corresponding
motion of the piston moving to the left, the gas on the left
compressing, and the lefthand weight being lowered slightly. Third
and fourth lines show the piston reaching the right end of the
Szilard box, and the weight reaching it's maximum height, before
the piston is reflected and starts to compress the gas while
lowering the weight. Finally the fifth and sixth lines represent
the left moving piston reaching maximum compression of the gas, on
the left of the piston, in the center of the box, with the weight
coming to a rest on the floor, before the piston reverses
direction under pressure from the gas, and starts to move to the
right again, with the expanding gas lifting the weight.

As Figure \ref{fg:airy}(b) shows, raising the weight can leave
substantial overlap between states, so that $\bk{A_m^\lambda
(Y_i)}{A_m^\lambda(Y_j)}\neq \delta_{ij}$ in general. However, as
in Equation \ref{eq:szdel2}, the orthogonality of the piston
states ensures that the operator is a permutation of orthonormal
states. Furthermore, for any given position $Y$ of piston, and so
by $h(Y)$ a given position of the pan under the weight, the
$\ket{A_m^\lambda(Y)}$ form a complete basis for the subspace of
the weight. The set of joint $(l,m,n,A)$ states
$\ket{A_m^\lambda(Y_n)\Psi_l^\lambda(Y_n)\Phi_A(Y_n)}$ therefore
spans the accessible space of the joint system, and the operator
is unitary.

We now, by symmetry, construct a similar operator for the one atom
gas located entirely to the right of the piston. Now we
temporarily ignore the lefthand weights, and obtain from Equation
\ref{eq:szqm2}

\begin{eqnarray*}
U_{W3}^\rho(\tau) &=& \sum_{l,m}
    \{\sum_{n=-j+1}^{-2}
    \kb{A_m^\rho(Y_{n+1})\Psi_l^\rho(Y_{n+1})\Phi_R(Y_{n+1})}
        {A_m^\rho(Y_n)\Psi_l^\rho(Y_n)\Phi_R(Y_n)}
\\ && +\sum_{n=-j+2}^{-1}
    \kb{A_m^\rho(Y_{n-1})\Psi_l^\rho(Y_{n-1})\Phi_L(Y_{n-1})}
        {A_m^\rho(Y_n)\Psi_l^\rho(Y_n)\Phi_L(Y_n)}
\\ &&+\kb{A_m^\rho(0)\Psi_l^\rho(0)\Phi(0)}
        {A_m^\rho(Y_{-1})\Psi_l^\rho(Y_{-1})\Phi_R(Y_{-1})}
\\ && +\kb{A_m^\rho(Y_{-1})\Psi_l^\rho(Y_{-1})\Phi_L(Y_{-1})}
        {A_m^\rho(0)\Psi_l^\rho(0)\Phi(0)}
\\ && +\kb{A_m^\rho(-1+p)\Psi_l^\rho(-1+p)\Phi(-1+p)}
        {A_m^\rho(Y_{-j+1})\Psi_l^\rho(Y_{-j+1})\Phi_L(Y_{-j+1})}
\\ && +\kb{A_m^\rho(Y_{-j+1})\Psi_l^\rho(Y_{-j+1})\Phi_R(Y_{-j+1})}
        {A_m^\rho(-1+p)\Psi_l^\rho(-1+p)\Phi(-1+p)}
\}
\end{eqnarray*}

We now need to combine this into a single unitary operator.
Denoting the identity operator upon the unraised lefthand weight
space by
\[
I^\lambda_W=\sum_m \kb{A_m^\lambda (0)}{A_m^\lambda (0)}
\]
and that on the unraised righthand weight by
\[
I^\rho_W=\sum_m \kb{A_m^\rho (0)}{A_m^\rho (0)}
\]
we have a combined operator
\begin{eqnarray}
U_{W4}(\tau)&=&
    \left[U^\lambda_{W3}(\tau) \otimes I^\rho_W \right]
    \oplus
    \left[I^\lambda_W \otimes U^\rho_{W3}(\tau) \right]
\label{eq:szqm3}
\end{eqnarray}

 This unitary operator may be associated with a Hamiltonian
$H_{W4}$, constructed from the subsystem interaction Hamiltonians,
in the same manner as discussed above in Section \ref{s:szqm3},
and the complete expansion of the system of gas, piston and
weights has the Hamiltonian

\[
H_{T4}=(1-h(t)) \left[ H_{G2}^\lambda +H_{W2}^\lambda +H_{G2}^\rho
+H_{W2}^\rho \right] +h(t) H_{W4}+H_P
\]

We now simplify Equation \ref{eq:szqm3}, by allowing the
interaction to run for exactly the time necessary for a complete
expansion, or compression, of the one atom gas, and include only
those states which can be obtained from an initial subspace in
which the piston is located in the center of the box ($Y=0$). This
gives us the unitary operation
\begin{eqnarray}
U_{W4}&=&\sum_{l,m,n}
    \kb{A_m^\lambda(0)A_n^\rho(h(-1+p))\Psi_l^\rho(-1+p)\Phi(-1+p)}
        {A_m^\lambda(0)A_n^\rho(0)\Psi_l^\rho(0)\Phi(0)} \nonumber \\
&&+\kb{A_m^\lambda(0)A_n^\rho(0)\Psi_l^\rho(0)\Phi(0)}
    {A_m^\lambda(0)A_n^\rho(h(-1+p))\Psi_l^\rho(-1+p)\Phi(-1+p)} \nonumber \\
&&+\kb{A_m^\lambda(h(1-p))A_n^\rho(0)\Psi_l^\lambda(1-p)\Phi(1-p)}
        {A_m^\lambda(0)A_n^\rho(0)\Psi_l^\lambda(0)\Phi(0)} \nonumber \\
&&+\kb{A_m^\lambda(0)A_n^\rho(0)\Psi_l^\lambda(0)\Phi(0)}
        {A_m^\lambda(h(1-p))A_n^\rho(0)\Psi_l^\lambda(1-p)\Phi(1-p)}
        \label{eq:expand}
\end{eqnarray}

This operator simply generalises the conclusions of Section
\ref{s:szqm3}, to include the two weights in the quantum
description of the Popper-Szilard Engine. With the initial state
\[
\ket{\chi_\mathrm{initial}}=\left(
    \alpha
    \ket{A_l^\lambda (0) A_m^\rho (0) \Psi_n^\rho(0)}
    +\beta
    \ket{A_l^\lambda (0) A_m^\rho (0) \Psi_n^\lambda(0)}
    \right)\ket{\Phi(0)}
\]
the system will evolve into
\begin{eqnarray*}
\ket{\chi_\mathrm{final}}&=&
    \alpha
    \ket{A_l^\lambda (0) A_m^\rho (-1+p)
    \Psi_n^\rho(-1+p)\Phi(-1+p)} \\
    &&+\beta
    \ket{A_l^\lambda (1-p) A_m^\rho (0) \Psi_n^\lambda(1-p)\Phi(1-p)}
\end{eqnarray*}

The internal energy of the one atom gas can apparently be
converted into the energy required to lift a quantum weight,
although it may leave the system of piston and weights in an
entangled superposition. This completes the analysis of the stage
of the Popper-Szilard Engine shown in Figure \ref{fg:raise}(b).

\section{Resetting the Engine}
\label{s:szqm5}

The previous two Sections have analysed the interaction of the one
atom gas, moveable piston and weights, using quantum mechanics. We
have seen that, contrary to the assertions of \cite{Zur84,BS95},
the piston is not stationary when the one atom gas is in a
superposition. Instead, the joint system evolves into an entangled
superposition. This has significance for the final problem that
must be addressed in this Chapter: the issue of restoring the
Popper-Szilard Engine to it's initial state before commencing a
second cycle. As we recall, it is this, according to \cite[pages
25-28]{LR90} that requires work to be performed upon the system.
The three stages identified in Section \ref{s:szmd3} associated
with resetting the piston position are shown in Figure
\ref{fg:raise}(c-e) and are dealt with in this Section.

First, for Stage (c), we must see what the effect of inserting a
shelf at height $h_T=h(1-p)$ has upon the weights. This stage is
significant as the weights are quantum systems and this leads to a
wavefunction where there is a probability of finding an unraised
weight {\em above} the shelf.

For Stage (d) we construct states to describe the piston when it
is outside the box, and a unitary operator that incorporates the
effect upon the gas of inserting and removing the piston.

In Stage (e) we will attempt to construct a unitary operator that
restores the piston to the center, ready for re-insertion. We will
find that correlating the position of the piston to the position
of the weights is necessary to attempt to return the piston to the
center, but even so, cannot be achieved without some error, due to
the quantum nature of the weights shown in Stage (c).

The effects of this error will be shown to lead to a possibility
of the Popper-Szilard Engine going into reverse. The consequences
of this will be evaluated in later Chapters.

\subsection{Inserting Shelves}

The insertion of the shelves on each side can be considered as the
raising of an infinitely high potential barrier at height
$h_T=h(1-p)$ in the Hamiltonians of {\em both} weights. For the
raised weight, this will have no effect upon the wavefunction, as
the quantum weight wavefunction $A_n(z,h(1-p))$ is non-zero only
above the height $h_T$.

For the unraised weight, however, the wavefunction $A_n(z,0)$ has
a 'tail' that, for large values of $z$, has the form
$\frac{e^{-\frac{2}{3}z^{2/3}}}{z^{1/4}}$. While this is small, it
is non-zero and so there is always some possibility of finding a
quantum weight above the height $h_T$. While we could attempt to
treat this by an adiabatic raising of the potential barrier, as we
did for the one atom gas, the form of the wavefunction below the
shelf does not have a simple solution. Instead we will proceed by
a rapid insertion of the potential barrier, and project out the
portions of the wavefunctions above and below the shelf height.

For a given state, $\ket{A_n(0)}$, the projected state on finding
the weight above the shelf height is given by:
\begin{eqnarray*}
\ket{RA_n(h_T)}&=&\frac {1}{\alpha_n(h_T)}
    \int_{h_T}^\infty \ket{z}\bk{z}{A_n(0)}dz \\
\magn{\alpha_n(h_T)} &=&
    \int_{h_T}^\infty \magn{A_n(z,0)} dz
\end{eqnarray*}
while the 'unraised' state (below the shelf height) is
\begin{eqnarray*}
\ket{UN_n(h_T)}&=&\frac {1}{\beta_n(h_T)}
    \int_0^{h_T}\ket{z}\bk{z}{A_n(0)}dz \\
\magn{\beta_n(h_T)} &=&
    \int_0^{h_T}\magn{A_n(z,0)} dz
\end{eqnarray*}
so that
\[
\ket{A_n(0)}=
    \alpha_n(h_T)\ket{RA_n(h_T)}
    +\beta_n(h_T)\ket{UN_n(h_T)}
\]

$\magn{\alpha_n(h)}$ is the probability of finding an unraised
weight above the height $h$. Unfortunately, the values of
$\alpha_n(h_T)$ and $\beta_n(h_T)$ do not generally have simple
expressions\footnote{Although as $A_n(z,0)$ is a real function,
$\alpha _n(h_T)$ and $\beta _n(h_T)$ will always be real
numbers.}. However, using the properties of Airy functions we are
able to calculate approximate values of these for large values of
$n$. \pict{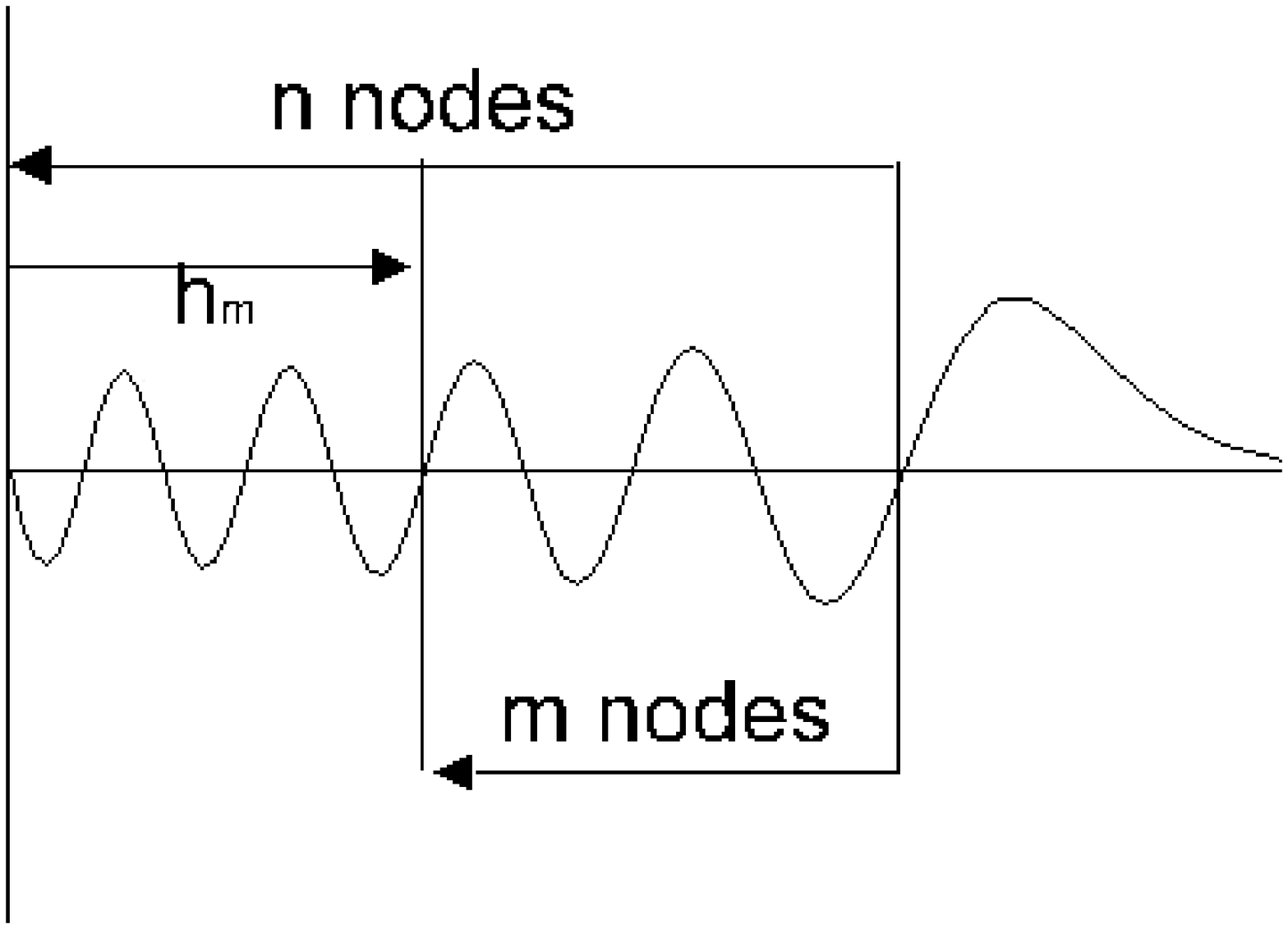}{Splitting Airy Function at Height $h$} The
wavefunction $A_n(z,0)$ has $n$ nodes above the floor at $z=0$,
which occur at heights $h_m=(a_m-a_n)H$, where $m<n$ (remembering
that the values $a_n,a_m<0$). This is shown in Figure
\ref{fg:airy2}. When the shelf is inserted at the height of a node
$a_m$, we can calculate the value of $\alpha_n(h_m)$ from Equation
\ref{eq:airy1}, and the properties of integrals of Airy functions
$A(z)$

\begin{eqnarray*}
\int_{h_m}^\infty \magn{A_n(z,0)}dz &=&
    \frac{1}{A^{\prime}(a_n)^2H}
        \int_{(a_n-a_m)H}^\infty A\left(\frac{z}{H}-a_n\right)^2dz \\
    &=&\frac{1}{A^{\prime}(a_n)^2}
        \int_{a_m}^\infty A(z)^2dz \\
    &=&\frac{1}{A^{\prime}(a_n)^2}
        \left[ -A^{\prime}(z)^2+zA(z)^2\right]_{a_m}^\infty \\
    &=&\left(\frac{A^{\prime}(a_m)}{A^{\prime}(a_n)}\right)^2
\end{eqnarray*}

If $m\gg 1$ the asymptotic value $A^{\prime }(a_m) \approx
\frac{(-)^{m-1}}{\sqrt{\pi}}\left(\frac{3\pi
m}2\right)^{\frac{1}{6}}$ leads to the result

\[
\alpha_n(h_m)=\left(\frac{m}{n}\right)^{\frac{1}{6}}
\]

If the shelf is not inserted at the position of a node, we must
interpolate between the nearest two nodes. As $\alpha_n(h_m)$
varies slowly for large $m$, this will be a reasonable
approximation. Using the asymptotic value $a_l=-\left(\frac{3\pi
l}{2}\right)^{\frac{2}{3}}$ and $h_m=(a_n-a_m)H$ to estimate an
interpolated (non-integer) value of $m$, we can approximate
$\alpha_n(h)$ for any shelf height from:

\begin{eqnarray}
h &=& \left(
     \left(\frac{3\pi n}{2}\right)^{\frac{2}{3}}
    -\left(\frac{3\pi m}{2}\right)^{\frac{2}{3}}
\right)H\nonumber \\
m &=&n\left(
    1-\left(\frac{2}{3\pi n}\right)^{\frac{2}{3}}
        \frac{h}{H}\right)^{\frac{3}{2}} \nonumber \\
\alpha_n(h)&\approx&
    \left(1-\left(\frac{2}{3\pi n}\right)^{\frac{2}{3}}
        \frac{h}{H}\right)^{\frac{1}{4}} \label{eq:alpha1}
\end{eqnarray}

This is valid whenever the height is lower than the final node
$(h<-a_nH)$. If $h>-a_nH$ the shelf is inserted into the 'tail' of
the wavefunction. To estimate the value of $\alpha_n(h)$ in this
case, we will evaluate the probability that the weight is located
anywhere above the height $-a_nH$, which must be larger than the
probability of the weight located above $h$

\begin{eqnarray*}
\alpha_n(-a_nH)^2 &=&
    \frac{1}{A^{\prime}(a_n)^2}\int_{a_n}^\infty A(z)^2dz \\
    &=&\left(\frac{A^{\prime}(0)}{A^{\prime}(a_n)}\right)^2
\end{eqnarray*}

Using $A^{\prime}(0) \approx -0.25$ and $n \gg 1$ as before, this
gives
\[
\alpha_n(h)^2 <
    \frac{\pi}{16}\left(\frac{2}{3\pi n}\right)^{\frac{1}{3}}
\]
which may be treated as negligible. In effect, we have shown that
if $h>\left(\frac{3 \pi n}{2}\right)^\frac{2}{3}H$, or,
equivalently,
\[
n<\frac {2}{3\pi }\left(\frac{h}{H}\right)^{\frac{3}{2}}
\]
then we can approximate
\begin{eqnarray}
\alpha_n(h)&=&0 \nonumber \\
\beta_n(h)&=&1 \label{eq:alpha2}
\end{eqnarray}

When
\[
n \geq \frac {2}{3\pi }\left(\frac{h}{H}\right)^{\frac{3}{2}}
\]
we calculate $\alpha_n(h)$ from Equation \ref{eq:alpha1} above,
and $\beta_n(h)$ from
\begin{equation}
\beta_n(h)=\sqrt{1-\alpha_n(h)^2} \label{eq:alpha3}
\end{equation}

This completes the calculation of the effect of inserting the
shelves at height $h$ in Stage (c) of the Popper-Szilard cycle.

\subsection{Removing the Piston}

We will now consider Stage (d) of the cycle. The piston state is
removed from the ends of the box, effectively 'switching off' the
interaction between the gas and the piston.

Firstly, we need to introduce quantum states to describe the
piston outside the box. These will be the orthonormal states, with
$\ket{\phi_L}$,$\ket{\phi_R}$ and $\ket{\phi_0}$ describing the
piston outside the box, but in the lefthand, righthand and central
positions, respectively. These states also include the pulley and
pan, and so the state $\ket{\phi_L}$ implies that the righthand
weight is raised, and so on.

We now need a general unitary operator to account for the
insertion and removal of the piston from the box. This will have
an effect upon the internal states of the gas. As noted in
Equation \ref{eq:gas}, when the piston is at one or the other end
of the box, the gas will be approximately in an unperturbed energy
eigenstate\footnote{There will be a slight expansion of the gas
states, of order $2p$ as the piston is removed. Technically this
could be used to perform work upon the piston during it's removal.
However, we shall ignore this effect as negligible.} and so will
be unaffected by the piston's removal. If the piston was in the
center of the box when it was removed, however, it's removal can
have a significant effect upon the state of the gas. This effect
is the adjoint operation to inserting the piston into the center
of the box, in Section \ref{s:szqm2}. The complete insertion and
removal operator is therefore
\begin{eqnarray}
U_{IR}&=&I_G \otimes
    \{ \kb{\phi_L}{\Phi(-1+p)}+\kb{\Phi(-1+p)}{\phi_L} \nonumber\\
        &&+\kb{\phi_R}{\Phi(1-p)}+\kb{\Phi(1-p)}{\phi_R} \} \nonumber\\
    &&+U_G \otimes \kb{\Phi(0)}{\phi_0}
        +U_G^\dag \otimes \kb{\phi_0}{\Phi(0)}
\label{eq:ir}
\end{eqnarray}
where $I_G$ is the identity operator upon the gas states, and
$U_G$ is from Equation \ref{eq:barrier} in the limit of the
infinitely high barrier.

\subsection{Resetting the Piston}

We now need to consider Stage (e). This is the critical stage to
the argument of Leff and Rex. They argue that Landauer's Principle
implies an expenditure of $kT_G\ln2$ energy to reset the piston
states. However, we have suggested that the piston may be returned
to $\ket{\phi_0}$ without such an expenditure, by correlating it
to the weights. We will now show that the piston may indeed by
returned in this way, but, due to the quantum nature of the
weights, there is always some possibility of error in the
resetting mechanism.

First, it will be useful to consider if we can reset the piston
without correlating to the weights. The ideal operation would
include
\begin{eqnarray*}
U_{R1}\ket{\phi_L}&=&\ket{\phi_0} \\
U_{R1}\ket{\phi_R}&=&\ket{\phi_0}
\end{eqnarray*}
but this is clearly non-unitary as orthogonal states are being
mapped to non-orthogonal states. The most general operation acting
only upon the piston states is
\begin{eqnarray*}
U_{R2}\ket{\phi_0}&=&
    a_1\ket{\phi_0}+b_1\ket{\phi_L}+c_1\ket{\phi_R} \\
U_{R2}\ket{\phi_L}&=&
    a_2\ket{\phi_0}+b_2\ket{\phi_L}+c_3\ket{\phi_R} \\
U_{R2}\ket{\phi_R}&=&
    a_3\ket{\phi_0}+b_3\ket{\phi_L}+c_3\ket{\phi_R} \\
\end{eqnarray*}
Unitarity requires that the vectors $a_i$,$b_i$ and $c_i$ (with
$i=1,2,3$) are orthonormal (or, equivalently, the vectors
$\alpha_1$, $\alpha_2$ and $\alpha_3$ with $\alpha=a,b,c$).

To maximise the probability of the piston being returned to the
center, we need to maximise $\magn{a_2}+\magn{a_3}$. This would
imply setting $a_1=0$. However, if we are not going to change the
state of the weights, the piston initially in the state
$\ket{\phi_0}$ cannot be moved to either $\ket{\phi_L}$ or
$\ket{\phi_R}$ as these states both imply one of the pans is
raised. We are therefore constrained to have $a_1=1$ and so there
is no possibility of resetting the piston. We must, therefore,
include the states of the weights.

After the piston is removed from the box, we will have combined
piston and weight states of:
\begin{eqnarray*}
\ket{A_m^\lambda(0)A_n^\rho(1-p)\phi_L} \\
\ket{A_m^\lambda(1-p)A_n^\rho(0)\phi_R}
\end{eqnarray*}
If we simply attempt to correlate the action on the piston with
the raised and unraised states, $\ket{A_m(1-p)},\ket{A_m(0)}$ we
would construct a resetting operator along the lines of
\begin{eqnarray*}
U_{R3}\ket{A_m^\lambda(0)A_n^\rho(1-p)\phi_L}
    &=&\ket{A_m^\lambda(0)A_n^\rho(1-p)\phi_0} \\
U_{R3}\ket{A_m^\lambda(1-p)A_n^\rho(0)\phi_R}
    &=&\ket{A_m^\lambda(1-p)A_n^\rho(0)\phi_0}
\end{eqnarray*}
However, the inner product of these input states is given by
\begin{eqnarray*}
\bk{A_m^\lambda(0)A_n^\rho(1-p)\phi_L}
    {A_m^\lambda(1-p)A_n^\rho(0)\phi_R}&=&
    \bk{A_m^\lambda(0)}{A_m^\lambda(1-p)}
    \bk{A_n^\rho(1-p)}{A_n^\rho(0)}
    \bk{\phi_L}{\phi_R} \\
    &=&0
\end{eqnarray*}
while the inner product of the output states is
\begin{eqnarray*}
\bk{A_m^\lambda(0)A_n^\rho(1-p)\phi_0}
    {A_m^\lambda(1-p)A_n^\rho(0)\phi_0}&=&
    \bk{A_m^\lambda(0)}{A_m^\lambda(1-p)}
    \bk{A_n^\rho(1-p)}{A_n^\rho(0)}
    \bk{\phi_0}{\phi_0} \\
&=& \bk{A_m^\lambda(0)}{A_m^\lambda(1-p)}
    \bk{A_n^\rho(1-p)}{A_n^\rho(0)} \\
& \neq & 0
\end{eqnarray*}
The output states are not orthogonal as the Airy functions of the
raised and unraised weight states overlap, as shown in Figure
\ref{fg:airy}. $U_{R3}$ is still not a unitary operator.

To construct a proper unitary operator we need to correlate the
movement of the piston to the projection of the weights above or
below the shelf. The relevant projection operators are
\begin{eqnarray*}
P(RA)&=&\int_{h_T}^\infty \kb{z}{z}dz \\
P(UN)&=&\int_0^{h_T} \kb{z}{z}dz
\end{eqnarray*}

However it is more useful to construct them from the raised
eigenstates:
\begin{eqnarray*}
P(RA)&=& \sum_n \kb{A_n(1-p)}{A_n(1-p)}
\end{eqnarray*}
or from the projections of the unraised eigenstates:
\begin{eqnarray*}
P(RA) &=&\sum_n\alpha_n(h_T)^2 \kb{RA_n}{RA_n} \\
    &=&\int \int_{h_T}^\infty \kb{z}{z}
        \sum_n\ket{A_n}\bk{A_n}{z^\prime}\bra{z^\prime}dzdz^\prime \\
    &=&\int_{h_T}^\infty \kb{z}{z}dz \\
P(UN) &=&\sum_n\beta_n(h_T)^2 \kb{UN_n}{UN_n} \\
    &=&\int \int_0^{h_T} \kb{z}{z}
        \sum_n\ket{A_n}\bk{A_n}{z^\prime}\bra{z^\prime}dzdz^\prime \\
    &=&\int_0^{h_T} \kb{z}{z}dz
\end{eqnarray*}
From these it follows that:
\begin{eqnarray*}
P(RA)\ket{A_n(0)}&=&\alpha_n\ket{RA_n} \\
P(UN)\ket{A_n(0)}&=&\beta_n\ket{UN_n} \\
P(RA)\ket{A_n(1-p)}&=&\ket{A_n(1-p)} \\
P(UN)\ket{A_n(1-p)}&=&0
\end{eqnarray*}
\begin{figure}[htb]
    \resizebox{\textwidth}{!}{
        \includegraphics{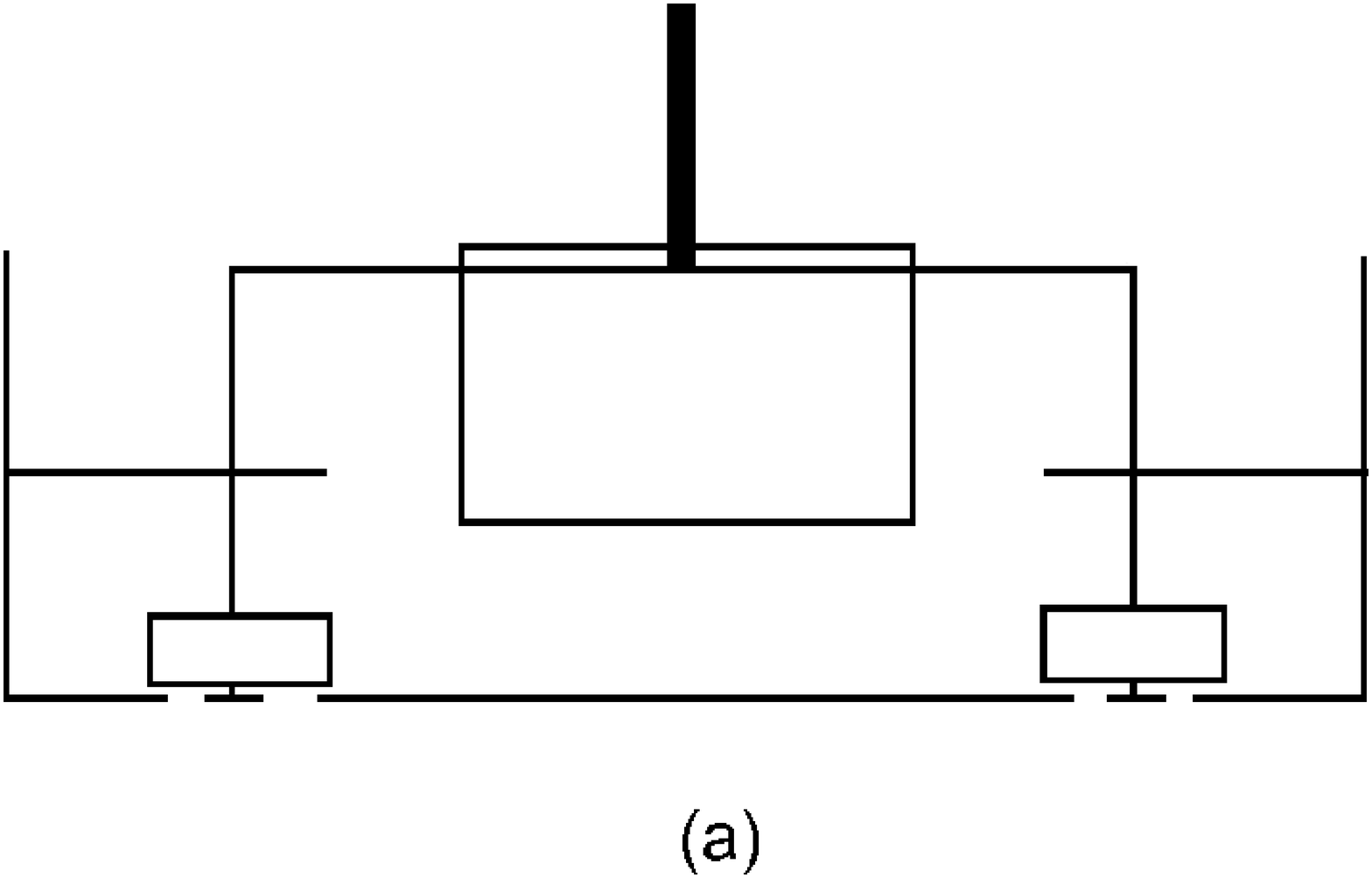}
        \includegraphics{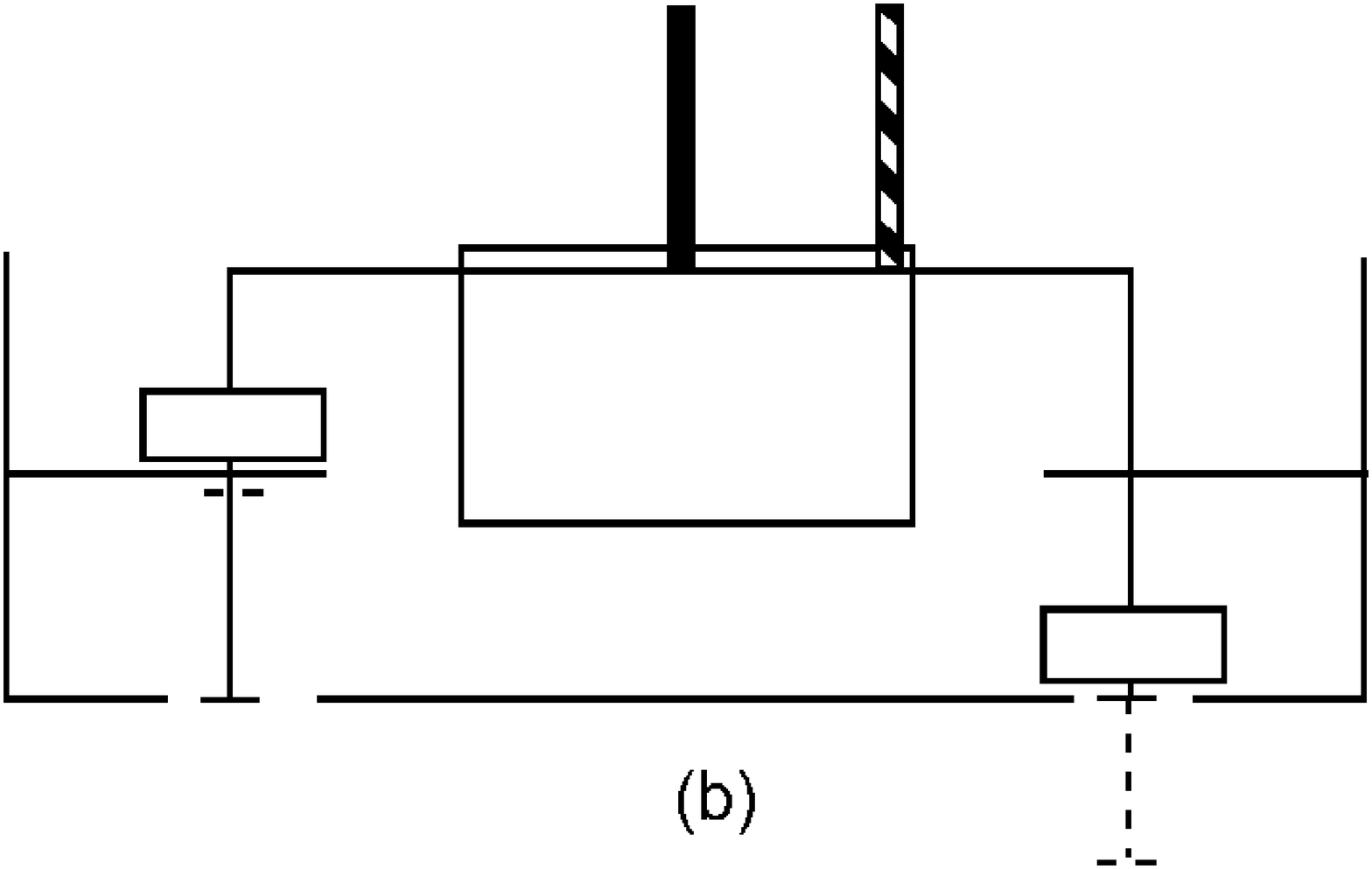}
        }
    \resizebox{\textwidth}{!}{
        \includegraphics{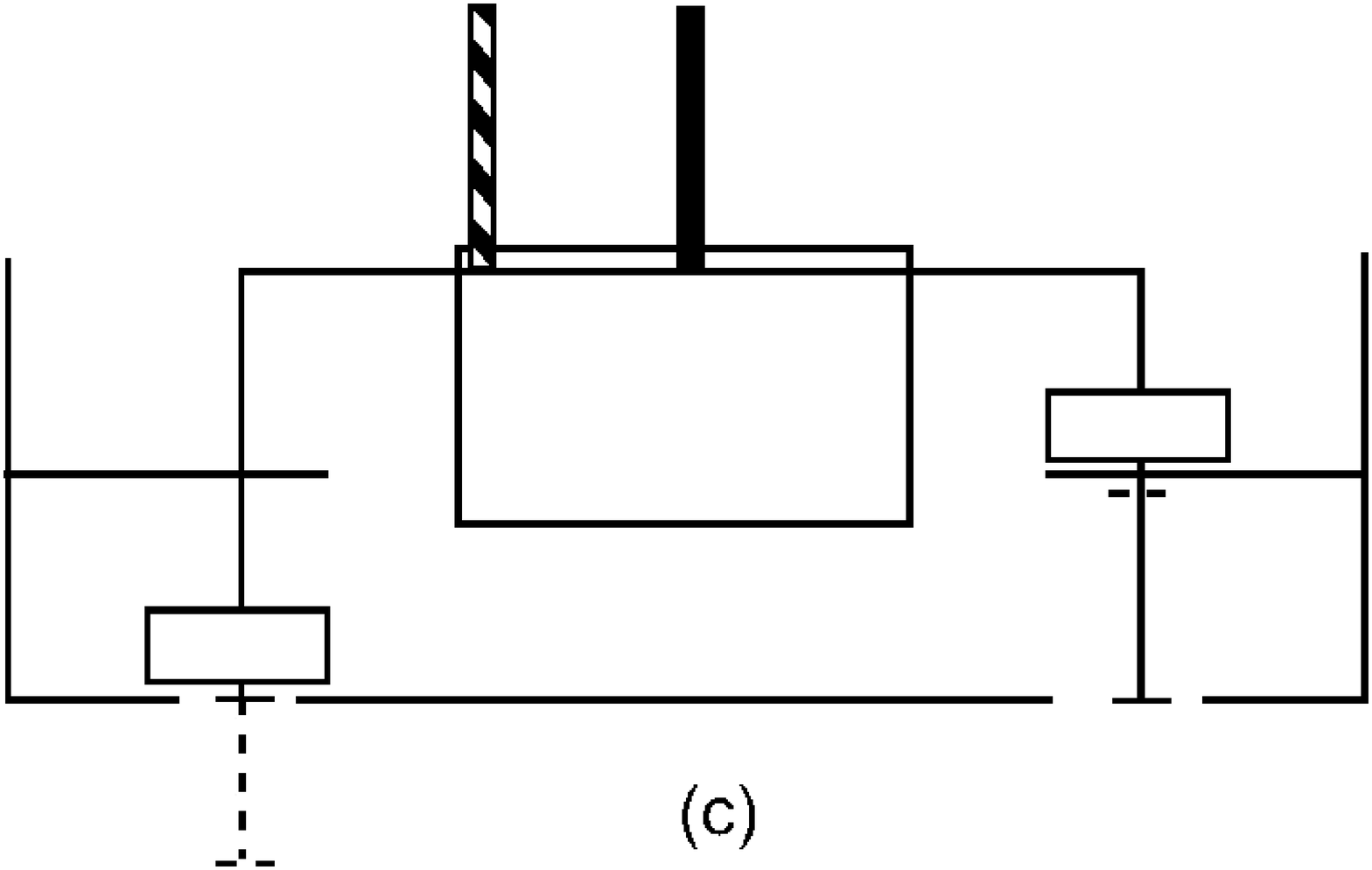}
        \includegraphics{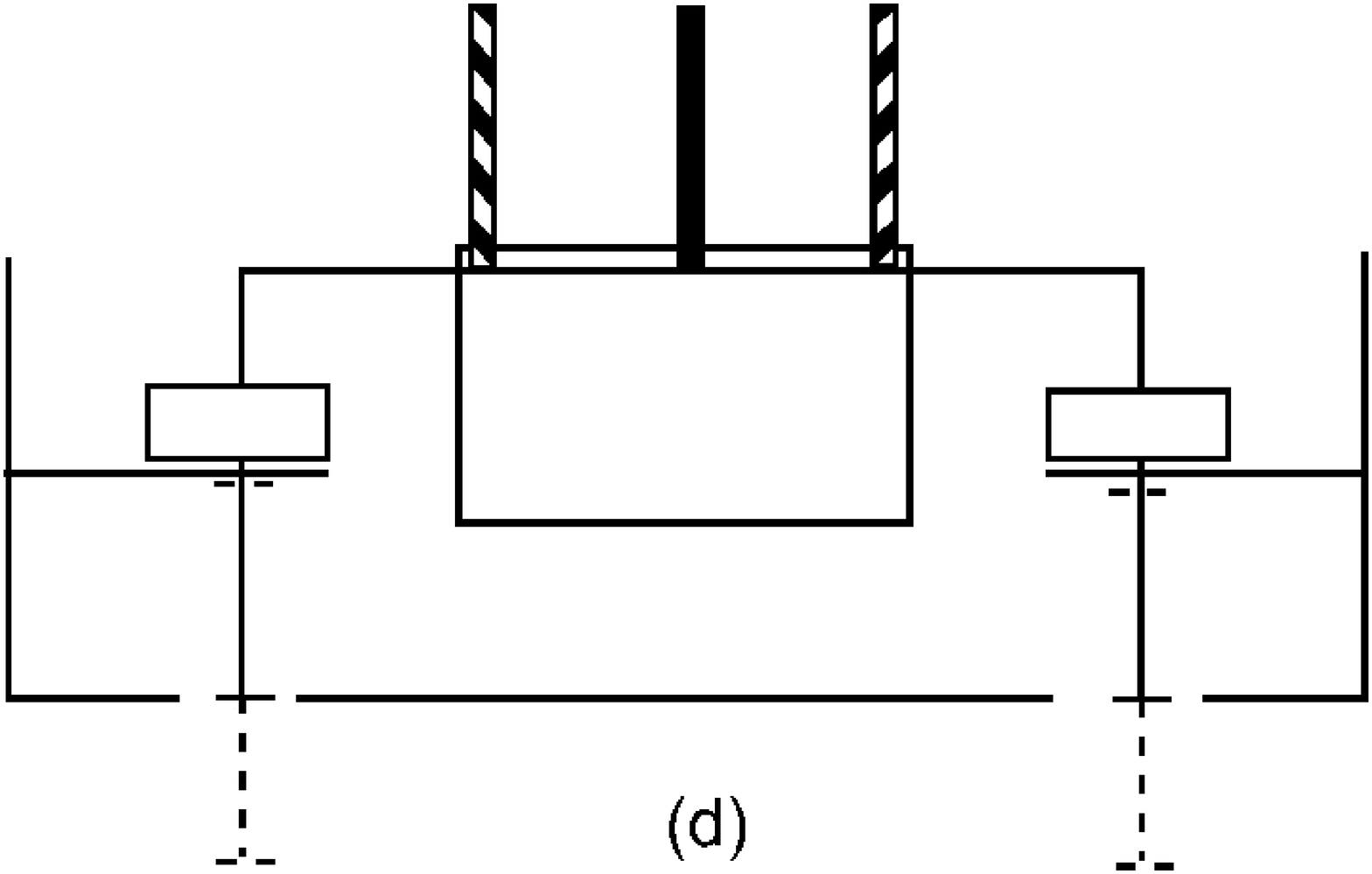}
        }
    \caption{Correlation of Weights and Piston Position
    \label{fg:weights}}
\end{figure}
We will now examine the correlation between the state of the
weights and the piston position. There are eight orthonormal sets
of states that are accessible for the combined system. These are
shown in Figure \ref{fg:weights}.
\begin{itemize}
\item (a) Both weights are resting upon the floor, below the shelf. The piston must
be located in the center of the Engine. The allowed state is:
\[
\ket{UN^\lambda(h_T) UN^\rho(h_T) \phi_0}
\]
\item (b) The left weight on the shelf and the right weight on the floor.
The piston can be in the center, or at the right of the engine.
Allowed states are:
\begin{eqnarray*}
\ket{RA^\lambda(h_T) UN^\rho(h_T) \phi_0} \\
\ket{A^\lambda(1-p) UN^\rho(h_T) \phi_R}
\end{eqnarray*}
\item (c) The left weight on the floor and the right weight on the shelf. The piston
may now be found either in the center, or at the left of the
engine. Allowed states are:
\begin{eqnarray*}
\ket{UN^\lambda(h_T) RA^\rho(h_T) \phi_0} \\
\ket{UN^\lambda(h_T) A^\rho(1-p) \phi_L}
\end{eqnarray*}
\item (d) Both weights are upon the shelves. The piston may be located at any of the
three locations:
\begin{eqnarray*}
\ket{RA^\lambda(h_T) RA^\rho(h_T) \phi_0}\\
\ket{RA^\lambda(h_T) A^\rho(1-p) \phi_L}\\
\ket{A^\lambda(1-p) RA^\rho(h_T) \phi_R}
\end{eqnarray*}
\end{itemize}
If the resetting interaction is not to change the location of the
weights, these must form four separate subspace under the
operation.

We can now state the most general form of the resetting operation,
consistent with the requirements of unitarity.

\begin{eqnarray}
U_{RES} &=& \kb{\phi_0}{\phi_0}P^\lambda(UN)P^\rho(UN) \nonumber\\
    &&+\left[\kb{\phi_R}{\phi_0}+\kb{\phi_0}{\phi_R}\right]
        P^\lambda(RA)P^\rho(UN) \nonumber\\
    &&+\left[\kb{\phi_L}{\phi_0}+\kb{\phi_0}{\phi_L}\right]
        P^\lambda(UN)P^\rho(RA) \nonumber\\
    &&+\left[\kb{\phi_1}{\phi_0}+\kb{\phi_2}{\phi_L}
        +\kb{\phi_3}{\phi_R}\right]P^\lambda(RA)P^\rho(RA)
    \label{eq:reset}
\end{eqnarray}

The first line represents the subspace where both weights are
located beneath the shelf height. The only possible location of
the piston is in the center.

The second and third lines represent one weight above and one
weight below the shelf. When the piston is located in the
corresponding left or right position, we want to reset the piston
by moving it to the center. To preserve unitarity with this, the
reset operator must also include a term moving the piston
initially located in the center to the appropriate left or right
position.

Finally, when both weights are located above the shelf height, in
line four, the weights do not correlate to the location of the
piston.  The most general transformation possible is given, where
the $\ket{\phi_j}$ states are superpositions of the
$\ket{\phi_0}$, $\ket{\phi_L}$ and $\ket{\phi_R}$ states:
\begin{eqnarray*}
\ket{\phi_1}&=&a_1\ket{\phi_0}+b_1\ket{\phi_L}+c_1\ket{\phi_R} \\
\ket{\phi_2}&=&a_2\ket{\phi_0}+b_2\ket{\phi_L}+c_2\ket{\phi_R} \\
\ket{\phi_3}&=&a_3\ket{\phi_0}+b_3\ket{\phi_L}+c_3\ket{\phi_R}
\end{eqnarray*}

For the operation to be unitary, orthonormal states must transform
into orthonormal states, so $\bk{\phi_i}{\phi_j}=\delta_{ij}$.
This leads to the conditions
\begin{eqnarray*}
a_1^{*}a_2+b_1^{*}b_2+c_1^{*}c_2 &=&0 \\
a_1^{*}a_3+b_1^{*}b_3+c_1^{*}c_3 &=&0 \\
a_2^{*}a_3+b_2^{*}b_3+c_2^{*}c_3 &=&0
\end{eqnarray*}
\begin{eqnarray}
a_1^{*}a_1+b_1^{*}b_1+c_1^{*}c_1 &=&1 \nonumber \\
a_2^{*}a_2+b_2^{*}b_2+c_2^{*}c_2 &=&1 \nonumber \\
a_3^{*}a_3+b_3^{*}b_3+c_3^{*}c_3 &=&1 \label{eq:unres}
\end{eqnarray}
Rearranging the expression
\begin{eqnarray*}
&&\left[\kb{\phi_1}{\phi_0}+\kb{\phi_2}{\phi_L}+\kb{\phi_3}{\phi_R}\right] \\
&=&\ket{\phi_0}\left\{a_1\bra{\phi_0}+a_2\bra{\phi_L}+a_3\bra{\phi_R}\right\} \\
&&+\ket{\phi_L}\left\{b_1\bra{\phi_0}+b_2\bra{\phi_L}+b_3\bra{\phi_R}\right\} \\
&&+\ket{\phi_R}\left\{c_1\bra{\phi_0}+c_2\bra{\phi_L}+c_3\bra{\phi_R}\right\}
\end{eqnarray*}
leads to an equivalent set of conditions
\begin{eqnarray*}
a_1^{*}a_1+a_2^{*}a_2+a_3^{*}a_3 &=&1 \\
b_1^{*}b_1+b_2^{*}b_2+b_3^{*}b_3 &=&1 \\
c_1^{*}c_1+c_2^{*}c_2+c_3^{*}c_3 &=&1
\end{eqnarray*}
\begin{eqnarray*}
a_1^{*}b_1+a_2^{*}b_2+a_3^{*}b_3 &=&0 \\
a_1^{*}c_1+a_2^{*}c_2+a_3^{*}c_3 &=&0 \\
b_1^{*}c_1+b_2^{*}c_2+b_3^{*}c_3 &=&0
\end{eqnarray*}

We can examine the effect of this operator by considering the
effect upon the state where the piston is to the left, before the
shelves are inserted
\[\ket{A_m^\lambda(0)A_n^\rho(1-p)\phi_L}\]
When the shelves are inserted this becomes separated into raised
and unraised portions of the lefthand weight
\[\alpha_m(h_T)\ket{RA_m^\lambda(h_T)A_n^\rho(1-p)\phi_L}+
\beta_m(h_T)\ket{UN_m^\lambda(h_T)A_n^\rho(1-p)\phi_L}\]. The
operation of $U_{RES}$ on the unraised portion of the wavefunction
moves the piston to the center. The effect of $U_{RES}$ on the
raised portion is to set the piston state to $\ket{\phi_2}$. This
makes the state
\begin{eqnarray*}
&&\alpha_m(h_T)\ket{RA_m^\lambda(h_T)A_n^\rho(1-p)\phi_2}
    +\beta_m(h_T)\ket{UN_m^\lambda(h_T)A_n^\rho(1-p)\phi_0} \\
&&=\alpha_m(h_T)b_2\ket{RA_m^\lambda(h_T)A_n^\rho(1-p)\phi_L} \\
&&+\alpha_m(h_T)c_2\ket{RA_m^\lambda(h_T)A_n^\rho(1-p)\phi_R}\\
&&+\left(\alpha_m(h_T)a_2\ket{RA_m^\lambda(h_T)}
    +\beta_m(h_T)\ket{UN_m^\lambda(h_T)}\right)\ket{A_n^\rho(1-p)\phi_0}
\end{eqnarray*}
Although the resetting operation has partially succeeded, there is
still some probability of finding the piston to the left or right
of the Engine, whatever choice we make for the values of $a_i$
etc. Selection of the optimum values of the $a_i$'s can only be
made once we include the full statistical mechanics in Chapter
\ref{ch:szsm}.

This completes the analysis of Stage (e) of the Popper-Szilard
Engine in this chapter. We have found that the quantum state of
the weight leads to the possibility of an unraised weight being
spontaneously located above the height $h_T$ through which the
raised weight has been lifted. This possibility, combined with the
requirement that the resetting operation be unitary, leads to an
imperfect resetting. This is clearly not sufficient to show that
the Popper-Szilard Engine does not work. The error in the
resetting is only partial, and it is not yet certain that an
optimal choice of resetting operation could not violate the second
law of thermodynamics.

\section{Conclusions}\label{s:szqm6}

We have examined the operation of the quantum Popper-Szilard
Engine given in Figure \ref{fg:raise} in detail, explicitly
constructing unitary operations for all relevant stages of the
cycle. We will now summarise this cycle, and consider the effects
of the errors in the resetting operation.

There is a final unitary operation we need to add to the ones
constructed. This is the act of inserting and removing the shelves
at height $h_T$, at Stages (c) and (f). This can be treated by
assuming a narrow potential barrier is inserted in the Hamiltonian
in Equation \ref{eq:airyh}. The result is a time dependant
perturbation of the Hamiltonian, exactly equivalent to the raising
or lowering of the potential barrier in the one atom gas, in
Section \ref{s:szqm2}. The unitary operator for this can be
constructed in the same manner as the operator $U_G$ in Equation
\ref{eq:barrier}. We will not explicitly do this, but will simply
describe the unitary operator corresponding to the insertion of
the shelves by $U_S$ and their removal by $U_S^\dag$. The complete
cycle of the Popper-Szilard Engine is now given by the unitary
operation:

\begin{equation}
U_T = U_S^\dag U_{RES} U_{RI} U_S U_{W4} U_{RI} \label{eq:cycle}
\end{equation}

Moving from right to left through $U_T$, the successive stages
are:
\begin{itemize}
\item $U_{RI}$ Stage (a) Equation \ref{eq:ir}
\item $U_{W4}$ Stage (b) Equation \ref{eq:expand}
\item $U_S$ Stage (c) above
\item $U_{RI}$ Stage (d) Equation \ref{eq:ir}
\item $U_{RES}$ Stage (e) Equation \ref{eq:reset}
\item $U_S^\dag$ Stage (f) above
\end{itemize}

We will now review the effect of $U_T$ on the system.

\subsection{Raising Cycle}
If we start from the state where the piston is in the center,
outside the box, and both weights are at rest upon the floor, the
state is
\[\ket{A_m^\lambda(0)A_n^\rho(0)\phi_0}\]
We can now see how the operation of $U_T$ attempts to reproduce
the cycle in Figure \ref{fg:raise}.

\begin{itemize}
\item $U_{RI}$ The insertion of the piston in the center
of the box (Section \ref{s:szqm2})
\item $U_{W4}$ The expansion of the one atom gas against
the piston, lifting one of the weights. This may leave the system
in an entangled superposition (Sections \ref{s:szqm3},
\ref{s:szqm4}).
\item $U_S$ Inserting shelves on both sides at height
$h_T$.
\item $U_{RI}$ Removing the piston from the box (Section
\ref{s:szqm5})
\item $U_{RES}$ Resetting the piston by correlating it's
state to the location of the raised or unraised weights (Section
\ref{s:szqm5})
\item $U_S^\dag$ Removing the shelves and allowing any
raised weights to fall to the floor
\end{itemize}
This will be described as a 'raising cycle'.

We saw in Section \ref{s:szqm5} above, that this leaves the Engine
in a superposition of states. To complete the cycle, we want the
Engine to be in state
\[\ket{A_m^\lambda(0)A_n^\rho(0)\phi_0}\]
at the end of Stage (f). However, due to the imperfect nature of
the resetting, the Engine is in a superposition with states such
as
\begin{eqnarray*}
\ket{A_m^\lambda(0)A_n^\rho(1-p)\phi_L} \\
\ket{A_m^\lambda(1-p)A_n^\rho(0)\phi_R}
\end{eqnarray*}
We must now consider the effect of starting a new cycle with these
states.
\subsection{Lowering Cycle}
If the Engine starts with a raised weight on the righthand side,
and the piston to the left side of the Engine, the state will be
\[\ket{A_m^\lambda(0)A_n^\rho(1-p)\phi_L}\]
We must now consider the effect of $U_T$ on this state.

\begin{figure}[htb]
    \resizebox{\textwidth}{!}{
        \includegraphics{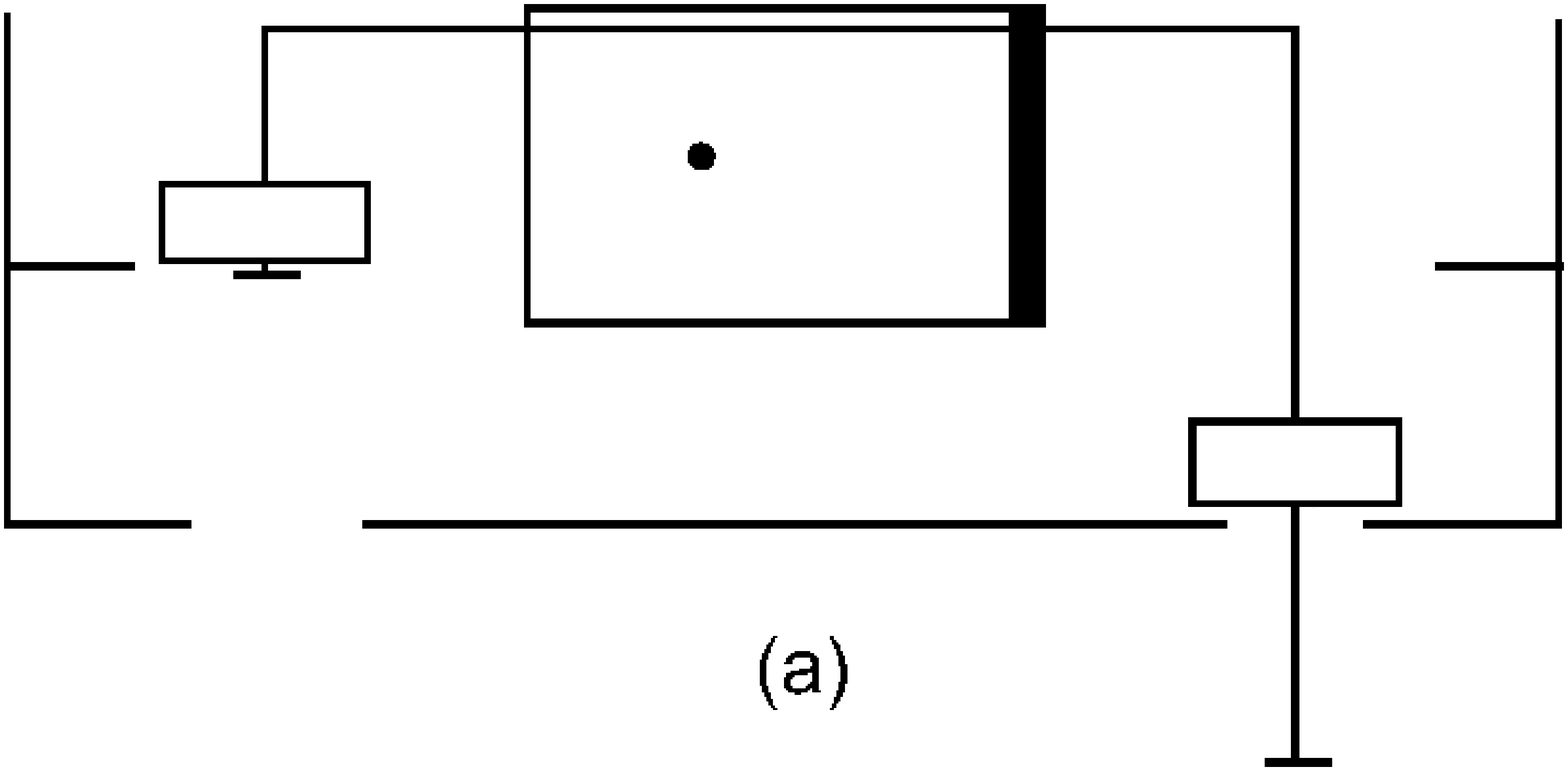}
        \includegraphics{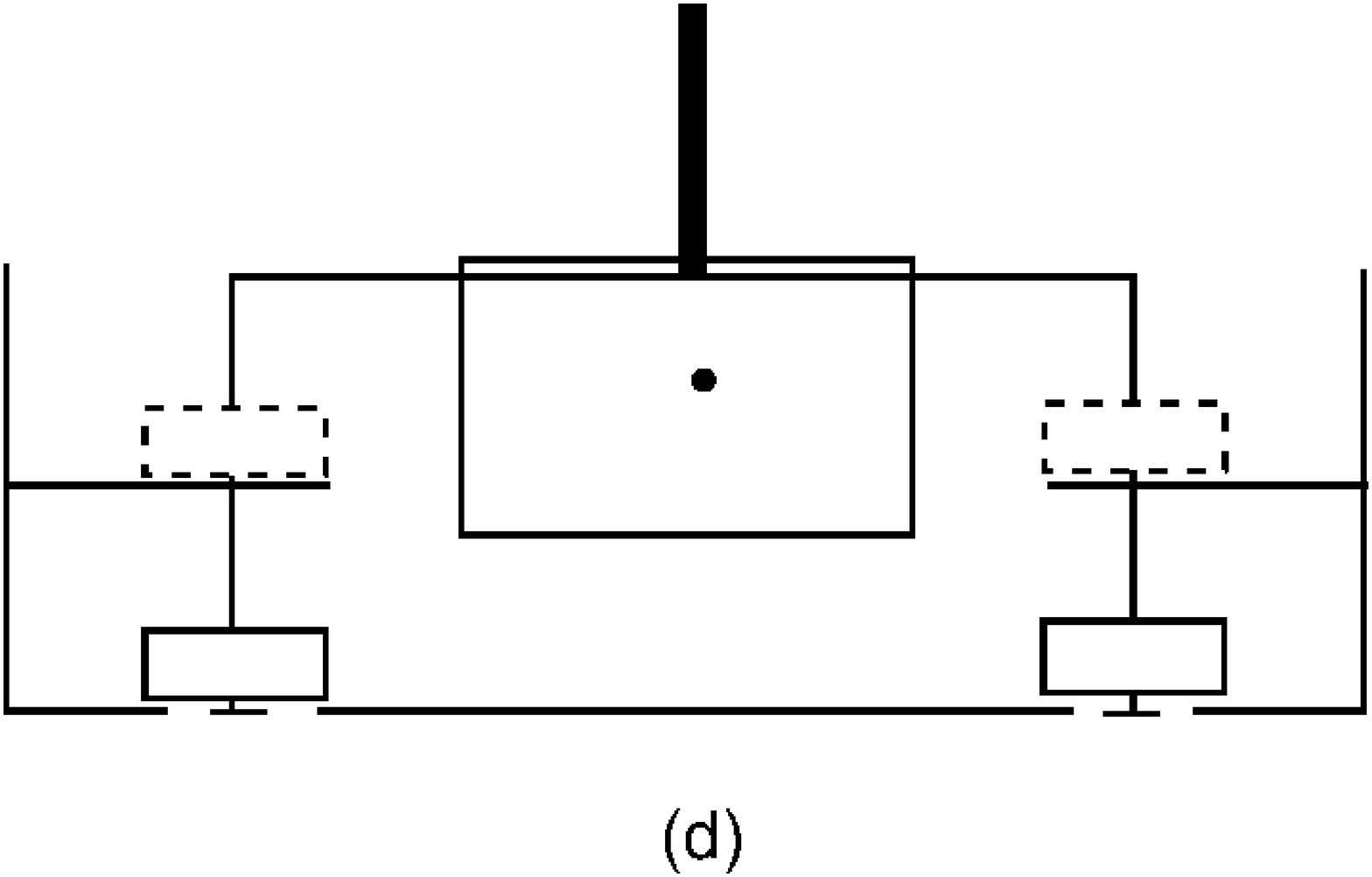}
        }
    \resizebox{\textwidth}{!}{
        \includegraphics{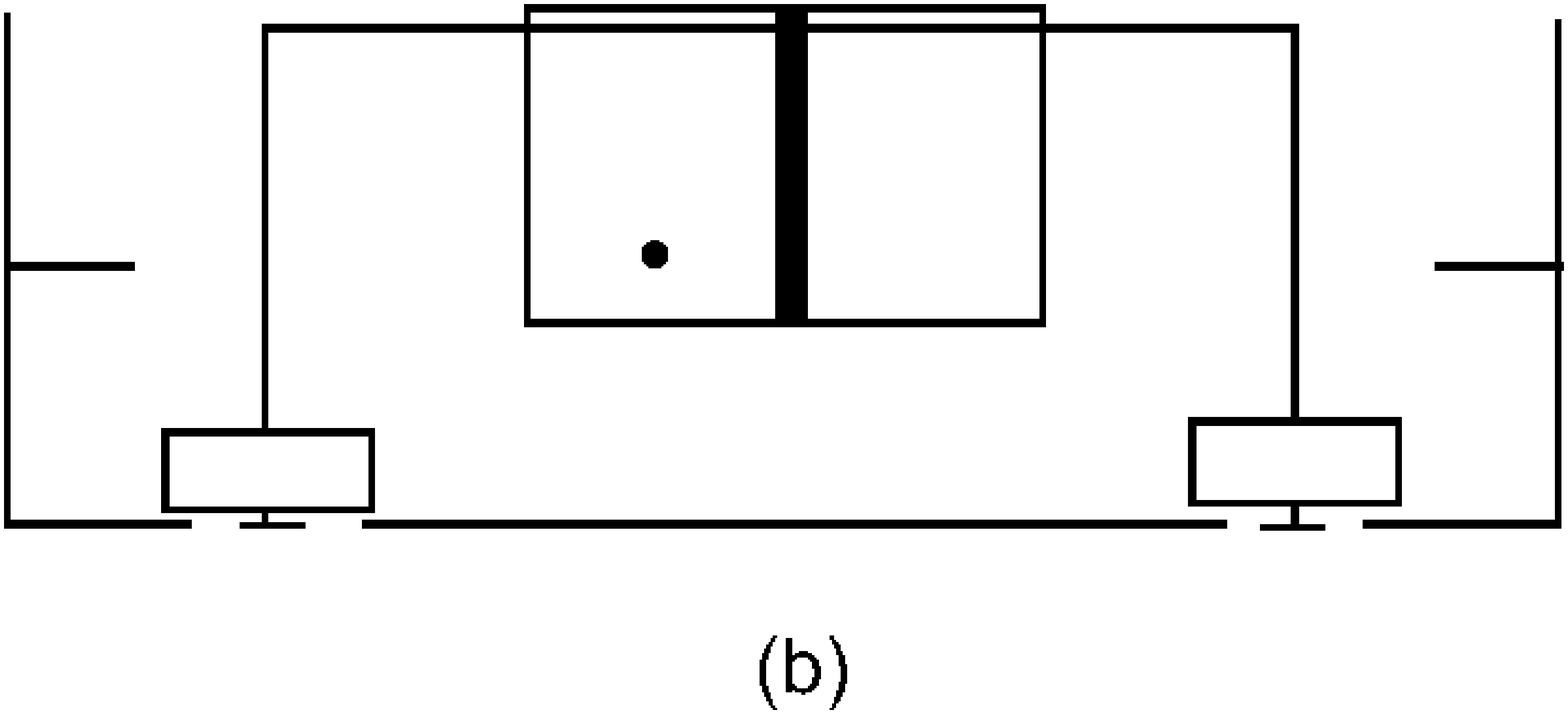}
        \includegraphics{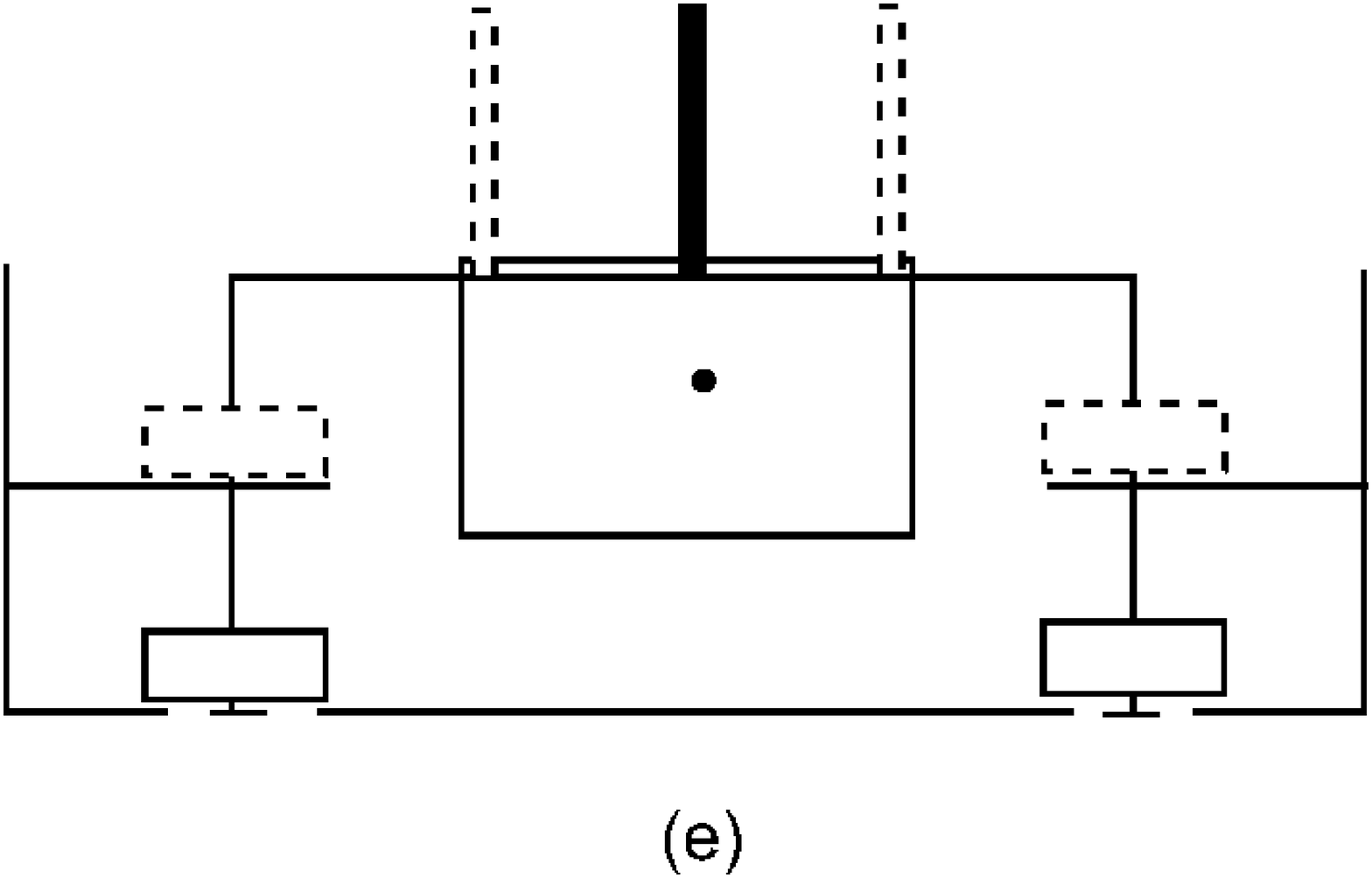}
    }
    \resizebox{\textwidth}{!}{
        \includegraphics{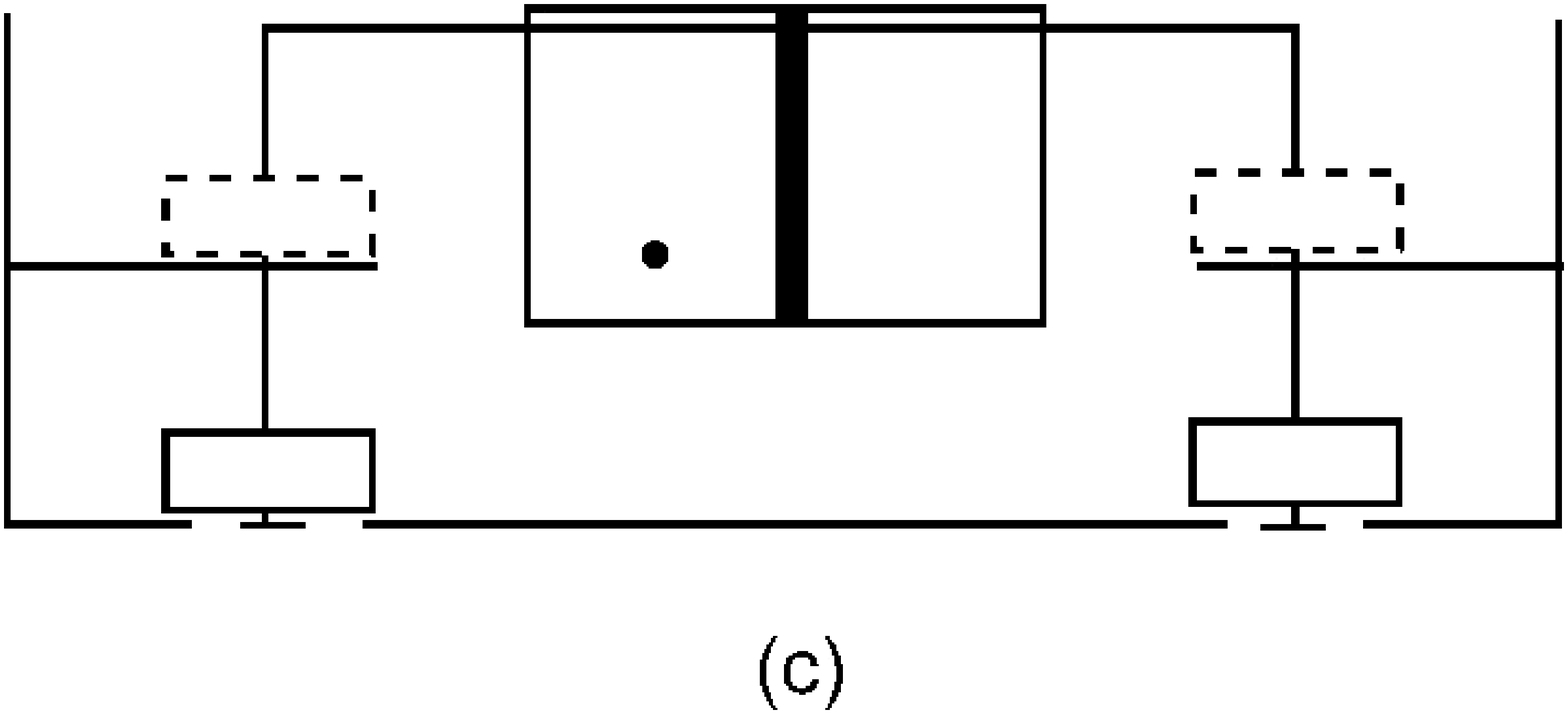}
        \includegraphics{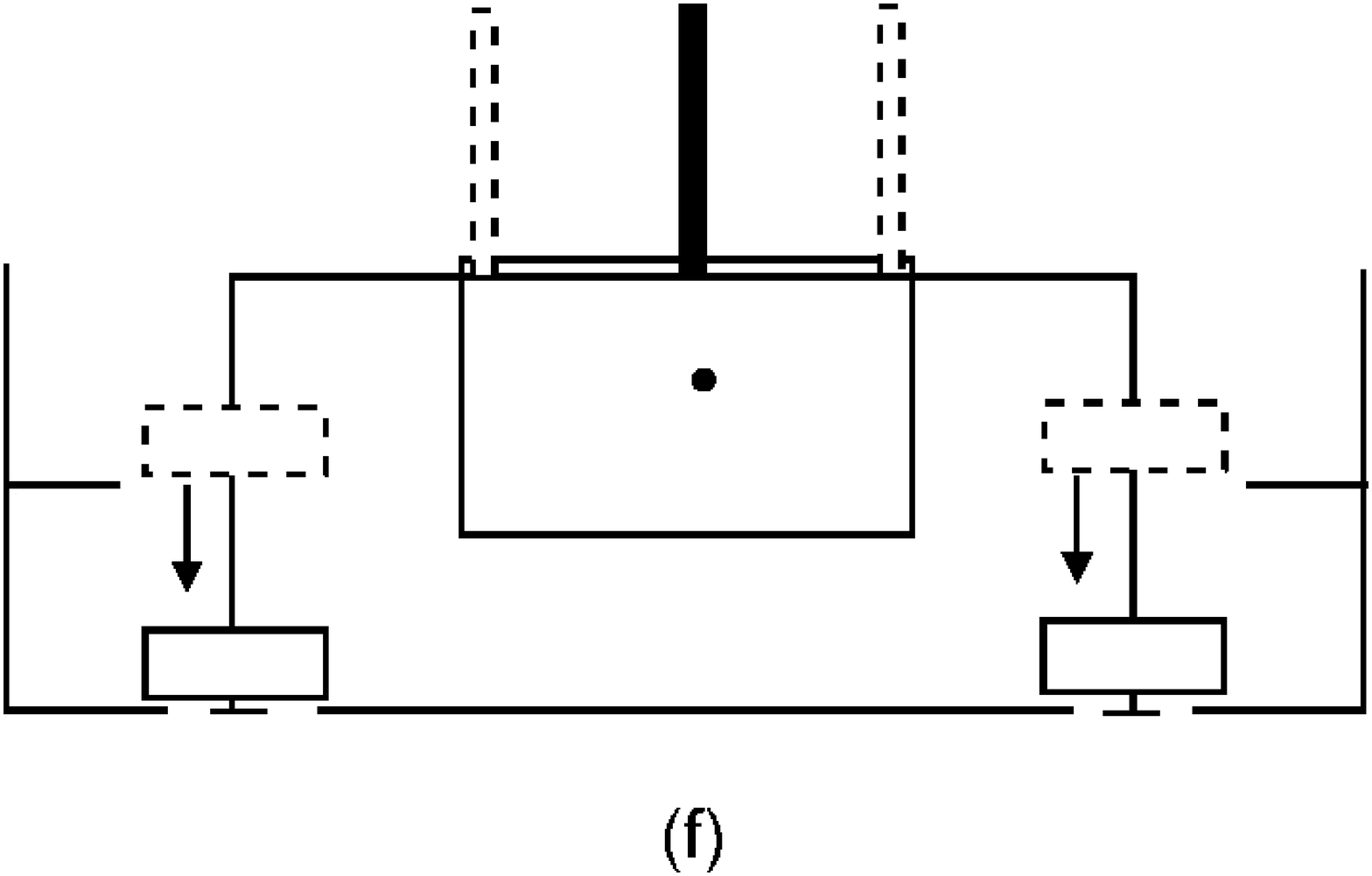}
    }
    \caption{The Lowering Cycle of the Popper-Szilard Engine
    \label{fg:lower}}
\end{figure}

\begin{itemize}
\item $U_{RI}$ The piston is inserted into the box on
the lefthand side. Negligible compression of the gas takes place.
The state is now
\[\ket{A_m^\lambda(0)A_n^\rho(1-p)\Psi_l^\rho(-1+p)\Phi(-1+p)}\]
\item $U_{W4}$ The combined gas, piston and weight
system now runs through a {\em compression} phase. The righthand
weight is lowered, and the piston moves from the left to the
center of the box, compressing the gas to the right. The energy of
the weight is reduced and the internal energy of the gas is
raised. The system is left in state
\[\ket{A_m^\lambda(0)A_n^\rho(0)\Psi_l^\rho(0)\Phi(0)}\]
\item $U_S$ At the end of Stage (b) both weights are
in the unraised state. When the shelves emerge there is a
possibility that either, or both, could be trapped above the shelf
height $h_T$. This involves rewriting
\begin{eqnarray*}
\ket{A_m\lambda(0)A_n^\rho(0)\Psi_l^\rho(0)\Phi(0)}&=&
    \left(\alpha_m(h_T)\alpha_n(h_T)
        \ket{RA_m^\lambda(h_T)RA_n^\rho(h_T)} \right.\\
&&+    \alpha_m(h_T)\beta_n(h_T)
        \ket{RA_m^\lambda(h_T)UN_n^\rho(h_T)}\\
&&+    \beta_m(h_T)\alpha_n(h_T)
        \ket{UN_m^\lambda(h_T)RA_n^\rho(h_T)}\\
&&\left. +    \beta_m(h_T)\beta_n(h_T)
        \ket{UN_m^\lambda(h_T)UN_n^\rho(h_T)}\right)\ket{\Psi_l^\rho(0)\Phi(0)}
\end{eqnarray*}
\item $U_{RI}$ The piston is removed from the {\em center} of the box. As the one
atom gas was confined to the right of the piston, this will have a
significant effect upon the gas state, as it is allowed to expand
to occupy the entire box. This involves replacing
$\ket{\Psi_l^\rho(0)}$ with $\frac{1}{\sqrt{2}}
\left(\ket{\Psi_l^{\mathrm{even}}}-\ket{\Psi_l^{\mathrm{odd}}}\right)$
and $\ket{\Phi(0)}$ with $\ket{\phi_0}$.
\item $U_{RES}$ The resetting operation moves the piston
according to the location of the weights. As noted in Stage (c),
all four combinations of weight states occur with some
probability. After this operation the piston may therefore be
found in the left, right or central position
\begin{eqnarray*}
&& \left(\alpha_m(h_T)\alpha_n(h_T)
    \ket{RA_m^\lambda(h_T)RA_n^\rho(h_T)\phi_1} \right.\\
&& +\alpha_m(h_T)\beta_n(h_T)
    \ket{RA_m^\lambda(h_T)UN_n^\rho(h_T)\phi_R}\\
&& +\beta_m(h_T)\alpha_n(h_T)
    \ket{UN_m^\lambda(h_T)RA_n^\rho(h_T)\phi_L}\\
&& \left. +\beta_m(h_T)\beta_n(h_T)
    \ket{UN_m^\lambda(h_T)UN_n^\rho(h_T)\phi_0}\right)
    \frac{1}{\sqrt{2}}\left(\ket{\Psi_l^{\mathrm{even}}}
    -\ket{\Psi_l^{\mathrm{odd}}}\right)
\end{eqnarray*}
\item $U_S$ The shelves are removed, allowing
unsupported weights to fall to the floor. If the piston state is
in the $\ket{\phi_L}$ or $\ket{\phi_R}$, then the corresponding
right or lefthand weight will be supported at height $h_T$.
However, if the piston state is $\ket{\phi_0}$ then both weights
will fall to the floor.
\end{itemize}
We will describe this as the 'lowering cycle' and it is shown in
Figure \ref{fg:lower}. The key point to this cycle is that energy
is transferred from the {\em weight} to the {\em gas} during Stage
(b) . This is in the opposite direction to the 'raising cycle'.
 At the end of the 'lowering cycle' the piston may again be
found, outside the box, in the lefthand, righthand or central
positions. If the piston is in the center, then the next cycle of
$U_T$ will result in a 'raising cycle'. If the piston is instead
in the left or right states, then a weight is trapped at the
height $h_T$ and the system will continue with another 'lowering
cycle'.

\subsection{Summary}

This completes the analysis of the quantum mechanics of the
Popper-Szilard Engine. We have demonstrated how the Engine
proceeds without the need for external measurements or
interventions from 'demons'. The arguments of \cite{Zur84,BS95} do
not appear to be sustained with respect to the quantum state of
the one atom gas.

With respect to the arguments of \cite{LR90} we have shown that an
imperfect resetting does appear to be possible, without the need
to perform work upon the system. However, the imperfect resetting
leads to the possibility of the cycle of the Popper-Szilard Engine
reversing from a 'raising cycle' to a 'lowering cycle'. However,
at the end of a lowering cycle, there is a possibility of
reversing back onto a raising cycle. The Engine therefore switches
between the two cycles.

On raising cycles, energy is transferred from the one atom gas to
the weight. On lowering cycles, the energy in pumped in the
opposite direction. To avoid violating the second law of
thermodynamics, the energy flow must go from the hotter to the
colder system. This requires a delicate balance of probabilities.
If the temperature of the gas heat bath is lower than the
temperature of the weight heat bath, then the Engine must spend
more time transferring heat from the weights to the gas, and so
must spend most of it's time on the lowering cycle. Conversely, if
the one atom gas is hotter than the weights, the Engine must spend
most of it's time on the raising cycle. This must continue to hold
true for {\em all possible} choices of the parameters for
$U_{RES}$ given in Equation \ref{eq:unres}. To verify that this is
the case, we must introduce the statistical mechanical properties
of the Engine. We will do this in the next Chapter.
\chapter{The Statistical Mechanics of Szilard's Engine}
\label{ch:szsm}

In Chapter \ref{ch:szqm} we examined the physical limitations
imposed by quantum theory upon the interactions of the microstates
of the Popper-Szilard Engine. This would be sufficient if we
wished to analyse the Engine as a closed system, initially in a
definite quantum state. However, this is not the problem for which
the thought experiment was designed. The purpose of the analysis
is to decide whether the Engine is capable of transferring energy
between heat baths in an anti-entropic manner. For this we need to
introduce statistical mechanical concepts. These concepts will be
introduced and applied in this Chapter, and will demonstrate that
such anti-entropic behaviour is {\em not} possible.

Section 1 summarises the statistical mechanical concepts which
will be used. This includes ensembles, heat baths and generalised
pressure. With the exception of the temperature of the heat baths,
we will avoid making use of any explicitly thermodynamic
quantities, such as entropy or free energy.

Sections 2 and 3 will apply these concepts to the gas and the
weight subsystems, respectively, paying particularly close
attention to the changes in pressure and internal energies of
these systems, for different piston positions. In Section 4 we
will use the results of the previous two sections to calculate the
optimum gearing ratio $h(Y)$ for the piston and pulley system (see
Section \ref{s:szqm4}).

In Sections 5 and 6 we will put together these results to describe
the behaviour of the Popper-Szilard Engine for the raising and
lowering cycles, respectively. Section 7 will finally analyse the
mean flow of energy between the gas and weight heat baths. It will
now be possible to show that, for any choice of temperatures of
the two heat baths, and for any choice of resetting operation
$U_{RES}$, the long term behaviour of the Engine is to produce a
flow of energy from the hotter to the colder heat bath. The
Popper-Szilard Engine is therefore unable to produce anti-entropic
heat flows.

\section{Statistical Mechanics} \label{s:szsm1}

\subsubsection{Statistical Ensemble}
Many textbooks (\cite{Pen70,Wal85}, for example) introduce
statistical mechanics as the study of systems which have a large
number of constituents. It has been argued \cite{Pop74,Cha73} that
this is part of the explanation of the Szilard Paradox. However,
is not necessary that a system be large for statistical mechanics
to be used. Statistical mechanical concepts can be applied
whenever the preparation of a system, however large or small, does
not uniquely specify the initial state of the system. Instead we
must specify the probabilities $p_i$ of the different possible
initial states $\ket{\Gamma_i}$.

We will describe such a system using the Gibbs ensemble, where we
conceive of an infinite number of equivalently prepared systems,
with the initial states $\ket{\Gamma_i}$ occurring with relative
frequencies $p_i$. The ensemble is represented by the density
matrix $\rho=\sum_i p_i \proj{\Gamma_i}$ (\cite{Tol79,BH96a}, for
example). Obviously such an ensemble does not actually exist.
However, if we use the preparation method to prepare a finite
number of systems, with no special ordering, then the statistics
of the outcomes of the real systems will approach the statistics
of the ensemble\footnote{In \cite{Per93} the large finite number
of systems is referred to as an 'assembly'. If instead the systems
can be considered as occurring in a particular order, it may be
more accurate to describe them as a 'string'\cite{Zur89a}.} as the
number of systems becomes large. The ensemble is a representation
of the mean behaviour when the same experiment is repeated a large
number of times, and applies even when each experiment is
performed upon a system which consists of only a few constituents.

In our case we are therefore supposing an infinite number of
Popper-Szilard Engines, each connected to their own heat baths and
each containing only a single atom. We will describe the behaviour
of this 'representative ensemble' of Engines as the mean behaviour
of the Popper-Szilard Engine.

\subsubsection{Generalised Pressure}
The mean energy of a system is given by $E=\trace{\rho H}$, where
$H$ is the Hamiltonian. If the $\ket{\Gamma_i}$ are energy
eigenstates, with eigenvalues $E_i$, then this leads to $E=\sum_i
p_i E_i$, as we would expect. Typically, these $E_i$ depend upon
both internal co-ordinates (such as the location of the atoms in a
gas) and external co-ordinates (such as the location of the walls
surrounding the gas). The energy is a property of the internal
co-ordinate (such as the kinetic energy of the motion of the atoms
in the gas), while the external parameters define the boundary
conditions upon the eigenstates.

If the system is in state $\ket{\Gamma_i}$ and an external
parameter ($X$ for example) is changed, this affects the
eigenstate, and through it the energy of the state. The force that
is required to change the parameter is given by $\frac{\partial
E_i}{\partial X}$. For the ensemble the mean force, or generalised
pressure, on co-ordinate X is
\[
P(X)=\sum_i p_i(X) \frac{\partial E_i}{\partial X}
\]
The work done, or mean energy required, to
change the co-ordinate from $X_1$ to $X_2$ is therefore
\[
W=\int_{X_1}^{X_2}\sum_i p_i(X) \frac{\partial E_i}{\partial X}dX
\]
\subsubsection{Heat Baths}
An infinitesimal change in the Energy of a system is given by
$dE=\sum_i p_i dE_i + \sum_i E_i dp_i$. As $dE_i=\frac{\partial
E_i}{\partial X}dX$ we can see the first term corresponds to the
work, $dW$, done upon the system. The second term corresponds to
the change in heat, $dQ=\sum_i E_i dp_i$, and requires the system
to be in contact with an environment (in an isolated system,
occupation probabilities do not change). The 'environment' system
we will use will be the canonical heat bath.

The canonical heat bath consists of a large assembly of weakly
interacting systems, parameterised by the temperature $T$. Each
system has an internal Hamiltonian $H_B$. The density matrix of
individual system $n$, removed from the assembly, is given by the
canonical ensemble:
\[
\rho_n=\frac{e^{-H_B(n)/kT}}{\trace{e^{-H_B(n)/kT}}}
\]
The ensemble of the heat bath is
\[
\rho_B=\prod_n\frac{e^{-H_B(n)/kT}}{\trace{e^{-H_B(n)/kT}}}
\]
This is the {\em most likely} distribution consistent with a given
mean energy.

The most significant property of the canonical heat bath is the
effect of bringing another system into temporary
contact\footnote{By 'temporary contact' we mean that for a short
period there is a non-zero interaction Hamiltonian affecting the
two systems} with one of the heat bath subsystems. It can be shown
that if a system which is {\em not} initially described by
canonical distribution, is brought into successive contact with
many systems, which are each in a canonical distribution with
temperature $T$, the first system will approach a canonical
distribution, also with temperature
$T$\cite{Tol79,Par89a,Par89b,Per93}.

When a system is brought into contact with a heat bath, we assume
that it is in effect brought sequentially into contact with
randomly selected subsystems of the heat bath. This will gradually
bring the system into a canonical distribution with the same
temperature as the heat bath, so the density matrix of the system
itself becomes
\[
\rho=\frac{e^{-H/kT}}{\trace{e^{-H/kT}}}
\]
where $H$ is the systems internal Hamiltonian. As the heat bath
subsystems are weakly interacting, and there is a large number of
them, we will assume that any energy transferred to or from the
heat bath does not significantly affect the state of the heat
bath, and that any correlations that develop between heat bath and
system states are rapidly lost. This process of {\em
thermalisation}, by which the system is brought into equilibrium
with the heat bath at temperature $T$, occurs with a
characteristic time $\tau$, the thermal relaxation time.

This property needs qualifying with regard to {\em accessible}
states. It may be the case that the Hamiltonian $H$ can be
subdivided into separate Hamiltonians $H=H_1+H_2+...$ where $H_1$,
$H_2$ correspond to disjoint subspaces, between which there are no
transitions, or transitions can only take place at a very slow
rate.

An example of this would be locating a particle in one of several
large boxes, with the separate Hamiltonians corresponding to the
states within each box. In this case, placing the boxes in contact
with the heat bath over a time period of order $\tau$ will cause a
particle to be thermalised with a given box but would not cause
transitions between boxes. The resulting thermalised density
matrix $\rho ^{\prime }$ will be
\begin{equation}
\rho^{\prime}=\trace{P_1\rho}
    \frac{e^{-H_1/kT}}{\trace{e^{-H_1/kT}}}
    +\trace{P_2\rho}
    \frac{e^{-H_2/kT}}{\trace{e^{-H_2/kT}}}+\ldots
\label{eq:therm1}
\end{equation}

where $\rho $ is the initial, unthermalised, density matrix and
$P_1$ is the projection operator onto the subspace of $H_1$ and so
forth. If the contact is maintained for a much longer period of
time $\tau^{\prime\prime}$, so that significant numbers of
transitions between the $H_i$ states can take place, the complete
thermalisation will occur and
\[
\rho^{\prime\prime}=\frac{e^{-H/kT}}{\trace{e^{-H/kT}}}
\]
It should be noted that this implies there can be more than one
thermal relaxation time associated with a given system.

Developing this further, we must consider {\em conditional}
Hamiltonians
\[
H=\Pi _1H_1+\Pi _2H_2+\ldots
\]
where the $\Pi _1$'s are orthogonal projection operators on states
of a {\em second} quantum system, or Hilbert space. An example of
this might be a situation where a system has {\em spin}, but the
interaction between the system and the heat bath does not allow
transitions between spin states (or these transitions are
suppressed) and the $H_i$ do not explicitly include the spin
states. In this case the thermalisation will take place separately
within the separate spin subspaces.

In this case the effect of contact with the heat bath will be to
thermalise the density matrix to
\begin{equation}
\rho^{\prime\prime\prime}=
    \partrace{\Pi_1\rho}{1}\frac{e^{-H_1/kT}}{\trace{e^{-H_1/kT}}}
    +\partrace{\Pi_2\rho}{1}\frac{e^{-H_2/kT}}{\trace{e^{-H_2/kT}}}+\ldots
    \label{eq:therm2}
\end{equation}
where the trace is taken only over the Hilbert space of the first
system. This produces a density matrix for the {\em joint} system,
which has the property of no interference terms between the
subspaces of the {\em second} system. However, we should be clear
that there has been no interaction between the heat bath and the
second Hilbert space. Again, if there is a process by which
transitions take place between the states of the second Hilbert
space, then the complete thermalisation of the joint system may
take place, with a second, longer thermal relaxation time.

Within the context of the Popper-Szilard Engine, Equation
\ref{eq:therm1} will apply to situations where a single Hilbert
space is divided into a tensor sum of subspaces. This includes the
one atom gas, when the partition is raised in the center of the
box, or the unraised weight when the shelf is inserted. The
Hamiltonian in Equation \ref{eq:szh1} shows how the gas Hilbert
space divides into the two disjoint subspaces. Equation
\ref{eq:therm2} applies when there is a joint Hilbert space
composed of a tensor product of two (or more) Hilbert spaces, only
one of which is in thermal contact with a heat bath. This will
apply to the joint systems of the gas and piston located in the
box, and to the joint system of a raised weight and the pan
located beneath it. Equations \ref{eq:szhg2} and \ref{eq:hw2} give
the relevant conditional Hamiltonians for these cases.

In general there may be many relaxation times associated with the
thermalisation of a system, depending upon the different subspaces
and interactions with the heat bath. We will assume all relaxation
times are either very short (or effectively instantaneous), or
very long (or effectively infinite), with respect to the time
period over which the Popper-Szilard Engine operates.

The following transitions will be assumed to have short thermal
relaxation times: \begin{itemize}
\item Transitions between one atom gas states when the partition is not inserted in the box.
\item Transitions between one atom gas states on the same side of the piston or
partition.
\item Transitions between quantum weight states when the shelves
are not present.
\item Transitions between quantum weight states on the same side
of the shelf.
\end{itemize}

Transitions with long thermal relaxation times are assumed to be:
\begin{itemize}
\item Transitions of the one atom gas states across the partition
or piston.
\item Transitions of the quantum weight states across the shelf.
\item All transitions of the piston states.
\end{itemize}

We will also always assume that temperatures $T$ are high enough
for us to approximate summations over energy eigenstates by
integrations of the form

\[
\sum_{n=1,\infty}e^{-\frac{E_n}{kT}}f(E_n)
     \approx \int_0^\infty e^{-\frac{E(n)}{kT}}f(E(n))dn
\]
where the eigenvalue relations for integer $n$ are replaced by the
corresponding functions of a continuous parameter $n$, so that
$E_n=E(n)$. This approximation is valid if $kT$ is much greater
than the spacing of the energy levels.

\section{Thermal state of gas} \label{s:szsm2}

In this Section we will analyse the effect on the one atom gas of
bringing it into contact with a heat bath at temperature $T_G$. It
is assumed that the thermal relaxation time is very short.

We will start by analysing the energy levels, and mean internal
energy of the one-atom gas, in equilibrium, before and after the
partition is inserted. Proceeding in a similar manner to Chapter
\ref{ch:szqm} we will then consider the situation where the one
atom gas is confined entirely to the left of the partition, at
some variable position $Y$. Finally we will consider the situation
where there is a moving piston in the box.

\subsection{No partition}

The initial Hamiltonian in Equation \ref{eq:szh0}, can be written
as
\[
H_{G0}=\sum_n\epsilon n^2 \proj{\psi_n}
\]
In contact with a heat bath at $T_G$, the gas will be in an
initial equilibrium ensemble of\footnote{In some situations the
normalisation constant $Z$ will coincide with the thermodynamic
partition function. However, this will {\em not} necessarily be
the case, so we will not make use of this fact in this Chapter.}
\begin{eqnarray}
\rho _{G0} &=&
    \frac{1}{Z_{G0}}\sum_n e^{-\frac{\epsilon n^2}{kT_G}}\proj{\psi_n} \\
Z_{G0} &=&\sum_n e^{-\frac{\epsilon n^2}{kT_G}} \nonumber \\
&\approx& \int e^{-\frac{\epsilon n^2}{kT_G}}dn
     = \frac{1}{2} \sqrt{\frac{\pi kT_G} \epsilon }\nonumber
     \label{eq:rg0}
\end{eqnarray}
The mean internal energy of the gas states is given by
\[
\mean{E_{G0}} \approx \frac{1}{Z_{G0}}
    \int \epsilon n^2 e^{-\frac{\epsilon n^2}{kT_G}}dn
    =\frac{1}{2}kT_G
\]
which confirms the usual formula for the internal energy of a gas
with a single degree of freedom.

\subsection{Partition raised}
Raising of the partition in the center of the box is equivalent to
applying the operator $U_G$, in Equation \ref{eq:barrier}. The
final Hamiltonian in Equation \ref{eq:szh1} from Section
\ref{s:szqm2} is
\[
H_{G1}=\frac{4\epsilon }{(1-p)^2}\sum_l l^2
    \left\{\proj{\Psi_l^\lambda}+\proj{\Psi_l^\rho}\right\}
\]
which, taking account a degeneracy factor of 2, leads to
\begin{eqnarray}
\rho_{G1} &=&\frac {1}{Z_{G1}}
    \sum_l e^{-\frac{\epsilon}{kT_G}\left( \frac{2l}{1-p}\right)^2}
    \left\{\proj{\Psi_l^\lambda}+\proj{\Psi_l^\rho}\right\} \\
Z_{G1} &=&\sum_n 2
    e^{-\frac{\epsilon}{kT_G}\left(\frac{2l}{1-p}\right)^2}
    \approx \frac{1-p}{2}\sqrt{\frac{\pi kT_G}{\epsilon}} \nonumber \\
\mean{E_{G1}} & \approx &
    \frac {1}{Z_{G1}}\int 2 \epsilon \left( \frac{2l}{1-p}\right)^2
        e^{-\frac{\epsilon}{kT_G}\left(\frac{2l}{1-p}\right)^2}dl
        =\frac{1}{2}kT_G \nonumber
        \label{eq:rg1}
\end{eqnarray}

The fact that the internal energy has not changed does not mean
that no work has been performed upon the system, only that any
energy that enters the gas while inserting the partition has been
transferred to the heat bath. We will now prove that the insertion
of the partition requires negligible work.

As the partition is inserted, the odd and even wavefunctions are
perturbed, leading to shifts in energy. There will also be a shift
in occupation probabilities, if the gas is kept in contact with a
heat bath. As the size of the energy change is small compared with
the initial energy, for all but the lowest eigenstates, we can
assume that the change in occupation probabilities is negligible.

For odd symmetry states, the change in energies is given by
\begin{eqnarray*}
W_l^{(odd)} &=&\epsilon \left(\frac{2l}{1-p}\right)^2 f(p)  \\
f(p)    &=& p(2-p)
\end{eqnarray*}
so the work done is
\begin{eqnarray*}
W^{(odd)} &=&
    \frac{\epsilon}{Z^{(odd)}} \sum_l\left(\frac{2l}{1-p}\right)^2
         f(p) e^{-\frac{\epsilon}{kT_G}(2l)^2} \\
Z^{(odd)} &=&\sum_l e^{-\frac{\epsilon}{kT_G}(2l)^2}
    \approx \frac{1}{4}\sqrt{\frac{\pi kT_G}{\epsilon}} \\
W^{(odd)} &\approx & \frac {\epsilon}{Z^{(odd)}}
     \left(\frac{2}{1-p}\right)^2 f(p)\frac{2Z_{odd}^3}{\pi}
         =\frac{\frac {1}{2}kT_G}{(1-p)^2}f(p)
\end{eqnarray*}

For even symmetry states, the energy shift is more complicated
\begin{eqnarray*}
W_l^{(even)} &=& \epsilon\left(\frac{1}{1-p}\right)^2
    \left[f(p)4l^2-(4l-1)(1-p)^2\right]  \\
W^{(even)} &=&\frac{\epsilon}{Z^{(even)}}\sum_l
    \left(\frac{1}{1-p}\right)^2
    \left[ f(p)4l^2-(4l-1)(1-p)^2\right]
    e^{-\frac{\epsilon}{kT_G}(2l-1)^2} \\
Z^{(even)} &=&\sum_l e^{-\frac{\epsilon}{kT_G}(2l-1)^2}
\end{eqnarray*}
This requires a substitution $2y=2l-1$ to give
\begin{eqnarray*}
W^{(even)} &=&\frac{\epsilon}{Z^{(even)}}
    \sum_y \left(\frac{1}{1-p}\right)^2
    \left[ f(p)4y^2+4y+1\right]
    e^{-\frac{\epsilon}{kT_G}(2y)^2} \\
Z^{(even)} & \approx &
    \frac{1}{4}\sqrt{\frac{\pi kT_G}{\epsilon}} \\
W^{(even)} &\approx &
    \frac{\epsilon}{Z^{(even)}}\left(\frac{2}{1-p}\right)^2
    \left[f(p) \frac{2\left(Z^{(even)}\right)^3}{\pi}
        +\frac{2\left(Z^{(even)}\right)^2}{\pi}+\frac{Z^{(even)}}{4}\right]  \\
&\approx& \frac{\frac{1}{2}kT_G}{(1-p)^2}
    \left[f(p)+4\sqrt{\frac{\epsilon}{kT_G}}+2\left(\frac{\epsilon}{kT_G}\right)\right]
\end{eqnarray*}
The mean work done is approximately
$W=\frac{1}{2}W^{(odd)}+\frac{1}{2}W^{(even)}$ . As can be seen,
when $p \ll 1$ and ground state energy $\epsilon \ll kT_G$, then
$W \ll \frac{1}{2}kT_G$. This confirms that the insertion of the
barrier does not require a significant amount of work, when the
barrier is narrow and the internal energy is high with respect to
the ground state.
\subsection{Confined Gas}
If we restrict the gas to be located on the lefthand side of the
partition, the density matrix only includes half the states
\begin{eqnarray}
\rho_{G2}^\lambda &=&
    \frac{1}{Z_{G2}}\sum_l e^{-\frac{\epsilon}{kT_G}
    \left(\frac{2l}{1-p}\right)^2}\proj{\Psi_l^\lambda} \\
Z_{G2} &=&\sum_n e^{-\frac{\epsilon}{kT_G}
    \left(\frac{2l}{1-p}\right)^2}
    \approx \frac{1-p}{4} \sqrt{\frac{\pi kT_G}{\epsilon}} \nonumber \\
\mean{E_{G2}^\lambda} &\approx& \frac{1}{Z_{G2}} \int \epsilon
     \left(\frac{2l}{1-p}\right)^2 e^{-\frac{\epsilon}{kT_G}
     \left(\frac{2l}{1-p}\right)^2}dl = \frac{1}{2}kT_G \nonumber
     \label{eq:rg2}
\end{eqnarray}
Similar expressions can be calculated from $\rho_{G2}^\rho$,
$Z_{G2}^\rho$ and $\mean{E_{G2}^\rho}$, where the gas is confined
entirely to the right of the partition.
\subsection{Moving partition}
We will now proceed with the gas located entirely on the left of
the piston, and consider the mean internal energy of the gas
states, and the pressure upon the piston, as the piston moves.

For the piston located at a position $Y$ we use the Hamiltonian
$H_{G2}^\lambda$ given in Equation \ref{eq:szhg2} for the internal
energy of the gas states. The energy and pressure of the
individual gas states are
\begin{eqnarray*}
E_l^\lambda(Y)&=&\frac{4\epsilon l^2}{(Y+1-p)^2} \\
\frac{\partial E_l(Y)}{\partial Y} &=&
    \frac{-8\epsilon l^2}{(Y+1-p)^3}
\end{eqnarray*}

The evaluation of the effect of the moving partition depends upon
how the probabilities of each state changes as the piston moves.
We will consider three cases: perfectly isolated, essentially
isolated and isothermal. The definition of these follows that
given in \cite[Chapter 12 B]{Tol79}\footnote{It will be seen that
essential isolation broadly corresponds to those processes that
are traditionally referred to as 'adiabatic' in thermodynamics. We
have not used this term to avoid confusion with the 'adiabatic
theorem' in quantum mechanics, which will be applicable to all
three of the above processes}.

\subsubsection{Perfect Isolation}
For this condition, we assume the gas is completely isolated, and
the expansion takes place sufficiently slowly, that the
probabilities are unchanged from their initial values,
proportional to
$e^{-\frac{\epsilon}{kT_G}\left(\frac{2l}{1-p}\right)^2}$

\begin{eqnarray*}
\rho_{G3}^\lambda(Y)&=&
    \frac{1}{Z_{G3}}\sum_le^{-\frac{\epsilon}{kT_G}
    \left(\frac{2l}{1-p}\right)^2}\proj{\Psi_l^\lambda(Y)} \\
Z_{G3}&=&\sum_l e^{-\frac{\epsilon}{kT_G}
    \left(\frac{2l}{1-p}\right)^2}
    \approx \frac{1-p}{4}\sqrt{\frac{\pi kT_G}{\epsilon}} \\
\mean{E_{G3}^\lambda(Y)}&=&
    \frac{1}{Z_{G3}}\int \epsilon \left(\frac{2l}{Y+1-p}\right)^2
    e^{-\frac{\epsilon}{kT_G}\left(\frac{2l}{1-p}\right)^2}dl
    =\frac{1}{2}kT_G\left(\frac{1-p}{Y+1-p}\right)^2 \\
P_{G3}^\lambda(Y) &=&
    \frac{1}{Z_{G3}}\int \frac{-8\epsilon l^2}{(Y+1-p)^3}
        e^{-\frac{\epsilon}{kT_G}\left(\frac{2l}{1-p}\right)^2}dl
            =-kT_G\frac{(1-p)^2}{(Y+1-p)^3}
\end{eqnarray*}

The pressure term is derived from the change in internal energies
of the gas, when the piston position $Y$ changes. Note, the piston
position is an {\em external} co-ordinate for the gas. The work
performed upon the piston by the gas, when the piston is initially
in the center of the box ($Y=0$) is
\[
W_{G3}^\lambda(Y)=\int_0^Y kT_G
    \frac{(1-p)^2}{(Y^{\prime}+1-p)^3}dY^{\prime}
    =\frac{1}{2}kT_G\frac{Y(Y+2(1-p))}{(Y+1-p)^2}
\]
As the system is completely isolated, the change in internal
energy must exactly equals work performed so that
$\mean{E_{G3}(Y)}+W_{G3}(Y) =\frac{1}{2}kT_G$.

After the expansion has ended at $Y=(1-p)$, the gas has internal
energy $\frac{1}{8}kT_G$, and the work extracted is
$\frac{3}{8}kT_G$. If the system is allowed to continue in perfect
isolation, the piston will now reverse direction and start to
compress the gas. This requires work to be performed by the piston
upon the gas
\[
W_{G3}^\lambda(Y)=\int_{1-p}^Y kT_G
    \frac{(1-p)^2}{(Y^{\prime}+1-p)^3}dY^{\prime}
\]
Again the total energy is constant, and when the piston has
reached the center, the gas has internal energy $\frac{1}{2}kT_G$
and the work performed upon the gas is $\frac{3}{8}kT_G$. As the
work extracted during the expansion is the same as that performed
during the compression, the cycle is reversible.

If, when the piston was at $Y=1-p$, instead of allowing the piston
to immediately return to the center, we brought the gas into
contact with the heat bath, it would return to the state
$\rho_{G0}$ above, absorbing $\frac{3}{8}kT_G$ heat from the bath
in the process. When the piston starts to compress the gas from
this state, different results occur, as the initial probabilities
are now proportional to $e^{-\frac \epsilon
{kT_G}\left(\frac{l}{1-p}\right) ^2}$

\begin{eqnarray*}
\rho _{G4}^\lambda(Y)&=& \frac{1}{Z_{G4}}\sum_l
    e^{-\frac{\epsilon}{kT_G}\left(\frac{l}{1-p}\right)^2}
        \proj{\Psi_l^\lambda(Y)} \\
Z_{G4} &=&\sum_l
    e^{-\frac{\epsilon}{kT_G}\left(\frac{l}{1-p}\right)^2}
    \approx \frac{1-p}{2} \sqrt{\frac{\pi kT_G}{\epsilon}} \\
\mean{E_{G4}^\lambda(Y)}&=&\frac{1}{Z_{G4}}
    \int \epsilon \left(\frac{2l}{Y+1-p}\right)^2
    e^{-\frac{\epsilon}{kT_G}\left(\frac{l}{1-p}\right)^2}dl
    =2kT_G\left(\frac{1-p}{Y+1-p}\right)^2 \\
P_{G4}^\lambda(Y)&=&\frac{1}{Z_{G4}}
    \int \frac{-8\epsilon l^2}{(Y+1-p)^3}
    e^{-\frac{\epsilon}{kT_G}\left(\frac{l}{1-p}\right)^2}dl
    =-4kT_G\frac{(1-p)^2}{(Y+1-p)^3} \\
W_{G4}^\lambda(Y)&=&\int_{1-p}^Y-P_{G4}(Y^{\prime})dY^{\prime}
    =-2kT_G\left(\left(\frac{1-p}{Y+1-p}\right)^2-\frac{1}{4}\right)
\end{eqnarray*}
Again,
$\mean{E_{G4}^\lambda(Y)}+W_{G4}^\lambda(Y)=\frac{1}{2}kT_G$, but
 after compression to $Y=0$, the gas has internal energy $2kT_G$.
The work performed upon the gas during the compression was
$\frac{3}{2}kT_G$. If we now bring the gas back into contact with
heat bath, it will be restored to the original state
$\rho_{G2}^\lambda$ with energy $\frac{1}{2}kT_G$, transferring
the $\frac{3}{2}kT_G$ to the heat bath. During the course of the
complete cycle, a total amount of work equal to $\frac{3}{2}kT_G
-\frac{3}{8}kT_G =\frac{9}{8}kT_G$ has been dissipated.

\subsubsection{Essential Isolation}

The perfect isolation assumed above is not achievable in practice.
The interactions with the surrounding environment will cause
transitions between eigenstates. As the energy levels change, the
system moves out of Boltzmann equilibrium, but the interactions
with the environment will cause the system to return to Boltzmann
equilibrium over a characteristic time $\tau_{G5}$. An essentially
isolated system is one for which this contact with the environment
takes place, but involves no net transfer of energy.

This can be considered as dividing the changes into a series of
infinitesimal changes in energy $dE=\sum_np_ndE_n+ \sum_nE_ndp_n$.
First, the system is in perfect isolation, so that $dp_n=0$, and
eigenstates are allowed to change. The work performed upon the
system is $dE=\sum_n p_n dE_n$. The next stage holds the
eigenstates constant, but brings the system into contact with a
heat bath, for a time $\tau_{G5}$. This will bring the system into
a new Boltzmann equilibrium. The key element to essential
isolation is that, at each point that the system is brought into
contact with a heat bath, the temperature of the heat bath is
chosen so that there is no net change in internal energy of the
system ($\sum_n E_n dp_n=0$) even though there is a change in
occupation probabilities ($dp_n \neq 0$).

A system which is essentially isolated is, therefore, always in
equilibrium with some notional heat bath at temperature $T$, but
this temperature is variable, and depends upon the external
parameters. Changes in internal energy of the system can only come
about through work extracted from, or performed upon the system.

For the Popper-Szilard Engine, the temperature of the gas is now a
function of the piston position $T=T(Y)$

\begin{eqnarray*}
\rho _{G5}^\lambda(Y)&=&\frac{1}{Z_{G5}^\lambda(Y)}
    \sum_l e^{-\frac{\epsilon}{kT}\left(\frac{2l}{Y+1-p}\right)^2}
        \proj{\Psi_l^\lambda(Y)} \\
Z_{G5}^\lambda(Y)&=&\sum_l e^{-\frac{\epsilon}{kT}\left(\frac{2l}{Y+1-p}\right)^2}
    \approx \frac{Y+1-p}{4}\sqrt{\frac{\pi kT}{\epsilon}} \\
\mean{E_{G5}^\lambda(Y)}&=&\frac{1}{Z_{G5}^\lambda(Y)}
    \int \epsilon \left(\frac{2l}{Y+1-p}\right)^2
        e^{-\frac{\epsilon}{kT}\left(\frac{2l}{Y+1-p}\right)^2}dl
        =\frac{1}{2}kT \\
P_{G5}^\lambda(Y)&=&\frac{1}{Z_{G5}^\lambda(Y)}
    \int \frac{-8\epsilon l^2}{(Y+1-p)^3}
        e^{-\frac{\epsilon}{kT}\left(\frac{l}{Y+1-p}\right)^2}dl
        =\frac{-kT}{Y+1-p}
\end{eqnarray*}

We cannot immediately evaluate $W=\int P_{G5}^\lambda(Y)dY$ as we
do not know the variation of $T$ with $Y$. We can solve this by
noting the essential isolation requires
\[P(Y)dY=dW=dE=\frac{1}{2}kdT\]
so
\[
\frac k2\frac{dT}{dY}=P\left( Y\right) =\frac{-kT}{Y+1-p}
\]
which has the solution (given the initial temperature is $T_G$)
\[
T=T_G\left( \frac{Y_0+1-p}{Y+1-p}\right) ^2
\]
For an expansion phase, $Y_0=0,$ while for a compression phase
$Y_0=1-p$. It can be readily verified that this gives the same
results as for perfect isolation above\footnote{This equivalence
between essential and perfect isolation occurs whenever the energy
eigenstates have the form $E_n=\alpha(V)n^\beta$, where
$\alpha(V)$ depends upon the varying external parameters, but
$\beta$ is a constant. This applies only to mean pressure. The
effect of fluctuations will still be different.}.

\subsubsection{Isothermal}

The third method we use is to keep the system in constant contact
with a heat bath at the initial temperature $T_G$. As the values
of the energy eigenvalues $E_n(Y)$ changes depending upon the
external parameters, the occupation probabilities continuously
adjust to be proportional to $e^{-\frac{E_n}{kT_G}}$. As this
means the infinitesimal change $\sum_n E_n dp_n \neq 0$ heat will
be drawn form or deposited in the heat bath.

\begin{eqnarray}
\rho _{G6}^\lambda(Y) &=&\frac 1{Z_{G6}\lambda(Y)}
    \sum_l e^{-\frac{\epsilon}{kT_G}\left(\frac{2l}{Y+1-p}\right)^2}
    \proj{\Psi_l^\lambda(Y)} \\
Z_{G6}^\lambda(Y)&=&\sum_n
    e^{-\frac{\epsilon}{kT_G}\left(\frac{2l}{Y+1-p}\right)^2}
    \approx \frac{Y+1-p}{4}\sqrt{\frac{\pi kT_G}{\epsilon}} \nonumber \\
\mean{E_{G6}^\lambda(Y)}&=&\frac{1}{Z_{G6}^\lambda(Y)}
    \int \epsilon \left(\frac{2l}{Y+1-p}\right)^2
        e^{-\frac{\epsilon}{kT_G}\left(\frac{2l}{Y+1-p}\right)^2}dl
        =\frac{1}{2}kT_G \nonumber \\
P_{G6}^\lambda(Y)&=&\frac{1}{Z_{G6}^\lambda(Y)}
    \int \frac{-8\epsilon l^2}{(Y+1-p)^3}
    e^{-\frac{\epsilon}{kT_G}\left(\frac{2l}{Y+1-p}\right)^2}dl
    =-\frac{kT_G}{(Y+1-p)} \nonumber
\label{eq:rg6}
\end{eqnarray}
Unlike in the isolated cases, the internal energy remains
constant, and the sum of internal energy and work is not constant,
as heat is drawn from, or deposited in the heat bath, to
compensate for work extracted or added by the moving piston. For
expansion we have
\[
W=\int_0^Y-\frac{kT_G}{Y^{\prime }+1-p}dY^{\prime}
    =kT_G\ln \left(\frac{1-p}{Y+1-p}\right)
\]
and compression gives
\[
W=\int_{1-p}^Y-\frac{kT_G}{Y^{\prime }+1-p}dY
    =kT_G\ln \left(\frac{2(1-p)}{Y+1-p}\right)
\]
The work extracted from expansion is $kT_G\ln2$ which equals the
work required for compression. The complete cycle therefore
requires no net work to be dissipated into the heat bath.

\begin{table}
\begin{tabular}{|l|c|c|} \hline
    & Expansion & Compression \\ \hline
Isolated & $\frac{3}{8}kT_G$ & $-\frac{3}{2}kT_G$ \\
Isothermal & $kT_G \ln 2$ & $-kT_G \ln 2$ \\ \hline
\end{tabular}
\label{tb:work} \caption{Work extracted from gas}
\end{table}

If we summarise the results of the three types of expansion in
Table \ref{tb:work}, we can see that the maximum energy extracted
from the expansion phase is under isothermal expansion, while the
minimum energy required during compression is also for isothermal
expansion. We will therefore assume that the gas is in isothermal
contact with a heat bath at temperature $T_G$ from now on.

\subsubsection{Fluctuations}

The mean values derived above are valid as an average over an ensemble.
However, that is no guarantee that the value for any individual case will be
close to the average. The usual formula for 'fluctuations' about the mean is
given by

\[
\frac{\mean{A^2}-\mean{A}^2}{\mean{A}^2}\approx \frac{1}{m}
\]

where $m$ is a large number of degrees of freedom in the system. However, in
this situation there is only one degree of freedom, and this suggests that
fluctuations in the pressure, and hence work done, may be very large.

Evaluation of the size of $\mean{E^2}$ and $\mean{P^2}$ for
perfect isolation gives

\begin{eqnarray*}
\mean{E_{G3}^2} &=&\frac{1}{Z_{G3}}
    \int \frac{16\epsilon ^2}{(Y+1-p)^4}l^4
    e^{-\frac{\epsilon}{kT_G}\left(\frac{2l}{1-p}\right)^2}dl
    =\frac{3}{4}(kT_G)^2\left(\frac{1-p}{Y+1-p}\right)^4 \\
&=& 3\mean{E_{G3}}^2 \\
\left\langle P_{G3}^2\right\rangle &=&\frac{1}{Z_{G3}}
    \int \frac{64\epsilon^2}{(Y+1-p)^6}l^4
    e^{-\frac{\epsilon}{kT_G}\left(\frac{2l}{1-p}\right)^2}dl
    =3(kT_G)^2\left(\frac{(1-p)^2}{(Y+1-p)^3}\right)^2 \\
&=&3\mean{P_{G3}}^2
\end{eqnarray*}

This gives substantial fractional deviations from the mean energy and
pressure. In the case of perfect isolation, the actual gas state will not
change during the course of the expansion, and the net energy transferred is
$\Delta W_n=\int \frac{\partial E_n}{\partial X}dX=\Delta E_n,$ which will
imply that over the ensemble we will have
\[
\frac{\mean{W^2}-\mean{W}^2}{\mean{W}^2}=2
\]
which corresponds to large fluctuations in the amount of energy
drawn from, or deposited in the work reservoir over each cycle.

Clearly the size of the fluctuation at any given time will be the same for
the essentially isolated expansion. For the isothermal expansion, we have
\begin{eqnarray*}
\mean{E_{G6}^2}&=&\frac{1}{Z_{G6}}
    \int \frac{16\epsilon^2}{(Y+1-p)^4}l^4
    e^{-\frac{\epsilon}{kT_G}\left(\frac{2l}{Y+1-p}\right)^2}dl
    =\frac{3}{4}(kT_G)^2 \\
&=&3\mean{E_{G6}}^2 \\
\mean{P_{G6}^2} &=&\frac{1}{Z_{G6}}
    \int \frac{64\epsilon^2}{(Y+1-p)^6}l^4
    e^{-\frac{\epsilon}{kT_G}\left(\frac{2l}{Y+1-p}\right)^2}dl
    =3(kT_G)^2\frac{1}{(Y+1-p)^2} \\
\  &=&3\mean{P_{G6}}^2
\end{eqnarray*}
so the fractional variation is still 2.

For the cases of essential isolation, or isothermal expansion,
however, we are assuming that, after each small expansion step,
the system is allowed to interact with an environment, so that it
is restored to a Boltzmann equilibrium. This contact, over a
characteristic thermal relaxation period $\tau_\theta $
effectively randomises the state of the system, in accord with the
probabilities of the Boltzmann distribution, from one expansion
step to the next. If we suppose the expansion takes place over a
time $ t=n\tau_\theta $ there will be $n$ such randomisations.
From this it can be shown (see Appendix \ref{ap:szfluc}), that,
although the fractional fluctuation in the energy transferred is
of order 2 on each small step, the fractional fluctuation in
energy transferred over the course of an entire expansion or
compression phase is of order $1/n=\tau_\theta /t$ . For
essentially isolated and isothermal expansions, as the expansion
takes place over a large time with respect to the thermal
relaxation time, the deviation from the mean work extracted from,
or deposited within, the work reservoir is negligible.
\subsubsection{Conclusion}
We have now examined the thermal state of the one atom gas, when
it is confined to the left side of the piston. The isothermal
expansion of this gas, as the piston moves from the center, to the
right end of the box, extracts $kT_G \ln 2$ energy from the gas.
Evidently, had we started with the gas confined to the right side
of the piston, we would have equally well extracted $kT_G \ln 2$
work.

Now, if we start with the gas occupying the entire box, and insert
the partition in the center, we would have the state
\[
\rho_{G1}=\frac{1}{2}\left(\rho_{G2}^\lambda+\rho_{G2}^\rho\right)
\]
Inserting the piston into the center, $\proj{\Phi_0}$, and
applying the expansion operators $U_{W4}$ leads to the state
\[
\frac{1}{2}\left(\rho_{G6}^\lambda(1-p)\proj{\Phi(1-p)}
    +\rho_{G6}^\rho(-1+p)\proj{\Phi(-1+p)}\right)
\]
In both cases the energy $kT_G \ln 2$ is extracted from the gas.
This confirms that the Szilard Paradox is still valid for quantum
systems, and the question of superposition of the wavefunction,
raised by Zurek, is irrelevant.

\section{Thermal State of Weights} \label{s:szsm3}
We now wish to describe the thermal states of the weights as they
are raised and lowered by the pulleys, and when a shelf is
inserted into an unraised weight at height $h$. The probability of
finding an unraised weight above the shelf height $h$ is also the
probability of an imperfect correlation between the location of
the weights and the piston states. This governs the tendency of
the Popper-Szilard Engine to switch between raising and lowering
cycles, and plays a critical role in the long term behaviour of
the Engine.

We will bring the weights into contact with a heat bath at
temperature $T_W$. It will be shown that, due to properties of the
quantum states, described by Airy functions, that there is no
difference between perfect isolation, essential isolation or
isothermal expansion, when raising or lowering a weight. We will
assume, for simplicity, that the weight is always in contact with
the heat bath. The initial density matrix, with the weights
resting upon the floor, is given by
\begin{eqnarray}
\rho_{W0} &=&\frac{1}{Z_{W0}}\sum_n
    e^{\frac{a_nM_wgH}{kT_W}}\proj{A_n(0)}\\
Z_{W0} &=&\sum_n e^{\frac{a_nM_wgH}{kT_W}} \nonumber
\label{eq:rw0}
\end{eqnarray}
(recall $a_n<0$)

\subsection{Raising and Lowering Weight}
We will consider the case of raising a weight, and then show that
the resulting density matrix describes a lowered weight as well.
If we start with the system in perfect isolation and the floor
beneath the weight is raised slowly from $0$ to a height $h(Y)$
then, by the adiabatic theorem, the new density matrix will
be\footnote{We have continued to use the notation developed in
Chapter \ref{ch:szqm} where the quantum wavefunction $A_n(z,h(Y))$
is represented by the Dirac ket $\ket{A_n(Y)}$.}
\[
\rho _{W1}^{\prime}(h)=\frac{1}{Z_{W0}}\sum_n
    e^{\frac{a_nM_wgH}{kT_W}}\proj{A_n(Y)}
\]
while the equilibrium density matrix, that results from bringing
$\rho_{W1}^\prime(h)$ into contact with the heat bath, will be
\begin{eqnarray}
\rho_{W1}(h)&=&\frac{1}{Z_{W1}}\sum_n
    e^{\frac{(Ha_n-h)M_wg}{kT_W}}\proj{A_n(Y)} \\
Z_{W1}(h) &=&\sum_n e^{\frac{(Ha_n-h)M_wg}{kT_W}} \nonumber
\label{eq:rw1}
\end{eqnarray}
Comparing these, it can be seen that the probability of a given
state $\ket{A_n(Y)}$ is the same in both cases
\[
p_n(h)=\frac{e^{\frac{(Ha_n-h)M_wg}{kT_W}}}
            {\sum_n e^{\frac{(Ha_n-h)M_wg}{kT_W}}}
    =\frac{e^{\frac{-hM_wg}{kT_W}}e^{\frac{a_nM_wgH}{kT_W}}}
            {e^{\frac{-hM_wg}{kT_W}}\sum_n e^{\frac{a_nM_wgH}{kT_W}}}
    =p_n^{\prime}(h)
\]
In other words, as
\[
\rho _{W1}^{\prime}(h)=\rho _{W1}(h)
\]
the density matrix resulting from perfect isolation is already in
equilibrium at $T_W$. By definition this will also apply to
essential isolation. As this holds for any height $h$, the three
processes are identical. It also follows that the density matrix
that arises from {\em starting} with a raising floor, and then
{\em lowering} it to a height $h$ will be the same.

One implication of this equivalence is that net exchange of heat
between the weight and the heat bath while it is being raised or
lowered isothermally will be zero. Any change in the internal
energy of the weight comes about through the work done upon the
weight. To examine this, we will now look at the generalised
pressure exerted upon the co-ordinate $h(Y)$.

The energy and pressure of the state $\ket{A_n(Y)}$ is given by
\begin{eqnarray*}
E_n &=&(h-a_nH) M_Wg \\
\frac{\partial E_n}{\partial h} &=&M_Wg
\end{eqnarray*}

The pressure $P_n(h)=\frac{\partial E_n}{\partial h}$ is
independant of both $n$ and $h$. This means we can evaluate the
average pressure for {\em any} ensemble as it is clearly simply
$\mean{P(h)}=M_Wg$. It should also be clear that
$\mean{P(h)^2}=\mean{P(h)}^2$ so there is zero fluctuation in the
pressure! From this it will also follow there is zero fluctuation
in the work required to raise the weight. This constancy of the
pressure gives the very pleasing result that if the weight is
raised slowly through a height of $h$ the work performed upon the
weight is always exactly $M_Wgh$. This makes a raised weight a
particularly useful system to use as a work reservoir.

As we know that no net flow of heat has entered or left the system
we can immediately state that the internal energy of the weight
must be of the form
\[
\mean{E(h,T_W)}=M_Wgh+f(T_W)
\]
We now use the asymptotic approximation
\[
a_n \approx -\left(\frac{3\pi n}{2}\right)^{\frac{2}{3}}
\]
valid for large $n$, to complete this equation.

\begin{eqnarray*}
Z_{W1}(h) &=&\sum_n
    e^{\frac{M_wg(Ha_n-h)}{kT_W}}
    \approx e^{-\frac{M_Wgh}{kT_W}}\int_0^\infty
     e^{-\left(\frac{3\pi n}2\right)^{\frac{2}{3}}\frac{M_WgH}{kT_W}}dn \\
&\approx &\frac{e^{-\frac{hM_Wg}{kT_W}}}{2\sqrt{\pi }}
    \left(\frac{kT_W}{M_WgH}\right)^{\frac{3}{2}} \\
\mean{E(h,T_W)}&=&\frac{1}{Z_{W1}(h)}\sum_n M_wg(h-Ha_n)
    e^{\frac{(Ha_n-h)M_wg}{kT_W}} \\
&=& M_Wgh-\frac{M_WgH}{Z_{W1}}
    e^{-\frac{M_WgH}{kT_W}}\sum_n
    a_n e^{\frac{M_wgHa_n}{kT_W}} \\
&\approx &M_Wgh+2\sqrt{\pi}M_WgH
    \left(\frac{M_wgH}{kT_W}\right)^{\frac{3}{2}}
    \int_0^\infty \left(\frac{3\pi n}{2}\right)^{\frac{2}{3}}
    e^{-\left(\frac{3\pi n}{2}\right)^{\frac{2}{3}}\frac{M_WgH}{kT_W}}dn \\
&\approx& M_Wgh+\frac{3}{2}kT_W
\end{eqnarray*}
Further analysis of the energy fluctuations gives
\begin{eqnarray*}
\mean{E^2} &=&(M_Wgh)^2+\frac{15}{4}(kT_W)^2+3M_WghkT_W \\
\mean{E^2} -\mean{E}^2 &=&\frac{3}{2}(kT_W)^2
\end{eqnarray*}
although, as noted above, there is no fluctuation in the pressure.

With regard to the internal energy term $\frac{3}{2}kT_W$, we can
break the Hamiltonian $H_W$ into two terms
\begin{eqnarray*}
H_{KE} &=&-\frac \hbar {2M_W}\frac{\partial ^2}{\partial z^2} \\
H_{PE} &=&M_Wgz
\end{eqnarray*}
representing kinetic and potential energies, and find they have
expectation values
\begin{eqnarray*}
\mean{H_{KE}} &=&\frac{1}{2}kT_W \\
\mean{H_{PE}} &=&kT_W
\end{eqnarray*}
The internal energy dividing in this ratio between kinetic and
potential energy is an example of the virial theorem.

\subsection{Inserting Shelf}
We now consider the effect of inserting a shelf at height $h$ into
an unraised thermal state $\rho_{W0}$. This projects out raised
and unraised portions of the wavefunction. The statistical weight
of these two portions gives the probability of locating the
unraised weight above or below the shelf height, and so determines
the reliability of the resetting mechanism at the end of a cycle
of the Popper-Szilard Engine.

For simplicity we will deal only with the projection of
$\rho_{W0}$ into raised and unraised density matrices. Although
there will, in general, be interference terms between the two
subspaces when the shelf is inserted using $U_S$, in the
situations we will be considering the contact with the $T_W$ heat
bath will destroy these coherence terms.

The projections of the unraised density matrix to below and above
the height $h$, respectively, are given by:
\begin{eqnarray*}
\rho_{W0}(0)^{\prime} &=& P(UN)\rho_{W0} P(UN) \\
    &=&\frac{1}{Z_{W0}}\sum_m
    e^{a_m\frac{M_WgH}{kT_W}}\beta_m^2(h)\proj{UN_m(h)} \\
\rho_{W0}(h)^{\prime} &=& P(RA)\rho_{W0} P(RA) \\
    &=&\frac{1}{Z_{W0}}\sum_m
    e^{a_m\frac{M_WgH}{kT_W}}\alpha_m^2(h)\proj{RA_m(h)}
\end{eqnarray*}
These have not been normalised. We must be careful when doing
this, as the $\ket{RA_m(h)}$ and $\ket{UN_m(h)}$ do not form an
orthonormal basis.
\begin{eqnarray*}
\trace{\rho_{W0}(0)^{\prime}} &=&
    \sum_n\ip{A_n(Y)}
    {\left\{\frac{1}{Z_{W0}}\sum_m e^{a_m\frac{M_WgH}{kT_W}}
        \beta_m^2(h)\proj{UN_m(h)}\right\}}
        {A_n(Y)} \\
&=&\frac{1}{Z_{W0}}\sum_m e^{a_m\frac{M_WgH}{kT_W}}\beta_m^2(h)
    \sum_n \beta_n^2(h)\bk{UN_n(h)}{UN_m(h)} \bk{UN_m(h)}{UN_n(h)}\\
&=&\frac{1}{Z_{W0}}\sum_m e^{a_m\frac{M_WgH}{kT_W}}\beta_m^2(h)
\end{eqnarray*}
In the last step we have used the fact that
$\sum_n\beta_n^2(h)\proj{UN_n(h)}$ is the identity operator for
the unraised subspace to substitute\footnote{This can be
generalised to the produce useful result $\trace{\sum_n
c_n\proj{UN_n(h)}}=\sum_n c_n$ despite the non-orthogonality of
the $\ket{UN_n(h)}$}
\[
\ip{UN_m(h)}
    {\left\{\sum_n\beta_n^2(h)\proj{UN_n(h)}\right\}}
    {UN_m(h)}
    =\bk{UN_m(h)}{UN_m(h)}=1
\]

We may similarly obtain the result
\[
\trace{\rho_{W0}(h)^{\prime}}
    =\frac {1}{Z_{W0}}\sum_m
    e^{a_m\frac{M_WgH}{kT_W}}\alpha_m^2(h)
\]

Using the asymptotic approximations for $a_m$ we get the high
temperature values
\begin{eqnarray*}
Z_{W0} &=&\sum_m e^{a_m\frac{M_WgH}{kT_W}}
    \approx \int_0^\infty e^{-\left(\frac{3\pi n}{2}\right)
        ^{\frac{2}{3}}\frac{M_wgH}{kT_W}} \\
&\approx &\frac{1}{2\sqrt{\pi}}
    \left(\frac{kT_W}{M_WgH}\right)^{\frac{3}{2}} \\
\end{eqnarray*}

Using the values of $\alpha_m(h)$ and $\beta_m(h)$ from Equations
\ref{eq:alpha1}, \ref{eq:alpha2} and \ref{eq:alpha3}, and in
particular noting that $\alpha_m(h)=0$, $\beta_m(h)=1$ for $m<
\frac{2}{3\pi }\left( \frac{h}{H}\right)^{3/2}$
\begin{eqnarray*}
\sum_m e^{a_m\frac{M_WgH}{kT_W}}\alpha_m^2(h)
     &\approx &\int_{\frac{2}{3\pi}\left(\frac{h}{H}\right)^{3/2}}^\infty
     e^{-\left(\frac{3\pi n}{2}\right)^{\frac{2}{3}}\frac{M_wgH}{kT_W}}
     \left(1-\left(\frac{2}{3\pi n}\right)^{\frac{2}{3}}\frac{h}{H}
        \right)^{\frac{1}{2}}dn \\
&\approx& \frac{1}{2\sqrt{\pi}}e^{-\frac{M_wgh}{kT_W}}
    \left(\frac{kT_W}{M_WgH}\right)^{\frac{3}{2}}
    =Z_{W0} e^{-\frac{M_wgh}{kT_W}} \\
\sum_m e^{a_m\frac{M_WgH}{kT_W}}\beta_m^2(h)
     &\approx& \int_0^\infty
     e^{-\left(\frac{3\pi n}{2}\right)^{\frac{2}{3}}
        \frac{M_wgH}{kT_W}}dn \\
     && -\int_{\frac{2}{3\pi}\left(\frac{h}{H}\right)^{3/2}}^\infty
     e^{-\left(\frac{3\pi n}{2}\right)^{\frac{2}{3}}\frac{M_wgH}{kT_W}}
     \left(1-\left(\frac{2}{3\pi n}\right)^{\frac{2}{3}}\frac{h}{H}
        \right)^{\frac{1}{2}}dn \\
&=&Z_{W0}-\frac{1}{2\sqrt{\pi}}e^{-\frac{M_wgh}{kT_W}}
    \left(\frac{kT_W}{M_WgH}\right)^{\frac{3}{2}}
    =Z_{W0}\left(1-e^{-\frac{M_wgh}{kT_W}}\right)
\end{eqnarray*}

These results give the probability of locating a weight at
temperature $T_W$ above or below the shelf at height $h$
\paragraph{Probability of Weight Above Shelf}
\begin{equation}
P_1(h,T_W)= e^{-\frac{M_wgh}{kT_W}} \label{eq:p1}
\end{equation}
\paragraph{Probability of Weight Below Shelf}
\begin{equation}
P_2(h,T_W)=1-e^{-\frac{M_wgh}{kT_W}} \label{eq:p2}
\end{equation}
(Before we can use these probabilities, we must calculate the
height at which the shelves are inserted. This will be undertaken
in the next Section).

We will represent the density operator for the thermal state of a
weight projected out above or below the shelf by
\begin{eqnarray}
\rho_{W0}(h)^{\prime\prime}
    =\frac{1}{P_1(h,T_W)}\rho_{W0}(h)^{\prime} \nonumber \\
\rho_{W0}(0)^{\prime\prime}
    =\frac{1}{P_2(h,T_W)}\rho_{W0}(0)^{\prime}
\label{eq:rwprime}
\end{eqnarray}

\subsection{Mean Energy of Projected Weights}
Now we shall calculate the mean internal energy of the weight when
it is trapped above or below the shelf. The mean energy of a
weight in the unraised state $\rho_{W0}$, conditional upon it
being above the height $h$, is given by:
\begin{eqnarray*}
E_W(z>h)&=&\frac{\int_{h}^\infty
    \ip{z}{H_{W1}(0)\rho_{W0}}{z}dz}
    {\int_{h}^\infty \ip{z}{\rho_{W0}}{z}dz} \\
&=&\frac{\sum_m
    e^{\frac{M_WgH}{kT_W}a_m}E_m\alpha_m^2(h)\bk{RA_m(h)}{RA_m(h)}}
    {\sum_m e^{\frac{M_WgH}{kT_W}a_m}\alpha_m^2(h)\bk{RA_m(h)}{RA_m(h)}} \\
&\approx & \frac{1}{P_1(h,T_W)Z_{W0}}
    \int_{\frac{2}{3\pi }\left(\frac{h}{H}\right)^{3/2}}^\infty
     e^{\frac{M_WgH}{kT_W}a_m}(-a_mM_WgH)
     \left(1-\left(\frac{2}{3\pi m}\right)^{\frac{2}{3}}\frac{h}H\right)
     ^{\frac{1}{2}} dm\\
&\approx &\frac {3}{2}kT_W+M_Wgh
\end{eqnarray*}
using the asymptotic value of $a_m$. This is the same energy as
for the equilibrium density matrix $\rho_{W1}(h)$.

We can likewise calculate for the weight trapped below the shelf:
\begin{eqnarray*}
E_W(z<h) &=&
    \frac{1}{(1-P_1(h,T_W))Z_{W0}}
    \sum_m e^{\frac{M_WgH}{kT_W}a_m}E_n\beta_m^2(h) \\
&\approx &\frac{3}{2}kT_W-M_Wgh
    \left(\frac{e^{-\frac{M_Wgh}{kT_W}}}{1-e^{-\frac{M_Wgh}{kT_W}}}\right)
\end{eqnarray*}

If we now calculate the mean height of the weight, conditional
upon it being above the shelf
\begin{eqnarray*}
\mean{z>h} &=&\frac{\int_{h}^\infty
    \ip{z}{z\,\rho_{W0}}{z}dz}
    {\int_{h}^\infty\ip{z}{\rho_{W0}}{z}dz} \\
&\approx &\frac{kT_W}{M_Wg}+h
\end{eqnarray*}
giving a mean potential energy
\begin{eqnarray*}
PE_W(z>h) &\approx & kT_W+M_Wgh \\
&=&E_W(z>h)-\frac{1}{2}kT_W
\end{eqnarray*}
and for below the shelf
\begin{eqnarray*}
\mean{z<h} &=&\frac{\int_0^{h}\ip{z}{z\,\rho_{W0}}{z}dz}
    {\int_0^{h}\ip{z}{\rho_{W0}}{z}dz} \\
\  &\approx &\frac{kT_W}{M_Wg}-
    h\left(\frac{e^{-\frac{M_Wgh}{kT_W}}}{1-e^{-\frac{M_Wgh}{kT_W}}}\right)
\end{eqnarray*}

\begin{eqnarray*}
PE_W(z<h) &\approx & kT_W-
    M_Wgh\left(\frac{e^{-\frac{M_Wgh}{kT_W}}}
        {1-e^{-\frac{M_Wgh}{kT_W}}}\right)  \\
\  &=&E_W(z<h)-\frac{1}{2}kT_W
\end{eqnarray*}
so the mean kinetic energy is still $\frac{1}{2}kT_W$ . This is an
important result, as it demonstrates that the mean kinetic energy
of a particle, in thermal equilibrium in a gravitational field, is
the same at any height.

It will be useful to note that
\begin{eqnarray*}
\mean{E(T_W)}=P_1(h,T_W) E_W(z>h)+P_2(h,T_W) E_W(z<h) \\
\mean{PE(T_W)}=P_1(h,T_W) PE_W(z>h)+P_2(h,T_W) PE_W(z<h)
\end{eqnarray*}

If the height of the shelf is large $\left( h \gg
\frac{kT_W}{M_Wg} \right)$ then the mean energy of the weight
below the shelf approaches $\frac{3}{2}kT_W$ - the same energy as
without the shelf. This corresponds to the case where there is
little probability of the weight being above the shelf, so
inserting it has no effect. If the shelf is low $\left(h \ll
\frac{kT_W}{M_Wg} \right)$ then the mean height below the shelf is
simply $\frac{1}{2}h$ . In this case the mean kinetic energy of
the particle is much higher than the gravitational potential below
the shelf and the probability distribution of the height is almost
flat. The mean energy becomes negligibly different from the mean
kinetic energy $\frac{1}{2}kT_W$. These are consistent with the
approximations for the perturbed Airy function eigenvalues derived
in Appendix \ref{ap:szai}.

When the potential barrier is raised in the center of the one-atom
gas, it was possible to show how the wavefunction deforms
continuously, and so we could demonstrate in Section \ref{s:szsm2}
that, for $kT_G$ much higher than the ground state energy,
negligible work is done by raising the potential. We would like to
show a similar result for the Airy functions, as the shelf is
inserted. Unfortunately, there is no simple solution for the
intermediate stages, or even for the weight confined between the
floor and the shelf. However, in Appendix \ref{ap:szai} it is
argued that, for high quantum numbers ($m \gg 1$) it is reasonable
to assume that there is negligible perturbation of the energy
eigenvalues as the shelf is inserted. For situations where the
weight's internal energy $kT_W$ is large in comparison to the
ground state energy of the weight, $-a_1M_WgH$, then the work done
inserting the shelves can be disregarded.

\section{Gearing Ratio of Piston to Pulley} \label{s:szsm4}
We now need to calculate the height $h_T$ at which the shelves are
inserted, to complete the calculation of the probability that an
unraised weight is trapped above the shelf. In Section
\ref{s:szqm4} it was noted that the height $h$ through which the
weight is raised is not necessarily proportional to the position
of the piston $Y$. Some frictionless gearing system is required to
provide a gearing ratio $h(Y)$. In this Section we calculate the
optimal gearing ratio, and use this to calculate the maximum
height $h_T$ through which the weight can be raised by the
expansion of the gas. This will be the height at which the shelves
must be inserted into the Popper-Szilard Engine.

We wish the mean energy given up by the expansion of the gas to
exactly match the energy gained by the raising of the weight, or
\begin{eqnarray*}
\int_0^{h(1-p)}P_W(h)dh &=&-\int_0^{1-p} P_G(Y)dY \\
\int_0^{1-p} P_W (h(Y))\frac{\partial h}{\partial Y}dY
    &=&-\int_0^{1-p} P_G(Y) dY \\
\frac{\partial h}{\partial Y} &=&-\frac{P_G(Y)}{P_W(h(Y))}
\end{eqnarray*}

For essential isolation of the gas, this would give
\begin{eqnarray*}
\frac{\partial h^{\prime }(Y)}{\partial Y}
    &=&\frac{kT_G(1-p)^2}{M_Wg(Y+1-p)^3} \\
h^{\prime}(Y)&=&\frac{kT_G}{2M_Wg}
    \left(1-\left(\frac{1-p}{Y+1-p}\right)^2\right)
\end{eqnarray*}
giving a maximum $h^{\prime}(1-p)=\frac{3kT_G}{8M_Wg}$

However, we can extract more energy from the gas per cycle if we use an
isothermal expansion, which requires a different gearing ratio

\begin{eqnarray*}
\frac{\partial h(Y)}{\partial Y}
     &=&\frac{kT_G}{M_Wg(Y+1-p)} \\
h(Y) &=&\frac{kT_G}{M_Wg}\ln\left(1+\frac{Y}{1-p}\right)
\end{eqnarray*}

giving $h_T=h(1-p)=\frac{kT_G}{M_Wg}\ln 2$.

This is the optimum gearing, based upon the mean energy transfer.
On average, the work extracted from the gas is equal to the work
done upon the weight, and vice versa. As noted in Sections
\ref{s:szsm2} and \ref{s:szsm3} above, there are fluctuations in
the pressure exerted upon the piston by the gas, but none in the
pressure exerted by the weight upon the floor. However, as
demonstrated in Appendix \ref{ap:szfluc}, the fluctuation about
the mean energy extracted from the gas becomes negligible, so we
have now justified our statement in Section \ref{s:szqm4} that the
amount of energy drawn from or deposited in the external work
reservoir is negligible.

\subsection{Location of Unraised Weight}
We now know the height at which the shelves are inserted, so we
can calculate the probability of locating the weight above or
below the shelf, as a function only of the temperatures of the gas
and the weight.

Substituting $h_T=\frac{kT_G}{M_Wg}\ln 2$ into Equations
\ref{eq:p1} and \ref{eq:p2} we obtain:
\subsubsection{Above Shelf at $h_T$}
\begin{equation}
P_1 = \left( \frac{1}{2}\right)^{\frac{T_G}{T_W}} \label{eq:p1a}
\end{equation}
\subsubsection{Below Shelf at $h_T$}
\begin{equation}
P_2 = 1-\left(\frac{1}{2}\right)^{\frac{T_G}{T_W}} \label{eq:p2a}
\end{equation}

The form of these results will be shown to play a critical role in
the failure of the Popper-Szilard Engine to produce anti-entropic
behaviour. We will be examining the origin of this relationship in
detail in Chapter \ref{ch:szsol}.

\section{The Raising Cycle}\label{s:szsm5}
We can now use the unitary operators in Equation \ref{eq:cycle} to
describe the complete operation of the engine. In this section we
will move through each step of the 'raising cycle' given in
Section \ref{s:szqm6}. We will confirm that the fully quantum
mechanical description of the Popper-Szilard Engine does not lead
to the conclusions of \cite{Zur84,BS95}, that the piston does not
move as the one atom gas is in a superposition. With regard to the
arguments of \cite{LR90}, we will show that the operation
$U_{RES}$ is capable of achieving a partial resetting of the
engine, without the requirement for external work. However, as
noted in Section \ref{s:szqm5}, there are inevitable errors in the
resetting operation. We will now be able to evaluate the effect of
these errors upon the state of the Engine at the end of the cycle.

\subsubsection{Extracting Energy from the $T_G$ Heath Bath}
For the 'raising cycle' (Figure \ref{fg:raise}) the initial
density matrix is given by
\begin{eqnarray*}
\rho_{T0}&=& \rho_{G0} \otimes \rho_{W0}^\lambda
        \otimes \rho_{W0}^\rho \otimes \proj{\phi_0}
\end{eqnarray*}
The internal energy of this state is
\[
E_{T0}=\frac{1}{2}kT_G+3kT_W
\]

During Stage (a), the operator $U_{RI}$ is applied. As the piston
is initially in state \ket{\phi_0} this corresponds to the raising
of a potential barrier in the center of the gas and the insertion
of the piston. The state of the system is now
\begin{eqnarray*}
\rho_{T1}(0) &=& \rho_{G1} \otimes \rho_{W0}^\lambda
        \otimes \rho_{W0}^\rho \otimes \proj{\Phi(0)} \\
&=&\frac{1}{2}\left(\rho_{G6}^\lambda(0)+\rho_{G6}^\rho(0) \right)
        \otimes \rho_{W0}^\lambda \otimes \rho_{W0}^\rho \otimes \proj{\Phi(0)}
\end{eqnarray*}
and the internal energy is unchanged. As the expansion and lifting
(operator $U_{W4}$) takes place in Stage (b) this evolves through
the $Y$ states
\begin{eqnarray}
\rho_{T1}(Y)&=&
    \frac{1}{2}\left(\rho_{G6}^\lambda(Y)\otimes \rho_{W1}^\lambda(h(Y))
        \otimes \rho_{W0}^\rho \otimes \proj{\Phi(Y)} \right. \nonumber \\
    &&\left. +\rho_{G6}^\rho(-Y)\otimes \rho_{W0}^\lambda
        \otimes \rho_{W1}^\rho(h(Y))\otimes \proj{\Phi(-Y)}
\right) \label{eq:rt1}
\end{eqnarray}
until the piston wavepackets reach the sides of the box at
$Y=1-p$. It is important to note how the parameter $Y$ has been
applied in this equation. For those states where the gas is to the
left of the piston, the value $Y$ represents the distance the
piston has moved to the right, from the center of the box. This
varies from $0$ to $1-p$ as the piston moves to the righthand side
of the box.

However, for the states where the gas is to the right of the
piston, the piston moves to the left. This would be represented by
a negative value of $Y$. To simplify the expression of this, we
have substituted $-Y$. The value of $Y$ goes from $0$ to $1-p$
again, but now represents the piston moving from position $0$ to
the lefthand side of the box, at position $-1+p$.

When $Y=1-p$, the state of the system is
\begin{eqnarray*}
\rho_{T1}(1-p)&=&
    \frac{1}{2}\left(\rho_{G6}^\lambda(1-p)\otimes \rho_{W1}^\lambda(h_T)
        \otimes \rho_{W0}^\rho \otimes \proj{\Phi(1-p)} \right.\\
    && \left.+\rho_{G6}^\rho(-1+p)\otimes \rho_{W0}^\lambda
        \otimes \rho_{W1}^\rho(h_T)\otimes \proj{\Phi(-1+p)}
\right)
\end{eqnarray*}
The internal energy is now
\[
E_{T1}(1-p)=\frac{1}{2}kT_G+3kT_W+M_Wgh_T
\]

This refutes the arguments of \cite{Zur84,BS95}, that the piston
cannot move because the quantum gas exerts an even pressure upon
it until an external measurement is performed. Clearly the piston
is not left in the center of the box. The gas expands, exerting
pressure upon the piston, and lifts one of the weights. This
extracts energy from the gas, but the isothermal contact with the
$T_G$ heat bath replaces this. At the end of the expansion, one of
the weights has been raised through the distance $h_T$. The energy
has increased by $M_Wgh_T=kT_G \ln 2$, which has been drawn from
the $T_G$ heat bath during the isothermal expansion. At this point
we appear to have proved the contention of Popper et al. that an
'information gathering measurement' is not necessary to extract
energy from the Szilard Engine.

The $M_Wgh_T$ energy is stored in the internal energy of the
raised weight. If we remove the support for the weight it will
start to fall to the floor. Contact with the $T_W$ heat bath will
then return it to the thermal equilibrium state $\rho_{W0}$. This
will have reduced it's energy by $M_Wgh_T$. The extra energy is
dissipated into the $T_W$ heat bath. As we argued in Section
\ref{s:szmd2}.3, we have encountered no reason, so far, that
prevents us from setting $T_W > T_G$. If we can reliably transfer
$M_Wgh_T$ energy per cycle from the $T_G$ to the $T_W$ heat baths,
we will then have violated the second law of thermodynamics.
However, we still have to address the problem of resetting the
Engine for the next cycle. Before we can allow the weight to fall
to the floor and dissipate the $M_Wgh_T$ energy into the $T_W$
heat bath we must correlate it's position to the location of the
piston. As we found in Section \ref{s:szqm5}, without this
correlation in the resetting stage we will be unable to start a
new cycle, or if we attempted to start a new cycle, the Engine
would automatically reverse into a lowering cycle.

\subsubsection{Resetting the Piston Position}
At this point, Stage (c), the shelves are inserted at a height
$h_T$, by the operator $U_S$ and then, Stage (d), the piston is
removed from the box by $U_{IR}$.

The effect of $U_S$ is to divide each of the unraised weight
wavefunctions $\ket{A_n(0)}$ into raised ($\ket{RA_n(h_T)}$) and
unraised ($\ket{UN_n(h_T)}$) portions. We will assume that contact
with the $T_W$ heat bath destroys interference terms between the
raised and unraised wavefunctions\footnote{Strictly, we can only
be certain this will have happened when the system is allowed to
thermalise, after the operation $U_{RES}$. However, it makes no
difference to the calculation, while simplifying the description,
if we also assume this happens after the shelves are inserted.}.
In terms of the projected density matrices in Equation
\ref{eq:rwprime}, the system is now:
\begin{eqnarray*}
\rho_{T2}&=&
    \frac{1}{2}\left(\rho_{G6}^\lambda(1-p)\otimes \rho_{W1}^\lambda(h_T)
        \otimes \left\{P_1\rho_{W0}^\rho(h_T)^{\prime\prime}
            +P_2\rho_{W0}^\rho(0)^{\prime\prime}\right\}
         \otimes \proj{\Phi(1-p)} \right. \\
    && \left. +\rho_{G6}^\rho(-1+p)\otimes
        \left\{P_1\rho_{W0}^\lambda(h_T)^{\prime\prime}
            +P_2\rho_{W0}^\lambda(0)^{\prime\prime}\right\}
        \otimes \rho_{W1}^\rho(h_T)\otimes \proj{\Phi(-1+p)}
\right)
\end{eqnarray*}
The operation of $U_{RI}$ upon $\rho_{T2}$, during Stage (d),
removes the piston states, and allows the gas state to return to
$\rho_{G0}$:
\begin{eqnarray*}
\rho_{T3}&=&
    \frac{1}{2}\rho_{G0} \otimes
    \left(\rho_{W1}^\lambda(h_T)
        \otimes \left\{P_1\rho_{W0}^\rho(h_T)^{\prime\prime}
            +P_2\rho_{W0}^\rho(0)^{\prime\prime}\right\}
         \otimes \proj{\phi_R} \right.\\
    && \left. +\left\{P_1\rho_{W0}^\lambda(h_T)^{\prime\prime}
            +P_2\rho_{W0}^\lambda(0)^{\prime\prime}\right\}
        \otimes \rho_{W1}^\rho(h_T)\otimes \proj{\phi_L}
\right)
\end{eqnarray*}
The density matrices $\rho_{W0}(h_T)^{\prime \prime}$ show the
possibility that the unraised weights have been trapped above the
shelf height $h_T$. This is a 'thermal fluctuation' in the
internal energy of the weights. It was shown in Section
\ref{s:szsm3} that the internal energy of the
$\rho_{W0}(h_T)^{\prime\prime}$ states is $M_Wgh_T$ higher than
the equilibrium state $\rho_{W0}$. The source of this energy is
the $T_W$ heat bath. Trapping the unraised weight does {\em not}
constitute energy drawn from the $T_G$ heat bath, in contrast to
the increase in internal energy of the raised weight
$\rho_{W1}(h_T)$.

If we calculate the mean internal energy of $\rho_{T3}$, we find
it is unchanged:
\begin{eqnarray*}
E_{T3} &=& \frac{1}{2}kT_G
    +\frac{1}{2}P_2\left(3kT_W+M_Wgh_T
            \left(1-\frac{e^{-\frac{M_Wgh_T}{kT_W}}}
            {1-e^{-\frac{M_Wgh_T}{kT_W}}}\right) \right) \\
&&  +\frac{1}{2}P_2\left(3kT_W+M_Wgh_T
        \left(1-\frac{e^{-\frac{M_Wgh_T}{kT_W}}}
            {1-e^{-\frac{M_Wgh_T}{kT_W}}}\right) \right)  \\
&&  +\frac{1}{2}P_1 (3kT_W+2M_Wgh_T)
        +\frac{1}{2}P_1 (3kT_W+2M_Wgh_T)  \\
&=& \frac{1}{2}kT_G+3kT_W+M_Wgh_T
        \left(P_2\left(1-\frac{P_1}{P_2}\right)+2P_1\right)  \\
&=&E_{T1}(1-p)
\end{eqnarray*}

Re-writing $\rho_{T3}$ in a form more suitable for applying
$U_{RES}$ in Stage (e) we get
\begin{eqnarray*}
\rho_{T3}=
    & \rho_{G0} \otimes & \left(
    \frac{1}{2}P_2 \rho_{W1}^\lambda(h_T) \otimes
        \rho_{W0}^\rho(0)^{\prime\prime} \otimes \proj{\phi_R}
    +\frac{1}{2}P_2 \rho_{W1}^\lambda(0)^{\prime\prime} \otimes
        \rho_{W0}^\rho(h_T) \otimes \proj{\phi_L}\right. \\
&&  \left.
    +\frac{1}{2}P_1 \rho_{W1}^\lambda(h_T) \otimes
        \rho_{W0}^\rho(h_T)^{\prime\prime} \otimes \proj{\phi_R}
    +\frac{1}{2}P_1 \rho_{W1}^\lambda(h_T)^{\prime\prime} \otimes
        \rho_{W0}^\rho(h_T) \otimes \proj{\phi_L}
\right)
\end{eqnarray*}
The first line of this represents the unraised weight trapped
below the shelf height. When this happens, the location of the
weight is correlated to the location of the piston, and can be
used to reset the piston. The second line corresponds to
situations where the unraised weight has been trapped above the
shelf height. It not possible to identify the location of the
piston from the location of the weights in this portion of the
density matrix.

Now applying $U_{RES}$ to $\rho_{T3}$ we are left with the state
\begin{eqnarray*}
\rho_{T4}=
    & \rho_{G0} \otimes & \left(
    \frac{1}{2}P_2 \rho_{W1}^\lambda(h_T) \otimes
        \rho_{W0}^\rho(0)^{\prime\prime} \otimes \proj{\phi_0}
    +\frac{1}{2}P_2 \rho_{W1}^\lambda(0)^{\prime\prime} \otimes
        \rho_{W0}^\rho(h_T) \otimes \proj{\phi_0} \right. \\
&&  \left.
    +\frac{1}{2}P_1 \rho_{W1}^\lambda(h_T) \otimes
        \rho_{W0}^\rho(h_T)^{\prime\prime} \otimes \proj{\phi_3}
    +\frac{1}{2}P_1 \rho_{W1}^\lambda(h_T)^{\prime\prime} \otimes
        \rho_{W0}^\rho(h_T) \otimes \proj{\phi_2} \right)
\end{eqnarray*}
Where the unraised weight is found below the shelf, in the first
line, the piston has been restored to the center. However, it is
left in states $\ket{\phi_2}$ and $\ket{\phi_3}$ on the second
line. These are in general superpositions of the piston states
$\ket{\phi_L}$, $\ket{\phi_R}$ and $\ket{\phi_0}$. As both weights
are above the shelf, the piston may be located anywhere. However,
as the probabilities of the locations of the weights have not
changed, the internal energy of the system is the same as
$E_{T3}$.

\subsubsection{Return to Equilibrium}
We now remove the shelves, in Stage (f), by the operation of
$U_S^\dag$, and allow the weights to come to a thermal equilibrium
at temperature $T_W$. The equilibrium states of the weights
depends upon the location of the piston and pulley system. The
piston states $\ket{\phi_L}$ and $\ket{\phi_R}$ will each support
one of the weights at a height $h_T$, while state $\ket{\phi_0}$
allows both weights to fall to the floor. This corresponds to an
conditional internal Hamiltonian for the weights of
\begin{eqnarray*}
H_{W3}&=&H_W^\lambda(0)H_W^\rho(0)\proj{\phi_0} \\
&&   +H_W^\lambda(h_T)H_W^\rho(0)\proj{\phi_R}
     +H_W^\lambda(0)H_W^\rho(h_T)\proj{\phi_L}
\end{eqnarray*}
As shown in Section \ref{s:szsm1}, thermalisation of a system with
conditional Hamiltonian leads to a canonical distribution within
each of the projected subspaces $\ket{\phi_L}$, $\ket{\phi_R}$ and
$\ket{\phi_0}$. The probability of each subspace is given by the
trace of the projection onto the subspaces in the original density
matrix:
\begin{eqnarray*}
\proj{\phi_L}\rho_{T3}\proj{\phi_L}&=&\rho_{G0}\otimes\left(
    \frac{1}{2}P_1 \magn{b_3} \rho_{W1}^\lambda(h_T) \otimes
        \rho_{W0}^\rho(h_T)^{\prime\prime} \right. \\
&&\left.
    +\frac{1}{2}P_1 \magn{b_2}\rho_{W1}^\lambda(h_T)^{\prime\prime} \otimes
        \rho_{W0}^\rho(h_T) \right) \otimes \proj{\phi_L} \\
\trace{\proj{\phi_L}\rho_{T3}\proj{\phi_L}}&=&
    \frac{1}{2}P_1 (\magn{b_2}+\magn{b_3}) \\
\proj{\phi_R}\rho_{T3}\proj{\phi_R}&=&\rho_{G0}\otimes\left(
    \frac{1}{2}P_1 \magn{c_3} \rho_{W1}^\lambda(h_T) \otimes
        \rho_{W0}^\rho(h_T)^{\prime\prime} \right. \\
&&\left.
    +\frac{1}{2}P_1 \magn{c_2}\rho_{W1}^\lambda(h_T)^{\prime\prime} \otimes
        \rho_{W0}^\rho(h_T) \right) \otimes \proj{\phi_R} \\
\trace{\proj{\phi_R}\rho_{T3}\proj{\phi_R}}&=&
    \frac{1}{2}P_1 (\magn{c_2}+\magn{c_3}) \\
\proj{\phi_0}\rho_{T3}\proj{\phi_0}&=&\rho_{G0}\otimes\left(
     \frac{1}{2}P_2 \rho_{W1}^\lambda(h_T) \otimes
        \rho_{W0}^\rho(0)^{\prime\prime}
    +\frac{1}{2}P_2 \rho_{W1}^\lambda(0)^{\prime\prime} \otimes
        \rho_{W0}^\rho(h_T) \right. \\
&&  +\frac{1}{2}P_1 \magn{a_3} \rho_{W1}^\lambda(h_T) \otimes
        \rho_{W0}^\rho(h_T)^{\prime\prime} \\
&&\left.
    +\frac{1}{2}P_1 \magn{a_2}\rho_{W1}^\lambda(h_T)^{\prime\prime} \otimes
        \rho_{W0}^\rho(h_T) \right)\otimes \proj{\phi_0} \\
\trace{\proj{\phi_0}\rho_{T3}\proj{\phi_0}}&=&
    P_2+\frac{1}{2}P_1 (\magn{a_2}+\magn{a_3})
\end{eqnarray*}

The weights now come into equilibrium on with the heat bath at
temperature $T_W$, with the final state of the weights conditional
upon the projected state of the piston. The canonical
distributions of the weights are:
\begin{eqnarray*}
\proj{\phi_0}& \rightarrow &
    \rho_{W1}^\lambda(0) \otimes \rho_{W1}^\rho(0) \\
\proj{\phi_R}& \rightarrow &
    \rho_{W1}^\lambda(h_T) \otimes \rho_{W1}^\rho(0) \\
\proj{\phi_L}& \rightarrow &
    \rho_{W1}^\lambda(0) \otimes \rho_{W1}^\rho(h_T)
\end{eqnarray*}

When the piston is in the center, the equilibrium consists of the
two weights in a thermal state on the floor. If the piston is in
the righthand position, the equilibrium thermal state has a raised
lefthand weight, with the righthand weight on the floor, and vice
versa.

\subsubsection{Conclusion}

We have now completed the 'Raising Cycle' of the Popper-Szilard
Engine. The final state of the density matrix of the system is:

\begin{eqnarray}
\rho_{T5}&=&
    \rho_{G0} \otimes \left(
    w_1 \rho_{W1}^\lambda(0) \otimes \rho_{W1}^\rho(0) \otimes
    \proj{\phi_0}
    + w_2 \rho_{W1}^\lambda(h_T) \otimes \rho_{W1}^\rho(0) \otimes
    \proj{\phi_R} \right. \nonumber \\
&& \left.
    + w_3 \rho_{W1}^\lambda(0) \otimes \rho_{W1}^\rho(h_T) \otimes
    \proj{\phi_L} \right)
\label{eq:rt5}
\end{eqnarray}

where the statistical weights $w_1$, $w_2$ and $w_3$ are
calculated from the projection onto the subspaces of
$\proj{\phi_0}$, $\proj{\phi_R}$ and $\proj{\phi_L}$ above.

\begin{eqnarray}
w_1 &=&P_2+\frac{1}{2} P_1 \left( \magn{a_2}+\magn{a_3}\right) \nonumber \\
    &=&1-\frac{1}{2} P_1 \left(1+\magn{a_1}\right) \nonumber \\
w_2 &=&\frac{1}{2}P_1 \left(\magn{b_2}+\magn{b_3}\right) \nonumber \\
    &=&\frac{1}{2}P_1 \left(1-\magn{b_1}\right) \nonumber \\
w_3 &=&\frac{1}{2}P_1 \left(\magn{c_2}+\magn{c_3}\right) \nonumber \\
    &=&\frac{1}{2}P_1 \left(1-\magn{c_1}\right)
\label{eq:w123}
\end{eqnarray}
and we have made use of the identities, from the unitarity of
$U_{RES}$, in Equation \ref{eq:unres}.

The internal energy of $\rho_{T5}$ is
\begin{eqnarray*}
E_{T5}&=&\frac{1}{2}kT_G+3kT_W+(w_2+w_3)M_Wgh_T \\
    &=& E_{T1}(1-p)-w_1M_Wgh_T
\end{eqnarray*}

In $w_1$ proportion of cycles, the piston is restored to the
center of the Engine. In these cases, the raised weight has been
allowed to fall back to the floor. This dissipates $M_Wgh_T$
energy into the $T_W$ heat bath. The system is then ready to
perform another raising cycle of the Popper-Szilard Engine.

However, with probability $(w_2+w_3)$, the piston will not be
restored to the center. On these cycles, the energy extracted from
the $T_G$ heat bath has been transferred to the weights, but it
has not been dissipated into the $T_W$ heat bath\footnote{Strictly
speaking, it is possible that the cycle has ended with the {\em
unraised} weight trapped in a thermal fluctuation, while the
raised weight {\em is} allowed to fall dissipatively. The result
of this, however, is still no net transfer of energy to the $T_W$
heat bath.}. Instead, one of the weights has been trapped by the
imperfect resetting of the piston leaving it on the left or right
of the Engine. The system will not be able to continue with a
raising cycle, but will instead 'reverse direction' and use the
trapped energy to start upon a lowering cycle.

\section{The Lowering Cycle}
\label{s:szsm6}

We will now repeat the analysis of Section \ref{s:szsm5}, but this
time we will consider the 'lowering cycle' described in Section
\ref{s:szqm6}. In this cycle, we start with the piston to one or
the other side of the Engine, and with the corresponding weight
trapped at the height $h_T$. We will then apply the stages of the
operator $U_T$, exactly as we did for the raising cycle. This will
be shown to take us through the steps in Figure \ref{fg:lower}.

\subsubsection{Pumping Energy into the $T_G$ Heath Bath}

We start with the initial density matrix corresponding to the
piston located on the right of the Engine:

\[
\rho_{T6}=\rho_{G0} \otimes \rho_{W1}^\lambda(h_T)
     \otimes \rho_{W0}^\rho \otimes \proj{\phi_R}
\]

This has internal energy
\[
E_{T6}=\frac{1}{2}kT_G+3kT_W+M_Wgh_T
\]

Stage (a) consists of the operation $U_{RI}$, which in this case
simply corresponds to inserting the piston in the right end of the
box, at $Y=(1-p)$. The gas will be entirely to the left of the
piston, and will be subject to a negligible compression. The state
is now
\[
\rho_{T7}(1-p)=\rho_{G6}^\lambda(1-p) \otimes
\rho_{W1}^\lambda(h_T)
     \otimes \rho_{W0}^\rho \otimes \proj{\Phi(1-p)}
\]
We now go through Stage (b), which involves the operation
$U_{W4}$. This causes the gas to compress, while the lefthand
weight is lowered. As the position of the piston moves from
$Y=1-p$ to $Y=0$, the system moves through
\[
\rho_{T7}(Y)=\rho_{G6}^\lambda(Y) \otimes \rho_{W1}^\lambda(h(Y))
     \otimes \rho_{W0}^\rho \otimes \proj{\Phi(Y)}
\]
until it reaches
\[
\rho_{T7}(0)=\rho_{G6}^\lambda(0) \otimes \rho_{W0}^\lambda
     \otimes \rho_{W0}^\rho \otimes \proj{\Phi(0)}
\]
at the end of Stage (b). This state has internal energy
\[
E_{T7}(0)=\frac{1}{2}kT_G+3kT_W
\]
The compression of the gas is isothermal, so the internal energy
of the gas remains constant throughout this stage at
$\frac{1}{2}kT_G$. The work performed upon the gas is passed into
the $T_G$ heat bath. The system has transferred $M_Wgh_T=kT_G\ln
2$ energy from the raised weight to the heat bath.

\subsubsection{Resetting the Piston Position}

Operation $U_S$, during Stage (c), inserts shelves at height $h_T$
into the space of the weights. As both of these weights are in the
unraised position, both of the weights will be projected out:
\begin{eqnarray*}
\rho_{T8}&=&\rho_{G6}^\lambda(0) \otimes
    \left\{P_1 \rho_{W0}^\lambda(h_T)^{\prime\prime}+
            P_2 \rho_{W0}^\lambda(0)^{\prime\prime}\right\} \\
&&  \otimes \left\{P_1 \rho_{W0}^\rho(h_T)^{\prime\prime}+
            P_2 \rho_{W0}^\rho(0)^{\prime\prime}\right\}
    \otimes \proj{\Phi(0)}
\end{eqnarray*}
(again, for convenience we have assumed that thermal contact with
the $T_W$ heat bath destroys coherence between the raised and
unraised density matrices). The mean energy is unaffected by this.

Stage (d) now removes the piston from the center of the box.
Unlike the raising cycle, this has a significant effect upon the
internal state of the one atom gas. In $\rho_{T8}$ the gas is
confined entirely to the left half of the box. When the piston is
removed, the internal Hamiltonian for the gas becomes $H_{G0}$.
With the full extent of the box accessible, the contact with the
$T_G$ heat bath allows the gas to expand to the equilibrium state
$\rho_{G0}$, leaving the system in the state
\begin{eqnarray*}
\rho_{T9}&=&\rho_{G0}\otimes
    \left(
        (P_1)^2 \rho_{W0}^\lambda(h_T)^{\prime\prime}
        \otimes \rho_{W0}^\rho(h_T)^{\prime\prime}
        +P_1P_2 \rho_{W0}^\lambda(0)^{\prime\prime}
        \otimes \rho_{W0}^\rho(h_T)^{\prime\prime}\right. \\
&&\left.
        +P_1P_2 \rho_{W0}^\lambda(h_T)^{\prime\prime}
        \otimes \rho_{W0}^\rho(0)^{\prime\prime}
        +(P_2)^2 \rho_{W0}^\lambda(0)^{\prime\prime}
        \otimes \rho_{W0}^\rho(0)^{\prime\prime}
        \right)
    \otimes \proj{\phi_0}
\end{eqnarray*}
However, the internal energy of the gas is still $\frac{1}{2}kT_G$
so the energy of the system has not been affected by the free
expansion of the one atom gas.

We can see all four of the possible configurations of the weights
are present. The resetting of the piston, $U_{RES}$, in Stage(e)
leads to the piston being in any of the possible locations,
including the superposition $\ket{\phi_1}$
\begin{eqnarray*}
\rho_{T10}&=&\rho_{G0}\otimes
    \left(
        (P_1)^2 \rho_{W0}^\lambda(h_T)^{\prime\prime}
        \otimes \rho_{W0}^\rho(h_T)^{\prime\prime}
        \otimes \proj{\phi_1} \right.\\
&&      +P_1P_2 \rho_{W0}^\lambda(0)^{\prime\prime}
        \otimes \rho_{W0}^\rho(h_T)^{\prime\prime}
        \otimes \proj{\phi_L} \\
&&      +P_1P_2 \rho_{W0}^\lambda(h_T)^{\prime\prime}
        \otimes \rho_{W0}^\rho(0)^{\prime\prime}
        \otimes \proj{\phi_R} \\
&&\left.+(P_2)^2 \rho_{W0}^\lambda(0)^{\prime\prime}
        \otimes \rho_{W0}^\rho(0)^{\prime\prime}
        \otimes \proj{\phi_0}
        \right)
\end{eqnarray*}
The second and third lines represent the situation where one
weight was trapped above the shelf, and one below. In this
situation, the piston is moved to the corresponding side of the
engine, to hold up the trapped weight. This allows the machine to
continue with a lowering cycle.

The fourth line gives the situation where both weights are trapped
below the shelf height. As neither weight is in a raised position,
the piston cannot be moved without changing the location of a
weight. $U_{RES}$ therefore leaves the piston in the central
position. This means that at the start of the next cycle, the
piston will be in the central position, and a raising cycle will
begin.

When both weights are trapped above the shelf height $h_T$, the
effect of $U_{RES}$ is to put the piston into the superposition of
states given by $\ket{\phi_1}$. This superposition is constrained
by the unitarity requirements on $U_{RES}$ given in Equation
\ref{eq:unres}.

\subsubsection{Return to Equilibrium}
As with the raising cycle, the shelves are removed by $U_S^\dag$
operation in Stage (f), and the weights come to a thermal
equilibrium with the $T_W$ heat bath.

The internal Hamiltonian for the weights is $H_{W3}$ as in the
raising cycle above. The process of thermalisation is therefore
exactly the same as for the raising cycle, requiring us to project
out each of the subspaces of the piston:

\begin{eqnarray*}
\proj{\phi_L}\rho_{T10}\proj{\phi_L}&=&\rho_{G0} \otimes
     \left(
        (P_1)^2 \magn{b_1} \rho_{W0}^\lambda(h_T)^{\prime\prime}
        \otimes \rho_{W0}^\rho(h_T)^{\prime\prime} \right.\\
&&\left.
        +P_1P_2 \rho_{W0}^\lambda(0)^{\prime\prime}
        \otimes \rho_{W0}^\rho(h_T)^{\prime\prime}\right)
        \otimes \proj{\phi_L} \\
\trace{\proj{\phi_L}\rho_{T10}\proj{\phi_L}}&=&
        (P_1)^2\magn{b_1}+P_1P_2 \\
\proj{\phi_R}\rho_{T10}\proj{\phi_R} &=& \rho_{G0} \otimes
    \left(
        (P_1)^2 \magn{c_1} \rho_{W0}^\lambda(h_T)^{\prime\prime}
        \otimes \rho_{W0}^\rho(h_T)^{\prime\prime} \right. \\
&&\left.+P_1P_2 \rho_{W0}^\lambda(h_T)^{\prime\prime}
        \otimes \rho_{W0}^\rho(0)^{\prime\prime} \right)
        \otimes \proj{\phi_R} \\
\trace{\proj{\phi_R}\rho_{T10}\proj{\phi_R}}&=&
        (P_1)^2\magn{c_1}+P_1P_2 \\
\proj{\phi_0}\rho_{T10}\proj{\phi_0}&=& \rho_{G0} \otimes
    \left(
        (P_1)^2\magn{a_1} \rho_{W0}^\lambda(h_T)^{\prime\prime}
        \otimes \rho_{W0}^\rho(h_T)^{\prime\prime}\right.\\
&&\left.+(P_2)^2 \rho_{W0}^\lambda(0)^{\prime\prime}
        \otimes \rho_{W0}^\rho(0)^{\prime\prime} \right)
        \otimes \proj{\phi_0} \\
\trace{\proj{\phi_0}\rho_{T10}\proj{\phi_0}}&=&
        (P_1)^2\magn{a_1}+(P_2)^2
\end{eqnarray*}

Contact with the $T_W$ heat bath will then bring the weights into
canonical equilibrium distributions, conditional upon the location
of the piston:
\begin{eqnarray*}
\proj{\phi_0}& \rightarrow &
    \rho_{W1}^\lambda(0) \otimes \rho_{W1}^\rho(0) \\
\proj{\phi_R}& \rightarrow &
    \rho_{W1}^\lambda(h_T) \otimes \rho_{W1}^\rho(0) \\
\proj{\phi_L}& \rightarrow &
    \rho_{W1}^\lambda(0) \otimes \rho_{W1}^\rho(h_T)
\end{eqnarray*}

\subsubsection{Conclusion}
The density matrix that results from the thermalisation in Stage
(f) is

\begin{eqnarray}
\rho_{T11}&=&
    \rho_{G0} \otimes \left(
    w_4 \rho_{W1}^\lambda(0) \otimes \rho_{W1}^\rho(0) \otimes
    \proj{\phi_0}
    + w_5 \rho_{W1}^\lambda(h_T) \otimes \rho_{W1}^\rho(0) \otimes
    \proj{\phi_R} \right. \nonumber \\
&& \left.
    + w_6 \rho_{W1}^\lambda(0) \otimes \rho_{W1}^\rho(h_T) \otimes
    \proj{\phi_L} \right)
\label{eq:rt11}
\end{eqnarray}
where the statistical weights $w_4$, $w_5$ and $w_6$ are
calculated from the projections onto the $\proj{\phi_0}$,
$\proj{\phi_R}$ and $\proj{\phi_L}$ subspaces, respectively.
Making use of the identities in Equation \ref{eq:unres} that come
from the unitarity of $U_{RES}$, we have:

\begin{eqnarray*}
w_4 &=& (P_2)^2+(P_1)^2\magn{a_1}\\
    &=& (1-2P_1)+(P_1)^2\left(1+\magn{a_1}\right) \\
w_5 &=& P_1\left(P_2+P_1\magn{b_1}\right) \\
    &=& P_1-(P_1)^2\left(1-\magn{b_1}\right) \\
w_6 &=& P_1\left(P_2+P_1\magn{c_1}\right) \\
    &=& P_1-(P_1)^2\left(1-\magn{c_1}\right)
\end{eqnarray*}

After thermal equilibrium has been established, the mean energy is
\[
E_{T11}=\frac{1}{2}kT_G+3kT_W+(w_5+w_6)M_Wgh_T
\]
In $(w_5+w_6)$ proportion of the cases, the cycle will complete
with one of the weights trapped at height $h_T$, gaining an energy
$M_Wgh_T$. This energy comes from thermal fluctuations of the
weight, and therefore is drawn from the $T_W$ heat bath. In these
cases, the piston is located to one side, or the other, of the
Engine, and when the next cycle starts it will be another lowering
cycle. This shows that the lowering cycle proceeds by capturing
thermal fluctuations from the $T_W$ heat bath, and using them to
compress the single atom gas. This transfers heat from the $T_W$
to the $T_G$ heat bath. We have confirmed that the flow of energy
in the lowering cycle is in the opposite direction to the flow of
energy in the raising cycle.

In $w_4$ proportion of the cases, however, both weights will be on
the floor at the end of a lowering cycle, and the piston will be
in the center. The next cycle of the Popper-Szilard Engine will
therefore be a raising cycle.

\section{Energy Flow in Popper-Szilard Engine} \label{s:szsm7}

We have now reached the conclusion of our analysis of the
behaviour of the quantum mechanical Popper-Szilard Engine. We
shall briefly review the situation, before calculating the long
term behaviour of the Engine. This will enable us to prove that,
for any choice of $U_{RES}$, the energy flow will be from the
hotter to the colder of $T_W$ and $T_G$. Thus we will show that
the Popper-Szilard Engine is incapable of producing anti-entropic
heat flows.

In Chapter \ref{ch:szqm} we analysed the detailed interactions
between the microstates of the Engine, restricting ourselves only
by the requirement that the evolution of the system be expressed
as a unitary operator. We found that it was possible to extract
energy from the quantum mechanical one atom gas, and use it to
lift a weight, without making a measurement upon the system. We
also found that we could try to reset the piston position, without
having to perform work upon it, albeit with some error. This error
leads to some probability of the Engine going into a reverse
lowering cycle. However, we found that there was also a
corresponding tendency for the Engine on the lowering cycle to
change back to a raising cycle.

An Engine which spends most of it's time on raising cycles will
transfer energy from the $T_G$ to the $T_W$ heat baths, while an
Engine which spends more time on lowering cycles will transfer
energy in the opposite direction. For the second law of
thermodynamics to hold, these tendencies must be balanced so that
the long term flow of energy is always in the direction of the
hotter to the colder heat bath.

In this Chapter we have added statistical mechanics to the
analysis. This allows us to optimise the energy transferred
between the one atom gas and the weights per cycle, and calculate
the probabilities that the Engine changes between the raising and
lowering cycles. We can now use these results to calculate the
long term energy flow between the two heat baths.

\subsubsection{Energy Transfer per Cycle}

On the raising cycle, the energy transfer is $kT_G \ln 2$ per
cycle, from the $T_G$ heat bath to the $T_W$ heat bath. We will
regard the energy of any raised weights at the end of the cycle as
part of the energy of the $T_W$ system, even though it has not
been dissipatively transferred to the $T_W$ heat bath itself.

\[
\Delta E_r=kT_G \ln 2
\]

On the lowering cycle, the energy transfer is from the raised
weight to the $T_G$ heat bath. Again, regarding the weights as
part of the $T_W$ system, this constitutes a transfer of $kT_G \ln
2$ energy, but now in the opposite direction

\[
\Delta E_l=-kT_G \ln 2
\]

\subsubsection{Length of Cycles}

If the probability of a cycle reversing is $p$, and of continuing
is $(1-p)$, then mean number of cycles before a reversal takes
place is $1/p$.

For raising cycle, the probability of the cycle continuing is
given by
\begin{eqnarray*}
1-P_r &=& w_1\\
    &=& 1-\frac{1}{2}P_1\left(1+\magn{a_1}\right)
\end{eqnarray*}
and of reversing
\begin{eqnarray*}
P_r &=& w_2+w_3\\
    &=& \frac{1}{2}P_1\left(1-\magn{b_1}\right)
        +\frac{1}{2}P_1\left(1-\magn{b_1}\right)\\
    &=& \frac{1}{2}P_1\left(1+\magn{a_1}\right)
\end{eqnarray*}
The mean number of raising cycles that takes place is therefore
\[
N_r=1/P_r=\frac{2}{P_1\left(1+\magn{a_1}\right)}
\]
The lowering cycle has continuation and reversal probabilities of
\begin{eqnarray*}
1-P_l &=& w_5+w_6 \\
    &=& P_1\left(2P_2+P_1\left(\magn{b_1}+\magn{c_1}\right)\right) \\
    &=& 2P_1-(P_1)^2\left(1+\magn{a_1}\right) \\
    &=& 2P_1(1-P_r) \\
P_l &=& w_4 \\
    &=& (P_2)^2+(P_1)^2\magn{a_1} \\
    &=& (1-2P_1)+(P_1)^2\left(1+\magn{a_1}\right) \\
    &=& 1-2P_1(1-P_r)
\end{eqnarray*}
respectively. The mean number of lowering cycles is
\[
N_l=1/P_l=\frac{1}{(1-2P_1)+(P_1)^2\left(1+\magn{a_1}\right)}
\]

\subsubsection{Mean Energy Flow}

As the Popper-Szilard Engine will alternate between series of
raising and lowering cycles, in the long term the net flow of
energy from the $T_G$ to the $T_W$ heat baths, per cycle, is given
by:

\[
\Delta E=\frac{N_r \Delta E_r + N_l \Delta E_l}
    {N_r+N_l}
\]

Substituting in the values and re-arranging leads to the final
equation for the flow of energy in the Popper-Szilard Engine

\begin{equation}
\Delta E= kT_G \ln 2 \left(
    \frac{(1-2P_1)\left(1-\frac{P_1}{2}\left(1+\magn{a_1}\right)\right)}
        {(1-2P_1)+(1+2P_1)\frac{P_1}{2}\left(1+\magn{a_1}\right)}
    \right)
\label{eq:energy}
\end{equation}

It is interesting to note that, of all the possible values that
could be chosen for the operation $U_{RES}$, in the long run it is
only the value $\magn{a_1}$ that has any effect. The value of
$\magn{a_1}$ is related to the probability of the lowering cycle
reversing direction when both weights are trapped above the shelf
height. The symmetry of the Popper-Szilard Engine between the
righthand and lefthand states, and the existence of the unitarity
constraints on $U_{RES}$, such as $\sum_i \magn{a_i}=1$, lead to
all relevant properties expressible in terms of $\magn{a_1}$.

The function
\[
f(P_1,\magn{a_1})=
\frac{(1-2P_1)\left(1-\frac{P_1}{2}\left(1+\magn{a_1}\right)\right)}
        {(1-2P_1)+(1+2P_1)\frac{P_1}{2}\left(1+\magn{a_1}\right)}
\]
is plotted in Figure \ref{fg:energy} as $P_1$ and $\magn{a_1}$
vary between the values of $0$ and $1$. \pict{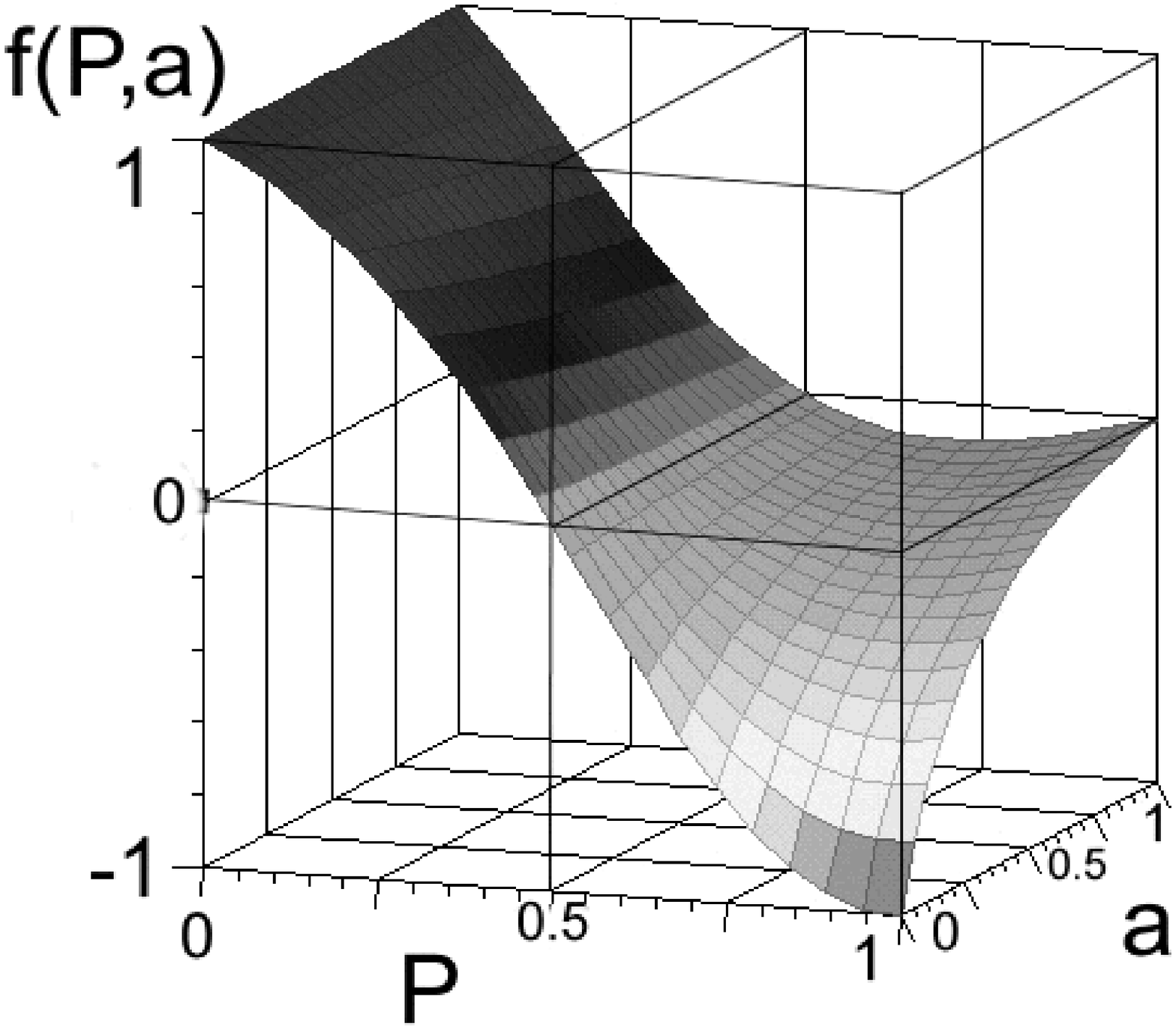}{Mean Flow of
Energy in Popper-Szilard Engine} This shows that
\begin{eqnarray*}
P_1<\frac{1}{2} & \Rightarrow & f(P_1,\magn{a_1})>0 \\
P_1=\frac{1}{2} & \Rightarrow & f(P_1,\magn{a_1})=0 \\
P_1>\frac{1}{2} & \Rightarrow & f(P_1,\magn{a_1})<0
\end{eqnarray*}
regardless of the value of $a_1$. The direction of the long term
flow of energy in the Popper-Szilard Engine is completely
independant of the choice of the resetting operation $U_{RES}$. It
depends only upon the size of $P_1$. When there {\em is} a mean
flow of energy, then the choice of $\magn{a_1}$, and thereby of
$U_{RES}$, does have an affect upon the {\em size} of mean energy
flow per cycle, but it cannot affect the direction of the flow.

If we now look at the form of $P_1$ in Equation \ref{eq:p1a}, we
find
\[
P_1=\left(\frac{1}{2}\right)^\frac{T_G}{T_W}
\]

From this, and the form of $f(P_1,\magn{a_1})$, we have the proof
of our central result, that the mean flow of heat is always in the
direction of hotter to colder:

\paragraph{Solution to Popper-Szilard Engine}

\begin{equation}
\begin{array}{ccccc}
T_G>T_W & \Rightarrow & P_1<\frac{1}{2} & \Rightarrow & \Delta E>0 \\
\\
T_G=T_W & \Rightarrow & P_1=\frac{1}{2} & \Rightarrow & \Delta E=0 \\
\\
T_G<T_W & \Rightarrow & P_1>\frac{1}{2} & \Rightarrow & \Delta E<0
\end{array} \label{eq:result}
\end{equation}

This proves that despite the arguments in Chapter \ref{ch:szmd},
the Popper-Szilard Engine is not, in the long run, capable of
violating the second law of thermodynamics, as defined by Clausius
\begin{quotation}
No process is possible whose sole result is the transfer of heat
from a colder to a hotter body
\end{quotation}
Although we have now achieved our primary goal, of providing a
complete analysis of the quantum mechanical Popper-Szilard Engine,
and demonstrating that it does not violate the second law of
thermodynamics, it will be useful to examine how the function
$f(P_1,\magn{a_1})$ varies with the choice of $\magn{a_i}$, $T_G$
and $T_W$.

\paragraph{$T_G \gg T_W$}
When $T_G \gg T_W$, then $P_1 \approx 0$. In this situation, the
gas is able to lift the weight through a very large distance,
compared with the mean thermal height of the weight. There is
correspondingly a vanishingly small probability that the unraised
weights will be found above the shelf height.

On the raising cycle, this leads to an unambiguous correlation
between the piston states and the location of the raised and
unraised weights, and the piston will be reset with negligible
error. The raising cycle will therefore continue almost
indefinitely.

Should the Engine find itself in a lowering cycle, however, at the
end of the cycle both weights will be found below the shelf
height. The operation of $U_{RES}$ will leave the piston in the
center. Lowering cycles will therefore immediately reverse into
raising cycles.

The result is that the Engine will switch to and reliably stay on
a raising cycle, and will transfer $kT_G \ln 2$ energy from the
hotter $T_G$ to the colder $T_W$ per cycle.

\paragraph{$T_G=T_W$}
If $P_1=\frac{1}{2}$, there is exactly $50\%$ probability of
finding an unraised weight above the shelf height. The
probabilities of continuing and reversing become
\[
P_r=P_l=\frac{1}{4}\left(1+\magn{a_1}\right)
\]
This varies between $1/4$ and $1/2$. The mean number of cycles
before a reversal takes place is between 2 and 4. As it is equal
for raising and lowering cycles, in the long term there is no mean
flow of energy between the two heat baths. However, the energy
transfer will fluctuate about this mean.

\paragraph{$T_G \ll T_W$}
When the gas temperature is much lower than the weight temperature
the situation is more complex, and the value of $\magn{a_i}$
becomes more significant. $P_1 \approx 1$ implies that unraised
weights will always be located above the shelf height. The only
part of $U_{RES}$ that will be relevant will be the projection
onto the $P^\lambda(RA)P^\rho(RA)$ subspace. This part of the
operation puts the piston state into a superposition, which is
dependant upon the values of the $a_i$ etc. parameters in
$U_{RES}$.

Let us first consider an operator for which $a_1=0$. On the
lowering cycle, the piston is in the center of the Engine, and
$U_{RES}$ will always move it to one of the lefthand or righthand
states. Lowering cycles will therefore continue indefinitely. For
the raising cycle, the piston comes out of the box in the lefthand
or righthand position, with equal probability, $\frac{1}{2}$. The
unitarity requirements then lead to $\magn{a_2}+\magn{a_3}=1$.
These are the probabilities of the raising cycle continuing, from
the lefthand and righthand piston positions, respectively. The
overall probability of the raising cycle continuing is therefore
$\frac{1}{2}\left(\magn{a_2}+\magn{a_3}\right)$. This gives only a
$50\%$ chance that a raising cycle will continue. On average, a
raising cycle will only perform two cycles before reversing into a
lowering cycle. The long term behaviour of this is to stay on the
lowering cycle, and transfer $kT_G \ln 2$ from the hotter $T_W$ to
the colder $T_G$ heat baths.

If we increase $a_1$, we start to introduce a possibility of the
lowering cycle reversing into a raising cycle. However, as we do
this, we simultaneously reduce $\magn{a_2}+\magn{a_3}$, reducing
the ability of the raising cycle to continue. If we reach $a_1=1$,
we guarantee that the lowering cycle will reverse into a raising
cycle. However, we have simultaneously removed all possibility of
the raising cycle continuing. The machine simply switches between
the two cycles, producing a net zero energy flow, despite the high
temperature of $T_W$.

If the value of $P_1<1$, though, there is some possibility of an
unraised weight being trapped below the shelf. This increases the
possibility of the machine staying on a lowering cycle, and allows
some flow of heat.

\subsubsection{Density Matrix}
We have derived these results in terms of the long term behaviour
of the Popper-Szilard Engine, implicity assuming that on each
cycle of the Engine it is in either a raising or lowering cycle.
We now wish to re-examine this in terms of the density matrix of
the system. For simplicity, we will make use of the symmetry of
the Engine, and set $\magn{b_1}=\magn{c_1}$, and use the lowering
cycle density matrix
\[
\rho_{T12}=\frac{1}{2}\rho_{G0} \otimes \left(
    \rho_{W1}^\lambda(h_T)\otimes\rho_{W1}^\rho(0)\otimes\proj{\phi_R}
   +\rho_{W1}^\lambda(0)\otimes\rho_{W1}^\rho(h_T)\otimes\proj{\phi_L}
    \right)
\]

If the Engine starts the cycle in a general state, with some
probability $w_r$ of being on a raising cycle, the density matrix
is:
\[
\rho_{T13}=w_r\rho_{T0}+(1-w_r)\rho_{T12}
\]
After one cycle, it will be left in the state
\[
\rho_{T14}=(w_4+w_r(w_1-w_4))\rho_{T0}+2(w_5+w_r(w_2-w_5))\rho_{T12}
\]
The Engine rapidly converges\footnote{Excluding the case where
$P_1=1$, $a_1=0$, which oscillates between $\rho_{T13}$ and
$(1-w_r)\rho_{T0}+w_r\rho_{T12}$} to a value of $w_r^{\prime}$ for
which $\rho_{T14}=\rho_{T13}$. This value is given by
\[
w_r^{\prime}=\frac{w_4}{2w_2+w_4}
\]
for which the density matrix can be shown to be
\[
\rho_{T15}=\frac{N_r}{N_r+N_l}\rho_{T0}+\frac{N_l}{N_r+N_l}\rho_{T12}
\]
This demonstrates that, even if we do not wish to interpret the
system as being in a determinate state, whose long run energy flow
is given by Equation \ref{eq:energy}, the system will still
rapidly settle into a density matrix for which the mean flow {\em
on each cycle} is given by $\Delta E$. Thus, for this system the
statistical state at a particular time rapidly produces the same
results as the average behaviour over a large number of cycles.

\section{Conclusion} \label{s:szsm8}
Let us step back from the detail by which the simple and expected
result was achieved, and try to understand why the attempt to
produce anti-entropic behaviour fails. As we saw, the essential
property of the Engine's long term behaviour is that it must spend
more time on the raising cycle when $T_G > T_W$, and more time on
the lowering cycle when $T_G < T_W$. This turns on the value of
$P_1$, and it's dependancy on the temperatures of the gas and
weights, and critically takes the value of $\frac{1}{2}$ when
$T_G=T_W$. It is the relationship
\[
P_1=\left(\frac{1}{2}\right)^{\frac{T_G}{T_W}}
\]
which determines the direction of the mean flow of energy.

We must now examine how the various features that go into the
derivation of $P_1$ produce this balance. The key relationship is
between the thermal states of the weights and the gas. The thermal
state of the weight gives it a height above the floor of the
Engine. This leads to a probability of the weight being located
above a given height. The thermal state of the gas, on the other
hand, allows energy to be extracted and used to raise the floor
beneath the weight, to some height (or the lowering of the floor
beneath the weight, from some height, can be use to compress the
gas).

The probability\footnote{ This is the same as the Boltzmann
distribution for a classical gas in a gravitational field.} of
finding the weight above a height $h$ is
$e^{-\frac{M_Wgh}{kT_W}}$. The median height of the weight is
$h_m=\frac{ kT_W}{M_Wg}\ln 2$, which gives the height above which
it is $50\%$ likely that the weight will spontaneously be found
(the mean height $\mean{h}=\frac{kT_W}{M_Wg}$, which confirms the
expectation value of the potential energy $kT_W$ in Section
\ref{s:szsm3}) This height may be reduced by increasing the mass
of the weight, or by reducing it's temperature.

However, the height through which the weight can be lifted, is set
by it's weight, and by the temperature of the gas $T_G.$ The
maximum height that can be achieved is using isothermal expansion,
which raises it by $h_T=\frac{kT_G}{ M_Wg}\ln 2$. This may be
increased by reducing the mass, or increasing the temperature of
the gas.

We want $h_m <h_T$ to be reliably transferring energy from $T_G$
to $T_W$. If we decrease the likelihood that an unraised weight is
found above the height $h_T$, we improve the probability that the
machine is properly reset to start the next cycle. Changing the
mass does not help, as any reduction in the median height of the
weight is offset by a reduction in the height through which it is
lifted. Instead, we are forced to reduce $T_W$ or increase $T_G$.

However, clearly, for $h_m < h_T$, then $T_W <T_G$. If we wish to
transfer energy from a cold to a hot heat bath we need $T_W >T_G$.
In more than $50\%$ of the cases, a shelf inserted at $h_T$ will
find the weight already lifted, without any action required by the
gas. We only start to reliably (more that $50\%$ of the time) find
the weight below the shelf height if the temperature of the weight
is below that of the gas - in which case we are simply arranging
for heat to flow from a hotter to a colder body, in agreement with
the second law.

If we try to run the machine in reverse, we need to be able to
reliably capture fluctuations in the height of the weights and use
them to compress the gas. To compress the gas, the weight must be
caught above the height $h_T$. To be reliably (ie. with
probability greater than $50\%$) caught above this height, then
$h_m>h_T$. Again, we find the balance between $h_m$ and $h_T$
implies $T_W>T_G$, so that the heat flows from the hotter to the
colder heat bath.

There are two key elements we have found. Firstly, unitarity
constrains the operation of the Engine. We are not able to ensure
the machine stays on one cycle (raising or lowering) because the
resetting operation $U_{RES}$ must be unitary and cannot map
orthogonal to non-orthogonal states. Furthermore, unitarity
requires we define the operation over the entire Hilbert space of
the Engine. Once we define the operation of the Engine for one
cycle, we find we have completely defined the operation of the
Engine on the reversed cycle. The way we attempt to extract Engine
in one direction automatically implies a flow of energy in the
opposite direction.

The second element is the subtle balance between the thermal
states of the two systems. When we try to capture a fluctuation in
the gas, and use it to lift the weight through some height, we
found that, unless the gas was hotter than the weight, then we
were at least as likely to find the weight already above that
height, due to it's own thermal state. Similarly, when we capture
a fluctuation in the height of the weight, and use the lowering of
it to compress the gas, we find that, unless the weight is hotter
than the gas, probability of capturing the weight above the height
is less than the probability of finding the gas spontaneously in
the compressed state.

In Chapter \ref{ch:szsol} we will show the general physical
principles which underly these two elements. This will enable us
to generalise the conclusion of our analysis of the Popper-Szilard
Engine.
\chapter{The Thermodynamics of Szilard's Engine} \label{ch:szth}
Chapters \ref{ch:szqm} and \ref{ch:szsm} present a detailed
analysis of the operation of the quantum Popper-Szilard Engine.
The conclusion showed that no operation of the Engine compatible
with unitary dynamics was capable of transferring energy from a
colder to a hotter heat bath. It was not found necessary to make
any reference to information theory to reach this conclusion.

However, little reference has been made to thermodynamics either,
so one might wonder if one could equally abandon the concepts of
entropy or free energy. In fact, the reason why we were able to
avoid referring to these is because the system studied is
sufficiently idealised that it was possible to explicitly
construct operators upon the microstates and analyse statistical
behaviour of an ensemble of microstates. The only thermodynamic
concept introduced was temperature, to describe the statistical
ensembles and the heat baths. This will not be possible for more
complex systems, involving many degrees of freedom. For such
systems it will only be possible to usefully describe them by
aggregate properties, associated with an ensemble. However, this
does {\em not} mean, as it is sometimes asserted, that these
ensemble properties are only {\em valid} for complex, many body
systems. The thermodynamic, ensemble properties can still be
defined for simple, single body systems.

In this Chapter we will analyse the thermodynamic properties of
the Szilard Engine, and show the extent to which they can be
considered valid. We will be principally concerned with the
properties of entropy and free energy. This will give us a deeper
understanding of the reason why the Popper-Szilard does not
operate in an anti-entropic manner, and will form the basis of the
general resolution of the problem in the next Chapter.

In Section \ref{s:szth1} the concepts of free energy and entropy
will be derived from the statistical ensemble mean energy and
pressure, for a system in thermal equilibrium at some temperature
$T$. This demonstrates that these concepts are quite valid for
single atom systems. We will then give some consideration to the
meaning of these terms for systems exhibiting non-equilibrium
mixing and for correlations between different systems. It will be
shown that in some circumstances the concept of free energy must
be modified, and in other circumstances cannot be applied at all.
Entropy, on the other hand, remains well defined at all times.

Section \ref{s:szth2} steps through the six stages of the raising
cycle, given in Sections \ref{s:szqm6} and \ref{s:szsm6}. The
entropy and free energy are tracked throughout the cycle. Section
\ref{s:szth3} then does the same for the lowering cycle (Sections
\ref{s:szqm6} and \ref{s:szsm7}). It will be shown here that the
entropy is always constant or increasing, at all stages of the
operation of the Engine. This conclusion is derived solely from
the principles of statistical mechanics, without reference to
information processing principles.

\section{Free Energy and Entropy}
\label{s:szth1}

In this section we will start by defining clearly what we mean by
free energy and entropy, in terms of mean energy and pressure.
This definition will apply to a single system in thermal
equilibrium at temperature $T$. We will apply these definitions to
the case of the single atom gas, and to the weight supported at
height $h$. We will use this to show how the pressure of the gas
on a moveable piston is used to lift the weight, in thermodynamic
terms. This will justify our argument that thermodynamic concepts
are applicable for single atom systems. Finally, we will examine
how the concepts must be modified to take into account the
non-equilibrium mixing of states, and the correlations between
states of different systems.

We recall from Section \ref{s:szsm1} that the mean pressure
exerted on a system parameter $x$ was defined by
\[
P(x)=\sum_n p_n(x) \frac{\partial E_n}{\partial x}
\]
In an isothermal system, the probabilities are given by
\[
p_n(x)= e^{-\frac{E_n}{kT}}/\sum_m e^{-\frac{E_m}{kT}}
\]
The work done when this parameter is changed isothermally and
reversibly from $x_1$ to $x_2$ is
\begin{eqnarray*}
W&=& \int_{x_1}^{x_2}P(x) dx \\
&=&\int_{x_1}^{x_2}\frac{1}{Z}
    \sum_n e^{-\frac{E_n}{kT}}\frac{\partial E_n}{\partial x}dx \\
&=&\int_{x_1}^{x_2}\frac{-kT}{Z}\frac{\partial}{\partial x}
    \left( \sum_ne^{-\frac{E_n}{kT}}\right) dx \\
&=&\left[ -kT\ln Z\right]_{x_1}^{x_2}
\end{eqnarray*}
where we have used the function $Z=\sum_n e^{-\frac{E_n}{kT}}=
\trace{e^{-H/kT}}$. As the path taken from $x_1$ to $x_2$ is
reversible, it does not matter which path is taken, so $W$ can be
regarded as the change in the function, $F=-kT\ln Z$. This defines
the free energy of the system - it is the energy that can be
extracted isothermally to do work upon another system.

The mean energy of the system is, of course,
\[
E=\frac{1}{Z}\sum_n e^{-\frac{E_n}{kT}}E_n
\]
so the difference between the mean and free energy is given by the 'heat'
\begin{eqnarray*}
Q &=&\frac 1Z\sum_ne^{-\frac{E_n}{kT}}E_n+kT\ln Z \\
&=&-kT\sum_n\frac{e^{-\frac{E_n}{kT}}}{Z}
    \ln \left(\frac{e^{-\frac{E_n}{kT}}}{Z}\right) \\
&=&-kT\trace{\rho \ln \rho}
\end{eqnarray*}
with $\rho =\frac{1}{Z}e^{-H/kT}$, as the density matrix of the
system in equilibrium, thus confirming that the Gibbs-von Neumann
entropy $S_{VN}=-k\trace{\rho \ln \rho}$ exactly satisfies the
{\em statistical} equation $E=F+TS_{VN}$, for systems in
equilibrium. We will therefore always use this to define the
quantum mechanical entropy of a system. This gives us a physical
basis for understanding the thermodynamic quantities $F$ and $S$.
These properties must be understood as properties of the
statistical ensemble itself, introduced at the start of Chapter
\ref{ch:szsm}. Unlike the mean energy and pressure, they do not
correspond to the average of any property of the individual
systems.

It should be carefully noted that the free energy and entropy have
been given significance only for ensembles of systems at a
specific temperature $T$. The entropy $S_{VN}$, however, is not
dependant upon the given temperature, and does not even require
the system to be in thermodynamic equilibrium to be calculated. We
will therefore assume that $S_{VN}$ is always valid.

Free energy, however, has been defined with respect to thermal
equilibrium at a particular temperature. In Appendix
\ref{ap:szfree} it is argued that the free energy can still be
defined where there is more than one temperature, but that it is
not conserved. When a quantity of entropy $S$ is transferred
reversibly, within a system, through a temperature difference
$\Delta T$, then the free energy changes by a quantity $-S \Delta
T$. This characteristic equation will occur at several points in
our understanding of the Popper-Szilard Engine.

\subsection{One Atom Gas}

We will now apply these concepts to the one atom gas, confined
within a box. We will consider here only the situation where the
one atom gas is confined entirely to the left of a piston at
location $Y$. The changes in thermodynamics properties of the
single atom gas will be shown to be consistent with an ideal gas,
even though there is a single particle involved.

\subsubsection{Free Energy}
The density matrix of the gas is given in Equation \ref{eq:rg6} by
$\rho_{G6}^\lambda(Y)$. This has function
\[
Z_{G6}^\lambda(Y)=\sum_n
    e^{-\frac{\epsilon}{kT_G}\left(\frac{2l}{Y+1-p}\right)^2}
    \approx \frac{Y+1-p}{4}\sqrt{\frac{\pi kT_G}{\epsilon}}
\]
giving a free energy
\begin{eqnarray*}
F_{G6}^\lambda(Y)&=&\frac{kT_G}{2}\left(
    4\ln 2-\ln \left(\frac{\pi kT_G}{\epsilon} \right)
    -2\ln (Y+1-p)\right)
\end{eqnarray*}

It will be convenient to also calculate the free energy for the
gas when there is no partition present at all. This has density
matrix $\rho_{G0}$, in Equation \ref{eq:rg0} with
\begin{eqnarray*}
Z_{G0} &=&\sum_n e^{-\frac{\epsilon n^2}{kT_G}}\\
&\approx& \int e^{-\frac{\epsilon n^2}{kT_G}}dn
     = \frac{1}{2} \sqrt{\frac{\pi kT_G} \epsilon }
\end{eqnarray*}
so has free energy
\begin{equation}
F_{G0}=\frac{kT_G}{2}\left(2\ln 2
    -\ln\left(\frac{\pi kT_G}{\epsilon}\right)\right)
    \label{eq:fg0}
\end{equation}
This gives
\begin{eqnarray}
F_{G6}^\lambda(Y)&=&F_{G0}+kT_G\ln \left(\frac{2}{Y+1-p}\right)
\label{eq:fg6}
\end{eqnarray}
If we neglect terms of order $k\ln(1-p)$, this gives us the
results
\begin{eqnarray*}
F_{G6}^\lambda(0) & \approx &F_{G0}+kT_G \ln 2 \\
F_{G6}^\lambda(1-p) & \approx & F_{G0}
\end{eqnarray*}
As we saw in Section \ref{s:szsm2}, the work performed upon the
piston by the expansion of the one atom gas is simply
\[
\Delta W=kT_G\ln \left(\frac{Y+1-p}{2}\right)
\]
so this confirms
\[
F_{G6}^\lambda(Y)+\Delta W = \mathrm{constant}
\]
or equivalently, the change in free energy of the system is equal
to the work performed upon the system.

\subsubsection{Entropy}
We calculate the entropies directly from the density matrix
\begin{eqnarray}
S_{G0} &=& \frac{k}{2}\left(
    1+\ln\left(\frac{\pi kT_G}{\epsilon}\right)
    -2\ln 2 \right) \nonumber \\
S_{G6}^\lambda(Y) &=&\frac{k}{2}\left(
    1+\ln \left(\frac{\pi kT_G}{\epsilon}\right)
    -4\ln 2+2\ln(Y+1-p) \right) \nonumber \\
    &=& S_{G0}-k\ln \left(\frac{2}{Y+1-p}\right)
    \label{eq:sg6}
\end{eqnarray}
which gives the approximate results for the piston in the center
and end of the box
\begin{eqnarray*}
S_{G6}^\lambda(0) & \approx & S_{G0}-k \ln 2 \\
S_{G6}^\lambda(1-p) & \approx & S_{G0}
\end{eqnarray*}
The entropy of the gas increases by $k \ln 2$ as it expands to
fill approximately twice it's initial volume.

\subsubsection{Heat Bath}
The internal energy of the gas, given in Equation \ref{eq:rg6}, is
constant at $\frac{1}{2}kT_G$. The free energy extracted from the
expansion must be drawn from the contact the gas has with the heat
bath. This means an energy of $kT_G \ln
\left(\frac{Y+1-p}{2}\right)$ comes out of the $T_G$ heat bath.

It can be readily be shown that when the energy change in the heat
bath is small compared to it's total energy, then the entropy
change in the heat bath is given by
\[
dS=\frac{dE}{T}
\]
We include this entropy change in the heat bath
\[
S_{T_G}(Y)=-k \ln \left(\frac{Y+1-p}{2}\right)
\]
to our analysis. This gives a combined entropy of
\[
S_{T_G}(Y)+S_{G6}^\lambda(Y)=\frac{k}{2}\left(
    1+\ln\left(\frac{\pi kT_G}{\epsilon}\right)-4\ln2\right)
\]
which is a constant. This confirms our expectations for a
reversible process.

We may also note that, in Section \ref{s:szsm2} the pressure obeys
the relationship
\[
P(Y)V(Y)=kT_G
\]
where we define the 'volume' of the gas as the length of the box
\[V(Y)=Y+1-p\]
that gas occupies. This relationship hold for isothermal expansion
and compression, where the temperature is constant. For isolated
expansion and compression, where the temperature is variable
\[
P(Y)V(Y)=kT
\]
still holds, but in addition, the one atom adiabatic relationship
\[
P(Y)V(Y)^3=\mathrm{constant}
\]
hold true (see also \cite{BBM00}). The single atom gas therefore
acts in exactly the manner we would expect from the thermodynamic
analysis of an ideal gas.

\subsection{Weight above height $h$}
We now calculate the thermodynamics properties of a single atom
weight, supported at a height $h$. Again, we will analyse how the
free energy and entropy changes as the height is changed, and we
will connect this to the thermodynamic state of the one atom gas,
being used to lift a weight through the pressure it exerts upon a
piston.

\subsubsection{Free Energy}
In Section \ref{s:szsm3} the thermal state of the weight is given
in Equation \ref{eq:rw1}. The free energy may be calculated
directly from $Z_{W1}(h)$ as
\begin{eqnarray*}
F_{W1}(h) &=&M_Wgh-kT_W\left( \frac{3}{2}\ln
    \left( \frac{kT_W}{M_WgH}\right) -\ln \left( 2\sqrt{\pi}\right)\right) \\
    &=&F_{W1}(0) +M_Wgh
\end{eqnarray*}
As was noted before, the work done in raising a weight through a
height $h$ is always $M_Wgh$, regardless of the ensemble, so again
we confirm the status of the free energy.

Substituting the isothermal gearing ratio
$h(Y)=\frac{kT_G}{M_Wg}\ln \left(1+\frac{Y}{1-p}\right)$ gives
\begin{equation}
F_{W1}\left(h(Y)\right)
    =F_{W1}(0)+kT_G\ln\left(1+\frac{Y}{1-p}\right)
        \label{eq:fw1}
\end{equation}
which produces
\[
F_{W1}(h_T)=F_{W1}(0) +kT_G\ln 2
\]

If we use the expansion of the one atom gas to lift the weight,
(or the compression of the weight lifting the gas) then
\[
F_{W1}\left(h(Y)\right)+F_{G6}^\lambda(Y)=\mathrm{constant}
\]

\subsubsection{Entropy}
Taking the density matrix $\rho_{W1}(h)$, we calculate the entropy
to be
\begin{eqnarray}
S_{W1}&=& \frac{3k}{2}\left(
    1+\ln \left(\frac{kT_W}{M_WgH}\right)
    -\frac{2}{3}\ln \left(2\sqrt{\pi }\right) \right)
    \label{eq:sw1}
\end{eqnarray}
This is independant of the height $h$ of the weight. As the
entropy of the weight does not change, it is easy to see from
$E=F+TS$ that the change in internal energy of a raised weight is
exactly equal to it's change in free energy, and therefore equal
to the work done upon the weight. This agrees with the conclusion
in Section \ref{s:szsm3} that no heat need be drawn from or
deposited within a heat bath, for a weight to be raised or lowered
in thermal equilibrium.

The combination of the one atom gas and the quantum weight behaves
exactly as we would expect for a reversible thermodynamic system.
The application of the thermodynamic concepts of free energy and
entropy to these systems have presented no special problems.

\subsection{Correlations and Mixing}
The systems considered in the previous Subsection are always
described by a product of density matrices
\[
\rho=\rho_{W1}(h(Y))\otimes\rho_{G6}^\lambda(Y)
\]
For the Popper-Szilard Engine, we will have to consider more
complex density matrices, were the subsystem density matrices are
not product density matrices, but instead have correlations
between their states. We must now address the behaviour of
thermodynamic properties where systems become correlated. To do
this we must consider two different features: the mixing of an
ensemble from two or more subensembles\footnote{Throughout we will
refer to the combination of subensembles as a 'mixture' or
'mixing'. Unfortunately this term is used in several different
ways when associated with entropy. Here we will use it exclusively
to refer to the relationship between an ensemble and it's
subensembles, that the density matrix of an ensemble is a 'mixed
state' of the density matrices of it's subensembles. This should
not be confused with the 'entropy of mixing' that occurs when
'mixtures' of more than one substance is considered
\cite{Tol79}[Chapter XIV] or the 'mixing' or 'mixing enhancement'
associated with coarse graining \cite{Weh78}.}, and the
correlation of two or more subsystems.

\subsubsection{Entropy}
The entropy of composite systems can be defined directly from the
properties of $S_{VN}$ \cite{Weh78}. If there are two independent
systems, with a total density matrix $\rho =\rho _1\otimes \rho
_2$ then the total entropy is additive, $S=S_1+S_2,$ where
$S_1=k\trace{\rho_1\ln \rho_1}$ etc. When the total density matrix
is given as the sum of two orthogonal subensembles, so that $\rho
=p_a\rho_a+p_b\rho_b$ where $p_a+p_b=1$ and $\rho_a\rho _b=0$,
then the total entropy is given by the formula
$S=p_aS_a+p_bS_b-kp_a\ln p_a-kp_b\ln p_b.$ This can be generalised
to

\begin{equation}
S=\sum p_iS_i-k\sum p_i\ln p_i \label{eq:smix}
\end{equation}

These two results may be combined to calculate the entropy of
correlated systems, such as $\rho =p_a\rho _{a1}\otimes \rho
_{a2}+p_b\rho _{b1}\otimes \rho _{b2}$ , which has an entropy of
$S=\sum p_i\left( S_{i1}+S_{i2}\right) -k\sum p_i\ln p_i$.

\subsubsection{Free Energy}
For free energy, the problem is more subtle. We can consistently
assume that the free energy of two independant systems are
additive, so that $F=F_1+F_2$. However, we must be careful when
considering a mixture, if it is not an equilibrium mixture. If we
suppose we have a system in equilibrium at temperature $T$, then
the free energy is given by
\[
F=-kT\ln \left( \sum e^{-\frac{E_i}{kT}}\right)
\]
Now let us consider the effect of splitting the system into two
orthogonal subspaces, with equilibrium density matrices $\rho _a$
and $\rho _b$. These density matrices have partition functions
\begin{eqnarray*}
Z_a &=& \sum_{i\subset a}e^{-\frac{E_i}{kT}} \\
Z_b &=& \sum_{i\subset b}e^{-\frac{E_i}{kT}} \\
Z   &=& Z_a+Z_b
\end{eqnarray*}

It can be readily shown that for the combined density matrix $\rho
=p_a\rho _a+p_b\rho _b$ to be in thermal equilibrium, then
$Z_a=p_aZ$ and $Z_b=p_bZ$. This allows us to calculate the free
energy of the subensembles using the formula

\begin{equation}
F_a=-kT\ln Z_a=F-kT\ln p_a \label{eq:freeprob}
\end{equation}
and similarly for $\rho _b$. This will turn out to be a key
relationship in understanding the thermodynamic explanation for
the failure of the Popper-Szilard Engine.

Using Equation \ref{eq:freeprob} we can re-write $F$ as

\begin{eqnarray}
F   &=&\sum p_i F_i+kT \sum p_i\ln p_i \label{eq:fmix}
\end{eqnarray}

or equivalently

\begin{eqnarray*}
F   &=& -kT\ln \left( \sum e^{-\frac{F_i}{kT}}\right)
\end{eqnarray*}

and we also find that

\begin{eqnarray}
p_a=& \frac{e^{-\frac{F_a}{kT}}}{\sum e^{-\frac{F_i}{kT}}} &=
e^{\frac{F-F_a}{kT}} \label{eq:probfree}
\end{eqnarray}

It is important to note that these relationships are no longer a
sum over the individual eigenstates. They are summations over the
orthogonal subspaces, or the subensembles. Rather than relating
the total free energy to the logarithmic averaging over the
individual energies, they relate the free energy to the
logarithmic averaging over the {\em free} energies of the
subensembles. Similarly, the probabilities are not those of the
individual eigenstates, depending upon the individual energies,
they are the probabilities of the subensemble, and they depend
upon the {\em free} energy of that subensemble.

Equation \ref{eq:freeprob} will turn out to be very important in
the next Chapter. The value of $-kT\ln p$ is always positive, so
the free energy of a subensemble is always greater than the free
energy of the ensemble from which it is taken. Despite the
similarity of the equations $S=\sum p_iS_i-k\sum p_i\ln p_i$ and
$F=\sum p_iF_i+kT\sum p_i\ln p_i,$ it should be noted that there
is no equivalent relationship to (\ref{eq:freeprob}) between the
entropy of an ensemble and the entropy of it's subensembles. While
the entropy of an ensemble must be greater than the {\em mean}
entropy of it's subensembles $\left( S\leq \sum p_iS_i\right)$ ,
there is no such restriction upon it's relationship to the
entropies of the individual subensembles.

While we have
\[
F\leq F_a
\]
for all $a$ for free energies, we only have
\[
\min \left( S_a\right) \leq S\leq \max \left( S_a\right) +\ln N
\]
where $N$ is the dimension of the Hilbert space of the system, for
entropy. It may be higher than all the subensemble entropies, but
may also be lower than any but the minimum entropy.

We now must understand how the free energy is affected when we
form the {\em non-equilibrium} density matrix $\rho^{\prime
}=p_a^{\prime }\rho _a+p_b^{\prime }\rho _b$ where $p_a^{\prime
}\neq p_a$ (we will assume that the subensembles $\rho _a$ and
$\rho _b$ are themselves in thermal equilibrium at temperature
$T$, and that it is only their mixing that is not in proportion).

This is a subtle problem and is addressed in Appendix
\ref{ap:noneq}. There it is shown that free energy can be
meaningful for such mixtures, and that the relation
\[F=\sum p_iF_i+kT\sum p_i\ln p_i \]
is still valid, but that the equations $ F_a=F-kT\ln p_a$ and
$F=-kT\ln \left( \sum e^{-\frac{F_i}{kT}}\right)$ cannot be used
directly\footnote{Combining the results for this non-equilibrium
mixing of $F$ and $S,$ it can be shown that the statistical
equation $E=F+TS$ is still valid}. We can therefore calculate the
free energy of a non-equilibrium mixture, at a given temperature,
but we cannot use the free energy of the subensemble to calculate
it's probability, in the manner Equation \ref{eq:probfree} allows.

While we have defined free energy for non-equilibrium mixtures at
a specific temperature, we should notice that the temperature
plays a key role in the change of the free energy with mixing. For
this equation to be valid, the relevant subensembles must
themselves be in thermal equilibrium at some temperature $T$. In
particular, when we have a correlated density matrix $\rho
=p_a\rho _{a1}\otimes \rho _{a2}+p_b\rho _{b1}\otimes \rho _{b2}$
and systems 1 and 2 are at different temperatures to each other,
there is clearly no well defined temperature $T$ for the mixture
between $p_a$ and $p_b$. In this situation it appears that the
concept of free energy has been stretched to it's limit and can no
longer be regarded as a well defined, or meaningful, quantity.
This is significant, as at several points in the cycle of the
Popper-Szilard Engine, the system will be described by precisely
such a correlated density matrix. We will not be able to assume
that the free energy remains well defined throughout the operation
of the Engine.

\section{Raising cycle} \label{s:szth2}

We will now apply these results to the raising cycle of the
Szilard Engine, to parallel the statistical mechanical analysis in
Section \ref{s:szsm5}. The density matrices $\rho_{T0}$ to
$\rho_{T5}$ are given in that Section. The raising cycle is shown
in Figure \ref{fg:raise}.

\paragraph{Stage a}

In the initial state of the raising cycle, the density matrix is
\[
\rho_{T0}= \rho_{G0} \otimes \rho_{W0}^\lambda
        \otimes \rho_{W0}^\rho \otimes \proj{\phi_0}
\]
To maintain a certain level of generality we will assume that the
piston states all have a notional internal free energy $F_P$ and
entropy $S_P$.

The initial entropy and free energy is given by
\begin{eqnarray*}
S_{T0} &=&S_P+S_{G0}+2S_{W1} \\
F_{T0} &=&F_P+F_{G0}+2F_{W1}
\end{eqnarray*}

On raising the partition and inserting the piston in the center of
the box, we have a new density matrix

\[
\rho_{T1}(0) =
\frac{1}{2}\left(\rho_{G6}^\lambda(0)+\rho_{G6}^\rho(0) \right)
        \otimes \rho_{W0}^\lambda \otimes \rho_{W0}^\rho \otimes \proj{\Phi(0)}
\]

Mixing the entropy and the free energies of the gas subensembles
$\rho _{G6}^\lambda(0)$ and $\rho_{G6}^\rho (0)$ at temperature
$T_G$ gives

\begin{eqnarray*}
S_{G1} &=&\left(
    \frac{1}{2}S_{G6}^\lambda(0)+\frac{1}{2}S_{G6}^\rho(0)\right)
    -k\left(\frac{1}{2}\ln\frac{1}{2}+\frac{1}{2}\ln\frac{1}{2}\right) \\
    &=&\frac{k}{2}\left(1+\ln\left(\frac{\pi kT_G}{\epsilon}\right)
     -2\ln 2 +2\ln(1-p) \right) \\
F_{G1} &=&\left(\frac{1}{2}F_{G6}^\lambda(0)
        +\frac{1}{2}F_{G6}^\rho(0)\right)
    +kT_G\left(\frac{1}{2}\ln\frac{1}{2}
        +\frac{1}{2}\ln\frac{1}{2}\right) \\
    &=&\frac{kT_G}{2}\left(2\ln 2 -\ln \left(
        \frac{\pi kT_G}{\epsilon} \right)-2\ln(1-p) \right)
\end{eqnarray*}

Neglecting terms of order $\ln (1-p)$ we have $S_{G1}\approx
S_{G0}$, $F_{G1}\approx F_{G0}$ so the total entropy $S_{T1}$ and
free energy $F_{T1}$ are unchanged from $S_{T0}$ and $F_{T0}$. The
insertion of the piston requires negligible work and is
reversible.

\paragraph{Stage b}

During the expansion phase of the raising cycle, the density
matrix of the system $\rho_{T1}(Y)$ is a correlated mixture of
subensembles at different temperatures $T_G$ and $T_W$. It follows
that the free energy is not well defined during this expansion
phase. At the end of the expansion the density matrix becomes
\begin{eqnarray*}
\rho_{T1}(1-p)&=&
    \frac{1}{2}\left(\rho_{G6}^\lambda(1-p)\otimes \rho_{W1}^\lambda(h_T)
        \otimes \rho_{W0}^\rho \otimes \proj{\Phi(1-p)} \right.\\
    && \left.+\rho_{G6}^\rho(-1+p)\otimes \rho_{W0}^\lambda
        \otimes \rho_{W1}^\rho(h_T)\otimes \proj{\Phi(-1+p)}
\right)
\end{eqnarray*}

Examining these terms we note that $\rho_{G6}^\lambda(1-p) \approx
\rho_{G6}^\rho(1-p) \approx \rho_{G0}$, so the gas can be factored
out of the correlation, and only the weight temperature $T_W$ is
involved in the mixing.

The raised weight subensemble $\rho_{W1}^\lambda(h_T)$ is not
orthogonal to the unraised $\rho_{W1}^\lambda(0)$, but the piston
states $\proj{\Phi(1-p)}$ and $\proj{\Phi(-1+p)}$ are orthogonal,
so we can use the mixing formula for the entropy and free energy,
to get
\begin{eqnarray*}
S_{T1} &=&S_{G0}+S_P+2S_{W1}+k\ln 2 \\
F_{T1} &=&F_{G0}+F_P+2F_{W1}+kT_G\ln 2-kT_W\ln 2 \\
    &=&F_{G0}+F_P+2F_{W1}-kT_W\ln (2P_1)
\end{eqnarray*}
where we have used the relationship $P_1=\left( \frac
{1}{2}\right)^{\frac{T_G}{T_W}}$ to substitute $kT_G\ln
2=-kT_W\ln(P_1)$.

During the course of the expansion, $kT_G\ln 2$ heat is drawn from
the $T_G$ heat bath, causing an decrease in entropy of $k\ln 2$.
This compensates for the increase in the entropy of the engine,
and confirms that the process so far has been thermodynamically
reversible.

During the expansion phase the free energy becomes undefined. At
the end of this phase, it has changed by an amount
$F_{T1}-F_{T0}=-kT_W \ln (2P_1) =-(T_W-T_G) k\ln 2$. This is just
a free energy change of $\Delta F=-S\Delta T$, where the entropy
$k\ln 2$ has been transferred from the $T_G$ heat bath to the
weights and piston at $T_W$. This is the occurrence of the
characteristic equation discussed in Appendix \ref{ap:szfree}.

\paragraph{Stage c}
Shelves now come out on both sides of the machine, at a height
$h_T$ to support a raised weight. This divides an unraised density
matrix into the subensembles for above and below the shelf. In
Sections \ref{s:szsm5} and \ref{s:szsm6} it was assumed that the
unraised density matrix divides into two orthogonal subensembles
\[
\rho_{W1}(0)=P_1\rho_{W0}(h_T)^{\prime \prime }
    +P_2\rho_{W0}(0)^{\prime \prime }
\]
without interference terms.

This implies the entropies and free energies combine according to
\begin{eqnarray}
S_{W1} &=&\left( P_1S_{W0}(h_T)^{\prime \prime }
    +P_2S_{W0}(0)^{\prime \prime }\right)
    -k\left( P_1\ln P_1+P_2\ln P_2\right) \nonumber \\
F_{W1}(0) &=&\left( P_1F_{W0}(h_T)^{\prime \prime }
    +P_2F_{W0}(0)^{\prime \prime }\right)
    +kT_W\left( P_1\ln P_1+P_2\ln P_2\right)
    \\ \label{eq:airyent}
\end{eqnarray}
and so inserting the shelves would be both reversible, and involve
negligible work.

Unfortunately, it is not possible to directly confirm these
relations. We can estimate the free energy and entropy of
$\rho_{W0}(h_T)^{\prime \prime }$ as the same as the free energy
and entropy of the raised weight $\rho_{W1}(h_T)$. However, as we
do not have suitable approximations for the wavefunctions trapped
below the shelf, we cannot calculate the entropy or free energy
for $\rho_{W0}(0)^{\prime \prime }$.

For the reasons given in Appendix \ref{ap:szai}, if $kT_W \gg
M_Wgh_T$ or $kT_W \ll M_Wgh_T$ the insertion of the shelf should
be reversible and involve negligible work, and it is reasonable to
assume that this will also be true at intermediate heights for
high temperature systems ($kT_W \gg M_wgH$, the characteristic
energy of the ground state). If this is the case, Equations
\ref{eq:airyent} will then be true.

This assumption simply allows us to continue to calculate entropy
and free energies during Stages (c-e) of the cycle. It does not
affect the behaviour of the Engine itself, as the interference
terms will disappear in Stage (f) of the cycle. The only part of
the assumption that is significant is that the insertion of the
shelf requires negligible work. This is similar to inserting the
narrow barrier into the one atom gas, which was proved to require
negligible work in Section \ref{s:szsm2}\footnote{It should also
be noted that if this assumption is false, it would imply a
difference between the quantum and classical thermodynamics of a
particle in a gravitational field, even in the high temperature
limit.}.

We will therefore assume that Equations \ref{eq:airyent} are true,
from which it can immediately be seen that the free energy and
entropy of $\rho_{T2}$ is the same as for $\rho_{T1}$.

\paragraph{Stage d}
The piston is now removed from the box. The only affect of this is
to change $\rho^\rho_{G6}(1-p)$ and $\rho^\lambda_{G6}(-1+p)$ into
$\rho_{G0}$. This has negligible effect upon the free energy or
entropy of the gas states, so the thermodynamic properties of
$\rho_{T3}$ are also unchanged from $\rho_{T1}$.

\paragraph{Stage e}
The operation of $U_{reset}$ then takes the density matrix on the
raising cycle to $\rho_{T4}$. Only the piston states are changed
by this, and so again, there is no change in entropy or free
energy.

\paragraph{Stage f}
The shelves are removed and the system is allowed to thermalise,
leading to a final density matrix of
\begin{eqnarray}
\rho_{T5}&=&
    \rho_{G0} \otimes \left(
    w_1 \rho_{W1}^\lambda(0) \otimes \rho_{W1}^\rho(0) \otimes
    \proj{\phi_0}
    + w_2 \rho_{W1}^\lambda(h_T) \otimes \rho_{W1}^\rho(0) \otimes
    \proj{\phi_R} \right. \nonumber \\
&& \left.
    + w_3 \rho_{W1}^\lambda(0) \otimes \rho_{W1}^\rho(h_T) \otimes
    \proj{\phi_L} \right)
\end{eqnarray}
from Equation \ref{eq:rt5}.

In the $w_1$ portion of the density matrix, $M_wgh_T$ energy is
dissipated into the $T_W$ heat bath, increasing it's entropy. The
total entropy is therefore
\begin{eqnarray*}
S_{T5}&=&S_{G0}+w_1 \left(2S_{W0}+S_P+\frac{M_wgh_T}{T_w} \right)+
    w_2\left(S_{W0}+S_{W1}(h_T)+S_P\right)\\
    &&+w_3\left(S_{W0}+S_{W1}(h_T)+S_P\right)-k\sum_{n=1,3} w_n
    \ln w_n -k\ln2 \\
    &=& S_{T0}-k\sum_{n=1,3} w_n \ln w_n -k\ln2 -k \ln P_1
\end{eqnarray*}
where we have included the $k \ln 2$ reduction in entropy of the
$T_G$ heat bath, and have used $M_W g h_T=-k T_W \ln P_1$

The free energy can similarly be calculated to be
\begin{eqnarray*}
F_{T5} &=&F_{G0}+F_P+2F_{W1}-kT_W\left((w_2+w_3) \ln P_1
     -\sum_{n=1,3}w_n\ln w_n\right) \\
    &=&F_{T0}-kT_W\left((w_2+w_3)\ln P_1 -\sum_{n=1,3}w_n\ln w_n\right)
\end{eqnarray*}
where the $(w_2+w_3)kT_W\ln P_1$ term comes from the free energy
of the raised weights in the $(w_2+w_3) $ portions of the density
matrix.

\begin{table}[t]
\resizebox{\textwidth}{!}{
    \begin{tabular}{||c|c|c|c|c|cc||}
    \hline\hline & $T_G$ & Gas & Piston & Weight 1 &
    \multicolumn{1}{|c|}{Weight 2} & $T_W$ \\ \hline
    \multicolumn{7}{||l||}{Stage a} \\ \hline
    \multicolumn{1}{||l||}{Energy} & / & $\frac{1}{2}kT_G$ & / &
        $\frac{3}{2}kT_W$ &
        \multicolumn{1}{|c|}{$\frac 32kT_W$} & / \\
    \multicolumn{1}{||l||}{Entropy} & / & $S_{G0}$ & $S_p$ & $S_{W1}$&
        \multicolumn{1}{|c|}{$S_{W1}$} & / \\
    \multicolumn{1}{||l||}{Free Energy} & / & $F_{G0}$ & $F_P$ &
        $F_{W1}$ & \multicolumn{1}{|c|}{$F_{W1}$} & / \\ \hline
    \multicolumn{7}{||l||}{Stage b} \\ \hline
    \multicolumn{1}{||l||}{Energy} & $-kT_G\ln 2$ & $\frac 12kT_G$ &/&
        \multicolumn{2}{|c|}{$3kT_W+M_Wgh_T$} & / \\
    \multicolumn{1}{||l||}{Entropy} & $-k\ln 2$ & $S_{G0}$ &
        \multicolumn{3}{|c|}{$S_p+2S_{W1}+k\ln 2$} & / \\
    \multicolumn{1}{||l||}{Free Energy} & / & $F_{G0}$ &
        \multicolumn{3}{|c|}{$F_P+2F_{W1}-kT_W\ln(2P_1)$} &/ \\ \hline
    \multicolumn{7}{||l||}{Stage f} \\ \hline
    \multicolumn{1}{||l||}{Energy} & $-kT_G\ln 2$ & $\frac{1}{2}kT_G$
        & / & \multicolumn{2}{|c|}{$3kT_W-(w_2+w_3)kT_W\ln P_1$} &
        $-k w_1 T_W\ln P_1$ \\
    \multicolumn{1}{||l||}{Entropy} & $-k\ln 2$ & $S_{G0}$ &
        \multicolumn{3}{|c|}{$S_p+2S_{W1}-k\sum w\ln w$} & $-k w_1\ln P_1 $ \\
    \multicolumn{1}{||l||}{Free Energy} & / & $F_{G0}$ &
        \multicolumn{3}{|c|}{$F_P+2F_{W1}+kT_W
            \left( \sum w\ln w-(w_2+w_3) \ln P_1 \right)$} & / \\
    \hline\hline
    \end{tabular}
    }
\caption{Thermodynamic Properties of the Raising Cycle
\label{tb:raise}}
\end{table}

\paragraph{Summary}
These results are summarised in Table \ref{tb:raise}, giving the
energy, entropy and free energy at the ends of Stages a, b and f.
The remaining stages are omitted as they are no different to Stage
b. Where the free energy or entropy is associated with correlated
subsystems, the quantity is spread across the relevant columns.

The total energy is constant. The total entropy remains constant
until the final stage, at which point it changes by
\[
\frac{\Delta S_R}{k}=- \ln 2 - w_1 \ln P_1
    - \sum_{n=1,2,3} w_n\ln w_n
\]
\begin{figure}[t]
    \resizebox{\textwidth}{!}{
        \includegraphics{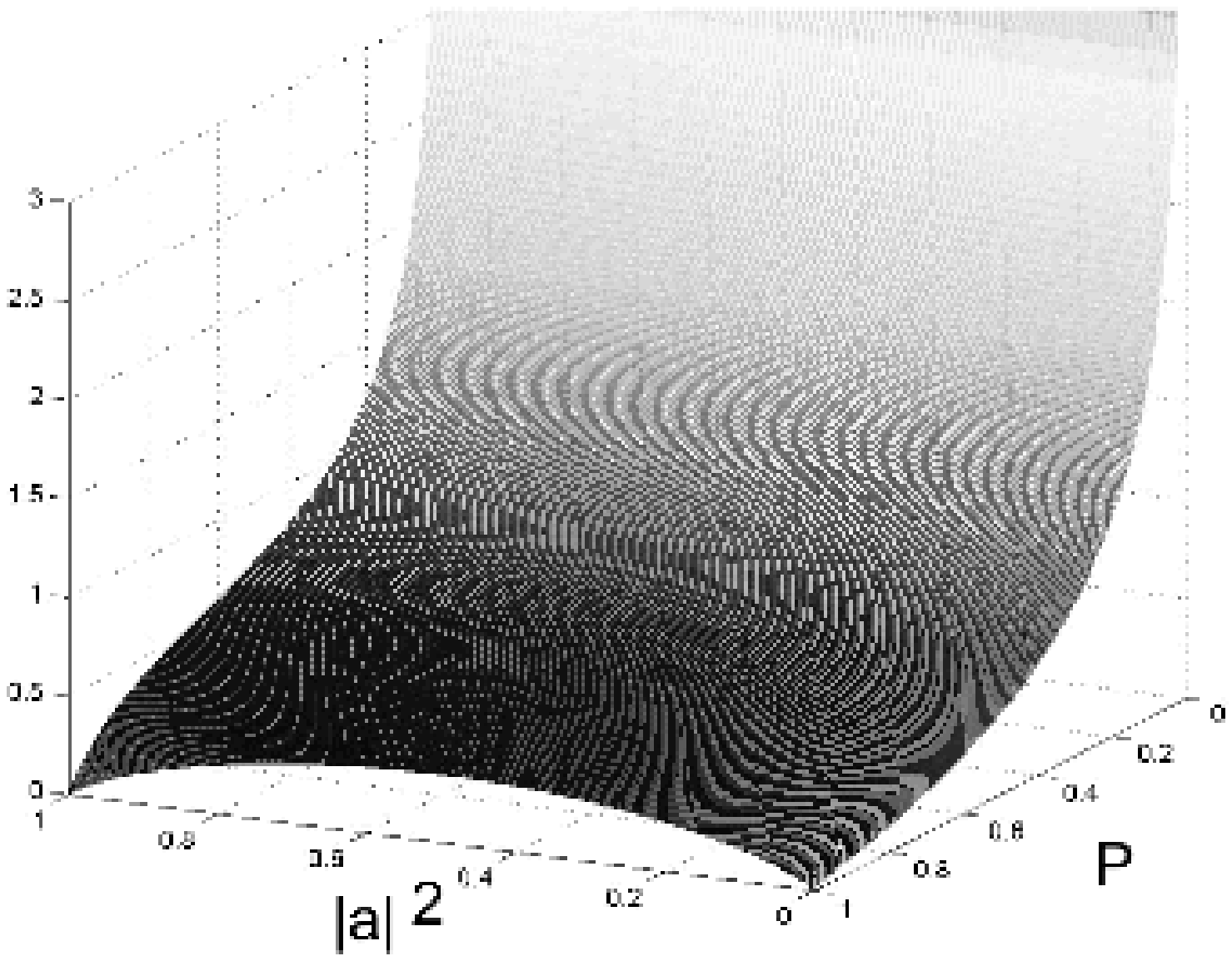}
        \includegraphics{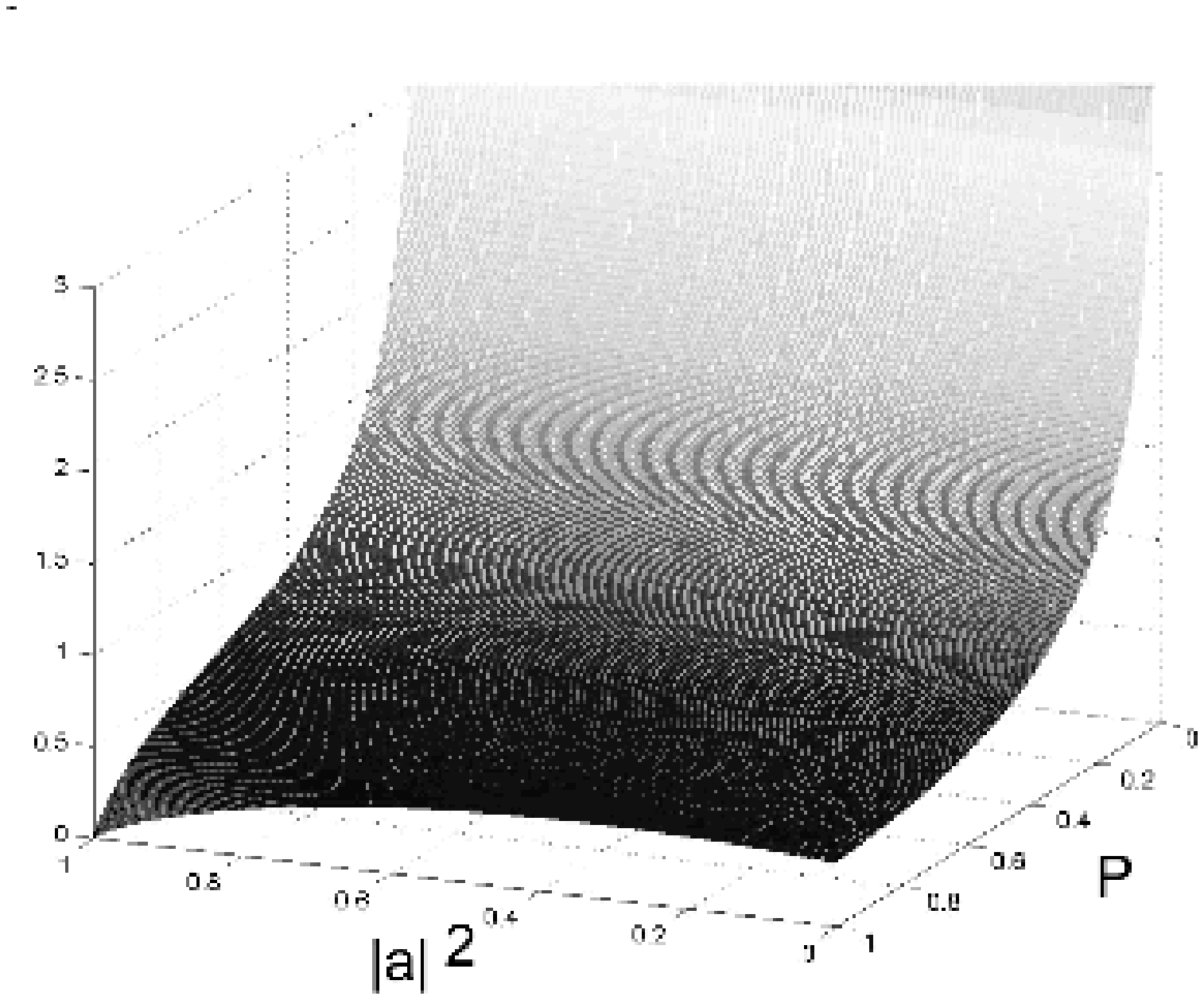}
        }
\caption[Change in Entropy on Raising Cycle]
    {Change in Entropy on Raising Cycle for
    (a) $\magn{c_1}=0$ and (b) $\magn{b_1}=\magn{c_1}$
    \label{fg:entraise}}
\end{figure}
This quantity has a complicated dependancy upon the values of
$P_1$, $\magn{a_1}$, $\magn{b_1}$ and $\magn{c_1}$, but is always
positive. In Figure \ref{fg:entraise}, the net change is plotted
for the two extreme cases of $\magn{c_1}=0$ and
$\magn{b_1}=\magn{c_1}$. As can be seen, this is always greater
than zero.

When the value of $P_1$ approaches 0, the entropy increase becomes
unbounded. This corresponds to the situation where $T_G \gg T_W$.
As the unraised weights will always be found upon the floor, there
is negligible increase in entropy due to mixing. However, the
entropy decrease when energy is extracted from the $T_G$ heat bath
is much less than the entropy increase when that same energy is
deposited in the $T_W$ heat bath.

In addition, it can be seen that when either $\magn{a_1}=1$ or
$\magn{b_1}=1$, and $P_1=1$ the net entropy increase is zero. In
this case $T_G \ll T_W$ and the unraised weights are always
located above the shelf height. The entropy increase here arises
only from the decoherence of the superposition of the piston
states, $\proj{\phi_2}$ and $\proj{\phi_3}$, after the operation
of $U_{RES}$. When any of $\magn{a_1},\magn{b_1},\magn{c_1}=1$,
the piston is not left in a superposition, so there is no increase
in entropy.

The free energy changes by
\[kT_W\ln(2P_1)=k(T_W-T_G)\ln 2\]
during Stage (b), as $k \ln 2$ entropy is transferred from the gas
and $T_G$ heat bath to the weights and $T_W$ heat bath. In the
final stage it changes again, alongside the entropy increase, to
give a net change of
\[
\frac{\Delta F_R}{kT_W}=\sum_{n=1,2,3} w_n\ln w_n -(w_2+w_3) \ln
P_1
\]
over the entire cycle. This can be shown to always be negative. We
should not be surprised by this, as our objective was to drop the
weight we had lifted, and so dissipate the energy used to raise
it.

\section{Lowering Cycle} \label{s:szth3}
The lowering cycle is shown in Figure \ref{fg:lower}. Following
the stages of this cycle given in Section \ref{s:szsm6}, where the
density matrices $\rho_{T6}$ to $\rho_{T11}$ are defined, we will
now calculate it's thermodynamic properties.

\paragraph{Stage a}
Assuming the piston starts initially on the right, the initial
density matrix is $\rho_{T6}$ and the entropy and free energies
are given by
\begin{eqnarray*}
S_{T6} &=&S_P+S_{G0}+2S_{W1} \\
F_{T6} &=&F_P+F_{G0}+2F_{W1}+kT_G\ln 2
\end{eqnarray*}
and will be negligibly affected by the piston being inserted into
one end of the box.

\paragraph{Stage b}
Under the operation of $U_{W4}$, the raised weight is lowered,
compressing the gas. During this stage, the density matrix is
\[
\rho _{T7}(Y)=\rho_{G6}^\lambda(Y) \otimes \rho_{W1}^\lambda(h(Y))
    \otimes \rho_{W1}^\rho(0) \otimes \proj{\Phi(Y)}
\]
giving entropies and free energies
\begin{eqnarray*}
S_{T7}(Y) &=&S_{G6}^\lambda(Y)+S_P+2S_{W1} \\
    &=&S_{T6}-k\ln \left(\frac{2}{Y+1-p}\right) \\
F_{T7}(Y) &=& F_{G6}^\lambda(Y) +F_p+F_{W1}(0)+F_{W1}(h(Y)) \\
    &=& F_{T6}
\end{eqnarray*}

During the compression, $kT_G\ln \left(\frac{2}{Y+1-p}\right)$
heat is transferred from the gas to the $T_G$ heat bath, giving a
compensating rise in entropy. At the end of this stage, the
entropy of the gas has reduced by approximately $k \ln 2$, having
halved in volume, and the entropy of the $T_G$ heat bath has
increased by the same amount. The total free energy remains
constant, as the work done by the weight in work done reversibly
upon the gas.

\paragraph{Stage c}
Shelves are inserted into the thermal state of the two weights at
height $h_T$. As explained in Stage c of the raising cycle above,
we must assume that this takes place reversibly and with
negligible work. The density matrix $\rho_{T8}$ will then have the
same entropy and free energy as $\rho_{T7}(0)$ at the end of Stage
b.

\paragraph{Stage d}
The operation of $U_{RI}$ now removes the piston from the center
of the box. The gas is now able to freely expand to occupy the
entire box, so that $\rho_{G6}^\lambda(0) \rightarrow \rho _{G0}$.
This leaves the system in state $\rho_{T9}$.

The internal energy of these two density matrices are both $\frac
{1}{2}kT_G$, and no work is done upon the gas, so no energy is
drawn from the $T_G$ heat bath by this free expansion. However,
the entropy of the gas increases by $k\ln 2$ and the free energy
decreases by a corresponding amount $ kT_G\ln 2$. There is no
compensating entropy decrease anywhere else in the system.

\paragraph{Stage e}
The application of $U_{RES}$ takes $\rho_{T9}$ to $\rho_{T10}$.
This changes only the state of the piston, and does not affect the
entropy or free energy.

\paragraph{Stage f}
Finally, the removal of the shelves and contact with the $T_W$
heat bath leaves the system in the state
\begin{eqnarray}
\rho_{T11}&=&
    \rho_{G0} \otimes \left(
    w_4 \rho_{W1}^\lambda(0) \otimes \rho_{W1}^\rho(0) \otimes
    \proj{\phi_0}
    + w_5 \rho_{W1}^\lambda(h_T) \otimes \rho_{W1}^\rho(0) \otimes
    \proj{\phi_R} \right. \nonumber \\
&& \left.
    + w_6 \rho_{W1}^\lambda(0) \otimes \rho_{W1}^\rho(h_T) \otimes
    \proj{\phi_L} \right)
\end{eqnarray}
from Equation \ref{eq:rt11}.

In the $(w_5+w_6)$ part of the density matrix, a thermal
fluctuation has caught a weight above one of the shelves. This
draws $M_Wgh_T$ energy from the $T_W$ heat bath, decreasing it's
entropy. The total entropy and free energy at the end of the
lowering cycle is therefore
\begin{eqnarray*}
S_{T11} &=&S_{G0}+S_P+2S_W-k\sum_{n=4,6}w_n\ln w_n
    +k(w_5+w_6)\ln P_1+ k\ln 2 \\
F_{T11} &=&F_{G0}+F_P+2F_W+kT_W\left((w_5+w_6) \ln P_1
    -\sum_{n=4,6}w_n\ln w_n\right)
\end{eqnarray*}
where we have explicitly included the entropy changes in the two
heat baths.

\paragraph{Summary}
Table \ref{tb:lower} summarises the changes in energy, entropy and
free energy for the lowering cycle. The values are shown at the
end of Stages a, b, d and f, and again, where subsystems are
correlated, the entropy and free energy are shown as a total
across the relevant columns.

\begin{table}[t]
\resizebox{\textwidth}{!}{
    \begin{tabular}{||l|c|c|c|c|c|c||}
    \hline\hline & $T_G$ & Gas & Piston & Weight 1 & Weight 2 & $T_W$
        \\ \hline
    \multicolumn{7}{||l||}{Stage a} \\
    \hline Energy & / & $\frac{1}{2}kT_G$ & / &$\frac{3}{2}kT_W+M_Wgh_T$
        & $\frac{3}{2}kT_W$ & / \\
    Entropy & / & $S_{G0}$ & $S_p$ & $S_{W1}$ & $S_{W1}$ & / \\
    Free Energy & / & $F_{G0}$ & $F_P$ & $
        F_{W1}+M_Wgh_T$ & $F_{W1}$ & / \\
    \hline \multicolumn{7}{||l||}{Stage b} \\
    \hline Energy & $kT_G\ln 2$ & $\frac{1}{2}kT_G$ & / &
        $\frac{3}{2}kT_W$ & $\frac{3}{2}kT_W$ & / \\
    Entropy & $k\ln 2$ & $S_{G0}-k\ln 2$ & $S_p$ & $S_{W1}$ & $S_{W1}$ & / \\
    Free Energy & / & $F_{G0}+kT_G\ln 2$ & $F_P$ & $F_{W1}$ & $F_{W1}$ & / \\
    \hline \multicolumn{7}{||l||}{Stage d} \\
    \hline Energy & $kT_G\ln 2$ & $\frac{1}{2}kT_G$ & / &
        $\frac{3}{2}kT_W$ & $\frac{3}{2}kT_W$ & / \\
    Entropy & $k\ln 2$ & $S_{G0}$ & $S_p$ & $S_{W1}$ & $S_{W1}$ & / \\
    Free Energy & / & $F_{G0}$ & $F_P$ & $F_{W1}$ & $F_{W1}$ & / \\
    \hline \multicolumn{7}{||l||}{Stage f} \\ \hline
    Energy & $kT_G\ln 2$ & $\frac{1}{2}kT_G$ & / &
        \multicolumn{2}{|c|}{$3kT_W-(w_5+w_6)kT_W\ln P_1$} &
            $(w_5+w_6) kT_W\ln P_1 $ \\
    Entropy & $k\ln 2$ & $S_{G0}$ &
        \multicolumn{3}{|c|}{$S_p+2S_{W1}-k\sum w\ln w$} &
        $(w_5+w_6) k\ln P_1 $ \\
    Free Energy & / & $F_{G0}$ &
        \multicolumn{3}{|c|}{$F_P+2F_{W1}+kT_W
            \left(\sum w\ln w-(w_5+w_6) \ln P_1 \right)$} & / \\
    \hline\hline
    \end{tabular}
    }
\caption{Thermodynamic Properties of Lowering Cycle
\label{tb:lower}}
\end{table}

Again, we see that the total energy is constant throughout the
operation. The entropy changes at two points. During Stage d, when
a free expansion of the one atom gas takes place, the entropy of
the gas increases by $k \ln 2$. At Stage f, there is a further
entropy change when the weights are allowed to thermalise through
contact with the $T_W$ heat bath. There is an entropy {\em
decrease} of $(w_5+w_6) \ln P_1$, where thermal energy from the
heat bath is trapped in a fluctuation of the weight, but an
increase of $-\sum_{n=4,5,6} w_n \ln w_n$. The change in entropy
at this stage is therefore
\[
\frac{\Delta S_L}{k}=(w_5+w_6) \ln P_1
    -\sum_{n=4,5,6} w_n \ln w_n
\]
which is always positive.
\begin{figure}[t]
    \resizebox{\textwidth}{!}{
        \includegraphics{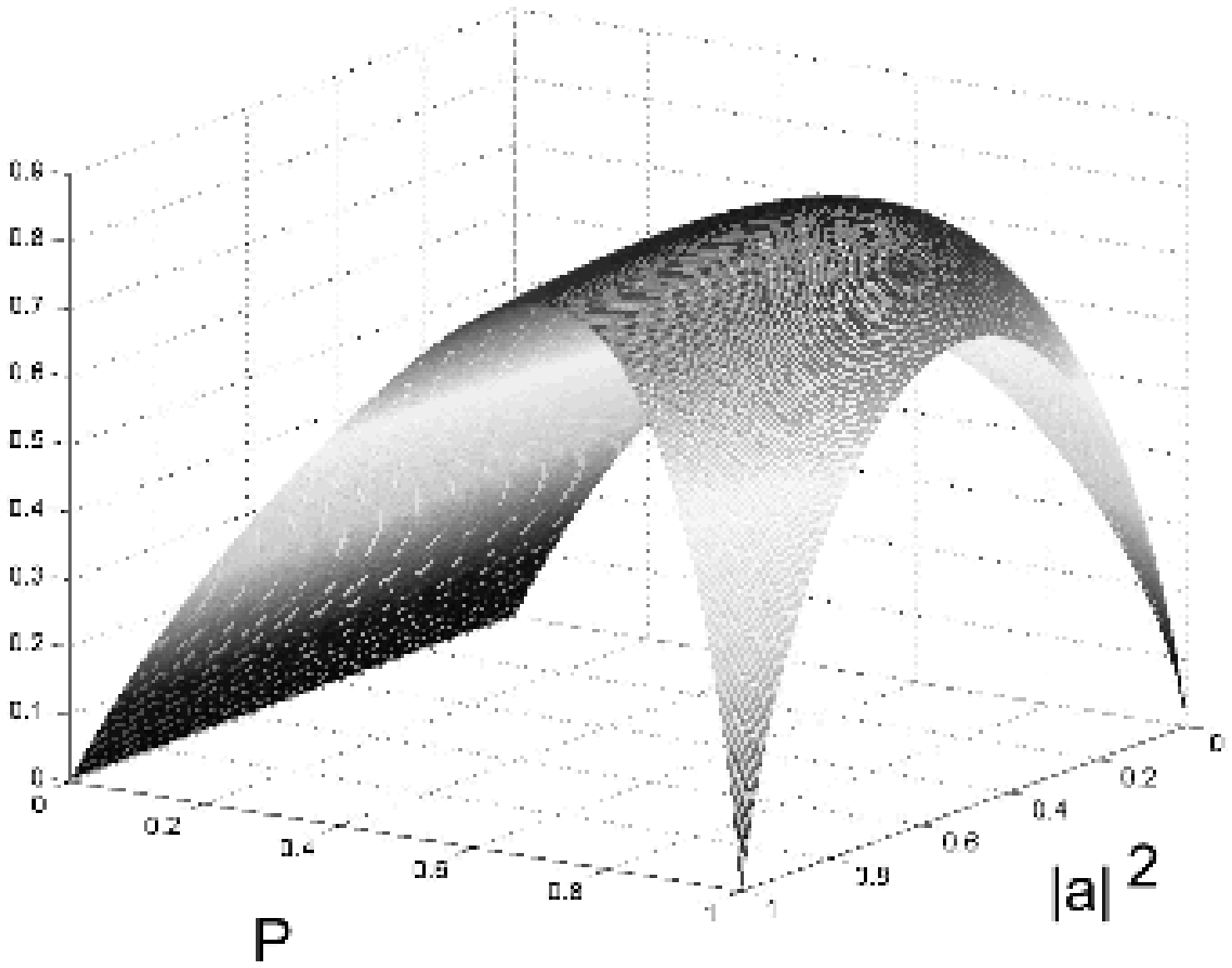}
        \includegraphics{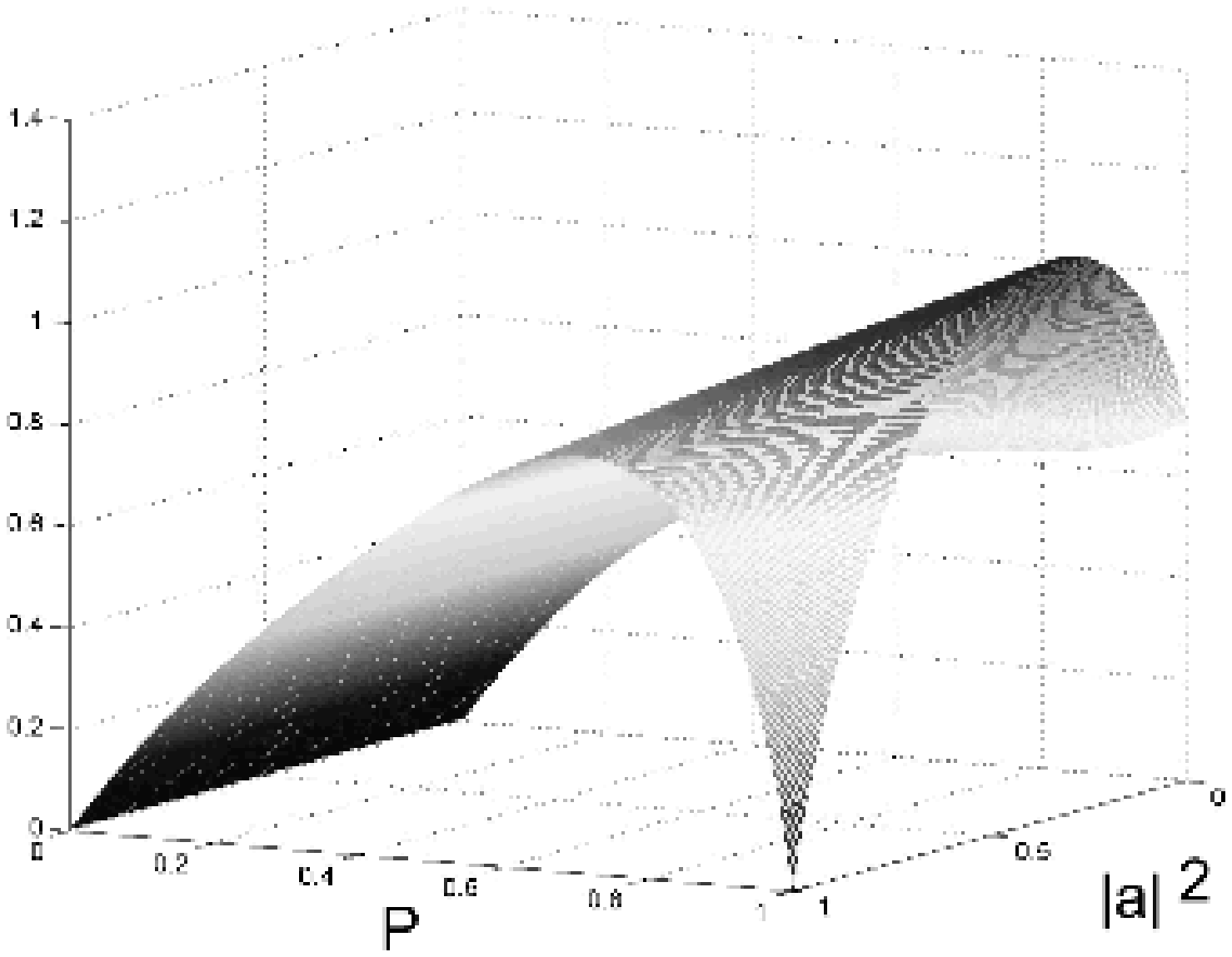}
        }
\caption[Change in Entropy on Lowering Cycle]
    {Change in Entropy on Lowering Cycle for
    (a) $\magn{c_1}=0$ and (b) $\magn{b_1}=\magn{c_1}$
    \label{fg:entlower}}
\end{figure}
This is shown in Figure \ref{fg:entlower}, for the two extremes,
where $\magn{c_1}=0$ and $\magn{b_1}=\magn{c_1}$. Notice that the
net change in entropy over the entire cycle includes an additional
increase of $k \ln 2$ from Stage d. The minimum entropy increase
on the lowering cycle is therefore $k \ln 2$.

The minimal increase in entropy occurs in two special cases. The
first case is the same as on the raising cycle, when $P_1=1$ the
weights are always located above the shelf height. The decoherence
of $\proj{\phi_1}$ when the weights are brought into contact with
the $T_W$ creates an entropy increase, unless the operation of
$U_{RES}$ is such that $\proj{\phi_1}$ is not a superposition.

The second case is when $P_1=0$, regardless of choice of
$U_{RES}$. In this case, at the end of Stage e, both weights will
be found unambiguously below the shelf height. The effect of
$U_{RES}$ must leave this unchanged, and only $\proj{\phi_0}$, the
piston in the center, is compatible with this state. No entropy
increase takes place at this stage, and the Engine cycle reverses.
However, there is still the $k \ln 2$ entropy increase that
occurred during Stage d.

The free energy similarly changes twice, both times as a direct
result of the change in entropy. At Stage d, the increase in the
gas entropy leads to a reduction in free energy of $kT_G \ln 2$,
while during Stage f, the it changes by $-kT_W((w_5+w_6) \ln
P_1-\sum_{n=4,5,6} w_n \ln w_n)$, giving a net change
\[
\frac{\Delta F_L}{kT_W}=w_4\ln P_1+\sum_{n=4,5,6} w_n \ln w_n
\]
over the complete cycle. All terms in this are negative. The free
energy must be reduced over the course of a lowering cycle.

\section{Conclusion} \label{s:szth4}
We have now completed a detailed analysis of the thermodynamic
quantities associated with the operation of the quantum Szilard
Engine.

The free energy becomes undefined at certain stages, and can
sometimes increase. However, when such an increase occurs it is
compatible with the characteristic equation (\ref{eq:free}), and
over the course of an entire cycle, the change in free energy will
be negative.

The entropy of the correlated systems also behaves as would be
expected. It is constant for all reversible processes, and
increases for irreversible processes. Regardless of the choice of
the resetting operation, or of the temperatures of the two heat
baths, it always increases over the course of a raising or
lowering cycle. There is an important subtlety to this result. In
Chapter \ref{ch:szsm} we accepted that an anti-entropic cycle
(such as a raising cycle when $T_W > T_G$) may continue, with some
probability, despite the fact that the energy flow would be from
colder to hotter. All we concluded was that the probability of the
anti-entropic flow reversing would ensure the {\em mean} energy
flow, over the long run, would be from hotter to colder. Now we
appear to be saying that, even so, the entropy must always
increase.

The answer to this apparent contradiction lies in the
interpretation of the entropy of the density matrix. In Chapter
\ref{ch:szsm} we assumed that the Engine was always either on a
raising or a lowering cycle, and we concerned ourselves with the
corresponding transfer of energy between the two heat baths.

To apply the concept of entropy, we must consider the density
matrices $\rho_{T5}$ and $\rho_{T11}$. In these, the Engine is
described by a mixture of states, and so is not determinately upon
a raising or lowering cycle. This implies an additional entropy of
mixing. The results of this Chapter demonstrate that, even when
the Engine starts on an anti-entropic cycle, at the completion of
that cycle the entropy due to mixing in the final state of the
Engine will always be larger than the reduction in entropy we may
have achieved from transferring heat between the two baths.
\chapter{Resolution of the Szilard Paradox}
\label{ch:szsol}

In Chapters \ref{ch:szqm}, \ref{ch:szsm} and \ref{ch:szth} we have
presented a detailed analysis of the operation of the
Popper-Szilard Engine. This has shown that, within certain
limitations, thermodynamic concepts are applicable to the single
atom systems, and that no operation of the Popper-Szilard Engine
was capable of violating the second law of thermodynamics.
However, we have not as yet gained any real insight into {\em why}
the Engine cannot work, nor why some further modification of the
Engine would not be successful. In this Chapter we will attempt to
address these issues by uncovering the essential properties of the
Engine, demonstrate that these properties are central to the
general problem of Maxwell's Demon, and explaining the
thermodynamics underlying them.

In Section \ref{s:szsol1}, we will consider first part of the role
played by the demon. The demon makes a measurement upon the system
of interest, and changes the state of the system, conditionally
upon the result of that measurement. This attempts to eliminate
the mixing entropy of the ensemble. However, the requirement of
unitary evolution leads to a change in the state of the demon
itself. We will show that the piston plays exactly the role of the
demon within the Popper-Szilard Engine. The first stage of the
resolution therefore rests in the consideration of the effect the
measurement has upon the demon itself.

The second stage of the resolution considers the consequences of
the change in the demons state, and the attempts to complete the
thermodynamic cycle. This problem is raised, but only partly
addressed, by advocates of Landauer's Principle as the resolution
to the problem. In Section \ref{s:szsol2}, it is shown that the
key thermodynamic relationship is  one relating the probabilities
of thermal fluctuations at different temperatures. This
relationship shows why the probabilistic attempt to reset must
fail, and why attempts to improve upon this, by performing work
upon the system, leads at best to the Carnot cycle efficiency.
This cycle differs from the phenomenological Carnot cycle,
however, as it operates through correlations in the statistical
states of the subsystems, to transfer {\em entropy}, rather than
energy, between subsystems at different temperatures. It is
further shown, from this relationship, that the attempt to capture
statistical fluctuations will always be an ineffective method of
extracting work from a thermal system.

This provides a comprehensive resolution to the general Maxwell's
demon problem. In Section \ref{s:szsol3} we will re-examine the
arguments offered in Chapter \ref{ch:szmd} and demonstrate they
are, at best, partial resolutions, each focussing upon one aspect
of the overall solution.

\section{The Role of the Demon}\label{s:szsol1}
We need to understand what are the essential features in the
system, that constrains the evolution of the Popper-Szilard Engine
in such a way that it fails to operate as intended. The essential
restriction placed upon it was that it must be described by a
unitary operator. The construction of an appropriate unitary
operator in Chapter \ref{ch:szqm} depended upon the moveable
piston in two particular ways. We will now examine this dependancy
and show that this captures the essential role played by the
Demon.

In Section \ref{s:szqm3} the unitarity of the expansion of the gas
states, in Equations \ref{eq:szqm1} and \ref{eq:szqm2}, is
guaranteed only through the orthonormality relationship, on the
gas and piston states, in Equation \ref{eq:szdel2}:
\begin{eqnarray}
\bk{\Psi_k^\alpha(Y_m)\Phi_A(Y_m)}
    {\Psi_l^\beta(Y_n)\Phi_B(Y_n)}
=\delta_{nm}\delta_{\alpha \beta }\delta_{kl}\delta_{AB}
\end{eqnarray}
However, this orthonormality does not come from the gas states
themselves, as the initially left and right gas states may become
overlapping under the action of the unitary operator $U_{T2}$. It
is the orthonormality of the different piston states, in Equation
\ref{eq:szdel}, that allows us to construct a suitable unitary
operator. However, it is also the orthonormality of the final
piston states that means we cannot construct a unitary operator to
reset the piston states and reliably start another cycle of the
Engine.

First we will examine precisely the role of the piston states.
This will show that the piston fulfils exactly the same role that
is required of a Maxwell's Demon. We will be able to characterise
the general role of Maxwell's Demon as an attempt to reverse the
mixing between subensembles in Equations \ref{eq:smix} and
\ref{eq:fmix}. It is then shown that the Demon can only achieve
such a reversal by increasing it's own entropy by at least as much
again.

\subsection{The Role of the Piston}
Let us examine the role of the piston, in the Popper-Szilard
Engine, in some detail. If we consider the raising cycle, the
insertion of the partition into the gas divides it into two
orthogonal subensembles
\[
\rho_{G1}=\frac{1}{2}\rho_{G6}^\lambda(0)
    +\frac{1}{2}\rho_{G6}^\rho(0)
\]
During the expansion Stage b, the correlated density matrix is
\begin{eqnarray*}
\rho_{T1}(Y)&=&\frac{1}{2} \rho_{G6}^\lambda(Y)\otimes
    \rho_{W1}^\lambda(h(Y)) \otimes \rho _{W1}^\rho(0)
    \otimes \proj{\Phi(Y)}\\
&&+\frac{1}{2}\rho_{G6}^\rho(Y) \otimes \rho_{W1}^\lambda(0)
    \otimes\rho_{W1}^\rho(h(Y)) \otimes \proj{\Phi(-Y)}
\end{eqnarray*}

None of the gas or weight subensembles are orthogonal in this
expansion. The left and right gas wavefunctions overlap, as do the
raised and unraised weight states. However, the piston states
$\proj{\Phi(Y)}$ and $\proj{\Phi(-Y)}$ are orthogonal. It is this
that maintains the orthogonality of the left and right
subensembles, and ensures the evolution is unitary.

As the expansion progresses, the overlap between the left and
right gas subensembles increases, until the piston reaches the end
of the box and is removed, at which point the overlap is complete.
The two, initially orthogonal, gas subensembles have been
isothermally expanded into the same density matrix. For the
weights, the overlap between $\rho_{W1}^\lambda(h(Y))$ and
$\rho_{W1}^\lambda(0)$ decreases, but never reaches zero (except
in the limit where $T_G\gg T_W$). Although the free energy from
the expansion of the gas is picked up by the weights, it is still
the piston states that ensures that the final density matrix has
orthogonal subensembles:

\begin{equation}
\frac{1}{2}\rho_{W1}^\lambda(h_T) \otimes \rho _{W1}^\rho(0)
    \otimes \proj{\phi_R}+\frac{1}{2}\rho_{W1}^\lambda(0)
    \otimes \rho_{W1}^\rho(h_T) \otimes \proj{\phi_L}
    \label{eq:mix1}
\end{equation}

When calculating the free energy and entropies in Chapter
\ref{ch:szth}, it was the orthogonality of the piston states that
allowed us to apply the mixing formulas. The entropy of mixing
between the two gas subensembles has been transferred to the
\textit{piston} states. The significance of the piston states can
be made clear by considering the density matrix:

\begin{equation}
\frac{1}{2}\rho_{W1}^\lambda(h_T) \otimes \rho_{W1}^\rho(0)
+\frac{1}{2}\rho_{W1}^\lambda(0) \otimes \rho_{W1}^\rho(h_T)
\label{eq:mix2}
\end{equation}

The correlated weight states in this matrix are not orthogonal, so
this density matrix has a lower entropy than the density matrix
that includes the piston states. If it were not for the
orthogonality of the piston states, the entropy of the Szilard
Engine would have been reduced at this stage. Only in the limit of
$T_G\gg T_W$ do the weights states become orthogonal, and the
entropy of (\ref{eq:mix2}) becomes equal to (\ref{eq:mix1}). In
this situation the different piston states \textit{can }both be
restored to the center (by correlating them to the position of the
weights), but this does not reduce the entropy of the Engine as it
only takes place where the transfer of heat is from the hotter to
the colder system.

For the lowering cycle, the stages described in Section
\ref{s:szsm6} do not show correlations. The reason for this is
that we started the lowering cycle by assuming the piston is
located on one particular side. In general, a lowering cycle can
start with the piston at either side of the Engine, and so will
have a density matrix of the form

\[
p_R \proj{\phi_R}\otimes \rho_{W1}^\lambda(h_T)\otimes
    \rho_{W1}^\rho(0)
+p_L \proj{\phi_L} \otimes \rho_{W1}^\lambda(0)
    \otimes \rho_{W1}^\rho(h_T)
\]
with $p_R+p_L=1$. This has an additional mixing entropy of
$-k\left(p_L\ln p_L+p_R\ln p_R\right)$, which has a maximum value
of $k\ln 2$, when $p_L=\frac 12$. Now we have a correlated states
with mixing entropy associated initially with the pistons.

The evolution following from this will be the reverse of the
raising cycle, and will transfer the entropy of mixing from the
piston states, to the gas subensembles. The gas will be left in
the state $p_L\rho_{G6}^\lambda(0)+p_R\rho_{G6}^\rho(0)$ just
before the removal of the piston from the center of the box.

After the removal of the piston, the gas returns to the uniform
distribution $\rho_{G0}$. This is an irreversible change, and the
entropy of the system increases by the difference between the
original entropy of mixing of the piston states, and $k\ln 2$. In
Section \ref{s:szth3}  then we have $p_L=0$ or $1$ and the maximum
entropy increase of $k\ln 2$ occurs. If $p_L=\frac {1}{2}$, then
no entropy increase occurs and we have the exact reverse of the
raising cycle\footnote{The net change in entropy over the cycle
will still be positive}.

The essential point is that the correlation between the orthogonal
piston and weight subensembles is transferred to the orthogonal
gas subensembles. This demonstrates the same features as the
raising cycle, which highlights the manner in which the Szilard
engine is intended to work.

The gas ensemble initially 'occupies' the entire box. When the
partition is inserted, it is divided into two orthogonal
subensembles. The intention of the engine is to extract useful
work from allowing each of these subensembles to expand back to
'occupy' the entire box again.

We have shown that this can be done, by inserting a freely moving
piston in the center of the box. The inclusion of the state of
this piston is an essential part of the evolution of the system,
as the required evolution is not unitary unless the orthogonality
of the piston states is taken into account. This transfers the
entropy of mixing from the gas subensembles to the piston and
weight subensembles. Now the same requirement of unitarity
prevents the piston from being restored to it's original position,
which, if successful would imply a reduction in the entropy of the
system.

\subsection{Maxwell's Demons}
It is the orthogonality of the pistons states that are essential
to the operation of the Szilard Engine. We will now show how this
relates to the Maxwell's Demon.

The original Maxwell's Demon thought experiments did not involve
an analysis of work or free energy. Maxwell described two systems,
a pressure demon and a temperature demon, using a trap door which
separates a gas into two portions. When an atom approaches, the
demon opens or closes the trapdoor, allowing the atom to pass or
not. We will present a very simplified analysis of the pressure
demon, to illustrate it's essential similarity to our analysis of
the Szilard Engine.

In the case of the pressure demon, if an atom approaches from the
left, it is allowed to pass, while if it approaches from the
right, it is reflected elastically. No work is performed upon the
system. We represent an atom on left by $\ket{L}$ and on the right
by $\ket{R}$.

If $U_1$ represents the unitary operator for the demon holding the
trapdoor open and $U_2$ the unitary operator for the demon holding
the trapdoor closed, we have
\begin{eqnarray*}
U_1\ket{L} &=& \ket{R} \\
U_2\ket{R} &=& \ket{R}
\end{eqnarray*}
These cannot be combined into a single unitary operator. To
operate the trapdoor the demon must involve it's own internal
states, or some auxiliary system.

The complete specification of the unitary operators is
\begin{eqnarray*}
U_1 &=&\kb{L}{R}+\kb{R}{L} \\
U_2 &=&\proj{L}+\proj{R}
\end{eqnarray*}
We now assume the demon has auxiliary states $\ket{\pi_0}$ and
$\ket{\pi_1}$, and uses these auxiliary states to produce a
combined unitary operation. There is some flexibility in choosing
this operator but this is not important, so we choose the fairly
simple form, assuming the demon initially in the state
$\ket{\pi_0}$ of

\begin{eqnarray*}
U_a &=&\kb{\pi_1L}{\pi_0L}+\kb{\pi_0R}{\pi_0R} \\
    &&+\kb{\pi_0L}{\pi_1L}+\kb{\pi_1R}{\pi_1R} \\
U_b &=&\kb{\pi_1R}{\pi_1L}+\kb{\pi_0R}{\pi_0R} \\
    &&+\kb{\pi_0L}{\pi_0L}+\kb{\pi_1L}{\pi_1R} \\
    &=&\proj{\pi_1}U_1+\proj{\pi_0}U_2
\end{eqnarray*}
The action of $U_a$ represents the Demon measuring the location of
the atom, and then $U_b$ represents the Demon holding the trapdoor
open or shut.

The atom may initially be on either side, so is described by
\[
\frac 12\proj{L}+\frac 12\proj{R}
\]
After the operation of $U_a$, the demon and atom are in a
correlated state
\[\frac 12\proj{L \pi_1}+\frac 12\proj{R \pi_0} \]
Under $U_b$, the atom then evolves into $\proj{R}$, but leaves the
demon in the state $\frac 12\proj{\pi_0}+\frac 12\proj{\pi_1}$.
Clearly the entropy of the atom has decreased, but the entropy of
the demon has correspondingly increased\footnote{If we now bring
in a second atom in the state $\frac 12\proj{L}+\frac 12\proj{R}$,
the demon fails to sort the atom at all. Having picked up the
mixing entropy of the atom, it is no longer able to function as
intended.}. The demon states play exactly the same role as the
piston states in the Popper-Szilard Engine. We will now consider
the thermodynamics of this.

\subsection{The Significance of Mixing}
What we have seen above is that the problem involves separating an
ensemble into subensembles. By correlating these subensembles to
an auxiliary system, such as a Demon or a piston, operations can
be performed upon the subensembles that cannot be performed upon
the overall ensemble. In other words, we are trying to reverse the
mixing of the subensembles. We will now have to consider the
physical origin of the mixing entropy, and the role it plays. We
will restrict the discussion to the case where there are only two
subensembles $\rho _1$ and $\rho _2,$ and focus upon the problem
of reversibly extracting work from the system.

To understand the significance of this requires us to explain the
physical origin of the mixing relationships
\begin{eqnarray*}
F_i &=&F-kT\ln p_i \\
S &=&\sum_ip_i\left( S_i-k\ln p_i\right)
\end{eqnarray*}
where an equilibrium density matrix may be decomposed into
orthogonal subensembles
\begin{eqnarray*}
\rho &=&\sum_ip_i\rho _i \\
\rho _i\rho _j &=&\left(\rho _i\right) ^2\delta _{ij}
\end{eqnarray*}

If we start with a system in the equilibrium state
$\rho=p_1\rho_1+p_2\rho_2$, we will be able to extract work from
the mean pressure exerted on some boundary parameter. This is
represented by the free energy $F$ which is the work that can be
isothermally extracted, when taking the density matrix $\rho$ to
some reference state $\rho_0$.

Let the free energy $F_1$ represents the isothermal work extracted
taking a density matrix $\rho_1$ to the reference state $\rho_0$.
This is given by $F_1=F-kT\ln p_1>F$.  Similarly for $\rho_2$ we
have $F_2=F-kT\ln p_2>F$. In both these cases, the free energy is
higher than is obtained by operating directly upon the ensemble,
by an amount $-kT\ln p_i$ so the mean gain in free energy from
operating upon the subensembles rather than the ensemble is simply
$-kT\sum p_i\ln p_i$. This is the free energy that is lost due to
the mixing.

In other words, by separating the ensemble into it's orthogonal
subensembles, we are attempting to avoid the loss of free energy
caused by the mixing. Although other versions of Maxwell's demon
do not address free energy directly (eg. creating pressure or
temperature gradients), they are all illustrated by being
connected to heat engines or turbines which extract work, so in
one way or another they are all implicitly concerned with
increasing the free energy of an ensemble by manipulating it's
subensembles.

We will now try to explain how mixing causes the free energy to be
lost. This will be shown to be a consequence of the unitarity of
the evolution operators.

\paragraph{Perfect Isolation}
First we will consider the situation of perfect isolation. In this
case there are no transitions between eigenstates, and the
evolution of a density matrix, initially $\rho ^{\prime }(0)$,
will be described by
\[
\rho ^{\prime }(t)=U(t)\rho^{\prime}(0)U^{\dagger}(t)
\]
where $U(t)$ is the solution to the operator \Sch equation.

Our first result to establish is that there is no operator that is
capable of separately operating upon $\rho_1$ and $\rho_2$ to take
them into the reference state $\rho_0$. This can be seen easily
from the fact that if we were to find an operator $U_1$ such that
\[
\rho_0=U_1\rho_1U_1^{\dagger}
\]
it cannot be also true that
\[
\rho_0=U_1\rho_2U_1^{\dagger }
\]
as this would mean
\[
\left(\rho_0\right)^2
    =U_1\rho_1U_1^{\dagger}U_1\rho_2U_1^{\dagger}
    =U_1\rho_1\rho_2U_1^{\dagger}=0
\]
and a density matrix such as $\rho_0$ cannot be nilpotent.

From this it follows that if we wish to perform an operation where
each of the two subensembles are taken to the same reference
state, we must involve a second system.

If we take a second operation, $U_2$, such that
\[
\rho_0=U_2\rho _2U_2^{\dagger }
\]
and introduce an auxiliary system, with orthogonal
states\footnote{We will always assume that eigenstates of the
auxiliary systems are at the same energy.} $\pi _1$ and $\pi _0$,
initially in the state $\pi _0,$ then we \textit{can }form two
unitary operators, containing the operations
\begin{eqnarray*}
U_a &=&\kb{\pi_1}{\pi _0}P_1+\kb{\pi_0}{\pi_0}P_2 \\
U_b &=&\proj{\pi_1}U_1+\proj{\pi_0}U_2
\end{eqnarray*}
where $P_1$ and $P_2$ are projectors onto the subspaces of
$\rho_1$ and $\rho_2$ respectively.

The effect of $U_a$ is to correlate the auxiliary system with the
subensembles. $U_b$ then acts as a conditional unitary operator.
If the auxiliary system is in $\pi_1$, then it switches on the
Hamiltonian necessary to take $\rho_1$ to $\rho_0,$ while if the
auxiliary system is in state $\pi_2$, the Hamiltonian for taking
$\rho_2$ to $\rho_0$ is switched on. This successfully takes each
of the subensembles to the reference state, extracting maximum
work in the process, but leaves the auxiliary system in the state
$p_1\proj{\pi_1}+p_2\proj{\pi_2}$. The entropy of mixing has been
transferred from the ensemble to the auxiliary. The $\pi_1$ and
$\pi_2$ are orthogonal, and so again there is no unitary operation
that is capable of restoring the auxiliary system to it's initial
state.

\paragraph{Contact with the environment}
The situation of perfect isolation, however, is too idealised. In
general, while the unitary operation is taking place, contact with
an environment will cause transitions between eigenstates. The
evolution of the  density matrix will not, in general, be
described by a unitary operation. We cannot assume that the final
and initial density matrices are unitarily equivalent, so the
proof given above, based upon the preservation of inner products,
is no longer valid.

As an example, let us consider the discussion of the Szilard box
with the partition raised, and the atom confined to the left. The
state is initially
\[
\ket{\psi_l^\rho}=\frac{1}{2}\left(
    \ket{\psi_l^\mathrm{even}}+\ket{\psi_l^\mathrm{odd}}\right)
\].

If the partition is removed, in perfect isolation, the free
evolution of the gas leads to the state
\[
\frac{1}{2}\left(
    e^{-i\frac{E_l^\mathrm{even}t}{\hbar}}\ket{\psi_l^\mathrm{even}}
    +e^{-i\frac{E_l^\mathrm{odd}t}{\hbar}}\ket{\psi_l^\mathrm{odd}}\right)
\]
where the energies are now the non-degenerate energies of the
unperturbed eigenstates. This leads to a time dependant factor in
the phase of the superposition. The state appears reasonably
uniformly spread most of the time, but when
\[
\frac{\left(E_l^\mathrm{even}-E_l^\mathrm{odd}\right)t}{\hbar}
    =n\pi
\]
for integer $n$, the atom will be located on a well defined side
of the box. If the piston is re-inserted at this time, the atom
will always be found on a specific side of the box.

If the atom had initially started confined to the right, it would
evolve to
\[
\frac{1}{2}\left(
    e^{-i\frac{E_l^\mathrm{even}t}{\hbar}}\ket{\psi_l^\mathrm{even}}
    -e^{-i\frac{E_l^\mathrm{odd}t}{\hbar}}\ket{\psi_l^\mathrm{odd}}\right)
\]
This will be found on the opposite side of the box at these same
well defined times. In fact, at all intervening times, the two
states are orthogonal. Although they are \textit{spatially}
overlapping most of the time, in principle the interference terms
maintain the distinguishability of the two states.

If we construct the density matrices $\rho_{G2}^\lambda$ and
$\rho_{G2}^\rho$ from the right and left wavefunctions, lowering
the partition causes these to evolve into states that are still
orthogonal to each other. The initially orthogonal subensembles
(of gas on the left or gas on the right) remain orthogonal at all
times.

If the box is in contact with an environment, however, decoherence
effects destroy the superposition between the even and odd
wavefunctions. Both $\ket{\psi^{\rho}_l}$ and
$\ket{\psi^{\lambda}_l}$ will now evolve into the density matrix
\[
\frac{1}{2}\left(\proj{\psi_l^\mathrm{even}}
    +\proj{\psi_l^\mathrm{odd}}\right)
\]
As the orthogonality between the $\rho_{G2}^\lambda$ and
$\rho_{G2}^\rho$ states depends upon the coherent phase of the
superpositions, when there is decoherence the left and right
subensembles evolve to the same equilibrium ensemble $\rho_{G0}$.
In this situation, the same unitary operation (lowering the
partition) leads to initially orthogonal subensembles evolving
into the same density matrix.

Although we must describe the evolution of the system with unitary
operators, contact with the environment can allow non-unitary
evolution of the system's density matrix. We must now analyse the
effect of this upon the mixing relationship.

\paragraph{Isothermal}
We must take into account the non-unitarity of the evolution, due
to interactions with the environment, when considering how to
extract the free energy. Our task is to see if the initially
orthogonal subensemble states can be taken into non-orthogonal
states, using contact with the heat bath, while extracting the
free energy that is lost due to mixing.

We will consider the situation where the environment is a heat
bath at temperature $T$. To extract the optimum free energy $F_1$,
from subensemble $\rho_1$, we need to apply a suitable time
dependant Hamiltonian (such as the one that leads to $U_1$) that
takes the subensemble to the reference state (at temperature $T$).
One of the properties of such a optimum path is that it is
thermodynamically reversible. The means that if we apply
$U_1^{\dagger}$ to the reference state, while in contact with a
heat bath at temperature $T$, we will obtain the original
subensemble $\rho_1$ (and will have to perform $F_1$ work upon the
system).

If we now try to extract the free energy $F_2$ from the
subensemble $\rho_2$, we clearly require a different time
dependant Hamiltonian as we need it to correspond to the adjoint
of that unitary operator $U_2^{\dagger }$ which, when isothermally
applied to the reference state, produces the subensemble $\rho
_2$. This leaves us in the same situation as with perfect
isolation - if we wish to combine the two unitary operations so
that the appropriate one is applied to the appropriate
subensemble, we need to include an auxiliary system. This
auxiliary system correlates itself to the subensemble, and is
itself left in a higher entropy state.

It appears that if we wish to extract the $-kT\ln p$ free energy
from the subensembles, we cannot combine the operations into a
single operator, but must employ an auxiliary. We know that there
{\em is} an operator that can take both the subensembles to the
same state, when in contact with a heat bath, but this operator
loses the free energy of mixing. We shall refer to this as a
'dissipation' of the mixing free energy $-kT \sum p \ln p$.

Let us try and understand more clearly the underlying reason why
the orthogonal subensembles can be decoherently transformed into
the same state using a single unitary operator, but if we wish to
extract the free energy rather than dissipate it, two different
unitary operators are required. We will consider the example of
the Szilard box, with a partition raised, where $\rho_1$ is the
atom confined to the left of the partition, $\rho_2$ the atom
confined to the right, and the reference state is the atom
unconfined with no partition.

When applying operator $U_{RI}$ to remove the partition, the
eigenstates deform continuously between the states $\Psi_l^{even}$
and $\Psi_l^{odd}$, and the corresponding unperturbed $\Psi_n$
states. If the atom is initially confined to the left, the initial
states are $\Psi_j^L$ which are superpositions of $\Psi_j^{even}$
and $\Psi_j^{odd}$. As the barrier is lowered, the initial states
evolve into a superposition of the unperturbed $\Psi_{2j}$ and
$\Psi_{2j-1}$ states. The $\Psi_j^R$ states, corresponding to an
atom initially confined to the right of the partition, will evolve
into an orthogonal superposition of the same states.

The most important feature of this is that the states into which
the $\Psi_l^L$ evolve span only half the Hilbert space - the
$\Psi_l^R$ evolve into states which span the other half. However,
once the barrier has been lowered, all the states are thermally
accessible to the atom, through interactions with the heat bath.
The evolution given by $U_{RI}$ does not cause the initially
confined atom to occupy the full space and become in the state
$\rho_{G0}$. It is the 'free energy dissipating' or decoherent
contact with the heat bath which allows the atom to expand to
occupy the entire state space.

Now let us consider the situation where the atom is confined to
the left, and we wish to extract the free energy of the expansion
to fill the entire box. Again, the atom starts in the $\Psi _l^L$
states. Now the evolution $U_1$, however it is implemented, to
extract the optimum work, must take the atom into $\rho_{G0}$,
occupying the complete set of the unperturbed $\Psi _l$ states -
which span the entire Hilbert space\footnote{This difference
between $U_1$ and $U_{RI}$, mapping the same initial states to
all, and one-half of the final Hilbert space, respectively, is
possible because there is a countable infinity of states
available.}.

Suppose the effect of $U_1$ left some of the final Hilbert space
unoccupied, but thermally accessible. Then, decoherence from
contact with the heat bath would lead to that portion of Hilbert
space becoming occupied, dissipating some free energy in the
process. To extract maximum work, or equivalently, to eliminate
the dissipation of free energy, the operation of $U_1$ must be a
one-to-one mapping of the $\Psi_l^L$ Hilbert space onto the
$\Psi_l$ Hilbert space.

Now, the same must also be true for the optimum extraction, using
$U_2$, of free energy from an atom initially confined on the
right. However, this means that $U_1$ and $U_2$ are attempting to
map initially orthogonal sets of eigenstates $\Psi_l^L$ and
$\Psi_l^R$ onto the same set of states $\Psi_l$. This is the
reason that $U_1$ and $U_2$ cannot be combined into a single
operator, as such a mapping cannot be unitary.

This significantly improves the result derived in the case of
perfect isolation above. For perfect isolation, we can rely upon
the unitary equivalence of the transformed \textit{density
matrices}, and the invariance of their inner product. This cannot
be relied upon when there are interactions with an environment.
Instead, we have used the properties of the unitary operation, as
a mapping upon the \textit{space of states} that the density
matrix occupies.

If we were to use a $U_1$ operator that mapped the $\Psi_l^L$ only
onto some subset of the $\Psi_l$ , then that would leave the
complementary subset available for some of the $\Psi_l^R$ under
$U_2$. This would allow some portion of $U_1$ and $U_2$ to be
combined. However, the atom initially confined to the left, would
come to occupy the entire Hilbert space, including that portion of
the Hilbert space left unoccupied by $U_1$ through decoherent
contact with the heat bath. The same would take place for the atom
initially confined to the right. In other words, the extent to
which the $U_1$ and $U_2$ operators may be combined is directly
linked to the amount of free energy that is dissipated rather than
extracted. The operator $U_{RI}$ maps the $\Psi_l^L$ and
$\Psi_l^R$ onto entirely orthogonal sets of states, but which are
accessible to the same set of states by a decoherent process. This
allows a single operator to take the left and right density
matrices into occupying the whole space, but at the cost of
dissipating the entire free energy of mixing.

The conclusion of this is that it is the requirement of unitarity
that prevents us from extracting the optimum free energy from the
subensembles. A unitary operator that acts upon both subensembles
will fall short of optimum by at least that amount of free energy
given by the mixing formula. We can use a different unitary
operator upon each subensemble only if we correlate an auxiliary
system to the subensembles. However, the consequence is that the
auxiliary system picks up precisely that entropy of mixing that
compensates for the increase in work we are now able to extract
from the subensembles.

\subsection{Generalised Demon}
We have argued that it is the relationship between the mixing and
correlations that both gives rise to, and resolves, the Maxwell's
Demon problem. Let us examine this in more detail, and greater
generality. Our intention here is to highlight the role of the
unitary operations upon the subspaces and the effect of
introducing an auxiliary system. Our argument is that the mixing
entropy is a consequence of unitarity. Reversing this mixing,
separating the ensemble into subensembles, can only be achieved by
introducing an auxiliary system. However, any gain in the free
energy or entropy due to this separation is offset by at least as
large an increase in the entropy of the auxiliary system.

We assume the initial Hilbert space is formed from two orthogonal
subspaces $\Gamma =\Gamma_1\oplus \Gamma_2$. The initial,
equilibrium ensemble may be written in terms of the orthogonal
subensembles $\rho =p_1\rho _1+p_2\rho _2. $ The subensemble
$\rho_1$ initially occupies\footnote{When we say a density matrix
'occupies' a subspace, we mean that those eigenvectors of the
density matrix which have non-zero eigenvalues, form a basis for
the subspace.} the subspace $\Gamma _1$ of the Hilbert space and
$\rho_2$ occupies the orthogonal subspace $\Gamma _2$. They occur
with probability $p_1$ and $p_2$ in the initial equilibrium
ensemble, and $p_1+p_2=1$. The unitary operator $U_1$ maps $\Gamma
_1$ to some subspace $\Gamma_1^{\prime}$ of $\Gamma$ and $U_2$
maps $\Gamma _2$ to $\Gamma _2^{\prime }$. We will assume that
contact with a thermal heat bath will cause an ensemble initially
localised in $\Gamma _1^{\prime }$ to decoherently spread
throughout $\Gamma$, returning the system to the initial
equilibrium ensemble $\rho$, and similarly for $\Gamma _2^{\prime
}$.

The probability of an equilibrium system $\rho$ being
spontaneously found in the $\Gamma _1^{\prime }$ subspace is
$p_1^{\prime }$ and the probability of the system being similarly
in $\Gamma _2^{\prime }$ is $p_2^{\prime }$. As we do not assume
that $\Gamma _1^{\prime }$ and $\Gamma _2^{\prime }$ are
orthogonal subspaces, there is no restriction on $p_1^{\prime
}+p_2^{\prime }$.

The free energy of the subensembles can be calculated from their
probabilities, and the free energy of the initial ensemble $F$
\begin{eqnarray*}
F_1 &=&F-kT\ln p_1 \\
F_1^{\prime } &=&F-kT\ln p_1^{\prime } \\
F_2 &=&F-kT\ln p_2 \\
F_2^{\prime } &=&F-kT\ln p_2^{\prime }
\end{eqnarray*}
We now wish to see how we can extract the extra free energy from
the subensembles.

In $p_1$ proportion of the cases, the system is in subensemble
$\rho _1$. Under the operation of $U_1$, it isothermally expands
to occupy $\Gamma _1^{\prime }$, becoming $\rho_1^{\prime}$. This
extracts $kT\ln \left( p_1^{\prime }/p_1\right)$ free energy. The
density matrix $\rho _1^{\prime } $ then expands freely into
$\rho$, and $-kT\ln \left( p_1^{\prime }\right) $ notional free
energy is dissipated.

In $p_2$ cases, the initial subensemble is $\rho _2$. Isothermally
expanding this with the operation of $U_2$ extracts $kT \ln
\left(p_2^{\prime}/p_2\right)$ and then dissipates the notional
free energy $-kT\ln \left( p_2^{\prime }\right)$.

The mean free energy gained is
\[
\frac{\Delta F_G}{kT}=p_1\ln \left(
    \frac{p_1^{\prime}}{p_1}\right)
    +p_2\ln \left( \frac{p_2^{\prime }}{p_2}\right)
\]
and the subensemble free energy which may be regarded as
dissipated is
\[
\frac{\Delta F_D}{kT}=-p_1\ln p_1^{\prime }-p_2\ln p_2^{\prime
}\geq 0
\]
giving
\[
\frac{\Delta F_G+\Delta F_D}{kT}=-p_1\ln p_1-p_2\ln p_2\geq 0
\]
which is equal to the entropy of mixing of the two subensembles.
As the free energy dissipated is never negative, it is immediately
apparent that the free energy gained cannot exceed the entropy of
mixing.

When we wish to distinguish between the actual free energy of an
ensemble, $F$, and the mean free energy of it's subensembles $\sum
p_iF_i$ we shall refer to the additional free energy $-kT \sum p_i
\ln p_i$ of the subensembles as a 'notional' free energy. This is
the free energy we would \textit{like }to be able to extract by
splitting the ensemble into subensembles. The sense in which this
'notional' free energy is 'dissipated' is simply that we have
failed to extract it. This is not the same as the situation where
the initial matrix is actually $\rho_1$ say, and it is allowed to
expand freely to $\rho$ in which case an actual, rather than
notional, free energy $-kT\ln p_1$ would have been lost.

\paragraph{No overlap in final subspaces}

In the case where $\Gamma _1^{\prime }$ and $\Gamma _2^{\prime }$
are complementary\footnote{If we were to use subensembles which
were orthogonal, but not complementary, then $p_1^{\prime
}+p_2^{\prime }<1.$ The only effect of this would be to reduce the
amount of free energy that could be extracted.} orthogonal
subspaces, then $U_1$ and $U_2$ may be combined into a single
unitary operator $U_3$ and $p_1^{\prime }+p_2^{\prime }=1$. This
yields a value of
\[
\frac{\Delta F_G}{kT}=p_1\ln \left(
    \frac{p_1^{\prime}}{p_1}\right)
    +(1-p_1) \ln \left( \frac{1-p_1^{\prime}}{1-p_1}\right) \leq 0
\]
with equality occurring only for $p_1=p_1^{\prime}$.

To understand this we must consider what is happening to the two
respective subensembles. As $p_1+p_2=p_1^{\prime }+p_2^{\prime}$
any 'expansion' of one subensemble is paid for by a 'compression'
of the other. What the relationship above shows, is that when we
divide an equilibrium ensemble into subensembles, the work
required to perform the compression on one will always outweigh
the work gained from the expansion on the other.

It is important to remember the values of $p_1^{\prime }$ and
$p_2^{\prime }$ are the equilibrium probabilities that initial
density matrix would have spontaneously been found in $\Gamma
_1^{\prime }$ or $\Gamma _2^{\prime }$, while $p_1$ and $p_2$ are
the probabilities of spontaneously finding the system in a
subensemble that is isothermally moved into those subspaces.
Unless these probabilities are the same, the final density matrix
will not be in equilibrium. This result tells us that any attempt
to rearrange an equilibrium distribution into a non-equilibrium
distribution requires work.

For the case of the Szilard Box, we divide the gas ensemble
$\rho_{G0}$ into the two subensembles $\rho_{G2}^\lambda$ and
$\rho_{G2}^\rho$ by inserting a partition. This gives us
$p_1=p_2=\frac{1}{2}$. If we simply remove the piston, we
'dissipate' the notional $kT \ln 2$ energy we could have extracted
from expanding either of the subensembles, as we do not have an
operator that, acting upon the gas alone, can extract this as
work.

\paragraph{Complete overlap in final subspaces}

Now let us consider the case where $\Gamma _1^{\prime }$ and
$\Gamma _2^{\prime }$ have an overlapping subspace
$\Gamma_{12}^{\prime }$. We are not restricted to $p_1^{\prime
}+p_2^{\prime }=1$ anymore, but we can no longer combine $U_1$ and
$U_2$ into a single operator, so must employ an auxiliary system.
The increase in entropy of the auxiliary system is
\[
\frac{\Delta S_{aux}}k=-p_1\ln p_1-p_2\ln p_2
\]
which is the same as the entropy of mixing of the subensembles,
and equal to the total free energy that is available to extraction
and dissipation.

As we have no restrictions upon $p_1^{\prime }$ and $p_2^{\prime
},$ we obtain minimum 'dissipation', and extract maximum free
energy, by setting $\Gamma _1^{\prime }=\Gamma _2^{\prime }=\Gamma
_{12}^{\prime }=\Gamma _1 \oplus \Gamma _2$ so that $p_1^{\prime
}=p_2^{\prime }=1$. This allows us to extract the free energy $-kT
\ln p_1$ with probability $p_1$ and $-kT\ln p_2$ with probability
$p_2$. Each subensemble has been allowed to expand to fill the
entire space, extracting maximum free energy. However, the
auxiliary system has had an equivalent increase in entropy.

This corresponds to the isothermal expansion of the Szilard box,
where the piston plays the role of the auxiliary system. The free
energy is extracted from each of the gas subensembles, but the
piston is left in a mixture of states.

\paragraph{Partial overlap in final subspaces}

We might now ask that if $\Gamma _1^{\prime }$ and $\Gamma
_2^{\prime }$ are not completely overlapping but not completely
orthogonal, is there some way we can avoid the auxiliary system
picking up the entire entropy of mixing. If we assume that
$p_2\leq p_1$, without loss of generality, we start by separating
$\Gamma _2^{\prime }$ into orthogonal subspaces $\Gamma
_{12}^{\prime }$ and $\Gamma _{2a}^{\prime },$ where $\Gamma
_{2a}^{\prime }$ does not overlap with $\Gamma _1^{\prime}$.

We now need to separate the initial density matrix $\rho _2$ into
the orthogonal subensembles $\rho _{2a} $ and $\rho _{2b}$, where
the subspace containing $\rho _{2a}$ is mapped onto
$\Gamma_{2a}^{\prime}$ and $\rho _{2b}$ onto
$\Gamma_{12}^{\prime}$ by $U_2$. The probabilities of these
subensembles will be $p_{2a}$ and $p_{2b}$ and the probabilities
associated with $\Gamma_{12}^{\prime}$ and $\Gamma_{2a}^{\prime}$
are $p_{12}^{\prime}$ and
$p_{2a}^{\prime}=p_2^{\prime}-p_{12}^{\prime}$. Finally, we split
$U_2$ into an operator $U_{2a}$ acting upon $\rho _{2a}$ and an
operator $U_{2b}$ acting on $\rho _{2b}$.

We are now able to combine $U_1$ with $U_{2a}$, as
$\Gamma_{2a}^{\prime }$ and $\Gamma _1^{\prime }$ do not overlap,
into a single operator $U_A=U_1\otimes U_{2a}$. This allows us to
reformulate the problem as involving the two complementary
orthogonal subspaces $\Gamma _A$ and $\Gamma _B$ with
\begin{eqnarray*}
\rho _A &=&\frac{p_1\rho _1+p_{2a}\rho _{2a}}{p_1+p_{2a}} \\
\rho _B &=&\rho _{2b} \\
\Gamma _A &=&\Gamma _1\oplus \Gamma _{2a} \\
p_A &=&p_1+p_{2a} \\
\Gamma _B &=&\Gamma _{2b} \\
p_B &=&p_{2b} \\
\Gamma _A^{\prime } &=&\Gamma _1^{\prime }\oplus \Gamma _{2a}^{\prime } \\
p_A^{\prime } &=&p_1^{\prime }+p_2^{\prime }-p_{12}^{\prime } \\
\Gamma _B^{\prime } &=&\Gamma _{12}^{\prime } \\
p_B^{\prime } &=&p_{12}^{\prime }
\end{eqnarray*}
Now the final entropy of the auxiliary system
\[
\frac{\Delta S_{aux}}k=-p_A\ln p_A-p_B\ln p_B
\]
is lower than the increase that would have occurred based upon
$p_1$ and $p_2$, so we have reduced it's increase in entropy.
However, now we still have a dissipation of
\[
\frac{\Delta F_D}{kT}=
    -p_A\ln p_A^{\prime}-p_B\ln p_B^{\prime}\geq 0
\]
notional free energy and an extraction of only
\[
\frac{\Delta F_G}{kT}=p_A\ln \left(
    \frac{p_A^{\prime}}{p_A}\right)
    +p_B\ln \left( \frac{p_B^{\prime }}{p_B}\right)
\]
so the gain in free energy is still less than the equivalent
increase in entropy of the auxiliary.

In the special case where $p_{2b}=p_{12}^{\prime }=0$, there is no
overlap between $\Gamma _1^{\prime }$ and $\Gamma _2^{\prime }$,
there is no increase in entropy of the auxiliary, but there is no
extraction of free energy. This is the case where we may write
$U_3=U_1\otimes U_2$.

If there is an overlap, however, unless $p_{2a}^{\prime }=0,$
(there is no portion of $\Gamma _2^{\prime }$ that is \textit{not
}overlapped by $\Gamma _1^{\prime }$) we cannot set $p_B^{\prime
}=1,$ and will always dissipate some of the free energy. We will
only be able to extract an amount of free energy equivalent to the
increase in entropy of the auxiliary when
$p_A^{\prime}=p_B^{\prime }=1$. So, although the case where the
final subspaces are partially overlapping may allow us to reduce
the entropy increase of the auxiliary system, it does not allow us
to do better than the case where the final subspaces are either
completely overlapping, or completely orthogonal.

\paragraph{Conclusion}

This now answers the question why we are unable to extract the
free energy of the subensembles. The optimum operators acting upon
the subensembles cannot be combined into a single unitary
operator. The only way of using a combined operator on the
subensembles is to allow processes that would dissipate the
notional free energy if applied to the individual subensembles.
This is the meaning of the reduction in free energy due to mixing.

We can try and avoid this, by correlating an auxiliary system to
the subensembles, and applying conditional unitary operators. This
will successfully extract the mean free energy from the expansion
of the system, without the loss of free energy due to mixing.
However, the cost of this is to leave an auxiliary system in a
higher entropy state, and this increase in entropy at least
matches the gain in free energy that results from separating the
system into it's orthogonal subensembles. So, through the
combination of dissipated free energy, and entropy transfer to an
auxiliary system, we are unable to improve our position.

It is important to note that the correlation between the auxiliary
and the subensembles must be carefully controlled. If we have
complete overlap in the final subspaces, then the operator $U_1$,
which maps $\Gamma_1$ onto $\Gamma$, will map $\Gamma _2$ onto a
space which occurs with $p=0$. If the auxiliary becomes correlated
to the wrong subensemble, the conditional operation may attempt to
apply $U_1$ to $\rho_2$. Instead of extracting free energy, this
will attempt to compress the system into a zero volume. This would
require an infinite amount of work. Obviously this is not
physically possible, and so would lead to the engine breaking down
in some way. If there is any possibility of the auxiliary being in
the wrong state, therefore, this imposes an additional constraint
upon the unitary operations that may be conditionalised upon it.
In the Szilard Engine, for example, this leads to the restriction
on the four subspaces of the piston and weights, for $U_{RES}$ in
Equation \ref{eq:reset}.

\subsection{Conclusion}

We believe this has brought out one of the essential features of
the general Maxwell's demon problem, and shown why it does not
constitute a problem for the second law of thermodynamics. In
essence, the problem arises from the increase in entropy that
comes about when subensembles are mixed. The demon Maxwell
proposed was able to examine each atom, and sort the ensemble into
it's subensembles. This reverses the entropy increase due to the
mixing, in apparent violation of the second law of thermodynamics.

However, we have seen that this sorting cannot be implemented by
any unitary operation acting only upon the space of the
gas\footnote{Maxwell argued that his demon proves the second law
of thermodynamics cannot be derived from Hamiltonian mechanics.
Clearly this is mistaken. The demon Maxwell envisages is able to
violate the second law only because it is a non-Hamiltonian
system.}. Instead, it must include an auxiliary system. This
auxiliary system increases in entropy to match the decrease in
entropy of the gas.

When we consider the change in free energy from mixing, we find
the same problem. To extract the free energy from each
subensemble, we must employ an auxiliary system, whose entropy
increases in direct relation to the gain in free energy. For the
Szilard Engine, this auxiliary system is clearly the piston
system.

This completes the first stage of the resolution to the Maxwell's
Demon problem. The 'measurement' of the system by the 'Demon' (or
equivalently, the correlation of the auxiliary to the system) does
not decrease entropy, as there is a compensating increase in
entropy of the auxiliary system.

However, this does not constitute the whole resolution. In the
Popper version of Szilard's Engine, there are also weights whose
state is imperfectly correlated to the auxiliary state. This
suggests that it is possible to imperfectly reset the auxiliary.
Although we have shown that, in the case of the Popper-Szilard
Engine, this resetting cannot succeed, we need to understand why
such a resetting mechanism cannot succeed in general, and how this
resetting relates to the $kT \ln 2$ energy that Landauer's
Principle suggests is necessary to reset the state of the
auxiliary.

\section{Restoring the Auxiliary} \label{s:szsol2}
We now must consider means by which the auxiliary system may be
restored to it's initial state. This would allow the system to
continue extracting energy in cyclic process. For the
Popper-Szilard Engine this involves attempting to reset the piston
state by correlating it to the location of the two weights.

The essential point to note here is that it was necessary to
include the quantum description of the weights as a thermodynamic
system at some temperature $T_W$, rather than simply as a 'work
reservoir'. Although we noted certain properties of the
thermodynamic weight\footnote{The equivalence of perfect
isolation, essential isolation and isothermal lifting, and also
the constancy of entropy as it is raised}, in Sections
\ref{s:szsm3} and \ref{s:szth1} that make the weight in a
gravitational field a very convenient system to use as a 'work
reservoir', our treatment of it was as an isothermal compression.

In the previous Section we showed how the correlation of an
auxiliary could be used to extract work from the mixing free
energy of the system. To complete the analysis we must also take
into account the effect of this work on a second system, and the
possible correlations this second system can have with the
auxiliary.

First we will derive a general relation, which we will refer to as
the 'fluctuation probability relation', which characterises the
effect upon one system that can be achieved from a thermal
fluctuation in a second. We will then apply this relation to the
generalisation of the Popper-Szilard Engine. The fluctuation
probability relation will be shown to govern the long term energy
flows in such a way as to ensure that any attempt to reset the
Engine must fail in exactly such a way as to ensure that the mean
flow of energy is always in an entropy increasing direction. We
will also show how, by performing work upon the system, the Engine
can be made to operate without error, but only at the efficiency
of the Carnot Cycle.

\subsection{Fluctuation Probability Relationship}
We will now calculate the key relationship governing the work that
may be extracted from a thermal fluctuation. We must first discuss
what we mean by a fluctuation within the context of the Gibbs
ensemble. Generally, the equilibrium density matrix
\[
\rho=\frac{e^{-\frac{H}{kT}}}{\trace{e^{-\frac{H}{kT}}}}
\]
may be interpreted as the system being in one of the eigenstates
of the Hamiltonian with probability
\[
p_i=\frac{e^{-\frac{E_i}{kT}}}{\trace{e^{-\frac{H}{kT}}}}
\]
and that contact with a heat bath at temperature $T$ completely
randomises the state of the system, on a timescale of order
$\tau$, the thermal relaxation time. The system jumps randomly
between the available states. These are the thermal
'fluctuations'.

If we had a macroscopic system, we could partition the Hilbert
space into macroscopically distinct subspaces. From the
perspective of the Gibbs ensemble, this is the separation of the
density matrix into subensembles
\[
\rho=\sum_\alpha p_\alpha \rho_\alpha
\]
where $\rho_\alpha$ is the equilibrium density matrix occupying
the subspace and $p_\alpha$ is the probability that the system
state is in the subspace.

For macroscopic systems, the majority of states will be in one
large subspace, which will have approximately the same entropy as
the ensemble. However, there will be some states in small
subspaces that correspond to situations with lower entropy, such
as the atoms of a macroscopic gas all located in one half of a
room. At any point there will be a small probability that the
thermal fluctuations will lead to such a subspace being occupied.
As we have seen in Equation \ref{eq:freeprob}, these fluctuations
will have a free energy given by
\[
F_i=F-kT \ln p_i
\]
If the fluctuation is very rare ($p_i \ll 1$) the increase in free
energy will be large in comparison to {\em macroscopic}
quantities.

For microscopic systems, such as the single atom Szilard Engine,
the ensemble free energy may well be of the order of $kT$. If this
is the case, reasonably common fluctuations may show an increase
in free energy comparable to the free energy of the ensemble
itself. We are now going to consider trying to harness this gain
in free energy, and put it to use on some other system, such as by
lifting a weight.

If we find a system at temperature $T_1$ in a subensemble which
spontaneously occurs with probability $p_1$, we can extract $-kT_1
\ln p_1$ work from allowing the subensemble to expand back to the
equilibrium. We wish to use this work to perform some action upon
a second system. If treat this as storing the energy in a work
reservoir, such as a weight, we have noted this is exactly
equivalent to isothermally compressing the second system (lifting
the weight).

The free energy $F_2^{\prime }$ of the compressed state of the
second system will differ from the free energy $F_2$ of it's
original state by
\[
F_2^{\prime }=F_2-kT_1\ln p_1
\]
Now, we know that the second system will spontaneously occur in a
fluctuation state with free energy $F_2^{\prime }$ with a
probability $p_2$, where
\[
F_2^{\prime }=F_2-kT_2\ln p_2
\]
and $T_2$ is the temperature of the second system.

\subsubsection{The Fluctuation Probability Relation}

Equating these we reach the essential result\footnote{ For the
Popper-Szilard Engine, this gives us $P_1=\left( \frac 12\right)
^{\left( T_G/T_W\right) }$, which we saw in Chapter \ref{ch:szsm}
was the key relationship in the failure of the Engine.} of this
section, the fluctuation probability relation:

\begin{eqnarray}
(p_1)^{T_1}=(p_2)^{T_2} \label{eq:flucprob}
\end{eqnarray}

We are now going to examine a key consequence of this result:

\[
p_1>p_2
\]

only if

\[
T_1>T_2
\]

The probability of the second system to be spontaneously found in
the desired state is less than the probability of the original
fluctuation occurring, only if the second system is at a lower
temperature.

Let us consider what this means. We have some system, at
temperature $T_2$, and we wish to perform some action upon it,
that requires work. We wish to obtain this work from a thermal
fluctuation in another system, at temperature $T_1$.

Now, if $T_1>T_2$, we could simply connect a heat engine between
the two and reliably compress the second system without having to
bother with identifying what fluctuations were occurring in system
one (remember - although we are not considering it here, we will
have to introduce an auxiliary system to determine which
fluctuation has taken place in system one, and this auxiliary
suffers an increase in entropy). Unfortunately, if system one is
not at a higher temperature than system two, then the probability
of system two spontaneously being found in the desired state is at
least as high as the probability that the fluctuation occurs in
system one.

The most effective way of obtaining a desired result from thermal
fluctuations is to wait for the fluctuation to occur in the system
of interest, rather than in any other system. Other systems will
only give a higher probability of being able to achieve the
desired result if they are at a higher temperature than the system
of interest, and so can achieve the result more reliably by more
conventional methods, and without involving auxiliaries. So the
most effective means of boiling a kettle by thermal fluctuations
is to unplug it and wait for it to spontaneously boil. This is an
important result, which is perhaps not well appreciated. In
\cite{Cav90}, for example, it is suggested that it may be possible
to build a demon capable of
\begin{quote}"violating" the second
law by waiting for rare thermal fluctuations
\end{quote}
while from the opposite point of view in \cite{EN99} it is argued
\begin{quote}
the result assures us that over the longer term, no \ldots demon
can exploit this fluctuation. But it can make no such assurance
for the shorter term. Short term and correspondingly improbable
violations of the Second Law remain.
\end{quote}
The result we have obtained here suggests that there is nothing to
be gained even from waiting for such improbable fluctuations to
occur - as any objective we could achieve by exploiting such a
rare fluctuation would be more likely to occur spontaneously than
the fluctuation itself!

\subsection{Imperfect Resetting} \label{s:szsol2:2}

We will now combine the results just obtained, with those of
Section \ref{s:szsol1}. This will demonstrate the significance of
the fluctuation probability relationship, completing our
understanding of why the Popper-Szilard Engine must fail.

Let us recall some of the key features of the resetting of the
piston in Chapter \ref{ch:szqm} and \ref{ch:szsm}. There are two
weights, but only one is raised, depending upon which side of the
piston that the gas is initially located. This leaves a
correlation between the position of the raised and unraised
weights and the position of the piston. We attempted to make use
of this correlation to reset the piston, but found that the
thermal state of the weights themselves defeated this attempt. The
result was that a mean flow of heat would occur only in the
direction of hot to cold.

When work was extracted from the expansion of the subensemble it
was assumed that this was simply absorbed by a suitable work
reservoir, such as a raised weight. Note, however, that this
raising of a weight can equally well be regarded as the isothermal
compression of the weight system, once we take into account the
fact that the weight must itself be at some temperature. Having
noted that the raising of the weight may be regarded as an
isothermal compression, we see that the fluctuation relation above
applies and
\[
\left( P_W\right) ^{T_W}=\left( P_G\right) ^{T_G}
\]
For the Popper-Szilard Engine, $P_W=P_1$ and $P_G=\frac 12$. This
leads directly to the relationship in Equation \ref{eq:p1a}
\[
P_1=\left(\frac{1}{2}\right)^{\frac{T_G}{T_W}}
\]
We saw in Section \ref{s:szsm7} that this equation plays the key
role in ensuring that the mean flow of energy in the
Popper-Szilard Engine is in an entropy increasing direction,
regardless of the choice of $T_W$ and $T_G$.

We must now try to understand how this relationship enters into
the attempt to reset a general Maxwell Demon. The key is the
additional feature that the arrangement of the weights makes to
the standard Szilard Engine. This feature is that the work
extracted from the gas is used to compress the weights in a {\em
different} manner, depending upon which subensemble of the gas is
selected. A different weight is lifted, depending upon which side
of the piston the one-atom gas is located. This produces the
correlation between weights and piston states at the end of the
raising cycle, and it is this correlation that enables an
imperfect resetting to be attempted. We need to understand how the
relationship between the fluctuation probabilities ensures that
this correlation is just sufficiently imperfect to prevent a mean
flow of energy from the colder to the hotter heat bath.

To do this we must add a second system, at a second temperature,
to the analysis of Section \ref{s:szsol1}. When the auxiliary
draws energy from the expansion of the subensembles of the first
system, it uses it to compress the second system in such a way
that there is a correlation between the final state of the second
system and the final state of the auxiliary. This correlation will
be used to reset the state of the auxiliary, in an attempt to
complete the engine cycle.

If the first system is at a higher temperature, we will see the
auxiliary can be reset by a correlation to the compression of the
second system, allowing the engine cycle to continue. However,
this is a flow of energy from a hotter to colder heat bath, so is
in an entropy increasing direction.

When the transfer of energy is in an anti-entropic direction, the
correlation between the second system and the auxiliary will be
shown to be imperfect. This leaves a mixture, whose entropy
offsets the transfer of energy between the heat baths. If we
attempt to reset the auxiliary imperfectly, the consequences of
the resetting failing are determined by the unitarity of the
evolution operators. It is shown that this leads inevitably to a
reversal of the direction of operation of the engine.

We will calculate general expressions for the mean number of
cycles the engine spends in each direction, and the mean energy
transferred between the heat baths per cycle. This will allow us
to show, quite generally, that the mean flow of energy will always
be in an entropy increasing direction.

\subsubsection{Expansion and Compression} We start with the system
from which we wish to extract free energy. Assuming this system to
be in thermal equilibrium at some temperature $T_G$, it's density
matrix is separated into orthogonal subensembles
\[
\rho _G=p_A\rho _{GA}+p_B\rho _{GB}
\]
which have free energies which differ from the ensemble free
energy by $kT_G\ln p_A$ and $kT_G\ln p_B$. We will not be assuming
that the two subensembles occur with equal probability. This
differs from the Szilard Engine, but is necessary to ensure the
generality of the results.

To extract the maximum amount of free energy, we need to expand
each subensemble to occupy the entire space, isothermally, leaving
it in the state $\rho _G$. We use the energy extracted from this
to compress a second system, at a temperature $T_W$ (if $p_A \neq
p_B$ then this second system will be compressed by different
amounts). If the equilibrium density matrix of the second system
is $\rho _W$, then $\rho _{WA}$ and $\rho _{WB}$ will represent
the density matrices it is isothermally compressed into by $\rho
_{GA}$ and $\rho _{GB}$, respectively. From the fluctuation
probability relationship, the $\rho _{WA}$ and $\rho _{WB}$
density matrices would occur{\em spontaneously} in $\rho _W$ with
probabilities $p_\alpha =\left( p_A\right) ^\tau $ and $p_\beta
=\left( p_B\right) ^\tau $ where $\tau =T_G/T_W$. We may write the
initial density matrix of the second system in two different ways:
\begin{eqnarray*}
\rho_W&=&p_\alpha \rho_{WA}+(1-p_\alpha)\rho_{W \overline{A}} \\
\rho_W&=&p_\beta \rho_{WB}+(1-p_\beta)\rho_{W \overline{B}}
\end{eqnarray*}

As shown in Section \ref{s:szsol1} above, we must also employ an
auxiliary system, which is initially in a state $\proj{\pi_0}$.
This system is required as the initially orthogonal states $\rho
_{GA}$ and $\rho _{GB}$ cannot be mapped to the same space $\rho
_G$, while extracting free energy. We cannot use the second system
as the auxiliary, as we do not yet know if the states $\rho _{WA}$
and $\rho _{WB}$ can be made orthogonal. It is also helpful to
regard the auxiliary as representing the state of the pistons,
pulleys, and other mechanisms (such as demons and memory
registers, if they are considered necessary) by which the
subensembles of the first system are selected, and used to
compress the second system.

The initial evolution of the system is from
\[
\rho_1=\left\{ p_A\rho_{GA}+p_B\rho _{GB}\right\}
    \otimes \rho_W\otimes \proj{\pi_0}
\]
to
\[
\rho _2=\rho_G \otimes \left\{
    p_A\rho_{WA}\otimes \proj{\pi_A}+p_B\rho_{WB}\otimes \proj{\pi_B}
    \right\}
\]
through intermediate stages
\[
\rho _1^{\prime }=p_A\rho_{GA}(Y)\rho_{WA}(Y) \proj{\pi_A(Y)}
    +p_B\rho_{GB}(Y) \rho_{WB}(Y) \otimes \proj{\pi _B(Y)}
\]
where $Y$ is a parameter varying from $0$ to $1$, and
\begin{eqnarray*}
\proj{\pi_A(0)}&=&\proj{\pi _B(0)}=\proj{\pi_0} \\
\proj{\pi_A(1)}&=&\proj{\pi_A}\\
\proj{\pi_B(1)}&=&\proj{\pi_B} \\
\rho _{GA}(1) &=&\rho _{GB}(1) =\rho _G \\
\rho _{GA}(0) &=&\rho _{GA} \\
\rho _{GB}(0) &=&\rho _{GB} \\
\rho _{WA}(0) &=&\rho _{WB}(0) =\rho _W \\
\rho _{WA}(1) &=&\rho _{WA} \\
\rho _{WB}(1) &=&\rho _{WB}
\end{eqnarray*}
In the process of this evolution, either $-kT_G\ln p_A$ or
$-kT_G\ln p_B$ energy is drawn from a heat bath at $T_G$.

The Hilbert space $\Gamma _G$ of the first system can be
partitioned into complementary subspaces as
\begin{eqnarray*}
\Gamma_G&=&\Gamma_{GA}(Y) \oplus \Gamma_{G \overline{A}}(Y) \\
&=&\Gamma _{GB}(Y) \oplus \Gamma _{G\overline{B}}(Y)
\end{eqnarray*}
where $\Gamma _{GA}(Y) $ is the space occupied by the density
matrix $\rho _{GA}(Y) $ etc.

The Hilbert space $\Gamma _W$ of the second system has a more
complicated partition. Let $\Gamma _{WA}(Y) $ be the subspace
occupied by the density matrix $\rho _{WA}(Y)$, $\Gamma _{WB}(Y)$
the subspace occupied by $\rho _{WB}(Y)$ and $\Gamma _{WAB}(Y)$ is
the subspace of the overlap between these two, then
\[
\Gamma_W=\Gamma_{WA}^{\prime}(Y) \oplus
    \Gamma_{WB}^{\prime}(Y)\oplus \Gamma _{WAB}(Y)\oplus
         \Gamma_{W \overline{AB}}(Y)
\]
where
\begin{eqnarray*}
\Gamma_{WA}(Y)&=&\Gamma_{WA}^{\prime}(Y) \oplus \Gamma_{WAB}(Y) \\
\Gamma_{WB}(Y)&=&\Gamma_{WB}^{\prime}(Y) \oplus \Gamma_{WAB}(Y)
\end{eqnarray*}
while $\Gamma _{W\overline{AB}}\left( Y\right) $ is the space
occupied by neither density matrix. The complementary subspaces
are
\begin{eqnarray*}
\Gamma_{W\overline{A}}(Y)&=&
    \Gamma_{WB}^{\prime}(Y) \oplus \Gamma_{W\overline{AB}}(Y) \\
\Gamma_{W\overline{B}}(Y)&=&
    \Gamma_{WA}^{\prime}(Y) \oplus \Gamma_{W\overline{AB}}(Y)
\end{eqnarray*}
When $Y=1$ we will simply refer to $\Gamma _{WA}$, $\Gamma
_{WA}^{\prime }$ etc. Projectors onto the subspaces are denoted by
$P_{WA}$, $P_{GA}$ and so forth.

To ensure the isothermal expansion is optimal, the systems have
internal Hamiltonians conditional upon discrete $Y_n$ states of
the auxiliary system
\begin{eqnarray*}
H_G &=&\sum_n\proj{\pi_A(Y_n)} \left\{
        H_{GA}(Y_n)+H_{G\overline{A}}(Y_n) \right\} \\
    &&+\proj{\pi_B(Y_n)}\left\{
        H_{GB}(Y_n)+H_{G\overline{B}}(Y_n) \right\} \\
H_W &=&\sum_n\proj{\pi_A(Y_n)}\left\{
        H_{WA}(Y_n)+H_{W\overline{A}}(Y_n) \right\} \\
    &&+\proj{\pi_B(Y_n)}\left\{
        H_{WB}(Y_n)+H_{W\overline{B}}(Y_n) \right\}
\end{eqnarray*}
where $H_{W\overline{A}}(Y_n)$ represents the Hamiltonian for the
subspace $\Gamma_{W\overline{A}}$ (complementary to the subspace
occupied by $\rho _{WA}\left( Y\right) $ ) and so on. When the
auxiliary is in the state $\proj{\pi_A(Y_n)}$, then transitions
between states in $H_{GA}\left( Y_n\right) $ and states in
$H_{G\overline{A}}\left( Y_n\right) $ are forbidden, and similarly
for $H_{WA}\left( Y\right)$, $H_{GB}\left( Y\right) $ and
$H_{WB}\left( Y\right)$. As compression and expansion takes place
isothermally, the subensembles are equilibrium density matrices
for their respective subspaces.

\subsubsection{Perfect Correlation}

If $T_G\geq T_W$ then
\[
p_\alpha +p_\beta \leq 1
\]
This means that $\Gamma _{WA}$ and $\Gamma _{WB}$ can be
non-overlapping, so that $\Gamma_{WAB}=0$, and the density
matrices $\rho_{WA}$ and $\rho_{WB}$ can be orthogonal.

If we use a reset operation which includes
\[
U_{r1}=\kb{\pi_0}{\pi_A}P_{WA}+\kb{\pi_0}{\pi_B}P_{WB}+\ldots
\]
where $P_{WA}$ is the projector onto $\Gamma_{WA}$, and $P_{WB}$
onto $ \Gamma _{WB}$, then we can reset the auxiliary state to
$\proj{\pi_0}$ and begin a new cycle, with perfect accuracy.

Restoring the auxiliary will make the second system internal
Hamiltonian $H_W(0)$, which has the equilibrium density matrix
$\rho_W$. This leads to a dissipation of the notional free energy,
$-kT_W\ln p_\alpha =-kT_G \ln p_A$ from $\rho _{WA}$, with
probability $p_A$, and dissipation of $-kT_W\ln p_\beta =-kT_G \ln
p_B $ from $\rho _{WB}$ with probability $p_B$. The mean
dissipation of notional free energy is then
\[Q=-kT_G(p_A\ln p_A+p_B\ln p_B)
\]
which equals the heat drawn from the $T_G$ heat bath. In other
words, a quantity of heat $Q$ can be reliably and continuously
drawn from one heat bath at $T_G$ and deposited at a colder heat
bath at $T_W$. This simply represents a flow of heat from the
hotter to the colder heat bath, and so presents no particular
problem for thermodynamics.

\subsubsection{Imperfect Correlation}

We now turn to the more interesting case, where the second system,
which is initially receiving energy, is at a higher temperature
than the first system, $T_W>T_G,$ and so
\[
p_\alpha +p_\beta >1
\]
In this case the subspace occupied by $\rho_{WA}$ and that
occupied by $\rho_{WB}$ will be overlapping. The projectors
$P_{WA}$ and $P_{WB}$ in $U_{r1}$ will not be orthogonal so the
operation $U_{r1}$ is no longer unitary.

To reduce the overlap, $\rho_{WA}$ and $\rho_{WB}$ should leave no
portion of the Hilbert space unoccupied, so that $\Gamma
_{W\overline{AB}}=0$ and
\[
\Gamma_W=\Gamma_{WA}^{\prime}\oplus
    \Gamma_{WB}^{\prime} \oplus \Gamma_{WAB}
\]
The probabilities of an equilibrium density matrix $\rho _W$ being
found in these subspaces are $p_\alpha ^{\prime }$, $ p_\beta
^{\prime }$ and $p_{\alpha \beta }$, with $p_\alpha ^{\prime
}+p_\beta ^{\prime }+p_{\alpha \beta }=1$, so that
\begin{eqnarray*}
\rho_{W}&=& p_\alpha^\prime \rho_{WA}^\prime
    +p_\beta^\prime \rho_{WB}^\prime+p_{\alpha \beta} \rho_{WAB}\\
\rho_{WA} &=&\left(1-\frac{p_{\alpha \beta}}{p_\alpha}
        \right)\rho_{WA}^{\prime}
    +\left(\frac{p_{\alpha \beta}}{p_\alpha}\right) \rho _{WAB} \\
\rho_{WB} &=&\left(1-\frac{p_{\alpha \beta}}{p_\beta}
        \right)\rho_{WB}^{\prime }
    +\left(\frac{p_{\alpha \beta}}{p_\beta}\right) \rho _{WAB}
\end{eqnarray*}

Using $\tau=T_G/T_W$, the probabilities are related by
\begin{eqnarray*}
p_\alpha &=&\left( p_A\right) ^\tau \\
p_\beta &=&\left( p_B\right) ^\tau \\
p_{\alpha \beta } &=&p_\alpha +p_\beta -1 \\
p_\alpha ^{\prime } &=&p_\alpha -p_{\alpha \beta }=1-p_\beta \\
p_\beta ^{\prime } &=&p_\beta -p_{\alpha \beta }=1-p_\alpha
\end{eqnarray*}

Now, if the second system is located in either $\Gamma
_{WA}^{\prime }$ or $\Gamma_{WB}^{\prime }$, then there is a
correlation between that system and the auxiliary system. The
auxiliary system may be restored to it's initial state
$\proj{\pi_0}$, by a correlated unitary operation.

However, if the second system is located in $\Gamma _{WAB}$, the
auxiliary may be in either position, and there is no correlation.
The resetting is now not possible. This is equivalent to the
situation in the Popper-Szilard Engine when both weights are
located above the shelf height.

As we can only unambiguously identify the state of the auxiliary
from the state of the second system when the second system is
located in a non-overlapping portion of the Hilbert space, we
choose to reset the auxiliary when the second system is in $\Gamma
_{WA}^{\prime}$ or $\Gamma_{WB}^{\prime}$, but perform no
resetting when the second system is in $\Gamma _{WAB}$. The
conditional unitary operation for this is
\[
U_{r2}=P_{WA}^{\prime }U_{RA}+P_{WB}^{\prime }U_{RB}+P_{WAB}U_{AB}
\]
where $P_{WA}^{\prime }$ etc. are projection operators onto the
relevant subspace of the second system, and the $U_{RA}$ are
unitary operators \footnote{Similar to the $U_{RES}$ in Section
\ref{s:szqm5} there is some flexibility in the choice of $U_{RA}$,
$U_{RB}$, and $U_{AB}$, so the ones chosen here are not the only
ones possible. However, they are the simplest choice, and a more
complicated expression would not essentially affect the outcome.}
on the auxiliary space of the form
\begin{eqnarray*}
U_{RA} &=&\kb{\pi_0}{\pi_A}+\kb{\pi_A}{\pi_0}+\proj{\pi_B} \\
U_{RB} &=&\kb{\pi_0}{\pi_B}+\kb{\pi_B}{\pi_0}+\proj{\pi_A} \\
U_{AB} &=&\proj{\pi_0}+\proj{\pi_A}+\proj{\pi_B}
\end{eqnarray*}

When the second system can be reliably correlated to the state of
the auxiliary, these operators will restore the auxiliary to its
initial state. Following this, the notional free energy of the
subensemble is dissipated, and a net transfer of heat from the
$T_G$ to the $T_W$ heat bath has taken place. However, in those
cases where the second system is found in $\Gamma _{WAB},$ the
system has not been restored to it's initial condition.

\subsubsection{Raising Cycle}

We can summarise the evolution so far, which we shall call the
'raising cycle' as it corresponds to the raising cycle of the
Szilard Engine:
\begin{eqnarray*}
\rho _1 &=&\rho _G\Pi _0\rho _W
    =\left\{ p_A\rho _{GA}+p_B\rho_{GB}\right\} \Pi _0\rho _W \\
\rho _2 &=&p_A\rho _G\Pi _A\rho _{WA}+p_B\rho _G\Pi _B\rho _{WB} \\
    &=&p_A\rho_G\Pi _A\left\{
         \left(1-\frac{p_{\alpha \beta}}{p_\alpha}\right)\rho_{WA}^{\prime}
         +\left( \frac{p_{\alpha \beta}}{p_\alpha}\right)\rho_{WAB}\right\} \\
    &&+p_B\rho_G\Pi_B\left\{
        \left(1-\frac{p_{\alpha \beta}}{p_\beta}\right)\rho_{WB}^{\prime}
        +\left(\frac{p_{\alpha \beta}}{p_\beta}\right)\rho_{WAB}\right\} \\
\rho_3 &=&\rho_G\Pi_0\left\{
    p_A\left(1-\frac{p_{\alpha \beta}}{p_\alpha}\right)\rho_{WA}^{\prime}
    +p_B\left(1-\frac{p_{\alpha \beta}}{p_\beta}\right)\rho_{WB}^{\prime}\right\} \\
    &&+\rho_G\left\{p_A\frac{p_{\alpha \beta}}{p_\alpha}\Pi_A
        +p_B\frac{p_{\alpha \beta}}{p_\beta}\Pi_B\right\}\rho_{WAB} \\
\rho_4 &=&\left\{
        p_A\left(1-\frac{p_{\alpha \beta}}{p_\alpha}\right)
        +p_B\left( 1-\frac{p_{\alpha \beta}}{p_\beta}\right)
    \right\} \rho _G\Pi_0\rho _W \\
    &&+\rho_G\left\{
        p_A\frac{p_{\alpha \beta}}{p_\alpha}\Pi_A\rho_{WA}
        +p_B\frac{p_{\alpha \beta}}{p_\beta}\Pi_B\rho_{WB}\right\}
\end{eqnarray*}

The initial density matrix is $\rho _1$, in equilibrium. The first
stage correlates the auxiliary to the subensembles of system one,
extracts free energy from their conditional expansion, and uses
the same free energy to compress the second system. However, the
compression of the second system is also conditional upon the
auxiliary, so that at the end of the expansion-compression stage
the auxiliary and the second system are correlated, in density
matrix $\rho_2$. An amount of heat equal to $Q=-kT_G\left( p_A\ln
p_A+p_B\ln p_B\right)$ has been drawn from the $T_G$ heat bath,
and used to compress the second system.

The next stage uses the operator $U_{r2}$. This utilises the
correlation between the auxiliary and the second system to restore
the auxiliary to it's initial state. When the second system is
located in the $\Gamma_{WAB}$ subspace, however, the imperfect
correlation does not allow the auxiliary to be reset. The final
state of the system is $\rho_3$.

Finally, the contact with the $T_W$ heat bath causes the second
system subensembles to thermally expand throughout their
accessible Hilbert space, leading to $\rho _4$.

With a probability given by
\[
p_C=\left\{p_A\left(1-\frac{p_{\alpha \beta }}{p_\alpha }\right)
    +p_B\left( 1-\frac{p_{\alpha \beta }}{p_\beta }\right) \right\}
\]
the system will be ready to start another raising cycle. However,
in the final line of $\rho _4$ we find that the system has a
probability of not being restored, with probability
\[
p_R=p_{\alpha
\beta}\left(\frac{p_A}{p_\alpha}+\frac{p_B}{p_\beta}\right)
\]

\subsubsection{Lowering Cycle}

We now need to consider what must happen to the unrestored system
at the start of a new cycle. We must be very careful when doing
this. As noted towards the end of Section \ref{s:szsol1}, if the
auxiliary is in the wrong state, the expansion/compression unitary
operation may attempt to compress a density matrix into a zero
volume. In such situations the operation of the engine would break
down. Avoiding such situations occurring constrains the form of
the operation upon the reversed cycle. We must always be sure that
the energy extracted from one system is equal to the energy added
to the other.

The conditional internal Hamiltonians $H_G$ and $H_W$ shows that
the states consistent with the different positions of the
auxiliary are
\begin{eqnarray*}
\rho_{GA}\Pi _0\rho _W && \rho _{GB}\Pi _0\rho _W \\
\rho _G\Pi _A\rho _{WA} && \rho _G\Pi _B\rho _{WB} \\
\rho _G\Pi _A\rho_{W\overline{A}}
    &&\rho _G\Pi _B\rho_{W\overline{B}}
\end{eqnarray*}
The expansion/compression operation must map the space of
$\rho_{GA}\Pi_0\rho_W$ to $\rho_G\Pi_A\rho_{WA}$ and
$\rho_{GB}\Pi_0\rho W$ to $\rho_G\Pi_B\rho_{WB}$.  The states
$\rho _G\Pi _A\rho _{W\overline{A}}$ and $\rho _G\Pi _A\rho
_{W\overline{B}}$ are inaccessible, and would lead to a breakdown
of the engine, should they occur.

The unitary operation for the expansion and compression phase must
therefore map the space $\rho _G\Pi _A\rho _{WA}$ onto $\rho
_{GA}\Pi _0\rho _W$ and $\rho _G\Pi _B\rho _{WB}$ onto $\rho
_{GB}\Pi_0\rho _W$, and then allow $\rho _{GA}$ and $\rho _{GB}$
to dissipate into $\rho _G$ (which corresponds to the piston being
removed from the Szilard box) when the auxiliary system is reset.
This is a 'lowering cycle' where the expansion of $\rho _{WA}$ or
$\rho _{WB}$ is used to compress $\rho _G$, in a reverse direction
to the 'raising cycle'.

The energy $Q_A=-kT_G\ln p_A$ is transferred to the first system,
on a lowering $A$-cycle' and $Q_B=-kT_B\ln p_B$ on a 'lowering
$B$-cycle'. If we follow the stages of the 'lowering $A$-cycle'
for a system initially in state $\rho _G\Pi _A\rho _{WA}$ we have
\begin{eqnarray*}
\rho_1^{\prime} &=& \rho_G\Pi_A\rho_{WA} \\
\rho_2^{\prime} &=& \rho_{GA}\Pi_0\rho_W \\
    &=&\rho_{GA}\Pi_0\left\{
        p_\alpha^{\prime}\rho_{WA}^{\prime}
        +p_\beta^{\prime}\rho_{WB}^{\prime}
        +p_{\alpha \beta}\rho_{WAB}\right\} \\
\rho_3^{\prime} &=& \rho_G\left\{
    p_\alpha^{\prime}\Pi_A\rho_{WA}^{\prime}
    +p_\beta^{\prime}\Pi_B\rho_{WB}^{\prime}\right\}
    +p_{\alpha \beta}\rho_{GA}\Pi_0\rho_{WAB} \\
\rho_4^{\prime} &=& \rho_G\left\{
    p_\alpha^{\prime}\Pi_A\rho_{WA}
    +p_\beta^{\prime}\Pi_B\rho_{WB}\right\}
    +p_{\alpha \beta}\rho_{GA}\Pi_0\rho_W
\end{eqnarray*}

These follow the same stages as the 'raising cycle' above.
Initially, the density matrix $\rho_1^\prime$ compresses the first
system, through the expansion of the second, leaving the system in
state $\rho_2^\prime$. Now we must apply the reset operation
$U_{r2}$, which leaves the system in state $\rho_3^\prime$.
Finally, contact with the $T_W$ heat bath leads to state
$\rho_4^\prime$.

Now the probability of a 'reversal' back onto the 'raising cycle'
is $p_{\alpha \beta }$. For a system initially in $\rho _G\Pi
_B\rho _{WB},$ the dissipation of $\rho _{GB}$ to $\rho _G$
between $\rho _2^{\prime }$ and $\rho _3^{\prime }$ leads to the
same probability of reversing, only now starting the raising cycle
on $\rho _{GB}\Pi _0\rho _W$.

This completes the optimal design for attempting to imperfectly
reset the auxiliary system, using correlations with the second
system, and the effect of the imperfect resetting. We have found
that, quite generally, the same considerations that constrained
the design of the Popper-Szilard Engine have arisen.

The compression of the second system, by expansion of subensembles
in the first system, is governed by the fluctuation probability
relation
\[
\left(p_G\right)^{T_G}=\left(p_W\right)^{T_W}
\]
When the flow of energy is in an anti-entropic direction, then
$\tau=\frac{T_G}{T_W}<1$. The compression of the second system is
into subensembles $\rho_{W\alpha}$ which would spontaneously occur
with probabilities $p_{W\alpha}$. This gives
\begin{equation}
\sum_\alpha p_{W\alpha}=
    \sum_\alpha \left(p_{G\alpha}\right)^\tau>1 \label{eq:pover}
\end{equation}
as $\left(p_{G\alpha}\right)^\tau>p_{G\alpha}$ and $\sum_\alpha
p_{G\alpha}=1$. There must be overlaps between the compressed
subensembles of the second system. Should the second system be in
one of the non-overlapping regions of the Hilbert space, then
there will be a correlation between the auxiliary and the second
system that allows the auxiliary to be reset. If, instead, the
second system is located in one of the overlapping regions, then
there is more than one auxiliary state possible, and a unitary
resetting operation does not exist.

The imperfect correlations lead to a failure to reset the
auxiliary, so we must consider the effect of starting a new cycle
with the auxiliary in the other states. The constraints upon this
is that the evolution of the system be described by a unitary
operation and no work is performed upon the system. When the
auxiliary has not been reset this forces the engine to reverse
direction.

\subsubsection{Average length of cycles}

We have shown that the engine must switch between 'raising' and
'lowering' cycles. We now need to demonstrate that this switching
will lead to a mean flow of heat in the entropy increasing
direction. There are two factors which need to be evaluated to
calculate this: the mean number of raising or lowering cycles
before a reversal takes place, and the average amount of energy
transferred per cycle.

The average length of a complete run of raising or lowering cycles
is simply given by the reciprocal of the probability of it
reversing. The total probability of reversal from a raising cycle
is
\begin{eqnarray*}
P_R &=&p_A\frac{p_{\alpha \beta }}{p_\alpha }
    +p_B\frac{p_{\alpha\beta }}{p_\beta } \\
    &=&p_{\alpha \beta }\left(
        \frac{p_A}{p_\alpha }+\frac{p_B}{p_\beta }\right) \\
    &=&\left(\left(p_A\right)^\tau+\left(p_B\right)^\tau-1\right)
        \left(\left(p_A\right)^{1-\tau}+\left(p_B\right)^{1-\tau}\right)
\end{eqnarray*}
while the probability of reversal from a lowering cycle is
\begin{eqnarray*}
P_L &=&p_{\alpha \beta } \\
    &=&\left(\left(p_A\right)^\tau+\left(p_B\right)^\tau-1\right)
\end{eqnarray*}

The mean number of cycles for the raising and lowering cycles,
$N_R$ and $N_L$ are then related by

\[
N_L=\left( \left( p_A\right) ^{1-\tau }+\left( p_B\right) ^{1-\tau
}\right) N_R
\]

This is the essential relationship between the relative
temperatures of the systems, and the mean length of time spent on
the raising and lowering cycles.

As $0 \leq 1-\tau \leq 1$ then we have
\[
1 \leq \left( \left( p_A\right) ^{1-\tau }+\left( p_B\right)
^{1-\tau }\right) \leq 2
\]

This produces the result that

\[
N_L \geq N_R
\]

so that the engine will, on average, spend more cycles
transferring energy from the hotter to the colder heat bath, on
the lowering cycle, than it will transferring energy in the from
the colder to the hotter, on the raising cycle. The engine spends
a proportion
\[
\frac{N_L}{N_L+N_R}=\frac{(p_A)^{1-\tau}+(p_B)^{1-\tau}}
    {(p_A)^{1-\tau}+(p_B)^{1-\tau}+1}
\]
of the time on the lowering cycle, and the remaining
\[
\frac{N_R}{N_L+N_R}=\frac{1}{(p_A)^{1-\tau}+(p_B)^{1-\tau}+1}
\]
of the time on the raising cycle. The limit that $T_G \approx T_W$
leads to $N_L=2N_R$. This spends one-third of the time on a
raising cycle, and two-thirds of the time on a lowering cycle In
the limit $T_G \ll T_W$, the engine approaches half the time on
each cycle. Surprisingly, as the temperature difference increases,
the proportion of the time on the anti-entropic cycle goes up.
This is because with large temperature differences, both cycles
are highly likely to go into reverse, until at the limit the
auxiliary is never reliably reset and the engine switches with
certainty between the two cycles.

It is interesting to note that if $T_G$ is only slightly lower
than $T_W$, the initial run of raising cycle can last for a very
long time (both $N_L$ and $N_R$ become very large). However, the
apparent entropy increase implied by this transfer of energy from
the colder to the hotter is very small, precisely because the
temperature difference is so small, and will be more than offset
by the increase in entropy that comes about from the small
probability of the cycle reversing, and the effect this has on the
mixing entropy of the auxiliary system. Once a reversal has
occurred, of course, the probability is that the Engine will stay
on the lowering cycle, for an even longer period of time.

\subsubsection{Mean energy per cycle}

To complete the analysis, we must calculate the mean energy per
cycle. It is not generally the case that the same mean amount of
energy is transferred on a lowering cycle as on a raising cycle.

On a raising cycle, the mean energy transfer is
\[
Q_R=-kT_G\left( p_A\ln p_A+p_B\ln p_B\right)
\]

On a lowering $A$-cycle, the energy transfer is $Q_A=-kT_G\ln p_A$
and on a lowering $B$-cycle it is $Q_B=-kT_G\ln p_B,$ but the
probabilities of a lowering cycle being an $A$ or $B$ cycle are
{\em not} $p_A$ and $p_B$. The mean energy transfer will therefore
be different to a raising cycle.

For the initial lowering cycle, which follows from a reversal from
the raising cycle, the probabilities of the $A$ or $B$ cycles are
\begin{eqnarray*}
p_{A1} &=&\frac{p_A\frac{p_{\alpha \beta}}{p_\alpha}}
    {p_A\frac{p_{\alpha \beta }}{p_\alpha}
    +p_B\frac{p_{\alpha \beta }}{p_\beta }} \\
    &=&\frac{p_A p_\beta }{p_Ap_\beta +p_Bp_\alpha } \\
    &=& \frac{\left(p_A\right)^{1-\tau}}
        {\left(p_A\right)^{1-\tau}+\left(p_B\right)^{1-\tau}} \\
p_{B1} &=&\frac{p_B\frac{p_{\alpha \beta}}{p_\beta}}
    {p_A\frac{p_{\alpha \beta }}{p_\alpha}
    +p_B\frac{p_{\alpha \beta }}{p_\beta}} \\
    &=&\frac{p_B p_\alpha }{p_A p_\beta+p_B p_\alpha } \\
    &=& \frac{\left(p_B\right)^{1-\tau}}
        {\left(p_A\right)^{1-\tau}+\left(p_B\right)^{1-\tau}}
\end{eqnarray*}
while a continuation of the lowering cycle will give probabilities
\begin{eqnarray*}
p_{A2} &=&\frac{p_\alpha^{\prime}}
    {p_\alpha^{\prime}+p_\beta^{\prime }} \\
p_{B2} &=&\frac{p_\beta ^{\prime }}
    {p_\alpha ^{\prime }+p_\beta^{\prime }}
\end{eqnarray*}

The mean energy transfer on the first lowering cycle is then
\[
Q_1=-kT_G\left( p_{A1}\ln p_A+p_{B1}\ln p_B\right)
\]
and on subsequent lowering cycles
\[
Q_2=-kT_G\left( p_{A2}\ln p_A+p_{B2}\ln p_B\right)
\]

To calculate the mean energy transfer, per cycle, over the course
for a complete run of lowering cycles, we need to include both
these results. Any run of lowering cycles starts with one $Q_1$
cycle. If it continues, with probability $\left( p_\alpha ^{\prime
}+p_\beta ^{\prime }\right)$, then the mean energy per cycle after
that is $Q_2$. The probability of reversal is the same on all
cycles, so, if we are given that it does continue beyond the $Q_1$
cycle, then the mean number of $Q_2$ cycles will be $N_L.$ The
mean energy transferred over the course of an entire run of
lowering cycles will be
\[
Q_1+\left(p_\alpha ^{\prime }+p_\beta^{\prime }\right)
     \left(N_LQ_2\right)
\]

As the mean number of cycles is still $N_L$, the mean energy
transfer, per cycle is
\begin{eqnarray*}
Q_L &=&\frac{Q_1+\left(p_\alpha^{\prime }
    +p_\beta ^{\prime}\right) \left(N_LQ_2\right) }{N_L} \\
    &=&p_{\alpha \beta }Q_1
        +\left( p_a^{\prime }+p_\beta ^{\prime}\right) Q_2 \\
\frac{Q_L}{-kT_G} &=&\left( p_{\alpha \beta }
    \frac{\left(p_A\right)^{1-\tau}}
        {\left(p_A\right)^{1-\tau}+\left(p_B\right)^{1-\tau}}
    +p_\alpha ^{\prime }\right) \ln p_A
    +\left( p_{\alpha \beta }
    \frac{\left(p_B\right)^{1-\tau}}
        {\left(p_A\right)^{1-\tau}+\left(p_B\right)^{1-\tau}}
    +p_\beta ^{\prime}\right) \ln p_B
\end{eqnarray*}
which can be rearranged to give
\[
Q_L=-\frac{kT_G \left(
    \left(p_A-p_B+(p_B)^{1-\tau}\right) \ln p_A
    +\left(p_B-p_A+(p_A)^{1-\tau}\right) \ln p_B \right)}
    {(p_A)^{1-\tau}+(p_B)^{1-\tau}}
\]

\paragraph{Long Term Mean}

We are now in a position to complete the analysis of the mean heat
flow for the imperfect resetting of the generalised Szilard
Engine. The mean flow of energy, per cycle, from the $T_G$ heat
bath to the $T_W$ heat bath is

\begin{eqnarray*}
Q &=&\frac{N_R Q_R-N_L Q_L}{N_R+N_L} \\
    &=&kT_G\frac{\left((p_B)^{1-\tau }-p_B\right) \ln p_A
        +\left((p_A)^{1-\tau }-p_A\right) \ln p_B}
    {(p_A)^{1-\tau}+(p_B)^{1-\tau}+1}
\end{eqnarray*}

We know that $(1-\tau) \leq 1$ so
\begin{eqnarray*}
\left(p_A\right)^{1-\tau }>p_A \\
\left(p_B\right)^{1-\tau }>p_B
\end{eqnarray*}

The value of $Q$ is always negative\footnote{In the limit of $T_G
\ll T_W$ the value approaches zero as the engine reverses between
cycles with certainty}. The mean flow of energy must go from the
hotter heat bath to the colder heat bath.

This generalises the conclusion to Chapters \ref{ch:szqm},
\ref{ch:szsm} and \ref{ch:szth} and is independant of any
particular physical model. We have demonstrated than, even when we
attempt to correlate an auxiliary to a second system, the
correlation must always fail sufficiently often to prevent a long
term anti-entropic energy flow.

\subsubsection{Summary}
We have seen that, when $T_G < T_W$ it is impossible to create a
perfect correlation between the auxiliary and the subensembles of
the $T_W$ system. The requirement that the resetting operation be
unitary then leads to the engine switching from a 'raising' to a
'lowering' cycle. However, this also leads to a 'lowering' cycle
switching back to a 'raising' cycle.

The key result we have shown here, is that the engine must, in the
long run, transfer more energy on the 'lowering' cycles, than on
the 'raising' cycles. The reason for this lies in the average
length of the cycles. On the entropic lowering cycle, the
probability of reversal is
\[p_{\alpha \beta}\]
which comes from the subspace $\Gamma_{WAB}$, representing the
overlap between the compressed subensembles. This is the
probability of finding an equilibrium system in the overlap
region, out of the entire Hilbert space $\Gamma_{W}$.

On the anti-entropic raising cycle, the probability of reversal
depends upon which subensemble was selected. With probability
$p_A$ the subensemble was $\rho_{GA}$. In this case the reversal
occurs if the second system is located within $\Gamma_{WAB}$, but
now it is out of the compressed subspace $\Gamma_{WA}$. The
probability
\[\frac{p_{\alpha \beta}}{p_\alpha}\]
must be higher than the probability of reversal from the raising
cycle.

The same will be true had the subensemble selected been
$\rho_{GB}$, which has probability $\frac{p_{\alpha
\beta}}{p_\beta}$. Clearly, therefore, the mean reversal
probability
\begin{eqnarray*}
p_A\left(\frac{p_{\alpha \beta}}{p_\alpha}\right)+
    p_B\left(\frac{p_{\alpha \beta}}{p_\beta}\right)
&=&    p_{\alpha \beta}\left(\frac{p_A}{p_\alpha}
    +\frac{p_B}{p_\beta}\right)
\end{eqnarray*}
will always be at least as large as the reversal probability for
the lowering cycle. It is therefore unavoidable that the engine
will spend more time, in the long run, on the lowering cycles, and
so will lead to a long term energy flow from the hotter to the
colder heat bath.

\subsection{The Carnot Cycle and the Entropy Engine} \label{s:szsol2:3}
We saw that when $T_G \geq T_W$ there was a perfect correlation
between the auxiliary and the second system, that could be used to
perfectly reset the auxiliary. However, this only leads to a
transfer of heat from the hotter to the colder heat bath.

In this Subsection we will see how we can extract work from the
second system, before the auxiliary is reset, without losing the
correlation. After the auxiliary is reset, we will discover that
this leads to heat engine operating at Carnot Cycle efficiency. We
will then apply the same method to the case where $T_G < T_W$. By
performing work upon the second system, we will show that the
imperfect correlation can be made perfect, allowing the auxiliary
to be reset without error. Again, when we take the complete cycle
of this, we will have a heat pump, operating at the Carnot Cycle
efficiency, so we still will not have succeeded in violating the
second law of thermodynamics. The resulting cycle is a form of the
Entropy Engine considered in Appendix \ref{ap:szfree}.

\subsubsection{$T_G \geq T_W$}

As $p_\alpha+p_\beta \leq 1$ there is no overlap between the
subspaces $\Gamma_{WA}$ and $\Gamma_{WB}$, so we can write
\[
\Gamma_{W}=\Gamma_{WA} \oplus \Gamma_{WB}
    \oplus \Gamma_{W\overline{AB}}
\]

The space $\Gamma_{W\overline{AB}}$ represents an unoccupied
portion of the Hilbert space. By allowing the second system to
isothermally expand into this space, we can extract some energy as
work, without creating an overlap and so without losing the
correlation with the auxiliary.

To do this, the two subensembles $\rho _{WA}$ and $\rho _{WB}$
must isothermally expand to $\rho _{WA}^{\prime \prime }$ and
$\rho _{WB}^{\prime \prime }$ respectively. These density matrices
spontaneously occur with probabilities $p_\alpha ^{\prime \prime
}$ and $p_\beta ^{\prime \prime }$ in the equilibrium density
matrix $\rho _W$.

Provided the expansion leaves $p_\alpha ^{\prime \prime }+p_\beta
^{\prime \prime }\leq 1$, we do not need to have any overlap
between $\rho _{WA}^{\prime \prime }$ and $\rho _{WB}^{\prime
\prime }$, and we will still have perfect correlation with the
auxiliary, and we will be able to reset the system. The expansion
of the system has allowed us to extract some of the heat flow from
the hotter to the colder bath, and turn it into useful work.

The most energy can be extracted, without allowing the density
matrices to overlap, will be when $p_\alpha ^{\prime \prime
}+p_\beta ^{\prime \prime }=1$, so that
\[
\rho _W=p_\alpha ^{\prime \prime }\rho _{WA}^{\prime \prime
}+p_\beta ^{\prime \prime }\rho _{WB}^{\prime \prime }
\]

After the second system expands and the auxiliary is reset, the
second system density matrix is
\[
\rho_W^{\prime \prime}=
    p_A \rho _{WA}^{\prime \prime }+p_B\rho _{WB}^{\prime \prime }
\]
The second system will then return to the equilibrium distribution
$\rho_W$.

Using the results in Section \ref{s:szsol1}, there is a
dissipation of notional free energy into the $T_W$ heat bath of
\[
\frac{\Delta F_D}{kT_W}=-\left( p_A\ln p_\alpha ^{\prime \prime
}+p_B\ln p_\beta ^{\prime \prime }\right)
\]
and mean work extracted of
\begin{eqnarray*}
\Delta F_G &=&kT_W\left( p_A\ln \left( \frac{p_\alpha ^{\prime \prime }}{%
p_\alpha }\right) +p_B\ln \left( \frac{p_\beta ^{\prime \prime }}{p_\beta }%
\right) \right) \\
&=&-kT_G\left( p_A\ln p_A+p_B\ln p_B\right) +kT_W\left( p_A\ln
\left( p_\alpha ^{\prime \prime }\right) +p_B\ln \left( p_\beta
^{\prime \prime }\right) \right)
\end{eqnarray*}
The first term in this is simply the heat extracted from the $T_G$
heat bath. The second term is the notional dissipation, and has a
minimum value (subject to $p_\alpha ^{\prime \prime }+p_\beta
^{\prime \prime }\leq 1$) when $p_\alpha ^{\prime \prime }=p_A,$
and $p_\beta ^{\prime \prime }=p_B$. This gives
\begin{eqnarray*}
\Delta F_G &\leq &k\left( T_W-T_G\right) \left( p_A\ln p_A+p_B\ln
p_B\right)
\\
&\leq &-S\Delta T
\end{eqnarray*}
where $S$ is the mixing entropy transferred from the system at
temperature $T_G$ to the system at temperature $T_W$.

This gives a heat engine efficiency of
\[
\frac{\Delta F_G}Q\leq 1-\frac{T_W}{T_G}
\]
which is in complete agreement with the efficiency of a Carnot
cycle.

\subsubsection{$T_G < T_W$}

We will now use the same approach for the case where the first
heat bath is colder than the second heat bath, and we have
extracted energy from the colder system to compress the hotter
system. As we saw above, the compression of the second system will
lead to an imperfect correlation with the auxiliary, as there will
be an overlap between the $\rho_{WA}$ and $\rho_{WB}$ density
matrices.

To remove the overlap, we must compress $\rho_{WA}$ and
$\rho_{WB}$ further, performing work upon the system, until they
are no longer overlapping. This will allow us to reset the
auxiliary system without error using $U_{r1}$ above. This will
lead to the density matrices $\rho _{WA}^{\prime \prime }$ and
$\rho _{WB}^{\prime \prime }$ as before, only now, as $p_\alpha
+p_\beta
>p_\alpha ^{\prime \prime }+p_\beta ^{\prime \prime }=1,$ the mean
work 'extracted'
\begin{eqnarray*}
\Delta F_G &=&kT_W\left( p_A\ln \left( \frac{p_\alpha ^{\prime \prime }}{%
p_\alpha }\right) +p_B\ln \left( \frac{p_\beta ^{\prime \prime }}{p_\beta }%
\right) \right) \\
\ &=&-kT_G\left( p_A\ln p_A+p_B\ln p_B\right) +kT_W\left( p_A\ln
\left( p_\alpha ^{\prime \prime }\right) +p_B\ln \left( p_\beta
^{\prime \prime }\right) \right)
\end{eqnarray*}
is negative, and is least negative when $p_\alpha ^{\prime \prime
}=p_A$ and $p_\beta ^{\prime \prime }=p_B$.

Re-expressing this as work, $W=-\Delta F_G$, required to pump heat
$Q=-kT_G\left( p_A\ln p_A+p_B\ln p_B\right)$ from a heat bath at
$T_G$ to a hotter heat bath at $T_W$, we have
\[
\frac WQ\geq \frac{T_W}{T_G}-1
\]
once again agreeing with the Carnot efficiency.

\subsection{Conclusion}

In Section \ref{s:szsol1} we examined how the mixing of
subensembles lead to an increase in entropy, and corresponding
reduction in free energy of the ensemble. We demonstrated that
this loss of free energy is because of the restriction of
unitarity upon the evolution operators. The optimal operations
cannot be applied to their respective subensembles, as this would
require mappings of orthogonal to non-orthogonal states. If an
auxiliary system is introduced, the optimal operators can by
applied, by a conditional interaction with the auxiliary system.
However, this leads to a compensating increase of the entropy of
the auxiliary system.

The two-weight Szilard Engine suggested that the work extracted
from the subensembles could be used to correlate a second system
to the auxiliary, and that this correlation could be used to reset
the auxiliary, if imperfectly. However, it was found that the
relationship $P_1=\left( \frac 12\right) ^{T_G/T_W}$ played a
critical role, preventing the correlation from being sufficient to
allow heat to flow in an anti-entropic direction. In this section
we have examined the origin of this, in terms of the free energy
subensemble formula (\ref{eq:freeprob})

\[
F_i=F-kT\ln p_i
\]

which leads to the probability fluctuation relationship
(\ref{eq:flucprob})

\[
\left(p_1\right)^{T_1}=\left(p_2\right)^{T_2}
\]

This relationship plays a key role in preventing the violation of
the statistical second law of thermodynamics. It is this
relationship that ensures that correlations are imperfect when the
heat flow would otherwise be anti-entropic. When we try to use an
imperfect resetting, this relationship then also guarantees that
the switching between raising and lowering cycles will always
prefer the lowering cycle.

The fluctuation probability relationship also ensures that thermal
fluctuations are ineffective as a means of performing work upon
other systems. Any objective, such as boiling a kettle, that could
be achieved through capturing a rare thermal fluctuation, will be
more likely to occur spontaneously, by unplugging it and leaving
it, or else could be achieved reliably without resort to
fluctuations.

Finally, when we attempt to improve the correlation with the
auxiliary, by performing work upon the second system, we find that
we recover a heat pump or heat engine operating at the Carnot
Cycle efficiency. It should be noted, however, that the cycle we
have here is not the same as the phenomenological Carnot Cycle,
using adiabatic and isothermal expansion and compression. At
several stages in this cycle we find key thermodynamic concepts,
such as the free energy, become undefined, as we have a correlated
mixture of systems at different temperatures. In fact, we have
here an example of the Entropy Engine, considered in Appendix
\ref{ap:szfree}. The origin of the work extracted is the transfer
of mixing entropy between systems at different temperatures.

\section{Alternative resolutions} \label{s:szsol3}

Having thoroughly investigated the physics of the quantum Szilard
Engine, we now wish to re-examine the arguments and resolutions
put forward by other authors, and explored in Chapter
\ref{ch:szmd}. We will use the simplest models possible to
demonstrate how these relate to our own conclusions. We will find
that, where these resolutions are not flawed, they are physically
equivalent to some aspect of our resolution, and so represent only
partial resolutions.

\subsection{Information Acquisition}\label{s:szsol3.1}

The first argument we will review will be that of Gabor and
Brillouin. We will examine this because, although, in it's
information theoretic form, it is no longer supported, it's
physical basis has been defended by opponents of the resolution
based upon Landauer's Principle. We will find that Gabor and
Brillouin did make unnecessary assumptions in their analysis, and
without these assumptions, their explanation of the resolution
does not hold. It will be instructive to examine the basis of this
when considering later arguments.

The key suggestion they made was that the demon was required "to
make some physical means of distinguishing between the gas
molecules" \cite{DD85} and that this physical means of acquiring
information inevitably lead to a dissipation of $kT\ln 2$ energy.
In the context of Szilard's Engine, it was the demon using a light
source to illuminate the location of the atom that would dissipate
the energy. Brillouin went on to argue that each elementary act of
information acquisition was associated with such a dissipation of
energy.

If we start by considering the physical connection between the
demon and the gas, we must consider three systems
\begin{itemize}
\item A gas, initially in a mixture of two subensembles $\rho
_G=\frac {1}{2}\left(\rho_G(A)+\rho_G(B) \right)$
\item A physical connection (such as a photon), initially in the unscattered state
$\rho_{Ph}(Un)$, but which will be scattered into a different
state, $\rho_{Ph}(Sc)$, if the gas is in the particular
subensemble $\rho _G(B)$.
\item the demon, initially in state $\rho_D(A)$, but which will
move into state $\rho_D(B)$ if it sees the photon in the scattered
state.
\end{itemize}
The system is initially in the state
\[
\rho_1=\frac{1}{2}\left(\rho_G(A)+\rho_G(B) \right)
     \rho_{Ph}(Un)\rho_D(A)
\]
If the photon encounters the state $\rho_G(B)$, it is scattered
into a new state, creating a correlation
\[
\rho_2=\frac{1}{2}\left(\rho_G(A) \rho_{Ph}(Un)
    +\rho_G(B)\rho_{Ph}(Sc) \right)\rho_D(A)
\]
and then the demon sees the photon, creating a correlation to it's
own state
\[
\rho_3=\frac{1}{2}\left(\rho_G(A) \rho_{Ph}(Un) \rho_D(A)
     +\rho_G(B) \rho_{Ph}(Sc) \rho_D(B) \right)
\]

Gabor and Brillouin now argue that the mean entropy of the gas has
been reduced by a mean factor of $k \ln 2$ on the basis that the
demon, by inspecting it's own state, knows which of the
subensembles the gas lies in. As a compensation, however, the
energy of the scattered photon is dissipated. They then argue that
the energy of the photon must be at least $kT \ln 2$, and this
completes the entropy balance.

There are two assumptions that they must make for this argument to
hold. Firstly, the demon must be able to identify the entropy
reduction only when the photon is scattered, otherwise the entropy
reduction would take place each time, while the dissipation of the
photon energy takes place only on the $50\%$ of occasions in which
it is scattered. Secondly, the energy of the scattered photon must
be dissipated.

There seems little real basis for either assumption. The demon's
actions are determined by it's state, so it can perform a
conditional unitary operation upon the gas, to produce
\[
\rho_4=\frac{1}{2}\rho_G(A)\left(\rho_{Ph}(Un) \rho_D(A)
     +\rho_{Ph}(Sc) \rho_D(B) \right)
\]
reducing the entropy of the gas for either outcome. Secondly,
there appears no reason why the detection of the scattered photon
must be dissipative. A suitably quick and idealised demon could
detect the photon through the recoil from it's deflection from a
mirror, rather than absorbtion by a photodetector, and by a rapid
adjustment of the apparatus effect a conditional operation upon
the photon to restore it to the unscattered state, giving
\[
\rho_5=\frac{1}{2}\rho_G(A)\rho_{Ph}(Un)
    \left(\rho_D(A)+\rho_D(B) \right)
\]
These operations are quite consistent with unitary evolution. The
entropy of the gas has been reduced, and the photon energy has not
been dissipated.

Finally, as the example of the piston in the Popper-Szilard Engine
above shows, there is no necessary reason why a physical
intermediary is even needed between the gas and the demon. The
essential issue, as we have seen, is not the energy of the photon,
but the fact that the demon itself, in $\rho_5$, is described by a
mixture, whose increase in entropy matches the reduction in
entropy of the gas.

We will now examine the conceptual difficulties this brings, and
where the error in thinking comes about. The problem lies in the
interpretation of the density matrix of the demon. The demon, of
course, does not regard itself as being in a mixture, as it should
be quite aware that it is in either the state $\rho_D(A)$ or the
state $\rho_D(B)$. This cuts to the heart of the statistical
nature of the problem. The density matrix $\rho_5$ is interpreted
as meaning that the state of the system, in reality, is either
\[
\rho_5^{\prime}=\rho_G(A)\rho_{Ph}(Un)\rho_D(A)
\]
or
\[
\rho_5^{\prime \prime}=\rho_G(A)\rho_{Ph}(Un)\rho_D(B)
\]
In each of these cases the entropy is reduced by $k \ln 2$ from
it's initial value.

The compensation is in the mixing entropy of the demon. However,
if we interpret this mixing entropy as a measure of ignorance, we
are left with the awkward fact that the demon is quite aware of
it's own state. From the perspective of the demon, the entropy
would have appeared to have decreased. Unfortunately the demon is
simply a particularly efficient observer, and there is nothing in
principle to stop us substituting a human being in it's place.
This brings us right back to Szilard's original problem - that the
intervention of an intelligent being, by making a measurement upon
a system, appears to be able to reduce it's entropy.

The error lies in the fact that we have abandoned the ensemble,
and with it the entropy of mixing, as soon as we correlate an
intelligent being to the system. We are led into this error by the
belief that the entropy of mixing represents ignorance about the
exact state of a system, and an intelligent being is certainly not
ignorant about it's own state. Thus we substitute for the ensemble
density matrix $\rho_5$ the particular subensemble
$\rho_5^{\prime}$ or $\rho_5^{\prime \prime}$ that the intelligent
being knows to be the case.

The flaw in this reasoning only comes about when we consider the
future behaviour of the demon, and the requirement of unitarity,
For example, we wish the demon to extract the energy from
expanding the one atom gas, and then start a new cycle. If we
think of the demon in state $\rho_5^{\prime}$, then it is a simple
matter to construct a unitary operation that achieves this. The
same holds true for $\rho_5^{\prime \prime}$. The problem lies in
the fact that these operations cannot be combined into a single
unitary operation. The unitary operator to complete the cycle {\em
must} be defined for the entire ensemble $\rho_5$. By implicitly
abandoning the description of the system in terms of ensembles, we
are led to construct unitary operations that do not, in fact,
exist. We will find ourselves returning to this point.

\subsection{Information Erasure}\label{s:szsol3.2}
We have found that, contrary to \cite{DD85,EN99}, Gabor and
Brillouin do not provide a resolution to the problem. Information
acquisition need not be dissipative. In this we are in agreement
with Landauer \cite{Lan61}. We must now examine how Bennett's
resolution \cite{Ben82} using Landauer's Principle of information
erasure relates to our analysis. It will be shown that Bennett's
analysis is a special case of the Entropy Engine discussed above
in Section \ref{s:szsol2:3} and in Appendix \ref{ap:szfree}. It is
therefore only a partial resolution.

Dispensing with the need for a physical intermediary between demon
and system, we have the simple process
\begin{eqnarray*}
\rho_1 &=&\frac 12\left(\rho_G(A)+\rho_G(B)\right)\rho_D(A) \\
\rho_2 &=&\frac 12\left(\rho_G(A)\rho_D(A)
    +\rho_G(B)\rho_D(B)\right) \\
\rho_3 &=&\rho_G(A)\frac 12\left(
    \rho_D(A)+\rho_D(B)\right)
\end{eqnarray*}

Bennett, in essence, accepts the argument that entropy represents
ignorance and the demon has reduced the entropy of the system, as
it is not ignorant of it's own state, but realises that the future
behaviour of the system depends upon the state the demon is left
in. The cycle must be completed.

The two different states $\rho_D(A)$ and $\rho_D(B)$ are taken to
represent the demon's own knowledge, or memory, of the measurement
outcome. To complete the cycle, and allow the Engine to extract
further energy, the demon must 'forget' this information. This
will return the demon to it's initial state and allow the cycle to
continue. It is the erasure of the information, Bennett argues,
that dissipates $k T \ln 2$ energy, and saves the second law of
thermodynamics.

This dissipation is based upon Landauer's Principle, that the
erasure of 1 bit of information requires the dissipation of $kT
\ln 2$ energy. The basis of Landauer's Principle may be summarised
as:
\begin{enumerate}
\item Information is physical. It must be stored and processed in
physical systems, and be subject to physical laws.

\item Distinct logical states must be represented within the
physical system by distinct (orthogonal) states.
\end{enumerate}
from which it is derived that the erasure of one bit of logical
information requires the dissipation of $kT\ln 2$ free energy, or
work.

There is an additional assumption, which is physically unnecessary
and usually unstated, which is also necessary to Landauer's
Principle

\begin{enumerate}
\item[3.] The physical states that are used represent the logical states
all have the same internal entropy, and mean energy.
\end{enumerate}

and the denial of this forms the basis of Fahn's
critique\cite{Fah96}\footnote{Fahn considers states with different
entropies, but neglects the possibility of different energies. In
other respects his resolution is equivalent to Bennett's.}.
Removing this assumption generalises the principle, and requires
taking note of the thermodynamic expansion and compression between
different states as part of the physical operations by which the
logical states are manipulated. As the effect of this is only to
make the relationship between information and thermodynamics more
complex, we will adopt Assumption 3 as a simplification.

It is an immediate consequence of these assumptions that the
physical storage of 1 bit of Shannon information requires a system
to have $k\ln 2$ entropy. The reason for this is simple. 1 bit of
Shannon information implies two logical states (such as true or
false), occurring with equal probability, so that the Shannon
information $I_{Sh}=\frac 12\log _2\frac 12+\frac 12\log _2\frac
12=1$. To store this in a physical system takes two orthogonal
physical states, which will be occupied with equal probability,
giving an ensemble mixing entropy of $S=k\left( \frac 12\ln \frac
12+\frac 12\ln \frac 12\right) =k\ln 2$. Now, to eliminate this
bit, the logical state must be restored to a single state. The
Shannon information of this is zero, and the mixing entropy is
zero. As Assumption 3 requires the mean energy to be unaffected by
this, a simple manipulation of the formula $E=F+TS$ demonstrates
that the reduction of entropy by $k \ln 2$ required to 'erase' the
bit of information isothermally requires $kT\ln 2$ work to be done
upon the system.

In this there is nothing controversial about Landauer's Principle.
However, it clearly rests upon the assumption that the second law
of thermodynamics is valid, which was precisely the point at
issue. To examine the Principles's relevance to the Szilard engine
we must consider how the erasure is to be achieved. Our demon will
be identified with the piston state, extracted from the box in a
mixed state.

As shown in Appendix \ref{ap:szfree}, there is a procedure by
which the piston may be restored to it's original state. This is
equivalent to inserting the piston into a second Szilard box at
some 'erasure' temperature $T_E$. This corresponds to the piston
alternating between a raising cycle, at temperature $T_G$ and a
lowering cycle at temperature $T_{E}$. The work extracted from the
$T_G$ heat bath on the raising cycle is $kT_G\ln 2$, and the work
dissipated into the $T_E$ heat bath is $kT_E\ln 2$. There is an
entropy increase of $k\ln 2$ in the $T_E$ heat bath, and decrease
of $k\ln 2$ in the $T_G$ heat bath. It should be immediately
apparent that this reversible cycle is equivalent to a Carnot
cycle, with efficiency
\[
\frac WQ=1-\frac{T_E}{T_G}
\]
Whether this cycle is acting as a heat pump or a heat engine
naturally depends upon which of $T_E$ or $T_G$ is the hotter.

Bennett assumes that the second heat bath is at $T_E=T_G$, so the
system acts as neither pump nor engine - the work extracted from
the raising cycle is used up on the lowering cycle. This cycle is
clearly the same as the Entropy Engine considered in Section
\ref{s:szsol2:3} and Appendix \ref{ap:szfree}, when restricted to
the case $T_W=T_G$. Removing this restriction, the Engine operates
at a Carnot cycle efficiency.

It is nevertheless operating on a quite different principle to the
more standard Carnot engine, which is based upon the isothermal
and adiabatic compression and expansion of a gas. No heat energy
actually flows directly between the two heat baths. Rather, it is
the piston (or 'demon') that transfers $S=k\ln 2$ entropy through
a temperature difference of $\Delta T=T_G-T_E$, and produces the
characteristic gain in free energy, $\Delta F=-S\Delta T$.

To obtain this gain, the temperature of erasure must be different
to the temperature at which the free energy is extracted from the
Szilard Box. This raises an issue that is not often addressed by
the information theoretic analysis of Maxwell's demon and
thermodynamics - there is no relationship between the entropy
involved in information storage and manipulation, and
thermodynamic temperature. Although Landauer's Principle is framed
in terms of an isothermal erasure process, such as that used for
the Szilard box above, the discussion of the 'fuel value' of blank
tapes \cite{Ben82,Fey99a} rarely makes clear how this temperature
is to be identified, as a purely information theoretical blank
tape has no temperature associated with it. For example, if we
represent the states by the spin up and spin down states of an
array of electrons, and there is no magnetic field, then all
possible logical states have the same energy, and the temperature
is undefined. By emphasising the role of information, the
additional role of temperature has been missed. An exception is
Schumacher\cite{Sch94} whose information theoretic heat engine may
be compared to the more physically explicit arrangement considered
here.

The information erasure argument can now be seen to be
insufficient to produce a complete resolution, and unnecessary
even where it is valid. It's physical basis is sound, but it is
not general enough, and information theory is not necessary to
understand it once the physical principles are correctly
understood.

Let us examine how it works as a resolution. First, we create the
problem by abandoning the ensemble of the states of the auxiliary
system. Then we characterise the different auxiliary states as
information. To quantify the information, however, we must use the
Shannon formula, and this just reintroduces the ensemble we
abandoned. We then try to connect the Shannon information back to
thermodynamics by appealing to the Landauer Principle, which is
itself derived from an assumption that the second law of
thermodynamics is universally valid. Had we not abandoned the
ensemble of auxiliary states in the first place, no reference to
information would have been necessary.

Finally, we note that information erasure has nothing to say about
the imperfect resetting considered in Section \ref{s:szsol2:2},
and so, as it does not apply to the Popper-Szilard Engine, it is
also insufficient to completely resolve the paradox.

\subsection{'Free will' and Computation}\label{s:szsol3.3}

There have recently been criticisms of the information erasure
resolution by Earman and Norton \cite{EN98,EN99}, and by
Shenker\cite{She99b}. Although we agree with the general tenor of
both papers, we believe that, unfortunately, both of them
misunderstand the nature of the Bennett-Landauer resolution. This
leads them to suspect that there are faults to be uncovered in the
Landauer principle, and to suggest that the true resolution should
be found in thermal fluctuations, with a similar physical basis to
Gabor and Brillouin's work, but that these fluctuations need not
be interpreted in any information theoretic manner. Thus, in
Earman and Norton we read

\begin{quotation}
[Bennett's] devices can only succeed in so far as we presume that
they are not canonical thermal systems. Thus Bennett's logic is
difficult to follow. Landauer's Principle is supported by
arguments that require memory devices to be canonical thermal
systems, but Szilard's Principle is defeated by the expedient of
ignoring the canonical thermal properties of the sensing device.
\end{quotation}

and in Shenker

\begin{quotation}
[The resolution] sacrifices basic ideas of statistical mechanics
in order to save the Second Law of Thermodynamics. Szilard and his
school claim that if we add the dissipation \ldots then the Demon
{\em never} reduces the entropy of the universe \ldots This way
the Second Law is invariably obeyed. The principles of statistical
mechanics, however, are violated. According to these principles,
entropy can decrease as well as increase, with some non-zero
probability.
\end{quotation}

\subsubsection{Thermal Fluctuations}

It is unclear what Earman and Norton mean when they suggest
Bennett ignores 'canonical thermal properties of the sensing
device'. It is clearly the case that the auxiliary starts in only
one of the states that is possible, so is not in a full thermal
equilibrium. However, this depends upon the thermal relaxation
times. There is no reason why selecting systems with large thermal
relaxation times, for transitions between some subspaces, and
preparing them initially in one of the subspaces, does not
constitute a 'canonical thermal system', or that use of such a
system is illegitimate.

In \cite{EN99}[Appendix 1] they claim to present a resolution,
equivalent to information theoretic arguments, in terms of thermal
fluctuations. However, their analysis rests upon the two equations
\begin{eqnarray*}
S[O,D]&=&S[O]+S[D] \\
\Delta S&=&0
\end{eqnarray*}
where $S[O]$ is the entropy of the object subsystem and $S[D]$ is
the entropy of the demon. From this they deduce $\Delta S[D]=-
\Delta S[O]$ and conclude that, as the entropy of the system is
reduced by the measurement, the entropy of the demon must have
increased.

The problem with this analysis is that these equations are simply
wrong when applied to correlated systems. The correct equation is
given in Equation \ref{eq:qinf} as
\[
S^{\prime}[O,D]=S[O]+S[D]+S[O:D]
\]
where $S[O:D]$ is the correlation between the subsystems. The
value of $S^{\prime}$ will be constant, while Earman and Norton's
$S$ will increase by $k \ln 2$ when the demon measures the state
of the gas, then {\em decrease} by the same amount when the demon
uses this correlation to change the state of the gas. Thus Earman
and Norton's argument that
\begin{quotation}
A demon closing the door at this moment has effected a reduction
in entropy. [$\Delta S[O]=-\Delta S[D]$] assures us that this
reduction must be compensated by a corresponding dissipation of
entropy in the demonic system
\end{quotation}
is incorrect, and it is unsurprising the they are unable to offer
an account of how this dissipation occurs. While it is true that
an increase in entropy of the demon system takes place, it does
not do so for the reason, or in the manner that Earman and Norton
appear to think.

Earman and Norton proceed to suggest that, if the demon can
non-dissipatively measure the location of the atom in the box,
then an erasure \textit{can} take place non-dissipatively,
allowing the second law to be violated. As this criticism would
seem to be applicable to our analysis of the Szilard Engine above,
we must consider it carefully below. It will be useful to examine
Shenker's arguments first, though.

\subsubsection{Free Will}
Shenker presents a different resolution, based upon the issue of
whether the demon may be considered to have 'free will'. If we
strip this of it's philosophical connotations, we find that the
specific property Shenker makes use of is more or less equivalent
to the absence of 'self-conditional' operations in unitary
dynamics, and that this is the same reason why Earman and Norton's
suggestion fails. Specifically, she refers to

\begin{quotation}
a system has free will if it is capable of choosing and
controlling its own trajectory in the state space
\end{quotation}

Now, to represent this in terms of unitary dynamics this would
correspond to an operation where
\begin{eqnarray*}
U\ket{0}&=&\ket{0} \\
U\ket{1}&=&\ket{0}
\end{eqnarray*}
and we have seen before, this is not a unitary operation. It will
be useful now to elaborate this with the help of the conditional
dynamics on an auxiliary system
\begin{eqnarray*}
U_a &=&\kb{\pi_1}{\pi_0}P_0+\proj{\pi_0}P_1 \\
    &&+\kb{\pi_0}{\pi_1}P_0+\proj{\pi_1}P_1 \\
U_b &=&\Pi_0 U_1+\Pi_1 U_2
\end{eqnarray*}
with $P_0$ and $P_1$ are projectors on the system of interest,
$\Pi_0$ and $\Pi_1$ are projectors onto the states of the
auxiliary system, and $U_1=\kb{1}{0}+\kb{0}{1}$,
$U_2=\proj{1}+\proj{0}$.

The system is initially in the state
$\rho=\frac{1}{2}\left(P_0+P_1\right)$ and the auxiliary is in the
state $\Pi_0$. The auxiliary examines the object, and goes into a
correlated state. It then refers to it's own state and sets the
object system to $P_0$. As noted before, this conditional
operation leaves the auxiliary system in a higher entropy state,
which compensates for the manner in which the entropy of the
system of interest has been reduced.

Shenker's characterisation of the absence of 'free will' amounts
to the statement that a system cannot refer to it's own state to
reset itself. A unitary operation cannot be conditionalised upon
the state of the system it acts upon. There are no
'self-conditional' unitary interactions. If we attempt to
construct such an operator, we must identify the auxiliary with
the system of interest. Terms such as $\kb{\pi_1}{\pi_0}P_0$ would
'collapse' as the operators act upon each other. Even assuming
such a 'collapse' is well defined, the two conditional operators
would become operators such as
\begin{eqnarray*}
U_a^{\prime } &=&\kb{1}{0}+\proj{1} \\
U_b^{\prime } &=&\kb{1}{0}+\proj{0}
\end{eqnarray*}
neither of which are unitary. A system which could exercise 'free
will', in this sense, would be able to violate the second law of
thermodynamics by resetting it's own state.

However, this is not the whole story. In \cite{ZZ92}, it is
demonstrated that there are classical, deterministic systems which
can be rigorously entropy decreasing. None of the elements in the
system can be regarded as exercising 'free will' in Shenker's
terminology. Nevertheless, the second law of thermodynamics is
broken. The reason for this is that the forces considered in
\cite{ZZ92} are Non-Hamiltonian. This is equivalent to a form of
non-unitary dynamics in quantum theory. In \cite[Chapter 9]{Per93}
Peres shows how such a non-unitary modification to quantum theory
will also lead to situations where entropy can decrease. Clearly,
the absence of free will is not enough to completely resolve the
problem.

\subsubsection{Computation}
Earman and Norton argue that a computer resetting
non-dissipatively should be possible. Their argument turns upon
the fact that there exists a non-dissipative program by means of
which a bit may be switched from one state to the other. This is
simply the operation $U_{1}$. There is a second program,
represented by operation $U_2$ which leaves the bit unchanged.
Neither of these operations are dissipative. They now propose a
program in which the bit is used to store the location of the atom
in the Szilard Engine. The computer then goes into one of two
subprograms, depending upon the state of the bit, which extracts
the energy from expanding the state of the atom.
\begin{quotation}
Programme-L leaves the memory register unaltered [$U_2$ is
applied] as it directs the expansion that yields a net reduction
of entropy. Programme-R proceeds similarly. However, at its end
Programme-R resets the memory register to L [$U_1$ is applied].
This last resetting is again not an erasure.
\end{quotation}

The flaw is that the choice of whether to execute Programme-R or
Programme-L (which are, of course, just unitary operations), is
made by a unitary operation that must be conditionalised upon the
state of the memory register itself. As we have seen, such an
operation \textit{cannot }include the $U_1$ or $U_2$ operations,
as this would be a 'self-conditionalisation' and would result in a
non-unitary operation. A similar confusion affects their later
argument, where they combine several Szilard Engines, and attempt
to extract energy only when 'highly favourable' (and
correspondingly rare) combinations of atom positions occur. In
this argument, they propose to only perform the 'erasure' when
those favourable combinations occur, thereby incurring a very
small mean erasure cost. Again, however, the choice of whether to
perform the 'erasure' operation or not cannot be made conditional
upon the state of the very bit it is required to erase, and their
argument fails. This is not some ''details of computerese'', but
due to the requirement that the evolution of any system be
described by a unitary operation.

\subsection{Quantum superposition}

We now return to the quantum mechanical arguments put forward by
Zurek\cite{Zur84} and Biedenharn and Solem\cite{BS95}. They argue
that the gas, being in a quantum superposition of both sides of
the partition, exerts no net pressure upon the piston, and so the
piston cannot move until the gas is localised by a quantum
measurement by the demon. Clearly, the piston arrangement
considered in Chapters \ref{ch:szqm} and \ref{ch:szsm} provides a
decisive counterexample to this argument. In fact, as we have
argued in Section \ref{s:szqm3:3}, the opposite conclusion, that
the piston must move, can be reached purely from consideration of
the linearity of quantum evolution.

However, it is now possible, and informative, to consider how such
a mistake could have been made. We believe that the reason for
this can be understood from the discussion of Section
\ref{s:szsol1}. This mistake, we will find, has been at the heart
of much of the confusion surrounding the operation of the Szilard
Engine, applies to the classical as well as the quantum
description and is responsible for making the information
theoretic analysis seem more plausible. By removing this mistake,
we can even apply this analysis of the Szilard Engine to the
expansion of a macroscopic N-atom gas, and we will find the same
issues are raised, and resolved, as for the one atom gas.

We start with the Hamiltonian in Section \ref{s:szqm1}, with an
infinitely high potential barrier. We now consider a modification
of this Hamiltonian, with the potential barrier displaced by a
distance $Y$
\[
H^{\prime }(Y) \Psi_n=\left(
    -\frac \hbar {2m}\frac{\partial^2}{\partial x^2}
    +V^{\prime }\left( x,Y\right) \right) \Psi _n
\]
with
\[
V^{\prime }\left( x,Y\right) =\left\{
\begin{array}{cc}
\infty & \left( x<-L\right) \\
0 & \left( -L<x<Y-d\right) \\
\infty & \left( Y-d<x<Y+d\right) \\
0 & \left( Y+d<x<L\right) \\
\infty & \left( x>L\right)
\end{array}
\right\}
\]

The eigenstates of this gas are the same as the internal
eigenstates of the gas, with a piston located at position $Y$,
denoted by $\ket{\Psi_l^\lambda(Y)}$ and $\ket{\Psi_l^\rho(Y)}$,
for states located entirely to the left or right of the partition,
respectively. The density matrix of the gas with $Y=0$ is
\begin{eqnarray*}
\rho_{P0}&=&\frac{1}{2}\left( \rho^\lambda+\rho^\rho\right) \\
\rho^\lambda &=&\frac 1{Z_{P0}}\sum_l e^{-\frac{\epsilon}{kT_G}
    \left( \frac{2l}{1-p}\right)^2}\proj{\Psi_l^\lambda(0)} \\
\rho^\rho &=& \frac 1{Z_{P0}}\sum_l e^{-\frac{\epsilon}{kT_G}
    \left( \frac{2l}{1-p}\right)^2}\proj{\Psi_l^\rho(0)} \\
Z_{P0}&=&\sum_l e^{-\frac{\epsilon}{kT_G}
    \left(\frac{2l}{1-p}\right)^2}
\end{eqnarray*}

If we now consider $H^{\prime }(Y)$ as a time dependant
Hamiltonian, with a changing parameter $Y$, we can apply the
analysis of Section \ref{s:szsm2} to the movement of the potential
barrier, rather than the movement of the piston (this will involve
ignoring or suppressing the piston states where they occur). As
$Y$ moves, the density matrix $\rho_{P0}$ will evolve into
\begin{eqnarray*}
\rho_{P1}^{\prime }(Y)&=&\frac 1{Z_{P1}^{\prime}}\{
    \sum_l e^{-\frac{\epsilon}{kT_G}\left( \frac{2l}{Y+1-p}\right)^2}
    \proj{\Psi_l^\lambda(Y)} \\
    && +e^{-\frac{\epsilon}{kT_G}\left( \frac{2l}{Y-1+p)}\right)^2}
    \proj{\Psi_l^\rho(Y)}\} \\
Z_{P1}^{\prime } &=&\sum_l\left\{e^{-\frac{\epsilon}{kT_G}
    \left( \frac{2l}{Y+1-p}\right)^2}
    +e^{-\frac \epsilon {kT_G}\left(\frac{2l}{Y-1+p}\right)^2}\right\}
\end{eqnarray*}

This is a significantly different density matrix to the density
matrix the gas evolves into when the moveable piston is present.
If we trace out the weight and piston states from $\rho_{T1}(Y)$
in Equation \ref{eq:rt1}, we find
\begin{eqnarray*}
\rho_{P1}(Y)&=&\frac 1{Z_{P1}}\{
    \sum_l e^{-\frac{\epsilon}{kT_G}\left( \frac{2l}{Y+1-p}\right)^2}
    \proj{\Psi_l^\lambda(Y)} \\
&&    +e^{-\frac{\epsilon}{kT_G}\left( \frac{2l}{Y+1-p)}\right)^2}
    \proj{\Psi_l^\rho(Y)}\} \\
Z_{P1}&=&\sum_l\left\{e^{-\frac{\epsilon}{kT_G}
    \left( \frac{2l}{Y+1-p}\right)^2}
    +e^{-\frac \epsilon {kT_G}\left(\frac{2l}{Y+1-p}\right)^2}\right\}
\end{eqnarray*}

Let us consider the behaviour of $\rho_{P1}^\prime$, supposing $Y$
has moved to the right. The $\ket{\Psi_l^\lambda(Y)}$ states will
have expanded, giving up energy as before, through pressure
exerted upon the potential barrier (this energy must be absorbed
by a work reservoir, as before). However, the $\ket{\Psi_l^\rho
(Y)}$ states have been compressed, which requires energy to be
extracted from the work reservoir. The pressure from the left is
$-\frac{kT_G}{Y+1-p}$ and that from the right
$-\frac{kT_G}{Y-1+p},$ giving a mean pressure on the co-ordinate
$Y $ of
\[
P_{P1}^{\prime }=-kT_G\left( \frac Y{Y^2-\left( 1-p\right)
^2}\right)
\]
Now, this pressure is zero when $Y=0$, is positive (pushing in the
positive $Y$ direction) when $Y$ is negative and vice versa. This
appears to be a restoring force, which if applied to a piston,
would keep it located in the center! Yet we saw from
$\rho_{T1}(Y)$ that the piston moves.

The reason for this apparent paradox is that $Y$ is used quite
differently in $\rho _{P1}^{\prime }(Y)$ compared to
$\rho_{P1}(Y)$. In $\rho _{P1}(Y)$, for the wavefunctions on the
right of the piston $Y$ represents the piston at a position $-Y$.
The result of this change of sign is that, when the pressure
exerted upon the moving piston is calculated from $\rho_{P1}(Y)$,
it is always in the direction of increasing $Y$ (which for the gas
on the right represents $-Y$ becoming more negative). The freely
moving piston represents a physically very different situation to
the constrained potential barrier.

Let us consider the difference between the two situations. The
density matrices are represented by
\begin{eqnarray*}
\rho_{P1}(Y) &=&\frac 12\rho^\lambda(Y)+\frac 12\rho^\rho(-Y)\\
\rho_{P1}^{\prime}(Y)&=&\frac 12\rho^\lambda(Y)+\frac 12\rho^\rho(Y)\\
\rho^\lambda(Y)&=&\frac 1{Z^\lambda(Y)}\sum_l
    e^{-\frac \epsilon {kT_G}\left(
        \frac{2l}{Y+1-p}\right) ^2}\proj{\Psi_l^\lambda(Y)} \\
Z^\lambda(Y)&=&\sum_l
    e^{-\frac \epsilon {kT_G}\left(\frac{2l}{Y+1-p}\right)^2} \\
\rho^\rho(Y)&=&\frac 1{Z^\rho(Y)}\sum_l
    e^{-\frac \epsilon {kT_G}\left(
    \frac{2l}{Y-1+p}\right)^2}\proj{\Psi_l^\rho(Y)}\\
Z^\rho(Y) &=&\sum_l
    e^{-\frac \epsilon{kT_G}\left(\frac{2l}{Y-1+p}\right)^2}
\end{eqnarray*}
Note that $\rho_{P1}(0)=\rho_{P1}^{\prime }(0)=\rho_{G1}$, so the
system starts in equilibrium

We represent the unitary evolution operator associated with
$H^{\prime }(Y)$ where $Y$ is moving slowly to the right by $U_R$
and where $Y$ is moving slowly to the left by $U_L$. Now $U_R$ is
the optimum operator for extracting energy from $\rho^\lambda(Y)$,
while $U_L$ is the optimum operator for extracting energy from
$\rho^\rho(Y)$. As discussed in Section \ref{s:szsol1}, these
cannot be combined into a single operator. The application of
either $U_R$ or $U_L$ to $\rho_{G1}$ will lead to
$\rho_{P1}^{\prime}(Y)$. This is not the equilibrium distribution
that would be reached had we started by inserting the potential
barrier at $Y$.

The equilibrium distribution of $\rho ^\lambda (Y) $ and $\rho
^\rho (Y)$ is
\[
\rho (Y) =p_1^{\prime }\rho ^\lambda \left( Y\right) +p_2^{\prime
}\rho ^\rho \left( Y\right)
\]
where $p_1^{\prime }+p_2^{^{\prime }}=1$, but $p_1^{\prime }\neq
\frac 12$ unless $Y=0$. This evolution moves the density matrix
away from equilibrium. As was shown in Section \ref{s:szsol1},
this requires a mean work expenditure. Note, however, that this
work expenditure is only expressed as an average. We are still
able to regard this as gaining energy on some attempts, but losing
more energy on others.

In order to gain energy reliably, we must employ an auxiliary
system, and correlate this to the application of $U_R$ or $U_L$,
depending upon the location of the one atom gas. This leads to the
density matrix of the gas to become $\rho_{P1}(Y)$, instead of
$\rho_{P1}^\prime(Y)$. The mistake is to assume that this
auxiliary requires the act of observation by an external 'demon'.
As we have noted, the piston itself constitutes an auxiliary
system, so no external observer is required to 'gather
information'.

The conditionalisation of the evolution operator upon the piston
is related to the conditionalisation of the internal Hamiltonian
of the gas. The constrained potential barrier Hamiltonian breaks
down into right and left subspaces $H^{\prime }\left( Y\right)
=H^\lambda \left( Y\right) \oplus H^\rho \left( Y\right) ,$
between which there are no transitions, with $Y$ as the externally
constrained parameter. The internal Hamiltonian for the gas, when
the piston is taken into account, however, is always a conditional
Hamiltonian
\[
H=\sum_n\Pi \left( Y_n\right) \left( H^\lambda \left( Y_n\right)
\oplus H^\rho \left( Y_n\right) \right)
\]
where $\Pi \left( Y_n\right) $ are projectors on the position of
the piston.

If we demand that the position of the piston is an externally
constrained parameter, then we find that \cite{Zur84,BS95} would
be correct. Nonetheless, this is not a quantum effect, \textit{as
the same result would also hold for a classical one-atom gas}.
Thus, even to the extent to which their contention is true, it is
nothing to do with quantum superpositions. However, the most
important conclusion is that this demand is simply unreasonable.
It does not correspond to any standard practice in thermodynamics.
This point Chambadal\cite{Cha73} argues is the key error in the
'paradox' of the Szilard Engine

\begin{quotation}
In all piston engines work is supplied by the movement of a piston
under the action of an expanding fluid. Here, though, it is the
operator who displaces the piston$\ldots $ It is clear that this
strange mode of operation was imagined only to make it necessary
to have information about the position of the molecule.
\end{quotation}

It is hard to disagree with this sentiment\footnote{Although we
must then disagree with Chambadal's conclusion that work can be
continuously extracted from the Engine.}. In fact, we can now go
further and consider how this 'mode of operation' would affect an
N-atom gas. Let us examine the situation where $\rho _N^\lambda
\left(Y\right)$ corresponds to N atoms confined to the left of a
piston at $Y$, and $\rho_N^\rho \left(Y\right) $ with them
confined to the right. Obviously such a situation would not be
likely to arise from the insertion of a piston into an N-atom gas,
but we can still consider a situation where there are two boxes,
one of which encloses a vacuum, and one contains an N-atom gas,
and some randomising process in the stacking of the boxes makes it
equally likely which box contains the gas.

In an ensemble of such situations, the mixing entropy is still
$k\ln 2$. If $N$ is large, this will be negligible compared to the
entropy of the gas. It is unsurprising that this negligible mixing
entropy will pass unnoticed by macroscopic experiments. However,
if we wish to place the two boxes side by side, and replace their
shared wall with a moveable piston, we can extract energy of
expansion by connecting the piston to some arrangement of weights,
similar to that considered for the Popper-Szilard Engine. No-one,
under such circumstances, could seriously believe that the piston
would not move, without an external observation to determine on
which side of the piston the N-atom gas is located, or that an
operator is required to know in which direction the piston should
be moved\footnote{Or even worse, Biedenharn and Solem's suggestion
that an observation may be required to 'localise' the N-atom gas
to one side or the other, and that this 'observation' involves the
thermal compression of the gas!}. The 'strange mode of operation'
is seen to be quite unnatural and unnecessary.

Nevertheless, if we consider the work we gain from the expansion,
$NkT\ln 2$, and the change in entropy of the gas $\Delta S=\left(
N-1\right) k\ln 2$, we find we have gained the tiny amount $kT\ln
2$ more than we should have done. No information gathering of any
kind has taken place, and no observation was necessary. The reason
for this gain is that the mixing entropy of $k\ln 2$ has been
eliminated from the gas. However, the piston is now in a mixture
of states, having increased it's own entropy by $k\ln 2$. As this
is a negligible quantity, compared to the dissipation of
macroscopic processes, it would naturally seem a simple matter to
restore the piston to it's original condition (though, of course,
with an N-atom gas, one could not start a new cycle by
re-inserting the piston). In fact such a restoration requires some
compression of the state of the piston as it's entropy must
decrease by $k \ln 2$, and so requires some tiny compensating
increase in entropy elsewhere. No paradox would ever be noticed
for such macroscopic objects, as both the free energy gain, and
entropy increase are negligible.

Nevertheless, the situation is otherwise identical, in principle,
to the Szilard Engine. No-one, we hope, would suggest that the
most sensible resolution is that $k\ln 2$ information must be
gathered about the location of the N-atom gas, by some dissipative
process, before the expansion can take place, or that thermal
fluctuations in the piston prevent it's operation! If such
interpretations seem absurdly contrived in the N-atom case, they
should be regarded as equally contrived in the single atom case.

\section{Comments and Conclusions} \label{s:szsol4}
The analysis and resolution of the Szilard Paradox presented in
this Chapter addresses all the problems raised in Chapter
\ref{ch:szmd}, and shows how the previous resolutions stand in
respect to one another. Rather than 'unseating' previous attempts
to resolve the problem, we have attempted to show how the
resulting partial resolutions fit into a more general structure.
Nevertheless, the analysis of this Chapter is not definitively
comprehensive. We will now briefly discuss the principal areas
where further analysis may be considered to be desirable. We will
then conclude by reviewing the reason for the occurrence of the
Szilard Paradox, and how our analysis shows this reason to be
mistaken.

\subsection{Criticisms of the Resolution}
There are four places in the analysis where we have made
assumptions about the physical processes involved, or where we
have not analysed the most general situation conceivable. These
represent situations where further work could be done to provide a
more comprehensive resolution.

These four areas may be summarised as:

\begin{itemize}
\item Non-orthogonality of subensembles;
\item More than two subensembles;
\item Pressure fluctuations;
\item Statistical Carnot Cycle.
\end{itemize}

We will now review each of these areas

\subsubsection{Non-orthogonality of subensembles}
Throughout Chapter \ref{ch:szsol} we have assumed that the density
matrix of a system is decomposed into orthogonal subensembles:
\[
\rho=p_1 \rho_1+p_2 \rho_2
\]
or if it is not, it can be decomposed into three orthogonal
subensembles, where the third is the overlap between the initial
two subspaces. This will always be the case for {\em classical}
ensembles.

However, for quantum systems, the problem is more subtle. Let us
consider the projection $\hat{P}$ of a density matrix $\rho$, onto
some subspace of the total Hilbert space, and onto it's complement
$\widehat{1-P}$.
\begin{eqnarray*}
\rho_1&=&\hat{P}\rho\hat{P} \\
\rho_2&=&(\widehat{1-P})\rho(\widehat{1-P})
\end{eqnarray*}
The decomposition
\[
\rho=\rho_1+\rho_2
\]
will only be true if $\rho$ was diagonalised in a basis for the
projected spaces. This can be seen in both the Szilard Box, and
the quantum weight. The insertion of the potential barrier, or
shelf, must deform the wavefunctions until previously
non-degenerate solutions become degenerate (which allows the
density matrix to diagonalise in a different basis). Until this
degeneracy occurs, there will be phase coherence between the
wavefunctions, that means we cannot simply divide the density
matrix into two.

For the situations considered here, we have argued that the work
required to create this degeneracy is negligible. Naturally there
will be situations where this will not be true. As long as this
work is applied slowly and isothermally, however, it should always
be recoverable at some other point in the cycle. This simply
represents an additional, if difficult, energy calculation and so
we do not believe it significantly affects our argument.

\subsubsection{More than two subensembles}
We have only considered situations where the ensemble is separated
into two. The most general solution is where the ensemble is
separated into a large number of subensembles, and the notional
free energy is extracted from each. It can be readily shown that
the increase in the entropy of the auxiliary must be at least as
large as $T$ times the gain in free energy. However, complications
arise when we attempt to consider an imperfect correlation between
the auxiliary and a compressed second system, as we must consider
all possible overlaps between the compressed states of the second
system. For $n$ initial subensembles, there will be $(2^n-1)$
different correlations between the auxiliary and the second
system. Demonstrating that the Engine must, in the long run, go
into reverse for all possible unitary operations, for all possible
values of $n$, remains a considerable task.

\subsubsection{Pressure fluctuations}
We have assumed that the piston moves with a constant speed, under
pressure from the gas and that, although the fluctuation in
pressure exerted by the gas upon the piston, at any one time, is
large, over the course of an entire cycle it is small. A more
rigorous approach would be to attribute a kinetic energy to the
piston, and allow the pressure fluctuations from the gas to cause
this to vary. The result would be a form of Brownian motion in the
piston. It might be argued that this is the 'fluctuations in the
detector' that should be seen as the real reason the Engine cannot
operate, similar to the fluctuating trapdoor. However we believe
this is false.

Although such motion would mean the piston would not reach the end
of the box at a specific time, we can be certain that it would
never reach the 'wrong' end of the box (as this would require
compressing the one atom gas to a zero volume). It is a simple
matter to create a new set of evolution operators, which, rather
than extract the piston at a given time, will extract the piston
at any time when it is in one of the three states: at the left
end; at the right end, and in the center of the box. This means
that sometimes the piston will be inserted and removed without
having any net effect, reducing the time it takes for the Engine
to operate. However, other than this, it would not affect the
conclusions above.

\subsubsection{Statistical Carnot Cycle}
Finally, in Section \ref{s:szsol2} we have only considered two
extremes: the Entropy Engine, where we perform work upon the
system to ensure a perfect correlation between the auxiliary and
the second system; and the imperfect correlation, where we perform
no work at all. In between there would be the situations where
some work is performed to improve the correlation, but not enough
to make the correlation perfect. It may be possible to use this to
produce a 'Statistical Carnot Cycle', in which the efficiency of
the Carnot Engine is exceeded, as long as the cycle continues, but
a probability of the Engine going into reverse is allowed. Any
initial gains in such an Engine are always more than offset in the
short run by the increase in entropy of the auxiliary, and in the
long run by the tendency of the machine to go into reverse.

\subsection{Summary}
In Chapter \ref{ch:szmd} we considered the arguments surrounding
the identification of information with entropy. Essentially, these
came from a dissatisfaction with the description of physical
systems using statistical mechanics, and in particular, the status
of entropy. At least part of the problem arises because of
confusion between the Boltzmann description of entropy, and the
Gibbs description, and how these two descriptions deal with
fluctuations.

The system is assumed to be in a particular state, at any one
time, but over a period of time comparable to the thermal
relaxation time, the state becomes randomly changed to any of the
other accessible states, with a probability proportional to
$e^{-E/kT}$. The Boltzmann entropy involves partitioning the phase
space into macroscopically distinct 'observational states', with
entropy $S_B=k \ln W$, where $W$ is the phase space volume of the
partition. The system will almost always be found in the high
entropy 'observational states', but has some small probability of
'fluctuating' into a low entropy state. Further, if the
'observational states' can be refined, then the entropy of the
system will decrease, until, with a completely fine grained
description, it appears to become zero!

For the Gibbs entropy, an ensemble of equivalently prepared states
must be considered, and the entropy is the average of $-k\ln p$
over this ensemble. A fluctuation is simply the division of the
ensemble into subensembles, only one of which will be actually
realized in any given system. However, by refining this to the
individual states, the entropy of the {\em subensembles} go to
zero. This is not a problem, so long as one does not abandon the
ensemble description, as the entropy is still present in the
mixing entropy.

The conceptual difficulty arises because the ensemble clearly does
not actually exist. Instead there is actually only a single
system, in a single state. It should seem that if we could
determine the actual state, we could reduce the entropy of the
system to zero. This is the origin of Maxwell's Demon and the
Szilard Paradox.

The resolution rests upon the fact that the Demon, as an active
participant within the system, must be described by the same laws
as the rest of the system. We find that, to be subject to a
unitary evolution, the Demon can only reduce the observed system's
entropy by increasing it's own. The fluctuation probability
relationship ensures that correlating a second system cannot
improve the situation.

Information theory would see the idea that the demon is an
intelligent being as central, and that this is different from the
'demonless' auxiliary, such as the fluctuating trapdoor. To
resolve this, it is necessary to supply principles to connect the
operation of intelligence to the physical system. What are the
principles required? No less than the Church-Turing thesis, that
\begin{quote}
What is human computable is Universal Turing Machine computable
\cite{Zur90a}
\end{quote}
to be sure that all intelligent creatures can be simulated as a
computer, and then Landauer's Principle, to connect the storage of
information to thermodynamics. However, if we consider what the
net effect of this is, we find it is simply to establish that we
must treat the 'intelligent being' as a physical system, subject
to unitary evolution and described by an ensemble. As we have
shown, the role played by an information processing demon is
nothing more or less than that of the auxiliary in the demonless
engine, for which no reference to information theory was
considered necessary.
\chapter{Information and Computation}\label{ch:comp}

In Chapters \ref{ch:szmd} and \ref{ch:szsol} we made reference to
Landauer's Principle, as a means of providing a link between
thermodynamics and information. Although we concluded that the
Principle was insufficient to provide a complete resolution to the
Szilard Paradox, we did not find a problem with the Principle
itself.

In this Chapter we will re-examine Landauer's Principle to see if,
on it's own, it provides a connection between information and
thermodynamics. In Section \ref{s:comp1} we will briefly review
the theory of reversible computation. We will show that classical
reversible computation can be made very efficient, or 'tidy', by a
procedure due to Bennett. However, we will also demonstrate that
Bennett's procedure does not work in general for quantum
computations. While these must be reversible, there exist quantum
computations that cannot be made 'tidy' and this has consequences
for the thermodynamics of distributed quantum computations.

Section \ref{s:comp2} will then consider the different meanings of
the information measure and the entropy measure. It will be
demonstrated that there are physical process that are logically
reversible but not thermodynamically reversible, and there are
physical processes that are thermodynamically reversible, but not
logically reversible. It is therefore demonstrated that, although
Shannon-Schumacher information and Gibbs-Von Neumann entropy share
the same mathematical form, they refer to different physical
concepts and are not equivalent.

\section{Reversible and tidy computations} \label{s:comp1}

The theory of reversible computation was developed following the
discovery of Landauer's Principle\cite{Lan61}, that only logically
irreversible operations implied an irretrievable loss of energy
(prior to that, it was thought that each logical operation
involved a dissipation of $kT\ln 2$ per bit). The amount of lost
energy is directly proportional to the Shannon measure of the
information that is lost in the irreversible operation.

We will now give a concrete physical example of how this Landauer
erasure operates, using the Szilard Box. It will be demonstrated
that the dissipation of $kT \ln 2$ work only occurs over a
complete cycle, and not during the actual process of erasing the
'information'. For understanding the thermodynamics of computation
we find that this distinction is unimportant, although in the
remainder of the Chapter we will see that the distinction can be
significant.

In Subsection \ref{s:comp1.2} we will then show how Landauer's
Principle is applied by Bennett to produce thermodynamically
efficient classical computations, but in Subsection
\ref{s:comp1.3} we will show that this approach cannot, in
general, be applied quantum computations\cite{Mar01}.

\subsection{Landauer Erasure}\label{s:comp1.1}
Landauer's Principle is typically formulated as:
\begin{quote}
to erase a bit of information in an environment at temperature $T$
requires dissipation of energy $\ge kT \ln 2$ \cite{Cav90}
\end{quote}

We will represent the storage of a bit of information by a Szilard
Box, with a potential barrier in the center. The atom on the
lefthand side of the barrier represents the logical state zero,
while the atom on the righthand side represents the logical one.
Landauer argues that RESTORE TO ZERO is the only logical operation
that must be thermodynamically irreversible\footnote{For a single
bit, the only other logical operation is NOT.}.

Firstly let us consider how much information is stored in the bit.
If the bit is always located in the logical one state, there is an
obvious procedure to RESTORE this to the logical zero state:
\begin{enumerate}
\item Isothermally move the barrier {\em and} the righthand wall to
the left at the same rate. The work performed upon the barrier by
the atom is equal to the work the wall performs upon the atom so
no net work is done.
\item When the wall has reached the original
location of the barrier, the barrier is by the lefthand wall. Now
lower the barrier from the lefthand wall, and raise it by the
righthand wall, confining the atom to the left of the barrier,
\item Return the righthand wall to it's original state.
\end{enumerate}
Naturally, if we have the bit in the logical zero state, an
operation required to RESTORE it to zero is simply: do nothing. At
first, this implies that Landauer's Principle is wrong - a bit may
always be RESTORED TO ZERO without any work being done. Of course,
we saw the fallacy in this argument in Section \ref{s:szsol3.3},
as the two procedures here cannot be combined into a single
operation.

What this tells us, however, is that if it {\em is} certain that
the bit is on one side or the other, it may be RESTORED TO ZERO
without any energy cost. It is only when the location of the bit
is uncertain that there is an energy cost. The information
represented by this is
\[
I_{Sh}=-\sum_a p_a \log p_a
\]
If the location of the bit is certain, it conveys no useful
information. It is only if there is a possibility of the bit being
in one state or the other that it represents information. In other
words, after the performing of some series of logical operations
the atom in the Szilard Box will be to the left of the barrier
with probability $p_0$ and to the right with probability $p_1$,
over an ensemble of such operations. $I_{Sh}$ represents the
information the person running the computation gains by measuring
which side of the box contains the atom.

We will now show how the RESTORE TO ZERO operation implies an
energy cost of $I_{Sh} kT \ln 2$. We are going to assume that the
probabilities $p_a$ are known. The information that is unknown is
the precise location of the atom in each individual case from the
ensemble.

First, let us note that we have already shown above that for
$p_0=1$ and $p_0=0$ we can perform the operation with zero energy
cost. These are situations where $I_{Sh}=0$.

Next, we follow this procedure if $p_0=p_1=\frac12$, for which
$I_{Sh}=1$:
\begin{enumerate}
\item Remove the barrier from the center of the box, and allow the
atom the thermalise.
\item Isothermally move the righthand wall to the center of the
box. This compresses the atom to the lefthand side, and requires
work $kT \ln 2$.
\item Re-insert the potential barrier by the righthand wall,
confining the atom to the left of the barrier
\item Return the righthand wall to it's initial location.
\end{enumerate}
This has required $kT \ln 2$ work to be performed upon the gas.
This energy is transferred into the heat bath, compensating for
the reduction in entropy of the atomic state.

If the probabilities are not evenly distributed the Shannon
information, $I_{Sh}< 1$ and we must follow a slightly different
procedure:
\begin{enumerate}
\item While keeping the central barrier raised, isothermally move
it's location to $Y=1-2p_1$. As shown in Section \ref{s:szsol1}
and Appendix \ref{ap:noneq}, this extracts a mean energy
$\left(1-I_{Sh}\right)kT \ln 2$.
\item Remove the barrier from the box and allow the atom to
thermalise.
\item Isothermally move the righthand wall to the center of the
box. This compresses the atom to the lefthand side, and requires
work $kT \ln 2$.
\item Re-insert the potential barrier by the righthand wall,
confining the atom to the left of the barrier
\item Return the righthand wall to it's initial location.
\end{enumerate}
The net work performed upon the gas is now $I_{Sh} kT \ln 2$.

This shows how the RESTORE TO ZERO operation comes with the work
requirement of $kT \ln 2$ per bit of Shannon information. This
work is transferred into an environmental heat bath, so represents
the heat emitted by a computer. Other logical operations do not
give off heat.

However, it is not clear that the work here has been lost, as the
key stage (compressions of the atom by the righthand wall) is
thermodynamically reversible. Although the energy may described as
dissipated into the heat bath, the entropy of the one atom gas has
decreased by $k \ln 2$ in compensation. The free energy of the
atom increases by $kT \ln 2$. The work performed upon the system
may, it appears, be recovered. The actual erasure of the
information occurs when the potential barrier lowered, and this
does not require any work to be performed.

The key to understanding the role of Landauer's Principle in the
thermodynamics of computation is to consider the entire
computational cycle. At the start of the computation, there will,
in general, be large numbers of memory registers. To perform
operations upon these, they must all be initially in a known
state, which we may by convention choose to be logical zero. So
the computation must start by initialising all the memory
registers that will be used. If we start with our Szilard Box
representing a Landauer Bit, then the atom will be equally likely
to be on either side of the box. To initialise it, we must
compress the atom to the left. This takes $kT \ln 2$ work. This
work has not been lost, as it has been stored as free energy of
the atom.

In other words, computation requires an {\it investment} of $kT\ln
2$ free energy, per bit of information that must be stored in the
system. At any time in the computation, any bit that is in a known
state can have this free energy recovered, by allowing it's state
to expand to fill the entire Szilard Box once more. A known state
is one that is in a particular value, regardless of the choice of
input state, (we may extend this to include the same state as an
initial input state).

When we examine a computational network, given the program and the
input state, we can recover all the free energy from the bits that
are known. Other bits may be in determinate states, well defined
functions of the input. It may be argued that these are,
therefore, 'known' but, as these states are non-trivially
dependant upon the input state (eg. (A OR NOT B) AND (C XOR D)),
to extract the energy requires one to find the value of the bit
from the input state ie. to recapitulate the calculation on a
second system. This requires an investment of an equivalent amount
of free energy into the second computation, so no gain is made in
terms of recoverable energy.

When a computation is reversible, we can recover all the free
energy initially invested in the system by completely reversing
the operation of the computation. However, if we have performed
the RESTORE TO ZERO operation, we cannot recover the original free
energy invested in the system, we only recover the $kT\ln2$ we
invested during the RESTORE TO ZERO operation. So we see that it
is only over the course of an entire cycle of computation that the
RESTORE TO ZERO operation has a thermodynamic cost. The objective
of reversible computing is to reduce the heat emitted during the
operation of a computer, and reduce the amount of the free energy
invested into the calculation that cannot be recovered at the end,
without losing the results of the computation. We will now look at
how this is achieved.

\subsection{Tidy classical computations}\label{s:comp1.2}

A reversible calculation may be defined as one which operates,
upon an input state $i$ and an auxiliary system, prepared in an
initial state $Aux0$ , to produce an output from the calculation
$O(i)$, and some additional 'junk' information $Aux(i)$:

\[
F:(i,Aux0)\rightarrow (O(i),Aux(i))
\]

in such a manner that there exists a complementary calculation:

\[
F^{\prime }:(O(i),Aux(i))\rightarrow (i,Aux0)
\]

The existence of the 'junk' information corresponds to a history
of the intervening steps in the computation, so allowing the
original input to be reconstructed. A computation that did not
keep such a history, would be irreversible, and would have lost
information on the way. The information lost would correspond to
an amount of free energy invested into the system that could not
be recovered.

However, $Aux(i)$ is not generally known, being non-trivially
dependant upon the input, $i,$ and so represents free energy that
cannot be recovered. A general procedure for discovering the
complementary calculation $F^{\prime }$ can be given like this:
\begin{itemize}
\item Take all the logical operations performed in $F$, and reverse
their operation and order.
\end{itemize}
As long as all the logical operations in $F$ are reversible logic
gates, this is possible. It is known that the reversible
Fredkin-Toffoli gates are capable of performing all classical
logical operations, so it is always possible to make a computation
logically reversible. However, this is not immediately very
useful: although we could recover the energy by reversing the
computation, we lose the output $O(i)$ in doing so.

Bennett\cite{Ben73,Ben82} showed that a better solution was to
find a different reverse calculation $F^{\prime\prime}$

\[
F^{\prime \prime }:(O(i),Aux(i),AuxO)\rightarrow (i,Aux0,O(i))
\]

Now the only additional unknown information is $O(i)$, which is
simply the output we desired (or extra information we needed to
know). A general procedure for $F^{\prime\prime}$, is:
\begin{itemize}
\item Copy $O(i)$ into a further auxiliary system $AuxO$ by means of
a Controlled-NOT gate;
\item Run $F^{\prime }$ on the original system.
\end{itemize}
This has also been shown to be the optimal
procedure\cite{LTV98,LV96} for $F^{\prime\prime}$. We call such a
calculation TIDY. All classical reversible computations can be
made TIDY.

\subsection{Tidy quantum computations}\label{s:comp1.3}
We will now show that when we try to apply this procedure to
quantum computations, it fails. This fact does not appear to be
widely appreciated\cite[for example]{BTV01}. The problem is that
the Controlled-NOT gate does not act as a universal copying gate
for quantum computers. In fact, the universal copying gate does
not exist, as a result of the 'no-cloning
theorem'\cite{WZ82,BH96b,GM97,BBBH97,Mar01}.

Clearly, in the case where the output states from a quantum
computer are in a known orthogonal set, then the quantum
computation can be made tidy. In fact, for other reasons, having
orthogonal output states was initially taken as a requirement on a
quantum computer, as it was deemed necessary for reading out the
output. This was suggestive not of a general quantum computation,
but of limited quantum algorithmic boxes: each connected by
classical communication. However, developments in quantum
information theory have suggested that distributed quantum
information may be desirable - in particular, a more general
conception of quantum computation may be required which takes
inputs from different sources, and/or at different
times.\pict{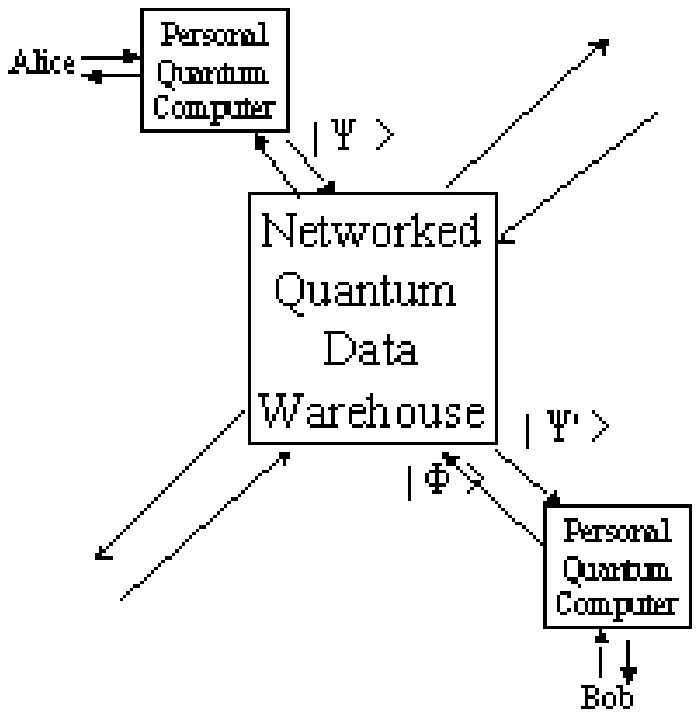}{Distributed quantum computing} In Figure
\ref{fg:qcomp} we see an example of this - Alice performs some
quantum computation, and stores the result of it in a 'quantum
data warehouse'. At some later time, Bob takes part of these
results as an input into his own computation.

We are going to take our definition of a quantum
computation\footnote{There is further complication when
entanglement enters the problem. When the output part of an
entangled state is non-recoverably transmitted, the loss of free
energy in the remainder is always at least equal to the entropy of
the reduced density matrix of the output. However, this minimum
loss of free energy requires knowledge of an accurate
representation of the resulting density matrix - which may not be
possible without explicitly calculating the output states.} as the
operation:

\[
U_C:\ket{i}\ket{Aux0})\rightarrow \ket{O(i)}\ket{Aux(i)}
\]

so that the output is always in a separable state (in other words,
we regard the 'output' of the computation as the subsection of the
Hilbert space that is interesting, and the 'auxiliary' as
everything that is uninteresting. If the 'output' were entangled
with the 'auxiliary' space, then there would be additional
information relevant to the 'output', contained in the
super-correlations between 'output' and 'auxiliary' spaces). As
any quantum computation must be performed by a unitary operation,
all quantum computers must be reversible. But are they TIDY?

If this model of computation is classical, then each time data is
sent to the central database, the local user can copy the data
before sending it, and tidy up their computer as they go along.
The only energy commitment is: total input, plus stored data. At
end of all processing - if it happens - reconstruction of
computation from stored input would allow tidying of any stored
data no longer needed. The difference between computation using
distributed classical algorithmic boxes and a single classical
computation is a trivial distinction, as the computation may be
tidied up along the way. However, this distinction depends upon
the classical nature of the information transferred between the
algorithmic boxes.

In our generalised quantum computation network, we can no longer
guarantee that the operations performed at separate locations are
connected by classical signals only. We now need to generalise the
definition of reversibility and tidiness to quantum computers.

Considering a general operation, unitarity requires that the inner
products between different input states and between the
corresponding output states is unchanged by the computation.
Reversibility must always hold. This leads to the conditions:
\paragraph{Reversible}
\begin{eqnarray*}
\bk{i}{j}\bk{Aux0}{Aux0}&=&\bk{O(i)}{O(j)}\bk{Aux(i)}{Aux(j)}
\end{eqnarray*}
\paragraph{Tidy}
\begin{eqnarray*}
\bk{i}{j}\bk{Aux0}{Aux0}\bk{AuxO}{AuxO}
    &=&\bk{i}{j}\bk{O(i)}{O(j)}\bk{Aux0}{Aux0}
\end{eqnarray*}

We can eliminate $\bk{Aux0}{Aux0}=1$ and $\bk{AuxO}{AuxO}=1$,
leaving only three cases.

\subsubsection{Orthogonal Outputs}

The output states are orthogonal set:

\[
\bk{O(i)}{O(j)}=\delta _{ij}
\]

Reversibility {\it requires} the input states to be an orthogonal
set $\kb{i}{j}=0$, and the TIDY condition will hold. This is not
too surprising, as an orthogonal set of outputs {\it can} be
cloned, and so can be tidied using Bennett's procedure.

\subsubsection{Orthogonal Inputs}

The input states are orthogonal set $\bk{i}{j}=\delta _{ij}$, but
the output states are not.

To satisfy unitarity, this requires the {\it auxiliary} output
states to be orthogonal.

\[
\bk{Aux(i)}{Aux(j)}=\delta _{ij}
\]

There does exist a unitary operator (and therefore a computable
procedure) for tidying the computation, without losing the output.
However, this tidying computation is not derivable from the
initial computation by Bennett's procedure. If we were to clone
the auxiliary output, and run the reverse operation, we would lose
the output, and be left with the 'junk'! Whether there is an
equivalent general procedure for obtaining $F^{\prime \prime }$ is
not known.

One obvious method is to examine the resulting auxiliary output
states, construct a unitary operator from

\[
U_G\ket{Aux\left( i\right) ,O\left( i\right)}
    =\ket{Aux0,O\left(i\right)}
\]
and decompose $U_G$ into a quantum logic circuit. However, it is
not clear whether the operator can be constructed without
explicitly computing each of the auxiliary output states - which
may entail running the computation itself, for each input, and
measuring the auxiliary output basis. Alternatively, examine the
form of the auxiliary output (eg. (A OR NOT B) AND (C XOR D)) )
and devise a logic circuit that reconstructs the input state from
this. However, these simply restates the problem: although some
such circuit (or $U_G$) must exist, is there a general procedure
for efficiently constructing it from only a knowledge of $U_C$?

\subsubsection{Non-orthogonal Inputs}

The input states are a non-orthogonal set. This corresponds to
Bob's position in the quantum distribution network of Figure
\ref{fg:qcomp}.

If we look at the requirements for a tidy computation, this leads
to:

\[
\bk{O(i)}{O(j)}=1
\]

The output is always the same, regardless of the input! Obviously
for a computation to be meaningful, or non-trivial, at least some
of the output states must depend in some way upon the particular
input state. So in this case we can say there are NO procedures
$F^{\prime \prime }$ that allow us to tidy our output from $F$. To
state this exactly:
\begin{quotation}
There does not exist any non-trivial $\left( \ket{O(i)} \neq
\ket{O(j)}\right) $ computations of the form

\[
G:\ket{i}\ket{Aux0}\ket{AuxO}
    \rightarrow \ket{i}\ket{Aux0}\ket{O(i)}
\]

for which $\ket{i}\ket{j} \neq \delta_{ij}$\footnote{It is
interesting to note that the 'no-cloning' theorem is a special
case of this theorem.}.
\end{quotation}

It should be made clear: this does NOT mean useful quantum
computations of the form

\[
F:\ket{i}\ket{Aux0}\rightarrow \ket{Aux(i)}\ket{O(i)}
\]

do not exist if $\ket{i}{j} \neq \delta_{ij}$ - simply that such
computations cannot be 'tidy'. For such computations, not only is
the free energy used to store the auxiliary output unrecoverable,
but also the input state cannot be recovered, except through
losing the output. For our distributed network, this means that
not only can Bob not 'tidy' his computation, but he cannot restore
Alice's data to the database.

\subsection{Conclusion}

We have now seen how Landauer's Principle arises within
computation. However we have seen that, strictly speaking, the
interpretation of Landauer's Principle as:
\begin{quote}
To erase information requires one to do $kT\ln2$ work per bit upon
the system
\end{quote}
is not strictly justified. A better use of language would be
\begin{quote}
To erase information requires the loss of $kT\ln2$ free energy per
bit
\end{quote}
This applies both in the classical computation (where the
information is measured in Shannon bits) and the quantum
computation (where information is measured in Schumacher bits).
However, the efficient tidying procedure due to Bennett is not
applicable to all quantum computations. Some quantum computations
may be tidied, but only by using some other procedure, and some
cannot be tidied at all.

\section{Thermodynamic and logical reversibility} \label{s:comp2}
We have clarified the significance of Landauer's Principle for the
thermodynamics of computation. However, we found that the logical
erasure step of the process is at a different stage to the stage
that involves the thermodynamic work of $kT \ln 2$ per bit of
information. Over the course of a computational cycle, this is of
little significance.

Nevertheless, when the interpreting the relationship between
information and entropy, this is very significant. We are now
going to briefly examine the relationship between thermodynamic
entropy and logical information. We will find that the two
concepts are quite distinct. There are processes that are
thermodynamically reversible but logically irreversible and
processes that are logically reversible but thermodynamically
irreversible.

\subsection{Thermodynamically irreversible computation}

Modern computers gives off heat well in excess of that suggested
by Landauer's Principle. They also use irreversible logic gates,
such as AND/OR gates. However, these two facts are not related in
the manner that Landauer's Principle would suggest.

While it is true that the development of quantum computing
requires the heat dissipation of computers to be minimised, the
desktop PC does not use anything approximating this kind of
technology. The computer gives off heat simply because it is very
inefficient.

Now, as Bennett has shown, any logically irreversible computation
could be implemented on a reversible computer. It would be
perfectly possible, using existing technology, to construct a
computer which was based upon reversible logic gates. Such a
computer would have to store more bits in it's memory while it was
making it's calculations, and would take approximately twice as
long to perform a calculation. The storing and reading of all
these extra bits would mean that {\em more} heat was given off
than in a corresponding irreversible computer. With current
technology, logically reversible computers are thermodynamically
{\em less} efficient than logically irreversible computers.

To put this another way: current computers are implemented using
irreversible logic gates {\em because} they are thermodynamically
inefficient, rather than the reverse. In the limit, where the
dissipation per bit stored, analysed or transmitted, is
significantly less than $kT\ln2$, a reversible computer would be
more thermodynamically efficient than an irreversible one.
However, if the technology is such that there is a dissipation per
bit stored, transmitted or analysed of more than $kT\ln2$ per bit,
then a logically irreversible computer will be thermodynamically
{\em more} efficient than a reversible one, as it has to store
less bits. With current technology, the desktop PC is far more
efficient if it is built from irreversible gates.

If we were to construct a desktop PC using reversible gates, they
would still give off heat. In short, they would be
thermodynamically irreversible, while logically reversible. This
demonstrates the first main point of this Section: logical
reversibility does not imply thermodynamic reversibility.

\subsection{Logically irreversible operations}

When we examined the Landauer Erasure, from the point of view of
the Szilard Box, we found that the logically irreversible stage
was distinct from the stage at which work is performed upon the
system. From the point of view of efficient computation these
distinctions are, perhaps, not very important. However, when we
are considering the relationship between information and entropy,
we will find this distinction becomes critical.

We are now going to consider very carefully what we mean by
logical reversibility, and demonstrate that there are operations
which are not logically reversible, but are thermodynamically
reversible. The computations will be taking place at the limiting
efficiency, where no dissipation takes place.

The information of the represented by the output states of the
computation is
\[
-\sum_a p_a \ln p_a
\]
Now we must ask, where do the $p_a$ come from? If the computation
is deterministic then, given a specific input there must be a
specific output, and the probabilities are all either zero or one.
This would imply that the information contained in the output is
zero.

Naturally this is not the case. The computation will typically
have a number of possible inputs, and a corresponding number of
possible outputs. For a reversible, deterministic computation
there will be a one-to-one correspondence between inputs and
outputs, and so the $p_a$ in the output bits are simply the
probabilities of the corresponding inputs being fed into the
computation.

This reminds us that the Shannon information is only defined over
an ensemble of possible states. To attempt to compare the Shannon
information of a computation to the thermodynamic entropy we must
consider an ensemble of computations run with different input
states.

Now let us consider how the logical reversibility comes into the
computation. The computation is fed an input state $I_a$. After
successive computation it produces the output state $O_a$. The
Shannon information of the ensemble is the same at the end of the
computation as at the start of the ensemble. This is only natural,
as we could equally well have considered the reverse computation.
This takes as it's input the states $O_a$ and produces the output
states $I_a$.

The definition of the logically reversible computation is
effectively one where, given the output state $O_a$ we can
determine exactly which input state ($I_a$) was fed into the start
of the computation.

Now, this is actually a much stronger condition that thermodynamic
reversibility. For a process to be thermodynamically reversible,
all that is required is that the entropy of the system, including
auxiliaries, is the same before and after the process.

We can now show the simple procedure that is thermodynamically
reversible but is not logically reversible. Let us return to our
Szilard Box, holding the output of some computation \footnote{As
there are only two possible outputs in this case we know there can
have only been only two possible inputs. It is a very simple
computation we are considering! However, this argument can easily
be generalised to computations with any size of output.}. We
suppose that the atom representing the outcome of the computation
is located on the left with probability $p_a$ and on the right
with probability $1-p_a$.

\begin{enumerate}
\item Move the partition, isothermally, from the center to
the location $Y=1-2p$, as described in Section \ref{s:comp1}
above.
\item The partition is removed completely from the Szilard
Box and the Box is left in contact with a heat bath for a period
of time long with respect to the thermal relaxation time.
\item The partition is reinserted in the box at the location $Y$. The
atom is again located upon the left with probability $p_a$ and on
the right with probability $1-p_a$.
\item The piston can now be isothermally returned to the center of the box, again in
connection to a work reservoir.
\end{enumerate}
This process we have described fulfils all the criteria of
thermodynamic reversibility.

In fact the thermodynamic description of the Szilard Box and the
heat bath is exactly the same at the end of this cycle as at the
start. However, there is also clearly no correspondence between
the location of the atom at the end of the cycle and the location
of the atom at the start of the cycle. If we were to now reverse
the cycle completely, and run the original computation in reverse,
there is no guarantee that the state we will end up with was the
original input state. The process is not logically reversible.

This demonstrates the second main point to this Section: that
thermodynamic reversibility does not imply logical reversibility.

\section{Conclusion}\label{s:comp3}

We have looked at the relationship between information and entropy
given by Landauer in some more detail in this Chapter. This has
lead to a better understanding of the thermodynamics of
computation but also has lead to a perhaps surprising conclusion:
\begin{itemize}
\item Logically reversible operations do not imply thermodynamic
reversibility.
\item Thermodynamically reversible operations do not
imply logical reversibility.
\end{itemize}

This pair of conclusions undermines any attempt to connect Shannon
information to Gibbs entropy\footnote{The arguments can be easily
generalised to Schumacher information and Von Neumann entropy in
quantum systems.} using Landauer's Principle and computation. We
will now see why this is so by considering the conceptual basis of
the two terms.

\subsubsection{Shannon Information}
Shannon information represents a situation where a system is in
one of a number of states $\rho_a$, and over an ensemble of such
situations occurs with probability $p_a$. Logically reversible
computations may be performed upon the system, where the state of
the system undergoes one-to-one transformations, and it is always
possible to reverse the computation and recover exactly the
initial state. For this to be possible, {\em there must be no
possibility of spontaneous transitions between the different
$\rho_a$ states}. The whole point of Shannon information is that
it quantifies the knowledge gained, on discovering that the state
is the particular $\rho_a$, out of the ensemble of possible
states.

When sending a signal, or performing a computation, any tendency
of the signal states to undergo transitions during transmission is
'noise'. This reduces the information that the receiver gains
about the signal sent, even if the effect of the noise is to leave
the {\em density matrix} over the ensemble unchanged. If the
system is allowed to completely randomise during transmission, so
that any input state $\rho_a$ leads to the density matrix $\sum_a
p_a \rho_a$ by the time it reaches the receiver, then {\em no}
information is conveyed.

\subsubsection{Entropy}
Thermodynamic entropy, on the other hand, is completely
insensitive to such transitions, {\em so long as} the ensemble
density matrix is unchanged. In a thermodynamic system the states
$\rho_a$ occur with probability $p_a$. Assuming the system is in
equilibrium at some temperature $T$, the system can be left in
contact with a heat bath at that temperature, and allowed to
undergo random transitions between all of the possible states. The
final density matrix will be the same as at the start and none of
the thermodynamic properties of the system will have changed.

In complete contrast to Shannon information, the exact individual
state $\rho_a$ that the system may be occupying has no
significance at all.
\paragraph{Summary}
The fact that signal information and entropy share the same
functional form, in both quantum and classical cases, is
remarkable. This means that many results derived in information
science will be applicable in thermodynamics, and vice versa. It
also means that, as information processing must take place on
physical systems, there are limiting cases where the two terms
will appear to coincide. However, despite their functional
similarity they refer to quite different concepts. They are not
the same thing.
\chapter{Active Information and Entropy}\label{ch:active}

In Chapters \ref{ch:szmd} and \ref{ch:szsol} we examined the
arguments surrounding the Szilard Engine thought experiment and
the role of information in it's resolution. We found that the
intrusion of information into the problem came about only because
of the failure to follow through with the ensemble description of
a thermodynamic system when that ensemble includes intelligent
beings. However, the reason for that failure can be traced, not to
a specific property of the intelligent beings, as such, but rather
a dissatisfaction with the ensemble description.

In this final Chapter we are going to briefly discuss this
dissatisfaction with the ensemble description. This has lead some
to suggest that the quantum density matrix should be treated as a
description applying to an individual system, rather than a
statistical ensemble of systems. We will argue that the attempt to
do this, rather than resolving the problem, simply imports the
quantum measurement problem into statistical mechanics.

However, we will then show that the Bohm approach to quantum
theory may be used to resolve this problem, by extending the
concept of active information to apply to the density matrix. This
resolves the tension in thermodynamics between the statistical
description and the individual system. We will construct a very
simple model suggesting how this approach could work, and how it
would be applied in the case of the interferometer and the Szilard
Engine.

\section{The Statistical Ensemble} \label{s:active1}
The statistical ensemble,
\[
\rho=\sum_a p_a \proj{a}
\]
as introduced in Chapters \ref{ch:info} and \ref{ch:szsm}, is a
description of the limiting case where an experiment is run an
infinitely large number of times, on a system that is prepared in
such a manner that state $\ket{a}$ occurs with the relative
frequency $p_a$. As noted before, if the \ket{a} do not form an
orthogonal basis then they do not diagonalise $\rho$, and the
Schumacher information of the ensemble is less than the Shannon
information
\[
S[\rho] < -\sum_a p_a \log_2 p_a
\]

In reality, of course, there is no such limiting case.  We never
have an infinite number of systems to act upon. The actual
physical situation should then be represented by a finite ensemble
or assembly\footnote{The terminology {\em assembly} is due to
Peres\cite{Per93}.}. This is a sequence of systems, $i$, each in a
particular state \ket{a_i}. The correct way to represent this
would be in a product of the Hilbert spaces of the individual
systems
\begin{eqnarray*}
\proj{\Psi}&=&\proj{a_1} \otimes \proj {a_2}
    \otimes \proj{a_3} \otimes \ldots \\
    &=& \Pi_i \proj{a_i}
\end{eqnarray*}

If there are $N$ such systems, and the state \ket{a} occurs $n_a$
times, the relative frequency of \ket{a} is
\[
f_a=\frac{n_a}{N}
\]
In the limit $N \rightarrow \infty$, then $f_a \rightarrow
p_a$\footnote{Although the probability that the relative
frequencies match the probabilities exactly, $f_a=p_a$, approaches
zero as $N$ becomes large!}.

The properties of an assembly differ from the statistical ensemble
in a number of ways.

\paragraph{Ordered systems}
The individual systems occur in a particular order, and this order
may display a pattern in the occurrence of the particular states.
It is generally assumed that the particular state \ket{a} is
randomly selected with probability $p_a$, and this will be
unlikely to produce a pattern in the appearance of the states.
Such patterned assemblies are less likely to occur the larger the
value of $N$, and become a set of measure zero as $N \rightarrow
\infty$, assuming that the states are indeed, probabilistically
generated. However, for a finite system, there is still a non-zero
probability of such order occurring. Of course, if the states are
not randomly generated (and it remains an open problem of how to
generate truly random states) then there may be an order in the
assembly even when $N$ becomes infinitely large.

An example of such a pattern is the assembly of spin-$\frac12$
particles, where the even numbered states are in the spin-up
state, while the odd numbered states are in the spin-down state.
This represents information, or a pattern, within the assembly,
that could be revealed by the appropriate measurements. Such
information is not represented in the statistical ensemble.

\paragraph{Joint measurements}
Measurements performed upon the system represented by the
statistical ensemble must be designed as a single POVM experiment.
This experiment is repeated for each system in turn, and the
relative frequencies of the POVM outcomes, $B_b$, occur. As the
value of $N$ gets large, these relative frequencies will approach
the values
\[
p_b=\trace{B_b \rho}
\]
However, this is {\em not} the most efficient method for gathering
information, given an assembly.

Firstly, one has the classically available option to correlate the
measurements performed upon a given system to the outcomes of
previous measurements. A given measurement is performed upon
system 1, then the outcome of this measurement is used to modify
the experiment performed upon system 2. The outcome of both
measurements can be used to perform an experiment upon system 3,
and so forth. It is even possible, if one performs measurements
that do not completely collapse the state of the system measured
('weak' measurements), to go back and perform further measurements
upon system 1, correlated to the outcomes of the measurements on
system 2 and 3. Such a scheme is referred to as 'Local operations
and classical communications' or LOCC measurements, as it can be
implemented by a separate experimentalist acting with locally
defined operations upon their own system, and communicating with
each other using classical information obtained from their
measurements.

Secondly, for quantum systems it is possible to improve upon LOCC
measurements by performing a joint measurement upon the combined
Hilbert space of the entire
assembly\cite{MP95,LPT98,BDE98,LPTV99,TV99}. Although joint
measurements have long been known to be required for entangled
systems, it has recently been discovered that such joint
measurements can have surprising consequences\cite[for
examples]{BDFMRSSW99,GP99,Mas00} even for systems constructed
entirely out of separable states, such as the assemblies
considered here.

\paragraph{Entropy of the universe}
The issues considered above arise because the assembly \proj{\Psi}
describes, not a statistical ensemble, but a {\em single state}
albeit one with a very large number of constituent subsystems.
This remains the case even if $N$ is allowed to become infinitely
large\footnote{Although if the universe is finite, then this will
not be possible.}.

When we consider the entropy of the {\em assembly}, we find
\[
S[\proj{\Psi}]=0
\]
as it is a pure state! Apparently, no matter how large we make the
assembly, it will have an entropy of zero. How do we reconcile the
entropy of the assembly with the entropy of the ensemble?

We have seen before that, for any given state \ket{a}, there
exists a unitary operator that will take it to a reference state
\ket{0}. A simple example of this is
\[
U_a=\kb{0}{a}+\kb{a}{0}+\sum_{\alpha \ne 0,a} \proj{\alpha}
\]
If we use $U^1$ to represent an operator acting on the Hilbert
space of the first subsystem in the assembly, then the combined
unitary operation
\begin{eqnarray*}
U_A&=&U_{a_1}^1 \otimes U_{a_2}^2 \otimes
    U_{a_3}^3 \otimes \ldots \\
&=& \Pi_i U_{a_i}^i
\end{eqnarray*}
will convert the entire assembly to the state \ket{0}. The
equivalent ensemble is now \proj{0}, which has an entropy of zero.
Thus, although there is no unitary operation which can act upon
the ensemble to reduce it's entropy, there do exist unitary
operations that can act upon assemblies, that reduce the entropy
of their equivalent ensembles.

What we have seen here is the 'global entropy problem'. The
universe does not occur as a statistical ensemble, it occurs once
only, and so has an entropy of zero. Naively, this might suggest
that we could exploit this to extract work from heat, somehow.
This is not the case. To implement an operation such as $U_A$, we
must apply the correct $U_a$ to each $i$ subsystem. This requires
a conditionally correlated system $B$ to the original assembly
$A$, and when we find the equivalent ensemble to the joint system,
the entropy we gain from the ensemble of the first system is just
the correlation entropy $-S[A:B]$, in
\[
S[A,B]=S[A]+S[B]+S[A:B]
\]
The overall entropy $S[A,B]$ of the joint ensemble remains
constant\footnote{The operation $U_A$ may also come about through
some fundamentally random process, that fortuitously happens to
apply the correct operator to each system. Such a situation is a
form of fluctuation, and the probability becomes negligible as $N$
becomes large.}.

\section{The Density Matrix} \label{s:active2}
Although we have seen that the finite assembly does not imply we
can violate the second law of thermodynamics, we are still left
with an uncomfortable situation. To express thermodynamic
properties, such as entropy and temperature, we must move from the
physically real assembly to a fictitious ensemble. This calls into
question whether the thermodynamic properties are physically real.

In addition to this, in Chapter \ref{ch:info} we saw that the
statistics of measurement outcomes were defined in terms of the
ensemble. The density matrix of the ensemble represents all the
information that can be gained from a measurement\footnote{This
may appear to contradict the joint measurements on the assembly
considered above. This is not the case. The statistics on the
outcomes of these measurements turns out to be defined in terms of
an ensemble of assemblies!}. There is no measurement that we can
perform that reveals the actual structure of the randomly
generated assembly, as opposed to the 'fictitious' ensemble, as
the statistics of measurements performed upon such an assembly can
only be expressed in terms of the ensemble density matrix.

As we cannot discover which states actually went into composing a
given density matrix, it is surely a matter of choice as to
whether we consider it to be constructed from individual pure
states, or not. Could we not abandon the idea that the density
matrix is composed of actual pure states? Can we treat the density
matrix as the fundamental description of a state, and the pure
states as simply representing the special cases of zero entropy?

If we could consistently make this assumption, then the density
matrix would no longer represent a 'fictitious' ensemble and
instead represents the actual state of a physically real system.
The thermodynamic quantities would then be undoubtedly physically
real properties rather than statistical properties. This would
significantly affect our discussion of Maxwell's Demon and the
Szilard Engine.

This question has been raised recently by \cite{AA98}. We will
find that their suggestion is only valid if the measurement
problem is assumed solved, and their suggestion does not provide a
solution to this. On the contrary, we find instead that the
general agreement that a measurement can be said to have taken
place when there has been a, for all practical purposes,
irreversible loss of phase coherence, can no longer be relied
upon.

\subsection{Szilard Box}
Let us be very clear what is being suggested here. Aharanov and
Anandan suggest taking the density matrix as the fundamental
expression of a {\em single} system with
\begin{quote} the same
ontological status as the wavefunction describing a pure
state\cite{AA98}\end{quote}

This is a very different situation to the statistical density
matrices in Chapter \ref{ch:info}. The density matrices there do
indeed represent an absence of knowledge of the exact state of the
system, while the system is actually in a definite state. To
distinguish between the two cases, we will continue to use $\rho$
to represent statistical ensembles, but will now use $\varrho$ to
represent the kind of ontological density matrices suggested by
\cite{AA98}.

The obvious situation to apply the ontological density matrix is
to thermodynamic systems. If we can do this, then the entropy
\[
S[\varrho]=\trace{\varrho \ln \varrho}
\]
can be associated with an individual system, rather then with a
representative, or fictitious, ensemble of equivalently prepared
systems. If the system is in a thermal equilibrium then it also
has a temperature $T$, and a free energy $F$, expressed as
physically real properties of the individual system, in much the
same manner as mass, or energy.

We will now consider the consequences of this by applying it to
the Szilard Box. We start with the one atom gas occupying the
entire box, with a density matrix
\[
\varrho_{G0}=\frac{1}{Z_{G0}}
    \sum_n e^{-\frac{\epsilon n^2}{kT_G}}\proj{\psi_n}
\]
as in Equation \ref{eq:rg0}. However, this no longer represents a
statistical mixture of \ket{\psi_n} states, with the atom in a
particular, but unknown state. Rather, it represents the actual
state of the individual atom. Clearly the probability distribution
of the particle throughout the box is given by
\begin{eqnarray*}
P_{G0}(x)&=& \ip{x}{\varrho_{G0}}{x} \\
    &=&\frac{1}{Z_{G0}}
    \sum_n e^{-\frac{\epsilon n^2}{kT_G}}\magn{\psi_n(x)} \\
    &=&\frac{1}{Z_{G0}}
    \sum_n e^{-\frac{\epsilon n^2}{kT_G}}R_n(x)^2
\end{eqnarray*}
where we have used the polar decomposition
$\psi_n(x)=\bk{x}{\psi_n}=R_n(x)e^{\imath S_n(x)}$, to emphasise
this is now just a real probability distribution. If we follow
standard quantum theory, this represents the probability of
finding the atom at a particular location $x$, if it is measured.
It is important to be clear that no possible measurement could
distinguish between this point of view and the statistical point
of view, where the probability density $P_{G0}$ represents the
probability of finding an atom at a location $x$ only over an {\em
ensemble} of measurements, as in each case the system would be in
a pure state.

If the partition is inserted into the center of the box, the
density matrix splits into two
\begin{eqnarray*}
\varrho_{G1}&=&\frac{1}{Z_{G1}}
    \sum_l e^{-\frac{\epsilon}{kT_G} \left(\frac{2l}{1-p}\right)^2}
        \left(\proj{\Psi_l^\lambda}+\proj{\Psi_l^\rho}\right) \\
        &=& \frac12 \left(\varrho_{G2}^\lambda
        +\varrho_{G2}^\rho \right)
\end{eqnarray*}
Now we cannot interpret this as the atom being on one side or the
other of the partition, any more than we could interpret the
wavefunction
\[
\frac{1}{\sqrt2}\left(\Psi_l^\rho(x)+\Psi_l^\rho(x)\right)
\]
as a statistical mixture. However, the reason for this is now
entirely interpretational: we are no longer assuming
$\varrho_{G1}$ represents a statistical mixture as a matter of
principle. Unlike interference in the wavefunction, there are {\em
no observable consequences} that tell us that the statistical
mixture is an untenable point of view.

\subsection{Correlations and Measurement}

Now let us suppose an auxiliary system (or Demon) attempts to
observe the box to determine on which side of the partition the
atom lies. The auxiliary is originally in the state
$\varrho_0(Aux)$. We wish an interaction so that, if the atom is
{\em actually} on the left, the auxiliary state changes to
$\varrho_L(Aux)$, and similarly $\varrho_R(Aux)$ if the atom is
{\em actually} on the right.

When we apply this interaction to the density matrix
$\varrho_{G1}$, the joint system evolves into:
\[
\varrho_2=\frac12\left(\varrho_{G2}^\lambda \otimes \varrho_L(Aux)
    +\varrho_{G2}^\rho \otimes \varrho_R(Aux) \right)
\]
How are we to understand this correlated matrix? For a statistical
ensemble $\rho_2$, the situation would be very clear. The ensemble
represents the situation where the system is either
\[
\rho_{G2}^\lambda \otimes \rho_L(Aux)
\]
or
\[
\rho_{G2}^\rho \otimes \rho_R(Aux)
\]
The demon is in a particular state, and observes the atom to be in
the correlated state.

However, \cite{AA98} cannot make use of this interpretation of the
correlated density matrix. To be consistent in the interpretation
of a density matrix $\varrho_2$, the correlated state simply
represents a joint probability density for finding the atom on one
side and the demon observing it, {\em when a measurement is
performed}. For the measurement to be brought to a closure, and a
{\em particular} outcome be observed, we must change from the
ontological density matrix $\varrho_2$ to the statistical ensemble
$\rho_2$
\[
\varrho_2 \rightarrow \rho_2
\]
and no process has been suggested through which this change will
occur.

Even if we include ourselves within the description, as Demon
states, we do not produce a well defined measurement procedure.
Instead we simply include ourselves in the quantum uncertainty,
exactly as if we were \Sch cats. Nevertheless, we know, from our
own experience, that specific outcomes of measurements {\em do}
occur. Even if we are able to interpret the density matrix as a
single system, at some point it {\em must} cease to be physically
real and become a statistical ensemble.

We notice that this new problem of measurement is even more
intractable than the old measurement problem of quantum theory! It
includes the old measurement problem, as a special case involving
pure states. The old problem consists of the fact that no unitary
transformation exists to convert the entangled pure state into the
physically real density matrix. On top of this, we then have the
fact that, even where we do not start with pure states, there is
no clear process by which the physically real density matrix
becomes a statistical ensemble.

In the case of the old measurement problem, there is at least
general agreement on when a measurement can, for all practical
purposes, be said to have taken place. When there has been a
practically irreversible loss of phase coherence between two
elements of a superposition, the wavefunction may be replaced by
\[
\frac{1}{2}\left(\proj{\Psi_l^\rho}+\proj{\Psi_l^\rho}\right)
\]
which is then interpreted as a statistical mixture $\rho$.

Now, even when the phase coherence has gone, we may still be left
with an ontological density matrix $\varrho$. A further process
appears necessary to complete the measurement, but this further
process, unlike the loss of phase coherence, has no observable
consequences\footnote{This is not strictly correct. Without such a
process, measurements cannot be said to actually have outcomes.
The fact that measurements actually {\em do} have outcomes is in
itself, therefore, an observable consequence of the existence of
this process.}!

\section{Active Information} \label{s:active3}

We saw in Chapter \ref{ch:interf} how the Bohm approach to quantum
theory resolves the measurement problem. In addition to the
wavefunction, there is an actual trajectory (whether 'particle' or
'center of activity'), and it is the location of the trajectory
within the wavepacket that determines which of the measurement
outcomes is realized.

We now find a similar {\em interpretational} problem in
thermodynamics. We would like to be able to apply thermodynamic
concepts to individual systems. However, the only way we know how
to do this would be to interpret the density matrix as applying to
individual systems, and this leads us into a similar dilemma as
with the quantum measurement problem \footnote{Although there is
no equivalent to interference effects or Bell Inequality
violations.}.

We can now consider an obvious resolution to both problems: if the
density matrix can be a description of an individual system,
rather than an ensemble, can we construct a Bohm trajectory model
for it, and will this resolve the problem in \cite{AA98}'s
approach? By explicitly developing a simple and tentative model of
Bohm trajectories for a density matrix, we will find the answer
appears to be, yes.

Firstly we must understand how we can construct a Bohm trajectory
model for a density matrix. This will not be the statistical
mechanics suggested by \cite{BH96a}, which constructs statistical
ensembles in the manner of $\rho$ above. Instead we will apply the
formalism recently developed by Brown and Hiley\cite{BH00}, who
develop the use of the Bohm approach within a purely algebraic
framework.

\subsection{The Algebraic Approach}

In \cite{BH00}, it is suggested that Bohm approach can be
generalised to the coupled algebraic equations \footnote{
\begin{eqnarray*}
\com{A}{B}&=& AB-BA \\
\acom{A}{B}&=& AB+BA
\end{eqnarray*}
}:

\begin{eqnarray}
\partime{\varrho}&=&
    \imath \com{\varrho}{H}  \label{eq:qliou}\\
\varrho \partime{\hat{S}} &=&
    -\frac12 \acom{\varrho}{H} \label{eq:aqhj}
\end{eqnarray}

Equation \ref{eq:qliou} is simply the quantum Liouville equation,
which represents the conservation of probability, and reduces to
the familiar form of
\[
\frac{\partial R(x)^2}{\partial t}+\mathbf{\nabla \cdot j}=0
\]
where $\mathbf{j}$ is the probability current
\[
\mathbf{j}=R(x)^2 \frac{\mathbf {\nabla} S(x)}{m}
\]
in the case where the system is in a pure state
$\varrho=\proj{\psi}$ and $\bk{x}{\psi}=R(x)e^{\imath S(x)}$

The second equation is the algebraic generalisation of the quantum
Hamilton-Jacobi, which reduces to Equation \ref{eq:qhj} for pure
states. The operator $\hat{S}$ is a phase operator, and this
equation can be taken to represent the energy of the quantum
system. The application of this to the Aharanov-Bohm,
Aharanov-Casher and Berry phase effects is demonstrated in
\cite{BH00}.

\cite{BH00} are concerned with the problem of symplectic symmetry,
so their paper deals mainly with constructing momentum
representations of the Bohm trajectories, for pure states, and
does not address the issue of when the density matrix is a mixed
state. Here we will be concentrating entirely upon the mixed state
properties of the density matrix, and so we will leave aside the
questions of symplectic symmetry and the interpretation of
Equation \ref{eq:aqhj}. Instead we will assume the Bohm
trajectories are defined using a position 'hidden variable' or
'beable', and will concentrate on Equation \ref{eq:qliou}.

The Brown-Hiley method, for our purposes, can be summarised by the
use of algebraic probability currents
\begin{eqnarray*}
J_X &=& \nabla_P \left(\varrho H \right) \\
J_P &=& \nabla_X \left(\varrho H \right)
\end{eqnarray*}
for which
\[
\imath \frac{\partial \varrho}{\partial t} +
    \com{J_X}{P}-\com{J_P}{X}=0
\]
To calculate trajectories in the position representation (which
Brown and Hiley refer to as constructing a 'shadow phase space')
from this we must project out the specific location $x$, in the
same manner as we project out the wavefunction from the Dirac ket
$\psi(x)=\bk{x}{\psi}$
\[
\imath \partime{\ip{x}{\varrho}{x}} +
    \ip{x}{\com{J_X}{P}}{x} - \ip{x}{\com{J_P}{X}}{x}=0
\]
The second commutator vanishes and the first commutator is
equivalent to the divergence of a probability current
\[
\mathbf{\nabla_x \cdot J(x)} =\ip{x}{\com{J_X}{P}}{x}
\]
leading to the conservation of probability equation
\[
\partime{P(x)}+\mathbf{\nabla_x \cdot J(x)} =0
\]

To see the general solution to this, we will note that the density
matrix of a system will always have a diagonal basis
\ket{\phi_a}(even if this basis is not the energy eigenstates),
for which
\[
\varrho=\sum_a w_a \proj{\phi_a}
\]
Note, the $w_a$ are {\em not} interpreted here as statistical
weights in an ensemble. There are physical properties of the state
$\varrho$, with a similar status to the probability amplitudes in
a superposition of states.

We can put each of the basis states into the polar form
\[
R_a(x)e^{\imath S_a(x)}=\bk{x}{\phi_a}
\]
so the probability density is just
\[
P(x)=\sum_a w_a R_a(x)^2
\]

The probability current now takes the more complex form
\[
\mathbf{J}(x)=\sum_a w_a R_a(x)^2 \mathbf{\nabla} S_a(x)
\]

So far we have not left standard quantum theory\footnote{The
probability current is a standard part of quantum theory, as it's
very existence is necessary to ensure the conservation of
probability.}. We may do this by now constructing trajectory
solutions $X(t)$, in the manner of the Bohm approach, by
integrating along the flow lines of this probability
current\cite{BH93,Hol93,BH00}. This leads to
\begin{equation}
m\partime{\mathbf{X}(t)}=\frac{\mathbf{J}(X(t))}{P(X(t))}=
    \frac{\sum_a w_a R_a(X(t))^2 \mathbf{\nabla} S_a(X(t))}
    {\sum_a w_a R_a(X(t))^2} \label{eq:trajrho}
\end{equation}
Notice the important fact that, when the density matrix represents
a pure state, this reduces to {\em exactly} the Bohm
interpretation in Chapter \ref{ch:interf}.

The most notable feature of Equation \ref{eq:trajrho} is that the
constructed particle velocity is {\em not} the statistical average
of the velocities \mean{V(t)}, that would have been calculated
from the interpretation of $\rho=\sum_a w_a \proj{\phi_a}$ as an
ensemble:
\[
\mean{\mathbf{V}(t)}=\sum_a w_a \mathbf{\nabla} S_a(X(t))
\]

This should not be too surprising however. We are interpreting the
density matrix as providing the activity of information necessary
to guide the particle motion. {\em All} the elements of the
density matrix are physically present, for a particle at $X(t)$,
and each state \ket{\phi_a} contributes a 'degree of activity',
given by $R_a(x)^2$ to the motion of the trajectory, in addition
to the weighting $w_a$. If a particular state has a probability
amplitude that is very low, in a given location, then even if its
weight $w_a$ is large, it may make very little contribution to the
active information when the trajectory passes through that
location.

Let us consider this with the simple example of a system which has
two states \ket{\phi_a} and \ket{\phi_b}. The probability
equations are
\begin{eqnarray*}
P(x) &=& w_a R_a(x)^2+w_b R_b(x)^2 \\
\mathbf{J}(x) &=& w_a R_a(x)^2 \mathbf{\nabla} S_a(x)
    +w_b R_b(x)^2 \mathbf{\nabla} S_a(x)
\end{eqnarray*}
Let us suppose that the two states \ket{\phi_a} and \ket{\phi_b}
are superorthogonal. This implies $\phi_a(X)\phi_b(X) \approx 0$
for all $X$. This must also hold for the probability amplitudes
$R_a(X)R_b(X) \approx 0$. If the particle trajectory $X(t)$ is
located in an area where $R_a(X)$ is non-zero, then now the value
of $R_b(X) \approx 0$. The probability equations become
\begin{eqnarray*}
P(X) & \approx & w_a R_a(X)^2 \\
\mathbf{J}(X) & \approx & w_a R_a(X)^2 \mathbf{\nabla} S_a(X)
\end{eqnarray*}
and so the particle trajectory
\[
m \partime{\mathbf{X}(t)} \approx \mathbf{\nabla} S_a(X(t))
\]
follows the path it would have taken if system was in the pure
state \ket{\phi_a}. In this situation, where there is no overlap
between the states, then the Bohm trajectories behave in exactly
the same manner as if the system had, in fact, been in a
statistical ensemble.

Now, if we make the assumption necessary to the Bohm
interpretation, that the initial co-ordinate of the particle
trajectory occurs at position $X(0)$, with a probability given by
$P(X(0))$, it is apparent that the trajectories, at time $t$ will
be distributed at positions $X(t)$ with probability $P(X(t))$. We
have therefore consistently extended the Bohm approach to treat
density matrices (and therefore thermal states) as a fundamental
property of individual systems, rather than statistical ensembles.
As we know that the statistics of the outcomes of experiments can
be expressed entirely in terms of the density matrix, we also know
that the results of any measurements in the approach will exactly
reproduce all the statistical results of standard quantum theory.

\subsection{Correlations and Measurement}
We will now look at how this extension of the Bohm interpretation
affects the discussion of correlations and measurements.

The general state of a quantum system consisting of two subsystems
will be a joint density matrix $\varrho_{1,2}$. This {\em joint}
density matrix must be diagonalised, before we project onto the
configuration space of {\em both} particle positions, using
\ket{x_1,x_2}. We can represent this projection by a 6 dimensional
vector, $x$, in the configuration space, incorporating the 3
dimensions of $x_1$ and the 3 dimensions of $x_2$. The probability
equations are simply
\begin{eqnarray*}
P(x_1,x_2) &=& \sum_a w_a R_a(x_1,x_2)^2 \\
\mathbf{J}(x_1,x_2) &=&
    \sum_a w_a R_a(x_1,x_2)^2 \mathbf{\nabla_x} S_a(x_1,x_2)
\end{eqnarray*}
The probability current can be divided into two
\[
\mathbf{J}(x_1,x_2) = \mathbf{J_1}(x_1,x_2)+\mathbf{J_2}(x_1,x_2)
\]
where
\begin{eqnarray*}
\mathbf{J_1}(x_1,x_2) &=& \sum_a w_a R_a(x_1,x_2)^2
    \mathbf{\nabla_{x_1}} S_a(x_1,x_2) \\
\mathbf{J_2}(x_1,x_2) &=& \sum_a w_a R_a(x_1,x_2)^2
    \mathbf{\nabla_{x_2}} S_a(x_1,x_2)
\end{eqnarray*}
The conservation of probability is expressed as
\[
\partime{P(x_1,x_2)}+\mathbf{\nabla_{x_1} \cdot J}(x_1,x_2)
    +\mathbf{\nabla_{x_2} \cdot J}(x_1,x_2)=0
\]
The particle trajectories must be described by a joint co-ordinate
$X(t)$ in the configuration space of both particles, which evolves
according to
\[
m\partime{\mathbf{X}(t)}=\frac{\mathbf{J}(X(t))}{P(X(t))}
\]
If we separate this into the trajectories of the two separate
particles $X_1(t)$ and $X_2(t)$, this becomes the coupled
equations
\begin{eqnarray*}
m\partime{\mathbf{X_1}(t)} &=&
    \frac{\mathbf{J_1}(X_1(t),X_2(t))}{P(X_1(t),X_2(t))} \\
m\partime{\mathbf{X_2}(t)} &=&
    \frac{\mathbf{J_2}(X_1(t),X_2(t))}{P(X_1(t),X_2(t))}
\end{eqnarray*}
We see, exactly as in the pure state situation, that the evolution
of one particle trajectory is dependant upon the instantaneous
location of the second particle, and vice versa.

The first special case to consider is when the density matrices
are uncorrelated
\[
\varrho_{1,2}=\varrho_1 \otimes \varrho_2
\]
The probability equations reduce to the form
\begin{eqnarray*}
P(x_1,x_2)=P(x_1)P(x_2)=
    \sum_a w_a R_a(x_1)^2 \sum_b w_b R_b(x_2)^2 \\
\mathbf{J}(x_1,x_2)=P(x_2)\mathbf{J_1}(x_1)
    +P(x_1)\mathbf{J_2}(x_2)
\end{eqnarray*}
where
\begin{eqnarray*}
\mathbf{J_1}(x_1) &=&
    \sum_a w_a R_a(x_1)^2 \mathbf{\nabla_{x_1}} S_a(x_1) \\
\mathbf{J_2}(x_2) &=&
    \sum_b w_b R_b(x_2)^2 \mathbf{\nabla_{x_2}} S_b(x_2)
\end{eqnarray*}
The resulting trajectories
\begin{eqnarray*}
m\partime{\mathbf{X_1}(t)} &=&
    \frac{\mathbf{J_1}(X_1(t))}{P(X_1(t))} \\
m\partime{\mathbf{X_2}(t)} &=&
    \frac{\mathbf{J_2}(X_2(t))}{P(X_2(t))}
\end{eqnarray*}
show the behaviour of the two systems are completely independant.

Now let us consider a correlated density matrix
\[
\varrho_{1,2}=\frac12 \left(
    \proj{\phi_a\chi_a} +\proj{\phi_b\chi_b} \right)
\]
where the \ket{\phi} states are for system 1 and the \ket{\chi}
states are for system 2. The polar decompositions
\begin{eqnarray*}
R_a(x_1)R_a(x_2)e^{\imath S_a(x_1)+S_a(x_2)}&=&
    \bk{x_1,x_2}{\phi_a\chi_a} \\
R_b(x_1)R_b(x_2)e^{\imath S_b(x_1)+S_b(x_2)}&=&
    \bk{x_1,x_2}{\phi_b\chi_b}
\end{eqnarray*}
lead to probability equations
\begin{eqnarray*}
P(x_1,x_2) &=& \frac12 \left(R_a(x_1)^2 R_a(x_2)^2
    +R_b(x_1)^2 R_b(x_2)^2\right) \\
\mathbf{J}(x_1,x_2) &=& \frac12 \left(R_a(x_1)^2 R_a(x_2)^2
    (\mathbf{\nabla_{x_1}} S_a(x_1)+\mathbf{\nabla_{x_2}} S_b(x_2))\right. \\
&&  +\left. R_b(x_1)^2 R_b(x_2)^2
        (\mathbf{\nabla_{x_1}} S_b(x_1)+\mathbf{\nabla_{x_2}} S_b(x_2))\right)
\end{eqnarray*}
The trajectories, $X(t)$, are then given by
\begin{eqnarray*}
m\partime{\mathbf{X_1}(t)} &=&
    \frac{R_a(X_1(t))^2 R_a(X_2(t))^2 \mathbf{\nabla_{X_1}} S_a(X_1(t))
        +R_b(X_1(t))^2 R_b(X_2(t))^2 \mathbf{\nabla_{X_1}} S_b(X_1(t))}
        {R_a(X_1(t))^2 R_a(X_2(t))^2+R_b(X_1(t))^2 R_b(X_2(t))^2} \\
m\partime{\mathbf{X_2}(t)} &=&
    \frac{R_a(X_1(t))^2 R_a(X_2(t))^2 \mathbf{\nabla_{X_2}} S_a(X_2(t))
        +R_b(X_1(t))^2 R_b(X_2(t))^2 \mathbf{\nabla_{X_2}} S_b(X_2(t))}
        {R_a(X_1(t))^2 R_a(X_2(t))^2+R_b(X_1(t))^2 R_b(X_2(t))^2}
\end{eqnarray*}
Now in general this will lead to a complex coupled behaviour.
However, if {\em either} of the states \ket{\phi} or \ket{\chi}
are superorthogonal, then relevant co-ordinate, $X_1$ or $X_2$
respectively, will be active for only one of the $R_a$ or $R_b$
states. For example, suppose the \ket{\chi} states are
superorthogonal
\[
R_a(X_2)R_b(X_2) \approx 0
\]
For a given location of $X_2$, only one of these probability
densities will be non-zero. If we suppose this is the \ket{\chi_a}
wavepacket, then $R_b(X_2)^2 \approx 0$. The trajectory equations
become
\begin{eqnarray*}
m\partime{\mathbf{X_1}(t)} &=&
    \frac{R_a(X_1(t))^2 R_a(X_2(t))^2 \mathbf{\nabla_{X_1}} S_a(X_1(t))}
         {R_a(X_1(t))^2 R_a(X_2(t))^2} \\
         &=& \mathbf{\nabla_{X_1}} S_a(X_1(t)) \\
m\partime{\mathbf{X_2}(t)} &=&
    \frac{R_a(X_1(t))^2 R_a(X_2(t))^2 \mathbf{\nabla_{X_2}} S_a(X_2(t))}
        {R_a(X_1(t))^2 R_a(X_2(t))^2} \\
        &=&\mathbf{\nabla_{X_2}} S_a(X_2(t))
\end{eqnarray*}
Both trajectories behave as if the system was in the pure state
\ket{\phi_a\chi_a}. If the location of $X_2$ had been within the
\ket{\chi_b} wavepacket, then the trajectories would behave
exactly as if the system were in the pure state
\ket{\phi_b\chi_b}. The trajectories, as a whole, behave as if the
system was in a statistical mixture of states, as long as at least
one of the subsystems has superorthogonal states.

The Bohm approach, by adding the trajectories to the quantum
description, is able to avoid the new measurement problem of the
density matrix above, by exactly the same method as it avoids the
old measurement problem of quantum theory. The loss of phase
coherence does not play a fundamental role in the Bohm theory of
measurement. It is the superorthogonality that is important, and
the principles of active and passive information implied by this.
These principles carry directly over into the density matrix
description. It is a simple matter to generalise the above
arguments to a general N-body system, or to consider states where
the diagonalised density matrix involves entangled states.

We will now briefly apply the analysis above to the Interferometer
considered in Chapter \ref{ch:interf} and the Szilard Engine in
Chapters \ref{ch:szmd} to \ref{ch:szsol}.

\subsubsection{Interferometer}
The experimental arrangement we will now be considering is not,
strictly speaking, the interferometer in Figure \ref{fg:inter1}.
In that arrangement we send a pure states into a beam splitter,
creating a superposition in the arms of the interferometer, and an
interference pattern emerges in the region $R$. Instead we will be
considering situations where the atomic state entering the arms of
the interferometer is the mixed state
\[
\frac12 \left(\proj{\phi_u(x,t_1)}+\proj{\phi_u(x,t_1)}\right)
\]
No interference effects are expected in the region $R$.

We will describe the Bohm trajectories for this in the cases
where:
\begin{enumerate}
\item The mixed state is a physically real density matrix
$\varrho$;
\item The mixed state is a statistical mixture $\rho$;
\item The mixed state is a physically real density matrix, and a
measurement of the atomic location is performed while the atom is
in the interferometer.
\end{enumerate}

\paragraph{Physically real density matrix}
While the atom is in the arms of the interferometer, the
wavepacket corresponding to \proj{\phi_u} and that corresponding
to \proj{\phi_d} are superorthogonal. The trajectories in the arms
of the interferometer are much as we would expect. However, when
the atomic trajectory enters the region $R$ the previously passive
information from the other arm of the interferometer becomes
active again.

No interference fringes occur in the region $R$, and if phase
shifters are placed in the arms of the interferometer, their
settings have no effect upon the trajectories\footnote{To observe
interference fringes we would need a density matrix that
diagonalises in a basis that includes non-isotropic superpositions
of \ket{\phi_u} and \ket{\phi_d}.}. However, the trajectories do
change in $R$. The symmetry of the arrangement, and the
'no-crossing principle' for the flow lines in a probability
current, ensures that no actual trajectories can cross the center
of the region $R$. The Bohm trajectories follow the 'surrealistic'
paths similar to those in Figure \ref{fg:inter4}, even in the
absence of phase coherence between the two arms of the
interferometer.

\paragraph{Statistical Ensemble}
We have seen that, even in the absence of phase coherence, the
Bohm trajectories for the density matrix show the surrealistic
behaviour. Does this represent an unacceptable flaw in the model?
To answer this, we now consider the situation where the density
matrix is a statistical ensemble of pure states. This situation
should more properly be described, for the point of view of the
Bohm approach, as an assembly.

First consider the assembly
\[
\rho_1=\Pi_i \proj{\phi_{a_i}}
\]
where $a_i=u$ or $d$ with a probability of one-half. As the
assembly consists entirely of product states, the behaviour in
each case is independant of the other cases.

If the state is \proj{\phi_u}, then the trajectories pass down the
u-branch, and go through the interference region without
deflection. Similarly, systems in the \proj{\phi_d} state pass
down the d-branch and are undeflected at $R$. These trajectories
are what we would expect from an incoherent mixture.

However, now let us consider the assembly
\[
\rho_2=\Pi_i \proj{\phi_{b_i}}
\]
where $b_i=+$ or $-$ occur with equal probability and
\begin{eqnarray*}
\ket{\phi_{+}} &=& \frac{1}{\sqrt2}
    \left(\ket{\phi_u}+\ket{\phi_d}\right) \\
\ket{\phi_{-}} &=& \frac{1}{\sqrt2}
    \left(\ket{\phi_u}-\ket{\phi_d}\right)
\end{eqnarray*}
This forms exactly the same statistical ensemble. Now, however, in
each individual case there will be interference effects within the
region $R$, it is just that the combination of these effects will
cancel out over the ensemble. If we were to measure the state in
the $(+,-)$ basis, then we would be able to correlate the
measurements of this to the location of the atom on the screen and
exhibit the interference fringes. The Bohm trajectories for the
assembly $\rho_2$ all reflect in the region $R$ and display the
supposed 'surrealistic' behaviour.

There are no observable consequences of the choice of the
different assemblies to construct the statistical
ensemble\footnote{It is interesting to note that if we were to
measure the assembly $\rho_1$ in the $(+,-)$ basis we would still
obtain interference fringes!}. Consequently, if we are only given
the density matrix of a statistical ensemble, we are unable to say
which assembly it is constructed from and cannot simply assume
that the underlying Bohm trajectories will follow the pattern in
Figure \ref{fg:inter2}. It is only legitimate to assume the
trajectories will pass through the interference region undeflected
if we know we have an assembly of \ket{\phi_u} and \ket{\phi_d}
states, in which case the Bohm trajectories agree. Thus we
conclude the behaviour of the trajectories for the physically real
density matrix cannot be ruled out as unacceptable on these
grounds.

\paragraph{Measuring the path}
Finally, we consider what happens when we have the physically real
density matrix
\[
\varrho=\frac12
    \left(\proj{\phi_u(x,t_1)}+\proj{\phi_u(x,t_1)}\right)
\]
and we include a conventional measuring device in the u-path. The
measuring device starts in the state \ket{\xi_0}. If the atom is
in the state \ket{\phi_u}, the measuring device moves into the
state \ket{\xi_1}. The states \ket{\xi_0} and \ket{\xi_1} are
superorthogonal.

If we now apply the interaction to the initial state
\[
\varrho \otimes \proj{\xi_0}
\]
the system becomes the correlated density matrix
\[
\frac12
    \left(\proj{\phi_u\xi_1}+\proj{\phi_u\xi_0}\right)
\]
As we saw above, as the measuring device states are
superorthogonal, the system behaves exactly as if it were the
statistical ensemble. This is true even when the atomic states
enter the region $R$. The Bohm trajectories of the atom pass
undeflected through in the manner of Figure \ref{fg:inter2}.

We conclude that the Bohm trajectories for the density matrix
cannot be considered any more or less acceptable than the
trajectories for the pure states.

\subsubsection{The Szilard Box}

We saw in Section \ref{s:active2} that the atom in the Szilard Box
can be represented by the physically real density matrix
\[
\varrho_{G0}=\frac{1}{Z_{G0}}
    \sum_n e^{-\frac{\epsilon n^2}{kT_G}}\proj{\psi_n}
\]
The probability density calculated from this is
\[
P_{G0}(x)=\frac{1}{Z_{G0}}
    \sum_n e^{-\frac{\epsilon n^2}{kT_G}}R_n(x)^2
\]
However, the probability current is zero,
($\mathbf{J_{G0}}(x)=0$). As a result, the Bohm trajectories for
the atom in the box represent it as stationary. This should not be
considered too surprising. A similar result occurs for pure
states, when the system is in an energy eigenstate. The state
$\varrho_{G0}$ is an equilibrium state. While we have a classical
picture of such a state as a fluctuating system, in the quantum
case we see the equilibrium state is simply stationary!

In reality, of course, the box will be weakly interacting with the
environment. This weak interaction will perturb the states of the
joint system, and joint density matrix will not be diagonalised
exactly in the basis of the joint Hamiltonian. The result will be
a complicated correlation of movements of the atom and the
environmental degrees of freedom that, in the long run, may
produce an equivalent effect to the classical picture of dynamic
fluctuations.

However, we will ignore this potential for environmentally induced
fluctuation. The potential barrier is inserted into the box and
the density matrix divides into
\[
\varrho_{G2}=\frac12 \left(\varrho_{G2}^\lambda
    +\varrho_{G2}^\rho \right)
\]
Now the atomic trajectory is actually located on one side or the
other of the potential barrier. The information in the other half
of the thermal state is rendered passive.

When we insert the moveable piston into the box, the joint density
matrix moves into the correlated state
\[
\varrho_{3}(Y)=\frac12 \left(
    \varrho_{G6}^\lambda(Y) \otimes \proj{\Phi(Y)}
    +\varrho_{G6}^\rho(-Y) \otimes \proj{\Phi(-Y)}\right)
\]
The changing boundary conditions and the interaction between the
piston and gas ensures that the $\varrho_{G6} \otimes \proj{\Phi}$
states are not diagonalised in eigenstates of the joint
Hamiltonian (we considered this in Section \ref{s:szqm3}), so now
the Bohm trajectories can move. If the atomic trajectory was
located on the left of the partition, then only the lefthand
branch of the state is active. The piston trajectory moves to the
right, and the atomic trajectory also moves to the right, as the
Bohm trajectories of the atom spread out to fill the expanding
space.

As the piston states move, the $\varrho_{G6}^\lambda$ and
$\varrho_{G6}^\rho$ states start to overlap. However, this can
only happen once the piston states have become superorthogonal.
The information in the passive atomic state does not become active
again.

So the Bohm trajectories for the thermal states, in this case,
confirm the naive classical picture of the Szilard Box. The atom
is indeed located on one side of the partition, and the piston can
move in the opposite direction, extracting heat from the expansion
of the gas. However, as we have seen, the Engine cannot violate
the second law of thermodynamics. We explained this in Chapter
\ref{ch:szsol} from the unitarity of the evolution. The unitary
operator must be defined upon the entire Hilbert space. This so
constrains the evolution that the Engine cannot operate without
either error or an input of work from outside (as a heat pump).

From the point of view of the Bohm theory, the need to define the
unitary operation upon the entire Hilbert space is not an abstract
issue. The portion of the Hilbert space that is not active is not
empty anymore. It is filled with the physically real, but passive,
alternate state. The passive information in this state cannot be
abandoned, anymore than the passive information from the second
arm of the interferometer can be abandoned. Attempting to reset
the piston at the end of the cycle fails because the previously
passive information, representing the piston state that moved to
the left in our example above, is still physically present, and
will combine with the active state containing the actual piston
trajectory.

What of the Szilard paradox? If the atom and piston have
physically real trajectories, does the correlation reduce the
entropy? The answer is that the entropy, as defined for the
complete density matrix, does not decrease. On the other hand, the
entropy of the active part of the density matrix {\em can} go
down, and does when a correlated measurement takes place. This
does not represent a conceptual problem, however, as the passive
part of the density matrix no longer represents a fictitious
possibility that did not occur. Instead it represents the
physically real thermal state, which just happens to be passive at
this point in time.

\section{Conclusion}
The classical conception of information, given by the Shannon
measure, represents the ignorance about an actually existing
property of a system. As measurements are performed, the state of
the observer becomes correlated to the state of the observed
system. The correlation, or mutual information, represents the
increase in knowledge the observer has about the actual state of
the system. With sufficiently refined measurements the observer
can gain a perfect knowledge of the exact state of the system and
over an ensemble of systems, can discover the ensemble probability
distribution.

In classical statistical mechanics, the Gibbs entropy shares the
same functional form as the Shannon information measure. This can
lead to the argument that entropy {\em is} simply the lack of
information about the system. Such an argument, however, directly
implies that, by performing a measurement upon the system, it's
entropy can be reduced. The flaw in this argument is that it fails
to include the observer as an active participant in the system.
This inclusion is necessary to understand why the second law of
thermodynamics cannot be broken by Maxwell's Demon. However, this
inclusion now makes it hard to interpret entropy as a lack of
information. Originally, we described the entropy of the system as
the lack of information possessed by the observer. However, as we
now have to include the entropy of the observer in the system, it
is unclear whose lack of information we are supposed to attribute
this to. It can no longer be the observer, who is fully aware of
which state he is in.

With quantum theory, the situation becomes more complex. The
Schumacher information measure shares the same form as the von
Neumann entropy. However, except in the case of communication,
where a receiver is in possession of a priori knowledge of which
signal states are being sent, it is no longer clear what the
'information' is referring to. It cannot be simply assumed that
the measurement reveals a pre-existing property of the measured
system. A given density matrix may be formed from many different
combinations of signal states, and there is no measurement
procedure that is able to uncover which is the correct one. When
the system is in a superposition of states, such as in the
interferometer, the information gathering measurement plays an
active role in the creation of the phenomena it is intended to
measure.

It has been suggested that the 'wavefunction collapse' involved in
the measurement process is a necessary part of understanding the
problem of Maxwell's Demon. However, we have shown that the
linearity of quantum mechanics proves the opposite: wavefunction
collapse plays no role in Szilard's Engine. The demon, in fact,
need perform no information processing at all and still fulfil
it's function as an auxiliary system. Nevertheless, the conceptual
problem remains, that the thermodynamic properties are possessed
only by the fictitious ensemble and not by the actual physical
system.

We now turn to the concept of active information in quantum
theory. This suggests that, in addition to the wavefunction, there
is a particle trajectory, or center of activity. The Hamiltonian
encodes the information about the system into the evolution of the
wavefunction, and this information guides the particle trajectory.
When a measurement occurs, the information in the unobserved
outcomes is no longer active, through the non-local correlation
between the system and the measuring device. The information
considered here is not simply a static correlation between two
systems, but is a dynamic principle, actively organising the
behaviour of the system.

By extending the Bohm interpretation to cover density matrices, we
showed it was possible to consistently treat the density matrix as
a property, not of an ensemble, but of an individual system. The
temperature and entropy of thermal systems can then be regarded as
physically real attributes. Again, when a measurement occurs, the
information in unobserved outcome is passive, but still physically
real. Although the entropy of the active branch of the system may
be reduced, the total entropy is constant.

It is interesting to note that it is only because the Bohm
interpretation is a no-collapse interpretation that this is
possible. Suppose we assumed the density matrix was physically
real, rather than an ensemble, and applied a wavefunction collapse
interpretation. As we performed our measurements, the density
matrix would rapidly become converted into a statistical ensemble
again. We would be forced to say that the physical entropy of the
system was decreasing. The total entropy would again become a
property only of the statistical ensemble.

In both statistical mechanics and quantum measurement it is
necessary to include the observer as an active participant in the
system if we are to avoid apparent paradoxes. The Bohm
interpretation and activity of information provides a unified
framework for understanding both.

{\appendix
\chapter{Quantum State Teleportation}\label{ap:telep}

Quantum state teleportation\footnote{The material in this Appendix
originally appeared in \cite{HM99} as a joint paper with B J
Hiley.} has focused attention on the role of quantum information.
Here we examine quantum teleportation through the Bohm
interpretation. This interpretation introduced the notion of
active information and we show  that it is this information that
is exchanged during teleportation.  We discuss the relation
between our notion of active information and the notion of quantum
information introduced by Schumacher.

\section{Introduction}
The recent discovery of quantum state teleportation
\cite{BBCJPW93} has re-focused attention on the nature of quantum
information and the role of quantum non-locality in the transfer
of information. Developments in this area have involved state
interchange teleportation \cite{Mou97}, as well as multi-particle
entanglement swapping \cite{BKV97}, and position/momentum state
teleportation \cite{Vai94}. Although these effects arise from a
straight forward application of the formalism, the nature of the
quantum information and its transfer still presents difficulties.
Attempts to address the issue from the perspective of information
theory \cite{HH96,AC95} and without invoking wave function
collapse \cite{Bra96} have clarified certain aspects of this
process but problems still remain.

In order to obtain a different perspective on these phenomena we
first review the  salient features of the Bohm interpretation that
are of direct relevance to these situations
\cite{Boh52a,Boh52b,BH93,Hol93,Bel87}, before applying its
techniques to the specific example of spin teleportation.  One of
the advantages of using this approach in the present context is
that to account for quantum processes it is necessary to introduce
of the notion of `active' information. This notion was introduced
by Bohm \& Hiley \cite{BH93} to account for the properties of the
quantum potential which cannot be consistently regarded as a {\em
mechanical potential} for reasons explained in Bohm \& Hiley
\cite{BH93}. There is also the added advantage that the approach
gives a clear physical picture of the process at all times, and,
therefore provides an unambiguous description of where and how the
`quantum information' is manifested.  In this paper we will
discuss how the three notions of active, passive and inactive
information are of relevance to the teleportation problem.

\section{Quantum Teleportation}

The basic structure of quantum teleportation can be expressed
using three spin- $\frac{1}{2}$ particles, with particles 2 and 3
initially in a maximally entangled EPRB state, and particle 1, in
an unknown superposition:

\[
\Psi_1=(a|\uparrow\rangle_1+b|\downarrow\rangle_1)
(|\uparrow\rangle_2|\downarrow\rangle_3-
|\downarrow\rangle_2|\uparrow\rangle_3)/\sqrt{2}
\]
By introducing the `Bell states'
\[
\begin{array}{cc}
\beta^{(ij)}_1=(|\uparrow\rangle_i|\uparrow\rangle_j+|\downarrow\rangle_i
|\downarrow\rangle_j)/\sqrt{2}&
\beta^{(ij)}_2=(|\uparrow\rangle_i|\uparrow\rangle_j-
|\downarrow\rangle_i|\downarrow\rangle_j)/\sqrt{2}\\
\beta^{(ij)}_3=(|\uparrow\rangle_i|\downarrow\rangle_j+|\downarrow\rangle
_i|\uparrow\rangle_j)/\sqrt{2}&
\beta^{(ij)}_4=(|\uparrow\rangle_i|\downarrow\rangle_j-
|\downarrow\rangle_i|\uparrow\rangle_j)/\sqrt{2}
\end{array}
\]
we can re-write $\Psi_1$ as

\[
\begin{array}{cccccc}
\Psi_2&= &(\beta^{(12)}_1&[-b|\uparrow\rangle_3+a|\downarrow\rangle_3]+
&\beta^{(12)}_2&[+b|\uparrow\rangle_3+a|\downarrow\rangle_3]+ \\
&&\beta^{(12)}_3&[-a|\uparrow\rangle_3+b|\downarrow\rangle_3]+
&\beta^{(12)}_4&[-a|\uparrow\rangle_3-b|\downarrow\rangle_3])/2 \end{array}
\]
If we now measure the Bell state of particles 1 and 2, and communicate the
result to the recipient
of particle 3 who will,using that information, then perform one of the local
unitary operations on
particle 3 given below

\[
\begin{array}{cccccc}
U_1&=&\left( \begin{array}{cc}0&1\\-1&0\end{array}\right),
& U_2&=&\left(\begin{array}{cc}0&1\\1&0\end{array}\right)\\
 U_3&=&\left(\begin{array}{cc}-1&0\\0&1\end{array}\right),
& U_4&=&\left(\begin{array}{cc}-1&0\\0&-1\end{array}\right).
\end{array}
\]
In this way we have disentangled particle 3 from particle 2 and
produced the state $(a|\uparrow\rangle_3+b|\downarrow\rangle_3)$
on particle 3.  Thus the information represented by [a,b] has been
perfectly `teleported' from  particle 1 to particle 3, without our
having measured a or b directly.  Furthermore, during the transfer
process we have only passed 2 classical bits of information
(corresponding only to the choice of $U$) between the remote
particles.  Note that as 'a' and 'b' are continuous parameters, it
would require an infinite number of classical bits to perfectly
specify the [a,b] state.  This ability to teleport accurately has
been shown to be critically dependant upon the degree of
entanglement of particles 2 and 3 \cite{HH96,Pop94}.

We may note that in the Bell state expansion, the information signified by the
coefficients
[a,b] appears on the particle 3 spin states before any actual measurement has
taken place
(although this information is encoded in a different way for each Bell state).
What are we to
make of this?

It would seem absurd to assume that the information
described by a and b was already attached to particle 3 as, at this stage,
particle 1 could be
any other particle in the universe.
Indeed all that has happened is that  $\Psi_1$ has been the re-written in a
different basis to
give $\Psi_2$.  Clearly this cannot be regarded as an actual physical effect.

Following Heisenberg \cite{Hei58} and Bohm \cite{Boh51}, we can
regard the wave function as describing potentialities.  At this
stage  $\Psi_2$ describes the potentiality that particle 3 could
carry the [a,b] information that would be actualised during the
measurement. However, here we have a problem as
Braunstein\cite{Bra96} has shown that a collapse of the
wavefunction (the usual mechanism by which such potentialities
become actualised) is unnecessary to the description of quantum
teleportation, by including the Bell state measuring device within
the quantum formalism.  Using this description, we find that the
attachment of the [a,b] information to particle 3, {\it after} the
Bell state interaction, is the same as in the $\Psi_2$ expansion
{\it prior} to the interaction.  While this is clearly necessary
to maintain the no-signalling theorem, it leaves ambiguous the
question of whether the [a,b] information has been transferred to
particle 3, at this stage, or not.

To resolve these issues, we need to give a clearer meaning to the
nature of the information contained in [a,b] and to understand how
and when this information becomes manifested at particle 3. We now
turn to the Bohm interpretation (Chapter \ref{ch:interf}) to
provide some new insights into these questions.

\section{Quantum State Teleportation and Active Information}

In order to examine how the idea of active and passive information
can be used in quantum teleportation, we must explain how spin is
discussed in the Bohm interpretation.  There have been several
different approaches to  spin \cite{BH93,Hol88,Alb92}, but this
ambiguity need not concern us here as we are trying to clarify the
principles involved.  Thus for the purpose of this article we will
adopt the simplest model that was introduced by Bohm, Schiller and
Tiomno \cite{BST55,DHK87}. We start by rewriting the polar
decomposition of the wave function as $\Psi=Re^{iS}\Phi$ where
$\Phi$ is a spinor with unit magnitude and zero average phase.  If
we write:
\[
\Phi=\left(\begin{array}{c}r_1e^{is_1}\\r_2e^{is_2}\\
\vdots\\r_ne^{is_n}\end{array}\right)
\]
where $n$ is the dimension of the spinor space, then $\sum_is_i=0$
and $\sum_i(r_i)^2=1$. The many-body Pauli equation then leads to
a modified quantum  Hamilton-Jacobi equation given by:

\[
\frac{\partial S}{\partial t}-i\Phi^{\dag}\frac{\partial \Phi}{\partial t} =-
\sum_i\left(
\frac{p_i^2}{2m}+Q_i+ 2\mu_i{\bf B}.{\bf s}_i \right)
\]

with a momentum $p_i=\nabla _iS+\Phi^{\dag}\nabla _i\Phi$, a
quantum  potential $Q_i=\frac{1}{2m} (-\nabla _i^2R+\nabla
_i\Phi^{\dag}\nabla _i\Phi+(\Phi^{\dag}\nabla _i\Phi)^2)$. $\bf
{B}$ is the magnetic field and $\mu_i$ is the magnetic dipole
moment associated with particle $i$. We can, in addition,
attribute a real physical angular momentum to each  particle $i$
given by ${\bf s}_i=\frac{1}{2}\Psi^{\dag}{\bf \sigma}_i\Psi$,
where  ${\bf \sigma}_i$ are the Pauli matrices operating solely in
the spinor subspace of  particle $i$.

The information contained in the spinor wave function  is again
encoded in the quantum potential, so that the trajectory of the
particle is guided by the evolution of the spinor states, in
addition to the classical interaction of the ${\bf B}$ field with
the magnetic dipole moment of the particle.  Contracting the Pauli
equation with $\Psi^{\dag}\sigma _i$ leads the equation of motion
for the particle $i$ spin vector: \[ \frac{d{\bf s}_i}{dt}={\bf
T}_i + 2\mu_i{\bf B}\times{\bf s}_i
\]
where ${\bf T}_i$ is a quantum torque.  The $k$  components of the torque are
given by
\[
[T_i]_k=\sum_j\frac{1}{2\rho m_j}
\epsilon _{klm}\{
[s_i]_l[\nabla _j]_n(\rho[\nabla _j]_n[s_i]_m)+
 s_{lr}[\nabla _j]_n(\rho[\nabla _j]_ns_{mr})
\}
\]
where $\rho=R^2$ and $s_{ij}$ is the non-local spin correlation
tensor formed from $\Psi^{\dag}\sigma _i\sigma _j\Psi$.  Equations
of motion for these tensors  can be derived by contracting the
Pauli equation with $\Psi^{\dag}\sigma _i\sigma _j$, and similarly
for higher dimension correlation tensors.  Detailed application of
these ideas to the entangled spin state problem has been
demonstrated in Dewdney et al. \cite{DHK87}.

To complete the description of the particles, we must attach
position wave functions to  each of the particles.  We do this by
assuming that each particle can be represented by a localised
wavepacket.  Thus, for the teleportation problem:
\[
\begin{array}{lcl}
\Psi &=&(a|\uparrow\rangle_1+b|\downarrow\rangle_1)
(|\uparrow\rangle_2|\downarrow\rangle_3-|\downarrow\rangle_2|\uparrow\rangle_3)
\rho(x_1)\phi(x_2)\xi(x_3)/\sqrt{2} \\
&=&
\{\beta^{(12)}_1[-b|\uparrow\rangle_3+a|\downarrow\rangle_3]+
\beta^{(12)}_2[+b|\uparrow\rangle_3+a|\downarrow\rangle_3]+\\
&&\beta^{(12)}_3[-a|\uparrow\rangle_3+b|\downarrow\rangle_3]+
\beta^{(12)}_4[-a|\uparrow\rangle_3-b|\downarrow\rangle_3]\}
\rho(x_1)\phi(x_2)\xi(x_3)/2
\end{array}
\]
Initially, the three position wave packets are separable, and the
particle trajectories will be determined by separate information
potentials although the spin properties of particles 2 and 3 will
be linked via the spin quantum potential.  The particle spins can
be shown to be
\begin{eqnarray*}
{\bf s}_1=\frac{1}{2}(a^*b+b^*a,ia^*b-ib^*a,a^*a-b^*b)& {\bf s}_2=(0,0,0)& {\bf
s}_3=(0,0,0)
\end{eqnarray*}

Note that each of the particles 2 and 3  in a maximally entangled
anti-symmetric state have zero spin angular momentum, a surprising
point that has already been noted and discussed by Dewdney et al.
\cite{DHK87} and by Bohm \& Hiley \cite{BH93}.  More significantly
for our problem is that at this stage, the information described
by a and b acts only through the quantum potential, $Q_1$, which
organises the spin of particle 1, but not the spin of particles 2
and 3.

Before discussing the measurement involved in the actual
teleportation experiment, let us first recall what happens when a
simple spin measurement is made on particle 2 alone. The
wavepacket  $\phi(x_2)$ would divide into two, and the particle
would enter one of these packets with equal probability.  Thus the
wave function becomes
\[
\Psi=(a|\uparrow\rangle_1+b|\downarrow\rangle_1)\rho(x_1)
(|\uparrow\rangle_2|\downarrow\rangle_3\phi_1(x_2)-
|\downarrow\rangle_2|\uparrow\rangle_3\phi_0(x_2))\xi(x_3)/2 \]
Particle 2 will enter one of the packets, say $\phi_1(x_2)$.    As
$\phi_1(x_2)$ and $\phi_0(x_2)$ separate, particles 2 and 3 will
develop non-zero spins, with opposite senses, and will be
described by $ |\downarrow\rangle_2|\uparrow\rangle_3 $. Any
subsequent measurement of the spin of particle 3, would divide
$\xi(x_3)$ into two, but particle 3 would always enter the
wavepacket on the same branch of the superposition as particle 2
had entered earlier, as only the information in that branch is
active. This has been beautifully illustrated by Dewdney et al.
\cite{DHK87}

As the particle 1 is in a separable state for both spin and
position, no local interactions on particle 2 or 3 will have any
effect on the trajectory and spin of particle 1. Neither will any
measurement on particle 1 produce any effect on particles 2 and 3.
The behaviour of the spins of particles 2 and 3 will be determined
by the pool of information common to them both, while only the
behaviour of particle 1 is determined by the [a,b] information,
regardless of the basis in which the spin states are expanded.

Now let us return to the main theme of this paper and consider the
measurement that produces teleportation.  Here we need to
introduce a Bell state measurement.  Let the instrument needed for
this measurement be described by the wavepacket $\eta(x_0)$ where
$x_0$ is a variable (or a set of variables) characterising the
state of this apparatus. The measurement is achieved via an
interaction Hamiltonian that can be written in the
 form $H={\bf O}^{(12)}\nabla _0$.

The interaction operator ${\bf O}^{(12)}=\lambda O^{(12)}_\lambda$ couples the
$x_0$ co-ordinate to the Bell state of particles 1 and 2 through the
Bell state projection operators $O_\lambda=\beta_{\lambda}\beta_{\lambda}^\dag$.
This creates the state
\[ \begin{array}{lccc}
\Psi_f&=&\{\eta_1(x_0)\beta^{(12)}_1[-
b|\uparrow\rangle_3+a|\downarrow\rangle_3]+
\eta_2(x_0)\beta^{(12)}_2[+b|\uparrow\rangle_3+a|\downarrow\rangle_3]+\\
&&\eta_3(x_0)\beta^{(12)}_3[-a|\uparrow\rangle_3+b|\downarrow\rangle_3]+
\eta_4(x_0)\beta^{(12)}_4[-a|\uparrow\rangle_3-b|\downarrow\rangle_3]\}\\
&&\rho(x_1)\phi(x_2)\xi(x_3)/2\\ \end{array} \] where
$\eta_1(x_0)$, $\eta_2(x_0)$, $\eta_3(x_0)$ and $\eta_4(x_0)$ are
the wavepackets of the four non-overlapping position states
corresponding to the four outcomes of the Bell state measuring
instrument. Initially all four systems become entangled and their
behaviour will be determined by the new common pool of
information. This includes the [a,b] information that was
initially associated {\em only} with particle 1.

As the position variable $x_0$ of the measuring device enters one
of the non- overlapping wavepackets $\eta_i(x_0)$, only one of the
branches of the superposition remains active, and the information
in the other branches will become passive.  As this happens,
particle 3 will develop a non-zero particle spin $\bf{s_3}$,
through the action of the quantum torque. The explicit
non-locality of this allows the affects of the Bell state
measurement to instantaneously have an effect upon the behaviour
of particle 3.  The significance of the $\Psi_2$ Bell state
expansion is now revealed as simply the appropriate basis for
which the [a,b] information will be transferred entirely onto the
behaviour of particle 3, if only a single branch of the
superposition were to remain active. The interaction with the Bell
state measuring device is required to bring about this change from
active to passive information in the other branches (and thereby
actualising the potentiality of the remaining branch).

However, no meaningful information on [a,b] may yet be uncovered
at particle 3 until it is known which branch is active, as the
average over all branches, occurring in an ensemble, will be
statistically indistinguishable from no Bell state measurement
having taken place. Simply by noting the actual position ($x_0$)
of the measuring device, the observer, near particles 1 and 2,
immediately knows which wavepacket $x_0$ has entered, and
therefore which state is active for particle 3.
   The observer then sends this classical information to the
observer at 3 who will then apply the appropriate unitary
transformation $U_1\cdots U_4$ so that the
initial spin state of particle 1 can be recovered at particle 3.

\section{Conclusion}

In the approach we have adopted here, the notion of active
information introduced by Bohm and Hiley \cite{BH93} has been
applied to the phenomenon of state teleportation. This gives rise
to a different perspective on this phenomenon and provides further
insight into the notion of quantum information.  To see more
clearly how teleportation arises in this approach let us
re-examine the above spin example in more general terms. The
essential features can be seen by examining the general structure
of the quantum potential.  Using the initial wave function,
$\Psi_i$ given above, the quantum potential takes the form
 \[
Q(x_{1}, x_{2}, x_{3}) = Q_{1}(x_{1}, a, b)Q_{23}(x_{2}, x_{3})
\] Here the coefficients a and b characterise the quantum
potential acting only on particle 1.  This means that initially
the information carried by the pair [a, b] actively operates on
particle 1 alone.  At this stage the behaviour of particle 3 is
independent of a and b, as we would expect.

 To perform a Bell State measurement we must couple particle 1 to particle 2 by
introducing the interaction Hamiltonian given above.  During this
process, a quantum potential will be generated that will couple
all three particles with the measuring apparatus. When the
interaction is over, the final wave function becomes $\Psi_f$.
This will produce a quantum potential that can be written in the
form
 \[
Q(x_{1}, x_{2}, x_{3}, x_{0}) = Q_{12}(x_{1}, x_{2}, x_{0})Q_{3}(x_{3}, x_{0},
a, b)
 \]
Thus after the measurement has been completed, the information
contained in a and b  has now been encoded in $Q_{3}$ which
provides the active information for particle 3. Thus we see that
the information that was active on particle 1 has been transferred
to particle 3.  In turn this particle has been decoupled from
particle 2.  Thus the subsequent spin behaviour of particle 3 will
be different  after the measurement.

What we see clearly emerging here is that it is {\em active
information} that has been transferred from particle 1 to particle
3 and that this transfer has been mediated by the non-local
quantum potential.  Let us stress once again that this information
is in-formation for the particle and, at this stage has nothing to
do with `information for us'.

Previous discussions involving quantum information have been in
terms of its relation to Shannon information theory \cite{Sch95}.
In classical information theory, the expression $H(A) = - \sum
p_alog_2p_a$ is regarded as the entropy of the source. Here $p_a$
is the probability that the message source produces the message
$a$.  This can be understood to provide a measure of the mean
number of bits, per signal, necessary to encode the output of a
source.  It can also be thought of as a capacity of the source to
carry potential information.  The interest here is in the transfer
of `information for us'.

 Schumacher\cite{Sch95} extended Shannon's ideas to the
quantum domain by introducing  the notion of a `qbit' (the number
of qbits per quantum system is $log_2(H)$, where $H$ is the
dimension of the system Hilbert space). A spin state with two
eigenvalues, say 0 and 1, can be used to encode 1 bit of
information.  To relate this to Shannon's source entropy,
Schumacher represents the signal source by a source density
operator
  \[
 \rho = \sum_{a}p(a)\pi_{a}
 \]
where $\pi_{a}=|a_i\rangle\langle a_i|$ is the set of orthogonal
operators relevant to the measurements that will be performed and
$p(a)$ is the probability of a given eigenvalue being found.  The
von Neumann information $S(\rho)=Tr(\rho log_2 \rho)$ corresponds
to the mean number of qbits, per signal, necessary for efficient
transmission.  The `information' in a quantum system, under this
definition, is therefore defined only in terms of its belonging to
a particular ensemble $\rho$.  It is not possible to speak of the
information of the individual system since the von Neumann
information of the individual pure state is zero (regardless of
the actual values of a and b).

In contrast, in the Bohm interpretation, the information given by
[a,b] has an objective significance for each quantum system, it
determines the trajectories of the individual particles. The
standard interpretation attributes significance only to the
quantum state, leaving the particle's position as somewhat
ambiguous and, in spite of the appearance of co- ordinate labels
in the wave function, there may be a temptation to think that it
is the particles themselves that are interchanged under
teleportation.  This of course is not what happens and the Bohm
approach confirms this conclusion, making it quite clear that no
particle is teleported. What it also shows is that it is the
objective active information contained in the wave function that
is transferred from particle 1 to particle 3.
\chapter{Consistent histories and the Bohm approach}\label{ap:chba}

In a recent paper Griffiths\footnote{The material in this Appendix
originally appeared on the Los Alamos e-print archive\cite{HM00}
as a joint paper with B J Hiley.} claims that the consistent
histories interpretation of quantum mechanics gives rise to
results that contradict those obtained from the Bohm
interpretation. This is in spite of the fact that both claim to
provide a realist interpretation of the formalism without the need
to add any new {\it mathematical} content and both always produce
exactly the same probability predictions of the outcome of
experiments. In contrasting the differences Griffiths argues that
the consistent histories interpretation provides a more physically
reasonable account of quantum phenomena. We examine this claim and
show that the consistent histories approach is not without its
difficulties.

\section{Introduction}

It is well known that realist interpretations of the quantum
formalism are known to be notoriously difficult to sustain and it
is only natural that the two competing approaches, the consistent
history interpretation (CH) \cite{Gri84} \cite{Gri96} and the Bohm
interpretation (BI)\cite{BH87,BH93}, should be carefully compared
and contrasted. Griffiths \cite{Gri99a} is right to explore how
the two approaches apply to interferometers of the type shown in
Figure \ref{fg:ch1}.

Although the predictions of experimental outcomes expressed in
terms of probabilities are identical, Griffiths argues that,
nevertheless, the two approaches actually give very different
accounts of how a particle is supposed to pass through such an
interferometer. After a detailed analysis of experiments based on
Figure \ref{fg:ch1}, he concludes that the CH approach gives a
behaviour that is `physically acceptable', whereas the Bohm
trajectories behave in a way that appears counter-intuitive and
therefore `unacceptable'. This behaviour has even been called
`surrealistic' by some authors\footnote{This original criticism
was made by Englert et al. \cite{ESSW92}. An extensive discussion
of this position has been presented by Hiley, Callaghan and
Maroney \cite{CHM00}.}. Griffiths concludes that a particle is
unlikely to actually behave in such a way so that one can conclude
that the CH interpretation gives a `more acceptable' account of
quantum phenomena.
\begin{figure}[t]
\includegraphics{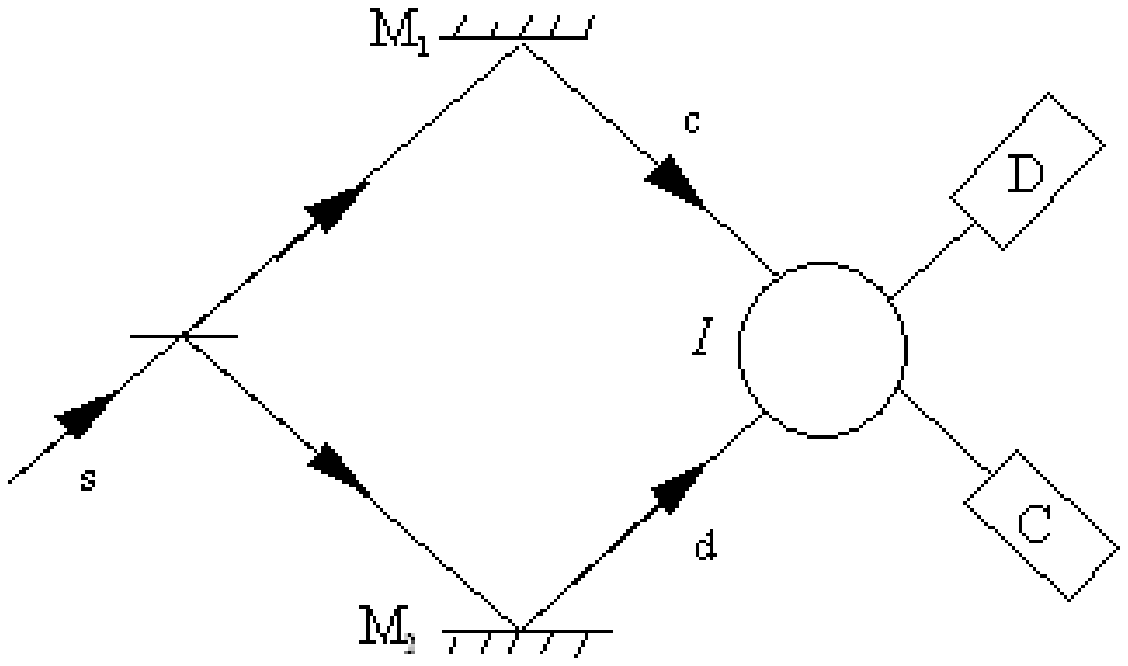}
\caption{Simple interferometer \label{fg:ch1}}
\end{figure}
Notice that these claims are being made in spite of the fact no
new mathematical structure whatsoever is added to the quantum
formalism in either CH or BI, and in consequence all the
experimental predictions of both CH and BI are identical to those
obtained from standard quantum mechanics. Clearly there is a
problem here and the purpose of our paper is to explore how this
difference arises. We will show that CH is not without its
difficulties.

We should remark here in passing that these difficulties have
already been brought out be Bassi and Ghirardi
\cite{BG99a,BG99b,BG99c} and an answer has been given by Griffiths
\cite{Gri00}. At this stage we will not take sides  in this
general debate. Instead will examine carefully how the analysis of
the particle behaviour in CH when applied to the interferometer
shown in Figure \ref{fg:ch1} leads to difficulties similar to
those highlighted by Bassi and Ghirardi \cite{BG99b}.

\section{Histories and trajectories}

The first problem we face in comparing the two approaches is that BI uses a
mathematically well defined concept of a trajectory, whereas CH does not use
such a notion, defining a more general notion of a history.

Let us first deal with the Bohm trajectory, which arises in the
following way. If the particle satisfies the \Sch equation then
the trajectories are identified with the one-parameter solutions
of the real part of the \Sch equation obtained under polar
decomposition of the wave function \cite{BH93}. Clearly these
one-parameter curves are mathematically well defined and
unambiguous.

CH does not use the notion of a trajectory. It uses instead the
concept of a history, which, again, is mathematically well defined
to be a series of projection operators linked by Schr\"{o}dinger
evolution and satisfying a certainty consistency condition
\cite{Gri84}. Although in general a history is not a trajectory,
in the particular example considered by Griffiths, certain
histories can be considered to provide approximate trajectories.
For example, when particles are described by narrow wave packets,
the history can be regarded as defining a kind of broad
`trajectory' or `channel'. It is assumed that in the experiment
shown in figure 1, this channel is narrow enough to allow
comparison with the Bohm trajectories.

To bring out the apparent difference in the predictions of the two
approaches, consider the interferometer shown in Figure
\ref{fg:ch1}. According to CH if we choose the correct framework,
we can say that if $C$ fires, the particle must have travelled
along the path $c$ to the detector and any other path is regarded
as ``dynamically impossible" because it violates the consistency
conditions. The type of trajectories that would be acceptable from
this point of view are sketched in Figure \ref{fg:ch2}. In
contrast a pair of typical Bohm trajectories \footnote{Detailed
examples of these trajectories will be found in Hiley, Callaghan
and Maroney \cite{CHM00}.} are shown in Figure \ref{fg:ch3} . Such
trajectories are clearly not what we would expect from our
experience in the classical world. Furthermore there appears, at
least at first sight, to be no visible structure present that
would `cause' the trajectories to be `reflected' in the region
$I$, although in this region interference between the two beams is
taking place.
\begin{figure}[t]
\includegraphics{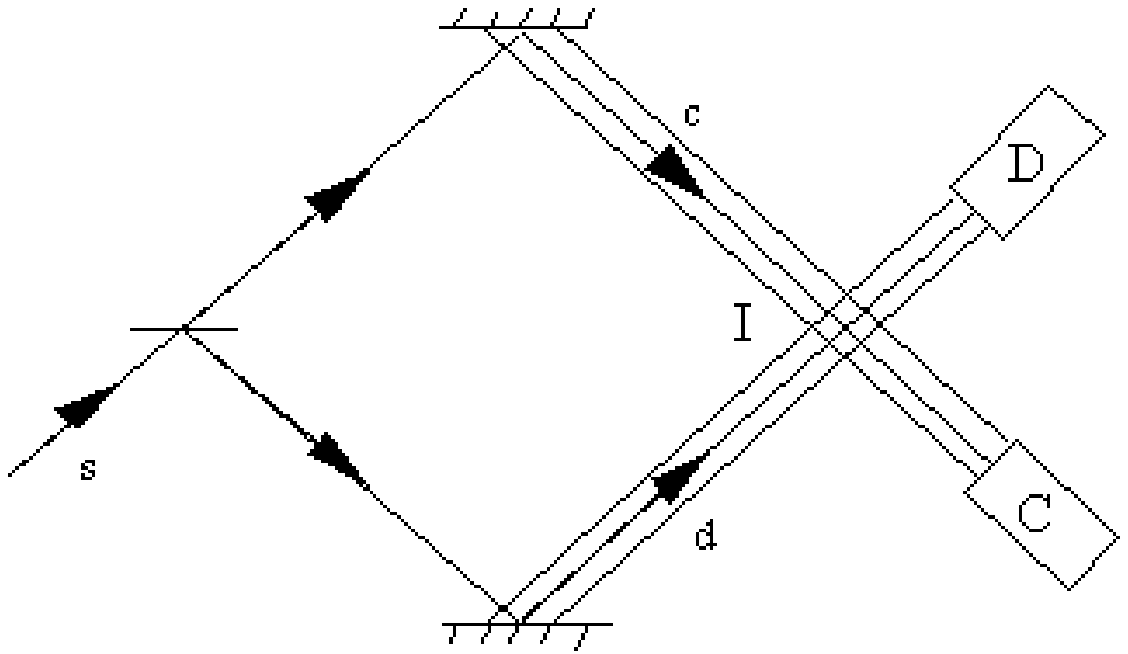}
\caption{The CH `trajectories'. \label{fg:ch2}}
\end{figure}
In the Bohm approach, an additional potential, the quantum
potential, appears in the region of interference and it is this
potential that has a structure which `reflects' the trajectories
as shown in Figure \ref{fg:ch3}. (See Hiley et al. \cite{CHM00}
for more details).
\begin{figure}[t]
\includegraphics{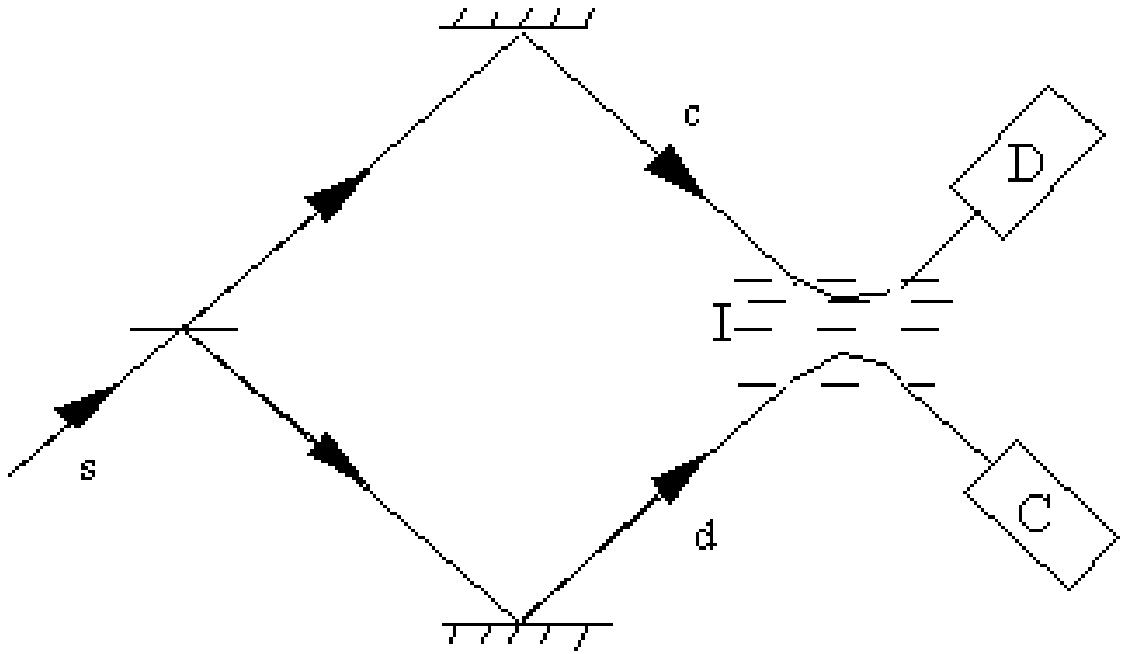}
\caption{The Bohm trajectories.\label{fg:ch3}}
\end{figure}
In this short note we will show that the conclusions reached by
Griffiths \cite{Gri99a} cannot be sustained and that it is not
possible to conclude that the Bohm `trajectories' must be
`unreliable' or `wrong'. We will show that CH cannot be used in
this way and the conclusions drawn by Griffiths are not sound.

\section{The interference experiment}

Let us analyse the experimental situation shown in figure 1 from the point
of view of CH.  A unitary transformation $U(t_{j+1}, t_{j})$ is used to
connect set of projection operators at various times. The times of interest
in this example will be $t_{0}, t_{1}$, and $t_{2}$. $t_{0}$ is a time
before the particle enters the beam splitter, $t_{2}$ is the time at which a
response occurs in one of the detectors $C$ or $D$ and $t_{1}$ is some
intermediary time when the particle is in the interferometer before the
region I is reached by the wave packets.

The transformation for $t_{0}\rightarrow t_{1}$ is
\begin{equation}
|\psi_{0}\rangle = |sCD\rangle_{0}\rightarrow \frac{1}{\surd{2}}[
|cC^{*}D\rangle_{1} + |dCD^{*}\rangle _{1}] \label{eq:ch1}
\end{equation}

The transformation for $t_{1}\rightarrow t_{2}$ is, according to
Griffiths \cite{Gri93a,Gri99a}
\begin{equation}
|cCD\rangle_{1} \rightarrow |C^{*}D\rangle_{2},\hspace{0.5in}
\mbox{and} \hspace{0.5in} |dCD\rangle_{1}\rightarrow
|CD^{*}\rangle_{2} \label{eq:ch2}
\end{equation}

These lead to the histories
\begin{equation}
\psi_{0} \otimes c_{1} \otimes C_{2}^{*},\hspace{0.5in} \mbox{and}
\hspace{0.5in} \psi_{0} \otimes d_{1}\otimes D_{2}^{*]}
\label{eq:ch3}
\end{equation}

Here $\psi_{0}$ is short hand for the projection operator
$\proj{\psi}$ at time $t_{0}$ etc.

These are not the only  possible consistent histories but only these two
histories are used by Griffiths to make judgements about the Bohm
trajectories. The two other possible histories
\begin{equation}
\psi_{0} \otimes d_{1} \otimes C_{2}^{*}, \hspace{0.5in}
\mbox{and} \hspace{0.5in} \psi_{0}\otimes c_{1} \otimes D_{2}^{*}
\label{eq:ch4}
\end{equation}
have zero weight and are therefore deemed to be {\em dynamically
impossible}.

The significance of the histories described by equation
\ref{eq:ch3} is that they give rise to new conditional
probabilities that {\em cannot} be obtained from the Born
probability rule \cite{Gri98}. These conditional probabilities are
\begin{equation}
Pr(c_{1}|\psi_{0}\wedge C_{2}^{*}) = 1,\hspace{0.5in}
Pr(d_{1}|\psi_{0} \wedge D_{2}^{*}) = 1. \label{eq:ch5}
\end{equation}

Starting from a given initial state, $\psi_{0}$, these probabilities are
interpreted as asserting that when the detector $C$ is triggered at $t_{2}$,
one can be certain that, at the time $t_{1}$, the particle was in the
channel {\em c} and not in the channel {\em d}. In other words when $C$
fires we know that the triggering particle must have travelled down path $c$
with certainty.

{\em This is the key new result from which the difference between
the predictions of CH and the Bohm approach arises}. Furthermore
it must be stressed that this result cannot be obtained from the
Born probability rule and is claimed by Griffiths \cite{Gri98} to
be a new result that does not appear in standard quantum
theory\footnote{It should be noted that the converse of
\ref{eq:ch5} must also hold. Namely, if $C$ does not fire then we
can conclude that at $t_{1}$ the particle was not in pathway $c$.
In other words $Pr(c_{1}|\psi_{0}\wedge C_{2}) = 0$}.

Looking again at Figure \ref{fg:ch1}, we notice that there is a
region $I$ where the wave packets travelling down $c$ and $d$
overlap. Here interference can and does take place. In fact
fringes will appear along any vertical plane in this region as can
be easily demonstrated. Indeed this interference is exactly the
same as that produced in a two-slit experiment. The only change is
that the two slits have been replaced by two mirrors. Once this is
realised alarm-bells should ring because the probabilities in
\ref{eq:ch5} imply that we know with certainty through which slit
the particle passed. Indeed equation \ref{eq:ch5} shows that the
particles passing through the lower slit will arrive in the upper
region of the fringe pattern, while those passing through the
upper slit will arrive in the lower half \footnote{Notice that in
criticising the Bohm approach, it is this consistent history
interpreted as a `particle trajectory' that is contrasted with the
Bohm trajectory. The Bohm approach reaches the opposite
conclusion, namely, the particle that goes through the top slit
stays in the top part of the interference pattern \cite{DHP79}}.

Recall that Griffiths claims CH provides a clear and consistent
account of standard quantum mechanics, but the standard theory
denies the possibility of knowing which path the particle took
when interference is present. Thus the interpretation of equation
\ref{eq:ch5} leads to a result that is not part of the standard
quantum theory and in fact contradicts it. Nevertheless CH uses
the authority of the standard approach to strengthen its case
against the Bohm approach. Surely this cannot be correct.

Indeed Griffiths has already discussed the two-slit experiment in
an earlier paper \cite{Gri94}. Here he argues that CH does not
allow us to infer through which slit the particle passes. He
writes; -

\begin{quote}
Given this choice at $t_{3}$ [whether $C$ or $D$ fires], it is inconsistent
to specify a decomposition at time $t_{2}$ [our $t_{1}$] which specifies
which slit the particle has passed through, i.e., by including the projector
corresponding to the particle being in the region of space just behind the $A
$ slit [our $c$], and in another region just behind the $B$ slit [our $d$].
That is (15) [the consistency condition] will not be satisfied if projectors
of this type at time $t_{2}$ [our $t_{1}$] are used along with those
mentioned earlier for time $t_{3}$.
\end{quote}

The only essential difference between the two-slit experiment and
the interferometer described by equation \ref{eq:ch3}  above is in
the position of the detectors. But according to CH measurement
merely reveals what is already there, so that the position of the
detector in the region $I$ or beyond should not affect anything.
Thus there appears to be a contradiction here.

To emphasise this difficulty we will spell out the contradiction
again. The interferometer in Figure \ref{fg:ch1} requires the
amplitude of the incident beam to be split into two before the
beams are brought back together again to overlap in the region
$I$. This is exactly the same process occurring in the two-slit
experiment. Yet in the two-slit experiment we are  not allowed to
infer through which slit the particle passed while retaining
interference, whereas according to Griffiths we are allowed to
talk about which mirror the particle is reflected off, presumably
without also destroying the interference in the region $I$. We
will return to this specific point again later.

One way of avoiding this contradiction is to assume the following: -

1. If we place our detectors in the arms $c$ and $d$ before the
interference region $I$ is reached then we have the consistent
histories described in equation \ref{eq:ch3}. Particles travelling
down $c$ will fire $C$, while those travelling down $d$ will fire
$D$. In this case we have an exact agreement with the Bohm
trajectories.

2. If we place our detectors in the region of interference $I$
then, according to Griffiths \cite{Gri94}, the histories described
by equation \ref{eq:ch3} are no longer consistent. In this case CH
can say nothing about trajectories.

3. If we place our detectors in the positions shown in Figure
\ref{fg:ch1}, then, according to Griffiths \cite{Gri99a}, the
consistent histories are described by equation \ref{eq:ch3} again.
Here the conditional probabilities imply that all the particles
travelling down $c$ will always fire $C$. Bohm trajectories
contradict this result and show that some of these particles will
cause $D$ to fire . These trajectories are shown in Figure
\ref{fg:ch3}.

It could be argued that this patchwork would violate the {\em
one-framework rule}. Namely that one must either use the
consistent histories described by equation \ref{eq:ch3} or use a
set of consistent histories that do not allow us to infer off
which mirror the particle was reflected. This latter would allow
us to account for the interference effects that must appear in the
region $I$.

A typical set of consistent histories that do not allow us to infer through
which slit the particle passed can be constructed in the following way.

Introduce a new set of projection operators $|(c + d)\rangle \langle (c + d)|
$ at $t_{3}$ where $t_{1} < t_{3} < t_{2}$. Then we have the following
possible histories
\begin{equation}
\psi_{0} \otimes (c + d)_{3} \otimes C_{2}^{*},\hspace{0.5in}
\mbox {and} \hspace{0.5in} \psi_{0} \otimes (c + d)_{3} \otimes
D_{2}^{*} \label{eq:ch6}
\end{equation}
Clearly from this set of histories we cannot infer any generalised
notion of a trajectory so that we cannot say from which mirror the
particle is reflected. What this means then is that if we want to
talk about trajectories we must, according to CH, use the
histories described by equation (3) to cover the whole region as,
in fact, Griffiths \cite{Gri99a} actually does. But then surely
the nodes in the interference pattern at $I$ will cause a problem.

To bring out this problem let us first forget about theory and consider what
actually happens experimentally as we move the  detector $C$ along a
straight line towards the mirror $M_{1}$. The detection rate will be
constant as we move it towards the region $I$. Once it enters this region,
we will find that its counting rate varies and will go through several zeros
corresponding to the nodes in the interference pattern. Here we will assume
that the detector is small enough to register these nodes.

Let us examine what happens to the conditional probabilities as
the detector crosses the interference region. Initially according
to \ref{eq:ch5}, the first history gives the conditional
probability $Pr(c_{1}|\psi_{0}\wedge C_{3}^{*}) = 1$. However, at
the nodes this conditional probability cannot even be defined as
$Pr(C_{3}^{*}) = 0$. Let us start again with the closely related
conditional probability, derived from the same history $Pr(
C_{3}^{*}|\psi_{0}\wedge c_{1}) = 1$. Now this probability clearly
cannot be continued across the interference region because
$Pr(C_{3}^{*}) = 0 $ at the nodes, while $Pr(\psi_{0}\wedge c_{1})
= 0.5$ regardless of where the detector is placed. In fact, there
is no consistent history that includes both $c_{1}$ and
$C_{3}^{*}$, when the detector is in the interference region. We
are thus forced to consider different consistent histories in
different regions as we discussed above.

If we follow this prescription then when the detector $C$ is placed on the
mirror side of path $c$, before the beams cross at $I$, we can talk about
trajectories and as stated above these trajectories agree with the
corresponding Bohm trajectories. When $C$ is moved right through and beyond
the region $I$, we can again talk about trajectories. However in the
intermediate region  CH does not allow us to talk about trajectories. This
means that we have no continuity across the region of interference and this
lack of continuity means that it is not possible to conclude that any
`trajectory' defined by $\psi_{0} \otimes c_{1} \otimes C^{*}$ before $C$
reaches the interference region is the same `trajectory' defined by the same
expression after $C$ has passed through the interference region. In other
words we cannot conclude that any particle travelling down $c$ will continue
to travel in the same direction through the region of interference and
emerge still travelling in the same direction to trigger detector $C$.

What this means is that CH cannot be used to draw any conclusions
on the validity or otherwise of the Bohm trajectories. These
latter trajectories are continuous throughout {\em all} regions.
They are straight lines from the mirror until they reach the
region $I$. They continue into the region of interference, but no
longer travel in straight lines parallel to the initial their
paths. They show `kinks' that are characteristic of
interference-type bunching that is needed to account for the
interference \cite{DHP79}. This bunching has the effect of
changing the direction of the paths in such a way that some of
them eventually end up travelling in straight lines towards
detector $D$ and not $C$ as Griffiths would like them to do.

Indeed it is clear that the existence of the interference pattern means that
any theory giving relevance to particle trajectories must give trajectories
that do not move in straight lines directly through the region $I$. The
particles must avoid the nodes in the interference pattern. CH offers us no
reason why the trajectories on the mirror side of $I$ should continue in the
same general direction towards $C$ on the other side of $I$. In order to
match up trajectories we have to make some assumption of how the particles
cross the region of interference. One cannot simply use classical intuition
to help us through this region because classical intuition will not give
interference fringes. Therefore we cannot conclude that the particles
following the trajectories before they enter the region $I$ are the same
particles that follow the trajectories after they have emerged from that
region. This requires a knowledge of how the particles cross the region $I$,
a knowledge that is not supplied by CH.

Where the consistent histories \ref{eq:ch3} could provide a
complete description is when the coherence between the two paths
is destroyed. This could happen if a measurement involving some
irreversible process was made in one of the beams. This would
ensure that there was no interference occurring in the region $I$.
In this case the trajectories would go straight through. This
would mean that the conditional probabilities given in equation
\ref{eq:ch5} would always be satisfied.

But in such a situation the Bohm trajectories would also go
straight through. The particles coming from Mirror $M_{1}$ would
trigger the detector $C$ no matter where it was placed. The reason
for this behaviour in this case is because the wave function is no
longer $\psi_{c} + \psi_{d}$, but we have two incoherent beams,
one described by $\psi_{c}$ and the other by $ \psi_{d}$. This
gives rise to a different quantum potential which does not cause
the particles to be `reflected' in the region $I$. So here there
is no disagreements with CH.

\section{Conclusion}

When coherence between the two beams is destroyed it is possible
to make meaningful inferences about trajectories in CH. These
trajectories imply that any particle reflected from the mirror
$M_{1}$ must end up in detector $C$. In the Bohm approach exactly
the same conclusion is reached so that where the two approaches
can be compared they predict exactly the same results.

When the coherence between the two beams is preserved then CH must
use the consistent histories described by equation \ref{eq:ch6}.
These histories do not allow any inferences about trajectories to
be drawn. Although the consistent histories described by equation
\ref{eq:ch3} enable us to make inferences about particle
trajectories because, as we have shown they lead to disagreement
with experiment. Unlike the situation in CH the Bohm approach can
define the notion of a trajectory which is calculated from the
real part of the Schr\"{o}dinger equation under polar
decomposition. These trajectories are well defined and continuous
throughout the experiment including the region of interference.
Since CH cannot make any meaningful statements about trajectories
in this case it cannot be used to draw any significant conclusions
concerning the validity or otherwise of the Bohm trajectories.
Thus the claim by Griffiths \cite{Gri99a}, namely, that the CH
gives a more reasonable account of the behaviour of particle
trajectories interference experiment shown in Figure \ref{fg:ch1}
than that provided by the Bohm approach cannot be sustained.
\chapter{Unitary Evolution Operators} \label{ap:szun}

The time evolution of a quantum system is usually calculated by
starting with a Hamiltonian energy operator $H$ and the \Sch
equation. When the Hamiltonian is time independant this leads to
the evolution, in the \Sch picture, of a quantum state \ket{\phi}
\[
\ket{\phi (t)}=e^{iHt}\ket{\phi (0)}
\]

The operator  $U=e^{iHt}$ is referred to as the unitary evolution
operator. When the Hamiltonian is not time independant, the
evolution of the system is still described by a unitary evolution
operator, but now $U$ is the solution to the more complex operator
\Sch equation

\begin{eqnarray} \label{eq:unop}
 i\hbar \frac{\partial U}{\partial t}=HU
\end{eqnarray}

$U$ is unitary if $H$ is hermitian and the integration constant is
such that at some given $t=t_0$, then $U(t_0)=I$, the unit matrix.
(We will assume $t_0=0$).

It would be normal practice to proceed by analysing the classical
interaction of a one-atom gas in a box, with a moveable partition,
replace the terms in the classical Hamiltonian with canonically
quantized operators, and then solve the operator \Sch equation.
However, this would tie our analysis to examining the properties
of a particular Hamiltonian. This is precisely the criticism that
was made of Brillouin and Gabor, that they generalised to a
conclusion from a specific form of interaction.

In order to avoid this, we will not attempt to start from a
specific Hamiltonian operator. Instead we will proceed by
constructing unitary time evolution operators, and assume that an
appropriate Hamiltonian can be defined by:

\[
H(t)=i\hbar \frac{\partial U(t)}{\partial t}U^{\dagger }(t)
\]

This Hamiltonian will be hermitian, if $U(t)$ is unitary\footnote{
We shall, nevertheless, present arguments as to the plausibility
of the existence of the necessary Hamiltonians, where it seems
appropriate to do so. According to the theory of quantum
computation \cite{Deu85,Deu89} any unitary operation can, in
principle, be efficiently simulated on a universal quantum
computer. This strongly suggests that any condition more
restrictive than unitarity would be too restrictive not to risk
coming under threat from developments in quantum computing.}.

The problem is therefore simplified to that of determining how the
evolution of the Szilard Engine is constrained by the requirement
of ensuring the evolution operator remains unitary. If the
appropriate transformations of the state of the Szilard Engine can
be expressed with a unitary time evolution operator, then there is
nothing, in principle, to prevent some physical system of being
constructed with an appropriate Hamiltonian. Such a system would
then perform all the necessary operations of the Szilard Engine
without needing an external 'demon' to make measurements or
process information about the system.

A unitary operator is defined by the conditions
\begin{eqnarray*}
    U^{\dagger }U =UU^{\dagger } &=& I
 \\ U\left( \alpha \ket{a}+\beta \ket{b} \right)
   &=& \alpha U\left( \ket{a} \right) +\beta U\left( \ket{b} \right)
\end{eqnarray*}
It can easily be shown that this is equivalent to the statement
that the unitary operator can be written in the form:
\[
U= \sum_n \kb{\phi_n}{\psi_n}
\]
where the \ket{\phi_n} and \ket{\psi_n} are two (usually
different) sets of orthonormal basis for the Hilbert space. If the
instantaneous eigenstates of the unitary operator at time $t$ are
given by the basis \ket{\varphi _n(t)} , then the unitary operator
will have eigenvalues $e^{-i\theta _n(t)}$ and the form

\[
U(t) = \sum_ne^{-i\theta_n(t)}
    \kb{\varphi_n(t)}{\varphi_n(t)}
\]

The associated Hamiltonian is given by

\begin{eqnarray*}
H(t) &=& \sum_n\hbar \frac{d\theta_n(t)}{dt}
    \kb{\varphi_n(t)}{\varphi_n(t)}
 \\ && +\sum_{m,n}e^{i\left(\theta_m(t)-\theta_n(t)\right)}
    \kb{\varphi_m(t)}{\varphi_m(t)}
    \frac{d\left(\kb{\varphi_n(t)}{\varphi_n(t)}\right)}{dt}
\end{eqnarray*}

For the Hamiltonian to be time independant, the eigenstates must
to be constant in time, and the eigenvalues must be of the form:

\[
\theta_n(t) =\frac{E_n}\hbar t
\]

An alternative formulation of this requirement is that the unitary
operator has the form

\[
U(t)U\left(t^{\prime}\right) =U\left(t+t^{\prime }\right)
\]

Instantaneous eigenstates of the time evolution operator are only
eigenstates of the Hamiltonian if they are also constant in time.
There are two special cases of the general time dependant
Hamiltonian:\ rapid transition and adiabatic transition
\cite[Chapter 17]{Mes62b}. These correspond to very fast and very
slow changes in the Hamiltonian, or alternatively, to the change
in the Hamiltonian taking place over a very short or very long
period $\tau $. In the first case (rapid transition) the
asymptotic evolution is given by:

\[
\lim_{\tau\rightarrow 0}U(\tau) =1
\]

while in the second case (adiabatic transition)

\[
\lim_{\tau \rightarrow \infty }U(\tau)=\sum_ne^{\frac i\hbar
\int^\tau E_n(t)dt}\kb{n(\tau)}{n(0)}
\]

where the $\ket{n(t)}>$ are the instantaneous eigenstates of the
Hamiltonian, and $E_n(t) $ are their instantaneous energy levels.

Time dependant Hamiltonians correspond to evolutions that do not
conserve the internal energy of a system. These will require
energy to be drawn from, and deposited in, a work reservoir -
corresponding to work done upon or extracted from the system -
through varying boundary conditions (or 'switching on'
potentials). Unitarity requires only that the variation in the
boundary condition (or potential) does not have any dependance
upon the specific internal state of the system\footnote{The use of
work reservoirs and their connection to time dependant
Hamiltonians is essential to the standard definition of a number
of thermodynamic entities, such as free energy.}. Instead, to
analyse the energy drawn from, or deposited in, the work reservoir
it is the necessary to calculate the change in the energy of the
system once the boundary conditions become fixed again (or the
potential is 'switched off') compared to the energy of the system
beforehand.

A more detailed approach separates the Hamiltonian into a
time-indepedant parts $H_i$, that refers to specific subsystems
$i$, and into a time-dependant part $V_{ij}(t) $, that refers to
the interaction between subsystems $ij$ or with the changing
external conditions.

\[
H(t) =\sum_{ij}(H_i+V_{ij}(t))
\]

If $V_{ij}$ does not commute with all the $H_i$, then the
eigenstates of $H(t)$ will involve superpositions of the
eigenstates of the $H_i$. Strictly speaking, this means there will
not be well-defined energies to the individual subsystems.
Nevertheless, it is usual practice to regard the change of
internal energy of subsystem $i$ as the expectation value of the
internal, time-independant Hamiltonian $\left\langle
H_i\right\rangle$, while the complete system evolves under the
influence of the full Hamiltonian $H(t)$. When the time-dependant
part is ''small'' this can be treated by perturbation theory, but
it is still meaningful when the time-dependant part is ''large'',
as $\left\langle H_i\right\rangle_t$ is still the expectation
value of measuring the internal energy of subsystem $i$ at time
$t$.

The Hamiltonian $H_i$ is also relevant as an internal energy where
a particular subsystem $i$ is in contact with a heat bath. The
interaction with a heat bath generally causes a subsystem density
matrix to diagonalise along the eigenstates of the subsystems
Hamiltonian $H_i$ (see Section \ref{s:szsm1}).
\chapter{Potential Barrier Solutions} \label{ap:szpb}
This Appendix contains a detailed analysis of the eigenstates of
the particle in a box, with a potential barrier of height $V$ and
width $2d$ raised in the centre of the box. We start with the
Hamiltonian given in Equation \ref{eq:szh1}
\[
 H\Psi=\left(-\frac{\hbar^2}{2m}\frac{\partial ^2}{\partial x^2}+V(x) \right) \Psi
\]
with
\begin{eqnarray*}
 V\left( x \right) = \left\{
    \begin{array}{cc}
       \infty & ( x<-L )
    \\      0 & ( -L<x<-d )
    \\      V & ( -d<x<d )
    \\      0 & ( d<x<L )
    \\ \infty & ( L<x )
    \end{array}
\right\}
\end{eqnarray*}
and substitute
\begin{eqnarray*}
    X      &=& \frac{x}{L}
\\  K_{al} &=& \frac{L \sqrt{2mE_l}}{\hbar}
\\  K_{bl} &=& \frac{L \sqrt{2m(E_l - V)}}{\hbar}
\\  K_{cl} &=& \frac{L \sqrt{2m(V - E_l)}}{\hbar}
\\  p      &=& \frac{d}{L}
\\  \epsilon &=& \frac{\hbar^2 \pi^2}{8mL^2}
\end{eqnarray*}
The solution is divided into three regions:
\begin{eqnarray*}
    \Psi _1( X) & -1<X<-p
\\  \Psi _2( X) & -p<X<p
\\  \Psi _3( X) &  p<X<1
\end{eqnarray*}
As the Hamiltonian is symmetric in $X$, then the solutions must be
of odd or even symmetry, imposing the additional conditions
\begin{eqnarray*}
ODD&&
\begin{array}{c}
    \Psi _1( X ) =-\Psi _3( -X )
\\  \Psi _2( X ) =-\Psi _2( -X )
\end{array}
\\
EVEN&&
\begin{array}{c}
    \Psi _1( X) =\Psi _3( -X)
 \\ \Psi _2( X) =\Psi _2( -X)
\end{array}
\end{eqnarray*}
Boundary conditions and continuity requires:
\begin{eqnarray*}
    \Psi_1( -1) &=& \Psi_3( 1) =0
 \\ \Psi_1( -p) &=& \Psi_2( -p)
 \\ \Psi_3(  p) &=& \Psi_2( p)
 \\  \frac{\partial \Psi_1 (X) }{\partial X}\mid_{X=-p}
 &=& \frac{\partial \Psi_2 (X) }{\partial X}\mid_{X=-p}
 \\  \frac{\partial \Psi_2 (X) }{\partial X}\mid_{X=p}
 &=& \frac{\partial \Psi_3 (X) }{\partial X}\mid_{X=p}
\end{eqnarray*}
The energy of the eigenstates are given by
\[
E_l=\frac{4 \epsilon}{\pi^2} (K_{al})^2
\]
\subsubsection{Outside Barrier}
The $l^{th}$ odd or even eigenstates have $\Psi_{1l}(X) $ and
$\Psi _{3l}(X)$ as sine functions of the form
\begin{eqnarray*}
    \Psi_{1l}(X)&=&     A_l\sin ( K_{al}(1+X) )
 \\ \Psi_{3l}(X)&=& \pm A_l\sin ( K_{al}(1-X) )
\end{eqnarray*}
with $\pm$ depending upon the odd or even symmetry.
\subsubsection{Within Barrier}
The form of $\Psi_{2l}(x) $ depends upon the height of the
barrier, $V$ relative to the energy of the eigenstate $E_l$. For
$E_l>V$ , $\Psi_{2l}(x) $ is a sine (odd symmetry) or cosine (even
symmetry) function, with wavenumber $K_{bl}$. When the barrier
height is higher than the energy, $E_l>V$, the wavefunction
becomes a hyperbolic function (sinh for odd symmetry, cosh for
even symmetry) of wavenumber $K_{cl}$. When the barrier height
$V=E$ , the Hamiltonian in the barrier region leads to:
\[
\frac{\partial ^2}{\partial X^2}\Psi = 0
\]
which has solutions
\[
\Psi_l=B_l X+C_l
\]
For odd functions, $C_l=0$ , while for even functions, $B_l=0$.

Two approximations will be made consistently: $p \ll 1$, and when,
for any $a,b$
\begin{eqnarray*}
    tan(a)& = b & \ll 1
 \\ a+l \pi & \approx b
\end{eqnarray*}
with $l=1,2,3 \ldots $ In addition, two further approximations
will be made, in the limit of a narrow, and a high potential
barrier.
\subsubsection{Narrow Barrier Approximation (NBA)}
The NBA is used whenever
\[
K_{bl}p < K_{al}p \ll 1
\]
The first inequality always holds when $E_l \geq V$, and the
second effectively states that the wavelength of the eigenstate is
much larger than the width of the potential barrier. Obviously for
very high quantum numbers this cannot be true. It will be
justified by the fact that we will later be using a thermal
wavefunction, and there will be exponentially little contribution
from high quantum number wavefunctions.

The NBA will also be used for $E<V$ if the energy eigenvalue is
only slightly lower than the barrier so that
\[
K_{cl}p < K_{al}p \ll 1
\]
\subsubsection{High Barrier Approximation (HBA)}
HBA can only be used where $V \gg E$, which approaches the limit
of an infinitely high potential. In this case we assume:
\[
K_{cl}p \gg 1 \gg K_{al}p
\]
where the second inequality is again assuming that very high
quantum numbers are thermodynamically suppressed. The main
approximations are:
\begin{eqnarray*}
    \tanh (K_{cl}p) &\approx& 1-2e^{-2 K_{cl}p}
\\  \sinh (K_{cl}p) &\approx& \frac{1}{2} e^{K_{cl}p}
\\  \cosh (K_{cl}p) &\approx& \frac{1}{2} e^{K_{cl}p}
\end{eqnarray*}

\section{Odd symmetry}

\subsection{$E > V$}

\begin{eqnarray*}
\Psi_l &=&
 \begin{array}{lc}
     A_l \sin (K_{al}(X+1))& -1<X<-p
 \\  B_l \sin (K_{bl} X)   & -p<X<p
 \\  A_l \sin (K_{al}(X-1))&  p<X<1
 \end{array}
\end{eqnarray*}
Continuity conditions lead to:
\begin{eqnarray*}
    A_l &=& B_l \frac{\sin(K_{bl}p)}{\sin (K_{al}(p-1))}
\\ \frac{\tan(K_{bl}p)}{K_{bl}} &=& \frac{\tan(K_{al}(p-1))}{K_{al}}
\end{eqnarray*}
and normalisation gives:
\begin{eqnarray*}
\left| A_l \right|^2 &=& \frac{2K_{al}K_{bl} \sin ^2(K_{bl}p)}
        {L \left( \begin{array}{l}
            K_{al} \sin^2(K_{al}(1-p))(2K_{bl}p    -\sin(2K_{bl}p))
        \\ +K_{bl} \sin^2(K_{bl}p)    (2K_{al}(1-p)-\sin(2K_{al}(1-p)))
        \end{array} \right) }
\end{eqnarray*}
\subsubsection{NBA}
Applying the NBA to $K_{bl}$ in the second continuity equation
leads to
\begin{eqnarray*}
    tan(K_{al}(p-1)) &\approx& K_{al}p
 \\ (K_{al}(p-1))+l \pi &\approx& K_{al}p
 \\ K_{al} &\approx& l \pi
 \\ E_l &\approx& \epsilon (2l)^2
\end{eqnarray*}
This corresponds to the energy of the $n=2l$ (symmetry odd)
solutions of the unperturbed wavefunction. For normalisation we
use
\begin{eqnarray*}
    \sin (K_{al}(p-1)) &\approx& \sin(K_{al}p-l\pi)=(-1)^l\sin(K_{al}p)
 \\ &\approx& (-1)^l K_{al}p
 \\ \sin (2K_{al}(p-1)) &\approx& \sin(2K_{al}p-l2\pi)=\sin(2K_{al}p)
 \\ &\approx &2K_{al}p
\end{eqnarray*}
to give
\begin{eqnarray*}
    A_l &\approx& \frac 1{\sqrt{L}}
 \\ B_l &\approx& (-1)^l \frac{K_{al}}{K_{bl}}\frac{1}{\sqrt{L}}
\end{eqnarray*}

The wavefunction in the region of the barrier approximates
\begin{eqnarray*}
    \Psi_l &=& B_l \sin(K_{bl}X)\approx B_l K_{bl}X
 \\   &\approx& (-1)^l \frac{K_{al}}{\sqrt{L}}
\end{eqnarray*}

\subsection{$E = V$}

\begin{eqnarray*}
\Psi_l &=&
 \begin{array}{lc}
     A_l \sin (K_{al}(X+1))& -1<X<-p
 \\  B_l X   & -p<X<p
 \\  A_l \sin (K_{al}(X-1))&  p<X<1
 \end{array}
\end{eqnarray*}
Continuity:
\begin{eqnarray*}
    A_l&=&B_l \frac {p}{\sin(K_{al}(p-1))}
 \\ \tan(K_{al}(p-1))&=&K_{al}p
\end{eqnarray*}
Normalisation:
\[
\left| A_l \right|^2=\frac{6K_{al}}
{L(4K_{al}p\sin^2(K_{al}(1-p))+6K_{al}(1-p)-3\sin(2K_{al}(1-p)))}
\]
\subsubsection{NBA}
\begin{eqnarray*}
    (K_{al}(p-1))+l\pi &\approx& K_{al}p
  \\ K_{al} &\approx& l\pi
  \\ E_l &\approx& \epsilon (2l)^2
  \\ A_l &\approx& \frac{1}{\sqrt{L}}\left(\frac {1}{1+\frac{8}{3}K_{al}p^2}\right)
       \approx \frac{1}{\sqrt{L}}
  \\ B_l &\approx& (-1)^l K_{al}\frac 1{\sqrt{L}}
\end{eqnarray*}
\subsection{$E < V$}
\begin{eqnarray*}
\Psi_l & = &
\begin{array}{lc}
   A_l \sin (K_{al}(X+1)) & -1<X<-p
\\ B_l \sinh(K_{cl}X)     & -p<X<p
\\ A_l \sin (K_{al}(X-1)) & p<X<1
\end{array}
\end{eqnarray*}
Continuity:
\begin{eqnarray*}
    A_l &=& B_l \frac{\sinh (K_{cl}p)}{\sin (K_{al}(p-1))}
 \\ \frac{\tanh (K_{cl}p)}{K_{cl}} &=& \frac{\tan (K_{al}(p-1))}{K_{al}}
\end{eqnarray*}
Normalisation:
\[
\left| A_l \right| ^2= \frac{2K_{al}K_{cl} \sinh ^2(K_{cl}p)}
    {L \left( \begin{array}{r}
         K_{cl}\sinh ^2(K_{cl}p)(2K_{al}(1-p)-\sin(2K_{al}(1-p)))
     \\ +K_{al} \sin ^2(K_{al}(1-p))(\sinh(2K_{al}p)+2K_{al}p)
     \end{array} \right)}
\]

\subsubsection{NBA}
When $E$ is only slightly larger than $V$ (ie. $K_{cl}p \ll 1$ ),
then the approximations for $sinh$ and $tanh$ match those made for
the NBA with $E<V$ and lead to the same approximate solutions.
\subsubsection{HBA}
\begin{eqnarray*}
    \tan (K_{al}(p-1)) &\approx&
        \frac{K_{al}}{K_{cl}}\left(1-2e^{-2K_{cl}p}\right) \ll 1
 \\ K_{al}(p-1)+l \pi &\approx& \frac{K_{al}}{K_{cl}}
        \left(1-2e^{-2K_{cl}p}\right)
 \\ K_{al} &\approx& \frac{l\pi }{(1-p)}\left( 1+\frac{(1-2e^{-2K_{cl}p})}{K_{cl}(1-p)}\right)^{-1}
 \\  &\approx& \frac{l\pi }{(1-p)}\left( 1-\frac{(1-2e^{-2K_{cl}p})}{K_{cl}(1-p)}\right)
 \\ E_l &\approx &\epsilon \left( \frac{2l}{(1-p)}\right)^2\left( 1-2\frac{(1-2e^{-2K_{cl}p})}{K_{cl}(1-p)}\right)
\end{eqnarray*}
which approaches $E_l\approx \epsilon
\left(\frac{2l}{(1-p)}\right)^2$.

Normalisation of the wavefunction is more complex, but dropping
terms of order $e^{-2K_{cl}p}$ we get
\begin{eqnarray*}
    K_{al}(p-1)+l\pi &\approx &\frac{K_{al}}{K_{cl}}
 \\ \sin (K_{al}(p-1)) &\approx &(-)^l\sin
    \left(\frac{K_{al}}{K_{cl}}\right) \approx (-)^l\left(\frac{K_{al}}{K_{cl}}\right)
 \\ \sin (2K_{al}(1-p)) &\approx & -2\frac{K_{al}}{K_{cl}}
 \\ A_l &\approx& \frac {1}{\sqrt{L(1-p)}}
 \\ B_l &\approx& 2(-)^l\left(\frac{K_{al}}{K_{cl}}\right) \frac{e^{-K_{cl}p}}{\sqrt{L(1-p)}}
\end{eqnarray*}
The wavefunction in the region of the barrier ($|X|<p$) is then:
\[
\Psi_l =B_l \sinh (K_{cl}X) \approx
    \frac{(-)^l}{\sqrt{L}}
    \left(\frac{K_{al}}{K_{cl}}\right)
    \frac{e^{-K_{cl}(p-X)}-e^{-K_{cl}(p+X)}}{\sqrt{(1-p)}}
\]
For large $K_{cl}$ this is non-negligible at the very edges of the
barrier $(|X| \approx p)$.

\subsection{Summary}
The wavefunction and eigenvalues undergo negligible perturbation
until $E>V$. As the potential barrier becomes large, the
wavefunction becomes zero inside the barrier and the wavenumber
increases by a factor of $\frac{1}{1-p}$, causing a minor increase
in the energy levels.
\section{Even symmetry}
\subsection{ $E>V$}
\begin{eqnarray*}
\Psi_l & =&
\begin{array}{lc}
    A_l \sin (K_{al}(1+X)) & -1<X<-p
 \\ C_l \cos(K_{bl}X)      & -p<X<p
 \\ A_l \sin (K_{al}(1-X)) & p<X<1
 \end{array}
\end{eqnarray*}
Continuity:
\begin{eqnarray*}
    A_l &=& C_1 \frac{\cos (K_{bl}p)}{\sin (K_{al}(1-p))}
 \\ \frac {1}{K_{bl}\tan (K_{bl}p)} &=& \frac{\tan (K_{al}(1-p))}{K_{al}}
\end{eqnarray*}

Normalisation:
\[
\left| A\right| ^2 = \frac{2K_{al}K_{bl}\cos ^2(K_{bl}p)}
 {L \left( \begin{array}{l}
    K_{al}\sin^2(K_{al}(1-p))(2K_{bl}p+\sin (2K_{bl}p))
 \\ +K_{bl}\cos ^2(K_{bl}p)(2K_{al}(1-p)-\sin (2K_{al}(1-p)))
 \end{array} \right)}
\]

\subsubsection{NBA}
\begin{eqnarray*}
    \cot (K_{al}(1-p)) &\approx& K_{bl}p \frac{K_{bl}}{K_{al}} \ll 1
 \\ -(K_{al}(1-p))+\frac{(2l-1)}{2}\pi &\approx& K_{bl}p \frac{K_{bl}}{K_{al}}
\end{eqnarray*}

If we assume the barrier is low ($K_{al}\approx K_{bl}$)
\begin{eqnarray*}
    K_{al} &\approx& \frac{(2l-1)}{2}\pi
 \\ E_l &\approx& \epsilon(2l-1)^2
\end{eqnarray*}
which gives the $n=(2l-1)$ unperturbed energies.

When the barrier rises to $V=E$, then $K_{bl}$ becomes small
enough to be negligible, and
\begin{eqnarray*}
    K_{al} &\approx& \frac{(2l-1)}{2(1-p)}\pi
 \\ E_l &=& \epsilon \left( \frac {2l-1}{1-p} \right)^2
\end{eqnarray*}
corresponding to a slightly perturbed ($p \ll 1$) energy of the
$n=(2l-1)$ solutions.
For normalisation
\begin{eqnarray*}
    \sin (K_{al}(1-p))&\approx& -(-1)^l
 \\ \sin (2K_{al}(1-p))&\approx& 2K_{bl}p \frac{K_{bl}}{K_{al}}
 \\ \left| A_l\right| ^2 & \approx& \frac {1}{L}\left(
     \frac {1}{1+p\left(1-\left( \frac{K_{bl}}{K_{al}}\right) ^2\right) }
     \right)
  \\ C_l&\approx&-(-1)^lA_l
\end{eqnarray*}
which gives the unperturbed values when $K_{bl} \approx K_{al}$.
When $K_{bl} \ll K_{al}$ it leads to
\[
\left| A_l\right| ^2 \approx \frac {1}{L}
    \left( \frac {1}{1+p}\right)
\]
 The wavefunction in the region of the
barrier approximates:
\[
\Psi =C_l \cos (K_{bl}X)\approx C_l
\]
\subsection{$E=V$}
\begin{eqnarray*}
\Psi_l  & = &
\begin{array}{lc}
     A_l \sin (K_{al}(1+X)) & -1<X<-p
 \\  C_l & -p<X<p
 \\  A_l \sin (K_{al}(1-X)) & p<X<1
\end{array}
\end{eqnarray*}
Continuity :
\begin{eqnarray*}
    A_l &=& \frac {C_l}{\sin (K_{al}(1-p))}
 \\ A_l K_{al}\cos (K_{al}(1-p))&=&0
\end{eqnarray*}
This has {\em exact} solutions
\begin{eqnarray*}
(K_{al}(1-p)) &=& \frac{(2l-1)\pi }{2}
 \\ K_{al} &=& \frac{(2l-1)}{2(1-p)}\pi
 \\ E_l &=& \epsilon \left( \frac{2l-1}{1-p} \right)^2
\end{eqnarray*}
Normalisation uses $\sin (K_{al}(1-p))=-(-1)^l$ and $\sin
(2K_{al}(1-p))=0$
\begin{eqnarray*}
    A_l=\frac {1}{\sqrt{L(1+p)}}
 \\ C_l=\frac{-(-1)^l}{\sqrt{(1+p)}}
\end{eqnarray*}
\subsection{$E < V$}
\begin{eqnarray*}
\Psi & = &
\begin{array}{lc}
    A_l \sin (K_{al}(1+X)) & -1<X<-p
 \\ C_l \cosh(K_{cl}X)     & -p<X<p
 \\ A_1 \sin (K_{al}(1-X)) & p<X<1
\end{array}
\end{eqnarray*}
Continuity:
\begin{eqnarray*}
    A_l=C_l \frac{\cosh (K_{cl}p)}{\sin (K_{al}(1-p))}
 \\ \frac {1}{K_{cl}\tanh (K_{cl}p)}=-\frac{\tan (K_{al}(1-p))}{K_{al}}
\end{eqnarray*}
Normalisation:
\[
\left| A_l\right| ^2= \frac{2K_{al}K_{cl}\cosh^2(K_{cl}p)}
 {L \left(\begin{array}{r}
    K_{cl}\cosh ^2(K_{cl}p)(2K_{al}(1-p)-\sin(2K_{al}(1-p)))
 \\ +K_{al}\sin ^2(K_{al}(1-p))(2K_{al}p-\sinh(2K_{al}p))
 \end{array}\right)}
\]
\subsubsection{NBA}
When $E$ is only slightly higher than $V$, these results
approximate to the same results as the approximation for NBA with
$E>V$.  These approximations match the exact solutions for $E=V$.
\subsubsection{HBA}
\begin{eqnarray*}
\tan (K_{al}(1-p)) &\approx&
    -\frac{K_{al}}{K_{cl}}\left(1+2e^{-2K_{cl}p}\right) \ll 1
 \\ K_{al}(1-p)-l\pi &\approx &-\frac{K_{al}}{K_{cl}}
        \left(1+2e^{-2K_{cl}p}\right)
 \\ K_{al} &\approx &\frac{l\pi }{(1-p)}
        \left( 1+\frac{(1+2e^{-2K_{cl}p})}{K_{cl}(1-p)}\right)^{-1}
 \\ &\approx& \frac{l\pi}{(1-p)}\left( 1-\frac{(1+2e^{-2K_{cl}p})}{K_{cl}(1-p)}\right)
 \\ E_l &\approx &\epsilon \left( \frac{2l}{(1-p)} \right)^2
        \left( 1-2\frac{(1+2e^{-2K_{cl}p})}{K_{cl}(1-p)}\right)
\end{eqnarray*}
which approaches $E_l\approx \epsilon \left( \frac{2l}{(1-p)}
\right)^2$. For normalisation, we drop terms involving
$e^{-2K_{cl}p}$ and get
\begin{eqnarray*}
    K_{al}(p-1)+l\pi &\approx &\frac{K_{al}}{K_{cl}}
 \\ \sin (K_{al}(p-1)) &\approx& (-)^l\sin \left(\frac{K_{al}}{K_{cl}}\right)
 \\ &\approx& (-)^l\left( \frac{K_{al}}{K_{cl}}\right)
 \\ \sin (2K_{al}(1-p)) &\approx& 2\frac{K_{al}}{K_{cl}}
 \\ A_l &\approx& \frac{1}{\sqrt{L}}\frac {1}{\sqrt{1-p}}
 \\ C_l &\approx& \frac{2(-)^l}{\sqrt{L}}
        \left(\frac{K_{al}}{K_{cl}}\right) \frac{e^{-K_{cl}p}}{\sqrt{1-p}}
\end{eqnarray*}
The wavefunction in the region of the barrier ($\left| X\right|
<p$) is then:
\begin{eqnarray*}
\Psi_l &=&C_l\cosh (K_{cl}X)
 \\ &\approx& \frac{(-)^l}{\sqrt{L}} \left( \frac{K_{al}}{K_{cl}}\right)
     \frac{e^{-K_{cl}(p-X)}+e^{-K_{cl}(p+X)}}{\sqrt{1-p}}
\end{eqnarray*}
For large $K_{cl}$ this is non-negligible at the very edges of the
barrier $(|X| \approx p)$
\subsection{Summary}
The even symmetry wavefunctions undergo a minor perturbation, of
order $p$, as the barrier rises to $V=E$. As the barrier rises
above the energy eigenvalue, the initial peak at $X=0$ becomes a
node, as the wavefunction is expelled from the potential barrier
region. The energy of the $l^{th}$ even eigenstate increases from
the unperturbed value $E_l=\epsilon (2l-1)^2$ to $E_l=\epsilon
\left( \frac{2l}{1-p} \right)^2$. The final energy level, in the
limit of an infinitely high barrier, becomes degenerate with the
$l^{th}$ odd symmetry eigenstate.

\section{Numerical Solutions to Energy Eigenvalues}\label{s:apszpb3}
Given the dependancies of $K_{al}$,
$K_{bl}$ and $K_{cl}$ on $E_l$ and $V$, each of the second of the
continuity equations can be rewritten in the form $f(E_n,V)=0$,
which defines a discrete set of eigenstates for a given $V$. These
eigenstates can be evaluated by numerically solving the
differential equation
\[
\frac{dE_l}{dV}=-\left( \frac{\partial f}{\partial V} \right)
\left/ \left( \frac{\partial f}{\partial E_l} \right) \right.
\]
with initial values given by solutions to $E_l$ for the
unperturbed eigenstates of $V=0$ given in Section \ref{s:szqm1}.
The solutions for $E_l$ can then be used to calculate $K_{al}$ and
so plot the wavefunction itself. \pict{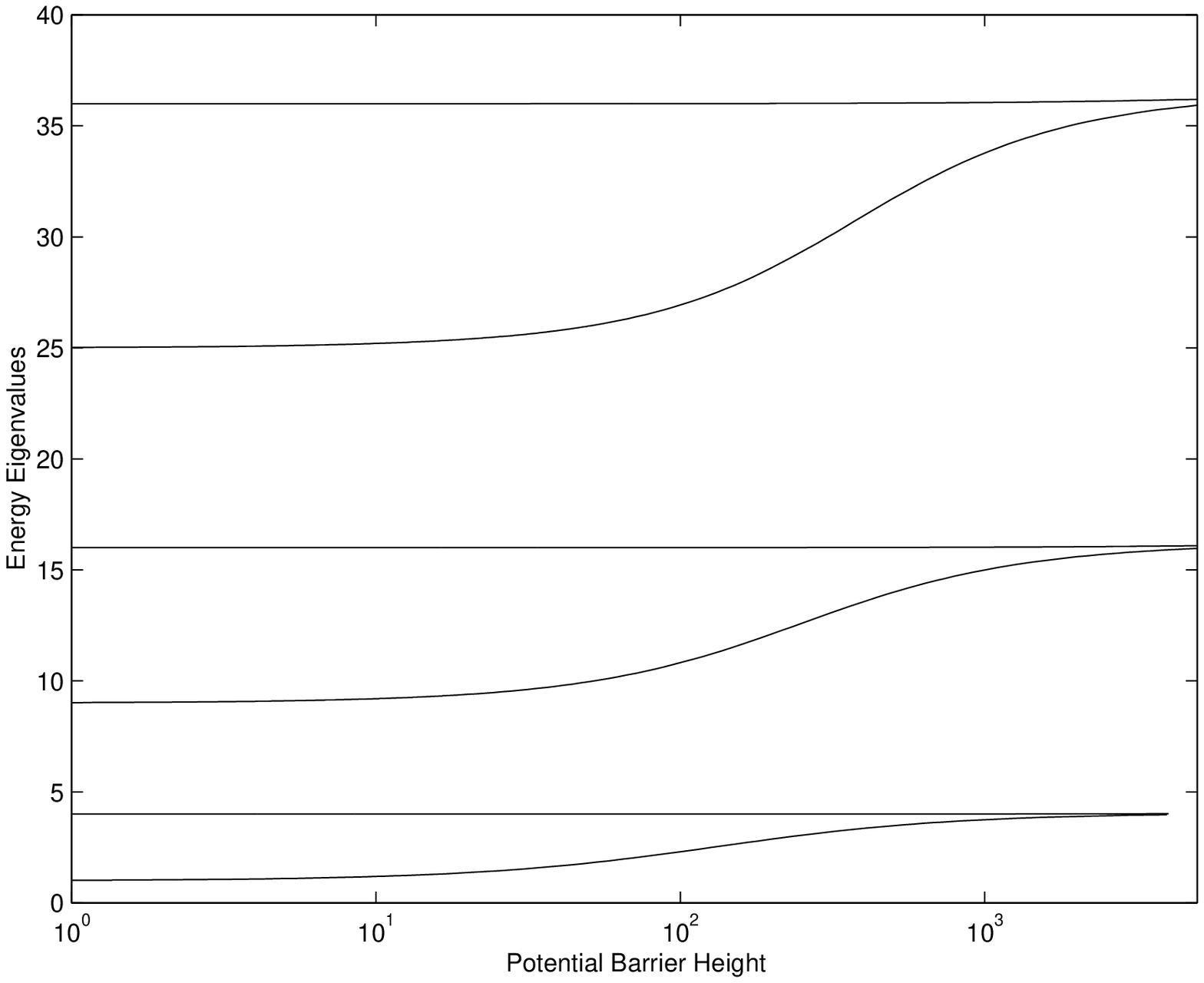}{First six energy
eigenvalues with potential barrier} Numerical solutions to these
equations were evaluated using the MATLAB\cite{MATLAB} analysis
package, and setting $\epsilon=L=1$, $p=0.01$  The results are
shown in Figures \ref{fg:szpb1},\ref{fg:szpb2} and \ref{fg:szpb3}.

Figure \ref{fg:szpb1} shows the changes in the eigenvalues of the
first three (odd and even symmetry) pairs of eigenstates as the
barrier height increases. The eigenvalues pass continuously from
the $V=0$ values, through $V=E$, to the $V \gg E$ values, becoming
degenerate only in the limit of the infinitely high barrier.
\begin{figure}[htb]
\resizebox{\textwidth}{!}{
    \includegraphics{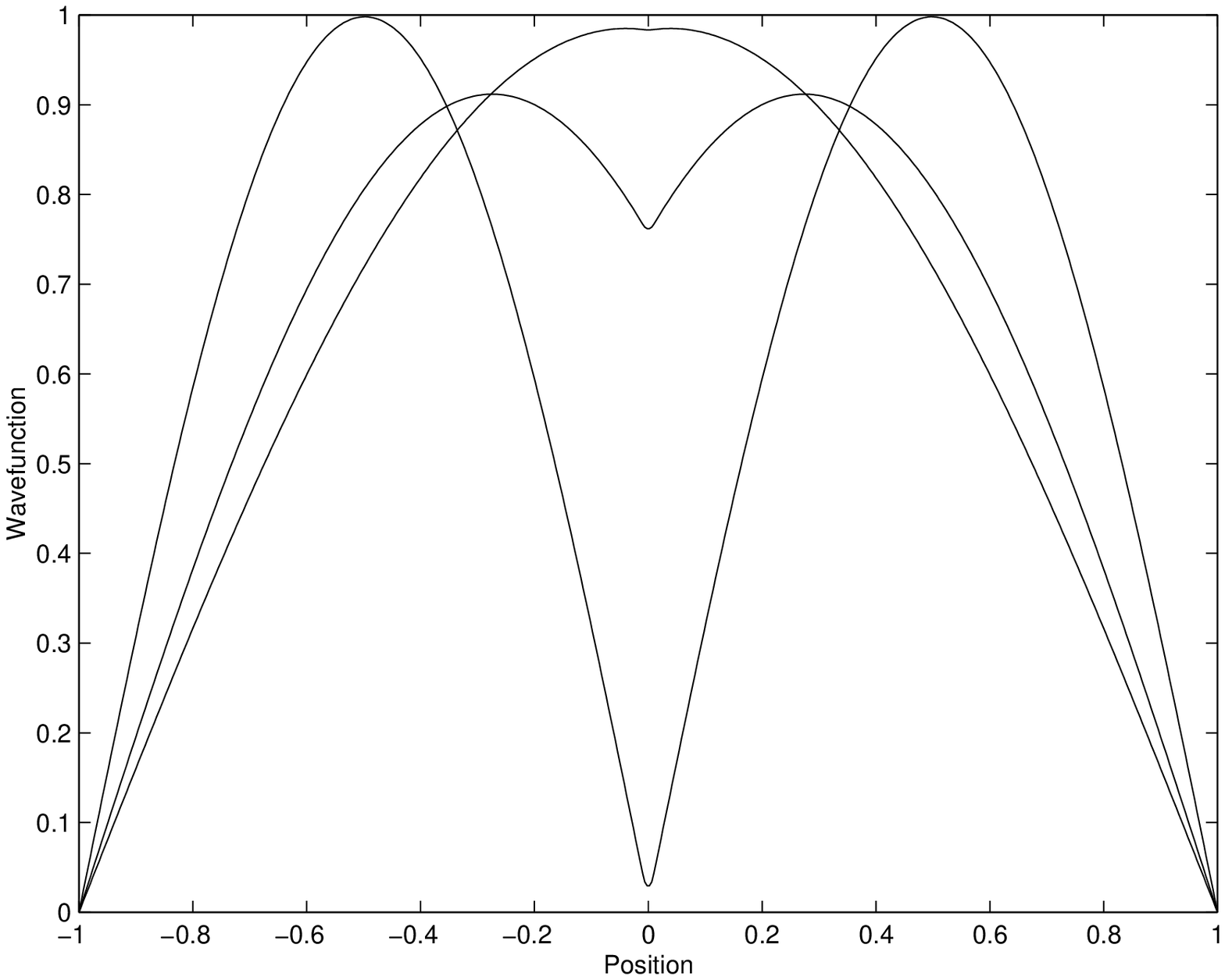}
    \includegraphics{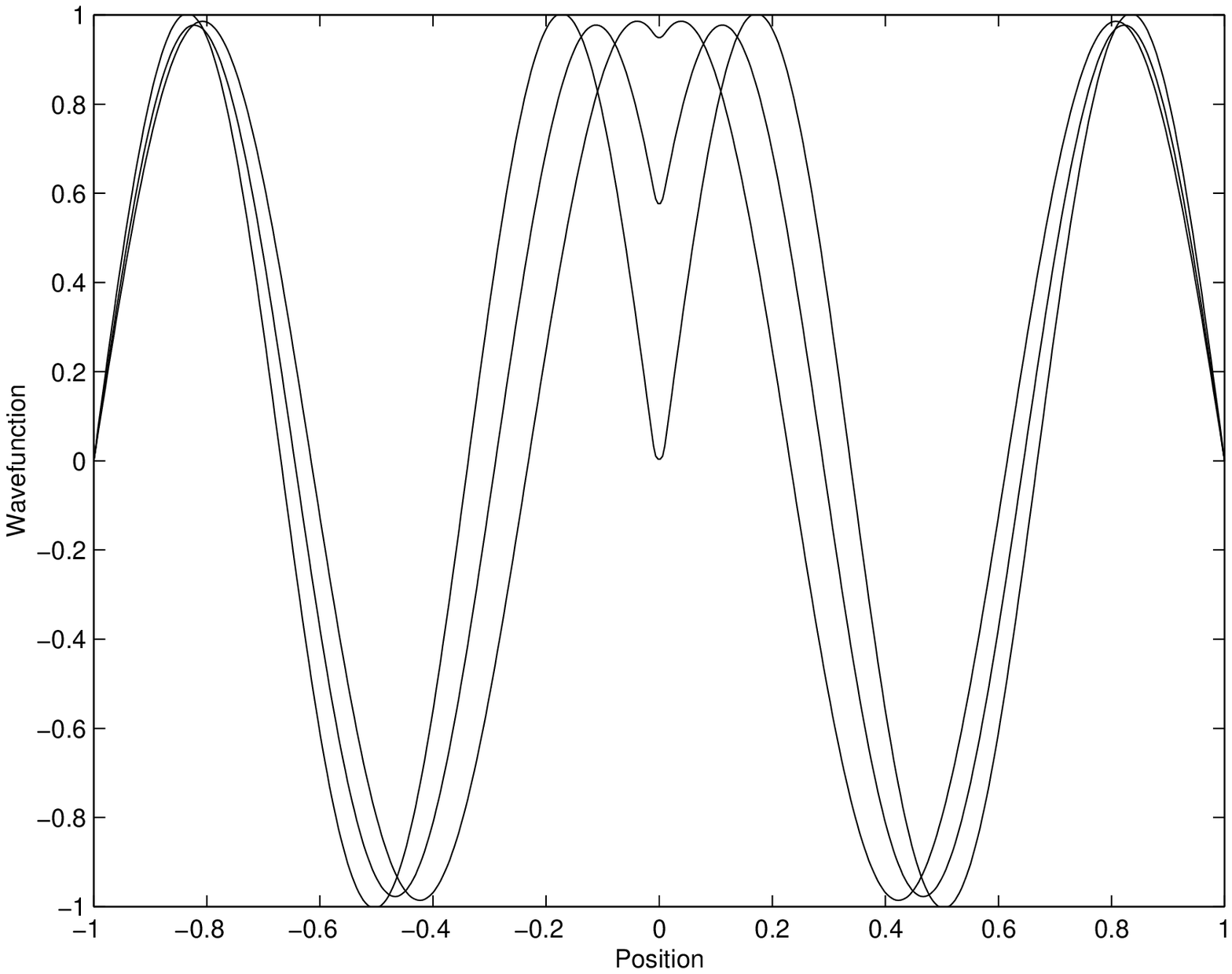}}
\caption{Perturbation of Even Symmetry
Eigenstates\label{fg:szpb2}}
\end{figure}
Figure \ref{fg:szpb2} shows the changes in the wavefunction of the
first and third even symmetry eigenstates, with barrier heights
starting at twice the energy eigenvalue. The eigenstates clearly
develop a node in the center, shortening their wavelengths, until
they reach the same wavelength as the corresponding odd symmetry
state. Finally, in the limit of the infinite potential barrier the
odd and even symmetry states differ only by a change of sign at
they pass through the origin, shown in Figure \ref{fg:szpb3}
\begin{figure}[htb]
\resizebox{\textwidth}{!}{
     \includegraphics{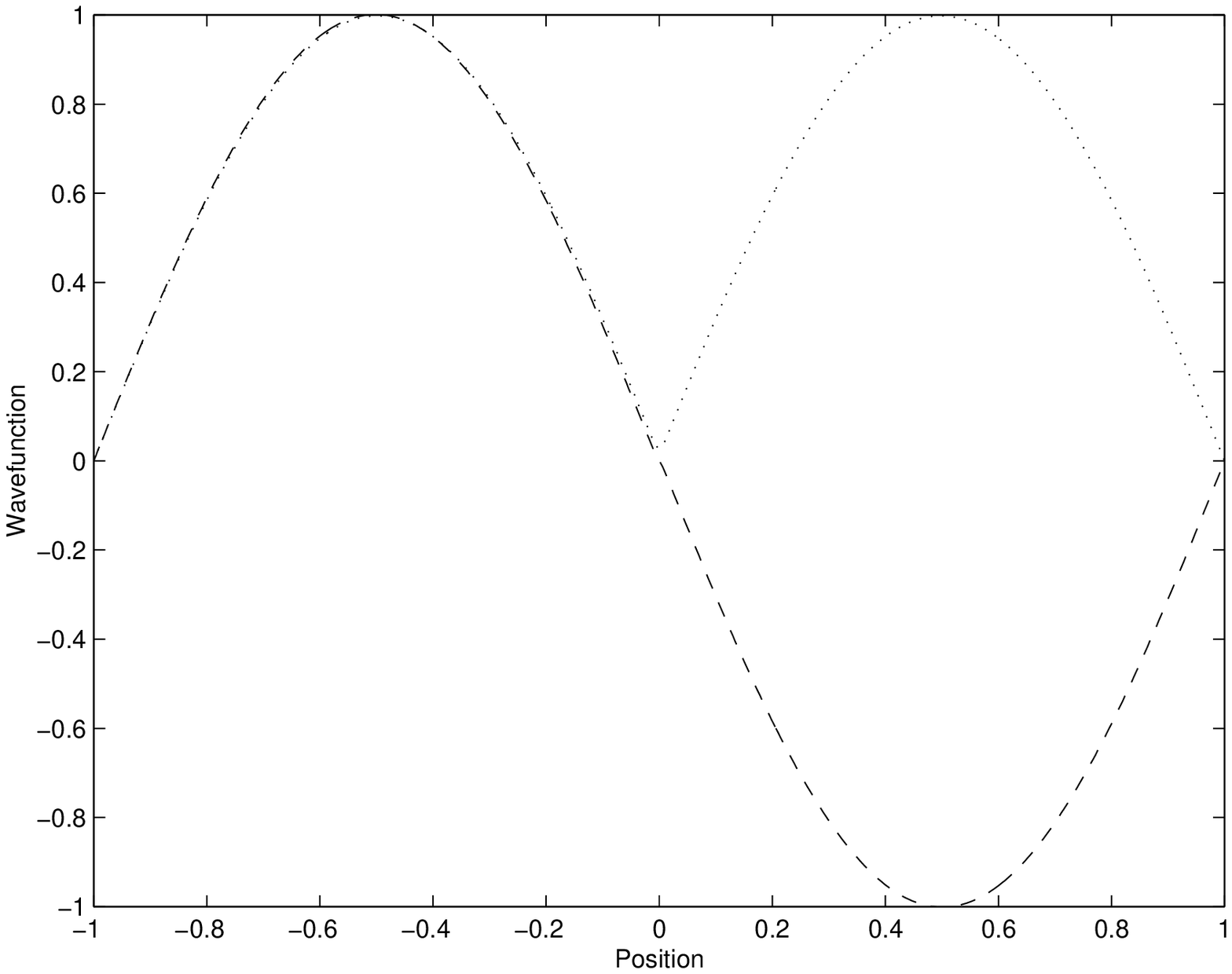}
     \includegraphics{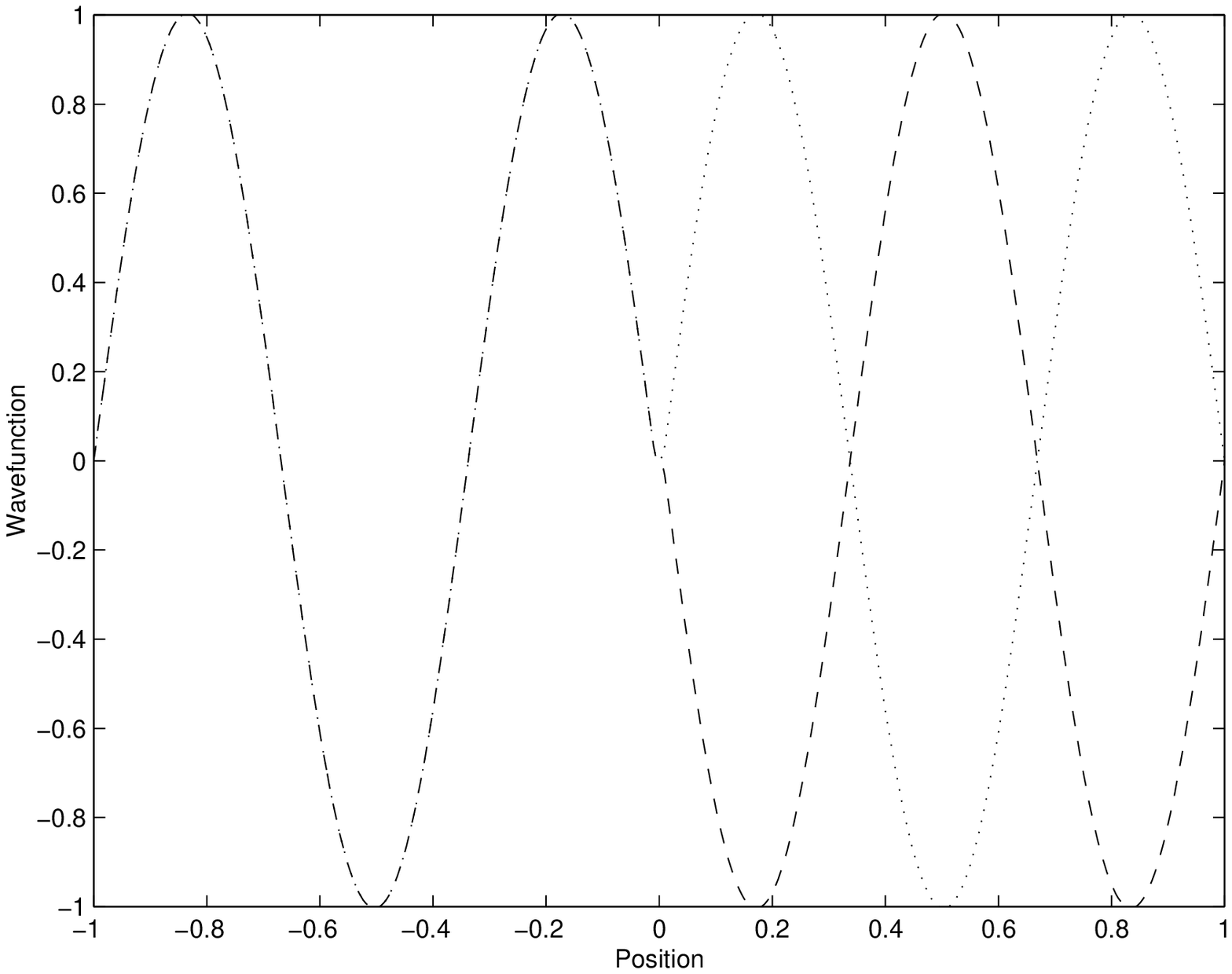}}
     \caption{Degeneracy of Even and Odd Symmetry Eigenstates\label{fg:szpb3}}
\end{figure}
\chapter{Energy of Perturbed Airy Functions}
\label{ap:szai}

The insertion of shelves at height $h$ into the wavefunction of a
quantum weight will cause a perturbation of the energy
eigenvalues. Due to the nature of the Airy functions, it is not
possible to calculate the effect of this perturbation exactly.
However, it can be estimated for two extremes, and be shown to
involve negligible energy changes for high quantum numbers. It is
argued that it is reasonable to assume that there is also
negligible energy changes between the two extremes.

This is based upon calculations in \cite{NIST} for the quantum
state of a particle in a linear potential between two barriers. We
will calculate the effect of inserting a potential barrier for
high quantum numbers $n$, where the shelf height is large and
small in comparison with the characteristic height of the
wavefunction $-a_nH$. The unperturbed energy of the state is
$E_n=-a_nMgH$. We will always use the asymptotic approximation
$a_n=-\left(\frac{3 \pi n}{2}\right)^{2/3}$.

\subsubsection{Large Shelf Height}
If the shelf is inserted at a height $h \gg -a_nH$ then there is
negligible perturbation of the wavefunction, as the potential
changes only in a region where the wavefunction is negligibly
small. The final energy is therefore approximately the same as the
unperturbed energy:
\[
E_n^{(1)}=\left(\frac{3 \pi n}{2}\right)^{2/3}MgH
\]

\subsubsection{Small Shelf Height}
If the shelf is inserted at a height $h \ll -a_nH$, the
wavefunction is split into two, above and below the shelf. We will
start by assuming that the shelf is inserted at a node, and that
$m$ nodes are above the shelf height (see Figure \ref{fg:airy2}).
The number of nodes below the shelf height is given by $k=n-m$,
and the shelf height is $h=(a_m-a_n)H$.

The low shelf height is equivalent to the assumption that $k \ll
n$. There are two subcases, depending upon whether $k$ itself is
large or small.

If $k$ is small, then $m \approx n$ and there is negligible
probability of the weight being located below the shelf, and we
only need to consider the wavefunction above. This is the same as
an unperturbed wavefunction with $m$ nodes, raised by a height
$h$, and so will have an energy
\begin{eqnarray*}
E_m^{(2)}&=&-a_mMgH+Mgh \\
    &=&-a_nMgH \\
    &=&E_n
\end{eqnarray*}

If $1 \ll k \ll n$ we need to estimate the energy of the
wavefunction above and below the barrier. Above the barrier, we
again have a wavefunction with energy $E_m^{(2)}=E_n$. Below the
barrier, it can be shown that the energy eigenstates approximate
those of a square well potential (in effect, the variation in
gravitational potential is negligible in comparison to the kinetic
energy). The energy of these states are
\begin{eqnarray*}
E_n^{(2)}&=&\frac{\hbar^2 \pi^2}{2Mh^2}k^2 \\
    &=& MgH \left(\frac{4 \pi^4}{9} \right)^{\frac{1}{3}}
        \left(\frac{k}{n^{2/3}-m^{2/3}}\right)^2 \\
    &\approx& MgH \left(\frac{2 \pi^2}{3} \right)^{\frac{2}{3}}
        n^{-\frac{4}{3}}k^2
        \left(1-\left(1-\frac{2k}{3n}\right)\right)^{-2} \\
    &\approx&\left(\frac{3 \pi n}{2}\right)^{2/3}MgH
\end{eqnarray*}
which approximates the unperturbed energy.

This shows if the shelf is inserted at a node, the energy values
are the same, regardless of whether the weight is trapped above or
below the shelf. Inserting the shelf adiabatically at some other
point will add, at most, one node to the wavefunction. As the
energies vary slowly with the quantum number (or number of nodes),
there will also be negligible change in energy states if the shelf
is inserted between nodes.

This demonstrates that for the three cases
\begin{eqnarray*}
k \ll m \leq n \\
1 \ll k \ll m < n \\
k=n
\end{eqnarray*}
the energy is negligibly affected by the insertion of the barrier.
There remains only the case where the shelf height is comparable
to the characteristic height of the eigenstate and $1 < m \ll k <
n$. Unfortunately this does not yield a simple solution. However,
as we have shown, the states above and below the shelf, for values
of $k$ both higher and lower than this region have the energy
eigenvalues $E \approx -a_nMgH$. As the energy values must be
monotonically increasing in the intermediate region, it is
reasonable to assume that they will also have this form. The
insertion of the potential barrier will then have negligible
effect upon the energy levels of the high quantum number states.
\chapter{Energy Fluctuations}
\label{ap:szfluc}

We suppose that the expansion of the gas, described in Section
\ref{s:szsm2} takes places in $n$ steps, and after each step the
gas is allowed to thermalise through interactions with an
environment. This thermalisation randomises the individual state
of the gas from step to step.

The energy transferred by the $i^\prime th$ state, on the
$m^\prime th$ step is denoted by $\delta E_{im}$ and the
probability of the gas being in the $i^\prime th$ state, on the
$m^\prime th$ step is $p_{im}.$ Clearly $ \sum_ip_{im}=1.$ The
randomisation of the state between steps means that the
probabilities at different steps can be treated as statistically
independant.

We describe the ordered set of states that the system passes
through on a given expansion by the array $\alpha=(ijk\ldots)$,
which means the system is in the $i^\prime th$ state on the first
step, $j^\prime th$ state on the second step, etc. We also write
this as $\alpha_1=i$, $\alpha_2=j$ or
$\alpha=(\alpha_1\alpha_2\ldots)$. The probability of $\alpha$
occurring is given by
\begin{eqnarray*}
P_\alpha  &=&\prod_{m=1,n} p_{\alpha_m m} \\
\sum_\alpha P_\alpha  &=&\sum_{\alpha_1,\alpha_2,\ldots}
    \left(\prod_{m=1,n} p_{\alpha_m m}\right)=1
\end{eqnarray*}
and the energy transferred on such an expansion is
\[
E_\alpha=\sum_m \delta E_{\alpha_m m}
\]
We also need to note the following identities

\begin{eqnarray*}
\sum_{\alpha_1,\alpha_2,\ldots}
    \left(\prod_{m=1,n}p_{\alpha_m m} f(\alpha_k) \right)
    &=&\sum_{\alpha_k} p_{\alpha_k k}f(\alpha_k)  \\
\sum_{\alpha_1,\alpha_2,\ldots}
    \left(\prod_{m=1,n} p_{\alpha_m m}f(\alpha_k,\alpha_l)\right)
    &=&\sum_{\alpha_k,\alpha_l}p_{\alpha_k k}p_{\alpha_l l}f(\alpha_k,\alpha_l)
\end{eqnarray*}
etc.

We can now write the following results

Mean energy transfer and fluctuation on $m^{\prime }th$ step:

\begin{eqnarray*}
\mean{\delta E_m} &=&\sum_ip_{im}\delta E_{im} \\
\mean{\delta E_m^2} &=&\sum_ip_{im}\left( \delta E_{im}\right) ^2
\end{eqnarray*}

Mean energy transfer and fluctuation of the overall expansion is:
\begin{eqnarray*}
\mean{E}&=&\sum_\alpha \left( \prod_{l=1,n}p_{\alpha_l l}\right)
    \left(\sum_m\delta E_{\alpha_m m}\right)
    =\sum_m\left(\sum_{\alpha_m}p_{\alpha_m m}\delta E_{\alpha_m m}\right)  \\
&=&\sum_{m=1,n}\mean{\delta E_m} \\
\mean{E}^2 &=&\sum_m\mean{\delta E_m}^2
    +2\sum_{l<m}\mean{\delta E_l}\mean{\delta E_m} \\
\mean{E^2}&=&\sum_\alpha\left(\prod_{l=1,n}p_{\alpha_l l}\right)
    \left(\sum_m\delta E_{\alpha_m m}\right)^2
    =\sum_\alpha \left(\prod_{k=1,n}p_{\alpha_k k}\right)
        \left(\sum_{l,m}\delta E_{\alpha_l l}\delta E_{\alpha_m m}\right)  \\
&=&\sum_m\left(\sum_{\alpha_m}p_{\alpha_m m}
    \left(\delta E_{\alpha_m m}\right)^2\right)
    +2\sum_{l<m}\sum_{\alpha_m,\alpha_l}
        p_{\alpha_m m}p_{\alpha _l l}\delta E_{\alpha_m m}\delta E_{\alpha_l l} \\
&=&\sum_m\mean{\delta E_m^2}+2\sum_{l<m}\mean{\delta E_l}\mean{\delta E_m} \\
\mean{E^2}-\mean{E}^2 &=&\sum_{m=1,n}
    \left(\mean{\delta E_m^2}-\mean{\delta E_m}^2\right)
\end{eqnarray*}

For the expansions in Section \ref{s:szsm2}, we have $\mean{\delta
E_m^2}-\mean{\delta E_m}^2=2\mean{\delta E_m}^2$. We may therefore
introduce the following inequalities:

\begin{eqnarray*}
\mean{E^2}-\mean{E}^2 & \leq & 2n\mean{\delta E_{\max}}^2 \\
\mean{E}^2 & \geq & (n\mean{\delta E_{\min}})^2
\end{eqnarray*}

and prove our required result that

\[
\frac{\mean{E^2}-\mean{E}^2}{\mean{E}^2} \leq
     \frac 2n\left(\frac{\mean{\delta E_{\max}}}
     {\mean{\delta E_{\min}}}\right) ^2
\]

The ratio $\frac{\mean{\delta E_{\max}}}{\mean{\delta E_{\min}}}$
approaches $\frac{P_{\max}}{P_{\min}}$, where $P_m=\frac{\partial
E_m}{\partial X}$ is the generalised pressure, as the size of the
step reduces, and so becomes independant of $n$. As
$n=t/\tau_\theta$, where $\tau_\theta$ is a characteristic thermal
relaxation time, and $t$ is the length of time of the expansion,
the size of fluctuations in the total energy transfer can be made
negligible if the expansion takes place sufficiently slowly with
respect to $\tau_\theta$.

It should be clear that the result obtained here is not the same
as, although it is similar to, the usual fluctuation formula. The
usual formula refers to the deviation from the mean value of the
thermodynamic variable at a given time, and is reciprocally
related to the number of constituents of the system. The formula
here refers to potentially large fluctuations at any particular
moment, for systems which may have only a few constituents, but
which, when integrated over a significant period of time, still
leads to negligible long term fluctuations.
\chapter{Free Energy and Temperature} \label{ap:szfree}

The free energy, $F$ is only one of a number of thermodynamics
potentials that may be associated with a system. For example, we
can also use the energy $E$, Gibbs function $G$ or enthalpy $H,$
defined by
\begin{eqnarray*}
E & & \\
F &=&E-TS \\
G &=&E-TS+PV \\
H &=&E+PV
\end{eqnarray*}
to describe the behaviour of a system. These terms can them be
generalised even further, when the number of particles is allowed
to vary. The choice of which thermodynamic potential to use
entirely is a question of which constraints are acting upon the
system, or which pair of the variables $S$, $T$, $P$ and $V$ are
controlled.

In Section \ref{s:szth1}, the significance of $F$ and from that
$S$ was derived from the work that can be extracted from an
isothermal expansion of a system. In terms of classical
thermodynamics potentials, this is derived from the infinitesimal
relationships
\[
dF=dE-TdS-SdT
\]
and the general relationship for heat and work acting upon a system
\[
dE=TdS-PdV
\]
which is equivalent to the statistical mechanical relationship
\[
dE=\sum_iE_idp_i+\sum_ip_idE_i
\]
This gives
\[
dF=-SdT-PdV
\]
and clearly, if the temperature is held fixed
\[
dF=-PdV
\]
so the change in free energy is equal to the negative of the work
extracted from the system, $dW=PdV$.

Now, if the temperature is not held fixed, then we clearly have
\[
dF+dW=-SdT
\]
If we can interpret the work as being the gain in free energy of a
second system (which has no change in entropy), such as a raised
weight, we can express this equation as being a net gain in free
energy $\Delta F=dF+dW,$ of a closed system, when a quantity of
entropy $S$ is taken through a temperature difference $\Delta
T=dT$. We will express this as

\begin{equation}
\Delta F=-S\Delta T \label{eq:free}
\end{equation}
and refer to this as the characteristic equation for free energy
in the presence of a temperature differences.

\paragraph{Adiabatic expansion}

The derivation above is essentially based upon the adiabatic
(essentially isolated) expansion of a gas. If we take a gas in
essential isolation, and extract work from it's expansion, the
free energy before and after is given by

\begin{eqnarray*}
F_1 &=&E_1-T_1S_1 \\
F_2 &=&E_2-T_2S_2
\end{eqnarray*}

As the expansion is reversible but thermally isolated we have
$\Delta W=E_1-E_2$ and $S_2=S_1=S$. This gives

\begin{eqnarray*}
\Delta F &=&F_2-F_1 \\
    &=&-\Delta W -(T_2-T_1) S \\
\Delta F+\Delta W &=&-S\Delta T
\end{eqnarray*}

\paragraph{Carnot Cycle}

The Carnot heat engine operates by drawing energy in the form of
heat $Q_1$ from a heat bath at temperature $T_1,$ extracting $W$
as work, and depositing $Q_2$ in a heat bath at temperature $T_2$.
The usual means of achieving this would be to have gas initially
in contact with the heat bath $T_1$. This is isothermally
expanded, drawing the $Q_1$ out as work. The gas is then removed
from contact with the heat bath, and adiabatically expanded, again
extracting work, until it's temperature falls to $T_2$. It is then
placed in contact with the $T_2$ heat bath, isothermally
compressed, depositing the $Q_2$ heat, and is then isolated again,
and adiabatically compressed further until it returns to it's
initial volume, at which point, on a reversible cycle, it will
have risen back to temperature $T_2$.

For a reversible process the entropy loss from the $T_1$ heat bath must
match the gain from the $T_2$ heat bath, so
\[
S=\frac{Q_1}{T_1}=\frac{Q_2}{T_2}
\]
and conservation of energy is
\[
Q_1=Q_2+W
\]
This is usually rearranged to give the Carnot efficiency
\[
\frac W{Q_1}=1-\frac{T_2}{T_1}
\]
However, there is an alternative way of expressing this
\[
W=-S\left( T_2-T_1\right)
\]
which is again the characteristic equation \ref{eq:free} for free
energy in the presence of two different temperatures.

\paragraph{Entropy Engine}

The two previous examples can be regarded as equations about the
movement of energy between, or within, systems, rather than an
equation about the gain in free energy from moving entropy between
different temperatures. We will now demonstrate a system, based
upon the Szilard Engine, and with some similarities to the heat
engines described in Chapter \ref{ch:szsol}, but which produces
this characteristic equation without any energy changes taking
place anywhere. This makes it very clear that the gain in free
energy is actually a consequence of the transferral of entropy
between temperatures.

First we start with two Szilard boxes, each containing a single
atom, and initially of length $L$. The boxes are initially at
temperatures $T_1$ and $T_2$, but are thermally isolated. A
partition is raised in the centre of the first box, dividing the
one atom gas into left and right subensembles, and a piston is
inserted between them.

Now, however, we modify the behaviour of the box, as shown in
Figure \ref{fg:free}. The piston is constrained so that it cannot
move to the right, even when the gas is located to the left. If
the gas is on the right, the piston moves to the left, as before.
However, regardless of the location of the gas, the right most
wall of the box starts to move to the left, at the same rate as a
left-moving piston would. When the wall of the box reaches the
initial center, it stops. \pict{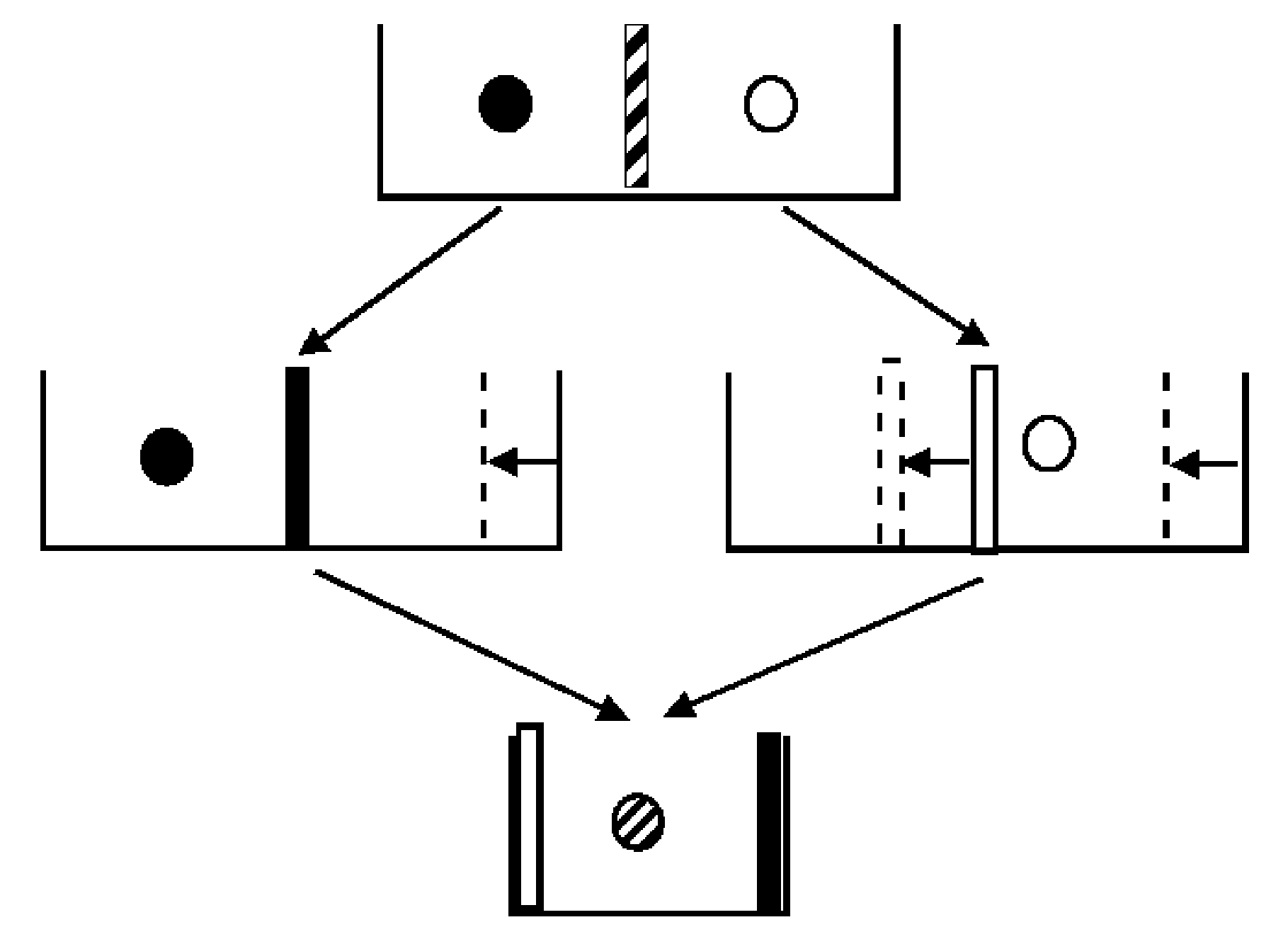}{The Entropy Engine}
 If the gas was initially located
to the left of the partition, the right wall simply moves in
through empty space on the right, until it comes against the
piston, still in the center. If, on the other hand, the gas was
initially located to the right, the piston and wall move leftwards
together. As long as this movement is sufficiently slow, any work
done upon the piston would be matched by work done by the wall. In
effect, no work is done upon the gas at all, as the right
subensemble keeps the same volume throughout. At the end of this
process, the wall is in the initial center, and the piston is
against the left wall. The initially left and right gas
subensembles are now entirely overlapping.

The remarkable consequence of this is that we have compressed the
gas to exactly half it's volume, but without performing any work
upon it, or changing it's energy in any other way. We have
succeeded in this by increasing the entropy of the piston, which
is now in a mixture of being on the left or the right. This effect
is possible only from statistical mechanics: there is no
equivalent process in phenomenological thermodynamics by which
such a compression can be achieved without any flow of energy.

We now remove the piston states from the ends of the first box and
insert them in the corresponding ends of the second. We can now
perform the same operation on the second Szilard box, in the
reverse direction. The second gas expands to twice it's volume,
while the piston is restored to it's initial state. Again, there
is no contact with a heat bath, no work is extracted from the gas,
and it's internal energy is constant throughout.

It is clear that we can continue this process indefinitely,
compressing the first gas to as small a fraction of it's initial
volume as we like, without ever performing any work upon it.
However, the cost is that we must proportionately increase the
volume occupied by the second. The only quantity that is
transferred between the two systems is the mixing entropy of the
piston, $S=k\ln 2$. However, by compressing the first gas we
increase it's free energy by $kT_1 \ln 2$, and by expanding the
second gas, reduce it's free energy by $kT_2 \ln 2$. The net
change in free energy is
\begin{eqnarray*}
\Delta F &=&k\ln 2 (T_1-T_2) \\
\ &=&-S\Delta T
\end{eqnarray*}
which corresponds to the entropy being transferred through the
temperature difference $\Delta T=T_2-T_1$ This provides an
'engine' by which the free energy of a system can be increased
indefinitely, by reversibly moving entropy between parts of the
system at different temperatures, yet without any energy flow
taking place.

Of course, when we attempt to extract this free energy by, for
example, isothermally restoring the system to it's initial
configuration, we simply recover the Carnot cycle efficiency.
Although this process produces the characteristic equation
\ref{eq:free} for the free energy change in the presence of
different temperatures, it should be clear that it's physical
basis is a purely statistical mechanical effect, and quite
different to the more commonly encountered manifestation in the
adiabatic expansion and Carnot Cycle.
\chapter{Free Energy and Non-Equilibrium Systems} \label{ap:noneq}

In Section \ref{s:szth1} the free energy of a system in a
canonical thermodynamical state $\rho
=\frac{1}{Z}e^{-\frac{H}{kT}}$ was derived in terms of it's
partition function $Z=\trace{e^{-\frac{H}{kT}}}=\sum_i
e^{-\frac{E_i}{kT}}$ as $F=-kT\ln Z$.

When the Hilbert space is split into subspaces with partition
functions $Z_a=\sum_{i\subset a}e^{-\frac{E_i}{kT}}$, the
equilibrium probability of the density matrix being in the
subspace is
\[
p_a=\sum_{i\subset a}p_i=\frac{\sum_{i\subset a}e^{-\frac{E_i}{kT}}}{%
\sum_ie^{-\frac{E_i}{kT}}}=\frac{Z_a}Z
\]

From this we can express the free energy of a density matrix in
equilibrium in the subspace by
\[
F_a=F-kT\ln p_a
\]

When the Hilbert space is divided into several orthogonal
subspaces, so that $Z=\sum_\alpha Z_\alpha $, we have
\begin{eqnarray*}
F &=&\sum_\alpha p_\alpha F_\alpha
    +kT\sum_\alpha p_\alpha \ln p_\alpha  \\
    &=&-kT \ln \left(
        \sum_\alpha e^{-\frac{F_\alpha}{kT}}\right)
\end{eqnarray*}
and also
\[
p_a=\frac{e^{-\frac{F_a}{kT}}}
    {\sum_\alpha e^{-\frac{F_\alpha}{kT}}}
\]

The equilibrium density matrix may be expressed as

\begin{eqnarray*}
\rho &=&\sum_\alpha p_\alpha \rho _\alpha \\
    &=&\frac{1}{Z}\sum_\alpha e^{-\frac{F_a}{kT}}\rho _\alpha
\end{eqnarray*}
Note that this is expressed in terms of the free energies of the
subensembles, rather than the energies of the microstates.

We now wish to consider what happens when an density matrix is
composed of the same equilibrium subensembles $\rho _\alpha $ but
for which the mixing probabilities $p_\alpha ^{\prime }$ are not
in thermal equilibrium\footnote{We may imagine that each of the
subspaces corresponds to a separate 'box', between which
transitions are inhibited. We can then easily prepare a system in
which the 'boxes' are each in equilibrium with some heat bath, but
the probabilities of the 'boxes' being occupied are not in an
equilibrium. As long as the thermal relaxation time for
transitions between boxes is very large, this will be stable.}
\[
\rho ^{\prime }=\sum_\alpha p_\alpha ^{\prime }\rho _\alpha
\]

We know the entropy of this matrix from the mixing equation
\[
S^{\prime }=\sum_\alpha p_\alpha ^{\prime }S_\alpha
    -k\sum_\alpha p_\alpha ^{\prime }\ln p_\alpha ^{\prime }
\]

However, it may seem unclear whether the free energy is at all
meaningful in this situation. We cannot simply use $F$ as the
equations would not agree. At the same time, there is a well
defined temperature associated with the system. We need to develop
a well defined generalisation of the equilibrium equations above.

We are going to proceed by proposing a non-equilibrium version of
the partition function
\[
Z^{\prime }=\sum_\alpha D_\alpha e^{-\frac{F_a}{kT}}
\]
where the $D_\alpha $ are a set of factors which determine the
extent to which the system is out of equilibrium. If all $D_\alpha
=1$ then the system is in equilibrium. We define the $D_a$ from
the constraint $\sum_\alpha p_\alpha ^{\prime }\ln D_\alpha =0$ to
give

\[
kT\ln D_a=\left(F_a+kT\ln p_a^{\prime}\right)
    -\sum_\alpha p_\alpha^{\prime}
        \left(F_\alpha +kT\ln p_\alpha^{\prime}\right)
\]
which allows us to write
\begin{eqnarray*}
p_a^{\prime} &=&\frac{1}{Z^{\prime}}D_a e^{-\frac{F_a}{kT}} \\
\rho^{\prime} &=&\frac{1}{Z^{\prime}}
    \sum_\alpha D_\alpha e^{-\frac{F_a}{kT}}\rho_\alpha
\end{eqnarray*}
in analogy to our equilibrium equations.

We would now like to express the non-equilibrium free energy as
just $F^{\prime}=-kT\ln Z^{\prime}$. Our primary justification for
believing this is because the mean energy $E$, the non-equilibrium
entropy $S^{\prime }$ and the subensemble temperature $T$ can be
shown to be related by
\[
E-TS^{\prime}=-kT\ln Z^{\prime}
\]
which is precisely the relationship we would like a free energy to
fulfil. However, the operational definition free energy, that
makes it a useful to use, is that it corresponds to the work
required to put the system into some reference state, by an
isothermal procedure. We must show the work required to change the
state matches the change in $F^{\prime}$. To be sure that this is
valid, the final reference state should be one in which the
subensembles occur with equilibrium probability.

Let us start with a particularly simple example, consisting of
Szilard box and a piston system. It is the piston system that we
are going to focus upon. The piston system is initially in one of
two states, which have the same internal entropies $S_P$, energies
$E_p$ and are in equilibrium temperature $T,$ which for simplicity
will be the same temperature as the Szilard box. The 'internal'
free energy of the piston states are therefore $F_p=E_p-TS_p$. In
a 'thermal equilibrium' each of the piston states would be equally
likely and in an equilibrium mixture of piston states, the free
energy would be $F=F_p-kT\ln 2$.

If we placed two piston states in opposite ends of the Szilard
box, and compress the gas until each piston state was found in the
center of the box, the isothermal work required is just $kT\ln 2$.
The piston is now no longer in the mixture, and has free energy
$F_P.$ When the piston is removed, the gas expands to refill the
entire box. This allows us to isothermally put the equilibrium
state into a reference state, with a work requirement of $kT\ln
2$. We could also reverse the procedure, and allow the piston
reference state to expand into an {\em equilibrium} mixture,
extracting $kT\ln 2$ work.

We now consider what happens if the initial piston states occur
with the more general probabilities of $p$ and $1-p$. We will
again place the two piston states at each end of the box, but now
we compress the two sides by different amounts, so the piston ends
up in some position $Y$, not necessarily the center. If the piston
is on the left, with probability $p$, we allow it to compress the
gas to the right of $Y$. This requires a work of $kT\ln \left(
\frac {2}{1-Y}\right)$. If the piston is on the right, with
probability $(1-p)$, the gas is compressed to the left of $Y,$ and
the work required is $kT\ln \left( \frac 2{1+Y}\right) .$ The mean
work requirement is therefore

\[
\frac{W}{kT}=
    p\ln \left(\frac{2}{1-Y}\right)
        +(1-p)\ln \left(\frac{2}{1+Y}\right)
\]
This has its smallest value when $p=\left(\frac{1-Y}{2}\right)$
and therefore
\[
W=-kT\left(p\ln p+(1-p)\ln (1-p) \right)
\]

This leaves the piston at position $Y=(1-2p)$, with the one atom
gas located to the left of the piston, with probability
$\left(\frac{1+Y}{2}\right)$ and to the right with probability
$\left(\frac{1-Y}{2}\right)$. Had we inserted a partition into the
box at position $Y$, we would have precisely these probabilities
for the location of the one atom gas. The piston can therefore be
reversibly removed from the box. Had the compression of the gas
left the piston at some other value of $Y^{\prime}$, removing and
reinserting the piston at $Y^{\prime}$ would lead to a
rearrangement of the probabilities of the one atom gas. This would
not be a reversible procedure. This demonstrates that the work
requirement to reversibly put the non-equilibrium mixture of
piston states into the reference state is exactly $-T\Delta S$,
where $\Delta S$ is just the mixing entropy of the non-equilibrium
state.

We consider this to be the required generalisation of isothermal
compression. For the change in free energy to be equal to the work
done, the initial free energy must be

\[
F^{\prime}=F_p+kT\left( p\ln p+(1-p)\ln (1-p) \right)
\]

This can be readily generalised to a situation with many different
subensembles and with different free energies in each subensemble,
but with all subensembles at the same temperature\footnote{If the
internal states of the piston are assumed to be thermally isolated
from the Szilard box, then the compression may take place at a
different temperature. While this complicates the process, it will
still be consistent with the free energy defined here, taking into
account the results of Appendix \ref{ap:szfree}, where there is
more than one temperature present.} to yield

\begin{eqnarray*}
F^{\prime}&=&\sum_\alpha p_\alpha^{\prime}
    \left(F_\alpha+kT\ln p_\alpha^{\prime}\right)  \\
    &=&E-TS^{\prime} \\
    &=&-kT\ln Z^{\prime}
\end{eqnarray*}
which is the desired result, and justifies the form of the
non-equilibrium partition function.

With regard to the other relationships involving the free energy,
we find these generalise to
\begin{eqnarray*}
F^{\prime} &=& -kT\ln \left(
    \sum_\alpha D_\alpha e^{-\frac{F_\alpha}{kT}}\right)  \\
p_a^{\prime} &=& \frac{D_a Z_a}{Z^{\prime}} \\
F_a &=& F^{\prime} -kT\ln \left(\frac{p_a^{\prime}}{D_a}\right)
\end{eqnarray*}

These relations are less useful than they might appear. We have
justified the existence of a free energy for situations where a
system is in a stable, non-equilibrium state, but has a well
defined temperature. However the dependance upon the values of
$D_\alpha$ makes the non-equilibrium partition function of limited
value when these are changeable (unless they can be constrained to
be changeable in a well defined way eg. when the system is not
isolated, the $D_\alpha$ will approach $1$, typically with an
exponential decay, and over a time period of the same order as the
thermal relaxation time). It should be noted, however, that the
non-equilibrium state will have a higher free energy than the
equivalent equilibrium state. As the system approaches equilibrium
this extra free energy will be lost in the process of
thermalisation.
}

\bibliographystyle{alpha}
\bibliography{phd}

\end{document}